\documentclass[12pt,oneside]{article}
\usepackage{breqn}
\begin{document}


\begin{center}
{\large\bf TeV-scale gravity in Ho\v{r}ava-Witten theory on a compact complex
hyperbolic threefold\\}
\vspace{0.14cm}
\vspace*{.05in}
{Chris Austin\footnote{Email: chris@chrisaustin.info}\\
\small 33 Collins Terrace, Maryport, Cumbria CA15 8DL, England\\
}
\end{center}
\begin{center}
{\bf Abstract}
\end{center}
\noindent The field equations and boundary conditions of Ho\v{r}ava-Witten
theory, compactified on a smooth compact spin quotient of $ \mathbf{CH}^3 $,
where $ \mathbf{CH}^3 $ denotes the hyperbolic cousin of $ \mathbf{CP}^3 $, are
studied in the presence of Casimir energy density terms.  If the Casimir energy
densities near one boundary result in a certain constant of integration taking
a value greater than around $ 10^5 $ in units of the $ d = 11 $ gravitational
length, a form of thick pipe geometry is found that realizes TeV-scale gravity
by the ADD mechanism, with that boundary becoming the inner surface of the
thick pipe, where we live.  Three alternative ways in which the outer surface
of the thick pipe might be stabilized consistent with the observed value of the
effective $ d = 4 $ cosmological constant are considered.  In the first
alternative, the outer surface is stabilized in the classical region and the
constant of integration is fixed at around $ 10^{13} $ in units of the $ d = 11
$ gravitational length for consistency with the observed cosmological constant.
In the second alternative, the four observed dimensions have reduced in size
down to the $ d = 11 $ gravitational length at the outer surface, and there are
Casimir effects near the outer surface. In the third alternative, the outer
surface is stabilized in the classical region by extra fluxes of the three-form
gauge field, whose four-form field strength wraps three-cycles of the compact
six-manifold times the radial dimension of the thick pipe. Some problems
related to fitting the strong/electroweak Standard Model are considered.

\vspace{3.7cm}

\tableofcontents

\vspace{3.0cm}

\section{Introduction}

The observed physical universe is a very stiff structure,
approximately flat up to distances larger, by a factor of $10^{61}$, than the
radius of curvature that would be expected on the basis of the Standard
Model, plus General Relativity, in $3 + 1$ dimensions.  Large two-dimensional
structures, such as the hull of an oil tanker, are often stiffened by
structures that extend a short distance into the third dimension.  So it is
natural to wonder whether compact additional spatial dimensions, not yet
observed, could play an active role in stiffening the universe.

To study the possibility of such a mechanism, I shall consider, in this paper,
the compactification of Ho\v{r}ava-Witten theory \cite{HW1, HW2} on a smooth
compact spin K\"{a}hler manifold,
that is obtained from $\mathbf{C} \mathbf{H}^3$, the hyperbolic cousin of
$\mathbf{C} \mathbf{P}^3$, by quotienting out the free, holomorphic action of
a cocompact, torsionless, discrete subgroup of the isometry group of
$ \mathbf{CH}^3 $, which is SU(3,1).  I shall look for solutions that
realize TeV-scale gravity by the ADD mechanism \cite{ADD1, AADD, ADD2} in
a form of thick pipe geometry \cite{Benakli, CM, NTSG}, such that the two
boundaries of the Ho\v{r}ava-Witten universe become the inner and outer
surfaces of the thick pipe, the eleventh dimension becomes
the radial direction of the thick pipe, and the diameter of the compact
six-manifold increases with increasing distance from the inner surface of
the thick pipe, where we live.

The choice of a smooth compact spin quotient of $\mathbf{C} \mathbf{H}^3$,
rather than a Calabi-Yau threefold \cite{CHSW}, as the compact six-manifold $
\mathcal{M}^6 $, means that all supersymmetries are broken by the
compactification.  By a fundamental theorem of Mostow, known as Mostow rigidity
\cite{Mostow}, the geometry of $ \mathcal{M}^6 $ is now completely determined
by its fundamental group, up to an overall scale factor, so that $
\mathcal{M}^6 $ has no shape moduli.  There are an infinite number of
topologically distinct smooth compact quotients of $ \mathbf{CH}^3 $, but only
a finite number with $ \left\vert\chi\left(\mathcal{M}^6\right)\right\vert $
up to a given value, where $ \chi\left(\mathcal{M}^6\right) $ denotes the Euler
number of $ \mathcal{M}^6 $, and only a small fraction of these are likely to
be spin manifolds.

The possible values of $ \left\vert\chi\left(\mathcal{M}^6\right)\right\vert $
are constrained by the fact that the squares of the Yang-Mills coupling
constants, at unification, are inversely proportional to $ \left\vert\chi\left(
\mathcal{M}^6\right)\right\vert $, and by combining an estimate of the
Yang-Mills coupling constants, at unification, with an estimate by Giudice,
Rattazzi, and Wells \cite{Giudice Rattazzi Wells} of the effective expansion
parameter for quantum gravity in eleven dimensions, the upper limit on $
\left\vert
\chi\left(\mathcal{M}^6\right)\right\vert $ is provisionally estimated in
subsection \ref{The expansion parameter}, on page \pageref{The expansion
parameter}, to be around $ 7 \times 10^4 $.  This upper limit might possibly
be slightly increased by an effect considered by Robinson and Wilczek
\cite{Robinson Wilczek}.

This limit on the value of $ \left\vert\chi\left(\mathcal{M}^6\right)\right
\vert $ means that TeV-scale gravity cannot be realized simply by choosing an
extremely large value of $ \left\vert\chi\left(\mathcal{M}^6\right)\right\vert
$.  Instead, it is necessary that the boundary conditions at the inner surface
of the thick pipe, and the Casimir energy density corrections to the
energy-momentum tensor on and near the inner surface, result
in a certain constant of integration taking a value greater than around $ 10^5
$ in units of the gravitational length in eleven dimensions.

Specifically, if $ y $ denotes the geodesic distance from the inner surface of
the thick pipe, up to an additive constant, and the $ d = 11 $ metric on $
\mathcal{M}^6 $ is $ b^2 h_{AB} dx^A dx^B $, where $ b $ depends only on $ y $,
and $ h_{AB} $ is the standard metric on $ \mathbf{CH}^3 $ introduced in
subsection~\ref{CH3}, on page \pageref{CH3}, then in the main part of the bulk,
where there are no significant source terms in the Einstein equations, we find
\begin{equation}
\label{d b by d y in the Introduction}
\frac{db}{dy} \simeq \left(\frac{B}{b}\right)^{1.8990},
\end{equation}
where $ B $ is a constant of integration, that for TeV-scale gravity has to
have a value greater than around $ 10^5 \kappa^{2/9} $, where $ \kappa
$ is the gravitational coupling constant in eleven dimensions.

The value of $ B $ is completely determined by the region close to the inner
surface of the thick pipe, because the only other physically significant
constant of integration, which is an overall constant multiplying the warp
factor that multiplies the metric in the four extended dimensions, does not
occur in any significant terms in the field equations or boundary conditions in
this region.  Thus the two boundary conditions at the inner surface fix $ B $
and $ b_1 $, the value of $ b $ at the inner surface.

A perturbative mechanism by which a large value of
$\frac{B}{\kappa^{2/9}}$ could occur is \mbox{identified} in subsection
\ref{The region near the inner surface of the thick pipe}, on page
\pageref{The region near the inner surface of the thick pipe}.  In essence,
the bulk power law (\ref{d b by d y in the Introduction}) holds only for
\mbox{$\frac{b}{\kappa^{2/9}} > \left(
\frac{B}{\kappa^{2/9}}\right)^{0.6551}$,} which is
greater than around $ 10^3 $, while for $1\sim\frac{b_1}{\kappa^{2/9}}
< \frac{b}{\kappa^{2/9}} < \left( \frac{B}{\kappa^{2/9}} \right)^{0.6551}$,
we find self-consistently that
\begin{equation}
  \label{d b by d y in the quantum region in the Introduction} \frac{db}{dy}
  \sim \frac{b}{\kappa^{2/9}},
\end{equation}
when the Casimir energy density corrections are taken into account, as
discussed in subsection \ref{The Casimir energy density corrections}, on page
\pageref{The Casimir energy density corrections}, and subsection \ref{Beyond
the proximity force approximation}, on page \pageref{Beyond the proximity
force approximation}.  Thus there is a quantum region of thickness greater than
around $ 8
\kappa^{2/9} $ adjacent to the inner surface, in which $ b $ increases
exponentially with $ y $.

The linear relation (\ref{d b by d y in the quantum region in the
Introduction}) starts to round off to a broad peak at $b \sim 10^3 \kappa^{
\frac{2}{9}}$, followed smoothly by the classical power law (\ref{d b
by d y in the Introduction}).  The only requirement for obtaining the linear
relation (\ref{d b by d y in the quantum region in the Introduction}) is that
a certain sign is positive rather than negative, so it seems possible that a
value of $ B $ significantly larger than $ \kappa^{2/9}$ could be found
for as many as fifty percent of the smooth compact quotients of $\mathbf{C}
\mathbf{H}^3$ that are spin manifolds.  The actual value of $b$ at which the
quantum relation (\ref{d b by d y in the quantum region in the Introduction})
transforms into the classical relation (\ref{d b by d y in the Introduction}),
and the corresponding value of $B$, will be determined by how close to the
self-consistent linear relation (\ref{d b by d y in the quantum region in the
Introduction}) the system is set by the boundary conditions at $b_1 \sim
\kappa^{2/9}$.

This mechanism is completely perturbative, and could be tested by one-loop
calculations, for smooth compact quotients of $\mathbf{C} \mathbf{H}^3$
that are spin manifolds.  The numerical coefficient in the linear relation
(\ref{d b by d y in the quantum region in the Introduction}) is found to be
$\sim 1$ if $b_1$ is at least a factor of $2$ or so larger than the minimum
value allowed by the Giudice, Rattazzi, and Wells estimate of the expansion
parameter, which is $b_1 \simeq 0.2 \kappa^{2/9}$.  Thus it seems
likely that $b_1$ will be somewhere in the range from $0.4
\kappa^{2/9}$, which corresponds to $\left| \chi \left( \mathcal{M}^6
\right) \right| \simeq 10^3$, to $0.8 \kappa^{2/9}$, which corresponds
to $\left| \chi \left( \mathcal{M}^6 \right) \right| \simeq 20$.

There are inevitably significant Casimir energy density terms in the
energy-momen-tum tensor on and near the inner surface of the thick pipe, due to
the Ho\v{r}ava-Witten relation $ \lambda \simeq 5.8 \kappa^{\frac{2}{3}} $
between the $ d = 10 $ Yang-Mills coupling constant $ \lambda $ and $ \kappa $
\cite{HW2}, and the fact that the $ d = 4 $ Yang-Mills coupling constants at
unification are not much smaller than $ 1 $, which implies that $ b_1 $ is
comparable to $ \kappa^{2/9} $.

Although the mechanism for realizing TeV-scale gravity considered in this
paper is completely perturbative, it would be desirable to be able to
calculate corrections beyond one loop, and the problem of the higher order
corrections to Ho\v{r}ava-Witten theory is considered in subsection \ref{The
higher order corrections to Horava-Witten theory}, on page \pageref{The higher
order corrections to Horava-Witten theory}.  The derivation of type IIA
superstring theory {\cite{type IIA superstring theory}} from the
Cremmer-Julia-Scherk (CJS) theory of $d = 11$ supergravity {\cite{Cremmer
Julia Scherk}} compactified on a small $\mathbf{S}^1$ {\cite{HW1}} is
reviewed, and $M$-theory on a smooth background is observed to be the same as
the CJS theory.

The superspace constructions of higher-derivative counterterms for the CJS
theory {\cite{Duff Toms, Howe Tsimpis}} are considered, and I suggest that
an obstruction might exist that prevents the geometrical transformations in
superspace {\cite{Cremmer Ferrara, Brink Howe}} from matching the CJS
supersymmetry variations for a general solution of the CJS field equations
beyond a certain power of $\theta$.  This would mean that with the exception
of the possibly unique counter-term constructible by the superform or ectoplasm
method {\cite{DAuria Fre Townsend van Nieuwenhuizen, Gates Ectoplasm 1, Gates
Grisaru Knutt Wehlau Siegel, Gates Ectoplasm 2, Howe Tsimpis, Howe R to the
fourth corrections}}, the superspace
counterterms do not result in locally supersymmetric deformations of the CJS
theory, so that since the coefficient of the unique dimension 8 counterterm
{\cite{Hyakutake}} is fixed by cancellation of the tangent bundle anomaly on
five-branes {\cite{Duff Liu Minasian, Witten Five Brane, Freed Harvey Minasian
Moore, Bilal Metzger 2, Harvey}}, it
might be possible to calculate the predictions of the CJS
theory and Ho\v{r}ava-Witten theory in the framework of effective field
theory, without the occurrence of undetermined parameters connected with
the short distance completion of the theory.

If the $d = 11$ metric on the four observed dimensions is $a^2 g_{\mu \nu}
dx^{\mu} dx^{\nu}$, where $a$ depends only on $y$, and $g_{\mu \nu}$ is a
metric on de Sitter space with de Sitter radius equal to $1$, then in the
classical region corresponding to (\ref{d b by d y in the Introduction}), we
find
\begin{equation}
  \label{classical dependence of a on b in the Introduction} a = A \left(
  \frac{\kappa^{2/9}}{b} \right)^{0.7753},
\end{equation}
where $A$ is a constant of integration whose value is determined by the region
close to the outer surface of the thick pipe.  And in the quantum region
corrresponding to (\ref{d b by d y in the quantum region in the
Introduction}), we find
\begin{equation}
  \label{quantum dependence of a on b in the Introduction} a = A_1 \left(
  \frac{b}{\kappa^{2/9}} \right)^{\tau},
\end{equation}
where the constant $A_1$ is determined by continuity with (\ref{classical
dependence of a on b in the Introduction}) at the transition between the
classical and quantum regions, and the exponent $\tau$ is determined by the
Casimir terms in the energy-momentum tensor for the self-consistent linear
relation (\ref{d b by d y in the quantum region in the Introduction}).

For $B \gg \kappa^{2/9}$, the existence of a solution of the boundary
conditions at the outer surface with $b_2 = \sqrt{2} a_2 \gg
\kappa^{2/9}$, where $b_2$ and $a_2$ are the values of $b$ and $a$ at
the outer surface, is demonstrated in subsection \ref{Solutions with both a
and b large at the outer surface}, on page \pageref{Solutions with both a and
b large at the outer surface}, and this type of solution is found in
subsection \ref{G sub N and Lambda for solutions with outer surface in
classical region}, on page \pageref{G sub N and Lambda for solutions with
outer surface in classical region}, to fit the observed values of Newton's
constant and the cosmological constant for TeV-scale gravity if $\tau \simeq -
3$ and $B \sim 10^{13} \kappa^{2/9} \sim 10^{- 5}$~metres.

This type of solution does not fully satisfy the condition for a valid
reduction to a four dimensional effective action, due to the fact that $a
\left( y \right)$ decreases from the observed de Sitter radius of around
$10^{26}$ metres at the inner surface of the thick pipe, to around $10^{- 5}$
metres at the outer surface.  The fact that Newton's law is recovered for the
gravitational force between point particles on the Planck brane {\cite{0412016
Arnowitt Dent}} of the first Randall-Sundrum model {\cite{Randall
Sundrum 1}} suggests there is a possibility that Newton's law might be
obtained between point particles on the inner surface of the thick pipe, but
this question is not resolved in this paper.

Solutions in which $a \left( y \right)$ has decreased to around
$\kappa^{2/9}$ at the outer surface, and there are Casimir effects
near the outer surface, are considered in subsection \ref{Solutions with a as
small as kappa to the two ninths at the outer surface}, on page
\pageref{Solutions with a as small as kappa to the two ninths at the outer
surface}.  The three observed spatial dimensions are in this case assumed to
be compactified to a smooth compact quotient of $\mathbf{H}^3$, whose
topology is significant for the Casimir effects near the outer surface.  There
is now an additional large constant of integration, $\tilde{A}$, which is the
analogue of $B$ for the quantum region near the outer surface, and by
increasing $\tau$ from around $- 3$ towards the exponent $- 0.7753$ in the
classical relation (\ref{classical dependence of a on b in the Introduction}),
the value of $\frac{B}{\kappa^{2/9}}$ can be reduced from around
$10^{13}$ towards a limiting value of around $10^5$, at a cost of rapidly
increasing the value of $\frac{\tilde{A}}{\kappa^{2/9}}$.

For the case when $\tau = - 0.7753$, this type of solution is demonstrated in
subsection \ref{Comparison with sub-millimetre tests of Newtons law}, on page
\pageref{Comparison with sub-millimetre tests of Newtons law}, to be
consistent with the precision sub-millimetre tests of Newton's law {\cite{Hoyle
et al}}, because most of the decrease of $a \left( y \right)$ takes place in a
very narrow region near the outer surface, so that only a fraction $\sim 10^{-
6}$ of the integral that determines Newton's constant comes from values of $y$
for which $a \left( y \right)$ is smaller than around $10^{18}$ metres.

Solutions with extra fluxes of the four-form field strength of the three-form
gauge field of $d = 11$ supergravity {\cite{Cremmer Julia Scherk}} wrapping
three-cycles of the compact six-manifold $\mathcal{M}^6$ times the radial
dimension are considered in subsection \ref{Stiffening by fluxes}, on page
\pageref{Stiffening by fluxes}.  The outer surface is in the classical region
$b_2 \gg \kappa^{2/9}$, $a_2 \gg \kappa^{2/9}$, and there is
an additional large constant of integration, $\tilde{G}$, whose square
corresponds to an average value of the energy-momentum tensor of the extra
fluxes.

The value of $\frac{B}{\kappa^{2/9}}$ can again be reduced from around
$10^{13}$ towards a limiting value of around $10^5$, by increasing $\tau$ from
around $- 3$ towards $- 0.7753$, at a cost now of rapidly increasing the value
of $\frac{\tilde{G}}{\kappa^{\frac{4}{3}}}$.  This results in greatly
increasing the value of $a_2$, so that $a_2$ is around $10^{22}$ metres for
$\tau = - 0.7753$, while $b_2$ remains $\sim B$.  These solutions are
therefore also consistent with the precision sub-millimetre tests of Newton's
law, for $\tau$ in a range including $- 0.7753$.

The value of $\tilde{G}$ in this type of solution does not appear to be
quantized, which suggests that cosmological models involving $\tilde{G}$ might
resemble quintessence models {\cite{Steinhardt quintessence}}.

Most of the results of this paper are also valid, with minor modifications, for
smooth compact spin quotients of $ \mathbf{H}^6 $, and the construction of an
infinite family of smooth compact quotients of $ \mathbf{CH}^3 $ and $
\mathbf{H}^6 $, called arithmetic quotients, which is due to Borel and
Harish-Chandra \cite{Borel Harish-Chandra}, is reviewed in subsection
\ref{Smooth compact arithmetic quotients of CHn and Hn}, on page
\pageref{Smooth compact arithmetic quotients of CHn and Hn}.  Non-arithmetic
smooth compact quotients of $ \mathbf{H}^6 $ have been constructed by Gromov
and Piatetski-Shapiro \cite{Gromov Piatetski-Shapiro}.  Non-arithmetic smooth
compact quotients of $ \mathbf{CH}^2 $ have been constructed by Mostow
\cite{Mostow SU 2 1}, and non-arithmetic smooth finite-volume, but non-compact,
quotients of $ \mathbf{CH}^3 $ have been constructed by Deligne and Mostow
\cite{Deligne Mostow}, but it does not at present seem to be known whether
there exist non-arithmetic smooth compact quotients of $ \mathbf{CH}^3 $.

The compact six-manifold $ \mathcal{M}^6 $ is required to be a spin manifold,
because the three-form gauge field \cite{Nahm} only enters the generalized spin
connection through its four-form field strength, which is well-defined
globally, so there is no possibility of defining an analogue of a $
\mathrm{spin}^c $ structure \cite{Hawking Pope} in the bulk.  I do not know
whether any of the arithmetic smooth compact quotients of $ \mathbf{CH}^3 $ or
$ \mathbf{H}^6 $ are spin manifolds, but the simplest known smooth compact
quotient of $ \mathbf{H}^4 $, which is called the Davis manifold \cite{Davis
manifold}, is both an arithmetic quotient and a spin manifold \cite{Ratcliffe
Tschantz, Everitt Maclachlan}.  A counting argument considered in section
\ref{Smooth compact
quotients of CH3 H6 H3 and S3}, on page \pageref{Smooth compact quotients of
CH3 H6 H3 and S3}, suggests that for sufficiently large $ \left\vert\chi\left(
\mathcal{M}^6\right)\right\vert $, non-arithmetic smooth compact quotients of
$ \mathbf{H}^6 $ will exist that are spin manifolds.  The value of the
integration constant $ B $ is likely to depend on the choice of the spin
structure on $ \mathcal{M}^6 $.

The value of $ B $ is also affected by the presence of topologically stabilized
vacuum Yang-Mills fields tangential to $ \mathcal{M}^6 $ on the inner surface
of the thick pipe, and the further Casimir energy density terms in the
energy-momentum tensor, to which they in turn give rise.  Such vacuum
Yang-Mills fields also affect the four-form field strength of the three-form
gauge field of $ d = 11 $ supergravity \cite{Nahm, Cremmer Julia Scherk}, due
to the boundary condition derived by Ho\v{r}ava and Witten \cite{HW2}, and this
also results in terms in the energy-momentum tensor that are significant near
the inner surface of the thick pipe, and thus affect the value of $ B $.

By considering certain Wilson lines formed from trees of hairpins, I
demonstrate in subsection~\ref{Dirac quantization condition for E8 vacuum gauge
fields}, on page \pageref{Dirac quantization condition for E8 vacuum gauge
fields}, that integrals over closed orientable
two-dimensional surfaces in $ \mathcal{M}^6
$, of the field strengths of Yang-Mills fields in the Cartan subalgebra of $ E8
$, whose field strengths are proportional to Hodge - de Rham harmonic
two-forms, are restricted by a form of Dirac quantization condition to lie on a
certain discrete lattice in the Cartan subalgebra of $ E8 $, and more
generally, that Abelian configurations of the $ E8 $ Yang-Mills fields, with
field strengths proportional to Hodge - de Rham harmonic two-forms, can be
topologically stabilized in magnitude, and partly also in orientation within $
E8 $, by a form of Dirac quantization condition.

Such topologically stabilized Abelian vacuum Yang-Mills fields are restricted
only by the requirements that they break $ E8 $ to the Standard Model
\cite{PDG, Rosner} in the correct way, as studied in subsection \ref{SU 3 cross
SU 2 squared cross U 1 to the fourth}, on page \pageref{SU 3 cross SU 2 squared
cross U 1 to the fourth}, and subsection \ref{SU 3 cross SU 2 cross U 1 to the
fifth}, on page \pageref{SU 3 cross SU 2 cross U 1 to the fifth}, and that a
topological constraint derived by Witten \cite{Witten Constraints on
compactification} is satisfied, and that the correct spectrum of chiral
fermions, namely three Standard Model generations, plus possible singlet
neutrinos, is obtained.  Witten's
topological constraint ensures that the effective field theory, in the four
extended dimensions, is free of chiral anomalies.

The first of these requirements leaves a substantial amount of flexibility in
the choice of the topologically stabilized Abelian vacuum Yang-Mills fields,
and Witten's topological constraint also leaves a substantial amount of
flexibility, unless it should happen that the symmetric trilinear form which
defines the topologically invariant cup product $ H^2 \times H^2 \to H^4 $ of
$ \mathcal{M}^6 $ is either positive definite or negative definite as a
bilinear form when one of its indices takes some fixed values, thus preventing
Witten's topological constraint from being satisfied by cancellations between
contributions from different elements of the Cartan subalgebra when the free
index takes one of those fixed values.  It seems reasonable to expect that this
is increasingly unlikely to occur, the larger the second Betti number of $
\mathcal{M}^6 $ is.

Now Mostow rigidity does not imply that $ \mathcal{M}^6 $ has no K\"{a}hler
shape moduli, so that $h^{ 1,1 }$, the dimension of the Dolbeault cohomology
group $H^{ 1,1 }$, is equal to 1, or that the second Betti number of
$ \mathcal{M}^6 $ is small.  Rather,
just as with any K\"{a}hler-Einstein metric with a nonvanishing Ricci scalar,
each K\"{a}hler modulus is equal to a fixed multiple of the corresponding
element of the first Chern class.  However, by a theorem of Gromov~\cite{Gromov
Bound on Betti numbers}, all the Betti numbers of $ \mathcal{M}^6
$ are bounded by a constant times $ \left\vert\chi\left(\mathcal{M}^6\right)
\right\vert $.  It seems reasonable to expect that the second Betti number of $
\mathcal{M}^6 $ will be comparable to $ \left\vert\chi\left(\mathcal{M}^6
\right)\right\vert $, and thus around $ 10^4 $.

If the embedding of the Standard Model in $ E8 $ is such that only
a small number of types of exotic fermion could occur, then the requirements of
anomaly cancellation, which are automatically satisfied when Witten's
topological constraint is satisfied, may already be sufficient to prevent the
occurrence of exotic chiral fermions.  This happens for the embeddings of the
Standard Model in $ E8 $ studied in subsection \ref{SU 3 cross SU 2 cross U 1
to the fifth}, on page \pageref{SU 3 cross SU 2 cross U 1 to the fifth}, where
there is only one type of exotic fermion, and the only solutions of the anomaly
cancellation constraints are an integer number of Standard Model generations.
In this case there would still be a substantial amount of flexibility in the
choice of the topologically stabilized Abelian vacuum Yang-Mills fields, when
all three requirements are satisfied.

It might also be possible to introduce partially topologically stabilized
Yang-Mills instantons in $ \mathrm{SU}\left(2\right) $ subgroups of $ E8 $,
associated with non-contractible closed four-dimensional surfaces in $
\mathcal{M}^6 $ \cite{Donaldson}, and this might be necessary for the more
complicated types of embedding of the Standard Model in $ E8 $ studied in
subsection \ref{SU 3 cross SU 2 squared cross U 1 to the fourth}, on page
\pageref{SU 3 cross SU 2 squared cross U 1 to the fourth}.  However, it is not
certain that this is possible, because it does not seem likely that
non-contractible closed four-dimensional surfaces in $ \mathcal{M}^6 $ will be
simply connected, and it is also unclear to what extent the orientation of such
$ \mathrm{SU}\left(2\right) $ subgroups in $ E8 $ could be topologically
stabilized \cite{Bernard Christ Guth Weinberg}.

The introduction of topologically stabilized Abelian vacuum Yang-Mills fields
of Hosotani type \cite{Hosotani 1, Hosotani 2, Hosotani 3}, with vanishing
field strength, is usually associated with a torsion element of the fundamental
group of the compact six-manifold, or in other words, a nontrivial element $ a
$ such that $ a^n = 1 $ for some finite integer $ n $ \cite{CHSW}.  A smooth
compact quotient of $ \mathbf{CH}^3 $ necessarily has torsionless fundamental
group, due to the fact that $ \mathbf{CH}^3 $ is the quotient of the isometry
group, $ \mathrm{SU}\left(3,1\right) $, by its maximal compact subgroup, $
\mathrm{SU}\left( 3 \right) \times \mathrm{U} \left( 1 \right) $, but examples
in three dimensions suggest that it might be
possible for $ H_1\left(\mathcal{M}^6,\mathbf{Z}\right) $ to have torsion even
though the fundamental group of $ \mathcal{M}^6 $ has no torsion, and I show in
subsection \ref{SU 3 cross SU 2 cross U 1 to the fifth}, on page \pageref{SU 3
cross SU 2 cross U 1 to the fifth}, that this would be sufficient to enable
Abelian vacuum Yang-Mills fields of Hosotani type to be topologically
stabilized.

The breakings of $ E8 $ to the Standard Model considered in subsections \ref{SU
3 cross SU 2 squared cross U 1 to the fourth} and \ref{SU 3 cross SU 2 cross U
1 to the fifth} partly suppress proton decay by a mechanism related to the
Aranda-Carone mechanism \cite{Aranda Carone}, but I do not know whether the
suppression is sufficient for consistency with current experimental limits
\cite{Limits on proton decay 1, Limits on proton decay 2, Limits on proton
decay 3}.  The breakings also produce natural candidates for light sterile
neutrinos \cite{Babu Seidl, Shaevitz} that might be relevant if the
forthcoming results of the MiniBooNE experiment \cite{MiniBooNE 1, MiniBooNE 2}
confirm the evidence for light sterile neutrinos from the LSND experiment
\cite{LSND}.  The possibility that the existence of multiple oscillation
channels involving light sterile neutrinos could improve the compatibility
between the KARMEN \cite{KARMEN} and LSND experiments was recently demonstrated
in \cite{Goldman Stephenson McKellar}.

$ \mathbf{C}\mathbf{H}^3 $ has previously been considered in the context of $M$
theory by Kehagias and Russo \cite{Kehagias Russo}.  Compact hyperbolic spaces
have been considered in the context of large extra dimensions by Kaloper,
March-Russell, Starkman, and Trodden \cite{Kaloper March-Russell Starkman
Trodden}, and by Tabbash \cite{Tabbash}.

\section{Thick pipe geometries}
\label{Thick pipe geometries}

I shall now briefly review Ho\v{r}ava-Witten theory, in Subsection
\ref{Horava-Witten theory}, on page \pageref{Horava-Witten theory}, then
summarize the relevant facts about
$ \mathbf{CH}^3 $, in Subsection \ref{CH3}, on page \pageref{CH3}.  The metric
ansatz is introduced, and the field equations and boundary conditions
derived, in the presence of assumed Casimir energy densities, in Subsection
\ref{The field equations and boundary conditions}, on page
\pageref{The field equations and boundary conditions}, and the equations
are studied in Subsection
\ref{Analysis of the Einstein equations and the boundary conditions}, on page
\pageref{Analysis of the Einstein equations and the boundary conditions}.

I use units such that $\hbar = c = 1$.  The metric signature is $\left( -, +,
+, \ldots, + \right)$.  The definitions of the Riemann and Ricci tensors are
chosen to agree with the conventions of Weinberg \cite{Weinberg}.  The Riemann
tensor is defined by:
\begin{equation}
\label{Riemann tensor definition}
\left[ D_{\mu}, D_{\nu} \right] V_{\sigma} = - R_{\mu \nu \sigma \tau}
   V^{\tau} = - R^{ \hspace{3.1ex} \tau}_{\mu \nu \sigma} V_{\tau}
\end{equation}
Hence:
\begin{equation}
\label{Riemann tensor}
R^{ \hspace{3.1ex} \tau}_{\mu \nu \sigma} = \partial_{\mu}
\Gamma_{\nu \sigma}^{\tau} -
   \partial_{\nu} \Gamma_{\mu \sigma}^{\tau} + \Gamma_{\mu \rho}^{\tau}
   \Gamma_{\nu \sigma}^{\rho} - \Gamma_{\nu \rho}^{\tau} \Gamma_{\mu
   \sigma}^{\rho}
\end{equation}
where $\Gamma_{\mu \nu}^{\tau}$, the Christoffel symbol of the second kind, is
defined by:
\begin{equation}
\label{Christoffel symbol}
\Gamma_{\mu \nu}^{\tau} = \frac{1}{2} g^{\tau \sigma} \left( \partial_{\mu}
   g_{\nu \sigma} + \partial_{\nu} g_{\mu \sigma} - \partial_{\sigma} g_{\mu
   \nu} \right)
\end{equation}
The Ricci tensor is defined by:
\[ R_{\mu \nu} = R^{ \hspace{3.1ex} \tau}_{\mu \tau \nu} = \partial_{\mu}
\Gamma_{\tau
   \nu}^{\tau} - \partial_{\tau} \Gamma_{\mu \nu}^{\tau} + \Gamma_{\mu
   \rho}^{\tau} \Gamma_{\tau \nu}^{\rho} - \Gamma_{\tau \rho}^{\tau}
   \Gamma_{\mu \nu}^{\rho} = \]
\begin{equation}
\label{Ricci tensor}
= \frac{1}{2} \partial_{\mu} \partial_{\nu} \ln \left| g \right| -
   \partial_{\tau} \Gamma_{\mu \nu}^{\tau} + \Gamma_{\mu \rho}^{\tau}
   \Gamma_{\tau \nu}^{\rho} - \frac{1}{2} \Gamma_{\mu \nu}^{\rho}
   \partial_{\rho} \ln \left| g \right|
\end{equation}
where $g$ is the determinant of the metric, $g_{\mu \nu}$.  These conventions
are consistent with references \cite{HW2, Lukas Ovrut Waldram,
Lukas Ovrut Stelle Waldram, Moss 1, Moss 2, Moss 3} on Ho\v{r}ava-Witten
theory, but the
Riemann and Ricci tensors, as defined here, have the opposite signs to those
used in Chapters 15 and 16 of \cite{Green Schwarz Witten}, and the Ricci tensor
also has the opposite sign, to that defined in Chapter 18 of \cite{PDG}.

Laboratory and astrophysical observations, excluding the hypothesized period
of inflation, in the very early universe, are consistent with an action
\begin{equation}
\label{S tot}
S_{\mathrm{tot}} = S_{\mathrm{Ein}} + S_{\mathrm{vac}} + S_{\mathrm{SM}} +
   S_{\mathrm{DM}}
\end{equation}
where
\begin{equation}
\label{Einstein action}
S_{\mathrm{Ein}} = - \frac{1}{16 \pi G_N} \int d^4 x \sqrt{- g} g^{\mu \nu}
   R_{\mu \nu}
\end{equation}
is the Einstein action,
\begin{equation}
\label{S vac}
S_{\mathrm{vac}} = - \rho_{\mathrm{vac}} \int d^4 x \sqrt{- g} = -
   \frac{\Lambda}{8 \pi G_N} \int d^4 x \sqrt{- g}
\end{equation}
is the vacuum energy, $S_{\mathrm{SM}}$ is the Standard Model matter action, and
$S_{\mathrm{DM}}$ is the action for the unknown dark matter, provided that the
metric, $g_{\mu \nu}$, is treated classically, rather than quantum
mechanically, and all contributions to the vacuum energy, other than
$\rho_{\mathrm{vac}}$, are discarded.  This means, in particular, that the
contributions to the vacuum energy from the VEV of the Standard Model Higgs
field, the chiral symmetry breaking condensate and possible other condensates
of QCD, and vacuum Feynman diagrams of the Standard Model fields and the dark
matter fields, in the metric $g_{\mu \nu}$, are all to be discarded.

$G_N$ is Newton's constant, with the value \cite{PDG}
\begin{equation}
\label{Newtons constant}
G_N = 6.7087 \times 10^{- 39} \hspace{0.8ex} \mathrm{GeV}^{- 2}
\end{equation}
so that $\sqrt{G_N} = 8.1907 \times 10^{- 20} \hspace{0.8ex} \mathrm{GeV}^{- 1}
= 1.6160 \times 10^{- 35} \hspace{0.8ex} \mathrm{metres}$.

Variation of $S_{\mathrm{tot}}$, with respect to the metric, gives Einstein's
field equations:
\begin{equation}
\label{Einsteins field equations}
R_{\mu \nu} - \frac{1}{2} Rg_{\mu \nu} - \Lambda g_{\mu \nu} + 8 \pi G_N
   T_{\mu \nu} = 0
\end{equation}
where the energy-momentum tensor, $T_{\mu \nu}$, is defined by:
\begin{equation}
\label{energy momentum tensor}
T^{\mu \nu} = \frac{2}{\sqrt{- g}} \left( \frac{\delta
   S_{\mathrm{SM}}}{\delta g_{\mu \nu}} + \frac{\delta S_{\mathrm{DM}}}{\delta
   g_{\mu \nu}} \right)
\end{equation}
The observed large-scale structure of the universe is consistent with a
Friedmann-Robertson-Walker metric
\begin{equation}
\label{FRW metric}
ds^2 = - dt^2 + R^2 \left( t \right) \tilde{g}_{ij} \left( x \right) dx^i
   dx^j
\end{equation}
where the spatial metric $\tilde{g}_{ij} \left( x \right)$ is maximally
symmetric, and satisfies $\tilde{R}_{ij} = - 2 k \tilde{g}_{ij}$, where $k = +
1$ for spherical spatial sections, $k = 0$ for flat spatial sections, and $k =
- 1$ for hyperbolic spatial sections.  The large-scale structure of $T_{\mu
\nu}$ is consistent with a perfect fluid form:
\begin{equation}
\label{perfect fluid}
T_{\mu \nu} = pg_{\mu \nu} + \left( p + \rho \right) u_{\mu} u_{\nu}
\end{equation}
with pressure $p$ and energy density $\rho$, where $u^{\mu} = \left( 1, 0, 0,
0 \right)$ is the velocity vector of the fluid in co-moving coordinates.
Einstein's equations then lead to the Friedmann-Lema\^{i}tre equation
\begin{equation}
\label{Friedmann Lemaitre equation}
\frac{k}{R^2} = \frac{8 \pi G_N \: \rho}{3} + \frac{\Lambda}{3} - H^2 = H^2
   \left( \frac{\: \rho}{\rho_c} + \frac{\rho_{\mathrm{vac}}}{\rho_c} - 1
   \right)
\end{equation}
where $H \left( t \right) = \frac{\dot{R}}{R}$ is the Hubble parameter,
$ \dot{R} = \frac{d R}{d t}$,
$\rho_c = \frac{3 H^2}{8 \pi G_N}$ is the critical value of
$\rho_{\mathrm{tot}}
= \rho + \rho_{\mathrm{vac}}$ for which $k$ vanishes, and $\Lambda = 8 \pi G_N
\rho_{\mathrm{vac}}$ is the cosmological constant.  The Hubble Space Telescope
Key Project \cite{HST} has given the value
\begin{equation}
\label{Hubble parameter}
H_0^{- 1} = 13.6 \pm 1.4 \hspace{0.8ex} \mathrm{Gyr} = \left( 1.29 \pm 0.13
   \right) \times 10^{26} \hspace{0.8ex} \mathrm{metres}
\end{equation}
for the present value of the Hubble parameter.  By combining WMAP data with
other astronomical data, Spergel et al \cite{0302209 Spergel et al}
give the value
\begin{equation}
\label{Omega tot}
\frac{\: \rho}{\rho_c} + \frac{\rho_{\mathrm{vac}}}{\rho_c} = 1.02 \pm 0.02
\end{equation}
However, there is no theoretical restriction on the magnitude of $HR =
\dot{R}$, so this value is consistent with any of the three possibilities $k =
+ 1$, $0$, or $- 1$, although $k = - 1$ is disfavoured by a standard
deviation.  In fact, visual inspection of the lower two panels, of Fig.~13 of
\cite{0302209 Spergel et al}, does not suggest any strong preference
for $k = + 1$, as opposed
to $k = - 1$.  It seems likely that the class of models considered in the
present paper will prefer $k = - 1$ to $k = + 1$, due to the infinitely
greater variety of the smooth compact quotients of $\mathbf{H}^3$, in
comparison to the smooth compact quotients of $\mathbf{S}^3$, and the
correspondingly improved chances of finding a quotient whose Casimir energy
densities are such that, in combination with a suitable quotient of
$\mathbf{C} \mathbf{H}^3$, the observed values of $G_N$ and $\Lambda$ can
be fitted.  There are, however, only 18 distinct topologies with $k = 0$, of
which ten are compact, and the remaining eight have one or more uncompactified
dimensions, \cite{flat topologies}.  This is far too small a number of distinct
topologies, for there to be any likelihood of any of them satisfying the
requirements on the Casimir energy densities, that will make it possible to
fit the observed values of $G_N$ and $\Lambda$, so I do not expect any model,
of the type studied in this paper, to have $k = 0$.  Furthermore, most of the
flat topologies have one or more shape moduli, unlike the hyperbolic
topologies, and possibly also unlike the spherical topologies.  I shall
therefore, for simplicity, assume $ k \neq 0 $.

The individual values of $\frac{\rho}{\rho_c}$, and
$\frac{\rho_{\mathrm{vac}}}{\rho_c}$, are not so precisely measured as their
sum.  Chapter 2 of \cite{PDG} quotes the values of
\cite{0302209 Spergel et al}: $\frac{
\rho}{\rho_c} = 0.27 \pm 0.04$, and $\frac{\rho_{\mathrm{vac}}}{\rho_c} = 0.73
\pm 0.04$.  The baryonic and dark matter contributions to $\frac{
\rho}{\rho_c}$ are quoted as $\frac{\rho_b}{\rho_c} = 0.044 \pm 0.004$, and
$\frac{\rho_{\mathrm{dm}}}{\rho_c} = 0.22 \pm 0.04$.  Chapter 19 of \cite{PDG}
quotes the values of Tonry et al \cite{Tonry et al}:
$\frac{\rho_{\mathrm{vac}}}{\rho_c} = 0.72 \pm
0.05$, and $\frac{\rho}{\rho_c} = 0.28 \pm 0.05$, if $k = 0$ is assumed.  And
Chapter 21 of \cite{PDG} quotes best-fit values from SNe Ia and CMB data of
$\frac{\rho}{\rho_c} \approx 0.3$ and $\frac{\rho_{\mathrm{vac}}}{\rho_c}
\approx 0.7$.  Using the middle value
$\frac{\rho_{\mathrm{vac}}}{\rho_c} = 0.72$,
and the above value of $H^{- 1}_0$, we have:
\[ \Lambda = 3 H_0^2 \frac{\rho_{\mathrm{vac}}}{\rho_c} = 0.012 \hspace{0.8ex}
   \mathrm{Gyr}^{- 2} = 1.3 \times 10^{- 52} \hspace{0.8ex}
\mathrm{metres}^{- 2} = \]
\begin{equation}
\label{Lambda}
= 3.4 \times 10^{- 122} G^{- 1}_N = 5.1 \times 10^{- 84} \hspace{0.8ex}
   \mathrm{GeV}^2
\end{equation}
Hence:
\begin{equation}
\label{rho vac}
\rho_{\mathrm{vac}} = \Lambda / \left( 8 \pi G_N \right) = 3.0 \times 10^{-
 47} \hspace{0.8ex} \mathrm{GeV}^4 = \left( 2.3 \times 10^{- 3} \hspace{0.8ex}
   \mathrm{eV} \right)^4
\end{equation}
If $\rho$ is set to zero, in the Friedmann-Lema\^{i}tre equation
(\ref{Friedmann Lemaitre equation}), then for
$\Lambda > 0$, the equation has the solutions $R = \sqrt{\frac{3}{\Lambda}}
\cosh \sqrt{\frac{\Lambda}{3}} \left( t - t_0 \right)$, for $k = + 1$, $R =
R_0 e^{\sqrt{\frac{\Lambda}{3}} t}$, for $k = 0$, and $R =
\sqrt{\frac{3}{\Lambda}} \sinh \sqrt{\frac{\Lambda}{3}} \left( t - t_0
\right)$, for $k = - 1$, $t > t_0$.  All three of these solutions satisfy
$R_{\mu \nu} = - \Lambda g_{\mu \nu}$, and all three are in fact pieces of the
maximally symmetric de Sitter space dS$_4$ \cite{de Sitter Les Houches}.  The
$k = + 1$ solution covers the full de Sitter hyperboloid, in the global
coordinates of \cite{de Sitter Les Houches}, the $k = 0$ solution covers the
future of a single point in the $t \rightarrow - \infty$ ``boundary'' of the
hyperboloid, in the planar coordinates of \cite{de Sitter Les Houches}, which
cover precisely half the hyperboloid, and the $k = - 1$ solution, for $t >
t_0$, covers the future of an ordinary point of the hyperboloid, in the
hyperbolic coordinates of \cite{de Sitter Les Houches}.
$\sqrt{\frac{3}{\Lambda}}$ is known as the de Sitter radius.  For the measured
value of $\Lambda$, the de Sitter radius is:
\begin{equation}
\label{de Sitter radius}
\sqrt{\frac{3}{\Lambda}} = 16.0 \hspace{0.8ex} \mathrm{Gyr} = 1.51 \times
   10^{26} \hspace{0.8ex} \mathrm{metres} = 0.94 \times 10^{61} \sqrt{G_N}
\end{equation}
For each of the cases $k = + 1$, $0$, and $- 1$, we can quotient the spatial
sections of the solutions by discrete subgroups of the isometry groups of the
spatial sections, that act freely, or in other words, without fixed points, on
the spatial sections, in order to obtain locally de Sitter solutions, with
$R_{\mu \nu} = - \Lambda g_{\mu \nu}$, and non-trivial spatial topology.  For
each of the cases $k = + 1$ and $k = - 1$, there are an infinite number of
distinct such topologies, so it seems plausible, especially for $k = - 1$,
that there will exist topologies for which Bose - Fermi cancellations occur in
the Casimir energy densities, for the compactifications of supergravity in
eleven dimensions, and supersymmetric Yang-Mills theory in ten dimensions, on
quotients with those topologies, with just the relative precisions I will show
are needed, in order for solutions involving those quotients, together with a
suitable quotient of $\mathbf{C} \mathbf{H}^3$, to fit the observed values
of $G_N$ and $\Lambda$.

The action (\ref{S tot}) is not applicable to the hypothesized period of
inflation, since according to section 19.3.5 of \cite{PDG},
most current models
of inflation are based on an unknown symmetry breaking involving a new scalar
field, the ``inflaton''.  Models of the type considered in the present paper
are expected to give very different behaviour from the standard hot big bang
model, at times earlier than the time, $t$, at which the hot big bang model
predicts the temperature, $T$, in units in which Boltzmann's constant is equal
to 1, to be comparable to the Planck mass in eleven dimensions, in the models
of the present paper.  According to \cite{Guth}, for temperatures higher than
all particle masses, the standard hot big bang model gives
\begin{equation}
\label{temperature time relation}
T^2 = \frac{1}{4 t} \sqrt{\frac{45}{\pi^3 \left( N_b + \frac{7}{8} N_f
   \right) G_N}}
\end{equation}
where $N_b$ denotes the number of bosonic degrees of freedom that are
effectively massless at temperature $T$, for example the photon contributes
two units to $N_b$, and $N_f$ is the corresponding number for fermions, for
example electrons and positrons together contribute four units to $N_f$.  If
we set $T = 1$ TeV, and count only the observed Standard Model particles plus
the graviton, so that $N_b = 26$, and $N_f = 90$, (if the neutrinos are
assumed left-handed, so that their masses are Majorana), then
\begin{equation}
\label{time when temperature is a TeV}
t = \left( 2.78 \times 10^{- 3} \hspace{0.8ex} \mathrm{eV} \right)^{- 1} =
   0.709 \times 10^{- 4} \hspace{0.8ex} \mathrm{metres} = 2.36 \times 10^{- 13}
   \hspace{0.8ex} \mathrm{seconds}
\end{equation}
which is comparable to the inverse one quarter power of the observed vacuum
energy density $\rho_{\mathrm{vac}}$, equation (\ref{rho vac}).  I do not
yet know whether models of the
type studied here have problems with initial conditions, analogous to the
horizon and flatness problems, that led to the hypothesis of inflation
\cite{Guth, Linde, Albrecht Steinhardt, Linde Linde, Olive, Lyth Riotto}.
To answer this question it will be necessary to study cosmological
versions of these models, which will involve partial differential equations,
with the time, and the radial coordinate of the thick pipe, as independent
variables.  In the present paper I shall only seek solutions such that the
metric in the four observed dimensions is locally maximally symmetric, with
the correct values of Newton's constant and the cosmological constant.  Thus
the metric in the four extended dimensions will be locally de Sitter, although
I will also consider whether or not flat and AdS solutions are possible.

The aim of this section is to determine the circumstances under which the
observed values of $G_N$ and $\Lambda$ can be fitted, in a certain class of
compactifications of Ho\v{r}ava-Witten theory.  Rather than seeking
supersymmetric
solutions, I shall seek solutions in which the universe is stiff and strong,
in the sense that the forces, that make it big and flat, are much stronger,
than the forces that occur in any other physical process.  In addition to
fitting $G_N$ and $\Lambda$, I shall also require that the gauge coupling
constants have approximately the correct values at unification, which
typically means that the $E_8$ fine structure constant, resulting from the
compactification to $3 + 1$ dimensions, is about $\frac{1}{10}$.

\subsection{Ho\v{r}ava-Witten theory}
\label{Horava-Witten theory}

Ho\v{r}ava-Witten theory \cite{HW1, HW2} is supergravity in eleven dimensions,
on a manifold with two boundaries, or, more precisely, on the orbifold $
\mathcal{M}^{10} \times \mathbf{S}^1 / \mathbf{Z}_2$, where $ \mathcal{M}^{10}
$ is a ten-dimensional manifold.  At one-loop
order in the Feynman diagram expansion, it is necessary to introduce a
supersymmetric Yang-Mills theory, with gauge group $E_8$, on each of the
ten-dimensional boundaries, in order to cancel anomalies.

The Ho\v{r}ava-Witten action in the bulk is the standard
Cremmer-Julia-Scherk (CJS) action \cite{Cremmer Julia Scherk}.
In the ``upstairs'' picture, working on the orbifold $ \mathcal{M}^{10} \times
\mathbf{S}^1 / \mathbf{Z}_2$, and omitting
terms quartic in the gravitino, this is:
\begin{eqnarray}
\label{upstairs bulk action}
  S_{ \mathrm{CJS} } & = & \frac{1}{\kappa^2} \int_{\mathcal{M}^{11}}
  d^{11} x \sqrt{-g}
  \left( - \frac{1}{2} R - \frac{1}{2} \bar{\psi}_I \Gamma^{IJK} D_J \psi_K -
  \frac{1}{48} G_{IJKL} G^{IJKL} \right. \nonumber\\
  &  & \hspace{2em} \hspace{2em} \hspace{2em} \hspace{2em} -
  \frac{\sqrt{2}}{192} \left( \bar{\psi}_I \Gamma^{IJKLMN} \psi_N + 12
  \bar{\psi}^J \Gamma^{KL} \psi^M \right) G_{JKLM} \nonumber\\
  &  & \hspace{2em} \hspace{2em} \hspace{2em} \hspace{2em} \hspace{2em}
  \hspace{2em} \left. - \frac{\sqrt{2}}{3456} g^{I_1 I_2 \ldots
  I_{11}} C_{I_1 I_2 I_3} G_{I_4 \ldots I_7} G_{I_8 \ldots I_{11}} \right)
\end{eqnarray}
where $ g^{I_1 I_2 \ldots I_{11}} $ is the tensor
$ g^{I_1 I_2 \ldots I_{11}} = \frac{ 1 }{ \sqrt{ -g } }
\epsilon^{I_1 I_2 \ldots I_{11}} $, $ \epsilon^{0 \, 1 \, 2 \, \ldots \, 9 \,
10} = 1 $, and $ G_{ I J K L } = 24 \partial_{ \left[ I \right. }
C_{ \left. J K L \right] } $.  Coordinate indices $ I, J, K, \ldots $ run
over all directions on $ \mathcal{M}^{ 11 } $.

The Dirac matrices $\Gamma_I$ are $32 \times 32$ real matrices satisfying
$\left\{ \Gamma_I, \Gamma_J \right\} = 2 g_{IJ}$.  A suitable representation
of the $\Gamma_a$, where $a, b, c, \ldots$ are local Lorentz indices, is
given, for example, in section 2.5 of {\cite{Miemiec Schnakenburg}}.  The
matrices $\Gamma^{I_1 I_2 \ldots I_n}$ are defined by $\Gamma^{I_1 I_2 \ldots
I_n} = \Gamma^{\left[ I_1 \right.} \Gamma^{I_2} \ldots \Gamma^{\left. I_n
\right]}$, so that when the indices are all different, $\Gamma^{I_1 I_2 \ldots
I_n} = \Gamma^{I_1} \Gamma^{I_2} \ldots \Gamma^{I_n}$.  Spinor indices are
written $\alpha, \beta, \gamma, \ldots$.  The matrices $\Gamma^0 \Gamma^{I_1
I_2 \ldots I_n}$, where the index of $\Gamma^0$ is a local Lorentz index, are
symmetric for $n = 1, 2, 5, 6, 9, 10$, and antisymmetric for $n = 0, 3, 4, 7,
8, 11$.  The charge conjugation matrix is the antisymmetric matrix
$\mathcal{C}= - \Gamma^0$, where the index of $\Gamma^0$ is a local Lorentz
index.

We note that the right-hand spinor index of a Dirac matrix transforms under
local Lorentz transformations by matrix multiplication on the right by $\left(
1 - \frac{1}{4} \lambda_{ab} \Gamma^{ab} \right)$, where $\lambda_{ab}$ are
the local Lorentz transformation parameters, and the left-hand spinor index of
$\mathcal{C} \Gamma^I$ transforms by matrix multiplication on the left by
$-\mathcal{C} \left( 1 + \frac{1}{4} \lambda_{ab} \Gamma^{ab} \right)
\mathcal{C}= \left( 1 - \frac{1}{4} \lambda_{ab} \Gamma^{ab} \right)^T$, which
is equivalent to acting on the left-hand spinor index of $\mathcal{C}
\Gamma^I$ by matrix multiplication on the right by $\left( 1 - \frac{1}{4}
\lambda_{ab} \Gamma^{ab} \right)$.  Thus the left-hand spinor index of
$\mathcal{C} \Gamma^I$ is an index with the Lorentz transformation properties
of the right-hand index of a Dirac matrix, so if spinor indices with the
Lorentz transformation properties of the left-hand and right-hand spinor
indices of a Dirac matrix are distinguished by writing them as upper and lower
spinor indices respectively, then $\mathcal{C}$ acts as a ``metric'', that
lowers a spinor index.  This is consistent with $\mathcal{C} \Gamma^a$ being
an invariant tensor under $\mathrm{SO} \left( 10, 1 \right)$ local Lorentz
transformations, since
\begin{equation}
  \label{Lorentz invariance of C Gamma up a} \left( \delta^a \, \!_b +
  \lambda^a \, \!_b \right) \left( \mathcal{C} \Gamma^b \right)_{\gamma
  \delta} \left( \delta^{\gamma} \, \!_{\alpha} - \frac{1}{4} \lambda_{cd}
  \left( \Gamma^{cd} \right)^{\gamma} \, \!_{\alpha} \right) \left(
  \delta^{\delta} \, \!_{\beta} - \frac{1}{4} \lambda_{ef} \left( \Gamma^{ef}
  \right)^{\delta} \, \!_{\beta} \right) = \left( \mathcal{C} \Gamma^a
  \right)_{\alpha \beta},
\end{equation}
up to terms quadratic in $ \lambda_{ab} $,
where the identity $\Gamma^a \Gamma^{cd} = \Gamma^{acd} + \eta^{ac} \Gamma^d -
\eta^{ad} \Gamma^c$ was used.  The inverse ``metric'' $\mathcal{C}^{\alpha
\beta}$ is defined in terms of $\mathcal{C}_{\alpha \beta}$ by
$\mathcal{C}^{\alpha \beta} \mathcal{C}_{\beta \gamma} = \delta^{\alpha} \,
\!_{\gamma}$.  The invariant tensor $\left( \mathcal{C} \Gamma^a
\right)_{\alpha \beta}$ can be written as $\Gamma_{\alpha \beta}^a$, since the
position of the first spinor index distinguishes it from $\left( \Gamma^a
\right)^{\alpha} \, \!_{\beta} = \Gamma^{a \alpha} \, \!_{\beta}$.  All
spinors in ten or eleven dimensions will be Majorana, which for a real
representation of the Dirac matrices, means real {\cite{Gliozzi Scherk
Olive}}.  The conjugate Majorana spinor is $\bar{\psi}_{\alpha} = -
\psi^{\beta} \mathcal{C}_{\beta \alpha} =\mathcal{C}_{\alpha \beta}
\psi^{\beta}$.

The manifold $ \mathcal{M}^{ 11 } $ is assumed to have the topology $
\mathcal{M}^{ 10 } \times \mathbf{S}^1 $.   Coordinate indices
$ U, V, W, \ldots $ will run over all directions on $ \mathcal{M}^{ 10 } $, and
I will use the lower-case letter
$ y $ for the coordinate in the $ \mathbf{S}^1 $ direction, and also
for the coordinate index in the $ \mathbf{S}^1 $
direction.  There is assumed to be an orbifold fixed point at $ y = y_1 $,
and another one at $ y = y_2 > y_1 $.  All fields are periodic in the $ y $
direction, with period $ 2 \left( y_2 - y_1 \right) $.  The bosonic fields
$ g_{ U V } $, $ g_{ y y } $, and $ C_{ U V y } $ are even under the
reflections $ y \to \left( 2 y_1 - y \right) $ and  $ y \to \left( 2 y_2 - y
\right) $, and $ g_{ U y } $ and $ C_{ U V W } $ are odd.  The gravitino
satisfies
\begin{equation}
\label{gravitino Z2 conditions}
\Psi_U \left( y \right) = \Gamma_y \Psi_U \left( 2 y_1 - y \right)
\hspace{1.5cm}
\Psi_y \left( y \right) = - \Gamma_y \Psi_y \left( 2 y_1 - y \right)
\end{equation}
together with the corresponding conditions, with $ y_1 $ replaced by
$ y_2 $.

The integral over $ \mathcal{M}^{ 11 } $, in (\ref{upstairs bulk action}),
includes two copies of the physical region $ y_1 \leq y \leq y_2 $, namely the
original region, and its reflection in one of the two fixed point sets.
I shall adopt the viewpoint of Ho\v{r}ava and Witten, that it should be
possible to switch, as convenient, between the ``upstairs'' viewpoint, of
working on the full $ \mathcal{M}^{ 11 } $, with these reflection symmetries
imposed on the fields, and the ``downstairs'' viewpoint, of working on a
manifold with boundary, with the topology $ \mathcal{M}^{ 10 } \times
\mathbf{I}^1 $, where
$ \mathbf{I}^1 $ denotes the interval $ y_1 \leq y \leq y_2 $.  For this to
work, it is essential, as noted in footnote 3 of \cite{HW2}, that
when working on the manifold with boundary, the factor $ \frac{1}{\kappa^2} $,
in (\ref{upstairs bulk action}), should be replaced by $ \frac{2}{\kappa^2} $.

The conditions (\ref{gravitino Z2 conditions}) imply that the gravitino is
chiral on the ten-dimensional orbifold fixed point sets, which results in a
gravitational anomaly, localized on the ten-dimensional fixed point sets.
Ho\v{r}ava and Witten argued, in \cite{HW1}, that this gravitational anomaly
could be cancelled by introducing an $ E8 $ supersymmetric Yang-Mills
multiplet, on each of the ten-dimensional fixed point sets, and they studied
the required couplings in \cite{HW2}.
The supersymmetric Yang-Mills action, on the orbifold fixed point set at
$ y_1 $, is
\begin{equation}
\label{Yang Mills action}
S_{\mathrm{YM}} = - \frac{ 1 }{ \lambda^2 } \int_{ \mathcal{M}^{ 10 }_{ 1 } }
d^{ 10 }
x \sqrt{ - g } \,\, \mathrm{tr} \left( \frac{ 1 }{ 4 } F_{ UV } F^{ UV } +
\frac{ 1 }{ 2 } \bar{\chi} \Gamma^{ U } D_{ U } \chi \right).
\end{equation}
and the action at $ y_2 $ is obtained from this by the substitution
$ \mathcal{M}^{ 10 }_{ 1 } \to \mathcal{M}^{ 10 }_{ 2 } $.  The action
(\ref{Yang Mills action})
is written in Ho\v{r}ava and Witten's notation, in which ``tr'', for
$ E8 $, denotes $ \frac{ 1 }{ 30 } $ of the trace in the adjoint
representation, which they denote by ``Tr''.  I will also use this notation.

Ho\v{r}ava and Witten do not explicitly specify the normalization of the $E_8$
generators they use, or, equivalently, their choice of normalization of the $E
8$ structure constants.  This needs to be determined for the present study,
because in Section \ref{E8 vacuum gauge fields and the Standard Model} I shall
use an SU(9) basis for $E_8$, rather
than an SO(16) basis, and the correct normalization of the generators, in the
SU(9) basis, has to be determined.  It is clear from (\ref{Yang Mills action})
that Ho\v{r}ava and Witten use hermitian $ E8 $ generators, and
I shall assume that their hermitian $E_8$ generators are given by $i$, or
alternatively $- i$, times
antihermitian generators, normalized so that, in the SO(16) basis, the $E_8$
Lie algebra is as given in Appendix 6.A of \cite{Green Schwarz Witten}.
Specifically, let $\gamma_i$, $1 \leq i \leq 16$, be a Majorana-Weyl
representation of the SO(16) gamma matrices, so the $\gamma_i$ are real and
off block diagonal, and let $\sigma_{ij} = \frac{1}{4} \left[ \gamma_i,
\gamma_j \right]$.  Then the $\sigma_{ij}$ are real, antisymmetric, and block
diagonal, with two $128 \times 128$ blocks, which are the two irreducible
spinor representations of SO(16).  Choose one of the two spinor
representations, say the first, and let $\bar{\sigma}_{ij}$ denote the
restriction of $\sigma_{ij}$ to the corresponding block.  Then the generators
of $E_8$ are the 120 generators $J_{ij}$ of SO(16), where $J_{ji} = - J_{ij}$,
together with 128 generators $Q_{\alpha}$, whose label, $\alpha$, runs over
the chosen spin representation of SO(16).  The commutation relations are:
\begin{equation}
  \label{JJ commutation relation} \left[ J_{ij}, J_{kl} \right] = J_{il}
  \delta_{jk} - J_{jl} \delta_{ik} - J_{ik} \delta_{jl} + J_{jk} \delta_{il}
\end{equation}
\begin{equation}
  \label{JQ commutation relation} \left[ J_{ij}, Q_{\alpha} \right] = \left(
  \bar{\sigma}_{ij} \right)_{\alpha \beta} Q_{\beta}
\end{equation}
\begin{equation}
  \label{QQ commutation relation} \left[ Q_{\alpha}, Q_{\beta} \right] =
  \left( \bar{\sigma}_{ij} \right)_{\alpha \beta} J_{ij}
\end{equation}
We therefore find that the matrix elements of the generators, in the adjoint
representation of $E_8$, which is also the fundamental, are given by:
\begin{equation}
  \label{matrix elements of J} \begin{array}{c}
    \\
    J_{ij} \quad =
  \end{array} \begin{array}{cc}
    & \begin{array}{cc}
      {\scriptstyle rs} \hspace{2.5ex} \quad & \quad
      \hspace{-0.2ex} {\scriptstyle \gamma}
    \end{array}\\
    \begin{array}{c}
      {\scriptstyle pq}\\
      {\scriptstyle \beta}
    \end{array} \hspace{-2.5ex} & \left(\begin{array}{cc}
      - f_{ij, pq, rs} & 0\\
      0 & - \left( \bar{\sigma}_{ij} \right)_{\beta \gamma}
    \end{array}\right)
  \end{array}
\end{equation}
\begin{equation}
  \label{matrix elements of Q} \begin{array}{c}
    \\
    Q_{\alpha} \quad =
  \end{array} \begin{array}{cc}
    & \begin{array}{cc}
      {\scriptstyle rs} \hspace{3.0ex} \quad & \hspace{0.2ex}
      {\scriptstyle \gamma}
    \end{array}\\
    \begin{array}{c}
      {\scriptstyle pq}\\
      {\scriptstyle \beta}
    \end{array} \hspace{-2.5ex} & \left(\begin{array}{cc}
      0 & \left( \bar{\sigma}_{pq} \right)_{\alpha \gamma}\\
      \left( \bar{\sigma}_{rs} \right)_{\beta \alpha} & 0
    \end{array}\right)
  \end{array}
\end{equation}
where
\[ f_{ij, pq, rs} = \frac{1}{2} \left( \delta_{iq} \delta_{ps} \delta_{rj} -
   \delta_{jq} \delta_{ps} \delta_{ri} - \delta_{ip} \delta_{qs} \delta_{rj} +
   \delta_{jp} \delta_{qs} \delta_{ri} - \delta_{iq} \delta_{pr} \delta_{sj} +
   \delta_{jq} \delta_{pr} \delta_{si} \right. \]
\begin{equation}
  \label{orthogonal group structure constants} \left. + \delta_{ip}
  \delta_{qr} \delta_{sj} - \delta_{jp} \delta_{qr} \delta_{si} \right)
\end{equation}
are the SO(16) structure constants, from (\ref{JJ commutation relation}).
These generators are correctly normalized so that, in doing matrix
multiplications with the generators (\ref{matrix elements of J}) and
(\ref{matrix elements of Q}), the vector index pairs $\left( p, q \right)$,
and $\left( r, s \right)$, are to be summed over the full ranges of all the
vector indices, without restrictions, so there is no restriction, for
example, to $p < q$.  In fact, $\left( J_{ij} \right)_{pq, rs} = - f_{ij, pq,
rs}$ are the correctly normalized generators of SO(16), in the adjoint
representation, with Young tableau shape $\left( 1, 1 \right)$, and can be
obtained, alternatively, by Young tableaux methods, starting from the
generators for the vector representation of SO(16), which are:
\begin{equation}
  \label{vector representation} \left( J_{ij} \right)_{ef} = \delta_{ie}
  \delta_{jf} - \delta_{je} \delta_{if}
\end{equation}
Using (\ref{matrix elements of J}) and (\ref{matrix elements of Q}), we find
that:
\begin{equation}
  \label{Tr JJ} \mathrm{Tr} \left( J_{ij} J_{kl} \right) = \left( 28 +
  \frac{128}{4} \right) \left( \delta_{il} \delta_{jk} - \delta_{ik}
  \delta_{jl} \right) = - 60 \left( \delta_{ik} \delta_{jl} - \delta_{il}
  \delta_{jk} \right) = - 120 \delta_{ij, kl}
\end{equation}
\begin{equation}
  \label{Tr QJ} \mathrm{Tr} \left( Q_{\alpha} J_{ij} \right) = 0
\end{equation}
\begin{equation}
  \label{Tr QQ} \mathrm{Tr} \left( Q_{\alpha} Q_{\beta} \right) = \left(
  \bar{\sigma}_{pq} \right)_{\alpha \gamma} \left( \bar{\sigma}_{pq}
  \right)_{\gamma \beta} + \left( \bar{\sigma}_{rs} \right)_{\delta \alpha}
  \left( \bar{\sigma}_{rs} \right)_{\beta \delta} = - 60 \delta_{\alpha \beta}
  - 60 \delta_{\beta \alpha} = - 120 \delta_{\alpha \beta}
\end{equation}
where in obtaining (\ref{Tr JJ}) I used that, for SO($d$), we have:
\begin{equation}
  \label{SO d adjoint trace} f_{ij, pq, rs} f_{kl, rs, pq} = \left( 2 d - 4
  \right) \left( \delta_{il} \delta_{jk} - \delta_{ik} \delta_{jl} \right)
\end{equation}
and $\delta_{ij, kl} = \frac{1}{2} \left( \delta_{ik} \delta_{jl} -
\delta_{il} \delta_{jk} \right)$ is the unit matrix, in the space of matrices
whose rows and columns are labelled by antisymmetrized pairs of vector
indices.  Hence, denoting the 248 generators $\left( J_{ij}, Q_{\alpha}
\right)$ collectively by $\Lambda_{\mathcal{A}}$, we have:
\begin{equation}
  \label{E8 trace} \mathrm{Tr} \left( \Lambda_{\mathcal{A}}
  \Lambda_{\mathcal{B}} \right) = - 120 \delta_{\mathcal{A} \mathcal{B}}
\end{equation}
On the other hand, for the vector representation (\ref{vector representation})
of SO(16), we have:
\begin{equation}
  \label{vector trace} \left( J_{ij} \right)_{ef} \left( J_{kl} \right)_{fe} =
  - 4 \delta_{ij, kl}
\end{equation}
Thus the trace of the square of a generator of SO(16), in the adjoint of $E
8$, is 30 times the trace of the square of the corresponding generator, in the
vector representation of SO(16).

Seeking to extend (\ref{Yang Mills action}) to a locally supersymmetric
action, coupled in a locally supersymmetric manner to the bulk supergravity
multiplet, Ho\v{r}ava and Witten found it necessary to modify the Bianchi identity
of the four-form gauge field, so that it reads:
\begin{equation}
  \label{Bianchi identity with FF only} dG_{yUVWX} = - 3 \sqrt{2}
  \frac{\kappa^2}{\lambda^2} \left( \delta \left( y - y_1 \right) \mathrm{tr}
  F^{\left( 1 \right)}_{\left[ UV \right.} F^{\left( 1 \right)}_{\left. WX
  \right]} + \delta \left( y - y_2 \right) \mathrm{tr} F^{\left( 2
  \right)}_{\left[ UV \right.} F^{\left( 2 \right)}_{\left. WX \right]}
  \right)
\end{equation}
where $dG_{IJKLM} = 5 \partial_{\left[ I \right.} G_{\left. JKLM \right]}$,
and $F^{\left( i \right)}_{UV}$ denotes the $E_8$ gauge fields at $y = y_i$.
This, in turn, implies that the three-form, $C_{IJK}$, is not invariant under
Yang-Mills gauge transformations.  It also implies, in the ``upstairs''
picture, that $G_{UVWX}$ has a discontinuity, at $y = y_1$, given by
\begin{equation}
  \label{G discontinuity} G_{UVWX} = - \frac{3}{\sqrt{2}}
  \frac{\kappa^2}{\lambda^2} \epsilon \left( y - y_1 \right) \mathrm{tr}
  F^{\left( 1 \right)}_{\left[ UV \right.} F^{\left( 1 \right)}_{\left. WX
  \right]} + \ldots
\end{equation}
where $\epsilon \left( x \right)$ is $1$ for $x > 0$, and $- 1$ for $x < 0$,
and $\ldots$ denotes terms that are regular near $y = y_1$, and thus vanish at
$y = y_1$.  While in the ``downstairs'' picture, on the interval $y_1 \leq y
\leq y_2$, (\ref{G discontinuity}) becomes a boundary condition:
\begin{equation}
  \label{G boundary condition} \left. G_{UVWX} \right|_{y = y_{1 +}} = -
  \frac{3}{\sqrt{2}} \frac{\kappa^2}{\lambda^2} \mathrm{tr} F^{\left( 1
  \right)}_{\left[ UV \right.} F^{\left( 1 \right)}_{\left. WX \right]}
\end{equation}
Corresponding results also hold in the region of $y = y_2$.

The non-vanishing variation of the three-form, $C_{IJK}$, under Yang-Mills
gauge transformations, now implies that the Chern-Simons term, $CGG$, in the
Cremmer-Julia-Scherk action (\ref{upstairs bulk action}), has a non-vanishing
variation, under Yang-Mills gauge transformations.  Ho\v{r}ava and Witten found
that this non-vanishing variation, under Yang-Mills gauge transformations, of
the Cremmer-Julia-Scherk Chern-Simons term, precisely cancels the one-loop
quantum gauge anomaly, of the Majorana-Weyl fermions in the supersymmetric
Yang-Mills multiplets on the orbifold fixed points, provided that
\begin{equation}
  \label{lambda kappa relation} \lambda^2 = 2 \pi \left( 4 \pi \kappa^2
  \right)^{\frac{2}{3}}
\end{equation}
A slightly different result was found by Conrad {\cite{Conrad}}, who found
$\lambda^2 = 2^{\frac{1}{3}} 2 \pi \left( 4 \pi \kappa^2 \right)^{\frac{2}{3}}
= 4 \pi \left( 2 \pi \kappa^2 \right)^{\frac{2}{3}}$.  This difference will
not have a major impact on the results of the present paper, so I shall use
the Ho\v{r}ava-Witten result (\ref{lambda kappa relation}), and not attempt to
resolve the issue here.  The relation (\ref{lambda kappa relation}) implies
that
\begin{equation}
  \label{one over lambda squared} \frac{1}{\lambda^2} = \frac{1}{2 \pi
  \kappa^2} \left( \frac{\kappa}{4 \pi} \right)^{\frac{2}{3}}
\end{equation}
so the Yang-Mills action is of relative order $\kappa^{\frac{2}{3}}$.

Having cancelled the Yang-Mills gauge anomalies, by relating the Yang-Mills
coupling constant to the gravitational coupling constant as just discussed,
Ho\v{r}ava and Witten returned to the original purpose of introducing the
Yang-Mills multiplets on the orbifold fixed points, which was to cancel the
gravitational anomalies of the gravitinos, on the orbifold fixed points.  As
explained in Section 2 {\emph{(i)}} of {\cite{HW1}}, the ``irreducible'' part
of the formal twelve-form, from which the gravitino anomaly in ten dimensions
is constructed, can only be cancelled by the introduction of 248 vector
multiplets on each of the orbifold fixed point hyperplanes.  This requirement
is fulfilled by the $E_8$ supersymmetric Yang-Mills multiplets.  Ho\v{r}ava and
Witten then argued that, in consequence of factorization properties of the
remaining terms in the full gravitational and mixed gravitational - gauge
anomalies, in ten dimensions, the remaining terms in the gravitational and
mixed anomalies can all be cancelled, provided that, in the equations
(\ref{Bianchi identity with FF only}), (\ref{G discontinuity}), and (\ref{G
boundary condition}), above, the substitutions
\begin{equation}
  \label{FF to FF RR substitutions} \mathrm{tr} F^{\left( i \right)}_{\left[ UV
  \right.} F^{\left( i \right)}_{\left. WX \right]} \rightarrow \mathrm{tr}
  F^{\left( i \right)}_{\left[ UV \right.} F^{\left( i \right)}_{\left. WX
  \right]} - \frac{1}{2} \mathrm{tr} R_{\left[ UV \right.} R_{\left. WX \right]}
\end{equation}
are made uniformly, where $R_{UV}$ is the curvature two-form, and $\mathrm{tr}
R_{\left[ UV \right.} R_{\left. WX \right]}$ must be defined, by analogy with
Section 16.1 of {\cite{Green Schwarz Witten}}, as $R^{\hspace{1.2em}
YZ}_{\left[ UV \right.}
R_{\left. WX \right] YZ}$, and provided that, in the quantum effective action,
or in other words, the generating functional of the proper vertices
{\cite{Nambu Jona Lasinio 1, Nambu Jona Lasinio 2}}, which is, in general, a
non-local functional of the fields, a certain local term, called the bulk
Green-Schwarz term, appears in the bulk, with an appropriate finite
coefficient.  The required form of the bulk Green-Schwarz term was in
agreement with the form already found from a one-loop calculation for Type IIA
superstrings {\cite{Vafa Witten}}, and from anomaly cancellation for
five-branes in eleven dimensions {\cite{Duff Liu Minasian, Witten 1}}, and its
coefficient was studied by de Alwis {\cite{de Alwis 1, de Alwis 2}} and Conrad
{\cite{Conrad}}.

Ho\v{r}ava and Witten then completed the calculation of the action at relative
order $\kappa^{\frac{2}{3}}$, and found a problem with a term in a
supersymmetry variation, proportional to $\delta \left( 0 \right)$.  This led
to a further problem, with a term in the action at relative order
$\kappa^{\frac{4}{3}}$, with a coefficient proportional to $\delta \left( 0
\right)$.  They suggested this implies that the full theory must have a
built-in cutoff, that would replace $\delta \left( 0 \right)$ by a finite
constant times $\kappa^{- \frac{2}{9}}$, for example, by having the gauge
fields propagate in a boundary layer, of thickness about
$\kappa^{2/9}$, rather than precisely on the orbifold fixed point
hyperplanes.  However Moss has presented an improved form of Ho\v{r}ava-Witten
theory {\cite{Moss 1, Moss 2, Moss 3}}, in which the $\delta \left( 0 \right)$
terms are absent.  The modifications introduced by Moss include the
introduction of a supersymmetrized Gibbons-Hawking boundary term {\cite{York,
Gibbons Hawking, Luckock Moss}}, additional terms bilinear in the gauginos in
(\ref{Bianchi identity with FF only}), (\ref{G discontinuity}), and (\ref{G
boundary condition}) above, and a modification to the chirality conditions
(\ref{gravitino Z2 conditions}) on the gravitino, in the neighbourhood
of an orbifold fixed point, which for the components $\psi_U$, in the upstairs
picture, amounts to introducing a step function term in the behaviour of
$\left( 1 - \Gamma_y \right) \psi_U$, near the fixed point, analogous to
(\ref{G discontinuity}) above.

The existence of Moss's improved form of the theory suggests it is reasonable
to assume that the Yang-Mills multiplets do not, after all, spread into a
boundary layer of nonzero thickness in the bulk, and do, indeed, stay in the
orbifold fixed point hyperplanes, of zero thickness.  The study of the
boundary conditions, and of the field equations in the bulk, near the
boundaries, in the present paper, depend on this assumption for their
validity, so the conclusions about the existence of thick pipe geometries, and
the possibility of fitting both Newton's constant and the cosmological
constant, for topologies such that the Casimir energy densities cancel to the
required relative precisions, depend on the existence of Moss's improved form
of the theory.  However, these studies do not involve the fermi fields, so I
will not need to use the explicit form of the modifications introduced by
Moss.

The assumption that the Yang-Mills multiplets do, indeed, stay in the orbifold
fixed-point hyperplanes, of zero thickness, means that, for the further
development of Ho\v{r}ava-Witten theory, it is essential to treat the step
functions, such as in (\ref{G discontinuity}) above, and their derivatives, by
a consistent limiting procedure, from properly regularized versions, as
discussed by Bilal and Metzger {\cite{Bilal Metzger 1, Bilal Metzger 2}}.
However this is not necessary in the present paper.

Lukas, Ovrut, and Waldram, {\cite{Lukas Ovrut Waldram}}, have pointed out
that, corresponding to the replacement (\ref{FF to FF RR substitutions}),
supersymmetry is likely to require that, in the Yang-Mills action (\ref{Yang
Mills action}), the corresponding replacement
\begin{eqnarray}
  \mathrm{tr} F^{\left( i \right)}_{UV} F^{\left( i \right) UV} & \rightarrow &
  \mathrm{tr} F^{\left( i \right)}_{UV} F^{\left( i \right) UV} - \frac{1}{2}
  \left( R_{UVWX} R^{UVWX} - 4 R_{UV} R^{UV} + R^2 \right) \nonumber\\
  \label{FF RR action substitutions} & = & \mathrm{tr} F^{\left( i \right)}_{UV}
  F^{\left( i \right) UV} - 3 R^{\hspace{1.2em} \left[ UV \right.}_{\left[ UV
  \right.} R^{\hspace{1.4em}  \left. WX \right]}_{\left. WX \right]}
\end{eqnarray}
is made, where $R_{UV}$ now denotes the Ricci tensor.  This would be analogous
to the situation for the $E 8 \times E 8$ heterotic superstring {\cite{Gross
Harvey Rohm Martinec 1, Gross Harvey Rohm Martinec 2}}, whose effective
low-energy field theory action contains the expression to the right of the
arrow in (\ref{FF RR action substitutions}), summed over both the $E_8$
groups.  In this case, the Lovelock-Gauss-Bonnet term {\cite{Lovelock 1,
Lovelock 2, Henneaux}} $R^{\hspace{1.2em} \left[ UV \right.}_{\left[ UV
\right.} R^{\hspace{1.4em} \left. WX \right]}_{\left. WX \right]}$ is stated,
in Section 16.1 of {\cite{Green Schwarz Witten}}, to be related, by
supersymmetry, to the Lorentz
Chern-Simons term that is included in the field strength of the $d = 10$, $N =
1$ supergravity two-form, by the original Green-Schwarz anomaly cancellation
mechanism {\cite{Green Schwarz}}.  The Lovelock-Gauss-Bonnet term, for the $E
8 \times E 8$ heterotic superstring, was found by Gross and Witten
{\cite{Gross Witten}}, by means of a low energy expansion of tree-level
superstring scattering amplitudes.  The relative coefficients of $R_{UVWX}
R^{UVWX}$, $R_{UV} R^{UV}$, and $R^2$ were fixed to the Lovelock-Gauss-Bonnet
form by Zwiebach \cite{Zwiebach}, who pointed out this linear combination
contains no terms
quadratic in the graviton, and thus does not lead to the occurrence of ghosts,
in the free graviton propagator.  An analogue of the positive energy theorem
{\cite{Schoen Yau, Witten positive energy theorem}} for the Einstein action,
together with the
Lovelock-Gauss-Bonnet term, as it occurs in the effective low-energy field
theory action for the $E 8 \times E 8$ heterotic superstring, was proved by
Kowalski-Glikman {\cite{Kowalski Glikman}}, and the Lovelock-Gauss-Bonnet term
was found by Candelas, Horowitz, Strominger, and Witten {\cite{CHSW}}, to make
it possible to circumvent the no-go theorem {\cite{no go theorem}}, for
compactifications of supersymmetric Yang-Mills theory coupled to $N = 1$
supergravity in ten dimensions {\cite{Bergshoeff de Roo de Wit van
Nieuwenhuizen, Chapline Manton}}.

To the best of my knowledge, the corresponding Lovelock-Gauss-Bonnet terms for
Ho\v{r}ava-Witten theory, given by making the substitutions
(\ref{FF RR action substitutions}) in the
Yang-Mills action (\ref{Yang Mills action}), have not yet been directly
derived, nor explicitly related by supersymmetry to the modified Bianchi
identity (\ref{Bianchi identity with FF only}), with the substitutions
(\ref{FF to FF RR substitutions}).  This would presumably require the
systematic study of Slavnov-Taylor identities {\cite{Slavnov, Taylor, Zinn
Justin}} for BRST quantized {\cite{Becchi Rouet Stora, Tyutin}} Ho\v{r}ava-Witten
theory, perhaps in the Batalin-Vilkovisky framework {\cite{Batalin
Vilkovisky 1, Batalin Vilkovisky 2, Batalin Vilkovisky 3, Batalin Vilkovisky 4,
Batalin Vilkovisky 5}}.  The Lovelock-Gauss-Bonnet terms would, then,
presumably be
found as local terms, in the generating functional of proper vertices, on the
orbifold fixed-point hyperplanes, with the expected finite coefficients, in a
similar manner to the bulk Green-Schwarz term.  I shall simply follow Lukas,
Ovrut, and Waldram {\cite{Lukas Ovrut Waldram}}, and assume these terms to be
present, with the coefficients implied by the substitutions (\ref{FF RR action
substitutions}).

\subsection{The complex hyperbolic space $ \mathbf{CH}^3 $}
\label{CH3}

I shall assume that six of the nine spatial dimensions, of $M^{10}$, are
compactified on a smooth compact spin quotient of $\mathbf{C} \mathbf{H}^3$,
the complex hyperbolic space with three complex dimensions.  A detailed
account of the geometry of complex hyperbolic space has been given by Goldman
{\cite{Goldman}}, but for the present study of the field equations and
boundary conditions, I shall only need the very simplest properties of
$\mathbf{C} \mathbf{H}^3$, which I shall now summarize.

The study of $\mathbf{C} \mathbf{H}^3$ is facilitated by the use of
complex coordinates.  I shall consider the transformation from Cartesian
coordinates to complex coordinates to be a special case of a general
coordinate transformation, and use the corresponding notation.  For $2 n$ real
dimensions, we define a $2 n$ by $2 n$ complex matrix $U_{\hspace{0.8ex}
\hspace{0.8ex} \nu}^{\mu}$ by:
\begin{equation}
\label{definition of U} 
  \begin{array}{cccc}
    U_{\hspace{1.2ex}  2 s - 1}^r = \frac{1}{\sqrt{2}}
    \delta_{\hspace{0.8ex}  s}^r, \:  \: & U_{\hspace{1.2ex}  2 s -
    1}^{\bar{r}} = \frac{1}{\sqrt{2}} \delta_{\hspace{1ex}  s}^r,
    \hspace{0.4ex}  & U_{\hspace{1.2ex}  2 s}^r = \frac{i}{\sqrt{2}}
    \delta_{\hspace{1ex}  s}^r, \hspace{0.4ex}  & U_{\hspace{1.2ex}  2
    s}^{\bar{r}} = - \frac{i}{\sqrt{2}}
  \end{array} \delta_{\hspace{1ex}  s}^r
\end{equation}
for $1 \leq r \leq n$ and $1 \leq s \leq n$.  Then we define complex
coordinates $z^{\mu}$ by a complex general linear transformation:
\begin{equation}
  \label{definition of z} z^{\mu} = U_{\hspace{0.4ex} \hspace{0.4ex}
  \hspace{0.4ex} \hspace{0.4ex} \nu}^{\mu} x^{\nu}
\end{equation}
Thus
\begin{equation}
  \label{z r} z^r = U_{\hspace{0.4ex} \hspace{0.4ex} \hspace{0.4ex} \nu}^r
  x^{\nu} = U_{\hspace{0.4ex} \hspace{0.4ex} \hspace{0.4ex} 2 s - 1}^r x^{2 s
  - 1} + U_{\hspace{0.4ex} \hspace{0.4ex} \hspace{0.4ex} 2 s}^r x^{2 s} =
  \frac{1}{\sqrt{2}} \left( x^{2 r - 1} + ix^{2 r} \right)
\end{equation}
and
\begin{equation}
  \label{z r bar} z^{\bar{r}} = U_{\hspace{0.4ex} \hspace{0.4ex}
  \hspace{0.4ex} \nu}^{\bar{r}} x^{\nu} = U_{\hspace{0.4ex} \hspace{0.4ex}
  \hspace{0.4ex} 2 s - 1}^{\bar{r}} x^{2 s - 1} + U_{\hspace{0.4ex}
  \hspace{0.4ex} \hspace{0.4ex} 2 s}^{\bar{r}} x^{2 s} = \frac{1}{\sqrt{2}}
  \left( x^{2 r - 1} - ix^{2 r} \right)
\end{equation}
for $1 \leq r \leq n$.  Thus $z^{\bar{r}} = \left( z^r \right)^{\ast}$, where
$^{\ast}$ denotes complex conjugation.  We define the inverse, $V$, of $U$,
by:
\begin{equation}
\label{definition of V} 
  \begin{array}{cccc}
    V_{\hspace{0.4ex} \hspace{0.4ex} \hspace{0.4ex}
    \hspace{0.4ex} \hspace{0.4ex} \hspace{0.4ex} \hspace{0.4ex} \hspace{0.4ex}
    \hspace{1ex} \, s}^{2 r - 1} = \frac{1}{\sqrt{2}} \delta_{\hspace{0.4ex}
    \hspace{0.4ex} \hspace{0.4ex} s}^r, & V_{\hspace{0.4ex} \hspace{0.4ex}
    \hspace{0.4ex} \hspace{0.4ex} \hspace{0.4ex} \hspace{0.4ex} \hspace{0.4ex}
  \hspace{0.4ex} \hspace{0.4ex} \hspace{0.4ex} \hspace{0.4ex} \, \bar{s}}^{2 r
    - 1} = \frac{1}{\sqrt{2}} \delta_{\hspace{0.4ex} \hspace{0.4ex}
    \hspace{0.4ex} s}^r, & V_{\hspace{0.4ex} \hspace{0.4ex} \hspace{0.4ex}
    \hspace{0.4ex} \hspace{0.4ex} \hspace{0.4ex} s}^{2 r} = -
    \frac{i}{\sqrt{2}} \delta_{\hspace{0.4ex} \hspace{0.4ex} \hspace{0.4ex}
    s}^r, & V_{\hspace{0.4ex} \hspace{0.4ex} \hspace{0.4ex} \hspace{0.4ex}
    \hspace{0.4ex} \hspace{0.4ex} \bar{s}}^{2 r} = \frac{i}{\sqrt{2}}
    \delta_{\hspace{0.4ex} \hspace{0.4ex} \hspace{0.4ex} s}^r
  \end{array}
\end{equation}
Thus
\begin{equation}
  \label{x in terms of z} x^{\mu} = V_{\hspace{0.4ex} \hspace{0.4ex}
  \hspace{0.4ex} \hspace{0.4ex} \hspace{0.4ex} \nu}^{\mu} z^{\nu}
\end{equation}
In general, on the change to complex coordinates, a contravariant index is
transformed by $U$, thus $x^{\mu} \to U_{\hspace{0.4ex} \hspace{0.4ex}
\hspace{0.4ex} \hspace{0.4ex} \nu}^{\mu} x^{\nu}$, and a covariant index is
transformed by $V$, thus $\partial_{\mu} \to V_{\hspace{0.4ex} \hspace{0.4ex}
\hspace{0.4ex} \hspace{0.4ex} \mu}^{\nu} \partial_{\nu}$.  The metric in flat
Cartesian coordinates, namely the Kronecker delta, $\delta_{\mu \nu}$, is not
preserved by the transformation to complex coordinates.  Its components in the
complex coordinate basis, which I will denote by $\tilde{\delta}_{\mu \nu}$,
are given by:
\begin{equation}
  \label{covariant delta tilde} \tilde{\delta}_{\mu \nu} = V_{\hspace{0.4ex}
  \hspace{0.4ex} \hspace{0.4ex} \hspace{0.4ex} \mu}^{\sigma} V_{\hspace{0.4ex}
  \hspace{0.4ex} \hspace{0.4ex} \hspace{0.4ex} \nu}^{\tau} \delta_{\sigma
  \tau}
\end{equation}
Explicitly:
\begin{equation}
  \label{components of covariant delta tilde} \begin{array}{cc}
    \tilde{\delta}_{rs} = \tilde{\delta}_{\bar{r} \bar{s}} = 0, \hspace{1ex} &
    \hspace{1ex} \tilde{\delta}_{r \bar{s}} =
  \end{array} \tilde{\delta}_{\bar{r} s} =  \delta_{rs}
\end{equation}
Thus $\tilde{\delta}_{\mu \nu} z^{\mu} z^{\nu} = 2 z^r z^{\bar{r}} =
\delta_{\sigma \tau} x^{\sigma} x^{\tau}$, where I have introduced a summation
convention specific to complex coordinates, namely that if a holomorpic
contravariant index, i.e. an unbarred contravariant index that runs from $1$
to $n$, has the same letter as an antiholomorphic contravariant index, i.e. a
barred contravariant index that runs from $1$ to $n$, then the formula is to
be summed over all values of that letter, from $1$ to $n$.  I shall also use
the corresponding convention when a holomorphic covariant index, and an
antiholomorphic covariant index, have the same letter, and summation from $1$
to $n$ also applies, when a holomorphic contravariant index, and a holomorphic
covariant index, have the same letter, and it also applies, when an
antiholomorphic contravariant index, and an antiholomorphic covariant index,
have the same letter.

Similarly:
\begin{equation}
  \label{contravariant delta tilde} \tilde{\delta}^{\mu \nu} =
  U_{\hspace{0.4ex} \hspace{0.4ex} \hspace{0.4ex} \hspace{0.4ex} \sigma}^{\mu}
  U_{\hspace{0.4ex} \hspace{0.4ex} \hspace{0.4ex} \hspace{0.4ex} \tau}^{\nu}
  \delta^{\sigma \tau}
\end{equation}
The components are:
\begin{equation}
  \label{components of contravariant delta tilde} \begin{array}{cc}
    \tilde{\delta}^{rs} = \tilde{\delta}^{\bar{r} \bar{s}} = 0, \hspace{1ex} &
    \hspace{1ex} \tilde{\delta}^{r \bar{s}} = \tilde{\delta}^{\bar{r} s} =
  \end{array} \delta^{rs}
\end{equation}
We then find, for example, that $\tilde{\delta}^{r \mu} \tilde{\delta}_{\mu s}
= \tilde{\delta}^{\bar{r} \mu} \tilde{\delta}_{\mu \bar{s}} =
\delta_{\hspace{0.4ex} \hspace{0.4ex} \hspace{0.4ex} s}^r$, $\tilde{\delta}^{r
\mu} \tilde{\delta}_{\mu \bar{s}} = \tilde{\delta}^{\bar{r} \mu}
\tilde{\delta}_{\mu s} = 0$.

It is convenient also to define $\delta_{r \bar{s}}$, $\delta_{\bar{r} s}$,
$\delta^{r \bar{s}}$, and $\delta^{\bar{r} s}$, (without tildes), by:
\begin{equation}
  \label{deltas without tildes} \delta_{r \bar{s}} = \delta_{\bar{r} s} =
  \delta^{r \bar{s}} = \delta^{\bar{r} s} = \delta_{rs} = \delta^{rs}
\end{equation}
I shall adopt the convention that, when using complex coordinates, the indices
of the {\emph{coordinates}} are lowered and raised by the {\emph{flat}} space
complex metric, $\tilde{\delta}_{\mu \nu}$ and $\tilde{\delta}^{\mu \nu}$, not
by whatever curved metric is under consideration.  We thus have:
\begin{equation}
  \label{covariant contravariant relation} z_r = z^{\bar{r}} = \left( z^r
  \right)^{\ast}, \hspace{4ex} z_{\bar{r}} = z^r = \left( z^{\bar{r}}
  \right)^{\ast}
\end{equation}
and
\begin{equation}
  \label{squared coordinates} z_r z_{\bar{r}} = z_r z^r = z^{\bar{r}}
  z_{\bar{r}} = z^{\bar{r}} z^r
\end{equation}
regardless of the curved metric under consideration.  This is convenient for
working with $\mathbf{C} \mathbf{H}^n$ and $\mathbf{C} \mathbf{P}^n$,
because it makes the SU($n$) properties of formulae manifest, and facilitates
the study of the transformation properties under SU($n, 1$) and SU($n + 1$),
respectively.

The metric on $\mathbf{C} \mathbf{H}^n$ is now defined by:
\begin{equation}
  \label{hermitian metric} g_{\mu \nu} = \left(\begin{array}{cc}
    0 & g_{r \bar{s}}\\
    g_{\bar{r} s} & 0
  \end{array}\right)
\end{equation}
where:
\begin{equation}
  \label{CHn metric} g_{r \bar{s}} = g_{\bar{s} r} = \frac{\delta_{r
  \bar{s}}}{ \left( 1 - z^{\bar{t}} z^t \right)} + \frac{z_r z_{\bar{s}}}{
  \left( 1 - z^{\bar{t}} z^t \right)^2} = \frac{1}{ \left( 1 - z^{\bar{t}} z^t
  \right)} \left( \delta_{r \bar{s}} + \frac{z_r z_{\bar{s}}}{\left( 1 -
  z^{\bar{u}} z^u \right)} \right)
\end{equation}
so that:
\begin{equation}
  \label{CHn squared line element} g_{\mu \nu} dz^{\mu} dz^{\nu} = g_{r
  \bar{s}} dz^r dz^{\bar{s}} + g_{\bar{r} s} dz^{\bar{r}} dz^s =
  \frac{2}{\left( 1 - z^{\bar{t}} z^t \right)} \left( \delta_{r \bar{s}} +
  \frac{z_r z_{\bar{s}}}{\left( 1 - z^{\bar{u}} z^u \right)} \right) dz^r
  dz^{\bar{s}}
\end{equation}
The complex hyperbolic space $\mathbf{C} \mathbf{H}^n$ corresponds to the
region $z^r z^{\bar{r}} < 1$.

We note that $g_{\bar{r} s}$ is the complex conjugate of $g_{r \bar{s}}$, or
in other words, $g_{\bar{r} s} = \left( g_{r \bar{s}} \right)^{\ast}$.  In
general, when working with complex coordinates, I shall choose definitions in
accordance with a convention such that if every index of a vector, tensor, or
matrix is of definite holonomic type, i.e. either holonomic or antiholonomic,
but not an index, such as $\mu$ in this section, which can be either, then
replacing every unbarred index by the corresponding barred index, and every
barred index by the corresponding unbarred index, is equivalent to complex
conjugation.

From (\ref{hermitian metric}) and (\ref{CHn metric}) we find:
\begin{equation}
  \label{determinant of CHn metric} g = \frac{1}{\left( 1 - z^{\bar{t}} z^t
  \right)^{2 n + 2}} = \frac{1}{\left( 1 - z^{\bar{t}} z^t \right)^{d + 2}}
\end{equation}
where $g$ denotes, as usual, the determinant of $g_{\mu \nu}$, and $d = 2 n$.
Also:
\begin{equation}
  \label{inverse hermitian metric} g^{\mu \nu} = \left(\begin{array}{cc}
    0 & g^{r \bar{s}}\\
    g^{\bar{r} s} & 0
  \end{array}\right)
\end{equation}
where
\begin{equation}
  \label{inverse CHn metric} g^{r \bar{s}} = g^{\bar{s} r} = \left( 1 -
  z^{\bar{v}} z^v \right) \left( \delta^{r \bar{s}} - z^r z^{\bar{s}} \right)
\end{equation}
We observe that
\begin{equation}
  \label{CHn metric is Kahler} g_{r \bar{s}} = - \partial_r \partial_{\bar{s}}
  \ln \left( 1 - z^{\bar{t}} z^t \right)
\end{equation}
so the metric is K\"ahler, with K\"ahler potential $- \ln \left( 1 - z^{\bar{t}}
z^t \right)$.  The K\"ahler form is
\begin{equation}
\label{Kahler form}
\omega_{r\bar{s}} = - i g_{r\bar{s}}, \hspace{2.0cm}
\omega_{\bar{s}r} = i g_{r\bar{s}}
\end{equation}
and has real components in a real coordinate system.
The nonvanishing Christoffel symbols of the second kind are:
\begin{equation}
  \label{CHn Christoffel symbols} \Gamma_{st}^r = \frac{\delta_{\hspace{0.8ex}
  \, t}^r z_s + \delta_{\hspace{0.8ex} \, s}^r z_t}{\left( 1 - z^{\bar{v}} z^v
  \right)}, \hspace{4ex} \Gamma_{\bar{s}  \bar{t}}^{\bar{r}} =
  \frac{\delta_{\hspace{0.8ex} \, \bar{t}}^{\bar{r}} z_{\bar{s}} +
  \delta_{\hspace{0.8ex} \, \bar{s}}^{\bar{r}} z_{\bar{t}}}{\left( 1 -
  z^{\bar{v}} z^v \right)}
\end{equation}
Hence, recalling the sign convention (\ref{Riemann tensor}) for the Riemann
tensor, we have:
\begin{equation}
  \label{CHn Riemann tensor} R^{\hspace{0.4ex} \hspace{0.4ex} \hspace{0.4ex}
  \hspace{1ex} u}_{r \bar{s} t} = - \partial_{\bar{s}} \Gamma_{rt}^u = -
  \delta_{\hspace{0.8ex} \, t}^u g_{r \bar{s}}
  - \delta_{\hspace{0.8ex} \, r}^u g_{t
  \bar{s}}, \hspace{1.5ex} R^{\hspace{0.4ex} \hspace{0.4ex} \hspace{0.4ex}
  \hspace{1ex} \bar{u}}_{r \bar{s} \bar{t}} = \partial_r \Gamma_{\bar{s}
  \bar{t}}^{\bar{u}} = \delta_{\hspace{0.8ex} \, \bar{t}}^{\bar{u}} g_{r \bar{s}
  } + \delta_{\hspace{0.8ex} \, \bar{s}}^{\bar{u}} g_{r \bar{t}}
\end{equation}
\begin{equation}
  \label{CHn covariant Riemann tensor} R_{r \bar{s} t \bar{u}} = - g_{r
  \bar{s}} g_{t \bar{u}} - g_{r \bar{u}} g_{t \bar{s}}
\end{equation}
and the Ricci tensor:
\begin{equation}
  \label{CHn Ricci tensor} R_{r \bar{s}} = \left( n + 1 \right) g_{r \bar{s}}
\end{equation}
so the metric is K\"ahler-Einstein.

To calculate the quantity $\epsilon_{\nu_1 \ldots \nu_{2 n}} \epsilon^{\mu_1
\ldots \mu_{2 n}} R^{\hspace{0.4ex} \hspace{0.4ex} \hspace{0.4ex}
\hspace{0.4ex} \hspace{0.4ex} \hspace{0.4ex} \hspace{0.4ex} \hspace{1.0ex}
\nu_1 \nu_2}_{\mu_1 \mu_2} \ldots R^{\hspace{0.4ex} \hspace{0.4ex}
\hspace{0.4ex} \hspace{0.4ex} \hspace{0.4ex} \hspace{0.4ex} \hspace{0.4ex}
\hspace{0.4ex} \hspace{0.4ex} \hspace{0.4ex} \hspace{0.4ex} \hspace{0.4ex}
\hspace{0.4ex} \hspace{0.4ex} \hspace{0.4ex} \hspace{0.4ex} \hspace{1.0ex}
\nu_{2 n - 1} \nu_{2 n}}_{\mu_{2 n - 1} \mu_{2 n}}$, which occurs in the
generalized Gauss-Bonnet formula {\cite{Allendoerfer Weil}}, we note that, if
we define the tensor, $g_{\nu_1 \ldots \nu_{2 n}}$, by $g_{\nu_1 \ldots \nu_{2
n}} \equiv \sqrt{g} \epsilon_{\nu_1 \ldots \nu_{2 n}}$, then on transforming
to complex coordinates, as described after (\ref{x in terms of z}), $g_{\nu_1
\ldots \nu_{2 n}}$ becomes $\det Vg_{\nu_1 \ldots \nu_{2 n}}$.  I shall assume
that the complex coordinates are taken in the order $1, \bar{1}, 2, \bar{2},
3, \bar{3}, \ldots$, so that $V$ is block diagonal.  Then $\det V = i^n$.
Thus $g_{1 \bar{1} 2 \bar{2} 3 \bar{3} \ldots n \bar{n}} = i^n \sqrt{g}
\epsilon_{123 \ldots n} \epsilon_{\bar{1} \bar{2} \bar{3} \ldots \bar{n}}$, so
$\epsilon_{1 \bar{1} 2 \bar{2} 3 \bar{3} \ldots n \bar{n}} = i^n \epsilon_{123
\ldots n} \epsilon_{\bar{1} \bar{2} \bar{3} \ldots \bar{n}}$.  We then find,
from (\ref{CHn covariant Riemann tensor}), that:
\begin{equation}
  \label{Euler number density for CHn} \epsilon_{\nu_1 \ldots \nu_{2 n}}
  \epsilon^{\mu_1 \ldots \mu_{2 n}} R^{\hspace{0.4ex} \hspace{0.4ex}
  \hspace{0.4ex} \hspace{0.4ex} \hspace{0.4ex} \hspace{0.4ex} \hspace{0.4ex}
  \hspace{1.0ex} \nu_1 \nu_2}_{\mu_1 \mu_2} \ldots R^{\hspace{0.4ex}
  \hspace{0.4ex} \hspace{0.4ex} \hspace{0.4ex} \hspace{0.4ex} \hspace{0.4ex}
  \hspace{0.4ex} \hspace{0.4ex} \hspace{0.4ex} \hspace{0.4ex} \hspace{0.4ex}
  \hspace{0.4ex} \hspace{0.4ex} \hspace{0.4ex} \hspace{0.4ex} \hspace{0.4ex}
  \hspace{1.0ex} \nu_{2 n - 1} \nu_{2 n}}_{\mu_{2 n - 1} \mu_{2 n}} = 2^{2 n}
  n! \left( n + 1 \right) !
\end{equation}

If we now formally introduce an $\left( n + 1 \right)$th coordinate, $z^{n +
1}$, that is actually set equal to 1, so that $dz^{n + 1} = 0$, let indices
$R$, $S$, $\ldots$, run from 1 to $n + 1$, and define $\eta_{R \bar{S}} =
\left( \delta_{r \bar{s}}, - 1 \right)$, then the above formula (\ref{CHn
squared line element}), for the squared line element of $\mathbf{C}
\mathbf{H}^n$, can be written:
\[ g_{\mu \nu} dz^{\mu} dz^{\nu} =
   \frac{2}{\left( - \eta_{\bar{T} Y} z^{\bar{T}} z^Y \right)} \left( \eta_{R
   \bar{S}} dz^R dz^{\bar{S}} + \frac{\left( z^{\bar{X}} \eta_{R \bar{X}} dz^R
   \right) \left( z^V \eta_{\bar{S} V} dz^{\bar{S}} \right)}{\left( -
   \eta_{\bar{U} W} z^{\bar{U}} z^W \right)} \right) = \]
\begin{equation}
  \label{homogeneous coordinates} = - \frac{2}{\eta_{\bar{T} Y} z^{\bar{T}}
  z^Y} \eta_{R \bar{S}} \left( dz^R - z^R  \frac{\eta_{Z \bar{X}} z^{\bar{X}}
  dz^Z}{\eta_{\bar{B} C} z^{\bar{B}} z^C} \right) \left( dz^{\bar{S}} -
  z^{\bar{S}}  \frac{\eta_{\bar{A} V} z^V dz^{\bar{A}}}{\eta_{\bar{U} W}
  z^{\bar{U} } z^W} \right)
\end{equation}
which, apart from the overall minus sign, and the factor of 2, which results
from my choice of normalization in (\ref{CHn metric}), would become equal to
the Fubini-Study metric on $\mathbf{C} \mathbf{P}^n$, as in equation
(15.3.14) of {\cite{Green Schwarz Witten}}, if each $\eta_{D \bar{E}}$ was now
replaced by $\delta_{D \bar{E}}$.  Conversely, if we now relax the condition
$z^{n + 1} = 1$, (\ref{homogeneous coordinates}) can be regarded as the metric
on $\mathbf{C} \mathbf{H}^n$ in homogeneous coordinates, since, just as
with the Fubini-Study metric in homogeneous coordinates, (\ref{homogeneous
coordinates}) is invariant under rescaling of the $z^R$, and also vanishes if
$dz^R$ is a multiple of $z^R$, or $dz^{\bar{S}}$ is a multiple of
$z^{\bar{S}}$.  Indeed, just as in the case of the Fubini-Study metric,
(\ref{homogeneous coordinates}) is invariant under
holomorphic {\emph{position-dependent}}
rescalings of the coordinates: $z^R \rightarrow f \left( z \right) z^R$.  For
(\ref{homogeneous coordinates}) is the square of the distance between $z$, and
a nearby point $z' = z + dz$.  And $z^R \rightarrow f \left( z \right) z^R$
implies:
\begin{equation}
  \label{position dependent rescaling} \left( z' \right)^R \rightarrow f
  \left( z' \right) \left( z' \right)^R = f \left( z \right) z^R + f \left( z
  \right) dz^R + \left( \partial_D f \left( z \right) \right) \left( dz^D
  \right) z^R
\end{equation}
Hence $dz^R \rightarrow f \left( z \right) dz^R + \left( \partial_D f \left( z
\right) \right) \left( dz^D \right) z^R$, and the second term here cancels
because it is a multiple of $z^R$.  $\mathbf{C} \mathbf{H}^n$ corresponds
to the region, in the space of the homogeneous coordinates, such that
\begin{equation}
  \label{CHn region of homogeneous coordinates} \eta_{R \bar{S}} z^R
  z^{\bar{S}} < 0
\end{equation}
Thus $z^{n + 1}$ does not vanish anywhere on $\mathbf{C} \mathbf{H}^n$,
so, unlike the case of $\mathbf{C} \mathbf{P}^n$, $\mathbf{C}
\mathbf{H}^n$ is naturally covered by a single coordinate patch.

The metric (\ref{homogeneous coordinates}) is manifestly invariant under
linear SU($n, 1$) transformations of the homogeneous coordinates:
\begin{equation}
  \label{linear SU n 1 transformation} z^R \rightarrow L_{\hspace{1.4ex}  S}^R
  z^S = L_{\hspace{1.4ex}  s}^R z^s + L_{\hspace{1.4ex} n + 1}^R z^{n + 1}
\end{equation}
where the matrix $L_{\hspace{1.4ex} S}^R$ satisfies:
\begin{equation}
  \label{condition on SU n 1 matrix L} \eta_{R \bar{S}} = \eta_{T \bar{U}}
  L_{\hspace{1.4ex} R}^T L_{\hspace{1.4ex} \bar{S}}^{\bar{U}}
\end{equation}
so that $\eta_{R \bar{S}} z^R z^{\bar{S}}$ is invariant.  The matrix
$L_{\hspace{1.4ex} \bar{S}}^{\bar{U}}$ in (\ref{condition on SU n 1 matrix L})
is by definition equal to $\left( L_{\hspace{1.4ex} S}^U \right)^{\ast}$, i.e.
the complex conjugate of $L_{\hspace{1.4ex} S}^U$, in accordance with the
convention stated above.

Now the hypersurface $z^{n + 1} = 1$, in the space of the homogeneous
coordinates, is equivalent to the whole of $\mathbf{C} \mathbf{H}^n$, in
the original coordinates.  The action of (\ref{linear SU n 1 transformation}),
on points in this hypersurface, is:
\[ z^r \rightarrow L_{\hspace{1.4ex} s}^r z^s + L_{\hspace{1.4ex} n + 1}^r  \]
\begin{equation}
\label{action of SU n 1 on hypersurface}
  z^{n + 1} \rightarrow L_{\hspace{1.4ex} \hspace{1.6ex} \, \, s}^{n + 1} z^s +
  L_{\hspace{1.4ex} \hspace{1.6ex}  n + 1}^{n + 1}
\end{equation}
Thus this hypersurface is not, in general, left invariant by (\ref{linear SU n
1 transformation}).  However, as noted above, (\ref{homogeneous coordinates})
is also invariant under holomorphic
position-dependent rescalings of the coordinates: $z^R
\rightarrow f \left( z \right) z^R$.   Hence (\ref{homogeneous coordinates})
is invariant under (\ref{linear SU n 1 transformation}), followed by division
by the new value of $z^{n + 1}$.  This compound transformation leaves the
hypersurface $z^{n + 1} = 1$ invariant, and transforms the points in this
hypersurface, which correspond to the points of $\mathbf{C} \mathbf{H}^n$,
by the projective transformations:
\begin{equation}
  \label{projective transformations} z^r \rightarrow \frac{L_{\hspace{1.4ex}
  s}^r z^s + L_{\hspace{1.4ex} n + 1}^r}{L_{\hspace{1.4ex} \hspace{1.6ex} \,
  \, s}^{n + 1} z^s + L_{\hspace{1.4ex} \hspace{1.6ex} n + 1}^{n + 1}}
\end{equation}
Thus, since the original squared line element, (\ref{CHn squared line
element}), is the restriction of (\ref{homogeneous coordinates}) to the
hypersurface $z^{n + 1} = 1$, the original squared line element, (\ref{CHn
squared line element}), is also invariant under the projective SU($n, 1$)
transformations, (\ref{projective transformations}).  This can also be
verified directly.

Returning now to the original coordinate system, or in other words, setting
$z^{n + 1} = 1$, and restricting $z$ to represent the $n$-vector, $\left( z^1,
\ldots, z^n \right)$, we can send the origin of $\mathbf{C} \mathbf{H}^n$
to an arbitrary point, $z$, of $\mathbf{C} \mathbf{H}^n$, by means of an
SU($n, 1$) projective transformation, (\ref{projective transformations}), by
choosing the matrix $L_{\hspace{1.4ex} S}^R$ to be the SU($n, 1$) ``boost'':
\begin{equation}
  \label{SU n 1 boost} L_{\hspace{1.4ex} S}^R = \left(\begin{array}{cc}
    L_{\hspace{1.4ex} s}^r & L_{\hspace{1.4ex} n + 1}^r\\
    L_{\hspace{1.4ex} \hspace{1.6ex} \, \, s}^{n + 1}  & L_{\hspace{1.4ex}
    \hspace{1.6ex} n + 1}^{n + 1}
  \end{array}\right) = \left(\begin{array}{cc}
    \delta_{\hspace{0.4ex} \hspace{0.4ex} s}^r + \left( \frac{ \gamma - 1}{z^t
    z^{\bar{t}}} \right) z^r z_s & \gamma z^r\\
    \gamma z_s & \gamma
  \end{array}\right)
\end{equation}
where
\begin{equation}
  \label{gamma in the SU n 1 boost} \gamma = \frac{1}{\sqrt{1 - z^t
  z^{\bar{t}}}}
\end{equation}

For $n \geq 2$, $\mathbf{C} \mathbf{H}^n$ is not maximally symmetric, and
the sectional curvature is not constant.  In general, for linearly independent
vectors $A^{\mu}$, $B^{\nu}$, the sectional curvature, at a point of a
Riemannian manifold, is defined, bearing in mind the sign convention
(\ref{Riemann tensor definition}), by:
\begin{equation}
  \label{definition of sectional curvature} K \left( S \right) = -
  \frac{R_{\mu \nu \sigma \tau} A^{\mu} B^{\nu} A^{\sigma} B^{\tau}}{\left(
  g_{\mu \sigma} g_{\nu \tau} - g_{\mu \tau} g_{\nu \sigma} \right) A^{\mu}
  B^{\nu} A^{\sigma} B^{\tau}}
\end{equation}
where $S$ denotes the linear space spanned by $A^{\mu}$ and $B^{\nu}$.  To
apply this to $\mathbf{C} \mathbf{H}^n$, in the complex coordinates, we
note that, if a vector, $A^{\mu}$, is real, in real coordinates, then after
transforming to complex coordinates, as described after (\ref{x in terms of
z}), its components satisfy $A^{\bar{r}} = \left( A^r \right)^{\ast}$, just as
for the complex coordinates themselves.  For the $\mathbf{C} \mathbf{H}^n$
metric, (\ref{hermitian metric}), (\ref{CHn metric}), with the Riemann tensor
components (\ref{CHn covariant Riemann tensor}), we find:
\[ R_{\mu \nu \sigma \tau} A^{\mu} B^{\nu} A^{\sigma} B^{\tau} = \hspace{1ex}
   \hspace{1ex} \hspace{1ex} \hspace{1ex} \hspace{1ex} \hspace{1ex}
   \hspace{1ex} \hspace{1ex} \hspace{1.6ex}
   \hspace{10cm} \]
\begin{equation}
  \label{R A B A B for CHn} = 2 \left( g_{t \bar{r}} A^t A^{\bar{r}} \right)
  \left( g_{s \bar{u}} B^s B^{\bar{u}} \right) + 2 \left( g_{t \bar{u}} A^t
  B^{\bar{u}} \right) \left( g_{s \bar{r}} B^s A^{\bar{r}} \right) - 2 \left(
  g_{r \bar{s}} A^r B^{\bar{s}} \right)^2 - 2 \left( g_{s \bar{r}} B^s
  A^{\bar{r}} \right)^2
\end{equation}
\[ \left( g_{\mu \sigma} g_{\nu \tau} - g_{\mu \tau} g_{\nu \sigma} \right)
   A^{\mu} B^{\nu} A^{\sigma} B^{\tau} = \hspace{1ex} \hspace{1ex}
   \hspace{1ex} \hspace{1ex} \hspace{1ex} \hspace{1ex} \hspace{1ex}
   \hspace{1ex} \hspace{1.6ex} \hspace{8cm} \]
\begin{equation}
  \label{g g A B A B for CHn} = 4 \left( g_{r \bar{s}} A^r A^{\bar{s}} \right)
  \left( g_{t \bar{u}} B^t B^{\bar{u}} \right) - 2 \left( g_{r \bar{s}} A^r
  B^{\bar{s}} \right) \left( g_{t \bar{u}} B^t A^{\bar{u}} \right) - \left(
  g_{s \bar{r}} B^s A^{\bar{r}} \right)^2 - \left( g_{r \bar{s}} A^r
  B^{\bar{s}} \right)^2
\end{equation}
If we now work at the origin of the complex coordinates, so that $g_{r
\bar{s}} = \delta_{r \bar{s}}$, we can define real magnitudes $\left| A
\right|$, $\left| B \right|$, and real angles, $\theta$, $\varphi$, by:
\[ \left| A \right| = \sqrt{g_{r \bar{s}} A^r A^{\bar{s}}}, \hspace{3cm}
   \left| B \right| = \sqrt{g_{r \bar{s}} B^r B^{\bar{s}}}, \]
\[ \theta = \arccos \left( \frac{\sqrt{\left( g_{r \bar{s}} A^r
   B^{\bar{s}} \right) \left( g_{t \bar{u}} B^t A^{\bar{u}} \right)}}{\left| A
   \right| \left| B \right|} \right), \]
\begin{equation}
  \label{magnitudes and angles for CHn sectional curvature} \varphi =
  \frac{1}{2} \arcsin \left( \frac{1}{2 i} \left( \frac{g_{r \bar{s}} A^r
  B^{\bar{s}}}{g_{t \bar{u}} B^t A^{\bar{u}}} - \frac{g_{t \bar{u}} B^t
  A^{\bar{u}}}{g_{r \bar{s}} A^r B^{\bar{s}}} \right) \right)
\end{equation}
Thus:
\begin{equation}
  - \label{CHn sectional curvature} \frac{R_{\mu \nu \sigma \tau} A^{\mu}
  B^{\nu} A^{\sigma} B^{\tau}}{\left( g_{\mu \sigma} g_{\nu \tau} - g_{\mu
  \tau} g_{\nu \sigma} \right) A^{\mu} B^{\nu} A^{\sigma} B^{\tau}} = -
  \frac{\left( 1 + \cos^2 \theta \left( 1 - 2 \cos 2 \varphi \right)
  \right)}{\left( 2 - \cos^2 \theta \left( 1 + \cos 2 \varphi \right) \right)}
\end{equation}
Now, if $n = 1$, the angle $\theta$ is $0$, so the sectional curvature is $-
2$.  For $n \geq 2$, and for all values of $\varphi$, such that $\cos 2
\varphi \neq 1$, the right-hand side of (\ref{CHn sectional curvature}) varies
between a minimum of $- 2$, when $\cos^2 \theta = 1$, and a maximum of $-
\frac{1}{2}$, when $\cos^2 \theta = 0$, since if we replace $\cos^2 \theta$ by
$x$, the function has no maximum or minimum between $x = 0$ and $x = 1$.  For
$\cos 2 \varphi = 1$, the right-hand side of (\ref{CHn sectional curvature})
is equal to $- \frac{1}{2}$, except for $\cos^2 \theta = 1$.  And when $\cos 2
\varphi$ and $\cos^2 \theta$ are both equal to $1$, $A^{\mu}$ and $B^{\nu}$
are no longer linearly independent.  Thus for $n \geq 2$, the sectional
curvature of $\mathbf{C} \mathbf{H}^n$, with the metric (\ref{hermitian
metric}), (\ref{CHn metric}), lies in the range $- 2$ to $- \frac{1}{2}$.

The equation for a geodesic is:
\begin{equation}
  \label{geodesic equation} \frac{d^2 z^r}{ds^2} + \Gamma_{st}^r
  \frac{dz^s}{ds} \frac{dz^t}{ds} \hspace{0.4ex} = \hspace{0.4ex} \frac{d^2
  z^r}{ds^2} + \frac{2 z_s}{\left( 1 - z^{\bar{v}} z^v \right)}
  \frac{dz^s}{ds} \frac{dz^r}{ds} \hspace{0.4ex} = \hspace{0.4ex} 0
\end{equation}
For the geodesics through the origin, we have $z^r = Z^r \tanh \left( \alpha s
\right)$, where $Z^r$ is a fixed complex $n$-vector such that $ Z_s Z^s = 1$,
and $\alpha$ is a real constant.  If we choose $\alpha = \frac{1}{\sqrt{2}}$,
then $s$ is the geodesic distance from the origin to $z$, in accordance with
(\ref{CHn squared line element}).  Hence the geodesic distance from the origin
to $z$, is
\begin{equation}
  \label{geodesic distance from origin} s = \frac{1}{\sqrt{2}} \ln \left(
  \frac{1 + \left| z \right|}{1 - \left| z \right|} \right)
\end{equation}
where $\left| z \right| \equiv \sqrt{z_r z^r}$.  To find the geodesic distance
from a point, $z$, to a point, $w$, we can use the invariance of the geometry
under the projective SU($n, 1$) transformations (\ref{projective
transformations}).  Sending $z$ to the origin, by the inverse of (\ref{SU n 1
boost}), sends $w$ to a point $\tilde{w}$, such that:
\begin{equation}
  \label{w tilde squared} \tilde{w}_r \tilde{w}^r = \left( \frac{\left( w_r -
  z_r  \right) \left( w^r - z^r \right) - z_r z^r w_s w^s + w_r z^r z_s
  w^s}{\left( 1 - z_t w^t \right) \left( 1 - w_u z^u \right)} \right)
\end{equation}
Hence:
\begin{equation}
  \label{cosh squared s over root 2} \cosh^2 \left( \frac{s}{\sqrt{2}} \right)
  = \frac{1}{1 - \tilde{w}_r \tilde{w}^r} = \frac{\left( 1 - z_r w^r \right)
  \left( 1 - w_s z^s \right)}{\left( 1 - z_t z^t \right) \left( 1 - w_u w^u
  \right)}
\end{equation}
where $s$ now denotes the geodesic distance from $z$ to $w$.  Thus in the
homogeneous coordinates, we have:
\begin{equation}
\label{cosh s over root 2}
\cosh \left( \frac{s}{\sqrt{2}} \right) = \frac{\left| \eta_{R \bar{S}} z^R
   w^{\bar{S}} \right|}{\sqrt{\left| \eta_{T \bar{U}} z^T z^{\bar{U}} \right|
   \left| \eta_{V \bar{X}} w^V w^{\bar{X}} \right|}}
\end{equation}
where $\left| \cdot \right|$ denotes the absolute value of a complex number.

\subsection{The field equations and boundary conditions}
\label{The field equations and boundary conditions}

I shall now assume that Ho\v{r}ava-Witten theory has been quantized in accordance
with standard procedures for quantizing supergravity in eleven dimensions
{\cite{de Wit van Nieuwenhuizen Van Proeyen, Bautier Deser Henneaux
Seminara}}, together with an appropriate treatment of the orbifold fixed-point
hyperplanes, and seek solutions of the field equations, and boundary
conditions, that follow from varying the quantum effective action, or in other
words, the generating functional of the proper vertices, $\Gamma$,
{\cite{Schwinger, DeWitt effective action, Nambu Jona Lasinio 1,
Nambu Jona Lasinio 2}}, with respect to the
fields.  The quantum effective action is expanded in terms of the number of
loops in Feynman diagrams, which in the bulk, is an expansion in powers of
$\kappa^2$, and on the orbifold fixed-point hyperplanes, is an expansion in
powers of $\lambda^2$, where $\lambda$ and $\kappa$ are related by
(\ref{lambda kappa relation}).  Since $\kappa$ and $\lambda$ are dimensional
constants, the actual expansion parameters have the form
$\frac{\kappa^2}{L^9}$, and $\frac{\lambda^2}{L^6}$, where $L$ will be, in
general, the smallest physically relevant distance, in a particular region of
the geometry.  I shall seek solutions with a ``thick pipe'' form of geometry,
so that, in particular, if $L$ denotes the geodesic distance between the two
orbifold fixed-point hyperplanes, then $\frac{\kappa^2}{L^9} \ll 1$.
Furthermore, if $L$ denotes a radius of curvature of the compact
six-manifold, which will in general be either smaller than, or comparable to,
its diameter, on account of
the hyperbolic nature of the manifold, then we will again have
$\frac{\kappa^2}{L^9} \ll 1$, throughout the main part of the bulk.

Thus, throughout the main part of the bulk, it will be a good approximation to
neglect all quantum corrections to $\Gamma$, and approximate $\Gamma$ as the
gauge-fixed classical action, together with the Fadeev-Popov terms.  We then
seek a solution, of the field equations that follow from varying $\Gamma$ with
respect to the fields, in which all the Fadeev-Popov fields, and also any
other fields introduced in the course of the gauge-fixing, vanish.  The field
equations then reduce to the classical field equations for the supergravity
multiplet, which are the Euler-Lagrange equations for the Cremmer-Julia-Scherk
action (\ref{upstairs bulk action}), together with gauge-fixing conditions.
We can always solve such equations by solving the Cremmer-Julia-Scherk field
equations in any convenient gauge we choose, then applying gauge
transformations to the solution, in order to satisfy the required gauge
conditions.

I shall now denote the full eleven-dimensional metric by $G_{IJ}$.  It will be
distinguished from $G_{IJKL}$ by context, and the number of indices.  Other
conventions are as in Subsection \ref{Horava-Witten theory}, on page
\pageref{Horava-Witten theory}.  Furthermore, coordinate indices $A, B, C,
\ldots$ will be tangent to the compact six-manifold, and coordinate indices
$\mu, \nu, \sigma, \ldots$ will be tangent to the four observed space-time
dimensions, which at the inner surface of the thick pipe, where we live, in
this type of model, are the extended dimensions.  I shall use the gauge
freedom of general coordinate invariance, in order to choose Gaussian normal
coordinates, such that $G_{yy} = 1$, and $G_{Uy} = 0$, and thus seek a
solution where the metric has the form:
\begin{equation}
  \label{metric ansatz} ds^2_{11} = G_{IJ} dx^I dx^J = a \left( y \right)^2
  g_{\mu \nu} dx^{\mu} dx^{\nu} + b \left( y \right)^2 h_{AB} dx^A dx^B + dy^2
\end{equation}
where $g_{\mu \nu}$ is the metric on a four-dimensional locally de Sitter
space, whose de Sitter radius I shall set equal to $1$, and whose spatial
sections may have been compactified, as discussed after equation
(\ref{de Sitter radius}), and
$h_{AB}$ is the metric on a smooth compact quotient of $\mathbf{C}
\mathbf{H}^3$, and is locally equal to the metric specified in
(\ref{hermitian metric}) and (\ref{CHn metric}).  Thus $R_{\mu \nu} \left( g
\right) = - 3 g_{\mu \nu}$, and $R_{AB} \left( h \right) = 4 h_{AB}$.  I shall
also consider the possibility of flat and AdS spacetimes, which would have
$R_{\mu \nu} \left( g \right)$ equal to zero, and a positive multiple of
$g_{\mu \nu}$, respectively.

I shall seek solutions such that the inner and outer surfaces of the thick
pipe, or, in other words, the orbifold fixed point hyperplanes, are at
$ y = y_1 $ and $ y = y_2 $, where $ y_1 $ and $ y_2 $ are determined by
the boundary conditions, and are independent of position in the four
observed dimensions, and on the compact six-manifold.
We are free to shift $y$ by a constant, and I shall use this freedom to obtain
the simplest formulae for the solution in the bulk, rather than to set $y_1$ or
$ y_2 $ to any particular value.

The inner surface of the thick pipe, where we
live, will be at $y = y_1$, so in the de Sitter case, it follows from (\ref{de
Sitter radius}), that we require $a \left( y_1 \right) = 16.0 \hspace{0.8ex}
\mathrm{Gyr} = 1.51 \times 10^{26} \hspace{0.8ex} \mathrm{metres} = 0.94 \times
10^{61} \sqrt{G_N}$.

\subsubsection{The Christoffel symbols, Riemann tensor, and Ricci tensor}
\label{The Christoffel symbols Riemann tensor and Ricci tensor}

The non-vanishing Christoffel symbols of the second kind, for the metric
ansatz (\ref{metric ansatz}), are:
\begin{eqnarray}
  \label{Christoffel symbols for the metric ansatz} \Gamma_{\nu \sigma}^{\mu}
  & = & \frac{1}{2} g^{\mu \tau} \left( \partial_{\nu} g_{\sigma \tau} +
  \partial_{\sigma} g_{\nu \tau} - \partial_{\tau} g_{\nu \sigma} \right)
  \nonumber\\
  \Gamma_{\nu y}^{\mu} & = & \Gamma_{y \nu}^{\mu} \hspace{0.4ex}
  \hspace{0.4ex} = \hspace{0.4ex} \hspace{0.4ex} \frac{\dot{a}}{a}
  \delta_{\hspace{0.4ex} \hspace{0.6ex} \nu}^{\mu} \nonumber\\
  \Gamma_{\mu \nu}^y & = & - a \dot{a} g_{\mu \nu} \hspace{0.4ex}
  \hspace{0.4ex} = \hspace{0.4ex} \hspace{0.4ex} - \frac{\dot{a}}{a} G_{\mu
  \nu} \nonumber\\
  \Gamma_{BC}^A & = & \frac{1}{2} h^{AD} \left( \partial_B h_{CD} + \partial_C
  h_{BD} - \partial_D h_{BC} \right) \nonumber\\
  \Gamma_{By}^A & = & \Gamma_{yB}^A \hspace{0.4ex} \hspace{0.4ex} =
  \hspace{0.4ex} \hspace{0.4ex} \frac{\dot{b}}{b} \delta_{\hspace{0.4ex}
  \hspace{0.8ex} B}^A \nonumber\\
  \Gamma_{AB}^y & = & - b \dot{b} h_{AB} \hspace{0.4ex} \hspace{0.4ex} =
  \hspace{0.4ex} \hspace{0.4ex} - \frac{\dot{b}}{b} G_{AB}
\end{eqnarray}
where a dot denotes differentiation with respect to $y$.  From this, and the
formula (\ref{Riemann tensor}), for the components of the Riemann tensor, it
follows that the only non-vanishing components of the form $R^{\hspace{0.4ex}
\hspace{0.4ex} \hspace{0.4ex} \hspace{0.4ex} \hspace{0.4ex} \hspace{1.5ex}
J}_{UVI}$, of the Riemann tensor, in eleven dimensions, are of the forms
$R^{\hspace{0.4ex} \hspace{0.4ex} \hspace{0.4ex} \hspace{0.4ex} \hspace{0.4ex}
\hspace{1.0ex} \tau}_{\mu \nu \sigma}$, $R^{\hspace{0.4ex} \hspace{0.4ex}
\hspace{0.4ex} \hspace{0.4ex} \hspace{0.4ex} \hspace{0.4ex} \hspace{0.4ex}
\hspace{1ex} D}_{ABC}$, $R^{\hspace{0.4ex} \hspace{0.4ex} \hspace{0.4ex}
\hspace{0.4ex} \hspace{0.4ex} \hspace{1.2ex} B}_{\mu A \nu}$,
$R^{\hspace{0.4ex} \hspace{0.4ex} \hspace{0.4ex} \hspace{0.4ex} \hspace{0.4ex}
\hspace{1.2ex} B}_{A \mu \nu}$, $R^{\hspace{0.4ex} \hspace{0.4ex}
\hspace{0.4ex} \hspace{0.4ex} \hspace{0.4ex} \hspace{0.4ex} \hspace{0.4ex}
\hspace{0.8ex} \nu}_{\mu AB}$, and $R^{\hspace{0.4ex} \hspace{0.4ex}
\hspace{0.4ex} \hspace{0.4ex} \hspace{0.4ex} \hspace{0.4ex} \hspace{0.4ex}
\hspace{0.8ex} \nu}_{A \mu B}$.  In particular, neither $I$, nor $J$, can be
$y$.  The non-vanishing components of the Riemann tensor, when one or more of
the indices is $y$, are $R^{\hspace{0.4ex} \hspace{0.4ex} \hspace{0.4ex}
\hspace{0.4ex} \hspace{0.4ex} \hspace{0.8ex} y}_{\mu y \nu}$,
$R^{\hspace{0.4ex} \hspace{0.4ex} \hspace{0.4ex} \hspace{0.4ex} \hspace{0.4ex}
\hspace{1.0ex} y}_{y \mu \nu}$, $R^{\hspace{0.4ex} \hspace{0.4ex}
\hspace{0.4ex} \hspace{0.4ex} \hspace{0.4ex} \hspace{1.0ex} \nu}_{\mu yy}$,
$R^{\hspace{0.4ex} \hspace{0.4ex} \hspace{0.4ex} \hspace{0.4ex} \hspace{0.4ex}
\hspace{0.8ex} \nu}_{y \mu y}$, $R^{\hspace{0.4ex} \hspace{0.4ex}
\hspace{0.4ex} \hspace{0.4ex} \hspace{0.4ex} \hspace{1.4ex} y}_{AyB}$,
$R^{\hspace{0.4ex} \hspace{0.4ex} \hspace{0.4ex} \hspace{0.4ex} \hspace{0.4ex}
\hspace{1.4ex} y}_{yAB}$, $R^{\hspace{0.4ex} \hspace{0.4ex} \hspace{0.4ex}
\hspace{0.4ex} \hspace{0.4ex} \hspace{0.4ex} \hspace{0.6ex} B}_{Ayy}$, and
$R^{\hspace{0.4ex} \hspace{0.4ex} \hspace{0.4ex} \hspace{0.4ex} \hspace{0.4ex}
\hspace{0.4ex} \hspace{0.6ex} B}_{yAy}$.  We find:
\begin{eqnarray}
  \label{Riemann tensor for the metric ansatz} R^{\hspace{0.4ex}
  \hspace{0.4ex} \hspace{0.4ex} \hspace{0.4ex} \hspace{0.4ex} \hspace{1.2ex}
  \tau}_{\mu \nu \sigma} & = & R^{\hspace{0.4ex} \hspace{0.4ex} \hspace{0.4ex}
  \hspace{0.4ex} \hspace{0.4ex} \hspace{1.0ex} \tau}_{\mu \nu \sigma} \left( g
  \right) + \hspace{0.4ex} \frac{\dot{a}^2}{a^2} \left( G_{\mu \sigma}
  \delta_{\nu}^{\hspace{0.4ex} \hspace{0.6ex} \tau} - \hspace{0.4ex} G_{\nu
  \sigma} \delta_{\mu}^{\hspace{0.4ex} \hspace{0.6ex} \tau} \right)
  \nonumber\\
  R^{\hspace{0.4ex} \hspace{0.4ex} \hspace{0.4ex} \hspace{0.4ex}
  \hspace{0.4ex} \hspace{0.4ex} \hspace{0.4ex} \hspace{0.4ex} \hspace{0.4ex}
  D}_{ABC} & = & R^{\hspace{0.4ex} \hspace{0.4ex} \hspace{0.4ex}
  \hspace{0.4ex} \hspace{0.4ex} \hspace{0.4ex} \hspace{0.4ex} \hspace{0.4ex}
  \hspace{0.4ex} D}_{ABC} \left( h \right) + \frac{\dot{b}^2}{b^2} \left(
  G_{AC} \delta_B^{\hspace{0.4ex} \hspace{0.8ex} D} - G_{BC}
  \delta_A^{\hspace{0.4ex} \hspace{0.8ex} D} \right) \nonumber\\
  R^{\hspace{0.4ex} \hspace{0.4ex} \hspace{0.4ex} \hspace{0.4ex}
  \hspace{0.4ex} \hspace{1.2ex} B}_{\mu A \nu} & = & \frac{\dot{a}
  \dot{b}}{ab} G_{\mu \nu} \delta^{\hspace{1.2ex} B}_A, \hspace{7.9ex}
  \hspace{0.4ex} \hspace{1.2ex} R^{\hspace{0.4ex} \hspace{0.4ex}
  \hspace{0.4ex} \hspace{0.4ex} \hspace{0.4ex} \hspace{0.4ex} \hspace{0.4ex}
  \hspace{0.6ex} \nu}_{A \mu B} \hspace{0.4ex} \hspace{0.4ex} = \hspace{0.4ex}
  \hspace{0.4ex} \frac{\dot{a} \dot{b}}{ab} G_{AB} \delta^{\hspace{0.4ex}
  \hspace{0.6ex} \nu}_{\mu} \nonumber\\
  R^{\hspace{0.4ex} \hspace{0.4ex} \hspace{0.4ex} \hspace{0.4ex}
  \hspace{0.4ex} \hspace{1.0ex} y}_{\mu y \nu} & = & \frac{\ddot{a} }{a}
  G_{\mu \nu}, \hspace{14.6ex} R^{\hspace{0.4ex} \hspace{0.4ex} \hspace{0.4ex}
  \hspace{0.4ex} \hspace{0.4ex} \hspace{1.0ex} \nu}_{y \mu y} \hspace{0.4ex}
  \hspace{0.4ex} = \hspace{0.4ex} \hspace{0.4ex} \frac{\ddot{a}}{a}
  \delta_{\mu}^{\hspace{0.4ex} \hspace{0.6ex} \nu} \nonumber\\
  R^{\hspace{0.4ex} \hspace{0.4ex} \hspace{0.4ex} \hspace{0.4ex}
  \hspace{0.4ex} \hspace{1.4ex} y}_{AyB} & = & \frac{\ddot{b}}{b} G_{AB},
  \hspace{14ex} R^{\hspace{0.4ex} \hspace{0.4ex} \hspace{0.4ex} \hspace{0.4ex}
  \hspace{0.4ex} \hspace{0.4ex} \hspace{0.6ex} B}_{yAy} \hspace{0.4ex}
  \hspace{0.4ex} = \hspace{0.4ex} \hspace{0.4ex} \frac{\ddot{b}}{b}
  \delta_A^{\hspace{0.4ex} \hspace{0.8ex} B}
\end{eqnarray}
where $R^{\hspace{0.4ex} \hspace{0.4ex} \hspace{0.4ex} \hspace{0.4ex}
\hspace{0.4ex} \hspace{0.8ex} \tau}_{\mu \nu \sigma} \left( g \right)$ denotes
the Riemann tensor calculated from the four-dimensional metric $g_{\mu \nu}$,
and $R^{\hspace{0.4ex} \hspace{0.4ex} \hspace{0.4ex} \hspace{0.4ex}
\hspace{0.4ex} \hspace{0.4ex} \hspace{0.4ex} \hspace{0.4ex} \hspace{0.4ex}
D}_{ABC} \left( h \right)$ denotes the Riemann tensor calculated from the
six-dimensional metric $h_{AB}$.  From (\ref{Riemann tensor for the metric
ansatz}), we find that the non-vanishing Ricci tensor components, in eleven
dimensions, are:
\begin{eqnarray}
  \label{Ricci tensor for the metric ansatz} R_{\mu \nu} & = & R_{\mu \nu}
  \left( g \right) + \left( \frac{\ddot{a} }{a} + 3 \hspace{0.4ex}
  \frac{\dot{a}^2}{a^2} + 6 \frac{\dot{a} \dot{b}}{ab} \right) G_{\mu \nu}
  \hspace{0.4ex} \hspace{0.4ex} = \hspace{0.4ex} \hspace{0.4ex} \left(
  \frac{\ddot{a} }{a} + 3 \hspace{0.4ex} \frac{\dot{a}^2}{a^2} + 6
  \frac{\dot{a} \dot{b}}{ab} - \frac{3}{a^2} \right) G_{\mu \nu} \nonumber\\
  R_{AB} & = & R_{AB} \left( h \right) + \left( \frac{\ddot{b}}{b} + 5
  \frac{\dot{b}^2}{b^2} + 4 \frac{\dot{a} \dot{b}}{ab} \right) G_{AB}
  \hspace{0.4ex} \hspace{0.4ex} = \hspace{0.4ex} \hspace{0.4ex} \left(
  \frac{\ddot{b}}{b} + 5 \frac{\dot{b}^2}{b^2} + 4 \frac{\dot{a} \dot{b}}{ab}
  + \frac{4}{b^2} \right) G_{AB} \nonumber\\
  R_{yy} & = & 4 \frac{\ddot{a}}{a} + 6 \frac{\ddot{b}}{b}
\end{eqnarray}
where I used the relations $R_{\mu \nu} \left( g \right) = - 3 g_{\mu \nu}$,
and $R_{AB} \left( h \right) = 4 h_{AB}$, from above.

For smooth compact quotients of $\mathbf{H}^6$, we choose the metric
$h_{AB}$ for $\mathbf{H}^6$ to have radius of curvature equal to $1$, so
that $R_{ABCD} \left( h \right) = h_{AC} h_{BD} - h_{AD} h_{BC}$, and $R_{AB}
\left( h \right) = 5 h_{AB}$.  Thus for a smooth compact quotient of
$\mathbf{H}^6$, the term $\frac{4}{b^2} G_{AB}$ in $R_{AB}$ is replaced by
$\frac{5}{b^2} G_{AB}$.

We also need the Riemann tensor components, on the orbifold fixed-point
hyperplanes, calculated from the ten-dimensional metric, on the orbifold
fixed-point hyperplanes.  The ten-dimensional metric is obtained from
(\ref{metric ansatz}), by setting $dy = 0$, and either $y = y_1$, or $y =
y_2$, as appropriate.  Then $\mathcal{M}^{10}$ is simply the Cartesian
product, of a four dimensional locally de Sitter space, with de Sitter radius
$a \left( y_1 \right)$, or $a \left( y_2 \right)$, as appropriate, and a
smooth compact quotient of $\mathbf{C} \mathbf{H}^3$, with the metric
(\ref{CHn metric}) multiplied by a factor $b^2 \left( y_1 \right)$, or $b^2
\left( y_2 \right)$, as appropriate.  All the Christoffel symbols and Riemann
tensor components with mixed indices now vanish, and the only non-vanishing
Riemann tensor components are now $R^{\hspace{0.4ex} \hspace{0.4ex}
\hspace{0.4ex} \hspace{0.4ex} \hspace{0.4ex} \hspace{1.0ex} \tau}_{\mu \nu
\sigma} \left( g \right)$ and $R^{\hspace{0.4ex} \hspace{0.4ex} \hspace{0.4ex}
\hspace{0.4ex} \hspace{0.4ex} \hspace{0.4ex} \hspace{0.4ex} \hspace{0.4ex}
\hspace{0.6ex} D}_{ABC} \left( h \right)$.

\subsubsection{The Yang-Mills coupling constants in four dimensions}
\label{The Yang-Mills coupling constants in four dimensions}

There are inevitably significant Casimir energy density terms in the
energy-momen-tum tensor on and near the inner surface of the thick pipe, due to
the Ho\v{r}ava-Witten relation $ \lambda \simeq 5.8 \kappa^{\frac{2}{3}} $
between the $ d = 10 $ Yang-Mills coupling constant $ \lambda $ and $ \kappa $
\cite{HW2}, and the fact that the $ d = 4 $ Yang-Mills coupling constants at
unification are not much smaller than $ 1 $, which implies that $ b_1 = b
\left( y_1 \right) $, the value of $b$ at the inner surface of the thick pipe,
is comparable to $ \kappa^{2/9} $.

The value of $b_1$ is fixed by the value of the Yang-Mills
fine structure constants in four dimensions at unification, $\alpha_U =
\frac{g^2_U}{4 \pi}$, which will be equal to the value of the QCD fine
structure constant at unification, and the magnitude $\left| \chi \left(
\mathcal{M}^6 \right) \right|$ of the Euler number of the compact six-manifold
$ \mathcal{M}^6 $.  For by the generalized Gauss-Bonnet theorem
{\cite{Allendoerfer Weil}}, the Euler characteristic, or Euler number, $\chi
\left( \mathcal{M}^{2 n} \right)$, of an arbitrary smooth $2 n$-manifold $
\mathcal{M}^{2 n}$, is given by:
\begin{equation}
  \label{generalized Gauss-Bonnet theorem} \chi \left( \mathcal{M}^{2 n}
  \right) = \frac{\left( - \right)^n}{\left( 8 \pi \right)^n n!}
  \int_{\mathcal{M}^{2 n}} d^{2 n} x \sqrt{g} \epsilon_{\nu_1 \ldots \nu_{2
  n}} \epsilon^{\mu_1 \ldots \mu_{2 n}} R^{\hspace{0.4ex} \hspace{0.4ex}
  \hspace{0.4ex} \hspace{0.4ex} \hspace{0.4ex} \hspace{0.4ex} \hspace{0.4ex}
  \hspace{1.0ex} \nu_1 \nu_2}_{\mu_1 \mu_2} \ldots R^{\hspace{0.4ex}
  \hspace{0.4ex} \hspace{0.4ex} \hspace{0.4ex} \hspace{0.4ex} \hspace{0.4ex}
  \hspace{0.4ex} \hspace{0.4ex} \hspace{0.4ex} \hspace{0.4ex} \hspace{0.4ex}
  \hspace{0.4ex} \hspace{0.4ex} \hspace{0.4ex} \hspace{0.4ex} \hspace{0.4ex}
  \hspace{1.0ex} \nu_{2 n - 1} \nu_{2 n}}_{\mu_{2 n - 1} \mu_{2 n}}.
\end{equation}
Thus defining $V \left( \mathcal{M}^6 \right) \equiv \int_{\mathcal{M}^6} d^6 z
\sqrt{h}$, we find from (\ref{Euler number density for CHn}), on page
\pageref{Euler number density for CHn}, that for a smooth compact quotient of
$\mathbf{C} \mathbf{H}^3$, with the standard metric (\ref{hermitian metric}),
(\ref{CHn metric}), as used in the metric ansatz (\ref{metric ansatz}):
\begin{equation}
  \label{volume in terms of Euler number for a smooth compact quotient of CH3}
  V \left( \mathcal{M}^6 \right) = - \frac{\pi^3}{3} \chi \left( \mathcal{M}^6
  \right) = - 10.3354 \chi \left( \mathcal{M}^6 \right)
\end{equation}
And for a smooth compact quotient of $\mathbf{H}^6$, with the metric
normalized such that $R_{ABCD} \left( h \right) = h_{AC} h_{BD} - h_{AD}
h_{BC}$, as stated after (\ref{Ricci tensor for the metric ansatz}), we have:
\begin{equation}
  \label{volume in terms of Euler number for a smooth compact quotient of H6}
  V \left( \mathcal{M}^6 \right) = - \frac{8 \pi^3}{15} \chi \left(
  \mathcal{M}^6 \right) = - 16.5367 \chi \left( \mathcal{M}^6 \right)
\end{equation}

Then on using (\ref{lambda kappa
relation}), and reducing (\ref{Yang Mills action}) to four dimensions, we
find {\cite{Witten Strong coupling expansion}} that when $ \mathcal{M}^6 $
is a smooth compact quotient of $\mathbf{C} \mathbf{H}^3$:
\begin{equation}
  \label{alpha sub U} \alpha_U = \frac{\left( 4 \pi \kappa^2
  \right)^{\frac{2}{3}}}{2 b^6_1 V \left( \mathcal{M}^6 \right)} =
  \frac{0.2615}{\left| \chi \left( \mathcal{M}^6 \right) \right|} \left(
  \frac{\kappa^{2/9}}{b_1} \right)^6
\end{equation}
And for a smooth compact quotient of $\mathbf{H}^6$, the same relation is
obtained, but with the coefficient $0.2615$ replaced by $\frac{5}{8} \times
0.2615 = 0.1634$.

The result (\ref{alpha sub U}) depends on the factor
$\frac{1}{30}$, in the definition of $\mathrm{tr}$ in (\ref{Yang Mills action}),
cancelling with a factor $30$, in the ratio of the trace of the square of a
generator of $\mathrm{SU} \left( 3 \right)$, naturally embedded in $E_8$, in the
adjoint of $E_8$, to the trace of the square of the corresponding generator,
in the fundamental representation of SU(3).  For standard Grand Unification,
this follows from the corresponding ratio for generators of SO(16), already
derived in subsection \ref{Horava-Witten theory}, via the natural embedding
$\mathrm{SU} ( 3 ) \subset \mathrm{SU} ( 5 ) \subset \mathrm{SO} ( 10 ) \subset
\mathrm{SO} ( 16 )$.  I will be using a different chain of
natural embeddings in
Section \ref{E8 vacuum gauge fields and the Standard Model},
namely $\mathrm{SU}
\left( 3 \right) \subset \mathrm{SU} \left( 9 \right) \subset E 8$, but the
embedding of SU(3) in $E_8$, by this chain, is equivalent to the embedding of
SU(3) in $E_8$, by the above SO(16) chain, as follows from, firstly, the
equivalence of the embedding of SU(3) in $E_8$ by the above SO(16) chain, and
the embedding of SU(3) in $E_8$ by the chain $\mathrm{SU} ( 3 ) \subset
\mathrm{SU} ( 5 ) \subset \mathrm{SO} ( 10 ) \subset E 6$, secondly, the
equivalence of the embeddings in $E_8$, of all four SU(3)'s, in the chain
$\mathrm{SU} \left( 3 \right) \times \mathrm{SU} \left( 3 \right) \times \mathrm{SU}
\left( 3 \right) \times \mathrm{SU} \left( 3 \right) \subset E 6 \times
\mathrm{SU} \left( 3 \right) \subset E 8$, and thirdly, the equivalence of the
embeddings in $E_8$, of any three of the four SU(3)'s in the preceding chain,
and the embeddings in $E_8$, of the three SU(3)'s in the chain $\mathrm{SU}
\left( 3 \right) \times \mathrm{SU} \left( 3 \right) \times \mathrm{SU} \left( 3
\right) \subset \mathrm{SU} ( 9 ) \subset E 8$.  Thus the required relation also
holds for the embedding of SU(3) in $E_8$, via the subgroup chain I will be
using in Section \ref{E8 vacuum gauge fields and the Standard Model}.  This
result will also be verified directly in Section \ref{E8 vacuum gauge fields and
the Standard Model}.  The subgroup chains just listed all follow simply by
identifying appropriate subsets of the roots of $E_8$, in the weight diagram
of $E_8$, without the need to project the roots to a subspace, and take linear
combinations of roots that coincide after the projection, as required, for
example, for embedding SO($n$) into SU($n$).

In Section \ref{E8 vacuum gauge fields and the Standard Model}, I shall consider
$E_8$ vacuum gauge fields, that break $E_8$ to the Standard Model $\mathrm{SU}
\left( 3 \right) \times \mathrm{SU} \left( 2 \right) \times U \left( 1 \right)$,
in such a way, that the values of the coupling constants, at unification, are
approximately equal to the observed values of the Standard Model coupling
constants, as evolved in the Standard Model, to around $142$ to 166 TeV.
However, if $\kappa^{- \frac{2}{9}}$ is around a TeV, it seems possible that
the higher dimensional accelerated unification mechanism of Dienes, Dudas, and
Gherghetta (DDG) {\cite{DDG1, DDG2}}, might perhaps reduce the unification
energy to not much larger than a TeV, since, provided $\left| \chi \left(
\mathcal{M}^6 \right) \right|$ is not too large, $b^{- 1}_1$ would also then
be not much larger than a TeV.  The supersymmetry in the higher dimensions,
required for the DDG mechanism to work, would of course automatically be
present, since it is only the compactification that breaks the supersymmetry,
in the models studied in the present paper.  The embedding of $\mathrm{SU}
\left( 3 \right)_{\mathrm{QCD}}$, in $E_8$, will be equivalent to the usual
embedding of $\mathrm{SU} \left( 3 \right)_{\mathrm{QCD}}$, in conventional Grand
Unification, as discussed in the preceding paragraph, so, assuming that the
DDG mechanism reduces the unification energy, without altering the unification
value of the coupling constants, I shall provisionally estimate $\alpha_U$ as
the value of the QCD fine structure constant, $\alpha_3$, as evolved to around
142 to 166 TeV, in the Standard Model, which gives the value:
\begin{equation}
  \label{alpha sub U in the Standard Model} \alpha_U \simeq 0.0602 \simeq
  \frac{1}{16.6}
\end{equation}

Robinson and Wilczek {\cite{Robinson Wilczek}} calculated the one loop
gravitational correction to the renormalization group running of the Standard
Model gauge coupling constants in a four-dimensional framework, and found
that, within the region of validity of their one loop result, the
gravitational correction reduces the magnitudes of the Yang-Mills gauge
coupling constants as energies are reached where quantum gravitational effects
become significant.  However this result is not directly applicable in the
present context, where higher dimensional effects and quantum gravitational
effects become significant together, so I shall not adjust the provisional
estimate (\ref{alpha sub U in the Standard Model}) for this effect.
Pietrykowski {\cite{Pietrykowski}} found that the Robinson-Wilczek effect is
gauge-dependent, and vanishes in a class of gauges different from the gauge
choice made by Robinson and Wilczek.

Thus when the compact six-manifold is a smooth compact quotient of
$\mathbf{C} \mathbf{H}^3$, we find:
\begin{equation}
  \label{b sub 1 in terms of chi} \frac{b_1}{\kappa^{2/9}} \simeq
  \frac{1.2772}{\left| \chi \left( \mathcal{M}^6 \right)
  \right|^{\frac{1}{6}}}
\end{equation}
And when the compact six-manifold is a smooth compact quotient of
$\mathbf{H}^6$, the coefficient $1.2772$ is replaced by $1.1809$.

\subsubsection{The problem of the higher order corrections to Ho\v{r}ava-Witten
theory}
\label{The higher order corrections to Horava-Witten theory}

At the inner surface of the thick pipe, where we live,
we must necessarily have $\lambda^2 \sim \left\vert \chi\left( \mathcal{M}^6
\right) \right\vert b_1^6$, up to factors of order 1, where $ b_1 = b\left( y_1
\right) $, and $ \chi\left( \mathcal{M}^6 \right) $ is the Euler number of the
compact six-manifold $ \mathcal{M}^6 $, which is an integer $ \leq -1 $.
Equivalently, from the Ho\v{r}ava-Witten relation (\ref{lambda kappa
relation}), we must have $ \kappa^{\frac{4}{3}} \sim \left\vert\chi\left(
\mathcal{M}^6 \right)\right\vert b_1^6 $, up to factors of order 1.  The
relation including all factors of order 1 is given in (\ref{b sub 1 in terms of
chi}), on page \pageref{b sub 1 in terms of chi}.  This
follows from the fact that the Yang-Mills coupling constant, $g$, in four
dimensions, at unification, is given by $g^2 \sim \frac{\lambda^2}{\left\vert
\chi \left( \mathcal{M}^6 \right) \right\vert b_1^6}$, up
to factors of order 1, and $g$, which is equal to the QCD coupling constant,
at unification, is of order 1.  Thus quantum effects must necessarily be
relevant, at the inner surface of the thick pipe, and, from (\ref{lambda
kappa relation}), also in the bulk, near the inner surface of the thick
pipe.

One type of quantum effect
has already been taken into account, namely the very existence of the
supersymmetric Yang-Mills multiplets, which, as discussed above, are required
to cancel the one-loop chiral anomalies of the gravitinos, which are chiral on
the orbifold fixed-point hyperplanes.  In order to consider what other relevant
quantum effects may occur, it is necessary to consider how Ho\v{r}ava-Witten
theory is defined, beyond the long distance limit.

Ho\v{r}ava-Witten theory was formally defined \cite{HW1, HW2} as
\emph{M}-theory on $ \mathcal{M}^{10} \times \mathbf{S}^1 / \mathbf{Z}_2 $,
where \emph{M}-theory is an unknown theory in eleven dimensions, whose defining
properties are that it is the strong coupling limit of type IIA superstring
theory \cite{Green Schwarz Witten}, and its low energy limit is supergravity in
eleven dimensions.  Ho\v{r}ava and Witten suggested that the theory would have
a built in short-distance cutoff, but left open the question of whether
the supermembrane in eleven dimensions \cite{supermembrane 1, supermembrane 2}
would play a role
in the physics of that short-distance cutoff, because at the time, it appeared
that, although the supermembrane contained the states of the
Cremmer-Julia-Scherk supergravity
multiplet in eleven dimensions \cite{Sezgin et al}, it could not
be consistently quantized.  The problem was that, due to supersymmetry, there
was no energy cost to deforming the shape of a membrane by drawing ``infinitely
thin'' tubes out from it, even when the zero point oscillations of the
thickness of the tubes were taken into account, and the spectrum was therefore
continuous \cite{de Wit Luscher Nicolai}.

However, the supermembrane has more recently been reinterpreted as a
second-quantized theory \cite{de Wit}, the idea being that little bubbles of
supermembrane, connected to one another by infinitely thin tubes, are like
independently moving single particles, with the sums over paths, of the
infinitely thin tubes connecting the bubble ``particles'', presumably building
up the eleven-dimensional analogue of the static Newtonian gravitational forces
between them.  Moreover, from section 12 of \cite{de Wit}, it is possible that
the supermembrane mass spectrum (in flat space) corresponds simply to the
single particle and multi-particle states of supergravity.

It is thus possible that the supermembrane is, in fact, a
kind of second quantized version of supergravity in eleven dimensions.  If this
is correct, there is then no known physical effect to provide the basis for any
difference, at the quantum level, between \emph{M}-theory and supergravity, on
a smooth background, in eleven uncompactified dimensions, because the classical
membrane \cite{classical membrane} and five-brane \cite{classical five brane}
solutions are infinitely massive, on a smooth background,
in eleven uncompactified dimensions, and thus do not take part in quantum
processes.

Now the classical membrane solution of $d = 11$ supergravity was reinterpreted
in {\cite{Duff Gibbons Townsend}} as a sourceless solitonic solution, with the
singularity at the origin found in {\cite{classical membrane}} being
reinterpreted as
a coordinate singularity at an event horizon, through which the solution can
be continued, although there is a curvature singularity hidden inside the
event horizon.  And by an argument of Hull and Townsend {\cite{Hull
Townsend}}, involving $U$-duality after toroidal compactification to four
dimensions, it is known that $M$-theory cannot contain a separate fundamental
supermembrane, distinct from the solitonic membrane of $d = 11$ supergravity.
This is consistent with the fact that the full dynamics of type IIA
superstring theory {\cite{type IIA superstring theory}} arises from the
solitonic membrane of the CJS theory, on toroidal compactification to ten
dimensions.  Let us recall how this works {\cite{HW1}}.

We first recall that, by a generalization of Dirac's argument {\cite{Dirac
magnetic monopole}}
for the quantization of the product of electric charge and the magnetic charge
of a magnetic monopole, the tensions $T_2$ and $T_5$, of a solitonic membrane
and a solitonic fivebrane, are constrained quantum mechanically to satisfy
{\cite{Nepomechie, Teitelboim, Duff Lu}}:
\begin{equation}
  \label{Dirac quantization condition for membrane and fivebrane} 2 \kappa^2
  T_2 T_5 = 2 \pi n, \quad n \in \mathbf{Z}
\end{equation}
Thus there must be a fundamental membrane tension, that is a numerical
multiple of $\kappa^{- \frac{2}{3}}$, and a fundamental fivebrane tension,
that is a numerical multiple of $\kappa^{- \frac{4}{3}}$, such that the
tensions of all solitonic membranes and solitonic fivebranes are constrained
quantum mechanically to be integer multiples of these fundamental tensions.  I
will confirm below, without reference to fivebranes, that the fundamental
membrane tension is $\frac{1}{2} \left( \frac{4 \pi}{\kappa}
\right)^{\frac{2}{3}}$, when the CJS action is $\frac{1}{\kappa^2}
\int_{\mathcal{M}^{11}} d^{11} x \sqrt{- g} \left( - \frac{1}{2} R + \ldots
\right)$, as in (\ref{upstairs bulk action}) for Ho\v{r}ava-Witten theory in the
``upstairs'' picture {\cite{Duff Liu Minasian, de Alwis 1}}.

Now since a solitonic membrane in eleven uncompactified dimensions is
infinitely extended and has a nonzero minimum tension, it is infinitely
massive, and cannot be produced in any physical process.  Solitonic membranes
of finite extent do not exist quantum mechanically in eleven uncompactified
dimensions, because a membrane of finite extent and tension $T_2$ would
contract into a region of size $T^{- \frac{1}{3}}_2$, the smallest size
allowed by the uncertainty principle, which for $T_2$ not smaller than the
fundamental membrane tension is comparable to or smaller than the thickness
$\sim \kappa^{\frac{1}{3}} T^{\frac{1}{6}}_2$ of the membrane {\cite{classical
membrane}}, so it would look like a lump rather than a membrane.  And while
such a lump could exist classically as a black hole, it has the wrong geometry
to be a source of $C_{IJK}$, so it cannot carry any charge to stabilize it, as
an extreme charged state of nonzero mass, against decay by Hawking radiation
\cite{Hawking Particle Creation by Black Holes},
so it will not lead to the existence of any massive single particle states in
the spectrum of the uncompactified CJS theory.

On the other hand, if a solitonic membrane of infinite extent, and tension
$T_2$, already exists in the vacuum, then its effective dynamics, at distances
$\gg$ both $\kappa^{2/9}$ and the thickness $\sim \kappa^{\frac{1}{3}}
T^{\frac{1}{6}}_2$ of the membrane, can be studied in terms of collective
coordinates, by deriving a worldvolume effective action for the membrane, by
the method of Callan, Harvey, and Strominger {\cite{Callan Harvey
Strominger}}.  The first step is the same as in studying a Kaluza-Klein
compactification, treating the dimensions parallel to the worldvolume of the
membrane as the ``extended'' dimensions, and the dimensions perpendicular to
the worldvolume of the membrane as the ``compact'' dimensions.  The CJS fields
are decomposed into blocks according to which of their tensor indices are
parallel to or perpendicular to the membrane, and the spinor index of the
gravitino is written as a pair of a two-valued $\mathrm{SO} \left( 2, 1 \right)$
spinor index and a sixteen-valued $\mathrm{SO} \left( 8 \right)$ spinor index.
Then all the fields are expanded in terms of a complete set of states on the
``compact'' dimensions, with coefficients that depend on position in the
``extended'' dimensions, or in other words, on position on the membrane
worldvolume.

The membrane thickness $\sim \kappa^{\frac{1}{3}} T^{\frac{1}{6}}_2$ now plays
the role of the size of the compact dimensions, and at distances large
compared to both $\kappa^{2/9}$ and $\kappa^{\frac{1}{3}}
T^{\frac{1}{6}}_2$, only the massless modes are dynamically significant.  The
massless modes correspond to the zero modes of the solitonic membrane, which
have been studied by Kaplan and Michelson {\cite{Kaplan Michelson}}.  The
solitonic membrane is a BPS solution of the CJS theory, so in accordance with
the general analysis of Callan, Harvey, and Strominger, 16 of the 32
supersymmetries of the CJS theory are realized linearly, as supersymmetries of
the world-sheet action, and the remaining 16 supersymmetries are realized
nonlinearly, as massless fermionic Goldstone modes.  Half of the fermionic
Goldstone modes vanish on the mass shell, so there are 8 bosonic Goldstone
modes, which are the 8 translational zero modes, corresponding to translations
of the membrane in the 8 directions perpendicular to the world sheet.

Choosing coordinates such that the membrane is in the $\left( 1, 2 \right)$
plane, let $x^{\mu}$, $0 \leq \mu \leq 2$ denote the coordinates on the
worldvolume, and $y^m$, $3 \leq m \leq 10$ denote the coordinates
perpendicular to the worldvolume.  Then an arbitrary diffeomorphism $y^m
\rightarrow y^m - \xi^m \left( y \right)$, with infinitesimal parameters
$\xi^m \left( y \right)$, generates a zero mode, by:
\begin{equation}
  \label{zero mode for g} \delta g_{IJ} = \xi^K \partial_K g_{IJ} + \left(
  \partial_I \xi^K \right) g_{KJ} + \left( \partial_J \xi^K \right) g_{IK}
\end{equation}
\begin{equation}
  \label{zero mode for C} \delta C_{IJK} = \xi^L \partial_L C_{IJK} + \left(
  \partial_I \xi^L \right) C_{LJK} + \left( \partial_J \xi^L \right) C_{ILK} +
  \left( \partial_K \xi^L \right) C_{IJL}
\end{equation}
In the right-hand sides, here, $g_{IJ}$ and $C_{IJK}$ are as given by the
classical membrane solution:
\begin{equation}
  \label{metric for classical membrane} ds^2 = \Lambda^{- \frac{2}{3}}
\eta_{\mu \nu} dx^{\mu} dx^{\nu} + \Lambda^{\frac{1}{3}} \delta_{m n} dy^m dy^n
\end{equation}
\begin{equation}
  \label{C for classical membrane} C_{\mu \nu \rho} = \pm \frac{\sqrt{2}}{12}
  \varepsilon_{\mu \nu \rho} \Lambda^{- 1}
\end{equation}
where
\begin{equation}
  \label{Lambda for classical membrane} \Lambda = 1 + \left( \frac{r_h}{\rho}
  \right)^6,
\end{equation}
$\rho \equiv \sqrt{y^m y^m}$, and the membrane thickness, $r_h$, is related to
the membrane tension $T_2$ by $r^6_h = \frac{\kappa^2 T_2}{3 \Omega_7}$, where
$\Omega_7$ is the volume of the unit seven-sphere $\mathbf{S}^7$.  The
horizon is located at $\rho = 0$ in these coordinates.

We note that the metric (\ref{metric for classical membrane}) tends to
Minkowski space as $\rho \rightarrow \infty$.  Let $\epsilon_{\left( i
\right)}^K \left( y \right)$ be a set of eight linearly independent vector
fields in the eight dimensions perpendicular to the membrane, (so
$\epsilon_{\left( i \right)}^{\mu} = 0$), such that $\lim_{\rho \rightarrow
\infty} \epsilon_{\left( i \right)}^m = \delta^m \, \!_i$, and such that when
any
of the $\epsilon_{\left( i \right)}^K$ is used as the diffeomorphism
parameter, $\xi^K$, in (\ref{zero mode for g}) and (\ref{zero mode for C}),
the corresponding zero modes $\delta g_{IJ}$ and $\delta C_{IJK}$ are
normalizable, in the sense that $\int d^8 y \sqrt{- g} g^{IK} g^{JM}
\delta g_{IJ} \delta g_{KM}$ and $\int d^8 y \sqrt{- g} g^{IL} g^{JM} g^{KN}
\delta C_{IJK} \delta C_{LMN}$ are finite,
where the integral extends over the region $\rho \geq 0$ outside the horizon.
Then when the small fluctuations of $g_{IJ}$ and $C_{IJK}$ are expanded as
\begin{equation}
  \label{expansion of small fluctuations for membrane} \delta g_{IJ} = \sum_i
  \lambda_{\left( i \right)} \left( x \right) \delta g_{IJ} \left(
  \epsilon_{\left( i \right)}, y \right), \hspace{2em} \delta C_{IJK} = \sum_i
  \lambda_{\left( i \right)} \left( x \right) \delta C_{IJK} \left(
  \epsilon_{\left( i \right)}, y \right)
\end{equation}
where $\delta g_{IJ} \left( \epsilon_{\left( i \right)}, y \right)$ and
$\delta C_{IJK} \left( \epsilon_{\left( i \right)}, y \right)$ are given by
(\ref{zero mode for g}) and (\ref{zero mode for C}), respectively, with
$\xi^K$ taken as $\epsilon_{\left( i \right)}^K \left( y \right)$, the
corresponding change of the CJS Lagrangian, defined as the integrand of the
Ho\v{r}ava-Witten bulk action in the ``upstairs'' picture, (\ref{upstairs bulk
action}), including the $\sqrt{- g}$ factor, has been calculated by Kaplan and
Michelson {\cite{Kaplan Michelson}} as:
\begin{equation}
  \label{change of L due to small fluctuations about membrane solution}
  \delta^2 \mathcal{L}= \quad \! \: \! \quad \! \: \! \quad \! \: \! \quad \!
  \: \! \quad \! \: \! \quad \! \: \! \quad \! \: \! \quad \! \: \! \quad \!
  \: \! \quad \! \: \! \quad \! \: \! \quad \! \: \! \quad \! \: \! \quad \!
  \: \! \quad \! \: \! \quad \! \: \! \quad \! \: \! \quad \! \: \! \quad \!
  \: \! \quad \! \: \! \quad \! \: \! \quad \! \: \! \quad \! \: \! \quad \!
  \: \! \quad \! \: \! \quad \! \: \! \quad \! \: \! \quad \! \: \! \quad \!
  \: \! \quad
\end{equation}
\[ = \frac{1}{\kappa^2} \left( \Lambda_{, m} \epsilon_{\left( i \right)}^m
   \partial_n \epsilon_{\left( j \right)}^n + \Lambda \partial_m
   \epsilon_{\left( i \right)}^m \partial_n \epsilon_{\left( j \right)}^n -
   \frac{1}{2} \Lambda \partial_m \epsilon_{\left( i \right)}^n \partial_m
   \epsilon_{\left( j \right)}^n - \frac{1}{2} \Lambda \partial_m
   \epsilon_{\left( i \right)}^n \partial_n \epsilon_{\left( j \right)}^m
   \right) \eta^{\mu \nu} \partial_{\mu} \lambda_{\left( i \right)}
   \partial_{\nu} \lambda_{\left( j \right)} \]
We see that $\delta^2 \mathcal{L}$ vanishes for constant $\epsilon_{\left( i
\right)}^m$, as expected, due to the global translation invariance of the
classical membrane solution, in the directions perpendicular to the membrane.
But constant $\epsilon_{\left( i \right)}^m$ do not lead to normalizable modes
$\delta g_{IJ} \left( \epsilon_{\left( i \right)}, y \right)$ and $\delta
C_{IJK} \left( \epsilon_{\left( i \right)}, y \right)$.  Let us try, instead,
$\epsilon_{\left( i \right)}^m = f \left(
\rho \right) \delta^m \, \!_i$, where $f
\left( \rho \right) \rightarrow 0$ as $\rho \rightarrow 0$, and $f \left( \rho
\right) \rightarrow 1$ as $\rho \rightarrow \infty$.  Then (\ref{change of L
due to small fluctuations about membrane solution}) becomes:
\begin{equation}
  \label{change of L with Kronecker delta ansatz} \delta^2 \mathcal{L}=
  \frac{1}{\kappa^2} \left( \frac{y_i y_j}{\rho^2} \left( \Lambda_{, \rho} f
  \partial_{\rho} f + \frac{1}{2} \Lambda \partial_{\rho} f \partial_{\rho} f
  \right) - \frac{1}{2} \Lambda \delta_{ij} \partial_{\rho} f \partial_{\rho}
  f \right) \eta^{\mu \nu} \partial_{\mu} \lambda_{\left( i \right)}
  \partial_{\nu} \lambda_{\left( j \right)}
\end{equation}
Thus after doing the angular integral over the $y_i$, we find:
\begin{equation}
  \label{change of CJS action due to small fluctuations about classical
  membrane} \delta^2 S_{\mathrm{CJS}} = - \frac{T_2}{8} \left( 1 + \frac{7}{6}
  \int^{\infty}_0 d \rho \rho \left( 1 + \frac{\rho^6}{r^6_h} \right)
  \partial_{\rho} f \partial_{\rho} f \right)
  \int d^3 x \hspace{0.4em} \eta^{\mu \nu} \partial_{\mu}
  \lambda_{\left( i \right)} \partial_{\nu} \lambda_{\left( i \right)}
\end{equation}
Thus with this ansatz for the $\epsilon_{\left( i \right)}^m \left( y
\right)$, the coefficient of $\eta^{\mu \nu} \partial_{\mu} \lambda_{\left( i
\right)} \partial_{\nu} \lambda_{\left( i \right)}$ in $\delta^2
S_{\mathrm{CJS}}$ is nonzero and has the correct sign.  Kaplan and Michelson
\cite{Kaplan Michelson} suggest that the uncertainty in the magnitude of the
coefficient should be absorbed into the definition of the $ \lambda_{\left( i
\right)} $.

Considering, now, the restrictions on the choice of $f \left( \rho \right)$
that result from the requirement that the zero modes are normalizable, so that
$\int d^8 y \sqrt{- g} g^{I K} g^{J M} \delta g_{I J} \delta g_{K M}$ and
$\int d^8 y \sqrt{- g} g^{I L} g^{J M} g^{K N} \delta C_{I J K} \delta C_{L M
N}$ are finite, we note that the $\rho$ integrals will certainly converge at
large $\rho$ if $f \left( \rho \right)$ tends to 1 rapidly enough as $\rho
\rightarrow \infty$. \ While for $\rho \rightarrow 0$, we find from (\ref{zero
mode for g}) and (\ref{zero mode for C}) that
$\int d^8 y \sqrt{- g} g^{\mu \sigma} g^{\nu \tau} \delta g_{\mu \nu} \delta
g_{\sigma \tau}$ and\\
$\int d^8 y \sqrt{- g} g^{\mu \rho} g^{\nu \tau}
g^{\sigma \lambda} \delta C_{\mu \nu \sigma} \delta C_{\rho \tau
\lambda}$\hfill lead\hfill to\hfill integrals\hfill of\hfill the\hfill
form\hfill $\int^{}_0 d \rho \rho^3 f^2$,\hfill while\linebreak
$\int d^8 y
\sqrt{- g} g^{j l} g^{k m} \delta g_{j k} \delta g_{l m}$ leads to integrals
of the forms $\int^{}_0 d \rho \rho^3 f^2$, $\int^{}_0 d \rho \rho^4 f
\partial_{\rho} f$, and $\int^{}_0 d \rho \rho^5 \left( \partial_{\rho} f
\right)^2$.

Thus we can choose $f = \left( \frac{\rho}{L} \right)^{1 / n}$ for $0 \leq
\rho \leq L$, and $f = 1$ for $\rho \geq L$, where $L > 0$ and $n \geq 1$. \
Then for $L \rightarrow 0$ and $n \rightarrow \infty$, $f \left( \rho \right)$
increases very rapidly from 0 to 1 in a small interval near $\rho = 0$, and
then stays equal to 1 for all larger $\rho$. \ In this limit, we see from
(\ref{zero mode for g}) and (\ref{zero mode for C}) that $\lambda_{\left( i
\right)}$ can be interpreted as $X^i$, the $x$-dependent transverse
displacement of the membrane. \ And with this choice of $f \left( \rho
\right)$, the integral in (\ref{change of CJS action
due to small fluctuations about classical membrane}) tends to $\frac{1}{2 n}$
as $L \rightarrow 0$, 
so for $L \rightarrow 0$ and $n \rightarrow \infty$, we find from (\ref{change
of CJS action due to small fluctuations about classical membrane}) that:
\begin{equation}
  \label{change of CJS action for L to 0 and n to infty} \delta^2
  S_{\mathrm{CJS}} = - \frac{T_2}{8} \int d^3 x \hspace{0.4em}
  \eta^{\mu \nu} \partial_{\mu} X^i \partial_{\nu} X^j \delta_{i j}
\end{equation}

We now use the fact
that the worldbrane effective action is completely determined by its
supersymmetries, up to an overall normalization factor.  From the general
principles discussed in {\cite{Hughes Polchinski}}, and the fact that the
classical solitonic membrane is a BPS solution that preserves half of the 32
supersymmetries, with the broken supersymmetries being realized nonlinearly as
Goldstone modes, it follows that the worldbrane effective action must be the
$d = 11$ supermembrane action found by Bergshoeff, Sezgin, and Townsend
{\cite{supermembrane 1, supermembrane 2}}.  However the supermembrane action
will be obtained in ``static'' gauge, as discussed in section 4 of
{\cite{supermembrane 2}}, such that Siegel's $\kappa$ symmetry {\cite{Siegel
kappa symmetry, Green Schwarz Covariant superstring action, Hughes Liu
Polchinski}} has been fixed by the gauge choice
$X^{\mu} = x^{\mu}$, where $X^I$ are the bosonic coordinates of the
supermembrane, and $x^{\mu}$ are the coordinates on the worldvolume, as above.

Reversing the gauge fixing of the $\kappa$ symmetry, and allowing a general
background satisfying the CJS field equations, the worldvolume effective
action of the infinitely extended classical solitonic membrane solution, that
describes its dynamics at distances large compared to both
$\kappa^{2/9}$ and the thickness $\sim \kappa^{\frac{1}{3}}
T^{\frac{1}{6}}_2$ of the solitonic membrane, is thus the $d = 11$
supermembrane action of {\cite{supermembrane 1}}.  The bosonic part of the
worldvolume action is then {\cite{classical membrane, Duff Liu Minasian}}:
\[ T_2 \int d^3 x \left( - \frac{1}{2} \sqrt{- \gamma} \gamma^{\mu \nu}
   \partial_{\mu} X^I \partial_{\nu} X^J G_{IJ} \left( X \right) + \frac{1}{2}
   \sqrt{- \gamma} \right. \quad \! \: \! \quad \! \: \! \quad \! \: \! \quad
   \! \: \! \quad \! \: \! \quad \! \: \! \quad \! \: \! \quad \! \: \! \quad
   \! \: \! \quad \]
\begin{equation}
  \label{bosonic part of supermembrane action} \quad \! \: \! \quad \! \: \!
  \quad \! \: \! \quad \! \: \! \quad \! \: \! \quad \! \: \! \quad \! \: \!
  \quad \! \: \! \quad \! \: \! \quad \! \: \! \quad \! \: \! \quad \! \: \!
  \quad \! \: \! \quad \! \: \! \quad \left. \pm \sqrt{2} \epsilon^{\mu \nu
  \sigma} \partial_{\mu} X^I \partial_{\nu} X^J \partial_{\sigma} X^K C_{IJK}
  \left( X \right) \right),
\end{equation}
where the sign choice in the third term in (\ref{bosonic part of supermembrane
action}), called the Wess-Zumino term, is the same as in (\ref{C for classical
membrane}) {\cite{classical membrane}}.  Here $\gamma^{\mu \nu}$ is the metric
on the worldvolume, and $X^I \left( x \right)$ are the bosonic coordinates of
the supermembrane, so that the $X^m \left( x \right)$, in (\ref{bosonic part of
supermembrane action}), should correspond to the $X^i \left( x \right)$
in (\ref{change of CJS action for L to 0 and n to infty}).  We note, however,
that there is a factor of $ \frac{1}{4} $ discrepancy between (\ref{change of
CJS action for L to 0 and n to infty}) and (\ref{bosonic part of supermembrane
action}), which I shall here leave unresolved.

We now note, following {\cite{Duff Liu Minasian}} and {\cite{de Alwis 1}},
that since $C_{IJK} \left( X \right)$ is not gauge-invariant, the worldvolume
action with the bosonic part (\ref{bosonic part of supermembrane action}) will
not lead to a well-defined quantum theory, unless changing the gauge of
$C_{IJK} \left( X \right)$ can only change the worldvolume action by an
integer multiple of $2 \pi$.  Let us consider a configuration such that $X^I
\left( x \right)$ sweeps out some closed three-dimensional surface
$\mathcal{S}_3$, as $x$ sweeps out some region $v$ of the worldvolume.  Then
the requirement that the worldvolume action leads to a well-defined quantum
theory implies that if $C'_{IJK}$ is any gauge transformation of $C_{IJK}$,
the integral
\begin{equation}
  \label{gauge change of Wess Zumino term over a three cycle} T_2 \int_v d^3 x
  \sqrt{2} \epsilon^{\mu \nu \sigma} \partial_{\mu} X^I \partial_{\nu} X^J
  \partial_{\sigma} X^K \left( C_{IJK} \left( X \right) - C'_{IJK} \left( X
  \right) \right) = 3! \sqrt{2} T_2 \int_{\mathcal{S}_3} \left( C - C' \right)
\end{equation}
must be an integer multiple of $2 \pi$.  Following {\cite{de Alwis 1}}, we now
apply this requirement with $\mathcal{S}_3$ being the equator of a topological
4-sphere $\mathcal{S}_4$, such that $C$ and $C'$ are the three-form gauge
field on coordinate patches that cover the north and south hemispheres of
$\mathcal{S}_4$ respectively.  Then from Stokes's theorem, and recalling that
the four-form $G$ was defined in section \ref{Horava-Witten theory} by
$G_{IJKL} = 24 \partial_{\left[ I \right.} C_{\left. JKL \right]}$, so that $G
= 6 dC$, it follows that $\sqrt{2} T_2 \int_{\mathcal{S}_4} G$ must be an
integer multiple of $2 \pi$.

On the other hand, by an argument of Witten {\cite{9609122 Witten}}, which I
review in subsection \ref{Stiffening by fluxes}, on page \pageref{Stiffening
by fluxes}, the vanishing of the Pontryagin number of $\mathcal{S}_4$ implies
that $\frac{\sqrt{2}}{4 \pi} \left( \frac{4 \pi}{\kappa} \right)^{\frac{2}{3}}
\int_{\mathcal{S}_4} G$ is an integer.  Thus the smallest possible nonzero
value of $\int_{\mathcal{S}_4} G$ is $\frac{4 \pi}{\sqrt{2}} \left(
\frac{\kappa}{4 \pi} \right)^{\frac{2}{3}}$.  The requirement that $\sqrt{2}
T_2 \int_{\mathcal{S}_4} G$ is an integer multiple of $2 \pi$ must be
satisfied, in particular, for this value of $\int_{\mathcal{S}_4} G$.  Hence
we find that
\begin{equation}
  \label{quantization condition for membrane tension} T_2 = \frac{n}{2} \left(
  \frac{4 \pi}{\kappa} \right)^{\frac{2}{3}}, \quad \! \: \! \quad n \in
  \mathbf{Z}
\end{equation}
which is in agreement with {\cite{Duff Liu Minasian}}, as amended by {\cite{de
Alwis 1}}.

Now the above argument for identifying the worldvolume effective action of the
infinitely extended classical solitonic membrane solution, that describes its
dynamics at distances large compared to both $\kappa^{2/9}$ and the
thickness $\sim \kappa^{\frac{1}{3}} T^{\frac{1}{6}}_2$ of the solitonic
membrane, as the $d = 11$ supermembrane action, also applies for
compactification on $\mathbf{S}^1$, when one dimension of the solitonic
membrane wraps the $\mathbf{S}^1$ and its other dimension extends
infinitely, and for compactification on $\mathbf{S}^1 \times
\mathbf{S}^1$, when the membrane wraps both $\mathbf{S}^1$'s, since these
are also BPS solutions of the CJS field equations, that preserve 16 of the 32
supersymmetries.  Furthermore, because these solutions are BPS, it is expected
that the semiclassical quantization of the worldvolume effective action will
be exact {\cite{Olive Witten}}, and will thus be valid even when the
circumference of one or both of the $\mathbf{S}^1$'s is small compared to
$\kappa^{2/9}$, which is a strong coupling limit for the CJS theory.
The semiclassical quantization of the supermembrane wrapping a two-torus was
studied in {\cite{Duff Inami Pope Sezgin Stelle}}.

Let us now consider compactification of the CJS theory on an $\mathbf{S}^1$
of circumference $L \ll \kappa^{2/9}$, such that a solitonic membrane
of the minimum tension $\frac{1}{2} \left( \frac{4 \pi}{\kappa}
\right)^{\frac{2}{3}}$ wraps once around the $\mathbf{S}^1$.  If we use the
same unit of length in ten dimensions as in eleven dimensions, the solitonic
membrane now looks like a string of tension $\frac{1}{\alpha'} = \frac{1}{2}
\left( \frac{4 \pi}{\kappa} \right)^{\frac{2}{3}} L$, while its thickness is
$\sim \kappa^{2/9}$, as in eleven dimensions.  The argument above for
the nonexistence quantum mechanically of solitonic membranes of finite extent
in eleven uncompactified dimensions no longer applies, because a string-like
solitonic membrane of finite extent would contract into a region of size $\sim
\sqrt{\alpha'} \sim \frac{\kappa^{\frac{1}{3}}}{\sqrt{L}}$, the smallest size
allowed by the uncertainty principle, which can be made arbitrarily large
compared to the string thickness $\sim \kappa^{2/9}$, by choosing $L$
sufficiently small compared to $\kappa^{2/9}$.  Furthermore, the
analogue of the Planck length, in ten dimensions, is $\left(
\frac{\kappa^2}{L} \right)^{\frac{1}{8}}$, and $\sqrt{\alpha'}$ can also be
made arbitrarily large compared to this, by choosing $L$ sufficiently small
compared to $\kappa^{2/9}$.  Thus the worldvolume effective action of
the classical solitonic membrane solution, namely the $d = 11$ supermembrane
action, can be made arbitrarily accurate at distances comparable to or larger
than $\sqrt{\alpha'}$, by choosing $L$ sufficiently small compared to
$\kappa^{2/9}$.

Now the compactification of the $d = 11$ supermembrane action on
$\mathbf{S}^1$, when the membrane also wraps the $\mathbf{S}^1$, was
calculated by Duff, Howe, Inami, and Stelle {\cite{Duff Howe Inami Stelle}},
and found to equal the covariant Green-Schwarz action for the superstring
{\cite{Green Schwarz Covariant superstring action}}.  Once the covariant
Green-Schwarz action is obtained, the standard spectrum of the type IIA
superstrings can be obtained in the light cone gauge {\cite{type IIA
superstring theory}}.

However, we still have to take account of the fact that the worldvolume
effective action of the classical solitonic membrane solution has only been
related to the $d = 11$ supermembrane action in the BPS configurations, so
each of the two dimensions of the solitonic membrane is either infinitely
extended or wraps a compactification $\mathbf{S}^1$.  So, following Ho\v{r}ava
and Witten {\cite{HW1}}, we consider compactification of the CJS theory on
$\mathbf{S}^1 \times \mathbf{S}^1$, such that a solitonic membrane of the
minimum tension $\frac{1}{2} \left( \frac{4 \pi}{\kappa}
\right)^{\frac{2}{3}}$ wraps once around each $\mathbf{S}^1$.  This is a BPS
solution of the CJS theory, so we can choose the circumference $L$ of one
$\mathbf{S}^1$ to be $\ll \kappa^{2/9}$.  Then the solitonic
membrane now looks like a closed string of tension $\frac{1}{\alpha'} =
\frac{1}{2} \left( \frac{4 \pi}{\kappa} \right)^{\frac{2}{3}} L$, wrapping
once around the second $\mathbf{S}^1$, whose radius $R$ we choose to be
comparable to or larger than $\sqrt{\alpha'} = \left( \frac{\kappa}{4 \pi}
\right)^{\frac{1}{3}} \sqrt{\frac{2}{L}}$.  Thus we now obtain the
Green-Schwarz action for the closed superstring wrapping once around the
second $\mathbf{S}^1$.

Going to light-cone gauge in the limit $L \ll \kappa^{2/9}$, the
oscillator degrees of freedom of the closed superstring completely decouple
from the wrapping degrees of freedom.  If the large $\mathbf{S}^1$ is in the
$X^9$ direction, then the bosonic coordinate $X^9$ of the superstring has the
standard expansion:
\begin{equation}
  \label{X 9 coordinate of closed superstring} X^9 = x^9 + \alpha' p^9 \tau +
  N_9 R \sigma + i \sqrt{\frac{\alpha'}{2}} \sum_{n \neq 0} \frac{1}{n} \left(
  \alpha_n^9 e^{- in \left( \tau - \sigma \right)} + \tilde{\alpha}_n^9 e^{-
  in \left( \tau + \sigma \right)} \right),
\end{equation}
where $\tau$ and $\sigma$, $0 \leq \sigma \leq 2 \pi$, are the timelike and
spacelike worldsheet coordinates of the closed superstring, $p^9 =
\frac{M_9}{R}$, for some integer $M_9$, and $N_9$ is the number of times the
closed superstring wraps the large $\mathbf{S}^1$, which is $1$ for the
configuration in which we have obtained the closed superstring.  The
oscillators $\alpha^n_{\mu}$ and $\tilde{\alpha}^n_{\mu}$, $n \neq 0$, satisfy
$\left[ \alpha_{\mu}^n, \alpha_{\nu}^m \right] = n \delta_{m, - n} \eta_{\mu
\nu}$ and $\left[ \tilde{\alpha}_{\mu}^n, \tilde{\alpha}_{\nu}^m \right] = n
\delta_{m, - n} \eta_{\mu \nu}$, because the semiclassical quantization of the
BPS solution is exact {\cite{Olive Witten}}.  And although we have only
obtained (\ref{X 9 coordinate of closed superstring}) in the case when $N_9 =
1$, the fact that the oscillators in (\ref{X 9 coordinate of closed
superstring}) are completely decoupled from the wrapping degrees of freedom
shows that these same oscillators also create the massive single particle
states of freely moving superstrings.  Applying the same treatment to the
fermionic collective coordinates of the solitonic membrane solution, we thus
obtain all the massive single-particle superstring states of the type IIA
supertring, while the massless single particle states arise from the
dimensional reduction of the $d = 11$ supergravity multiplet.

Finally, since $\sqrt{\alpha'} \sim \frac{\kappa^{\frac{1}{3}}}{\sqrt{L}}$
when we measure distances in ten dimensions in the same units as in eleven
dimensions, we should instead use a unit of length in ten dimensions that is
longer than the unit of length used in eleven dimensions by a factor
$\frac{\kappa^{\frac{1}{9}}}{\sqrt{L}}$, if we want to keep $\sqrt{\alpha'}$,
as measured in the new unit of length introduced for ten dimensions, fixed as
$\frac{\kappa^{2/9}}{L} \rightarrow \infty$ with $\kappa$ fixed.  This
can be implemented by writing the Kaluza-Klein ansatz for the $d = 11$ metric
as
\begin{equation}
  \label{Kaluza Klein ansatz to keep alpha primed fixed} ds^2 =
  \frac{\kappa^{2/9}}{L} g_{UV} dx^U dx^V + dy^2
\end{equation}
and interpreting the case where $g_{UV} = \eta_{UV}$ as ten-dimensional
Minkowski space.  Here $U$ and $V$ run from $0$ to $9$ as in subsection
\ref{Horava-Witten theory}, and $y$, the coordinate along the small
$\mathbf{S}^1$, runs from $0$ to $L$.  The metric $g_{UV}$ in (\ref{Kaluza
Klein ansatz to keep alpha primed fixed}) is called the string metric
{\cite{Witten Various dimensions}}, because it corresponds to choosing a unit
of length, in ten dimensions, with respect to which $\alpha' = \left(
\frac{\kappa}{4 \pi} \right)^{\frac{2}{3}} \frac{2}{L} \left(
\frac{\sqrt{L}}{\kappa^{\frac{1}{9}}} \right)^2 = \left( 2 \pi^2 \right)^{-
\frac{1}{3}} \kappa^{\frac{4}{9}}$ is independent of $L$.

Now since the Green-Schwarz action, which describes free superstrings, becomes
exact in the limit $\frac{L}{\kappa^{2/9}} \rightarrow 0$ with
$\kappa$ fixed, we expect that the string coupling constant $\lambda =
e^{\phi}$, where $\phi$ is the dilaton, should tend to zero in this limit.
This was demonstrated by Witten {\cite{Witten Various dimensions}}, by showing
that the string metric $g_{UV}$ in (\ref{Kaluza Klein ansatz to keep alpha
primed fixed}) is the correct metric to use for comparison with the low energy
effective action of the type IIA superstring, written in a standard form such
that the kinetic terms for the massless fields from the NS-NS sector include a
factor $\frac{1}{\lambda^2} = e^{- 2 \phi}$, and the kinetic terms for the
massless fields from the RR sector are independent of the dilaton.

A dynamical dilaton corresponds to the possibility that $L$, the circumference
of the small $\mathbf{S}^1$, can depend on position in the ten large
dimensions.  To allow for this possibility, we define $L =
\kappa^{2/9} e^{\gamma}$, where $\gamma$ can depend on $x^U$, and $y =
\frac{L}{\kappa^{2/9}} \tilde{y}$, so that $\tilde{y}$ runs from $0$
to $\kappa^{2/9}$.  The RR vector field $A_U$ is the Kaluza-Klein
vector field, and allowing also for a possible nonvanishing $A_U$, the $d =
11$ metric ansatz (\ref{Kaluza Klein ansatz to keep alpha primed fixed})
becomes:
\begin{equation}
  \label{KK ansatz with dilaton and KK vector} ds^2 = e^{- \gamma} g_{UV} dx^U
  dx^V + e^{2 \gamma} \left( d \tilde{y} - A_U dx^U \right)^2
\end{equation}
The massless NS-NS fields are the graviton, the dilaton, and $B_{UV} =
C_{UVy}$, and the other massless RR field, besides $A_U$, is $C_{UVW}$.
Substituting (\ref{KK ansatz with dilaton and KK vector}) into the CJS action
(\ref{upstairs bulk action}), the bosonic kinetic terms in the CJS action then
become schematically:
\begin{equation}
  \label{boson kinetic terms for KK ansatz} \sim
  \frac{1}{\kappa^{\frac{16}{9}}} \int d^{10} x \sqrt{- g} \left( e^{- 3
  \gamma} \left( R + \left( \partial \gamma \right)^2 + \left| dB \right|^2
  \right) + \left| dA \right|^2 + \left| dC \right|^2 \right)
\end{equation}
where this expression shows only the dependence on $\kappa$ and $\gamma$, not
the correct numerical coefficients of the terms.  Comparison with the low
energy effective action of the type IIA superstring, in the standard form
described above, then shows that the string coupling constant is given by
$\lambda = e^{\phi} \sim e^{\frac{3}{2} \gamma} =
\frac{L^{\frac{3}{2}}}{\kappa^{\frac{1}{3}}}$, and thus does, indeed, tend to
$0$ as $L \rightarrow 0$ with $\kappa$ fixed.

Thus the full dynamics of type IIA superstring theory is already contained in
the CJS theory of supergravity in eleven dimensions.  But since the defining
properties of $M$-theory are that it is the strong coupling limit of type IIA
superstring theory, and its low energy limit is supergravity in eleven
dimensions, there is then no detectable difference, on a smooth background,
between $M$-theory, and the CJS theory of supergravity in eleven dimensions.

In section 1.2 of {\cite{Green Russo Vanhove 0807}}, Green, Russo, and Vanhove
noted that on compactification of $d = 11$ supergravity on a 2-torus of radii
$r_A$ and $r_B$, terms of the form $e^{- cr_B}$ that arise in the string
theory 4-graviton amplitude are not reproduced by Feynman diagrams at any
number of loops. \ However for the case $r_B \ll \kappa^{2 / 9}$ that they
consider, the $d = 11$ solitonic membrane wrapping $r_B$ can form finite mass
solitonic closed strings with mass proportional to $r_B$, that would give
terms of this form by propagating as internal lines of the Feynman diagrams. \
The contribution of these solitonic closed strings to the 4-graviton amplitude
could presumably be calculated, for example, by the collective coordinate
techniques developed by Gervais, Jevicki, and Sakita {\cite{Gervais Jevicki
Sakita 1, Gervais Jevicki Sakita 2, Gervais Jevicki 1, Gervais Jevicki 2}}.

Type IIA superstring theory is thought to be UV complete {\cite{DHoker
Phong 1, DHoker Phong 2, DHoker Phong 3, DHoker Phong 4, DHoker Phong 5,
DHoker Phong 6, DHoker Phong 7, DHoker Phong 8}}, so apart from the factor of
$\frac{1}{4}$ discrepancy between (\ref{change of CJS action for L to 0 and n
to infty}) and (\ref{bosonic part of supermembrane action}) that I left
unresolved, the CJS theory of $d = 11$ supergravity, with the non-perturbative
effects of the classical membrane and 5-brane solutions properly included where
appropriate, appears to contain the full dynamics of the UV complete type IIA
superstring theory. \ However the CJS theory has been argued to be UV
incomplete {\cite{Bern Dixon Dunbar Julia Perelstein Rozowsky Seminara
Trigiante, Deser nonrenormalizability arguments 1, Deser nonrenormalizability
arguments 2, Stelle}}, on the basis of the existence of the linearized 4-field
counterterms of dimensions $8, 12, 14, 16, \ldots,$ constructed by Deser and
Seminara {\cite{Deser Seminara 1, Deser Seminara 2, Deser Seminara 3}}, which
have been proved by Metsaev {\cite{Metsaev}} to be the complete set of
linearized 4-field
counterterms, and the existence of an infinite set of counterterms
{\cite{Duff Toms}} constructed as integrals over the full $d = 11$ superspace
{\cite{Cremmer Ferrara, Brink Howe}}, together with a 2-loop dimensional
regularization calculation {\cite{Bern Dixon Dunbar Julia Perelstein Rozowsky
Seminara Trigiante}}, using the methods developed earlier in {\cite{Bern Dixon
Dunbar Perelstein Rozowsky, Bern Dixon Dunbar Grant Perelstein Rozowsky}},
which found that the dimension 20 Deser-Seminara linearized 4-field
counterterm would occur with an infinite coefficient.

Green, Vanhove, Kwon, and Russo have found that the coefficients of some local
counterterms of dimensions \(\geq 12\) in the $d = 11$ theory are fixed by
calculations in the type II superstring theories {\cite{Green Kwon Vanhove,
Green Vanhove 2, Green Russo Vanhove 1, Green
Russo Vanhove 0807}}, so the paradox of the UV-incomplete CJS theory
containing the full dynamics of the UV-complete type IIA superstring theory
cannot be resolved by ambiguities in the UV completion of the CJS theory,
which would arise as undetermined coefficients of the Deser-Seminara
and superspace counterterms of dimension \(\geq 12\) in the quantum effective
action of the CJS theory,
somehow disappearing during the compactification of the CJS theory on a small
circle to obtain the type IIA theory.

A possible resolution of half of the paradox, that appears to be consistent
with all known results, follows from noting that the Noether completion of the
Deser-Seminara linearized 4-field invariants, to fully non-linear
counterterms, invariant under the full non-linear CJS supersymmetry
variations, up to terms which vanish when the CJS field equations are
satisfied, and can thus be cancelled by the addition of higher dimension terms
to the CJS supersymmetry variations {\cite{Hyakutake Ogushi 1,
Hyakutake Ogushi 2}}, has never been carried out, and with the exception of
the unique dimension 8 invariant {\cite{Metsaev, Hyakutake}}, whose Noether
completion must exist, if $M$-theory is consistent, because it occurs in the
quantum effective action of $ d = 11 $ supergravity with a non-zero coefficient
that is fixed by the tangent bundle anomaly cancellation on five-branes
{\cite{Duff Liu Minasian, Witten Five Brane, Freed Harvey
Minasian Moore, Bilal Metzger 2, Harvey}}, and confirmed by anomaly
cancellation in Ho\v{r}ava-Witten theory {\cite{de Alwis 2,Conrad,Faux Lust
Ovrut,Lu,Bilal Derendinger Sauser,Harmark,Bilal Metzger 2,Meissner
Olechowski,Moss 3}}, and by comparison with types IIA and IIB superstring
theory \cite{Green Vanhove 1, Green Gutperle Vanhove}, it is possible that
their Noether completions do not exist.

In the case of $d = 4$, $N = 1$ supergravity {\cite{Freedman van Nieuwenhuizen
Ferrara, Deser Zumino}}, Noether completions were always found to exist, but
this follows from the existence of the auxiliary field formulations
{\cite{Breitenlohner, Stelle West Auxiliary fields, Ferrara van Nieuwenhuizen
Auxiliary fields, Ferrara van Nieuwenhuizen Tensor calculus, Sohnius West 1,
Sohnius West 2}}. \ However for the CJS theory in 11 dimensions, Rivelles and
Taylor showed that no similar auxiliary field formulation can exist
{\cite{Rivelles Taylor}}. \ An example of an obstruction to Noether completion
in 11 dimensions was found by Nicolai, Townsend, and van Nieuwenhuizen, when
they tried to construct an analogue of the CJS theory using a 6-form gauge
field instead of a 3-form gauge field {\cite{Nicolai Townsend van
Nieuwenhuizen}}.

A possible resolution to the other half of the paradox would be obtained if
the candidate counterterms constructed as integrals over the full $ d = 11 $
superspace \cite{Duff Toms, Howe Tsimpis} all vanished identically, or
alternatively, if an obstruction existed that prevented the geometrical
transformations in superspace {\cite{Cremmer Ferrara, Brink Howe}} from
matching the CJS supersymmetry variations, for a general solution of the CJS
field equations, beyond a certain power of $\theta$.
The mapping of the component fields and supersymmetry variations of a
supersymmetric theory into superspace, such that the geometrical
transformations in superspace match the supersymmetry variations of the
component fields, is called gauge completion \cite{Brink Gell-Mann Ramond
Schwarz, van Nieuwenhuizen Ferrara}, and for the CJS theory, this was 
initially carried out only to leading order in $\theta$ \cite{Cremmer Ferrara}.

The first terms beyond leading order in the gauge completion mapping of the
CJS theory into superspace were studied by de Wit, Peeters, and Plefka
{\cite{de Wit Peeters Plefka}}, and to consider whether an obstruction to
gauge completion appears in their results, I shall temporarily adopt their
notation.  Thus for the following discussion of {\cite{de Wit Peeters
Plefka}}, coordinate indices $\mu, \nu, \rho, \ldots$ will temporarily run
over all eleven bosonic coordinate directions, $r$ and $s$ are bosonic tangent
space indices, and $\alpha, \beta, \gamma, \ldots$ are fermionic coordinate
indices.  The relations between the normalizations of the fields are
$\psi^{\mathrm{HW}}_{\mu} = 2 \psi^{\mathrm{dWPP}}_{\mu}$, $C^{\mathrm{HW}}_{
\mu \nu \rho} = \frac{1}{6 \sqrt{2}} C^{\mathrm{dWPP}}_{\mu \nu \rho}$, and $
G_{\mu \nu \rho \sigma} = \frac{1}{\sqrt{2}} F_{\mu \nu \rho \sigma}$.

The first place to look for an obstruction is equation (4.5) of {\cite{de Wit
Peeters Plefka}}, which must be satisfied by the terms proportional to the
supercovariant field strength $\hat{F}_{\mu \nu \rho \sigma} = 4
\partial_{\left[ \mu \right.} C_{\left. \nu \rho \sigma \right]} + 12
\bar{\psi}_{\left[ \mu \right.} \Gamma_{\nu \rho} \psi_{\left. \sigma
\right]}$, in the conventions of {\cite{de Wit Peeters Plefka}}, at order
$\theta^2$ in the $\theta$ expansion of the superspace diffeomorphism
parameter.  Denoting these terms by $\epsilon^{\beta} N_{\beta} \,
\!^{\alpha}$, where $\epsilon^{\beta}$ is the parameter of a CJS local
supersymmetry variation, that is to be matched by a combination of superspace
diffeomorphisms, local Lorentz transformations, and possibly also gauge
transformations of a superspace three-form superfield, if one is included,
equation (4.5) of {\cite{de Wit Peeters Plefka}} reads:
\[ \epsilon^{\beta}_2 \partial_{\beta} N_{\gamma} \, \!^{\alpha}
   \epsilon^{\gamma}_1 - \left( \bar{\theta} \Gamma^{\mu} \epsilon_2 \right)
   \left( T_{\mu} \, \!^{\nu \rho \sigma \lambda} \epsilon_1 \right)^{\alpha}
   \hat{F}_{\nu \rho \sigma \lambda} - \left( 1 \leftrightarrow 2 \right) =
   \quad \! \: \! \quad \! \: \! \quad \! \: \! \quad \! \: \! \quad \! \: \!
   \quad \! \: \! \quad \! \: \! \quad \! \: \! \quad \]
\begin{equation}
  \label{equation for N gamma alpha} \quad \! \: \! \quad \! \: \! \quad \! \:
  \! \quad \! \: \! \quad \! \: \! \quad \! \: \! \quad \! \: \! \quad = -
  \frac{1}{288} \left( \Gamma_{rs} \theta \right)^{\alpha} \bar{\epsilon}_2
  \left( \Gamma^{rs} \, \!_{\nu \rho \sigma \lambda} + 24 e_{\nu} \, \!^r
  e_{\rho} \, \!^s \Gamma_{\sigma \lambda} \right) \epsilon_1 \hat{F}^{\nu
  \rho \sigma \lambda}
\end{equation}
Here $\partial_{\beta} = \frac{\partial}{\partial \theta^{\beta}}$, and
$T_{\mu} \, \!^{\nu \rho \sigma \lambda} = \frac{1}{288} \left( \Gamma_{\mu}
\, \!^{\nu \rho \sigma \lambda} - 8 \delta^{\left[ \nu \right.}_{\mu}
\Gamma^{\left. \rho \sigma \lambda \right]} \right)$.  This equation is to be
satisfied for arbitrary local supersymmetry variation parameters $\epsilon_1$
and $\epsilon_2$, and is thus a three-index equation for a two-index quantity.
Thus it will have no solution, unless the ``source'' terms satisfy an
appropriate integrability condition.  In fact, from the identity:
\begin{equation}
  \label{integrability identity} \partial_{\varepsilon} \left(
  \partial_{\beta} N_{\gamma} \, \!^{\alpha} + \partial_{\gamma} N_{\beta} \,
  \!^{\alpha} \right) + \partial_{\beta} \left( \partial_{\gamma}
  N_{\varepsilon} \, \!^{\alpha} + \partial_{\varepsilon} N_{\gamma} \,
  \!^{\alpha} \right) + \partial_{\gamma} \left( \partial_{\varepsilon}
  N_{\beta} \, \!^{\alpha} + \partial_{\beta} N_{\varepsilon} \, \!^{\alpha}
  \right) = 0,
\end{equation}
which follows from the fact that the spinor derivatives anticommute, we find
that integrability of (\ref{equation for N gamma alpha}) requires that the
following expression vanish for arbitrary $\hat{F}_{\nu \rho \sigma \lambda}$:
\[ - 2 \left( \Gamma^0 \Gamma^{\mu} \right)_{\varepsilon \beta} \left( T_{\mu
   \nu \rho \sigma \lambda} \right)^{\alpha} \, \!_{\gamma} \hat{F}^{\nu \rho
   \sigma \lambda} - 2 \left( \Gamma^0 \Gamma^{\mu} \right)_{\gamma
   \varepsilon} \left( T_{\mu \nu \rho \sigma \lambda} \right)^{\alpha} \,
   \!_{\beta} \hat{F}^{\nu \rho \sigma \lambda} \]
\[ - 2 \left( \Gamma^0 \Gamma^{\mu} \right)_{\beta \gamma} \left( T_{\mu \nu
   \rho \sigma \lambda} \right)^{\alpha} \, \!_{\varepsilon} \hat{F}^{\nu \rho
   \sigma \lambda} + \frac{1}{288} \left( \Gamma^{\mu \kappa} \right)^{\alpha}
   \, \!_{\varepsilon} \left( \Gamma^0 \left( \Gamma_{\mu \kappa \nu \rho
   \sigma \lambda} + 24 g_{\mu \nu} g_{\kappa \rho} \Gamma_{\sigma \lambda}
   \right) \right)_{\beta \gamma} \hat{F}^{\nu \rho \sigma \lambda} \]
\[ + \frac{1}{288} \left( \Gamma^{\mu \kappa} \right)^{\alpha} \, \!_{\beta}
   \left( \Gamma^0 \left( \Gamma_{\mu \kappa \nu \rho \sigma \lambda} + 24
   g_{\mu \nu} g_{\kappa \rho} \Gamma_{\sigma \lambda} \right) \right)_{\gamma
   \varepsilon} \hat{F}^{\nu \rho \sigma \lambda} \]
\begin{equation}
  \label{expression required to vanish for arbitrary F} + \frac{1}{288} \left(
  \Gamma^{\mu \kappa} \right)^{\alpha} \, \!_{\gamma} \left( \Gamma^0 \left(
  \Gamma_{\mu \kappa \nu \rho \sigma \lambda} + 24 g_{\mu \nu} g_{\kappa \rho}
  \Gamma_{\sigma \lambda} \right) \right)_{\varepsilon \beta} \hat{F}^{\nu
  \rho \sigma \lambda}
\end{equation}
Thus integrability of (\ref{equation for N gamma alpha}) requires the
following expression, which is antisymmetric in $\nu$, $\rho$, $\sigma$, and
$\lambda$, and symmetric in $\varepsilon$, $\beta$, and $\gamma$, to vanish
identically:
\[ - 2 \left( \Gamma^0 \Gamma^{\mu} \right)_{\varepsilon \beta} \left(
   \Gamma_{\mu \nu \rho \sigma \lambda} - 2 g_{\mu \nu} \Gamma_{\rho \sigma
   \lambda} + 2 g_{\mu \rho} \Gamma_{\sigma \lambda \nu} - 2 g_{\mu \sigma}
   \Gamma_{\lambda \nu \rho} + 2 g_{\mu \lambda} \Gamma_{\nu \rho \sigma}
   \right)^{\alpha} \, \!_{\gamma} \]
\[ - 2 \left( \Gamma^0 \Gamma^{\mu} \right)_{\gamma \varepsilon} \left(
   \Gamma_{\mu \nu \rho \sigma \lambda} - 2 g_{\mu \nu} \Gamma_{\rho \sigma
   \lambda} + 2 g_{\mu \rho} \Gamma_{\sigma \lambda \nu} - 2 g_{\mu \sigma}
   \Gamma_{\lambda \nu \rho} + 2 g_{\mu \lambda} \Gamma_{\nu \rho \sigma}
   \right)^{\alpha} \, \!_{\beta} \]
\[ - 2 \left( \Gamma^0 \Gamma^{\mu} \right)_{\beta \gamma} \left( \Gamma_{\mu
   \nu \rho \sigma \lambda} - 2 g_{\mu \nu} \Gamma_{\rho \sigma \lambda} + 2
   g_{\mu \rho} \Gamma_{\sigma \lambda \nu} - 2 g_{\mu \sigma} \Gamma_{\lambda
   \nu \rho} + 2 g_{\mu \lambda} \Gamma_{\nu \rho \sigma} \right)^{\alpha} \,
   \!_{\varepsilon} \]
\[ + \left( \Gamma^{\mu \kappa} \right)^{\alpha} \, \!_{\varepsilon} \left(
   \Gamma^0 \left( \Gamma_{\mu \kappa \nu \rho \sigma \lambda} + 4 g_{\mu \nu}
   g_{\kappa \rho} \Gamma_{\sigma \lambda} + 4 g_{\mu \sigma} g_{\kappa
   \lambda} \Gamma_{\nu \rho} + 4 g_{\mu \rho} g_{\kappa \sigma} \Gamma_{\nu
   \lambda} + 4 g_{\mu \nu} g_{\kappa \lambda} \Gamma_{\rho \sigma} \right.
   \right. \]
\[ \left. \left. + 4 g_{\mu \sigma} g_{\kappa \nu} \Gamma_{\rho \lambda} + 4
   g_{\mu \rho} g_{\kappa \lambda} \Gamma_{\sigma \nu} \right) \right)_{\beta
   \gamma} + \left( \Gamma^{\mu \kappa} \right)^{\alpha} \, \!_{\beta} \left(
   \Gamma^0 \left( \Gamma_{\mu \kappa \nu \rho \sigma \lambda} + 4 g_{\mu \nu}
   g_{\kappa \rho} \Gamma_{\sigma \lambda} + 4 g_{\mu \sigma} g_{\kappa
   \lambda} \Gamma_{\nu \rho} \right. \right. \]
\[ \left. \left. + 4 g_{\mu \rho} g_{\kappa \sigma} \Gamma_{\nu \lambda} + 4
   g_{\mu \nu} g_{\kappa \lambda} \Gamma_{\rho \sigma} + 4 g_{\mu \sigma}
   g_{\kappa \nu} \Gamma_{\rho \lambda} + 4 g_{\mu \rho} g_{\kappa \lambda}
   \Gamma_{\sigma \nu} \right) \right)_{\gamma \varepsilon} + \left(
   \Gamma^{\mu \kappa} \right)^{\alpha} \, \!_{\gamma} \left( \Gamma^0 \left(
   \Gamma_{\mu \kappa \nu \rho \sigma \lambda} \right. \right. \]
\begin{equation}
  \label{expression built from Gamma matrices that is required to vanish}
  \left. \left. + 4 g_{\mu \nu} g_{\kappa \rho} \Gamma_{\sigma \lambda} + 4
  g_{\mu \sigma} g_{\kappa \lambda} \Gamma_{\nu \rho} + 4 g_{\mu \rho}
  g_{\kappa \sigma} \Gamma_{\nu \lambda} + 4 g_{\mu \nu} g_{\kappa \lambda}
  \Gamma_{\rho \sigma} + 4 g_{\mu \sigma} g_{\kappa \nu} \Gamma_{\rho \lambda}
  + 4 g_{\mu \rho} g_{\kappa \lambda} \Gamma_{\sigma \nu} \right)
  \right)_{\varepsilon \beta}
\end{equation}
This expression (\ref{expression built from Gamma matrices that is required to
vanish}) is the type of expression that might vanish by a Fierz identity.  To
find out whether or not it vanished, I used the fact, reviewed for example in
{\cite{Miemiec Schnakenburg}}, that for a real representation of the $d = 11$
Dirac matrices, as assumed here, the 1024 matrices $\left( \Gamma_{\tau_1
\ldots \tau_n} \right)^{\gamma} \, \!_{\alpha}$, $0 \leq n \leq 5$, form a
complete basis for real $32 \times 32$ matrices.  We can therefore find out
whether or not (\ref{expression built from Gamma matrices that is required to
vanish}) vanishes, by contracting it with a general matrix $X^{\gamma} \,
\!_{\alpha}$, which turns it into an ordinary sum of matrices, with indices
$\varepsilon \beta$ or $\beta \varepsilon$, multiplied, in the case of the
first five terms and the last seven terms, by a trace, and then taking
$X^{\gamma} \, \!_{\alpha}$ to be each of these 1024 matrices in turn.

However, due to Lorentz invariance, it is not necessary to take $X^{\gamma} \,
\!_{\alpha}$ to be all 1024 of these matrices.  Instead, we first note that
(\ref{expression built from Gamma matrices that is required to vanish})
vanishes by antisymmetry, unless $\nu$, $\rho$, $\sigma$, and $\lambda$ are
all different.  Thus it is sufficient to evaluate (\ref{expression built from
Gamma matrices that is required to vanish}) for a fixed choice of $\nu$,
$\rho$, $\sigma$, and $\lambda$, all different from each other.  I chose $\nu
= 0$, $\rho = 8$, $\sigma = 9$, and $\lambda = y$, where, as throughout this
section, $y$ denotes the tenth spatial direction.  We then find, when we
choose $X^{\gamma} \, \!_{\alpha}$ to be a matrix $\left( \Gamma_{\tau_1
\ldots \tau_n} \right)^{\gamma} \, \!_{\alpha}$, for any specific value of
$n$, and any specific values for the indices $\tau_1, \tau_2, \ldots, \tau_n$,
that each term in (\ref{expression built from Gamma matrices that is required
to vanish}) is equal to a coefficient, times either the matrix $\left(
\Gamma^0 \Gamma_{\kappa_1 \ldots \kappa_m} \right)_{\varepsilon \beta}$ or the
matrix $\left( \Gamma^0 \Gamma_{\kappa_1 \ldots \kappa_m} \right)_{\beta
\varepsilon}$, where $\kappa_1, \ldots, \kappa_m$ are the indices in $\left\{
\nu, \rho, \sigma, \lambda \right\}$, that are not in $\left\{ \tau_1, \tau_2,
\ldots, \tau_n \right\}$, and the indices in $\left\{ \tau_1, \tau_2, \ldots,
\tau_n \right\}$ that are not in $\left\{ \nu, \rho, \sigma, \lambda
\right\}$, and may be taken in ascending order.

Furthermore, due to the symmetry of (\ref{expression built from Gamma matrices
that is required to vanish}) in $\beta$ and $\varepsilon$, the result vanishes
automatically, unless $m$ is one of the numbers for which the matrix $\left(
\Gamma^0 \Gamma_{\kappa_1 \ldots \kappa_m} \right)_{\beta \varepsilon}$ is
symmetric, namely 1, 2, 5, 6, 9, and 10.  Furthermore, for each value of $n$,
$0 \leq n \leq 5$, it is sufficient to consider just one choice of the indices
$\left\{ \tau_1, \tau_2, \ldots, \tau_n \right\}$ that gives each of these
values of $m$, since if the result vanishes for one choice, it will also
vanish for any other choice that gives the same value of $m$.

I chose $X$ to be the ten matrices $\Gamma_1$, $\Gamma_{08}$, $\Gamma_{12}$,
$\Gamma_{89 y}$, $\Gamma_{12 y}$, $\Gamma_{189 y}$, $\Gamma_{123 y}$,
$\Gamma_{1089 y}$, $\Gamma_{1239 y}$, and $\Gamma_{12345}$.  For each of these
ten choices of $X$, the contraction of (\ref{expression built from Gamma
matrices that is required to vanish}) with $X^{\gamma} \, \!_{\alpha}$ was
found to vanish.  The expression (\ref{expression built from Gamma matrices
that is required to vanish}) is therefore identically zero, so this potential
obstruction to the completion of the gauge completion procedure, at order
$\theta^2$, in fact vanishes.

However, this does not yet imply that there is no obstruction to completion of
the gauge completion procedure at order $\theta^2$, because the spin-spin
components of the supervielbein also contain terms of order $\hat{F}
\theta^2$, and these terms are required to satisfy equation (4.9) of {\cite{de
Wit Peeters Plefka}}, which is again a three-index equation for a two-index
quantity, and thus will have a nontrivial integrability condition, since it is
required to be satisfied for arbitrary supersymmetry variation parameter
$\epsilon$.  The ``source'' terms of equation (4.9) of {\cite{de Wit Peeters
Plefka}} include terms similar in structure, although different in detail,
from the source terms in equation (4.5) of {\cite{de Wit Peeters Plefka}},
reproduced as equation (\ref{equation for N gamma alpha}) above, and also a
term involving the solution $N_{\beta} \, \!^{\alpha}$ of equation (4.5) of
{\cite{de Wit Peeters Plefka}}, which cannot be eliminated by use of equation
(4.5) of {\cite{de Wit Peeters Plefka}}, because it does not occur in the
combination $\left( \partial_{\gamma} N_{\beta} \, \!^{\alpha} +
\partial_{\beta} N_{\gamma} \, \!^{\alpha} \right) \!$.  However $N_{\beta} \,
\!^{\alpha}$ does occur in the combination $\left( \partial_{\gamma} N_{\beta}
\, \!^{\alpha} + \partial_{\beta} N_{\gamma} \, \!^{\alpha} \right) $ in the
integrability condition for equation (4.9) of {\cite{de Wit Peeters Plefka}},
so that integrability condition could be checked by substituting for $\left(
\partial_{\gamma} N_{\beta} \, \!^{\alpha} + \partial_{\beta} N_{\gamma} \,
\!^{\alpha} \right) \!$ from equation (4.5) of {\cite{de Wit Peeters Plefka}},
without actually solving equation (4.5) of {\cite{de Wit Peeters Plefka}}, but
I will not do that in this paper.

The evaluation of (\ref{expression built from Gamma matrices that is required
to vanish}), contracted with each of the ten choices of $X^{\gamma} \,
\!_{\alpha}$ listed above, was speeded up by use of the well-known identities
{\cite{Miemiec Schnakenburg}}:
\begin{equation}
  \label{first identity from Miemiec Schnakenburg} \Gamma^{\mu} \Gamma_{\nu_1
  \ldots \nu_n} \Gamma_{\mu} = \left( - 1 \right)^n \left( d - 2 n \right)
  \Gamma_{\nu_1 \ldots \nu_n}
\end{equation}
\begin{equation}
  \label{second identity from Miemiec Schnakenburg} \Gamma^{\mu \sigma}
  \Gamma_{\nu_1 \ldots \nu_n} \Gamma_{\mu \sigma} = - \left( \left( d - 2 n
  \right)^2 - d \right)^{} \Gamma_{\nu_1 \ldots \nu_n}
\end{equation}
valid in $d$ dimensions.  For example, to evaluate the term $\left( \Gamma^0
\Gamma_{\mu \kappa 089 y} \Gamma_{1239 y} \Gamma^{\mu \kappa} \right)_{\beta
\varepsilon}$, which arises for the choice $X = \Gamma_{1239 y}$, we note that
we can treat $\mu$ and $\kappa$ here as summed only over the seven dimensions
different from $0$, $8$, $9$, and $y$.  So we split $\Gamma_{1239 y}$ as
$\Gamma_{123} \Gamma_{9 y}$ and commute the $\Gamma_{9 y}$ to the right, and,
with the understanding that $\mu$ and $\kappa$ are summed only over the range
$1$ to $7$, we also split $\Gamma_{\mu \kappa 089 y}$ as $\Gamma_{089 y}
\Gamma_{\mu \kappa}$.  We then use the identity (\ref{second identity from
Miemiec Schnakenburg}) above, with $d = 7$ and $n = 3$, to obtain:
\begin{equation}
  \label{example use of Miemiec Schnakenburg identity} \left( \Gamma^0
  \Gamma_{\mu \kappa 089 y} \Gamma_{1239 y} \Gamma^{\mu \kappa} \right)_{\beta
  \varepsilon} = - \left( \left( 7 - 6 \right)^2 - 7 \right) \left( \Gamma^0
  \Gamma_{089 y} \Gamma_{123} \Gamma_{9 y} \right)_{\beta \varepsilon} = 6
  \left( \Gamma^0 \Gamma_{01238} \right)_{\beta \varepsilon}.
\end{equation}

We see from above that already at order $\theta^2$, the possibility of mapping
the CJS theory into superspace such that the geometrical transformations in
superspace match the CJS supersymmetry variations, for a general solution of
the CJS field equations, requires that nontrivial integrability conditions be
satisfied.  Thus it is not possible to conclude, from the construction of a
counterterm in standard $d = 11$ superspace, that there exists a corresponding
higher derivative term, local in the CJS component fields, whose variation
under the CJS supersymmetry transformations is a total derivative when the CJS
field equations are satisfied, without explicitly checking that there are no
nonvanishing obstructions to the gauge completion mapping of the CJS theory
into superspace, up to the highest power of $\theta$ that occurs in the
superspace counterterm.  For the Duff-Toms superspace counterterms {\cite{Duff
Toms}} that would be $\theta^{32}$.  In the pure spinor framework of Berkovits
{\cite{0201151 Berkovits, 0612021 Berkovits}}, there are superspace invariants
involving an integration over only nine components of $\theta$, but it would
be necessary to check that there are no nonvanishing obstructions when the
pure spinor constraint is satisfied, at least through order $\theta^9$.

Turning now to the occurrence of fractional powers of $ \kappa $, in the
expansion of the quantum effective action $ \Gamma $,
$ \Gamma $ is formally given by an expansion in powers of $ \kappa^2 $,
starting with the classical action, of order $ \kappa^{ -2 } $, followed by
the one-loop term, which is formally independent of $ \kappa $.  However, it
is inevitable that other powers of $ \kappa $ will occur, especially if
$ \Gamma $, which is a non-local functional of the fields, is developed in a
low energy expansion, as a series of local terms, with increasing numbers of
derivatives on the fields.  Indeed, the bulk Green-Schwarz term, mentioned
at the end of subsection \ref{Horava-Witten theory}, which occurs in such an
expansion, is a sum of terms formed from a three-form gauge field, and four
Riemann tensors, with their indices contracted in various ways, using the
metric, and one antisymmetric eleven-tensor, and is of order $ \kappa^{
-\frac{ 2 }{ 3 } } $.  As already noted, if there had been a built-in
short distance
cutoff, of order $ \kappa^{2/9} $, then such fractional powers of
$ \kappa $ would have been interpreted as arising from powers of the short
distance cutoff.  But we now need to understand where they come from, when
there is no short distance cutoff.

\begin{figure}[t]
\setlength{\unitlength}{1cm}
\begin{picture}(11.85,7.86)(-3.0,1.0)
\put(6.28,7.81){\line(-1,0){3.89}}
\put(2.39,7.81){\line(-2,-5){1.25}}
\put(1.13,4.67){\line(5,-4){3.42}}
\put(4.55,1.94){\line(1,1){2.91}}
\put(7.45,4.84){\line(-2,5){1.18}}
\put(7.66,4.69){$ \varphi_{I_2 J_2} $}
\put(6.47,7.99){$ \varphi_{I_3 J_3} $}
\put(1.99,8.06){$ \varphi_{I_4 J_4} $}
\put(0.134,4.54){$ \varphi_{I_5 J_5} $}
\put(6.15,3.00){$ -11 $}
\put(7.07,6.24){$ -11 $}
\put(3.97,8.09){$ -11 $}
\put(0.87,6.29){$ -11 $}
\put(1.99,2.90){$ -11 $}
\put(1.73,4.75){$ x_5 $}
\put(4.29,2.69){$ x_1 $}
\put(6.50,4.87){$ x_2 $}
\put(5.61,7.08){$ x_3 $}
\put(2.66,7.10){$ x_4 $}
\put(4.34,1.34){$ C_{I_1 J_1 K_1} $}
\end{picture}
\caption{A pentagon contributing to the bulk Green-Schwarz term}
\label{pentagon for bulk Green-Schwarz term}
\end{figure}

Figure \ref{pentagon for bulk Green-Schwarz term} shows a typical Feynman
diagram, in the loop expansion of $ \Gamma $,
that can contribute to the bulk Green-Schwarz term.  It has a three-form
gauge field propagating in the loop, and the  $ C_{ I_1 J_1 K_1 } $
vertex comes from the Cremmer-Julia-Scherk Chern-Simons term in
(\ref{upstairs bulk action}), while the $ \varphi_{ I_i J_i } $ vertices
come from the three-form gauge field kinetic term, with the metric expanded
as $ G_{ I J } = \eta_{ I J } + \varphi_{ I J } $.  Each propagator has two
derivatives acting on it, one from the vertex at each end of it, so that,
for purposes of power counting, the line between two neighbouring vertices,
say $ x_i $ and $ x_j $, behaves as $ \left| x_i - x_j \right|^{ -11 } $.
On the basis of power counting, there is a logarithmic divergence whenever
any $ n $ consecutive vertices, such that $ 2 \leq n \leq 4 $, cluster
together, but the position space integral is in fact conditionally convergent
in these regions, and these apparent divergences, associated with tree
subdiagrams, can be dealt with by the method used to prove Theorem 2 of
\cite{BPHZ in EPS}.  However, the diagram as a whole has degree of divergence
11, so that, if we choose the three-form gauge field vertex, $ x_1 $, as the
contraction point of the diagram, then, in the BPHZ framework \cite{Dyson 1,
Dyson 2, Salam 1, Salam 2, Bogoliubov Parasiuk 1, Bogoliubov
Parasiuk 2, Bogoliubov Shirkov, Hepp, Weinberg convergence, Hahn
Zimmermann, Zimmermann Minkowski, Zimmermann convergence},
we have to subtract a counterterm, which, in this instance, has the form of
the ``internal function'' of the diagram, namely the propagators, with the
derivatives acting on them out of the vertices, times the terms, of degree
up to and including degree 11, of
the Taylor expansion of the ``external function'' of the diagram, namely the
function $ C_{I_1J_1K_1}\left(x_1\right)\varphi_{I_2J_2}\left(x_2\right)
\varphi_{I_3J_3}\left(x_3\right)\varphi_{I_4J_4}\left(x_4\right)
\varphi_{I_5J_5}\left(x_5\right)$, about the point $ \left(x_1,x_2,x_3,
x_4,x_5\right)=\left(x_1,x_1,x_1,x_1,x_1\right) $.  After integrating over
$ x_2 $, $ x_3 $, $ x_4 $, and $ x_5 $, this counterterm includes
terms with the structure of the bulk Green-Schwarz term, although with a
divergent coefficient, as well as many other terms.

Let us now consider a term of degree 11 in this counterterm, which has a total
of eleven derivatives acting on $ \varphi_{I_2J_2}\left(x_1\right)
\varphi_{I_3J_3}\left(x_1\right)\varphi_{I_4J_4}\left(x_1\right)
\varphi_{I_5J_5}\left(x_1\right) $, and a total of eleven factors, in the
counterterm integrand, of the form $ \left(x_{r}-x_1\right)_{I_r} $, where
the index $ r $ runs from 1 to 11, and $ x_{r} $ is one of $ x_2 $, $ x_3 $,
$ x_4 $, or $ x_5 $.  Suppose we now integrate over the vertex positions, in
the sequence $ x_2 $, $ x_3 $, $ x_4 $, then $ x_5 $.  We see that, when we
come to integrate over the position of $ x_5 $, the counterterm has an
uncancelled logarithmic divergence at \emph{large} distances, in consequence
of the
masslessness of the propagators.  There was no such large distance divergence
at all, in the original diagram, if the classical fields, $ C_{IJK} $, and
$ \varphi_{IJ} $, are assumed to fall off sufficiently rapidly, at large
distances.

Such large distance divergences, occurring in BPHZ counterterms,
but not in the original diagrams, are a well-known problem of BPHZ
renormalization, when there are massless particles.  Traditionally, the
problem has beeen dealt with by the BPHZL method \cite{Lowenstein 1,
Lowenstein 2, Lowenstein Zimmermann, Lowenstein 3}, which involves
the introduction of regulator masses for the massless particles, performing
additional infra-red subtractions, in addition to the short-distance
subtractions, then letting the regulator masses tend to zero, at the end of
the calculation.  An alternative method was presented in \cite{BPHZ in EPS},
where a generalized BPHZ convergence proof was presented, in Euclidean
position space, that allowed the propagators in the counterterms to differ,
at large distances, from the propagators in the original diagram, without
altering the propagators in the uncontracted diagram.  This
enables massless propagators, in the counterterms, to be cut off smoothly,
at large distances, so that the large distance divergences are eliminated
from the counterterms, without spoiling the convergence proof, and without
altering the propagators in the uncontracted diagram.  The proof
in \cite{BPHZ in EPS} applies only in Euclidean signature position space,
but it seems plausible that Hepp's convergence proof \cite{Hepp}, which can
be applied in Minkowski signature, could be
generalized in an analogous way, allowing the parameter integrals, of the
exponentiated propagators, to be cut off at large values of the
exponentiation parameters, in the counterterms, without altering them in
the terms coming from the uncontracted diagram.

When this method is used for a theory such as massless QCD, with no dimensional
parameters in the classical action, the distance at which the smooth
long distance cutoffs of the propagators in the counterterms begin, becomes
the distance that provides the basis for dimensional transmutation
\cite{Gross Neveu}.  In the
case of supergravity in eleven dimensions, the classical action has precisely
one parameter with the dimension of length, namely $ \kappa^{ \frac{2}{9} } $,
so the distance,
at which the smooth long distance cutoffs of the propagators in the
counterterms begin, will be a numerical multiple of $ \kappa^{ \frac{2}{9} } $.

Now the convergence proof in \cite{BPHZ in EPS} assumed that the same modified
propagators, differing at long distances from the propagators in the terms
coming from the uncontracted diagram, are used in all the terms of the Taylor
expansions that occur in the counterterms, so we will also be using these same
modified propagators, with a long distance cutoff commencing at some fixed
numerical multiple of $ \kappa^{ \frac{2}{9} } $, in those terms in the Taylor
expansions in the counterterms, where this is not actually needed, to ensure
convergence at large distances.  However, there is not, a priori, any reason
to choose any \emph{particular} numerical multiple of $ \kappa^{ \frac{2}{9} }
$, as the distance at which the smooth long distance cutoffs of the propagators
in the counterterms begin, and if we choose a
different numerical multiple of $ \kappa^{ \frac{2}{9} } $, the result will
change by the addition of local finite counterterms, whose coefficients will
involve powers of $ \kappa $, as determined by dimensional analysis.  In
particular, the term with the structure of the bulk Green-Schwarz term, which
contains eight derivatives, will include a factor of $ \kappa^{ -\frac{2}{3} }
$.  Then, when we require that the Slavnov-Taylor identities \cite{Slavnov,
Taylor, Zinn Justin}, which follow from local supersymmetry, in the BRST-BV
framework \cite{Becchi Rouet Stora, Tyutin, Batalin Vilkovisky 1, Batalin
Vilkovisky 2, Batalin Vilkovisky 3, Batalin Vilkovisky 4, Batalin Vilkovisky
5}, are satisfied,
and impose appropriate gauge-fixing conditions,
and use the freedom to redefine the fields, in order to set to zero the
coefficients of terms that vanish, when the classical field equations are
satisified, the coefficients of the possible finite counterterms will be fixed,
up to the addition of linear combinations of terms, that correspond to
nontrivial locally supersymmetric higher-derivative deformations, of the
CJS theory.

And as I discussed above, it is possible, and consistent with all
known results, that the only non-trivial higher-derivative deformation of the
CJS theory, that is locally supersymmetric at the full non-linear level, might
be the deformation whose lowest-dimension term is the unique dimension-8
CJS on-shell invariant \cite{Metsaev, Hyakutake} that contains the bulk
Green-Schwarz term.

The numerical coefficient of the unique dimension-8 CJS on-shell invariant
{\cite{Metsaev, Hyakutake}}, in the quantum effective action of $ d = 11 $
supergravity, is fixed by the tangent bundle anomaly cancellation on
five-branes {\cite{Duff Liu Minasian, Witten Five Brane, Freed Harvey
Minasian Moore, Bilal Metzger 2, Harvey}}, and confirmed by anomaly
cancellation in Ho\v{r}ava-Witten theory {\cite{de Alwis 2,Conrad,Faux Lust
Ovrut,Lu,Bilal Derendinger Sauser,Harmark,Bilal Metzger 2,Meissner
Olechowski,Moss 3}}, and by comparison with types IIA and IIB superstring
theory \cite{Green Vanhove 1, Green Gutperle Vanhove}.  This in turn depends
on the Dirac quantization of the two-brane and five-brane tensions
\cite{Dirac magnetic monopole, Nepomechie, Teitelboim, Duff Lu, Duff Khuri Lu,
Duff Liu Minasian, Schwarz 9509148, Schwarz 9510086, de Alwis 1, Lu}.

Thus if the unique dimension-8 CJS on-shell invariant is the only non-trivial
higher-derivative CJS on-shell invariant that is locally supersymmetric at the
full non-linear level, it might be possible to calculate the predictions of the
CJS theory and Ho\v{r}ava-Witten theory in the framework of effective field
theory, without the occurrence of undetermined parameters connected with
the short distance completion of the theory.

\subsubsection{The Casimir energy density corrections to the energy-momentum
tensor}
\label{The Casimir energy density corrections}

Having now considered some of the problems involved in the definition of
Ho\v{r}ava-Witten theory, or more specifically, the bulk \emph{M}-theory aspect
of it, beyond the long-wavelength limit, I shall now consider the Casimir-type
effects resulting from the compactification on the compact six-manifold.

The Casimir
corrections to the energy-momentum tensor, in Einstein's equations,
arise from the variation of the one loop, and higher loop terms, in the
quantum effective action, $\Gamma$, with respect to the classical metric,
$G_{IJ}$.  In general, these terms give corrections to the classical Einstein
equations, that are non-local functionals of the classical metric, $G_{IJ}$.
However, for a given classical metric $G_{IJ}$, the Casimir terms in the
energy-momentum tensor will be specific functions of position.  We can
therefore adopt an iterative approach to solving the quantum-corrected Einstein
equations, calculating the Casimir terms in a trial classical metric $G_{IJ}$,
then solving the Einstein equations with these Casimir terms, and if the
resulting ``output'' metric differs from the ``input'' metric, repeating the
process with an improved ``input'' metric, until agreement is reached.  This
method will be used, at the level of rough order of magnitude estimates, in
subsection \ref{The region near the inner surface of
the thick pipe}, on page \pageref{The region near the inner surface of the
thick pipe}.

The classical metric $G_{IJ}$ will not, in general, be a solution of the
classical field equations, in regions where the Casimir corrections to the
energy-momentum tensor are significant.  Nevertheless the gauge-fixed quantum
effective action, $\Gamma$, is still well defined, up to possible ultraviolet
divergences, as the generating functional of proper vertices \cite{Schwinger,
DeWitt effective action, Nambu Jona Lasinio 1, Nambu Jona Lasinio 2}.  Moreover
it can be calculated, for a classical action $A \left( \varphi \right)$, and
for an arbitrary classical field configuration $\Phi$, as the sum of all the
one-line-irreducible vacuum diagrams, calculated from the action $A \left(\Phi
+ \varphi \right)$, with the term linear in $\varphi$ deleted, where $\varphi$
denotes the quantum fields.  In other words, using DeWitt's compact index
notation {\cite{DeWitt}}, where a single index, $i$, runs over all
combinations of type of field, space-time position, and coordinate and other
indices, the quantum effective action, as a function of the classical fields,
$\Phi$, is given by the sum of all the one line irreducible vacuum diagrams,
calculated with the action:
\begin{equation}
  \label{action for calculating quantum effective action} A \left( \Phi +
  \varphi \right) - \varphi_i \frac{\delta A \left( \Phi \right)}{\delta
  \Phi_i}
\end{equation}
where the summation convention is applied to the index $i$.  The derivation of
this result is reviewed in section \ref{The Casimir energy densities}, on page
\pageref{The Casimir energy densities}.

I shall look for solutions such that all physical quantities are covariantly
constant in directions tangential to the four observed dimensions, which is
consistent with the choice of the de Sitter metric for the four observed
dimensions, in the metric ansatz (\ref{metric ansatz}).  The compactification
of $ \mathbf{CH}^3 $ or $ \mathbf{H}^6 $ to the compact six-manifold $
\mathcal{M}^6 $ usually breaks the homogeneity of the hyperbolic space, so the
Casimir terms in the energy-momentum tensor will not, in general, be
covariantly constant in directions tangential to $ \mathcal{M}^6 $.
Furthermore, in the models considered in section \ref{E8 vacuum gauge fields
and the Standard Model}, on page \pageref{E8 vacuum gauge fields and the
Standard Model}, there are topologically stabilized vacuum Yang-Mills fields on
the inner surface of the thick pipe, with non-vanishing field strengths, whose
contributions to the energy-momentum tensor explicitly break covariant
constancy in directions tangential to $ \mathcal{M}^6 $.  However, following
Lukas, Ovrut, and Waldram {\cite{Lukas Ovrut Waldram}}, we can introduce a
harmonic expansion on the compact six-manifold.  I shall work throughout this
section at the level of the leading term in such a harmonic expansion of the
energy-momentum tensor, which I shall assume has the form:
\begin{equation}
  \label{T IJ block diagonal structure} T_{\mu \nu} = t^{\left( 1 \right)}\!
  \left( y \right) G_{\mu \nu}, \hspace{6.0ex} T_{AB} = t^{\left( 2 \right)}\!
  \left( y \right) G_{AB}, \hspace{6.0ex} T_{yy} = t^{\left( 3 \right)}\!
  \left( y \right)
\end{equation}

Using the expressions (\ref{Christoffel symbols for the metric ansatz}), on
page \pageref{Christoffel symbols for the metric ansatz}, for the
non-vanishing Christoffel symbols of the second kind, the conservation
equation, $ D_I T^{ I J } = 0 $, for the energy-momentum tensor, now
reduces to:
\[
0 = D_I T^{ I y } = \partial_y T^{ y y } + \left( \Gamma^{ \mu }_{ \mu y }
+ \Gamma^{ A }_{ A y } \right) T^{ y y } + \Gamma^{ y }_{ \mu \nu }
T^{ \mu \nu } + \Gamma^{ y }_{ A B } T^{ A B } =
\]
\begin{equation}
\label{conservation equation for the t i}
 = \partial_y t^{\left(3\right)} + \left( 4 \frac{\dot{a}}{a} + 6 \frac{
\dot{b}}{b} \right)t^{\left(3\right)} - 4 \frac{\dot{a}}{a}t^{\left(1\right)}
- 6 \frac{\dot{b}}{b}t^{\left(2\right)}
\end{equation}

We will find that for thick pipe geometries that realize TeV-scale gravity by
the ADD mechanism \cite{ADD1, ADD2}, the energy-momentum tensor, including the
contributions of the four-form field strength $ G_{IJKL} $ of the three-form
gauge field, is negligible in the main part of the bulk, well away from the
boundaries.  Thus the Einstein equations in the main part of the bulk will,
indeed, be consistent with all physical quantities being covariantly constant
on $ \mathcal{M}^6 $.  We note that when $ \mathcal{M}^6 $ is a smooth compact
quotient of $ \mathbf{CH}^3 $, there will be $ h^{1,1} - 1 $ additional
harmonic $ \left(1,1\right) $-forms on $ \mathcal{M}^6 $ besides the K\"ahler
form, but only the K\"ahler form will be covariantly constant.  The K\"ahler
moduli do not correspond to massless modes because, just as for any
K\"ahler-Einstein metric with a nonvanishing Ricci scalar, each K\"ahler
modulus is equal to a fixed multiple of the corresponding element of the first
Chern class.  It seems reasonable to expect that the effects of the higher
harmonics in the Lukas-Ovrut-Waldram harmonic expansion will decrease rapidly
relative to the effects of the leading harmonic, as the distance from the
nearest boundary increases, so that the effects of the higher harmonics
will not be significant, in the main part of the bulk.

The functions $t^{\left( i \right)} \left( y \right)$ in (\ref{T IJ block
diagonal structure}) will be significant near the inner surface of the thick
pipe, where $b \left( y \right)$ is $\sim \kappa^{2/9}$.  I shall
consider three alternative ways in which the outer surface of the thick pipe
might be stabilized, consistent with the observed value (\ref{Lambda}) of the
effective $d = 4$ cosmological constant, and in one of the three alternatives,
$a \left( y \right)$ is $\sim \kappa^{2/9}$ near the outer surface, so
in that case, which is studied in subsection \ref{Solutions with a as small as
kappa to the two ninths at the outer surface}, on page \pageref{Solutions with
a as small as kappa to the two ninths at the outer surface}, the
$t^{\left( i \right)} \left( y \right)$ will also be significant near the
outer surface.

To calculate the quantum effective action $\Gamma$, and the
functions $t^{\left( i \right)} \left( y \right)$, for a particular classical
metric (\ref{metric ansatz}), the propagators and heat kernels for the $d =
11$ supergravity fields, and also for the Fadeev-Popov ghosts
\cite{Feynman, DeWitt ghosts, Faddeev Popov, t Hooft ghosts}, the ghosts for
ghosts for the three-form gauge field \cite{Namazie Storey, Townsend ghosts
for ghosts, de Wit van Nieuwenhuizen Van Proeyen}, and possible
Nielsen-Kallosh ghosts \cite{Nielsen, Kallosh ghosts}, are needed for that
metric.  These can be obtained from the corresponding propagators and heat
kernels on an uncompactified $\mathbf{C} \mathbf{H}^3$ or $\mathbf{H}^6$
background, as appropriate, with the same $a \left( y \right)$ and $b \left( y
\right)$, by the sum over images method of M\"uller, Fagundes, and Opher
\cite{Muller Fagundes Opher 1, Muller Fagundes, Muller Fagundes Opher 2},
provided the sum over images converges.

For the case of a massless scalar,
the sum over images marginally converges when the action of the massless
scalar is as simple as possible, with no ``conformal improvement'' term, but
diverges exponentially, due to the exponential growth of volume with distance,
when a ``conformal improvement term'' is added to the action, to make the
classical energy-momentum tensor traceless.  If the sum over images diverges
for any of the required propagators or heat kernels, it might be possible to
obtain the result by a resummation method \cite{Berry Keating, Agam
Fishman}, or a theta function method \cite{Poincare, Ewald, Crandall}.

The propagators and heat kernels on a flat $\mathbf{R}^5$ times uncompactified
$\mathbf{C} \mathbf{H}^3$ or $\mathbf{H}^6$ background can be obtained from
the corresponding propagators and heat kernels on a flat $\mathbf{R}^5$ times
$\mathbf{C} \mathbf{P}^3$ or $\mathbf{S}^6$ background, which can be
calculated by using the Salam-Strathdee harmonic expansion method {\cite{Salam
Strathdee}}, and summing the expansions by means of a generating function.
This calculation is currently in progress for $\mathbf{C} \mathbf{H}^3$, and
the scalar heat kernel on $\mathbf{C} \mathbf{H}^3$, obtained by this method,
is presented in subsection \ref{The Salam Strathdee harmonic expansion
method}, on page \pageref{The Salam Strathdee harmonic expansion method}.
The leading terms at short distances in the propagators and heat kernels have
been calculated for all the relevant fields on general smooth backgrounds by
Burgess and Hoover {\cite{Burgess Hoover 1, Burgess Hoover 2}}, using the heat
kernel expansion {\cite{Schwinger proper time method, DeWitt}}.
Casimir effects for compactification on hyperbolic quotients have also been
studied in \cite{Bytsenko Odintsov Zerbini, Bytsenko Cognola Vanzo Zerbini}.

Considering, now, the form of the functions $t^{\left( i \right)} \left( y
\right)$ near the inner surface of the thick pipe, where $b \sim
\kappa^{2/9}$, we note that the low energy expansion of the $M$-theory
quantum effective action, $\Gamma$, is known to contain local terms formed
from the Riemann tensor and its covariant derivatives.  The first such term is
formed from four Riemann tensors, and usually referred to as the $ t_8 t_8 R^4
$ term \cite{Green Vanhove 1, Green Gutperle Vanhove, Green Gutperle Kwon,
Green Kwon Vanhove, Peeters Vanhove Westerberg, Hyakutake Ogushi 1, Green
Vanhove 2, Hyakutake Ogushi 2, Green Russo Vanhove 1, Green Russo Vanhove 2},
where $ t^{I_1 I_2 J_1 J_2 K_1 K_2 L_1 L_2}_8 $ denotes the tensor obtained
from $\left( - 6 g^{I_1 J_2} g^{J_1 I_2} g^{K_1 L_2} g^{L_1 K_2} + 24 g^{I_1
J_2} g^{J_1 K_2} g^{K_1 L_2} g^{L_1 I_2} \right)$ by antisymmetrizing under
$I_1 \rightleftharpoons I_2$, and symmetrizing under all permutations of $
\left(I,J,K,L\right) $, with total weight one.

Recalling the definition (\ref{energy momentum tensor}), on page
\pageref{energy momentum tensor}, of the energy-momentum tensor, and looking
at the Riemann tensor components (\ref{Riemann tensor for the metric ansatz}),
on page \pageref{Riemann tensor for the metric ansatz}, for the metric ansatz
(\ref{metric ansatz}), we see that near the inner surface of the thick pipe,
the $t_8 t_8 R^4$ term will result in terms in the $t^{\left( i \right)}$
functions that are numerical multiples of $\frac{\kappa^{-
\frac{2}{3}}}{b^8}$, where the origin of the non-integer power of $\kappa$, in
the framework of effective field theory, was explained in the preceding
subsection, and there will also be terms where $\frac{\kappa^{-
\frac{2}{3}}}{b^8}$ is multiplied by up to four powers of $\dot{b}^2$ or $b
\ddot{b}$.

We will find in subsection \ref{The three form gauge field}, on page
\pageref{The three form gauge field}, that the vacuum configurations of the
three-form gauge field $ C_{IJK} $, that result, due to the Ho\v{r}ava-Witten
modified Bianchi identity (\ref{Bianchi identity with FF only}), on page
\pageref{Bianchi identity with FF only}, from the presence of
topologically stabilized vacuum Yang-Mills fields on the Ho\v{r}ava-Witten
orbifold hyperplanes, with non-vanishing field strengths tangential to the
compact six-manifold $ \mathcal{M}^6 $, also produce terms in the $t^{\left(
i \right)}$ functions that are numerical multiples of $\frac{\kappa^{-\frac{2}{
3}}}{b^8}$, and in this case, there are no additional terms involving
derivatives of $ b $ with respect to $ y $.

Calculations of Casimir energy effects often make use of the proximity force
approximation {\cite{proximity force 1, proximity force 2}}, which in the
present case would correspond to treating $b$ as independent of $y$, so that
all terms with factors of $\dot{b}$, $\ddot{b}$, or higher derivatives of $b$
with respect to $y$, could be neglected.  Thus in this approximation the $y$
direction would effectively be uncompactified, so that $t^{\left( 3 \right)}$
would be equal to $t^{\left( 1 \right)}$, and $t^{\left( 1 \right)}$ and
$t^{\left( 2 \right)}$ would correspond to $d = 11$ supergravity on flat
$\mathbf{R}^5$ times $\mathcal{M}^6$.  In this case, the first terms
dependent on the topology of $\mathcal{M}^6$ would be the one-loop
contributions from the terms in the sum over images other than the identity
term.  None of these terms contain short-distance divergences, so their
contributions to $\Gamma$ are independent of $\kappa$.  The corresponding
terms in the $t^{\left( i \right)}$ functions are thus numerical multiples of
$\frac{1}{b^{11}}$.

Several different indications have been found \cite{Green Vanhove 1, Green
Gutperle Vanhove, Green Gutperle Kwon, Green Kwon Vanhove, Green Vanhove
2, Green Russo Vanhove 1, E10, E11}, that suggest that the canonical dimensions
of non-vanishing terms, in the low-energy expansion of the \emph{M}-theory
quantum effective action $ \Gamma $, in eleven uncompactified dimensions, will
have the form $ 2\left(3k+1\right) $, for integer $ k $, or in other words, $
2,8,14, \ldots \hspace{0.4ex}$.  In that case, the next powers of $ b $, whose
coefficients, in the $t^{ \left( i \right) }$ functions, can get contributions
from local terms in the low-energy expansion of the quantum effective action,
in the context of the proximity force approximation, will be $ b^{-14} $ and $
b^{-20} $.  Neither of these terms would be expected to get contributions at
one loop, but both could get contributions at two loops.

It thus seems reasonable to assume that, within the context of the proximity
force approximation, the functions $t^{ \left( i \right) }$, near the inner
surface of the thick pipe, have expansions of the form:
\begin{equation}
  \label{the t i as functions of b} t^{\left( i \right)} = C^{\left( i
  \right)}_0 \frac{\kappa^{- \frac{2}{3}}}{b^8} + C^{\left( i \right)}_1
  \frac{1}{b^{11}} + C^{\left( i \right)}_2
  \frac{\kappa^{\frac{2}{3}}}{b^{14}} + \ldots
\end{equation}
where the $C^{ \left( i \right) }_n$ are numerical constants, that depend only
on the topology and spin structure of the compact six-manifold $ \mathcal{M}^6
$, and on the topologically stabilized configurations of the Yang-Mills fields
on the Ho\v{r}ava-Witten orbifold hyperplanes, and of the three-form gauge
field $ C_{IJK} $ in the bulk.

The conservation equation (\ref{conservation equation for the t i}) takes a
particularly simple form, near the inner surface of the thick pipe, when
$t^{\left( 3 \right)} = t^{\left( 1 \right)}$, and the $ t^{\left( i \right)} $
depend only on $ b $, as in the context of the proximity force approximation.
Specifically, when $t^{\left( 3 \right)} = t^{\left( 1 \right)}$:
\begin{equation}
  \label{conservation equation when t 3 equals t 1} \frac{dt^{\left( 1
  \right)}}{db} + \frac{6}{b} t^{\left( 1 \right)} - \frac{6}{b} t^{\left( 2
  \right)} = 0
\end{equation}
Hence, in this case:
\begin{equation}
  \label{C 2 and C 3 when t 3 equals t 1} C_n^{\left( 2 \right)} = -
  \frac{\left( 2 + 3 n \right)}{6} C_n^{\left( 1 \right)}, \hspace{10.0ex}
  C_n^{\left( 3 \right)} = C_n^{\left( 1 \right)} \hspace{10.0ex} n \geq 0
\end{equation}
For $n = 1$, this implies that, in the context of the proximity force
approximation, the topology dependent part of the one-loop Casimir
energy-momentum tensor, which is the $b^{- 11}$ terms in (\ref{the t i as
functions of b}), is traceless \cite{Fabinger Horava}.  This is presumably
connected with the formal relation between the trace of the energy-momentum
tensor, and the divergence of the ``dilation current'', and the fact that the
$b^{- 11}$ terms in (\ref{the t i as functions of b}) are independent of $
\kappa $.

The limitations of the proximity force approximation are discussed, for
example, in subsection 4.3 of {\cite{0106045 Bordag Mohideen Mostepanenko}}.
In the present case, the proximity force approximation would not be valid
unless $\dot{b}$, $b \ddot{b}$, and similar dimensionless quantities formed
from $b$ and its higher derivatives with respect to $y$, all had magnitude
small compared to $1$, and we will find in subsection \ref{Analysis of the
Einstein equations and the boundary conditions}, on page \pageref{Analysis of
the Einstein equations and the boundary conditions}, that this is not the
case.  It will therefore be necessary to go beyond the proximity force
approximation, as I will discuss in subsection \ref{Beyond the proximity
force approximation}, on page \pageref{Beyond the proximity force
approximation}.

However, for a given trial classical metric $G_{IJ}$, and in the approximation
of dropping all but the leading terms in the Lukas-Ovrut-Waldram harmonic
expansion of $T_{IJ}$ on $\mathcal{M}^6$, we can still assume that the
$t^{\left( i \right)}$ functions have an expansion of the form (\ref{the t i
as functions of b}) near the inner surface of the thick pipe, provided that
$b$ depends monotonically on $y$ in this region, except that other powers of
$b$, not included in (\ref{the t i as functions of b}), may occur, and we have
to check that when the boundary conditions are satisfied, the ``input''
$t^{\left( i \right)}$ functions lead self-consistently to a metric that
results in ``output'' $t^{\left( i \right)}$ functions equal to the ``input''
$t^{\left( i \right)}$ functions.

Considering, now, the energy-momentum tensor on the Ho\v{r}ava-Witten orbifold
fixed-point hyperplanes, let $\tilde{T}^{\left[ i \right] UV}$, $i = 1, 2$, be
defined by (\ref{energy momentum tensor}), on page \pageref{energy momentum
tensor}, with $\left( S_{\mathrm{SM}} + S_{\mathrm{DM}} \right)$ replaced by
the boundary action at $y = y_i$, and the metric $g_{\mu \nu}$ replaced by the
induced metric, $G_{UV}$, on the boundary at $y = y_i$.  This is a change of
notation from earlier sections, where the fixed-point hyperplanes were
distinguished by a superscript in round parentheses.  Then in the
approximation of dropping all but the leading terms in the Lukas-Ovrut-Waldram
harmonic expansions of the $\tilde{T}^{\left[ i \right] UV}$ on $ \mathcal{M}^6
$, I shall assume that the $\tilde{T}_{UV}^{\left[ i \right]}$ have the block
diagonal structure:
\begin{equation}
  \label{T tilde i UV block diagonal structure} \tilde{T}_{\mu \nu}^{\left[ i
  \right]} = \tilde{t}^{\left[ i \right] \left( 1 \right)} G_{\mu \nu},
  \hspace{1.8cm}
  \tilde{T}_{AB}^{\left[ i \right]} = \tilde{t}^{\left[ i \right]
  \left( 2 \right)} G_{AB}
\end{equation}

The coefficients $ \tilde{t}^{\left[ 1 \right] \left( j \right)} $ will
receive contributions that are numerical multiples of $ \frac{\kappa^{-\frac{4
}{3}}}{b^4} $, from the leading terms in the Lukas-Ovrut-Waldram harmonic
expansion of the energy-momentum tensor of topologically stabilized vacuum
Yang-Mills fields on the inner surface of the thick pipe.  It would seem
reasonable to expect these contributions to be roughly a positive numerical
multiple of the energy-momentum tensor that results from embedding the spin
connection in the gauge group for $ \mathbf{CH}^3 $, which will be calculated
in subsection \ref{The field equations and boundary conditions for the metric},
on page \pageref{The field equations and boundary conditions for the metric}.
The Lovelock-Gauss-Bonnet terms in the quantum effective action on the
Ho\v{r}ava-Witten orbifold hyperplanes, discussed in connection with (\ref{FF
RR action substitutions}), on page \pageref{FF RR action substitutions}, also
result in terms in the $ \tilde{t}^{\left[ 1 \right] \left( j \right)} $
coefficients that are numerical multiples of $ \frac{\kappa^{-\frac{4}{3}}}{b^4
} $, which will also be calculated in subsection \ref{The field equations and
boundary conditions for the metric}.

Thus by analogy with (\ref{the t i as
functions of b}), I shall assume that within the context of the proximity force
approximation, the coefficients $\tilde{t}^{\left[ 1 \right] \left( i \right)}$
can be expanded as:
\begin{equation}
  \label{surface Casimir action density} \tilde{t}^{\left[ 1 \right] \left(
  i \right)} = D^{\left( i \right)}_{- 1} \frac{\kappa^{-
  \frac{4}{3}}}{b^4} + D^{\left( i \right)}_{1} \frac{1}{b^{10}} +
  D^{\left( i \right)}_{2} \frac{\kappa^{\frac{2}{3}}}{b^{13}} + \ldots
\end{equation}
where the $D^{ \left( i \right) }_n$ are numerical constants, that depend only
on the topology and spin structure of $ \mathcal{M}^6 $, and on the
topologically stabilized vacuum configurations of the Yang-Mills fields on the
Ho\v{r}ava-Witten orbifold hyperplanes, and at higher orders, on the vacuum
configuration of the three-form gauge field in the bulk.  We note that in
consequence of the Ho\v{r}ava-Witten relation (\ref{lambda kappa relation}),
on page \pageref{lambda kappa relation}, between $ \kappa $, and the Yang-Mills
coupling constant $ \lambda $ on the orbifold hyperplanes, the expansion
(\ref{surface Casimir action density}) is equivalent to an expansion in
integer powers of $ \lambda $.

For the solutions I shall consider in subsection \ref{Solutions with a as small
as kappa to the two ninths at the outer surface}, on page \pageref{Solutions
with a as small as kappa to the two ninths at the outer surface}, where $ a
\left( y \right) $ becomes as small as $ \kappa^{2/9} $ at the outer
surface of the thick pipe, and the three observed spatial dimensions are
assumed to be compactified to a smooth compact quotient $ \mathcal{M}^3 $ of $
\mathbf{H}^3 $, the expansion analogous to (\ref{the t i as functions of b}) is
\begin{equation}
  \label{bulk Casimir action density for small a} t^{ \left( i \right) } =
  \tilde{C}^{ \left( i \right) }_0 \frac{\kappa^{-\frac{2}{3}}}{a^{8}} +
  \tilde{C}^{ \left( i \right) }_1 \frac{1}{a^{11}} +
  \tilde{C}^{ \left( i \right) }_2 \frac{\kappa^{\frac{2}{3}}}{a^{14}} +
  \ldots,
\end{equation}
and the expansion analogous to (\ref{surface Casimir action density}) is
\begin{equation}
  \label{surface Casimir action density for small a} \tilde{t}^{\left[ 2
  \right] \left( i \right)} = \tilde{D}^{\left( i \right)}_{- 1} \frac{\kappa^{
  -\frac{4}{3}}}{a^4} + \tilde{D}^{\left( i \right)}_{1} \frac{1}{a^{10}} +
  \tilde{D}^{\left( i \right)}_{2} \frac{\kappa^{\frac{2}{3}}}{a^{13}} +
  \ldots.
\end{equation}

The situation where $ a \left( y \right) $ becomes as small as $ \kappa^{\frac{
2}{9}} $ at the outer surface differs from the situation near
the inner surface, in that one of the four dimensions scaled by $a \left( y
\right)$ is the time dimension, and only the three spatial dimensions scaled by
$a \left( y \right)$ are assumed to be compactified.  The compactification
of the three observed spatial dimensions to $ \mathcal{M}^3 $ breaks $ d = 4 $
Lorentz invariance globally, although not locally, so the Casimir effects near
the outer surface will not, in general, be Lorentz invariant.

Thus for the solutions where $ a \left( y \right) $ becomes as small as $
\kappa^{2/9} $ at the outer surface, the $ G_{\mu\nu} $ form of $
T_{\mu\nu} $, in (\ref{T IJ block diagonal structure}), would in general have
to be replaced, near the outer surface, by a more general Robertson-Walker
form, and the $ G_{\mu\nu} $ form of $ \tilde{T}^{\left[ 2 \right] }_{\mu\nu}
$, in (\ref{T tilde i UV block diagonal structure}), would also have to be
replaced by a Robertson-Walker form.  However in subsection \ref{Solutions with
a as small as kappa to the two ninths at the outer surface} of this paper, I
shall consider the case where the Casimir effects near the outer surface
are, to sufficient accuracy, consistent with (\ref{T IJ block diagonal
structure}) and (\ref{T tilde i UV block diagonal structure}).  The
coefficients $ \tilde{C}^{ \left( i \right) }_n $ in (\ref{bulk Casimir action
density for small a}) and $ \tilde{D}^{\left( i \right)}_n $ in (\ref{surface
Casimir action density for small a}) are then numerical constants that depend
only on the topology and spin structure of $ \mathcal{M}^3 $.

\subsubsection{The orders of perturbation theory that the terms in the Casimir
energy densities occur at}
\label{The orders of perturbation theory that the terms occur at}

We recall that in subsection \ref{The higher order corrections to
Horava-Witten theory}, on page \pageref{The higher order corrections to
Horava-Witten theory}, we defined the homogeneity number, of a local monomial
in the CJS fields and their derivatives, to be the number of derivatives, plus
half the number of gravitinos.  Let us now extend this definition to an
arbitrary product of the CJS fields and their derivatives, not necessarily all
at the same point, and denote the homogeneity number by $h$.  Then the overall
degree of divergence of an $L$ loop Feynman diagram contributing to a term in
the quantum effective action, or in other words, the generating function of
proper vertices, in eleven dimensions, corresponding to a product of the CJS
fields and their derivatives, with homogeneity number $h$, is
\begin{equation}
  \label{overall degree of divergence in eleven dimensions} 9 b + 10 f +
  \left( 2 - 11 \right) v_0 + \left( 1 - 11 \right) v_1 - 11 v_2 + 11 - N = 9
  L + 2 - h,
\end{equation}
where in the left-hand side of (\ref{overall degree of divergence in eleven
dimensions}), $b$ denotes the number of boson propagators, $f$ denotes the
number of fermion propagators, $v_q$ denotes the number of vertices with $2 q$
fermion legs, and $N$ denotes the number of derivatives acting on the CJS
fields, which are here the ``background'' fields, and we noted that the number
of fermion propagator ends is $2 f = 2 v_1 + 4 v_2 - F$, where $F$ is the
number of gravitinos among the ``background'' fields, and $L = b + f + 1 - v_0
- v_1 - v_2$.

The maximum power of $\kappa$ that can occur for an $L$ loop Feynman diagram
contributing to the quantum effective action, in eleven dimensions, is $2$ for
each propagator, minus $2$ for each vertex, hence $2 \left( L - 1 \right)$.
However, as discussed in the second part of subsection \ref{The higher order
corrections to Horava-Witten theory}, starting around page \pageref{pentagon
for bulk Green-Schwarz term}, when we use BPHZ renormalization, with
propagators in the counterterms that differ from the propagators in the direct
terms, by being cut off at large distances, as allowed by the convergence
proofs in {\cite{BPHZ in EPS}}, so as to avoid the occurrence of divergences
at large distances in the BPHZ counterterms, due to the presence of massless
particles, terms involving lower powers of $\kappa$ also arise naturally at
$L$ loops.

Specifically, according to the prescription in {\cite{BPHZ in EPS}}, the same
modification of the propagator, at long distances, is used in all the internal
lines of a counterterm part.  Since a unit of distance, namely
$\kappa^{2/9}$, occurs in the CJS action (\ref{upstairs bulk action}),
it is natural to cut off the propagators, in the counterterms, at distances
greater than $\kappa^{2/9}$, where the numerical multiple of
$\kappa^{2/9}$, at which the cutoff occurs, is likely to get modified
later, in effect, when finite counterterms are added so as to satisfy
Slavnov-Taylor identities.  The position-space integral for the BPHZ
counterterm that has $p$ extra derivatives acting on the CJS ``background''
fields, and contributes to cancelling the short-distance divergence of a
direct term of overall ultraviolet degree of divergence $D$, where $0 \leq p
\leq D$, then has the schematic form $\int^{\kappa^{2/9}} \frac{x^p
dx}{x^{D + 1}}$, where the divergence at small $x$ cancels against a
corresponding ultraviolet divergence in the direct terms.  Thus this integral
gives $\sim \kappa^{\frac{2}{9} \left( p - D \right)}$, which is a power $\leq
0$ of $\kappa$.  The CJS fields, mostly at separated points in the direct
term, are now collected into a local monomial, in the CJS fields and their
derivatives, at a single point, in the counterterm, whose homogeneity number
is $h_f \equiv h + p$.  The total power of $\kappa$, including the overall
factor $\kappa^{2 \left( L - 1 \right)}$, is
\begin{equation}
  \label{total power of kappa} \kappa^{2 \left( L - 1 \right)}
  \kappa^{\frac{2}{9} \left( p - D \right)} = \kappa^{\frac{2}{9} \left( h_f -
  11 \right)},
\end{equation}
by (\ref{overall degree of divergence in eleven dimensions}).  This is the
correct power of $\kappa$ to multiply a local monomial, in the CJS fields and
their derivatives, of homogeneity number $h_f$, in order for the quantum
effective action to be dimensionless.

Thus we see that the terms of index $n$ in the expansions (\ref{the t i as
functions of b}), namely $C^{\left( i \right)}_n \frac{\kappa^{\frac{2}{3}
\left( n - 1 \right)}}{b^{8 + 3 n}} = \kappa^{- \frac{22}{9}} C^{\left( i
\right)}_n \left( \frac{\kappa^{2/9}}{b} \right)^{8 + 3 n}$, first
occur at a number of loops $L$, where $L$ is the smallest integer $\geq
\frac{n + 2}{3}$, and we also find the corresponding conclusion, for the terms
in the expansions (\ref{bulk Casimir action density for small a}).  Now for
the local terms in the low energy expansion of the quantum effective action,
such as terms formed from products of Riemann tensors, possibly with covariant
derivatives acting on them, and their indices contracted in various ways, a
number of indications have been found that, for at least some terms, their
coefficients, which will be independent of the topology of the background
field configuration, do not receive any further modifications, beyond certain
finite orders of perturbation theory {\cite{Duff Liu Minasian, Vafa Witten,
Green Vanhove 1, Green Gutperle Vanhove, Green Gutperle Kwon, Green Kwon
Vanhove, Green Vanhove 2, Green Russo Vanhove 1, Green Russo Vanhove 2}}.
However, for smooth compact quotients of $\mathbf{C} \mathbf{H}^3$ or
$\mathbf{H}^6$, the coefficients in (\ref{the t i as functions of b}) and
(\ref{bulk Casimir action density for small a}) also receive nonlocal
contributions, for example via the sums over images in the propagators, if
these converge, so we would expect the coefficients $C^{\left( i \right)}_n$
and $\tilde{C}^{\left( i \right)}_n$ to receive contributions from all loop
orders $L$ such that $L \geq \frac{n + 2}{3}$.

Considering, now, the terms of index $m$ in the expansions (\ref{surface
Casimir action density}), namely $D^{\left( i \right)}_m \frac{\lambda^{m -
1}}{b^{7 + 3 m}}$, an analogous argument, using power counting as appropriate
for Feynman diagrams in ten dimensions, indicates that $D^{\left( i
\right)}_m$ first receives contributions at a number of loops $L$, where $L$
is the smallest integer $\geq \frac{m + 1}{2}$, with a corresponding
conclusion, for the coefficients $\tilde{D}^{\left( i \right)}_m$ in
(\ref{surface Casimir action density for small a}).  However, Ho\v{r}ava-Witten
theory is fundamentally defined in eleven dimensions, and from the
Ho\v{r}ava-Witten relation (\ref{lambda kappa relation}), we see that
$\lambda^{\left( m - 1 \right)}$ is a numerical multiple of
$\kappa^{\frac{2}{3} \left( m - 1 \right)}$, so by analogy with the bulk case,
it seems likely that the coefficients $D^{\left( i \right)}_m$ and
$\tilde{D}^{\left( i \right)}_m$ will, in fact, receive contributions from all
loop orders $L$ such that $L \geq \frac{m + 2}{3}$.  For $m = - 1$, this is in
agreement with the fact that, in Ho\v{r}ava-Witten theory, the Yang-Mills actions,
on the orbifold ten-manifolds, first arise as a one-loop effect, while for $m
\geq 0$, it gives an onset value of $L$ that is less than or equal to that
given by the ``$d = 10$'' estimate.

\subsubsection{The expansion parameter}
\label{The expansion parameter}

Now we found in subsection \ref{The Yang-Mills coupling constants in four
dimensions}, on page \pageref{The Yang-Mills coupling
constants in four dimensions},
that for a reasonable estimate, (\ref{alpha sub U in the Standard Model}), on
page \pageref{alpha sub U in the Standard Model}, of the $d = 4$ Yang-Mills
fine structure constant at unification, the value $b_1 = b \left( y_1 \right)$
of $b \left( y \right)$, at the inner surface of the thick pipe, is
related to $\left| \chi \left( \mathcal{M}^6 \right) \right|$, the magnitude
of the Euler number of the compact six-manifold, by (\ref{b sub 1 in terms of
chi}), on page \pageref{b sub 1 in terms of chi}, which states that
$\frac{b_1}{\kappa^{2/9}} \simeq \frac{1.28}{\left| \chi \left(
\mathcal{M}^6 \right) \right|^{\frac{1}{6}}}$, when $\mathcal{M}^6$ is a
smooth compact quotient of $\mathbf{C} \mathbf{H}^3$, and we also find
that $\frac{b_1}{\kappa^{2/9}} \simeq \frac{1.18}{\left| \chi \left(
\mathcal{M}^6 \right) \right|^{\frac{1}{6}}}$, when $\mathcal{M}^6$ is a
smooth compact quotient of $\mathbf{H}^6$.  Thus to find out whether a
particular value of $\left| \chi \left( \mathcal{M}^6 \right) \right|$ is
possible, and indeed, whether $\left| \chi \left( \mathcal{M}^6 \right) \right|
\geq 1 $ is possible,
we need to know whether the expansions (\ref{the t i as functions of b}),
for the bulk Casimir energy density coefficients near the inner
surface of the thick pipe, and the expansions (\ref{surface Casimir action
density}), for the Casimir energy density coefficients on the inner surface of
the thick pipe, allow $b_1$ to be as small as the value given by (\ref{b sub 1
in terms of chi}), for that value of $\left| \chi \left( \mathcal{M}^6 \right)
\right|$, or whether the expansions (\ref{the t i as functions of b}) and
(\ref{surface Casimir action density}) already become infinite, for a value of
$b$ larger than that value of $b_1$.

We assume that the expansion coefficients in (\ref{the t i as functions of b}),
from $C^{\left( i \right)}_1$ onwards, and the expansion
coefficients in (\ref{surface Casimir action density}), from $D^{\left( i
\right)}_1$ onwards, depend on the topology of $\mathcal{M}^6$,
and in particular, that their signs depend on the topology of $\mathcal{M}^6$.
Kenneth and Klich {\cite{Kenneth Klich}} and Bachas {\cite{Bachas}} have
recently discovered that Casimir forces are always attractive in certain
circumstances, but their result does not apply in the present context because
$ \mathcal{M}^6 $ has no shape moduli, so that regions of $ \mathcal{M}^6 $
cannot be moved closer together without also being squeezed at the same time.

Now we know that the Casimir energy densities have local contributions,
independent of the topology of $\mathcal{M}^6$, such as the terms quartic in
the Riemann tensor, discussed in {\cite{Peeters Vanhove Westerberg, Hyakutake
Ogushi 1, Hyakutake Ogushi 2}}, that would contribute terms $C^{\left( i
\right)}_0 \frac{\kappa^{- \frac{2}{3}}}{b^8}$ in (\ref{the t i as functions of
b}), and the terms on the boundaries, quadratic in the Riemann tensor,
discussed in {\cite{Lukas Ovrut Waldram}}, and mentioned in subsection
\ref{Horava-Witten theory} above, in connection with equation (\ref{FF RR
action substitutions}), on page \pageref{FF RR action substitutions}, that
would contribute terms $D^{\left( i \right)}_{- 1} \frac{\lambda^{- 2}}{b^4}$
in (\ref{surface Casimir action density}), which will be calculated in
(\ref{Lovelock Gauss Bonnet t i j}), on page \pageref{Lovelock Gauss Bonnet t
i j}, when $\mathcal{M}^6$ is a smooth compact quotient of $\mathbf{C}
\mathbf{H}^3$, and in (\ref{Lovelock Gauss Bonnet t i j for H6}), on page
\pageref{Lovelock Gauss Bonnet t i j for H6}, when $\mathcal{M}^6$ is a smooth
compact quotient of $\mathbf{H}^6$.  There will also be local terms built
from more covariant derivatives and powers of the Riemann tensor {\cite{Green
Vanhove 1, Green Gutperle Vanhove, Green Gutperle Kwon, Green Kwon Vanhove,
Green Vanhove 2, Green Russo Vanhove 1, Green Russo Vanhove 2, E10, E11}}, that
will contribute to the
$C^{\left( i \right)}_n$ and $D^{\left( i \right)}_n$ with larger $n$.

Thus for the phenomenological estimates in this paper, I shall assume that the
signs of the $C^{\left( i \right)}_n$ and $D^{\left( i \right)}_n$, $n \geq
1$, depend on the topology of $\mathcal{M}^6$, and that their magnitudes can
depend on the topology of $\mathcal{M}^6$ though a factor of order $1$, but
that, apart from this factor of order $1$, the magnitudes of the $C^{\left( i
\right)}_n$ and $D^{\left( i \right)}_n$, $n \geq 1$, are determined by their
typical values, for a geometry of roughly constant curvature.  We therefore
need to know what those typical values are.

According to Giudice, Rattazzi, and Wells (GRW) {\cite{Giudice Rattazzi
Wells}}, the expansion parameter for graviton loop corrections in $D$
dimensions, in the sense that perturbation theory is reliable when the
expansion parameter is less than $1$, is $\frac{S_{D - 1}}{2 \left( 2 \pi
\right)^D} \left( \frac{E}{M_D} \right)^{D - 2}$, where $S_{D - 1} = \frac{2
\pi^{\frac{D}{2}}}{\Gamma \left( \frac{D}{2} \right)}$ is the $\left( D - 1
\right)$-volume of a unit radius $\mathbf{S}^{D - 1}$, $E$ is the relevant
energy of the process, and $M_D$ is defined such that the Einstein equation,
in $D$ dimensions, is $R_{AB} - \frac{1}{2} g_{AB} R = - \frac{\left( 2 \pi
\right)^{\left( D - 4 \right)}}{M_D^{D - 2}} T_{AB}$.  Thus from
(\ref{upstairs bulk action}), with $\frac{1}{\kappa^2}$ replaced by
$\frac{2}{\kappa^2}$, so as to work in the downstairs picture, we find that
for Ho\v{r}ava-Witten theory, $2^{\frac{1}{9}} M_{11} = 2 \pi \left( \frac{1}{\pi
\kappa} \right)^{\frac{2}{9}}$, and the GRW estimate of the expansion
parameter for graviton loop corrections is
\begin{equation}
  \label{Giudice Rattazzi Wells estimate of expansion parameter}
  \frac{\kappa^2}{1890 \left( 2 \pi \right)^4} \left( \frac{E}{2 \pi}
  \right)^9 = \left( 0.0304 \kappa^{2/9} E \right)^9
\end{equation}

Considering, now, the value of $E$ that would apply for the expansions
(\ref{the t i as functions of b}) and (\ref{surface Casimir action
density}), we note, from the discussion after (\ref{CHn sectional curvature}),
on page \pageref{CHn sectional curvature}, that with the metric
(\ref{hermitian metric}), (\ref{CHn metric}), the sectional curvature of
$\mathbf{C} \mathbf{H}^3$, at each point of $\mathbf{C} \mathbf{H}^3$,
lies in the range $- 2$ to $- \frac{1}{2}$, with the actual value depending on
the choice of the two-dimensional section through the point, so that the
magnitude of the corresponding ``radius of curvature'' lies in the range
$\frac{1}{\sqrt{2}}$ to $\sqrt{2}$.  Thus when $\mathcal{M}^6$ is a smooth
compact quotient of $\mathbf{C} \mathbf{H}^3$, its ``radius of
curvature'', at the inner surface of the thick pipe, lies in the range
$\frac{1}{\sqrt{2}} b_1$ to $\sqrt{2} b_1$.  And if $\mathcal{M}^6$ is a
smooth compact quotient of $\mathbf{H}^6$, and $h_{AB}$ is in that case
normalized so that $R_{ABCD} \left( h \right) = h_{AC} h_{BD} - h_{AD}
h_{BC}$, as assumed after (\ref{Lovelock Gauss Bonnet t i j}), on page
\pageref{Lovelock Gauss Bonnet t i j}, then its ``radius of curvature'', at
the inner surface of the thick pipe, has the fixed value $b_1$.  Thus for both
cases, it is reasonable to take $b_1$ as the typical ``radius of curvature''
of $\mathcal{M}^6$, at the inner surface of the thick pipe.

Now for the related cases of $\mathbf{C} \mathbf{P}^3$ and
$\mathbf{S}^6$, $b_1$ would be the actual radius of curvature, so the
corresponding ``wavelength'' would be $\lambda = 2 \pi b_1$, and the
corresponding energy would be $E = \frac{2 \pi}{\lambda} = \frac{1}{b_1}$.
Thus if we also use this estimate of $E$ for the cases of smooth compact
quotients of $\mathbf{C} \mathbf{H}^3$ and $\mathbf{H}^6$, the minimum
value of $\frac{b_1}{\kappa^{2/9}}$, allowed by the requirement that
the GRW estimate of the expansion parameter be $\leq 1$, would be:
\begin{equation}
  \label{first estimate of minimum value of b sub 1 over kappa to the two
  ninths} \frac{b_1}{\kappa^{2/9}} \simeq 0.03
\end{equation}
which by (\ref{b sub 1 in terms of chi}), implies that $\left| \chi \left(
\mathcal{M}^6 \right) \right|$ could not be larger than around $6 \times
10^9$.

On the other hand, since $E$ occurs in the combination $\frac{E}{2 \pi}$ in
(\ref{Giudice Rattazzi Wells estimate of expansion parameter}), it seems
possible that the appropriate value of $E$ should, in fact, be $\frac{2
\pi}{b_1}$, in which case the minimum value of
$\frac{b_1}{\kappa^{2/9}}$, allowed by the requirement that the GRW
estimate of the expansion parameter be $\leq 1$, would be:
\begin{equation}
  \label{second estimate of minimum value of b sub 1 over kappa to the two
  ninths} \frac{b_1}{\kappa^{2/9}} \simeq 0.2
\end{equation}
which by (\ref{b sub 1 in terms of chi}), implies that $\left| \chi \left(
\mathcal{M}^6 \right) \right|$ could not be larger than around $7 \times
10^4$.

As a first check of the GRW estimate of the expansion parameter, we note that,
for $D = 4$, their estimate of the expansion parameter becomes $\frac{G_N}{2
\pi} E^2$, where $G_N$ is Newton's constant, (\ref{Newtons constant}).
Looking now at Donoghue and Torma's formula for the one-loop graviton-graviton
scattering cross section in $D = 4$, equation (29) in their paper
{\cite{9901156 Donoghue Torma}}, and noting their convention for the coupling
constant, from their equation (2), or just after their equations (1) or (2),
we see that the expansion parameter is $\frac{2 G_N}{\pi^{}} E^2$, where $E$
is the square of the centre of mass energy, times a sum of terms, the first of
which is $\ln \frac{- t}{s} \ln \frac{- u}{s}$, where $s$, $t$, and $u$ are
the Mandelstam invariants of the scattering process.  Thus in a kinematic
region where this sum of terms is $\sim 1$, the GRW estimate of the expansion
parameter is, in this case, smaller than the actual parameter, by a factor
$\sim \frac{1}{4}$.

And looking at equation (15) of Donoghue's calculation of one-loop corrections
to the gravitational scattering of two heavy masses, for $D = 4$
{\cite{9310024 Donoghue}}, and noting that his convention for the coupling
constant is the same as Donoghue and Torma's, we see that the expansion
parameter is $\frac{G_N}{\pi} \left| q^2 \right|$, where $q$ is the momentum
transfer, times a sum of two terms, one of which is $- \frac{3}{4} \ln \left(
- q^2 \right)$, and the other of which, with a heavy mass in the numerator, is
identified, by considering the non-relativistic limit, as a post-Newtonian
correction of classical general relativity, rather than a quantum correction.
Thus, in this case, the expansion parameter for quantum gravitational
corrections is $\frac{3 G_N}{4 \pi} \left| q^2 \right| \ln \left( - q^2
\right)$, where $- q^2$ would be multiplied, in the argument of the logarithm,
by an undetermined multiple of $G_N$, that would have to have to be fixed by
an experimental measurement, due to the non-renormalizability of quantum
gravity for $D = 4$, although it might be determined in a resummation of
quantum gravity, for $D = 4$, recently developed by Ward {\cite{Ward
Resummation of quantum gravity}}.  So if we identify $\left| q^2 \right|$ as
the GRW $E^2$, we see that, in the kinematic region where the argument of the
logarithm is $\sim 1$, the GRW estimate of the expansion parameter is, in this
case, smaller than the actual parameter, by a factor $\sim \frac{2}{3}$.  So
it appears that, for $D = 4$, the GRW estimate of the expansion parameter is
reasonable, in kinematic regions where the logarithmic factors it omits are
not too large.

Considering, now, how the GRW estimate of the expansion parameter might be
understood in $D$ dimensions, let us choose the Ho\v{r}ava-Witten downstairs
convention for the gravitational action in $D$ dimensions, so that the
Einstein term in the action is $\frac{1}{\kappa^2} \int d^D x \sqrt{- g} R$.
The GRW estimate of the expansion parameter is then $\frac{\pi^2 S_{D -
1}}{\left( 2 \pi \right)^D} \kappa^2 \left( \frac{E}{2 \pi} \right)^{D - 2}$.
Working, now, in Euclidean signature momentum space, there is a factor
$\frac{\kappa^2}{\left( 2 \pi \right)^D}$ for each loop, a kinematic factor
$\frac{1}{k^2}$ for each propagator, where $k_{\mu}$ is the momentum in the
propagator, and two numerator momentum factors for each vertex.

Considering, now, a ladder diagram formed from graviton propagators, with an
external momentum $p_{\mu}$, with $p^2 = E^2$, running along the ladder, the
momentum integral for each loop of the ladder will be $\sim \int \frac{ d^D k
\left| k \right|^4}{\left( k^2 \right)^3}$, which we would expect to be cut
off for $\left| k \right|$ larger than around $E$, by BPHZ counterterms, and
thus to give around $\frac{S_{D - 1}}{D - 2} E^{D - 2}$, for $D > 2$.  Thus,
without considering sums over diagrams with a given number of loops, and the
Lorentz index structure of the
graviton propagator and vertices, the GRW estimate of the expansion parameter
is obtained for $D > 2$, up to a factor $\frac{\left( D - 2 \right)
\pi^2}{\left( 2 \pi \right)^{D - 2}}$.  This factor is $\frac{1}{2}$ for $D =
4$, so, in view of the two examples above, the estimate so far is as good as
the GRW estimate, for $D = 4$.

It is not clear, without further investigation, why the magnitude of the
Euclidean loop momentum would tend to be cut off, by BPHZ counterterms, at
around $\frac{E}{2 \pi}$, rather than at around $E$, as suggested by the GRW
estimate of the expansion parameter, and it is also not clear where the extra
factor of $\pi^2$, in the GRW estimate, comes from.  This seems to suggest
that, in applying the GRW estimate to the expansions (\ref{the t i as functions
of b}) and (\ref{surface Casimir action density}), $E$ should have been
taken as $\frac{2 \pi}{b_1}$, resulting in the estimate (\ref{second estimate
of minimum value of b sub 1 over kappa to the two ninths}), above, for the
minimum possible value of $\frac{b_1}{\kappa^{2/9}}$, rather than the
estimate (\ref{first estimate of minimum value of b sub 1 over kappa to the
two ninths}), above, except that the estimate (\ref{second estimate of minimum
value of b sub 1 over kappa to the two ninths}), above, could possibly be
reduced by a factor of $\pi^{- \frac{2}{9}}$, to around $0.15$, with a
corresponding increase in the maximum possible value of $\left| \chi \left(
\mathcal{M}^6 \right) \right|$, to around $4 \times 10^5$.

Considering, now, the effects of sums over diagrams with a given number of
loops, and the Lorentz index
structure of propagators and vertices, the fact that the estimate so far
includes a factor $\frac{1}{D - 2}$, which is absent from the GRW estimate,
and is thus presumably cancelled by Lorentz index contractions, for some
diagrams, suggests considering the limit $D \rightarrow \infty$.  The $D
\rightarrow \infty$ limit of the Feynman diagram expansion of quantum gravity
was considered by Strominger {\cite{Strominger}}, and recently reconsidered by
Bjerrum-Bohr {\cite{Bjerrum-Bohr}}, and the $D \rightarrow \infty$ limit of
quantum gravity was also considered, in the context of a lattice
regularization, by Hamber and Williams {\cite{Hamber Williams}}.

The graviton propagator has two $D$-vector indices at each end, and includes
terms in which two index-contraction lines run along it, so is in this respect
similar to the gluon propagator at large $N_c$, when the $\mathrm{SU} \left( N_c
\right)$ adjoint indices, of the gluon propagator, are written as pairs of an
$\mathrm{SU} \left( N_c \right)$ fundamental index and an $\mathrm{SU} \left( N_c
\right)$ antifundamental index {\cite{t Hooft Planar diagram theory}}, but
that is as far as the similarity with large-$N_c$ Yang-Mills theory goes.  One
difference is that the graviton interaction vertices all include two factors of
momentum, and in terms where the $D$-vector indices of these two momentum
factors are not contracted with each other, a $D$-vector index line ends on
each of them.  But the main difference is that the three-graviton vertex,
$V_{\mu_1 \nu_1, \mu_2 \nu_2, \mu_3 \nu_3}$, in an expansion about flat space,
includes, in Euclidean signature momentum space, terms of structure
$\delta_{\mu_1 \mu_2} \delta_{\nu_1 \nu_2} p_{\mu_3} q_{\nu_3}$, which allow
both index lines from one propagator ending at the vertex, to pass through the
vertex ``in parallel'', like a railway track, and leave the vertex together
along another
propagator, without getting separated.  There are no such terms in the
vertices of $\mathrm{SU} \left( N_c \right)$ Yang-Mills theory, with its usual
action, since they could only arise from Lagrangian terms with at least two
traces, such as $\mathrm{tr} \left( F_{\mu \nu} F_{\sigma \tau} \right)
\mathrm{tr} \left( F_{\mu \nu} F_{\sigma \tau} \right)$.

The presence of such ``railway track'' terms, in the three-graviton vertex,
means that for some diagrams, there are two factors of $D$ per loop, at large
$D$, and these are therefore the leading diagrams at large $D$, so far as
index contractions go.  For diagrams built from propagators and three-graviton
vertices only, the loops have to be separated from each other, as one-loop
propagator insertions, or one-loop vertex insertions, in order to be able to
have two factors of $D$ per loop, so in this respect, the large-$D$ limit of
quantum gravity is much simpler than the large-$N_c$ limit of $\mathrm{SU}
\left( N_c \right)$ Yang-Mills theory.  When $\left( 2 n + m \right)$-graviton
vertices, containing terms with $n$ ``railway tracks'' through them, with $n
\geq 2$, $m \geq 1$, are included, loops with two factors of $D$ per loop can
now touch one another, and the leading terms at large $D$, in the quantum
effective action, so far as index contractions go, are ``trees'' built from
one-loop bubbles, that meet one another at $\left( 2 n + m \right)$-graviton
vertices, $n \geq 2$, $m \geq 1$, that have $n$ ``railway tracks'' through
them.

Thus from considering the index contractions from the diagrams that are
leading at large $D$, so far as index contractions go, the estimate of the
expansion parameter now gets an additional factor $D^2$, so for large $D$, our
estimate of the expansion parameter is now larger than the GRW estimate, by a
factor $\frac{D \left( 2 \pi \right)^{D - 2}}{\pi^2}$.  However, for $D = 11$,
the factor $\frac{D}{\pi^2}$ is approximately $1$, and the factor $\left( 2
\pi \right)^{D - 2}$ is accomodated by taking $E$ as $\frac{2 \pi}{b_1}$ in
the GRW estimate (\ref{Giudice Rattazzi Wells estimate of expansion
parameter}), rather than $\frac{1}{b_1}$, as we would initially have expected.
Thus it appears that the second estimate, (\ref{second estimate of minimum
value of b sub 1 over kappa to the two ninths}), is at present the best rough
estimate of the minimum value of $\frac{b_1}{\kappa^{2/9}}$, and the
best rough estimate of the upper bound on $\left| \chi \left( \mathcal{M}^6
\right) \right|$ is
therefore around $7 \times 10^4$.  To check this estimate further, it would be
necessary to consider diagrams involving the gravitino and the three-form
gauge field, but that will not be done in this paper.

I shall seek solutions of the Casimir energy density corrected field equations
and boundary conditions, such that all the fermionic fields vanish.  Thus, in
the bulk, the only non-vanishing fields will be the metric, and the three-form
gauge field.  I shall now consider the implications of a topological
constraint, that was discussed by Witten in the context of superstrings, then
consider the field equation, and boundary conditions, for the three-form gauge
field.

\subsubsection{Witten's topological constraint}
\label{Wittens topological constraint}

By analogy with a constraint on the compactification of superstrings,
discussed by Witten {\cite{Witten Constraints on compactification}}, the fact
that the gauge-invariant field strength, $G_{IJKL}$, is globally well-defined,
implies that for any closed five-dimensional surface, $\mathcal{S}$, we must
have $\int_{\mathcal{S}} dG = 0$.  If we now work in the ``upstairs'' picture,
so that $\mathcal{M}^{11}$ is $\mathcal{M}^{10} \times \mathbf{S}^1$, and
the fields transform under reflection in the orbifold fixed-point hyperplanes
$y = y_1$, and $y = y_2$, as discussed in Subsection \ref{Horava-Witten
theory}, on page \pageref{Horava-Witten theory}, and choose $\mathcal{S}$ to
be the Cartesian product of the circle $\mathbf{S}^1$, and a closed
four-dimensional surface, $\mathcal{Q}$, in $\mathcal{M}^{10}$, then this
relation, together with (\ref{Bianchi identity with FF only}), after making
the substitutions (\ref{FF to FF RR substitutions}),
implies that the sum, over the two orbifold fixed-point hyperplanes, of the
integral:
\begin{equation}
  \label{summand in Wittens topological constraint} \int_{\mathcal{Q}} \left(
  \mathrm{tr} F^{\left( i \right)} \wedge F^{\left( i \right)} - \frac{1}{2}
  \mathrm{tr} R \wedge R \right)
\end{equation}
must be equal to zero.  We recall, from the discussion after (\ref{Yang Mills
action}), that for $E_8$, ``tr'' denotes $\frac{1}{30}$ of the trace in the
adjoint representation, and from the discussion after (\ref{FF to FF RR
substitutions}), that $\mathrm{tr} R_{\left[ UV \right.}
R_{\left. WX \right]}$ is defined as $R^{\hspace{0.4ex} \hspace{0.4ex}
\hspace{0.4ex} \hspace{0.4ex} \hspace{0.4ex} \hspace{0.4ex} \hspace{0.6ex}
YZ}_{\left[ UV \right.} R_{\left. WX \right] YZ}$, so that the trace is
effectively in the vector representation, of the SO(10) tangent space group of
$\mathcal{M}^{10}$.

I will show that, for the metric ansatz (\ref{metric ansatz}), the
implications of the topological constraint are the same, regardless of whether
the Riemann tensors, in $\mathrm{tr} R_{\left[ UV \right.} R_{\left. WX
\right]}$, are calculated entirely in ten dimensions, from the restriction of
the metric to the appropriate orbifold fixed-point hyperplane, or,
alternatively, treated as the restriction to the orbifold fixed-point
hyperplane, of the Riemann tensors calculated from the metric in eleven
dimensions.

Now in the problem studied here, $\mathcal{M}^{10}$ is the Cartesian product,
of a four dimensional locally de Sitter space, whose three spatial dimensions
may have been compactified, and a smooth compact quotient of $\mathbf{C}
\mathbf{H}^3$.  Suppose, first, that $\mathcal{Q}$ is the Cartesian product
of a topologically non-trivial closed four-dimensional surface, in
$\mathbf{C} \mathbf{H}^3$, and a point of the locally de Sitter space.
And suppose, first, that the Riemann tensors, in $\mathrm{tr} R_{\left[ UV
\right.} R_{\left. WX \right]}$, are calculated entirely in ten dimensions,
from the restriction of the metric to the appropriate orbifold fixed-point
hyperplane.  Then $\int_{\mathcal{Q}} \mathrm{tr} R \wedge R$ is generically
non-zero, and, moreover, is a topological invariant, specifically a
Pontrjagin number, so both the orbifold fixed points give the same
contribution, to the quantity that is required to vanish.  Thus it is necessary
to choose nonvanishing $E_8$ vacuum gauge fields, on one or
both of the orbifold fixed-point hyperplanes, in order to cancel the
contributions, to the sum, from $\int_{\mathcal{Q}} \mathrm{tr} R \wedge R$.  We
recall, from the discussion after (\ref{Tr JJ}), that the trace of the square
of a generator of SO(16), in the adjoint representation of $E_8$, is 30 times
the trace of
the square of the corresponding generator, in the vector representation of
SO(16).  Thus we can satisfy the topological constraint, for all $\mathcal{Q}$
of this type, by choosing {\emph{one}} of the two orbifold fixed-point
hyperplanes, and choosing an SO(6) subgroup of the $E_8$ gauge group on that
orbifold fixed-point hyperplane, embedded in that $E_8$ gauge group by the
natural embedding $\mathrm{SO} \left( 6 \right) \subset \mathrm{SO} \left( 16
\right) \subset E 8$, and setting the $E_8$ Yang-Mills gauge fields, in that
SO(6) subgroup of that $E_8$, equal to the spin connection, while the $E_8$
gauge fields, in the $E_8$ on the other orbifold fixed-point hyperplane, are
zero.  Furthermore, the classical Yang-Mills field equation is automatically
satisfied for such a configuration, in consequence of the fact that the
compact six-manifold is locally symmetric, so that the covariant derivative of
the Riemann tensor, $D_U R_{VWXY}$, vanishes identically, which implies that
$ D_U R^{\hspace{3.1ex} x y}_{VW} $ vanishes identically, where $ x $ and $ y $
are local Lorentz indices.  More generally, the
topological constraint, for this type of $\mathcal{Q}$, will also be
satisfied, for arbitrary Yang-Mills field configurations, in the same
topological class, as the configuration just described.  This is known as the
standard embedding of the spin connection of the compact six-manifold, in one
of the two $E_8$ gauge groups.  In the present case, $\mathbf{C}
\mathbf{H}^3$ is a K\"ahler manifold, so the spin connection will lie in a
U(3) subgroup of the SO(6).

Now suppose, instead, that the Riemann tensors, in $\mathrm{tr} R_{\left[ UV
\right.} R_{\left. WX \right]}$, are treated as the restriction to the
orbifold fixed-point hyperplane, of the Riemann tensors calculated from the
metric in eleven dimensions.  In that case, we find, from (\ref{Riemann tensor
for the metric ansatz}), that:
\begin{eqnarray}
\label{RR from eleven dimensions}
  \hspace{-0.4ex} \hspace{-0.8ex} \hspace{-0.8ex}
  \hspace{-0.4ex} \hspace{-0.8ex} \hspace{-0.8ex}
  \hspace{-0.4ex} \hspace{-0.8ex} \hspace{-0.8ex}
  \hspace{-0.4ex} \hspace{-0.8ex} \hspace{-0.8ex}
  R^{\hspace{0.4ex}
  \hspace{0.4ex} \hspace{0.4ex} \hspace{0.4ex} \hspace{0.4ex} \hspace{0.4ex}
  \hspace{0.4ex} \hspace{0.8ex} J}_{ABI} R^{\hspace{0.4ex} \hspace{0.4ex}
  \hspace{0.4ex} \hspace{0.4ex} \hspace{0.4ex} \hspace{0.4ex} \hspace{0.4ex}
  \hspace{0.4ex} \hspace{0.8ex} I}_{CDJ} & = & R^{\hspace{0.4ex}
  \hspace{0.4ex} \hspace{0.4ex} \hspace{0.4ex} \hspace{0.4ex} \hspace{0.4ex}
  \hspace{0.4ex} \hspace{0.4ex} \hspace{0.6ex} F}_{ABE} R^{\hspace{0.4ex}
  \hspace{0.4ex} \hspace{0.4ex} \hspace{0.4ex} \hspace{0.4ex} \hspace{0.4ex}
  \hspace{0.4ex} \hspace{0.4ex} \hspace{0.8ex} E}_{CDF} \hspace{0.4ex}
  \hspace{0.4ex} = \nonumber\\
  & = & R^{\hspace{0.4ex} \hspace{0.4ex} \hspace{0.4ex} \hspace{0.4ex}
  \hspace{0.4ex} \hspace{0.4ex} \hspace{0.4ex} \hspace{0.4ex} \hspace{0.6ex}
  F}_{ABE} \! \left( h \right) R^{\hspace{0.4ex} \hspace{0.4ex} \hspace{0.4ex}
  \hspace{0.4ex} \hspace{0.4ex} \hspace{0.4ex} \hspace{0.4ex} \hspace{0.4ex}
  \hspace{0.8ex} E}_{CDF} \! \left( h \right) \! + \! 4 \dot{b}^2 R_{ABDC}
  \left( h \right) \! + \! 2 \dot{b}^4 \left( h_{AD} h_{BC} \! - \! h_{AC}
  h_{BD} \right) \hspace{-1.6ex} \hspace{-2ex}
\end{eqnarray}
Thus $R^{\hspace{0.4ex} \hspace{0.4ex} \hspace{0.4ex} \hspace{0.4ex}
\hspace{0.4ex} \hspace{0.4ex} \hspace{0.6ex} IJ}_{\left[ AB \right.} R_{\left.
CD \right] IJ} = R^{\hspace{0.4ex} \hspace{0.4ex} \hspace{0.4ex}
\hspace{0.4ex} \hspace{0.4ex} \hspace{0.4ex} \hspace{0.4ex} EF}_{\left[ AB
\right.} R_{\left. CD \right] EF} = R^{\hspace{0.4ex} \hspace{0.4ex}
\hspace{0.4ex} \hspace{0.4ex} \hspace{0.4ex} \hspace{0.4ex} \hspace{0.4ex}
EF}_{\left[ AB \right.} \left( h \right) R_{\left. CD \right] EF} \left( h
\right)$, so for the metric ansatz (\ref{metric ansatz}), the topological
constraint, for a closed four-surface $\mathcal{Q}$, that has the form of the
Cartesian product of a topologically non-trivial four-dimensional closed
surface in the compact six-manifold, and a point in the locally de Sitter
space, has exactly the same form, regardless of whether the Riemann tensors,
in $\mathrm{tr} R \wedge R$, are calculated entirely in ten dimensions, from the
restriction of the metric to the appropriate orbifold fixed-point hyperplane,
or are the components, in the orbifold fixed-point hyperplane, of the
eleven-dimensional Riemann tensor, and, indeed, it still has the same form,
even if ``$\mathrm{tr}$'' sums the contracted indices, over all eleven
dimensions.

Suppose, now, that $\mathcal{Q}$ is the Cartesian product, of a topologically
non-trivial $n$-dimensional closed surface, in $\mathbf{C} \mathbf{H}^3$,
such that $1 \leq n \leq 3$, and a topologically non-trivial $\left( 4 - n
\right)$-dimensional closed surface, in the locally de Sitter space.  And as
before, suppose, first, that the Riemann tensors, in $\mathrm{tr} R_{\left[ UV
\right.} R_{\left. WX \right]}$, are calculated entirely in ten dimensions,
from the restriction of the metric to the appropriate orbifold fixed-point
hyperplane.  Then all the Riemann tensor components, with mixed indices,
vanish identically, so $\int_{\mathcal{Q}} \mathrm{tr} R \wedge R$ vanishes
identically, for any such $\mathcal{Q}$.  Furthermore, the $E_8$ vacuum gauge
fields already introduced, have no components tangential to the locally de
Sitter space, so $\int_{\mathcal{Q}} \mathrm{tr} F^{\left( i \right)} \wedge
F^{\left( i \right)}$ also vanishes identically, for both $E_8$ gauge groups,
for all such $\mathcal{Q}$.  So no nontrivial topological constraint arises
from any such $\mathcal{Q}$.

Now suppose, instead, that the Riemann tensors, in $\mathrm{tr} R_{\left[ UV
\right.} R_{\left. WX \right]}$, are treated as the restriction to the
orbifold fixed-point hyperplane, of the Riemann tensors calculated from the
metric in eleven dimensions.  Then it follows from the list, in subsection
\ref{The Christoffel symbols Riemann tensor and Ricci tensor}, of the
components of the Riemann tensor, of the form $R^{\hspace{0.4ex}
\hspace{0.4ex} \hspace{0.4ex} \hspace{0.4ex} \hspace{0.4ex} \hspace{1.4ex}
J}_{UVI}$, that do not vanish automatically, for the metric ansatz
(\ref{metric ansatz}), that $\int_{\mathcal{Q}} \mathrm{tr} R \wedge R$ vanishes
identically, unless $n = 2$.  We then find, from (\ref{Riemann tensor for the
metric ansatz}), that:
\begin{equation}
  \label{mixed Riemann Riemann} R^{\hspace{0.4ex} \hspace{0.4ex}
  \hspace{0.4ex} \hspace{0.4ex} \hspace{0.8ex} IJ}_{\mu A} R_{\nu BIJ} =
  R^{\hspace{0.4ex} \hspace{0.4ex} \hspace{0.4ex} \hspace{0.4ex}
  \hspace{0.8ex} UV}_{\mu A} R_{\nu BUV} = 2 R^{\hspace{0.4ex} \hspace{0.4ex}
  \hspace{0.4ex} \hspace{0.4ex} \hspace{0.8ex} \sigma C}_{\mu A} R_{\nu B
  \sigma C} = 2 \left( \frac{\dot{a} \dot{b}}{ab} \right)^2 G_{\mu \nu} G_{AB}
\end{equation}
Hence $R^{\hspace{0.4ex} \hspace{0.4ex} \hspace{0.4ex} \hspace{0.4ex}
\hspace{1.2ex} IJ}_{\left[ \mu A \right.} R_{\left. \nu B \right] IJ} =
R^{\hspace{0.4ex} \hspace{0.4ex} \hspace{0.4ex} \hspace{0.4ex} \hspace{1.2ex}
UV}_{\left[ \mu A \right.} R_{\left. \nu B \right] UV} = 0$, hence
$\int_{\mathcal{Q}} \mathrm{tr} R \wedge R$ vanishes identically, for the
metric
ansatz (\ref{metric ansatz}), regardless of whether the Riemann tensors, in
$\mathrm{tr} R \wedge R$, are calculated entirely in ten dimensions, from the
restriction of the metric to the appropriate orbifold fixed-point hyperplane,
or are the components, in the orbifold fixed-point hyperplane, of the
eleven-dimensional Riemann tensor, and, moreover, this is still true,
even if ``$\mathrm{tr}$'' sums the contracted indices, over all eleven
dimensions.

Finally, there are no topologically non-trivial $4$-dimensional closed
surfaces, in the locally de Sitter space, since the time dimension has not
been compactified.

Thus Witten's topological constraint is completely satisfied, by the standard
embedding of the spin connection of the compact six-manifold, in the $E_8$
gauge group, on one of the two orbifold fixed-point hyperplanes, as just
described, and this is true, for the metric ansatz (\ref{metric ansatz}),
regardless of whether the Riemann tensors, in $\mathrm{tr} R \wedge R$, are
calculated entirely in ten dimensions, from the restriction of the metric to
the appropriate orbifold fixed-point hyperplane, or are the components, in the
orbifold fixed-point hyperplane, of the eleven-dimensional Riemann tensor,
and, furthermore, this is still true, even if ``$\mathrm{tr}$'' sums the
contracted indices, over all eleven dimensions.

The fact that the spin connection is embedded in the $E_8$ gauge group, on
just {\emph{one}} of the two orbifold fixed-point hyperplanes, breaks the
symmetry between the two orbifold fixed-point hyperplanes, and it is known
from calculations by Witten {\cite{Witten Strong coupling expansion}}, and by
Lukas, Ovrut, Stelle, and Waldram {\cite{Lukas Ovrut Stelle Waldram}}, that
when the compact six-manifold is a Calabi-Yau manifold, the volume of the
compact six-manifold is greater, on the orbifold hyperplane that has the spin
connection embedded in its $E_8$ gauge group, than it is on the other orbifold
hyperplane.  I will show that this is also true, when the compact six-manifold
is a smooth compact quotient of $\mathbf{C} \mathbf{H}^3$, so the spin
connection will be embedded in the $E_8$ gauge group, on the {\emph{outer}}
surface of the thick pipe.  This is fortunate, because we must expect that, in
order to find smooth compact quotients of $\mathbf{C} \mathbf{H}^3$, such
that Fermi-Bose cancellations occur in the Casimir energy densities, to the
precisions required for thick pipe geometries to exist, the Euler number of
the compact quotient will have to be of larger order of magnitude than $1$.
Thus, if the spin connection was embedded in the $E_8$ gauge group on the
inner surface of the thick pipe, where we live, the number of generations of
chiral fermions would be of larger order of magnitude than $1$, in
contradiction with experiment.

In fact, as studied by Pilch and Schellekens {\cite{Pilch Schellekens}}, the
fact that the holonomy group of the compact six-manifold, in the present case,
is $\mathrm{SU} \left( 3 \right) \times U \left( 1 \right)$, rather than SO(6)
or SU(3), implies that there exist additional ways in which the spin
connection could be embedded in $E 8 \times E 8$, such that the topological
constraint is satisfied, and in some of these ways, part of the U(1) part of
the spin connection, is embedded in the $E_8$ on the inner surface of the
thick pipe.  However, it would seem likely that, for any uniform embedding of
part of the U(1) part of the spin connection, in the $E_8$ on the inner
surface of the thick pipe, in a manner that is independent of position on the
compact six-manifold, the number of chiral fermion modes, on the inner surface
of the thick pipe, would still be comparable, in order of magnitude, to the
Euler number of the compact six-manifold.  Thus I shall assume that the entire
spin connection is embedded in the $E_8$ gauge group, on the outer surface of
the thick pipe.

In Section \ref{E8 vacuum gauge fields and the Standard Model}, on page
\pageref{E8 vacuum gauge fields and the Standard Model}, I
shall introduce some $E_8$ vacuum gauge fields, on the inner surface of the
thick pipe, localized on Hodge - de Rham harmonic two-forms, and partly
topologically stabilized by a form of Dirac quantization condition, in order
to break $E_8$ to the Standard Model at around 140 TeV, if the couplings are
evolved in the Standard Model up to unification, and produce a small number of
chiral fermions, on the inner surface of the thick pipe, where we live.  This
has to be done without spoiling the satisfaction of Witten's topological
constraint, and I shall also require that, in the context of Lukas, Ovrut, and
Waldram's harmonic expansion, as discussed above, the modification to the
leading term in each harmonic expansion, resulting from the introduction of
these localized $E_8$ vacuum gauge fields, on the inner surface of the thick
pipe, is a small perturbation of the value which the leading term had, in the
absence of these localized $E_8$ vacuum gauge fields.  Thus the analysis of
the Casimir energy density corrected field equations and boundary conditions,
in the present Section, should still be a good first approximation, when the
localized $E_8$ vacuum gauge fields are introduced, in Section
\ref{E8 vacuum gauge fields and the
Standard Model}.  The idea is that the compactifications studied in the
present Section, should provide a strong, stiff, robust ``platform'', that
will only be slightly perturbed, by the interesting physics of the Standard
Model, taking place on the ``platform''.  However, it is necessary to note
that, since Fermi-Bose cancellations will be required to take place, to a
certain precision, in the leading term in the harmonic expansions of the
Casimir energy density contributions to the energy-momentum tensor, on the
inner surface of the thick pipe, even the small changes to this leading term,
resulting from the introduction of the localized $E_8$ vacuum gauge fields, on
the inner surface of the thick pipe, might imply that a smooth compact
quotient of $\mathbf{C} \mathbf{H}^3$, for which the cancellations occur
to the required precision, in the absence of the localized $E_8$ gauge fields,
on the inner surface of the thick pipe, might have to be replaced by a
different smooth compact quotient of $\mathbf{C} \mathbf{H}^3$, when those
localized $E_8$ gauge fields, are introduced.  I will not be able to
determine, in the present paper, whether such a substitution would be likely
to be necessary, and I will simply assume that, if such a substitution is
necessary, then it is made.

If I had chosen the compact six-manifold to be a smooth compact quotient of
the real hyperbolic space $\mathbf{H}^6$, rather than of $\mathbf{C}
\mathbf{H}^3$, then $R_{ABCD} \left( h \right)$ would have been a constant
multiple of $\left( h_{AC} h_{BD} - h_{AC} h_{BC} \right)$, and $\mathrm{tr}
R_{\left[ AB \right.} R_{\left. CD \right]}$ would have vanished identically,
so that Witten's topological constraint would not have given any nontrivial
constraints, and there would not have been any need to embed the spin
connection in the gauge group.  The symmetry between $y_1$ and $y_2$ would, in
that case, have remained unbroken, at this stage.  Nevertheless, we will see
that, in this case, thick pipe solutions, very similar to those obtained for
suitable quotients of $\mathbf{C} \mathbf{H}^3$, will still exist,
provided that the Casimir energy-momentum tensor coefficients satisfy
relations similar to those required for quotients of $\mathbf{C}
\mathbf{H}^3$.  The reason for this is that the terms in the energy-momentum
tensor, quadratic in $G_{ABCD}$, the field strength of the three-form gauge
field, as determined by the Ho\v{r}ava-Witten boundary conditions, are only
significant, for the $\mathbf{C} \mathbf{H}^3$ thick pipe solutions, in at
most a very small fraction of the bulk, and, in fact, at most, only for a
small fraction of the region $y_1 < y < \kappa^{2/9}$, while for the
case of TeV-scale gravity, we will find that $\left( y_2 - y_1 \right) \sim
10^{15} \kappa^{2/9}$.  Typical solutions of the Einstein equations
break the symmetry between $y_1$ and $y_2$, even when it is unbroken to start
with, because either $b \left( y \right)$ increases monotonically with
increasing $y$, while $a \left( y \right)$ decreases, or vice versa.  I always
choose the solutions for which $b \left( y \right)$ increases with increasing
$y$, while the warp factor, $a \left( y \right)$, decreases, since, by
assumption, we live at $y_1$, with $y_1 < y_2$.

Although I mainly consider compactification on quotients of $\mathbf{C}
\mathbf{H}^3$, in this paper, there are two reasons why compactification on
quotients of $\mathbf{H}^6$ might turn out to be preferable.  Firstly, on
the basis of existing knowledge, the number of smooth compact quotients of
$\mathbf{H}^6$, up to a given value of the modulus of the Euler number,
might be very much larger than the number of smooth compact quotients of
$\mathbf{C} \mathbf{H}^3$, up to the same value of the modulus of the
Euler number, as I shall discuss in Section \ref{Smooth compact quotients of
CH3 H6 H3 and S3}, on page \pageref{Smooth compact quotients of CH3 H6 H3 and
S3}.  And secondly, if the large number of
chiral fermion modes, on the outer surface of the thick pipe, for smooth
compact quotients of $\mathbf{C} \mathbf{H}^3$, should turn out to be a
phenomenological problem, it might be preferable to look for suitable smooth
compact quotients of $\mathbf{H}^6$, since there is no need to embed the
spin connection of the compact six-manifold in the gauge group, for smooth
compact quotients of $\mathbf{H}^6$.  However, it seems possible that the
most important criterion, that might favour either $\mathbf{C}
\mathbf{H}^3$ or $\mathbf{H}^6$, is that the local contributions to the
coefficients $D^{\left( i \right)}_1$, in (\ref{surface Casimir action
density}), should vanish, if this is necessary, in order to have an infinite
number of smooth compact quotients, with arbitrarily small, but nonvanishing,
values of the $D^{\left( i \right)}_1$.

\subsubsection{The field equations and boundary conditions for the
three-form \\
gauge field}
\label{The three form gauge field}

When the compact six-manifold is a smooth compact quotient of $\mathbf{C}
\mathbf{H}^3$, we can use the
ansatz of Lukas, Ovrut, Stelle, and Waldram {\cite{Lukas Ovrut Stelle
Waldram}}, (LOSW), for the four-form field strength in the bulk, namely that
$G_{IJKL}$ vanishes unless all four indices are on the compact six-manifold,
and:
\begin{equation}
  \label{four form field strength} G_{ABCD} = \frac{1}{6} \alpha h_{ABCDEF}
  h^{EG} h^{FH} \omega_{GH}
\end{equation}
for $y_1 < y < y_2$, where $\alpha$ is a fixed number, to be determined by the
boundary conditions, $h_{ABCDEF}$ is the tensor $\sqrt{h} \epsilon_{ABCDEF}$,
where $h$ is given by (\ref{determinant of CHn metric}), and
$\epsilon_{123456} =
1$, and $\omega_{GH}$ is the K\"ahler form, given by (\ref{Kahler form}).  This
satisfies the Bianchi identities, and field equations, in the bulk, due to its
independence from $y$, and from position in the four-dimensional locally de
Sitter space, the covariant constancy of $h_{ABCDEF}$, $h^{EG}$, and
$\omega_{FH}$, and the fact that there are not enough
non-vanishing components of $ G_{IJKL} $, for the $ G_{I_1\dots I_{11}}
G^{I_4\dots I_7} G^{I_8\dots I_{11}} $ term in the field equations, to be
nonzero.  Here $ G_{I_1\dots I_{11}} $ denotes the tensor $ \sqrt{-G}
\epsilon_{I_1\dots I_{11}} $.

To confirm that the vanishing of $G_{IJKL}$, unless all four indices are on
the compact six-manifold, is consistent with the boundary conditions (\ref{G
discontinuity}) or (\ref{G boundary condition}), after making the
substitutions (\ref{FF to FF RR substitutions}), we recall, from the preceding
subsection, that for the metric ansatz (\ref{metric ansatz}), and for all
cases of $\mathrm{tr} R_{\left[ UV \right.} R_{\left. WX \right]}$, other than
$\mathrm{tr} R_{\left[ \mu \nu \right.} R_{\left. \sigma \tau \right]}$, which
was not considered there, the value of $\mathrm{tr} R_{\left[ UV \right.}
R_{\left. WX \right]} = R^{\hspace{0.8ex} \hspace{0.4ex} \hspace{0.4ex}
\hspace{0.4ex} \hspace{0.4ex} \hspace{0.4ex} \hspace{0.4ex} YZ}_{\left[ UV
\right.} R_{\left. WX \right] YZ}$ is the same, regardless of whether the
Riemann tensors are calculated from the restriction of the metric to the
ten-dimensional orbifold hyperplanes, or are taken to be the components on the
orbifold hyperplanes, of the Riemann tensor in eleven dimensions.
Furthermore, all cases of $\mathrm{tr} R_{\left[ UV \right.} R_{\left. WX
\right]}$ with mixed components vanish identically, and $\mathrm{tr} R_{\left[
AB \right.} R_{\left. CD \right]} = R^{\hspace{0.6ex} \hspace{0.4ex}
\hspace{0.4ex} \hspace{0.4ex} \hspace{0.4ex} \hspace{0.4ex} \hspace{0.4ex}
EF}_{\left[ AB \right.} \left( h \right) R_{\left. CD \right] EF} \left( h
\right)$.  For the case of $\mathrm{tr} R_{\left[ \mu \nu \right.} R_{\left.
\sigma \tau \right]}$, we find, by a calculation precisely analogous to the
case of $\mathrm{tr} R_{\left[ AB \right.} R_{\left. CD \right]}$, that
$R^{\hspace{0.4ex} \hspace{0.4ex} \hspace{0.4ex} \hspace{0.4ex} \hspace{0.4ex}
\hspace{0.4ex} IJ}_{\left[ \mu \nu \right.} R_{\left. \sigma \tau \right] IJ}
= R^{\hspace{0.4ex} \hspace{0.4ex} \hspace{0.4ex} \hspace{0.4ex}
\hspace{0.4ex} \hspace{0.4ex} \rho \eta}_{\left[ \mu \nu \right.} R_{\left.
\sigma \tau \right] \rho \eta} = R^{\hspace{0.4ex} \hspace{0.4ex}
\hspace{0.4ex} \hspace{0.4ex} \hspace{0.4ex} \hspace{0.4ex} \rho \eta}_{\left[
\mu \nu \right.} \left( g \right) R_{\left. \sigma \tau \right] \rho \eta}
\left( g \right)$, so, as with the other cases, the result is the same,
regardless of whether the Riemann tensors are calculated from the restriction
of the metric to the ten-dimensional orbifold hyperplanes, or are taken to be
the components on the orbifold hyperplanes, of the Riemann tensor in eleven
dimensions, and this remains true, even if ``$\mathrm{tr}$'' sums the contracted
indices, over all eleven dimensions.  However, the metric $g_{\mu \nu}$ is
locally de Sitter, specifically dS$_4$, with de Sitter radius equal to $1$, so
we have $R_{\mu \nu \rho \eta} \left( g \right) = g_{\mu \eta} g_{\nu \rho} -
g_{\mu \rho} g_{\nu \eta}$.  Hence $R^{\hspace{0.4ex} \hspace{0.4ex}
\hspace{0.4ex} \hspace{0.4ex} \hspace{0.4ex} \rho \eta}_{\mu \nu} \left( g
\right) R_{\sigma \tau \rho \eta} \left( g \right) = 2 \left( g_{\mu \sigma}
g_{\nu \tau} - g_{\mu \tau} g_{\nu \sigma} \right)$, hence $R^{\hspace{0.4ex}
\hspace{0.4ex} \hspace{0.4ex} \hspace{0.4ex} \hspace{0.4ex} \hspace{0.4ex}
\rho \eta}_{\left[ \mu \nu \right.} \left( g \right) R_{\left. \sigma \tau
\right] \rho \eta} \left( g \right) = 0$.  Thus the boundary conditions are,
indeed, consistent with the vanishing of $G_{IJKL}$, unless all four indices
are on the compact six-manifold.

To determine $\alpha$, we note that, in consequence of the decision to embed
the spin connection, of the compact six-manifold, in the $E_8$ at the outer
surface of the thick pipe, it follows from (\ref{G discontinuity}), after
making the substitutions (\ref{FF to FF RR substitutions}), that near $y =
y_1$, we have:
\begin{equation}
  \label{G near y1} G_{ABCD} = \frac{3}{2 \sqrt{2}} \frac{\kappa^2}{\lambda^2}
  \epsilon \left( y - y_1 \right) \mathrm{tr} R_{\left[ AB \right.} R_{\left. CD
  \right]} + \ldots
\end{equation}
while near $y = y_2$, we have:
\begin{equation}
  \label{G near y2} G_{ABCD} = - \frac{3}{2 \sqrt{2}}
  \frac{\kappa^2}{\lambda^2} \epsilon \left( y - y_2 \right) \mathrm{tr}
  R_{\left[ AB \right.} R_{\left. CD \right]} + \ldots
\end{equation}
Thus, setting $y = y_{1 +}$, in (\ref{G near y1}), and $y = y_{2 -}$, in
(\ref{G near y2}), we see that the boundary conditions are consistent with
$G_{ABCD}$ taking the constant value:
\begin{equation}
  \label{G for y1 lt y lt y2} G_{ABCD} = \frac{3}{2 \sqrt{2}}
  \frac{\kappa^2}{\lambda^2} \mathrm{tr} R_{\left[ AB \right.} R_{\left. CD
  \right]} = \frac{3}{2 \sqrt{2}} \frac{\kappa^2}{\lambda^2} R^{\hspace{0.4ex}
  \hspace{0.4ex} \hspace{0.4ex} \hspace{0.4ex} \hspace{0.4ex} \hspace{0.4ex}
  \hspace{0.4ex} EF}_{\left[ AB \right.} \left( h \right) R_{\left. CD \right]
  EF} \left( h \right)
\end{equation}
for $ y_1 < y < y_2 $.

Now in the complex coordinate system of subsection \ref{CH3}, for the compact
six-manifold, we have, from (\ref{CHn Riemann
tensor}), and (\ref{CHn covariant Riemann tensor}), that:
\begin{equation}
  \label{CH3 RR} R^{\hspace{0.4ex} \hspace{0.4ex} \hspace{0.4ex}
  \hspace{0.4ex} EF}_{r \bar{s}} \left( h \right) R_{t \bar{u} EF} \left( h
  \right) = - 10 h_{r \bar{s}} h_{t \bar{u}} - 2 h_{r \bar{u}} h_{t \bar{s}}
\end{equation}
Hence:
\begin{equation}
  \label{antisymmetrized CH3 RR} R^{\hspace{0.4ex}
  \hspace{0.6ex} \hspace{0.4ex} \hspace{0.6ex} EF}_{\left[ r \bar{s} \right.}
  \left( h \right) R_{\left. t \bar{u} \right] EF} \left( h \right) = - 4
  \left( h_{r \bar{s}} h_{t \bar{u}} - h_{r \bar{u}} h_{t \bar{s}} \right)
\end{equation}
On the other hand, in the complex coordinate system, we have:
\begin{equation}
  \label{antisymmetric tensor h} h_{rst \bar{u} \bar{v} \bar{w}} = i \sqrt{h}
  \epsilon_{rst} \epsilon_{\bar{u} \bar{v} \bar{w}} = 6 i \sqrt{h}
  \mathcal{A}_{rst} \delta_{r \bar{u}} \delta_{s \bar{v}} \delta_{t \bar{w}}
\end{equation}
where the factor of $i$ is present because $h_{ABCDEF}$ is a tensor, which is
real in a real coordinate system, and on transforming to complex coordinates,
for example, by the matrix $U$, in (\ref{definition of U}), $h_{1 \bar{1} 2
\bar{2} 3 \bar{3}}$ acquires a factor $\left( \det U \right)^{- 1} = \left( -
i \right)^{- 3} = - i$, $h$ is given by (\ref{determinant of CHn metric}),
with $n = 3$, and $g$ rewritten as $h$, and the symbol $\mathcal{A}$, with a
list of indices underneath it, denotes the antisymmetrization of the
expression that follows it, under permutations of those indices.  Thus, from
(\ref{four form field strength}), we have:
\[ G_{rs \bar{t} \bar{u}} = \frac{1}{6} \alpha h_{rs \bar{t} \bar{u} EF}
   h^{EG} h^{FH} \omega_{GH} = \frac{1}{3} i \alpha h_{rs \bar{t} \bar{u} v
   \bar{w}} h^{v \bar{x}} h^{\bar{w} z} h_{\bar{x} z} = - \frac{1}{3} \alpha
   \left( \sqrt{h} \epsilon_{rsv} \epsilon_{\bar{t} \bar{u} \bar{w}} \right)
   h^{\bar{w} v} = \]
\begin{equation}
  \label{G in complex coordinates} = - \frac{1}{3} \alpha \left( h_{r \bar{x}}
  h_{s \bar{z}} h_{v \bar{q}} \epsilon_{xzq} \epsilon_{\bar{t} \bar{u}
  \bar{w}} \right) h^{\bar{w} v} = - \frac{1}{3} \alpha \left( h_{r \bar{t}}
  h_{s \bar{u}} - h_{r \bar{u}} h_{s \bar{t}} \right)
\end{equation}
Comparing with (\ref{G for y1 lt y lt y2}), and (\ref{antisymmetrized CH3
RR}), we see that the ansatz (\ref{four form field strength}) is, indeed,
consistent with the boundary conditions, and that
\begin{equation}
  \label{alpha in terms of kappa and lambda} \alpha = - 9 \sqrt{2}
  \frac{\kappa^2}{\lambda^2} = - \frac{9}{\sqrt{2} \pi} \left( \frac{\kappa}{4
  \pi} \right)^{\frac{2}{3}}
\end{equation}
where I used (\ref{one over lambda squared}), at the last step.  However, it
is also interesting to consider compactification on smooth compact quotients
of $\mathbf{H}^6$, for which there is no need to embed the spin connection
in the gauge group, so that we can set $G_{IJKL} = 0$.  I shall therefore
often leave the above coefficient, $\alpha$, in the Einstein equations, so
that the results for $\mathbf{C} \mathbf{H}^3$ can be obtained by setting
$\alpha = - \frac{9}{\sqrt{2} \pi} \left( \frac{\kappa}{4 \pi}
\right)^{\frac{2}{3}}$, and the results for $\mathbf{H}^6$ obtained by
setting $\alpha = 0$.

Making use of the K\"ahler geometry identity $\omega_{AB} h^{BC} \omega_{CD}
h^{DE} = - \delta^{\hspace{0.4ex} \hspace{0.6ex} E}_A$, and the relation
$G^{AB} = \frac{1}{b^2} h^{AB}$, we find, from (\ref{four form field
strength}), that:
\begin{equation}
  \label{quadratic in four form field strength 1} G^{BF} G^{CG} G^{DH}
  G_{ABCD} G_{EFGH} = \frac{4 \alpha^2}{3 b^8} G_{AE}
\end{equation}
and
\begin{equation}
  \label{quadratic in four form field strength 2} G^{AE} G^{BF} G^{CG} G^{DH}
  G_{ABCD} G_{EFGH} = \frac{8 \alpha^2}{b^8}
\end{equation}
Now, in the upstairs framework, the contribution of the three-form gauge
field, to the energy-momentum tensor, (\ref{energy momentum tensor}), for the
bulk action (\ref{upstairs bulk action}), in eleven dimensions, is:
\begin{equation}
  \label{three form energy momentum tensor} T^{\left( 3 f ) \right.}_{IJ} \!
  \! = \! \! \frac{1}{\kappa^2} \left( \frac{1}{6} G^{KN} G^{LO} G^{MP}
  G_{IKLM} G_{JNOP} - \frac{1}{48} G_{IJ} G^{QR} G^{KN} G^{LO} G^{MP} G_{QKLM}
  G_{RNOP} \right)
\end{equation}
Hence the non-vanishing components of $T^{\left( 3 f \right)}_{IJ}$ are:
\begin{equation}
  \label{three form energy momentum tensor components} T^{\left( 3 f
  \right)}_{\mu \nu} = - \frac{\alpha^2}{6 \kappa^2 b^8} G_{\mu \nu},
  \hspace{2.8em} T^{\left( 3 f \right)}_{AB} = \frac{\alpha^2}{18 \kappa^2
  b^8} G_{AB}, \hspace{2.8em} T^{\left( 3 f \right)}_{yy} = -
  \frac{\alpha^2}{6 \kappa^2 b^8}
\end{equation}
These contributions to the energy-momentum tensor are of the form (\ref{C 2
and C 3 when t 3 equals t 1}), on page~\pageref{C 2 and C 3 when t 3 equals t
1}, for $n = 0$, with $C^{\left( 1 \right)}_0$ negative.  They have been
calculated here, for nonzero $ \alpha $, only for the special case of the
standard embedding of the spin connection in the gauge group on the outer
surface of the thick pipe, when the compact six-manifold $ \mathcal{M}^6 $ is a
smooth compact quotient of $ \mathbf{CH}^3 $.  However it seems reasonable to
expect that in the approximation of restricting the energy-momentum tensor to
the leading term in the Lukas-Ovrut-Waldram harmonic expansion on $
\mathcal{M}^6 $ \cite{Lukas Ovrut Waldram}, as done throughout this section,
the same result would be obtained, but with a different value of $ \alpha $,
for the contributions to the energy-momentum tensor from the vacuum
configurations of the three-form gauge field that result, due to
the Ho\v{r}ava-Witten modified Bianchi identity (\ref{Bianchi identity with FF
only}), from the presence of
general topologically stabilized vacuum Yang-Mills fields on the
Ho\v{r}ava-Witten orbifold hyperplanes, with non-vanishing field strengths
tangential to $ \mathcal{M}^6 $.

It seems unlikely that the bulk Green-Schwarz term \cite{Vafa Witten, Duff Liu
Minasian} would have a significant
effect, near the inner surface of the thick pipe, because the extended
dimensions can to a good approximation be treated as flat, in this region,
and the bulk Green-Schwarz term includes an antisymmetric tensor, with
eleven indices, and the expression contracted with this antisymmetric tensor
would, in the approximation that the extended dimensions are treated as flat,
not have any nonvanishing components with enough different indices, to give a
nonvanishing result.  I shall assume that the bulk Green-Schwarz term does
not have any significant effect on the field equations of either the
three-form gauge field or the metric, for the geometries considered in the
present paper.

\subsubsection{The field equations and boundary conditions for the metric}

\label{The field equations and boundary conditions for the metric}

By analogy with (\ref{Einsteins field equations}), the field equations for
the gravitational field $G_{IJ}$, in the upstairs picture, in eleven
dimensions, are:
\begin{equation}
  \label{Einsteins field equations in eleven dimensions} R_{IJ} - \frac{1}{2}
  RG_{IJ} + \kappa^2 T_{IJ} = 0
\end{equation}
where $T_{IJ}$ is now defined by (\ref{energy momentum tensor}), with $\left(
S_{\mathrm{SM}} + S_{\mathrm{DM}} \right)$ replaced by the sum of all terms in the
quantum effective action $\Gamma$, in the upstairs picture in eleven
dimensions, except for the Ricci scalar term in (\ref{upstairs bulk action}).
We note that, due to the incompatibility of a cosmological constant in eleven
dimensions with local supersymmetry in eleven dimensions \cite{Nicolai Townsend
van Nieuwenhuizen, Sagnotti Tomaras, Bautier Deser Henneaux Seminara}, there is
not expected to be any $ d = 11 $ cosmological constant term, in the low energy
expansion of $\Gamma$.

The Einstein equations (\ref{Einsteins field equations in eleven dimensions})
can alternatively be written:
\begin{equation}
  \label{Einsteins equations in terms of T IJ} R_{IJ} + \kappa^2 \left( T_{IJ}
  - \frac{1}{9} G_{IJ} G^{KL} T_{KL} \right) = 0
\end{equation}
Now subject to the assumptions and approximations discussed in subsections
\ref{The Casimir energy density corrections} and \ref{The three form gauge
field}, $T_{IJ}$ will have the block diagonal structure (\ref{T IJ block
diagonal structure}), on page \pageref{T IJ block diagonal structure}.
Thus, using the Ricci tensor components (\ref{Ricci tensor for the metric
ansatz}), on page \pageref{Ricci tensor for the metric ansatz}, the Einstein
equations (\ref{Einsteins equations in terms of T IJ}) become:
\begin{equation}
  \label{first Einstein equation} \frac{\ddot{a} }{a} + 3 \hspace{0.4ex}
  \frac{\dot{a}^2}{a^2} + 6 \frac{\dot{a} \dot{b}}{ab} - \frac{3}{a^2} +
  \frac{\kappa^2}{9}\left( 5t^{\left( 1 \right)} \left( y \right)
  -6t^{\left( 2 \right)} \left( y \right)
  -t^{\left( 3 \right)} \left( y \right) \right) = 0
\end{equation}
\begin{equation}
  \label{second Einstein equation} \frac{\ddot{b}}{b} + 5
  \frac{\dot{b}^2}{b^2} + 4 \frac{\dot{a} \dot{b}}{ab} + \frac{4}{b^2} +
  \frac{\kappa^2}{9}\left( -4t^{\left( 1 \right)} \left( y \right)
  +3t^{\left( 2 \right)} \left( y \right)
  -t^{\left( 3 \right)} \left( y \right) \right) = 0
\end{equation}
\begin{equation}
  \label{third Einstein equation} 4 \frac{\ddot{a}}{a} + 6 \frac{\ddot{b}}{b}
  + \frac{\kappa^2}{9}\left( -4t^{\left( 1 \right)} \left( y \right)
  -6t^{\left( 2 \right)} \left( y \right)
  +8t^{\left( 3 \right)} \left( y \right) \right) = 0
\end{equation}
where the $ t^{\left(i\right)}\left(y\right) $ satisfy the conservation
equation (\ref{conservation equation for the t i}), on page
\pageref{conservation equation for the t i}.

We next need the boundary conditions for the metric, at $y_1$ and $y_2$.
Because of the simple structure of the metric ansatz (\ref{metric ansatz}), we
can obtain these either directly from the above Einstein equations, with
appropriate delta function terms in the $t^{\left( i \right)} \left( y
\right)$, located on the orbifold fixed point ten-manifolds, or alternatively,
from the Israel matching conditions {\cite{Israel, Chamblin Reall}}, which are
obtained by including a Gibbons-Hawking term {\cite{York, Gibbons Hawking}} in
the action on the boundary.  We recall from subsection \ref{Horava-Witten
theory}, that Moss's improved form of Ho\v{r}ava-Witten theory, which for the
purposes of the present paper I assume to be valid, includes a
supersymmetrized Gibbons-Hawking boundary term.

Considering first the direct approach,
the energy-momentum tensor $\tilde{T}_{UV}^{\left[ i \right]}$, $i = 1, 2$, on
the Ho\v{r}ava-Witten orbifold hyperplane at $y = y_i$, has the block-diagonal
structure (\ref{T tilde i UV block diagonal structure}), on page \pageref{T
tilde i UV block diagonal structure}, by assumption.
Hence $G^{KL} \tilde{T}^{\left[ i \right]}_{KL} = 4 \tilde{t}^{\left[ i \right]
\left( 1 \right)} + 6 \tilde{t}^{\left[ i \right] \left( 2 \right)}$.  Thus, by
(\ref{Einsteins equations in terms of T IJ}) and (\ref{T IJ block diagonal
structure}),
the first Einstein equation (\ref{first Einstein equation}) will include
delta function terms $\frac{\kappa^2}{9} \left( 5 \tilde{t}^{\left[ i \right]
\left( 1 \right)} - 6 \tilde{t}^{\left[ i \right] \left( 2 \right)} \right)
\delta \left( y - y_i \right)$, the second Einstein equation (\ref{second
Einstein equation}) will include delta function terms $\frac{\kappa^2}{9}
\left( - 4 \tilde{t}^{\left[ i \right] \left( 1 \right)} + 3 \tilde{t}^{
\left[ i \right] \left( 2 \right)} \right) \delta \left( y - y_i \right)$,
and the third Einstein equation (\ref{third Einstein equation}) will include
delta function terms \\
$- \frac{\kappa^2}{9} \left( 4 \tilde{t}^{\left[ i \right] \left( 1 \right)} +
6 \tilde{t}^{\left[ i \right] \left(
2 \right)} \right) \delta \left( y - y_i \right)$.  To match these delta
function terms, the slopes of $a \left( y \right)$, and $b \left( y \right)$,
must be discontinuous, at $y_1$, and $y_2$.  Furthermore, by the orbifold
conditions, $a \left( y \right)$, and $b \left( y \right)$, are to be
symmetric, under reflection about $y_1$, and under reflection about $y_2$.
Thus, near $y = y_1$, we must have, for example:
\begin{equation}
  \label{a near y1} a \left( y \right) = a_1 + \sigma \left| y - y_1 \right| +
  O \left( y - y_1 \right)^2
\end{equation}
If we now consider the Einstein equations, (\ref{first Einstein equation}),
(\ref{second Einstein equation}), and (\ref{third Einstein equation}), in the
vicinity of $y_1$ and $y_2$, and drop all terms except the delta function
terms, we find:
\begin{equation}
  \label{first Einstein equation delta function terms} \frac{\ddot{a} }{a} +
  \frac{\kappa^2}{9} \left( 5 \tilde{t}^{\left[ 1 \right] \left( 1 \right)} - 6
  \tilde{t}^{\left[ 1 \right] \left( 2 \right)} \right) \delta \left( y - y_1
  \right) + \frac{\kappa^2}{9} \left( 5 \tilde{t}^{\left[ 2 \right] \left( 1
  \right)} - 6 \tilde{t}^{\left[ 2 \right] \left( 2 \right)} \right) \delta
  \left( y - y_2 \right) = 0
\end{equation}
\begin{equation}
  \label{second Einstein equation delta function terms} \frac{\ddot{b}}{b} +
  \frac{\kappa^2}{9} \left( - 4 \tilde{t}^{\left[ 1 \right] \left( 1 \right)} +
  3 \tilde{t}^{\left[ 1 \right] \left( 2 \right)} \right) \delta \left( y - y_1
  \right) + \frac{\kappa^2}{9} \left( - 4 \tilde{t}^{\left[ 2 \right] \left( 1
  \right)} + 3 \tilde{t}^{\left[ 2 \right] \left( 2 \right)} \right) \delta
  \left( y - y_2 \right) = 0
\end{equation}
\begin{equation}
  \label{third Einstein equation delta function terms} 4 \frac{\ddot{a}}{a} +
  6 \frac{\ddot{b}}{b} - \frac{\kappa^2}{9} \left( 4 \tilde{t}^{\left[ 1
  \right] \left( 1 \right)} + 6 \tilde{t}^{\left[ 1 \right] \left( 2 \right)}
  \right) \delta \left( y - y_1 \right) - \frac{\kappa^2}{9} \left( 4
  \tilde{t}^{\left[ 2 \right] \left( 1 \right)} + 6 \tilde{t}^{\left[ 2 \right]
  \left( 2 \right)} \right) \delta \left( y - y_2 \right) = 0
\end{equation}
The third of these three equations follows from the first two, so we only need
to consider the first two.  Considering the first equation, near $y = y_1$, we
find that $\sigma$, in (\ref{a near y1}), is given by $\sigma = -
\frac{\kappa^2}{18} \left( 5 \tilde{t}^{\left[ 1 \right] \left( 1 \right)} - 6
\tilde{t}^{\left[ 1 \right] \left( 2 \right)} \right) a \left( y_1 \right)$.
Thus we find $\left. \frac{\dot{a}}{a} \right|_{y = y_{1 +}} = - \frac{
\kappa^2}{18} \left( 5 \tilde{t}^{\left[ 1 \right] \left( 1 \right)} - 6
\tilde{t}^{\left[ 1 \right] \left(
2 \right)} \right)$.  The other boundary conditions follow similarly, and we
find:
\begin{equation}
  \left. \left. \label{boundary conditions at y1} \frac{\dot{a}}{a} \right|_{y
  = y_{1 +}} = \frac{\kappa^2}{18} \left( - 5 \tilde{t}^{\left[ 1 \right]
  \left( 1 \right)} + 6 \tilde{t}^{\left[ 1 \right] \left( 2 \right)} \right),
  \hspace{6.0ex}
  \frac{\dot{b}}{b} \right|_{y = y_{1 +}} = \frac{\kappa^2}{18} \left( 4
  \tilde{t}^{\left[ 1 \right] \left( 1 \right)} - 3 \tilde{t}^{\left[ 1 \right]
  \left( 2 \right)} \right)
\end{equation}
\begin{equation}
  \left. \left. \label{boundary conditions at y2} \frac{\dot{a}}{a} \right|_{y
  = y_{2 -}} = \frac{\kappa^2}{18} \left( 5 \tilde{t}^{\left[ 2 \right]
  \left( 1 \right)} - 6 \tilde{t}^{\left[ 2 \right] \left( 2 \right)} \right),
  \hspace{6.0ex}
  \frac{\dot{b}}{b} \right|_{y = y_{2 -}} = \frac{\kappa^2}{18} \left( - 4
  \tilde{t}^{\left[ 2 \right] \left( 1 \right)} + 3 \tilde{t}^{\left[ 2 \right]
  \left( 2 \right)} \right)
\end{equation}
Alternatively, we can obtain the boundary conditions from the Israel matching
conditions {\cite{Israel, Chamblin Reall}}, which read:
\begin{equation}
  \label{Israel matching conditions} \left\{ K_{UV} - KH_{UV} \right\} = -
  \kappa^2 \tilde{T}_{UV}
\end{equation}
Here $H_{UV}$ is defined to be the components tangential to the orbifold
fixed-point hyperplane, of the projection tensor $H_{IJ} = G_{IJ} - n_I n_J$,
where $n_I$ is the unit normal pointing out of the fixed-point hyperplane, on
one side.  The curly braces denote summation over both sides of the
fixed-point hyperplane.  $K_{UV}$ is the extrinsic curvature of the
fixed-point hyperplane, defined by $K_{UV} = H^{\hspace{0.4ex} \hspace{0.4ex}
\hspace{0.4ex} I}_U H^{\hspace{0.4ex} \hspace{0.4ex} \hspace{0.4ex} J}_V D_I
n_J$, which is symmetric under swapping $U$ and $V$, because $n_J$ will be the
gradient of a scalar function, that takes a fixed value on the fixed-point
hyperplane, and whose gradient is normalized, at each point on the fixed-point
hyperplane, so that $G^{IJ} n_I n_J = 1$ there.  $K = H^{UV} K_{UV}$.  And
$\tilde{T}_{UV}$ is the energy-momentum tensor on the fixed-point hyperplane,
as above.

In the present case, if we first consider the $y = y_{1 +}$ side of the fixed
point hyperplane at $y = y_1$, we have $n_y = 1$, and all other components of
$n_I$ vanish, and $H_{UV}$ is simply the components $G_{UV}$ of $G_{IJ}$.
Furthermore, $K_{UV} = - \Gamma_{UV}^y$, hence, from (\ref{Christoffel symbols
for the metric ansatz}),
\begin{equation}
  \label{extrinsic curvature} \left. K_{\mu \nu} \right|_{y = y_{1 +}} =
  \left. \frac{\dot{a}}{a} \right|_{y = y_{1 +}} G_{\mu \nu}, \hspace{8.0ex}
  \left. K_{AB} \right|_{y = y_{1 +}} = \left. \frac{\dot{b}}{b} \right|_{y =
  y_{1 +}} G_{AB} \hspace{2.0ex}
\end{equation}
\begin{equation}
  \label{trace of extrinsic curvature} \left. K \right|_{y = y_{1 +}} = 4
  \left. \frac{\dot{a}}{a} \right|_{y = y_{1 +}} + 6 \left. \frac{\dot{b}}{b}
  \right|_{y = y_{1 +}}
\end{equation}
At $y = y_{1 -}$, $n_y = - 1$, and $\dot{a}$ and $\dot{b}$ have also been
multiplied by $- 1$, so we recover the boundary conditions (\ref{boundary
conditions at y1}), from the Israel matching conditions (\ref{Israel matching
conditions}).  And we also recover the boundary conditions (\ref{boundary
conditions at y2}), in a similar manner.

The energy-momentum tensors $\tilde{T}_{UV}^{\left[ i \right]}$, corresponding
to the bosonic part of the Yang-Mills action (\ref{Yang Mills action}), are
given by:
\begin{equation}
  \label{Yang Mills T AB} \tilde{T}^{\left[ i \right] \mathrm{YM}}_{AB} =
  \frac{1}{\lambda^2}\left( G^{CD}
  \mathrm{tr} F^{\left[ i \right]}_{AC} F^{\left[ i \right]}_{BD} - \frac{1}{4}
  G_{AB} G^{CD} G^{EF} \mathrm{tr} F^{\left[ i \right]}_{CE} F^{\left[ i
  \right]}_{DF} \right)
\end{equation}
\begin{equation}
  \label{Yang Mills T mu nu} \tilde{T}^{\left[ i \right] \mathrm{YM}}_{\mu \nu}
  = - \frac{1}{4 \lambda^2} G_{\mu \nu} G^{CD} G^{EF} \mathrm{tr} F^{\left[ i
  \right]}_{CE} F^{\left[ i \right]}_{DF}
\end{equation}
Now for compact quotients of $\mathbf{C} \mathbf{H}^3$, the spin
connection has been embedded in the $E_8$ at the outer surface of the thick
pipe, while $F^{\left[ 1 \right]}_{AB}$, and consequently
$\tilde{T}_{UV}^{\left[ 1 \right]}$, is zero.  And for compact quotients of
$\mathbf{H}^6$, the Yang-Mills fields are zero on both surfaces of the thick
pipe, and consequently $\tilde{T}_{UV}^{\left[ i \right]} = 0$, for both $i =
1$ and $i = 2$.

For the case of $\mathbf{C} \mathbf{H}^3$, and $i = 2$, we recall, from
subsections \ref{Horava-Witten theory} and \ref{Wittens topological
constraint}, that for $E_8$, ``$\mathrm{tr}$'' means $\frac{1}{30}$ of the trace
in the adjoint representation, and that the trace of the square of a generator
of SO(16), in the adjoint representation of $E_8$, is $30$ times the trace of
the square of the corresponding generator, in the vector representation of
SO(16).  Furthermore, the $E_8$ generators being used, are hermitian.  Thus we
have:
\begin{equation}
  \label{tr F 2 F 2} \mathrm{tr} F^{\left[ 2 \right]}_{AC} F^{\left[ 2
  \right]}_{BD} = R^{\hspace{0.4ex} \hspace{0.4ex} \hspace{0.4ex}
  \hspace{0.4ex} \hspace{0.4ex} \hspace{0.4ex} EF}_{AC} R_{BDEF}
\end{equation}
I shall now assume that the Riemann tensor $R^{\hspace{0.4ex} \hspace{0.4ex}
\hspace{0.4ex} \hspace{0.4ex} \hspace{0.4ex} \hspace{0.4ex} \hspace{0.4ex}
\hspace{0.4ex} \hspace{0.6ex} F}_{ACE}$, that is embedded in the $E_8$ on the
outer surface of the thick pipe, is the Riemann tensor $R^{\hspace{0.4ex}
\hspace{0.4ex} \hspace{0.4ex} \hspace{0.4ex} \hspace{0.4ex} \hspace{0.4ex}
\hspace{0.4ex} \hspace{0.4ex} \hspace{0.6ex} F}_{ACE} \left( h \right)$,
calculated from the induced metric $G_{UV}$, on the outer surface of the thick
pipe, and not the restriction to the outer surface of the thick pipe, of the
eleven-dimensional Riemann tensor.  We then have $R^{\hspace{0.4ex}
\hspace{0.4ex} \hspace{0.4ex} \hspace{0.4ex} \hspace{0.4ex} \hspace{0.6ex}
EF}_{AC} R_{BDEF} = R^{\hspace{0.4ex} \hspace{0.4ex} \hspace{0.4ex}
\hspace{0.4ex} \hspace{0.4ex} \hspace{0.6ex} EF}_{AC} \left( h \right)
R_{BDEF} \left( h \right)$, hence, from (\ref{CH3 RR}), we have, in the
complex coordinate system, that:
\begin{equation}
  \label{tr F 2 F 2 in terms of metric} \mathrm{tr} F^{\left[ 2 \right]}_{r
  \bar{s}} F^{\left[ 2 \right]}_{t \bar{u}} = - 10 h_{r \bar{s}} h_{t \bar{u}}
  - 2 h_{r \bar{u}} h_{t \bar{s}}
\end{equation}
\begin{equation}
  \label{G tr F 2 F 2 in complex coordinates} G^{CD} \mathrm{tr} F^{\left[ 2
  \right]}_{rC} F^{\left[ 2 \right]}_{\bar{u} D} = 16 \frac{1}{b^2} h_{r
  \bar{u}}
\end{equation}
We also have $G^{CD} \mathrm{tr} F^{\left[ 2 \right]}_{rC} F^{\left[ 2
\right]}_{uD} = G^{CD} \mathrm{tr} F^{\left[ 2 \right]}_{\bar{r} C} F^{\left[ 2
\right]}_{\bar{u} D} = 0$.  Hence:
\begin{equation}
  \label{G tr F 2 F 2 in general coordinates} G^{CD} \mathrm{tr} F^{\left[ 2
  \right]}_{AC} F^{\left[ 2 \right]}_{BD} = 16 \frac{1}{b^2} h_{AB} = 16
  \frac{1}{b^4} G_{AB}
\end{equation}
\begin{equation}
  \label{Yang Mills T 2 UV} \tilde{T}^{\left[ 2 \right] \mathrm{YM}}_{\mu \nu} =
  - \frac{24}{\lambda^2} G_{\mu \nu} \frac{1}{b^4}, \hspace{9.0ex}
  \tilde{T}^{\left[ 2 \right] \mathrm{YM}}_{AB} = - \frac{8}{\lambda^2} G_{AB}
  \frac{1}{b^4} \hspace{2.0ex}
\end{equation}
\begin{equation}
  \label{Yang Mills t 2 i} \tilde{t}^{\left[ 2 \right] \left( 1 \right)
  \mathrm{YM}} = - \frac{24}{\lambda^2 b^4}, \hspace{9.0ex} \tilde{t}^{\left[
  2 \right] \left( 2 \right)
  \mathrm{YM}} = - \frac{8}{\lambda^2 b^4} \hspace{2.0ex}
\end{equation}
When the functions $b \left( y \right)$, or $a \left( y \right)$, are shown
without arguments, they are evaluated at the appropriate value of $y$, which
for $\tilde{T}_{UV}^{\left[ 2 \right]}$, $F_{UV}^{\left[ 2 \right]}$, and
$\tilde{t}^{\left[ 2 \right] \left( i \right)}$, is at $y_2$.

Now, as discussed in subsection \ref{Horava-Witten theory}, the low energy
expansion of the quantum effective action, $\Gamma$, on the orbifold fixed
point hyperplanes, is believed to contain terms quadratic in the Riemann
tensor, of the Lovelock-Gauss-Bonnet form, obtained from the Yang-Mills
actions (\ref{Yang Mills action}), by the substitutions (\ref{FF RR action
substitutions}).  The corresponding term in $\Gamma$, at $y_i$, is:
\begin{equation}
  \label{Gamma i LGB} \Gamma_{\mathrm{LGB}}^{\left[ i \right]} = \frac{3}{4
  \lambda^2} \int_{\mathcal{M}^{10}_i} d^{10} x \sqrt{- G} R^{\hspace{0.4ex}
  \hspace{0.4ex} \hspace{0.4ex} \hspace{0.4ex} \hspace{0.4ex} \hspace{0.4ex}
  \hspace{0.4ex} \left[ UV \right.}_{\left[ UV \right.} R^{\left.
  \hspace{0.4ex} \hspace{0.4ex} \hspace{0.4ex} \hspace{0.4ex} \hspace{0.4ex}
  \hspace{0.4ex} \hspace{0.4ex} \hspace{0.4ex} \left. WX \right]
  \right.}_{\left. WX \right]}
\end{equation}
I shall now assume, as in the calculation above, of the Yang-Mills
energy-momentum tensor, when the spin connection is embedded in the gauge
group, that the Riemann tensor $R^{\hspace{0.4ex} \hspace{0.4ex}
\hspace{0.4ex} \hspace{0.4ex} \hspace{0.4ex} \hspace{0.4ex} \hspace{0.4ex}
\hspace{0.4ex} \hspace{1.3ex} X}_{UVW}$, in (\ref{Gamma i LGB}), is the
Riemann tensor calculated from the induced metric $G_{UV}$, on the orbifold
fixed point ten-manifold $\mathcal{M}^{10}_i$, and not the restriction to the
outer surface of the thick pipe, of the eleven-dimensional Riemann tensor.
Then, since $\mathcal{M}^{10}_i$ is the Cartesian product of the four observed
dimensions, and the compact six-manifold, all the Riemann tensor components,
with mixed indices, vanish identically.  The energy-momentum tensors
$\tilde{T}_{UV}^{\left[ i \right]}$, corresponding to (\ref{Gamma i LGB}),
are:
\[ \tilde{T}^{\left[ i \right] \mathrm{LGB}}_{UV} = - \frac{1}{2 \lambda^2}
   \left( \rule[-1.0ex]{0pt}{4.0ex}
   \tilde{R}_{UWXY} \tilde{R}^{\hspace{0.4ex} \hspace{0.4ex}
   \hspace{0.7ex} WXY}_V - 2 \tilde{R}_{UWVX} \tilde{R}^{WX} - 2
   \tilde{R}_{UW} \tilde{R}^{\hspace{0.4ex} \hspace{0.4ex} \hspace{0.5ex} W}_V
   + \tilde{R}_{UV} \tilde{R} \right. \]
\begin{equation}
  \label{Lovelock Gauss Bonnet T i UV} \hspace{18.0ex} \left. - \frac{1}{4}
  G_{UV} \tilde{R}^{\hspace{0.4ex} \hspace{0.4ex} \hspace{0.4ex}
  \hspace{0.4ex} \hspace{0.4ex} \hspace{0.4ex} \hspace{0.4ex} \hspace{0.4ex}
  \hspace{0.4ex}}_{WXYZ} \tilde{R}^{WXYZ} + G_{UV} \tilde{R}^{\hspace{0.4ex}
  \hspace{0.4ex} \hspace{0.4ex} \hspace{0.4ex}}_{WX} \tilde{R}^{WX}_{} -
  \frac{1}{4} G_{UV} \tilde{R}^2 \right)
\end{equation}
where I have now denoted curvatures calculated from the induced metric
$G_{UV}$, on the orbifold fixed point ten-manifold $\mathcal{M}^{10}_i$, by a
tilde.

To evaluate (\ref{Lovelock Gauss Bonnet T i UV}), we note that when the
compact six-manifold is a quotient of $\mathbf{C} \mathbf{H}^3$, we have,
from (\ref{CHn covariant Riemann tensor}), that for the metric induced on a
fixed point ten-manifold, by the metric ansatz (\ref{metric ansatz}):
\begin{equation}
  \label{two versions of RR for CH3} \tilde{R}_{ACDE}
  \tilde{R}^{\hspace{0.4ex} \hspace{0.4ex} \hspace{0.7ex} CDE}_B = 16
  \frac{1}{b^4} G_{AB}, \hspace{7.0ex} \tilde{R}_{ACBD} \tilde{R}^{CD} = 16
  \frac{1}{b^4} G_{AB} \hspace{2.0ex}
\end{equation}
Hence, recalling that $g_{\mu \nu}$, in the metric ansatz (\ref{metric
ansatz}), is normalized such that $R_{\mu \nu} \left( g \right) = - 3 g_{\mu
\nu}$, and when the compact six-manifold is a quotient of $\mathbf{C}
\mathbf{H}^3$, $h_{AB}$ is normalized such that $R_{AB} \left( h \right) = 4
h_{AB}$, we find, when the compact six-manifold is a quotient of $\mathbf{C}
\mathbf{H}^3$, that:
\begin{equation}
  \label{scalars formed from R tilde for CH3} \tilde{R}^{\hspace{0.4ex}
  \hspace{0.4ex} \hspace{0.4ex} \hspace{0.4ex} \hspace{0.4ex} \hspace{0.4ex}
  \hspace{0.4ex} \hspace{0.4ex} \hspace{0.4ex}}_{WXYZ} \tilde{R}^{WXYZ} =
  \frac{96}{b^4} + \frac{24}{a^4}, \hspace{4.0ex} \tilde{R}_{WX} \tilde{R}^{WX}
  = \frac{96}{b^4} + \frac{36}{a^4}, \hspace{4.0ex} \tilde{R} = \frac{24}{b^2} -
  \frac{12}{a^2}
\end{equation}
\begin{equation}
  \label{Lovelock Gauss Bonnet T i AB} \tilde{T}^{\left[ i \right]
  \mathrm{LGB}}_{AB} = \frac{3}{\lambda^2} \left( \frac{4}{b^4} - \frac{16}{a^2
  b^2} + \frac{1}{a^4} \right) G_{AB}
\end{equation}
\begin{equation}
  \label{Lovelock Gauss Bonnet T i mu nu} \tilde{T}^{\left[ i \right]
  \mathrm{LGB}}_{\mu \nu} = \frac{36}{\lambda^2} \left( \frac{1}{b^4} -
  \frac{1}{a^2 b^2} \right) G_{\mu \nu}
\end{equation}
\begin{equation}
  \label{Lovelock Gauss Bonnet t i j} \tilde{t}^{\left[ i \right] \left( 1
  \right)
  \mathrm{LGB}} = \frac{36}{\lambda^2} \left( \frac{1}{b^4} - \frac{1}{a^2 b^2}
  \right), \hspace{6.0ex} \tilde{t}^{\left[ i \right] \left( 2 \right)
  \mathrm{LGB}} =
  \frac{3}{\lambda^2} \left( \frac{4}{b^4} - \frac{16}{a^2 b^2} +
  \frac{1}{a^4} \right)
\end{equation}
To evaluate (\ref{Lovelock Gauss Bonnet T i UV}) when the compact six-manifold
is a quotient of $\mathbf{H}^6$, we recall that in this case we have chosen $
h_{AB}$, in the metric ansatz (\ref{metric ansatz}), to be normalized such that
$R_{ABCD} \left( h \right) = h_{AC} h_{BD} - h_{AD} h_{BC}$, so that $R_{AB}
\left( h \right) = 5 h_{AB}$, as stated after (\ref{Ricci tensor for the metric
ansatz}), on page \pageref{Ricci tensor for the metric ansatz}.  We then find,
when the compact six-manifold is a quotient of $\mathbf{H}^6$, that:
\begin{equation}
  \label{scalars formed from R tilde for H6} \tilde{R}_{WXYZ} \tilde{R}^{WXYZ}
  = \frac{60}{b^4} + \frac{24}{a^4}, \hspace{4.0ex} \tilde{R}_{WX}
  \tilde{R}^{WX} = \frac{150}{b^4} + \frac{36}{a^4}, \hspace{4.0ex} \tilde{R} =
  \frac{30}{b^2} - \frac{12}{a^2}
\end{equation}
\begin{equation}
  \label{Lovelock Gauss Bonnet T i AB for H6} \tilde{T}^{\left[ i \right]
  \mathrm{LGB}}_{AB} = \frac{3}{\lambda^2} \left( \frac{5}{b^4} - \frac{20}{a^2
  b^2} + \frac{1}{a^4} \right) G_{AB}
\end{equation}
\begin{equation}
  \label{Lovelock Gauss Bonnet T i mu nu for H6} \tilde{T}^{\left[ i \right]
  \mathrm{LGB}}_{\mu \nu} = \frac{45}{\lambda^2} \left( \frac{1}{b^4} -
  \frac{1}{a^2 b^2} \right) G_{\mu \nu}
\end{equation}
\begin{equation}
  \label{Lovelock Gauss Bonnet t i j for H6} \tilde{t}^{\left[ i \right] \left(
  1 \right) \mathrm{LGB}} = \frac{45}{\lambda^2} \left( \frac{1}{b^4} -
  \frac{1}{a^2 b^2} \right), \hspace{6.0ex} \tilde{t}^{\left[ i \right] \left(
  2 \right)
  \mathrm{LGB}} = \frac{3}{\lambda^2} \left( \frac{5}{b^4} - \frac{20}{a^2 b^2}
  + \frac{1}{a^4} \right)
\end{equation}
Now at the inner surface of the thick pipe, we will have $a \left( y_1 \right)
\sim 10^{26}$ metres, while $b \left( y_1 \right)$ will be less than about
$10^{- 19}$ metres, so for $i = 1$, we can neglect the terms with negative
powers of $a$, in (\ref{Lovelock Gauss Bonnet t i j}) and (\ref{Lovelock Gauss
Bonnet t i j for H6}).  On the other hand, we will find solutions where $a$ is
comparable to $b$, at the outer surface of the thick pipe, but these solutions
will not be able to fit the observed values of Newton's constant and the
cosmological constant, and other solutions where $a$ is small compared to $b$,
at the outer surface of the thick pipe, some of which will be able to fit the
observed values of Newton's constant and the cosmological constant.

\subsection{Analysis of the Einstein equations and the boundary conditions for
the metric}
\label{Analysis of the Einstein equations and the boundary conditions}

The Einstein equations (\ref{first Einstein equation}), (\ref{second Einstein
equation}), and (\ref{third Einstein equation}), with the range of $y$
restricted to $y_1 < y < y_2$, together with the boundary conditions
(\ref{boundary conditions at y1}) and (\ref{boundary conditions at y2}), now
constitute a system of coupled ordinary differential equations, and boundary
conditions, for the functions $a \left( y \right)$ and $b \left( y \right)$.

The functions $t^{\left( i \right)} \left( y \right)$, defined by (\ref{T IJ
block diagonal structure}), on page \pageref{T IJ block diagonal structure},
receive contributions from the energy-momentum
tensor of the three-form gauge field, given by (\ref{three form energy
momentum tensor components}) for quotients of $\mathbf{C} \mathbf{H}^3$,
and $0$ for quotients of $\mathbf{H}^6$, and from Casimir effects in the bulk,
near the inner surface of the thick pipe, and, for solutions such that $a
\left( y \right)$ becomes sufficiently small near the outer surface of the
thick pipe, also from Casimir effects in the bulk, near
the outer surface of the thick pipe.

The coefficients
$\tilde{t}^{\left[ i \right] \left( j \right)}$, defined by (\ref{T tilde i UV
block diagonal structure}), receive contributions from the energy-momentum
tensor of the Yang-Mills fields on the outer surface of the thick pipe, given
by (\ref{Yang Mills t 2 i}) for quotients of $\mathbf{C} \mathbf{H}^3$,
and $0$ for quotients of $\mathbf{H}^6$; from the leading terms in the
Lukas-Ovrut-Waldram harmonic expansion, on the compact six-manifold $
\mathcal{M}^6 $, of the energy-momentum tensor of topologically stabilized
vacuum Yang-Mills fields on the inner surface of the thick pipe;
from the Lovelock-Gauss-Bonnet
energy-momentum tensor on the surfaces of the thick pipe, given by
(\ref{Lovelock Gauss Bonnet t i j}) for quotients of $\mathbf{C}
\mathbf{H}^3$, and by (\ref{Lovelock Gauss Bonnet t i j for H6}) for
quotients of $\mathbf{H}^6$; and from Casimir effects on the inner surface of
the thick pipe, and, for solutions such that $a \left( y \right)$ becomes
sufficiently small at the outer surface of the thick pipe, also from Casimir
effects on the outer surface of the thick pipe.

The functions $t^{\left( i \right)} \left( y \right)$, and the coefficients
$\tilde{t}^{\left[ i \right] \left( j \right)}$, are required to be recovered
self-consistently, when they are recalculated for the solution of the Einstein
equations and the boundary conditions.

The equations are invariant under a uniform shift of $y$, $y_1$, and $y_2$,
but as already noted, in the discussion following (\ref{metric ansatz}), I
shall use this freedom to obtain the simplest form of the solution in the
bulk, near the inner surface of the thick pipe, rather than to set $y_1$ or
$y_2$ to any particular value.

Eliminating the double derivatives between the three Einstein equations, we
find:
\begin{equation}
  \label{Einstein equation without double derivatives} \frac{\dot{a}^2}{a^2} +
  4 \frac{\dot{a} \dot{b}}{ab} + \frac{5 \dot{b}^2}{2 b^2} + \frac{2}{b^2} -
  \frac{1}{a^2} - \frac{1}{6} \kappa^2 t^{\left( 3 \right)} = 0
\end{equation}
When the functions $t^{\left( i \right)} \left( y \right)$ are shown without
arguments, they are evaluated at $y$.  From (\ref{Einstein equation without
double derivatives}), we find:
\begin{equation}
  \label{a dot over a} \frac{\dot{a}}{a} = - 2 \frac{\dot{b}}{b} \pm
  \frac{1}{2} \sqrt{6 \frac{\dot{b}^2}{b^2} - \frac{8}{b^2} + \frac{4}{a^2} +
  \frac{2}{3} \kappa^2 t^{\left( 3 \right)}}
\end{equation}
The second Einstein equation, (\ref{second Einstein equation}), now becomes:
\begin{equation}
  \label{second Einstein equation without a dot} \frac{\ddot{b}}{b} - 3
  \frac{\dot{b}^2}{b^2} \pm 2 \frac{\dot{b}}{b} \sqrt{6 \frac{\dot{b}^2}{b^2}
  - \frac{8}{b^2} + \frac{4}{a^2} + \frac{2}{3} \kappa^2 t^{\left( 3 \right)}}
  + \frac{4}{b^2} + \frac{\kappa^2}{9} \left( - 4 t^{\left( 1 \right)} + 3
  t^{\left( 2 \right)} - t^{\left( 3 \right)} \right) = 0
\end{equation}
Now differentiating (\ref{a dot over a}) with respect to $y$, we find:
\begin{equation}
  \label{differentiated a dot over a} \frac{\ddot{a}}{a} -
  \frac{\dot{a}^2}{a^2} = - 2 \frac{\ddot{b}}{b} + 2 \frac{\dot{b}^2}{b^2} \pm
  \frac{1}{4 R} \left( 12 \frac{\ddot{b} \dot{b}}{b^2} - 12
  \frac{\dot{b}^3}{b^3} + 16 \frac{\dot{b}}{b^3} - 8 \frac{\dot{a}}{a^3} +
  \frac{2}{3} \kappa^2 \dot{t}^{\left( 3 \right)} \right)
\end{equation}
where I defined
\begin{equation}
  \label{definition of the square root R} R \equiv \sqrt{6
  \frac{\dot{b}^2}{b^2} - \frac{8}{b^2} + \frac{4}{a^2} + \frac{2}{3} \kappa^2
  t^{\left( 3 \right)}}
\end{equation}
Now, using the formula (\ref{differentiated a dot over a}) for $\ddot{a}$, the
left-hand side of the first Einstein equation, (\ref{first Einstein
equation}), becomes:
\[ \left( - 2 \pm \frac{3 \dot{b}}{Rb} \right) \left( \frac{\ddot{b}}{b} - 3
   \frac{\dot{b}^2}{b^2} \pm 2 \frac{\dot{b}}{b} R + \frac{4}{b^2} +
   \frac{\kappa^2}{9} \left( - 4 t^{\left( 1 \right)} + 3 t^{\left( 2 \right)}
   - t^{\left( 3 \right)} \right) \right) \]
\[ + 4 \hspace{0.4ex} \left( \frac{\dot{a}^2}{a^2} + 4 \frac{\dot{a}
   \dot{b}}{ab} + \frac{5 \dot{b}^2}{2 b^2} + \frac{2}{b^2} - \frac{1}{a^2} -
   \frac{1}{6} \kappa^2 t^{\left( 3 \right)} \right) \]
\[ - \left( 10 \frac{\dot{b}}{b} \pm \frac{2}{3 R} \kappa^2 \left( t^{\left( 3
   \right)} - t^{\left( 1 \right)} \right) \pm \frac{2}{Ra^2} \right) \left(
   \frac{\dot{a}}{a} - \left( - 2 \frac{\dot{b}}{b} \pm \frac{1}{2} R \right)
   \right) \]
\begin{equation}
  \label{first Einstein equation in terms of other equations} \pm \frac{1}{6
  R} \kappa^2 \left( \dot{t}^{\left( 3 \right)} + \left( 4 \frac{\dot{a}}{a} +
  6 \frac{\dot{b}}{b} \right) t^{\left( 3 \right)} - 4 \frac{\dot{a}}{a}
  t^{\left( 1 \right)} - 6 \frac{\dot{b}}{b} t^{\left( 2 \right)} \right)
\end{equation}
and thus vanishes when (\ref{a dot over a}) and (\ref{second Einstein equation
without a dot}) and the conservation equation (\ref{conservation equation for
the t i}) are satisfied, provided that the square root, (\ref{definition of
the square root R}), is nonvanishing.

Now the third Einstein equation, (\ref{third Einstein equation}), is
equivalent to (\ref{a dot over a}), provided that the first two Einstein
equations are satisfied.  Thus (\ref{a dot over a}) and (\ref{second Einstein
equation without a dot}), taken together, imply that all three Einstein
equations are satisfied, provided that the conservation equation
(\ref{conservation equation for the t i}) is satisfied, and the square root,
(\ref{definition of the square root R}), is nonvanishing.  This is true
whichever choice of sign we take in (\ref{a dot over a}) and (\ref{second
Einstein equation without a dot}), provided that we choose either the upper
sign in both equations, or the lower sign in both equations.

Now we are seeking solutions in the region $y_1 \leq y \leq y_2$, such that
for $y$ close to $y_1$, $a \left( y \right)$ is very large, and $b \left( y
\right)$ is very small.  Thus we may neglect the term $\frac{4}{a^2}$, in the
square root, for $y$ close to $y_1$, in this region.  In that case,
(\ref{second Einstein equation without a dot}) becomes an ordinary
differential equation for $b \left( y \right)$, since, in the approximations
discussed above, the $t^{\left( i \right)} \left( y \right)$ only depend on
$y$, through $b \left( y \right)$, in this region.  Moreover, we are looking
for solutions that realize the ADD mechanism {\cite{ADD1, ADD2}}, by a form of
thick pipe geometry, so we require $\dot{b} > 0$, for $y$ greater than $y_1$,
and close to $y_1$.

It is convenient to define $c \left( y \right) \equiv \dot{b}$, so that
$\ddot{b} = c \, \frac{dc}{db}$.  Then (\ref{second Einstein equation without a
dot}) reduces to a first order differential equation, for $c$ as a function of
$b$:
\begin{equation}
  \label{second Einstein equation for c in terms of b} \frac{c}{b}
  \frac{dc}{db} - 3 \frac{c^2}{b^2} \pm 2 \frac{c}{b} \sqrt{6 \frac{c^2}{b^2}
  - \frac{8}{b^2} + \frac{2}{3} \kappa^2 t^{\left( 3 \right)}} + \frac{4}{b^2}
  + \frac{\kappa^2}{9} \left( - 4 t^{\left( 1 \right)} + 3 t^{\left( 2
  \right)} - t^{\left( 3 \right)} \right) = 0
\end{equation}
We can now carry out a qualitative analysis of the differential equation
(\ref{second Einstein equation for c in terms of b}), in the $\left( b, c
\right)$ plane.  We are interested in the quadrant $b > 0$, $c > 0$.  For a
fixed choice of the sign of the square root, (\ref{second Einstein equation
for c in terms of b}) defines a unique curve through each point in the
quadrant $b > 0$, $c > 0$, such that the argument of the square root is
non-negative.  We can follow such a curve from the inner surface of the thick
pipe, where $b$ is very small.

Suppose, first, we choose the {\emph{lower}} sign of the square root, so
(\ref{second Einstein equation for c in terms of b}) becomes:
\begin{equation}
  \label{second Einstein equation for c in terms of b with lower sign}
  \frac{c}{b}  \frac{dc}{db} - 3 \frac{c^2}{b^2} - 2 \frac{c}{b} \sqrt{6
  \frac{c^2}{b^2} - \frac{8}{b^2} + \frac{2}{3} \kappa^2 t^{\left( 3 \right)}}
  + \frac{4}{b^2} + \frac{\kappa^2}{9} \left( - 4 t^{\left( 1 \right)} + 3
  t^{\left( 2 \right)} - t^{\left( 3 \right)} \right) = 0
\end{equation}
Now the functions $t^{\left( i \right)}$ all decrease rapidly in magnitude,
with increasing $b$, and become negligible as soon as $b$ is large compared to
$\kappa^{2/9}$.  In that case, (\ref{second Einstein equation for c in
terms of b with lower sign}) reduces to:
\begin{equation}
  \label{approximate second Einstein equation for c in terms of b with lower
  sign} \frac{dc}{db} = \frac{3 c^2 - 4}{bc} + \frac{2}{b} \sqrt{2 \left( 3
  c^2 - 4 \right)}
\end{equation}
We require $c \geq \sqrt{\frac{4}{3}}$, in order for the square root to be
real.  Then $\frac{dc}{db} \geq 0$, and will typically be $\sim \kappa^{-
\frac{2}{9}}$, or larger, once $b$ is $\sim \kappa^{2/9}$.  Then once
$b$ has
increased by a few multiples of $\kappa^{2/9}$, $c$ will be large
enough that we can to a reasonable approximation replace $3 c^2 - 4$ by $3
c^2$, and this becomes a better approximation as $c$ increases further.  Then
(\ref{approximate second Einstein equation for c in terms of b with lower
sign}) becomes:
\begin{equation}
  \label{lower sign at large c} \frac{dc}{db} = \left( 3 + 2 \sqrt{6} \right)
  \frac{c}{b} \simeq 7.8990 \frac{c}{b}
\end{equation}
Thus as soon as $b$ is as large as a few multiples of $\kappa^{2/9}$,
we have $c = \frac{db}{dy} \simeq \left( \frac{b}{B} \right)^{7.8990}$, for
some constant $B$, that cannot be much larger than $\kappa^{2/9}$, but
could be smaller, because we could be on a trajectory which starts out with a
large value of $c$, near the inner surface of the thick pipe.  Then $\left(
\frac{b}{B} \right)^{} \simeq 0.7559 \left( \frac{B}{\left( y_3 - y \right)}
\right)^{0.1449}$, where $y_3$ is some constant greater than $y_1$, but such
that $y_3 - y_1$ cannot be large compared to $\kappa^{2/9}$, unless
$b$ somehow remains smaller than $B$, all the way from $y = y_1$ to $y \sim
\left( y_3 - B \right)$, which would require $c$ to be smaller than around
$\frac{\kappa^{2/9}}{\left( y_3 - y_1 \right)}$ for most of this
range.  However, even if the functions $t^{\left( i \right)}$ were such that
this was possible, and the boundary conditions could be satisfied, such a
solution has no classical bulk, because as soon as a value of $y$ is reached,
such that $b$ is larger than $B$, $b$ starts increasing very rapidly, and
would reach infinity, if $y$ increased further by more than $B$.  Thus it is
not possible to find solutions with a thick pipe form of geometry, that can
realize the ADD mechanism in a simple way, without considering the upper
choice of sign, in (\ref{a dot over a}) and (\ref{second Einstein equation
without a dot}).

We now, therefore, choose the upper sign of the square root, so (\ref{second
Einstein equation for c in terms of b}) becomes:
\begin{equation}
  \label{second Einstein equation for c in terms of b with upper sign}
  \frac{c}{b}  \frac{dc}{db} - 3 \frac{c^2}{b^2} + 2 \frac{c}{b} \sqrt{6
  \frac{c^2}{b^2} - \frac{8}{b^2} + \frac{2}{3} \kappa^2 t^{\left( 3 \right)}}
  + \frac{4}{b^2} + \frac{\kappa^2}{9} \left( - 4 t^{\left( 1 \right)} + 3
  t^{\left( 2 \right)} - t^{\left( 3 \right)} \right) = 0
\end{equation}
We start again, at the inner surface of the thick pipe, where $b$ is very
small, and follow a curve in the $\left( b, c \right)$ plane as before, but
defined, this time, by (\ref{second Einstein equation for c in terms of b with
upper sign}).  The functions $t^{\left( i \right)}$ all become negligible, as
before, as soon as $b$ is large compared to $\kappa^{2/9}$.  Then
(\ref{second Einstein equation for c in terms of b with upper sign}) reduces
to:
\begin{equation}
  \label{approximate second Einstein equation for c in terms of b with upper
  sign} \frac{dc}{db} = - \frac{1}{b} \sqrt{3 c^2 - 4} \left( 2 \sqrt{2} -
  \sqrt{3 - \frac{4}{c^2}} \right)
\end{equation}
We again require $c \geq \sqrt{\frac{4}{3}}$, in order for the square root to
be real.  The right hand side of (\ref{approximate second Einstein equation
for c in terms of b with upper sign}) is $< 0$ for all $c >
\sqrt{\frac{4}{3}}$.  The simple dependence on $b$, of the right hand side of
(\ref{approximate second Einstein equation for c in terms of b with upper
sign}), means that the general solution of (\ref{approximate second Einstein
equation for c in terms of b with upper sign}) has the form $c = f \left(
\frac{b}{B} \right)$, for some function $f$, where $B$ is the constant of
integration.  Thus all trajectories, in this region, are related to one
another, by rescaling $b$.  I shall call the solutions of (\ref{approximate
second Einstein equation for c in terms of b with upper sign}) the bulk-type
trajectories.

Now for $c$ large compared to $\sqrt{\frac{4}{3}}$, (\ref{approximate second
Einstein equation for c in terms of b with upper sign}) reduces to
\begin{equation}
  \label{upper sign at large c} \frac{dc}{db} = - \left( 2 \sqrt{6} - 3
  \right) \frac{c}{b} \simeq - 1.8990 \frac{c}{b}
\end{equation}
Thus when $b$ is large compared to $\kappa^{2/9}$, and $c$ large
compared to $\sqrt{\frac{4}{3}}$, we have
\begin{equation}
  \label{bulk type power law for upper sign} c = \frac{db}{dy} \simeq \left(
  \frac{B}{b} \right)^{1.8990},
\end{equation}
for some constant $B$.  There is now no upper limit to how large $B$ can be,
but it cannot be much smaller than $\kappa^{2/9}$.  And for large $B$,
this approximate solution will be valid, throughout the range from $b$
somewhat larger than $\kappa^{2/9}$, to $b$ somewhat smaller than $B$,
and this range of $b$ can be made arbitrarily large, by choosing a
sufficiently large value of $B$.

The above approximate form (\ref{bulk type power law for upper sign}) of $c$,
as a function of $b$, corresponds to
\begin{equation}
  \label{bulk type power law dependence of b on y for upper sign} \left(
  \frac{b}{B} \right) \simeq 1.4436 \left( \frac{ y - y_0 }{B}
  \right)^{0.3449}
\end{equation}
for some $y_0$, which we could choose to set to $0$, by using the invariance
of the equations and boundary conditions, under a uniform shift of $y$, $y_1$,
and $y_2$.

It is convenient to regard $a$ as a function of $b$, in the same way as $c =
\frac{db}{dy}$ is being treated as a function of $b$.  Then in the region where
all the $ t^{\left(i\right)} $ are negligible, the equation (\ref{a dot over
a}) for $\frac{\dot{a}}{a}$, with the upper choice of sign, and dropping the
term $\frac{4}{a^2}$ in the square root, becomes:
\begin{equation}
  \label{a dot over a for a in terms of b in bulk} \frac{c}{a}  \frac{da}{db} =
  - 2 \frac{c}{b} + \frac{1}{2} \sqrt{6 \frac{c^2}{b^2} - \frac{8}{b^2}}
\end{equation}
When $c$ is sufficiently large, that we are on a bulk type power law
trajectory, this becomes:
\begin{equation}
  \label{bulk power law trajectory for a with upper sign} \frac{da}{db} = -
  \left( 2 - \frac{1}{2} \sqrt{6} \right) \frac{a}{b} = - 0.7753 \frac{a}{b}
\end{equation}
Hence:
\begin{equation}
  \label{bulk power law for a in terms of b} a = A \left(
  \frac{\kappa^{2/9}}{b} \right)^{0.7753}
\end{equation}
where $A$ is a constant of integration.

Now we will find, in subsection \ref{Newtons constant and the cosmological
constant}, on page \pageref{Newtons constant and the cosmological constant},
and subsection \ref{Stiffening by fluxes}, on page \pageref{Stiffening by
fluxes}, that for TeV-scale gravity, $ \frac{B}{\kappa^{2/9}}$,
whose value is determined by the boundary conditions at
the inner surface of the thick pipe, is required to have a value
around $\frac{1.5 \times 10^4}{\left| \chi \left(
\mathcal{M}^6 \right) \right|^{0.1715}}$.
Thus with our best rough estimate, (\ref{second estimate of minimum value of b
sub 1 over kappa to the two ninths}), on page \pageref{second estimate of
minimum value of b sub 1 over kappa to the two ninths}, of the minimum value
of $\frac{b_1}{\kappa^{2/9}}$, and the corresponding best rough
estimate of the upper bound on $\left| \chi \left( \mathcal{M}^6 \right)
\right|$ as around $7 \times 10^4$, we see that
$\frac{B}{\kappa^{2/9}}$ will be around $10^4$.  Thus if the
bulk power law, (\ref{bulk type power law for upper sign}), was valid down to
the inner surface of the thick pipe, the value of $c = \frac{db}{dy}$, at the
inner surface of the thick pipe, would be around $10^{8}$.  Thus the
proximity force approximation will certainly not be an adequate approximation
for the Casimir energy densities near the inner surface of the thick pipe, and
it is necessary to consider the effects of going beyond the proximity force
approximation.

\subsubsection{Beyond the proximity force approximation}
\label{Beyond the proximity force approximation}

I shall now consider the effects of including, in the expansions
(\ref{the t i as functions of b}), of the $c^{\left( i \right)}$ near the
inner surface of the thick pipe, and the expansions (\ref{bulk Casimir action
density for small a}), of the $\tilde{c}^{\left( i \right)}$ near the outer
surface of the thick
pipe, terms depending on $c = \dot{b}$, $\ddot{b}$, and higher derivatives of
$b$, with respect to $y$, in the case of (\ref{the t i as functions of b}),
and terms depending on $\dot{a}$, $\ddot{a}$, and higher derivatives of $a$,
with respect to $y$, in the case of (\ref{bulk Casimir action density for small
a}).  For
definiteness, I shall consider (\ref{the t i as functions of b}), near the
inner surface of the thick pipe, with similar considerations applying to
(\ref{bulk Casimir action density for small a}), near the outer surface of
the thick pipe.

Now the terms proportional to $b^{- 8}$, in (\ref{the t i as functions of b}),
get contributions (\ref{three form energy momentum tensor components}), from
the classical energy-momentum tensor, (\ref{three form energy momentum
tensor}), of the three-form gauge field, $C_{IJK}$, for the case of smooth
compact quotients of $\mathbf{C} \mathbf{H}^3$, and contributions from the
$t_8 t_8 R^4$ term, in the low energy expansion of the quantum effective
action of supergravity in eleven dimensions.  The proximity force
approximation is in fact exact, for the three-form gauge field configuration
(\ref{G for y1 lt y lt y2}), but we expect there to be contributions involving
$c$, $\ddot{b}$, $\frac{d^3 b}{dy^3}$, and $\frac{d^4 b}{dy^4}$, coming from
the metric variation of the $t_8 t_8 R^4$ term.

As a guide to the derivatives of $b$ with respect to $y$ that might be
expected, and the powers to which they might occur, at higher orders in the
expansion in $\kappa^{\frac{2}{3}}$, in (\ref{the t i as functions of b}), we
note that for even $n \geq 0$, the term $C^{\left( i \right)}_n
\frac{\kappa^{\frac{2}{3} \left( n - 1 \right)}}{b^{8 + 3 n}}$, in (\ref{the t
i as functions of b}), could come from terms built from $\left( 4 +
\frac{3}{2} n \right)$ Riemann tensors, in the low energy expansion of the
quantum effective action.  Considering, first, just the powers of $c$ that
might occur, we see, from the Riemann tensor components, (\ref{Riemann tensor
for the metric ansatz}), that each power of $\frac{1}{b}$, can bring in up to
one power of $c$.  If we extend this to odd $n \geq 1$ as well, and bear in
mind that for the bulk power-law solution, (\ref{bulk type power law for upper
sign}), $c$ will be very large compared to $1$, near the inner surface of the
thick pipe, the strongest dependence on $c$, that we expect at order
$\kappa^{\frac{2}{3} \left( n - 1 \right)}$, is $\kappa^{\frac{2}{3} \left( n
- 1 \right)}  \frac{c^{8 + 3 n}}{b^{8 + 3 n}}$.

We now need to determine the
range of values of $b$, and of $y$, where such a term could significantly
alter the results of the study of the Einstein equations, and the boundary
conditions for the metric, in the preceding subsections.  If we consider the
second Einstein equation, in the form (\ref{second Einstein equation for c in
terms of b with upper sign}), the ratio of $\frac{c^2}{b^2}$, to
the new term, will be $\left( \kappa^{- \frac{2}{9}}  \frac{b}{c} \right)^{6 +
3 n}$.  And for $b$ small compared to $B$, in (\ref{bulk type power law for
upper sign}), we have $c \simeq \left( \frac{B}{b} \right)^{1.8990}$, almost
right up to the inner surface of the thick pipe, according to subsection
\ref{The boundary conditions at the inner surface of the thick pipe}.  Thus
the ratio of $\frac{c^2}{b^2}$, to the new term, will be $\left( \left(
\frac{b}{\kappa^{2/9}} \right)^{2.8990} \left(
\frac{\kappa^{2/9}}{B} \right)^{1.8990} \right)^{6 + 3 n} = \left(
\left( \frac{b}{\kappa^{2/9}} \right) /^{} \left(
\frac{B}{\kappa^{2/9}} \right)^{0.6551} \right)^{2.8990 \times \left(
6 + 3 n \right)}$.  This is larger than $1$, for $\left(
\frac{b}{\kappa^{2/9}} \right) > \left( \frac{B}{\kappa^{2/9}}
\right)^{0.6551}$.  And, for $\frac{B}{\kappa^{2/9}} \gg 1$, this will
be for most of the range $\kappa^{2/9} < b < B$.  And by (\ref{bulk
type power law dependence of b on y for upper sign}), ignoring factors of
order $1$, $y > \kappa^{2/9}$ implies $b > B \left(
\frac{\kappa^{2/9}}{B} \right)^{0.3449} = \kappa^{2/9} \left(
\frac{B}{\kappa^{2/9}} \right)^{0.6551}$.  Thus the new terms,
involving $c$, will be significant for $y < \kappa^{2/9}$, and will be
likely to alter the conclusions of subsection \ref{The boundary conditions at
the inner surface of the thick pipe}, about this region, but they will be
negligible for $y \gg \kappa^{2/9}$, which for
$\frac{B}{\kappa^{2/9}} \gg 1$, will be most of the bulk.
We note that the point where $\frac{b}{\kappa^{2/9}} \simeq \left(
\frac{B}{\kappa^{2/9}} \right)^{0.6551}$, and $y \sim
\kappa^{2/9}$, is the point where $\frac{b}{\kappa^{2/9}}
\simeq c$.

Considering, now, terms involving higher derivatives of $b$, with respect to
$y$, we see, from (\ref{Riemann tensor for the metric ansatz}), that in
addition to terms proportional to $\frac{c^2}{b^2}$, one Riemann tensor can
also bring in terms proportional to $\frac{\ddot{b}}{b} = \frac{c}{b}
\frac{dc}{db}$, which, by (\ref{upper sign at large c}), is $\sim
\frac{c^2}{b^2}$ in the first bulk power law region, to the extent that
(\ref{approximate second Einstein equation for c in terms of b with upper
sign}), and (\ref{upper sign at large c}), are not significantly altered by
the new terms.  In general, from terms in the low energy expansion of the
effective action, built from polynomials in the Riemann tensor and its
covariant derivatives, we expect terms involving products of expressions
$\frac{c}{b} = \frac{1}{b}  \frac{db}{dy}$, $\frac{1}{b}  \frac{d^2 b}{dy^2}$,
$\frac{1}{b}  \frac{d^3 b}{dy^3}$, $\ldots$, and non-negative powers of
$\frac{1}{b}$.  But by repeated use of (\ref{upper sign at large c}), we find
that $\frac{1}{b}  \frac{d^n b}{dy^n} \sim \frac{c^n}{b^n}$, where $\sim$
means up to constant factors of order $1$.  Thus, at each mass dimension
$\left( 8 + 3 n \right)$, the largest terms, in the first bulk power law
region, where (\ref{upper sign at large c}) and (\ref{bulk type power law for
upper sign}) are approximately valid, that we can build by use of factors
involving higher derivatives of $b$ with respect to $y$, are no larger than
the terms $\kappa^{\frac{2}{3} \left( n - 1 \right)} \frac{c^{8 + 3 n}}{b^{8 +
3 n}}$, whose effect has already been considered.

Thus the effect of going beyond the proximity force approximation, is that the
bulk power law solutions (\ref{bulk type power law for upper sign}), (\ref{bulk
type power law dependence of b on y for upper sign}), and (\ref{bulk power law
for a in terms of b}), are no longer expected to be approximately valid
throughout the whole range from $ b $ somewhat larger than $ \kappa^{\frac{2}
{9}} $, to $ b $ somewhat smaller than $ B $, but rather, only over the
slightly smaller range, from where $y \sim \kappa^{2/9}$, and $b \sim
\kappa^{2/9} \left( \frac{B}{\kappa^{2/9}} \right)^{0.6551}$,
to $ b $ somewhat smaller than $ B $.

Now as I mentioned just before the start of this subsection, we will find, in
subsections \ref{Newtons constant
and the cosmological constant} and \ref{Stiffening by fluxes}, that for
TeV-scale gravity, we require $\frac{B}{\kappa^{ \frac{2}{9}}} \sim \frac{1.5
\times 10^{4}}{\left\vert\chi\left(\mathcal{M}^6\right)\right\vert^{0.1715}}
$.  And from subsection \ref{The expansion parameter}, on page \pageref{The
expansion parameter}, the minimum value of $ b_1 $ allowed by the
Giudice-Rattazzi-Wells estimate of the effective expansion parameter in quantum
gravity in eleven dimensions \cite{Giudice Rattazzi Wells} is $ b_1 \simeq 0.2
\kappa^{2/9} $, which means that $ \left\vert\chi\left(\mathcal{M}^6
\right)\right\vert $ cannot be larger than around $ 7 \times 10^4 $.  Thus $
\frac{B}{\kappa^{2/9}} \sim 10^4 $, so if the bulk power law (\ref{bulk
type power law for upper sign}) continued to be valid until very close to the
inner surface of the thick pipe, we would find $ c = \frac{db}{dy} \sim
10^{8} $ near the inner surface of the thick pipe.

Thus it is clear that the proximity force approximation, in which the Casimir
coefficients in (\ref{the t i as functions of b}) are approximated by their
values as calculated on flat $ \mathbf{R}^5 $ times the compact six-manifold,
will not, in fact, be an adequate approximation.  One way to take account of
this would be to generalize the expansions (\ref{the t i as functions of b}),
so as to include explicit dependence also on $ c = \frac{db}{dy} $, and on
higher derivatives of $ b $ with respect to $ y $, as in the order of magnitude
estimates above.  However this is not an appropriate way to study the detailed
form of the quantum corrections, just as it is not appropriate to
study the relativistic corrections to the Schr\"odinger equation for atoms, by
expanding $ \sqrt{m^2 + \vec{p}^2} \simeq m + \frac{\vec{p}^2}{2m} + \ldots $,
with $ \vec{p} $ interpreted as $ -i\vec{\partial} $, to higher orders in
$ \vec{p} $, because this results in differential equations of higher and
higher order, and correspondingly, larger and larger numbers of constants of
integration, making it difficult to single out the particular solution of
physical interest.

Instead, the appropriate way to study the quantum corrections is to use an
iterative approach, calculating the Casimir corrections for a trial form of
$ b\left(y\right) $ in the quantum region $ b_1 \sim \kappa^{2/9} \leq
b \leq \kappa^{2/9}\left(\frac{B}{\kappa^{2/9}}\right)^{0.6551}
$, and expressing the results, in the approximation of neglecting dependence on
position in the compact six-manifold, or in other words, of neglecting all but
the leading term, in the harmonic expansions of Lukas, Ovrut, and Waldram
\cite{Lukas Ovrut Waldram}, as expansions of the form (\ref{the t i as
functions of b}), depending only on $ b $, and
not on $ c $, or any higher derivatives of $ b $ with respect to $ y $, but
with coefficients that now depend on the trial form of $ b\left(y\right) $ in
the quantum region, and possibly, also, intermediate powers of $ b $, not
present in (\ref{the t i as functions of b}), then solving the field
equations and boundary conditions with these
Casimir coefficients, and if the resulting $ b\left(y\right) $ differs from the
trial $ b\left(y\right) $, repeating the process with a new trial $ b\left(y
\right) $, until a self-consistent solution is found for $ b\left(y\right) $ in
the quantum region, that joins smoothly onto the bulk power law (\ref{bulk type
power law for upper sign}), with the required value of $ \frac{B}{\kappa^{
\frac{2}{9}}} \sim 10^4 $ for TeV-scale gravity, at $ b \gg \kappa^{\frac{2}{9}
}\left(\frac{B}{\kappa^{2/9}}\right)^{0.6551} $.  Of course, the
possibility of finding such a self-consistent $ b\left(y\right) $, in the
quantum region, is likely to depend on the choice of the compact six-manifold
$ \mathcal{M}^6 $.

\subsubsection{The region near the inner surface of the thick pipe}
\label{The region near the inner surface of the thick pipe}

We now consider the region near the inner surface of the thick pipe, to find
out whether a value of $B$ greater than around $10^5 \kappa^{2/9}$
could occur, as required for TeV-scale gravity.  From the discussion above, we
know that the proximity force approximation will not be adequate.  However, we
can start by assuming that the $t^{\left( i \right)}$ functions have
expansions of the form (\ref{the t i as functions of b}), on page \pageref{the
t i as functions of b}, near the inner surface, and see whether the solution
of the Einstein equations can self-consistently reproduce the $t^{\left( i
\right)}$ functions that we started with, and also produce the required large
value of $\frac{B}{\kappa^{2/9}}$.  I shall consider first the case
where $t^{\left( 3 \right)} = t^{\left( 1 \right)}$, near the inner surface of
the thick pipe, so the $C_n^{\left( 2 \right)}$ and $C_n^{\left( 3 \right)}$
will be given by (\ref{C 2 and C 3 when t 3 equals t 1}), in terms of the $
C_n^{\left( 1 \right)}$.

We recall from subsection \ref{The Yang-Mills coupling constants in four
dimensions}, on page \pageref{The Yang-Mills coupling constants in four
dimensions}, that the value $b_1$ of $b$ at the inner surface of the thick
pipe cannot be larger than around $1.2 \kappa^{2/9}$, which
corresponds to $\left| \chi \left( \mathcal{M}^6 \right) \right| \simeq 1$,
and from subsection \ref{The expansion parameter}, on page \pageref{The
expansion parameter}, that it cannot be smaller than around $0.2
\kappa^{2/9}$, which corresponds to $\left| \chi \left( \mathcal{M}^6
\right) \right| \simeq 7 \times 10^4$.

Let us consider, first, the case where all the $C_n^{\left( 1 \right)}$ are
zero, except for a single value of $n$.  Then (\ref{second Einstein equation
for c in terms of b with upper sign}) becomes:
\begin{equation}
  \label{second Einstein equation for c in terms of b with upper sign and
  specific C n} \frac{c}{b}  \frac{dc}{db} - 3 \frac{c^2}{b^2} + 2 \frac{c}{b}
  \sqrt{6 \frac{c^2}{b^2} - \frac{8}{b^2} + \frac{2}{3} C^{\left( 1 \right)}_n
  \frac{\kappa^{\frac{2}{3} \left( n + 2 \right)}}{b^{8 + 3 n}}} +
  \frac{4}{b^2} - \left( \frac{4 + n}{6} \right) C_n^{\left( 1 \right)}
  \frac{\kappa^{\frac{2}{3} \left( n + 2 \right)}}{b^{8 + 3 n}} = 0
\end{equation}
for some fixed value of $n \geq 0$.  Let us now consider $b$ in the region
\begin{equation}
  \label{small b region} b \ll \left( \frac{1}{12} \left\vert C^{\left( 1
  \right)}_n \right\vert \right)^{\frac{1}{6 + 3 n}} \kappa^{2/9}
\end{equation}
so that we can neglect the term $- \frac{8}{b^2}$ in the square root, and the
term $\frac{4}{b^2}$, in (\ref{second Einstein equation for c in terms of b
with upper sign and specific C n}). Then (\ref{second Einstein equation for c
in terms of b with upper sign and specific C n}) becomes:
\begin{equation}
  \label{second Einstein equation for small b} \frac{c}{b}  \frac{dc}{db} - 3
  \frac{c^2}{b^2} + 2 \frac{c}{b} \sqrt{6 \frac{c^2}{b^2} + \frac{2}{3}
  C^{\left( 1 \right)}_n \frac{\kappa^{\frac{2}{3} \left( n + 2 \right)}}{b^{8
  + 3 n}}} - \left( \frac{4 + n}{6} \right) C_n^{\left( 1 \right)}
  \frac{\kappa^{\frac{2}{3} \left( n + 2 \right)}}{b^{8 + 3 n}} = 0
\end{equation}
Let us try for a power law trajectory,
\begin{equation}
  \label{small b power law} c = \sigma \left( \frac{b}{\kappa^{2/9}}
  \right)^{\rho},
\end{equation}
for some numerical constant $\sigma$, and exponent $\rho$.  We then find that
$C^{\left( 1 \right)}_n$ must be negative, which is satisfied for the
energy-momentum tensor of the three-form gauge field, (\ref{three form energy
momentum tensor components}), and corresponds to a positive contribution to
the energy density, $T_{00}$, and that:
\begin{equation}
  \label{power law trajectory for small b} \rho = - \frac{\left( 6 + 3 n
  \right)}{2}, \hspace{6.0ex} \hspace{6.0ex} \sigma = \sqrt{- \frac{C^{\left( 1
  \right)}_n \left( 4 + n \right)^2}{3 \left( 16 + 24 n + 3 n^2 \right)}}
  \hspace{4.0ex}
\end{equation}
Thus we see that, in contrast to the situation for the bulk-type trajectories,
where every trajectory is approximately a power law trajectory, for a certain
range of $b$, that depends on the trajectory, there is now just a single power
law trajectory.  If we now try for a solution of the form $c = c_0 \left( 1 +
s \right)$, where $c_0$ is the power law trajectory, and $s \left( b \right)$
is a small perturbation, we find that
\begin{equation}
  \label{perturbation of power law trajectory for small b} s = Sb^{\eta},
  \hspace{12.0ex} \eta = \frac{16 + 24 n + 3 n^2}{2 \left( 4 + n \right)}
  \hspace{4.0ex}
\end{equation}
where $S$ is a constant of integration.  Now $ \eta \geq 2 $ for $ n \geq 0 $,
so as we follow trajectories near
the power law trajectory, in the direction of decreasing $b$, they tend to
converge towards the power law trajectory, in the sense that $s$ decreases in
magnitude, so in this sense, the power law trajectory is an attractor, in the
direction of decreasing $b$.

Now in the proximity force case, using $t^{\left( 3 \right)} = t^{\left( 1
\right)}$ and (\ref{conservation equation when t 3 equals t 1}), the equation
(\ref{second Einstein equation for c in terms of b with upper sign}) can be
written, in the region where $- \kappa^2 t^{\left( 1 \right)} \gg
\frac{1}{b^2}$, as:
\begin{equation}
  \label{second Einstein equation in terms of C at small b} \frac{c}{b}
  \frac{dc}{db} - 3 \frac{c^2}{b^2} + 2 \frac{c}{b} \sqrt{6 \frac{c^2}{b^2}
  + \frac{2}{3} \kappa^2 t^{\left( 1 \right)}} + \frac{\kappa^2}{9} \left( - 2
  t^{\left( 1 \right)} + \frac{b}{2}  \frac{dt^{\left( 1 \right)}}{db} \right)
  = 0
\end{equation}
The small $b$ power law solutions,
(\ref{small b power law}) and (\ref{power law trajectory for small b}), can
all be written as:
\begin{equation}
  \label{small b attractor trajectory} c^2 = - \frac{b^2}{9} \kappa^2 t^{\left(
  1 \right)} \left( b \right) - \frac{4}{3 \sqrt{6}} b^6  \int^{\infty}_b
  \frac{dx}{x^5} \left( \left( \frac{x}{b} \right)^{4 \sqrt{6}} - \left(
  \frac{b}{x} \right)^{4 \sqrt{6}} \right) \kappa^2 t^{\left( 1 \right)}
  \left( x \right)
\end{equation}
The integral is convergent at $x \rightarrow \infty$, because $t^{\left( 1
\right)} \left( x \right)$ decreases at least as rapidly as $x^{- 8}$, as $x
\rightarrow \infty$.  Now (\ref{small b attractor trajectory}) does not give
an exact solution of (\ref{second Einstein equation in terms of C at small
b}), except when $t^{\left( 1 \right)} \left( b \right)$ is a pure power of
$b$.  In fact, on substituting (\ref{small b attractor trajectory}) into
(\ref{second Einstein equation in terms of C at small b}), the left-hand side
of (\ref{second Einstein equation in terms of C at small b}) reduces to:
\begin{equation}
  \label{remainder for general t 1} 2 \frac{c}{b} \sqrt{6 \frac{c^2}{b^2}
  + \frac{2}{3} \kappa^2 t^{\left( 1 \right)}} + \frac{8}{3} b^4
  \int^{\infty}_b \frac{dx}{x^5} \left( \left( \frac{x}{b} \right)^{4
  \sqrt{6}} + \left( \frac{b}{x} \right)^{4 \sqrt{6}} \right) \kappa^2
  t^{\left( 1 \right)} \left( x \right)
\end{equation}
When $t^{\left( 1 \right)} \left( b \right)$ is not a pure power of $b$, there
are cross terms between different powers of $b$, that do not cancel out of the
first term in (\ref{remainder for general t 1}), but the second term in
(\ref{remainder for general t 1}) is a linear combination of the contributions
from different powers of $b$.  However, if $t^{\left( 1 \right)} \left( b
\right)$ is a linear combination of two different pure powers, say $b^{-
\left( 8 + 3 n \right)}$ and $b^{- \left( 8 + 3 m \right)}$, with $n$ and $m$
large, then the remainder term, (\ref{remainder for general t 1}), is $\sim
\frac{1}{\sqrt{nm}}$, while the leading terms, in (\ref{second Einstein
equation in terms of C at small b}), are $\sim n$ or $m$.  Furthermore, for a
pure power $b^{- \left( 8 + 3 n \right)}$, with large $n$, the integral term,
in (\ref{small b attractor trajectory}), is of order $\frac{1}{n^2}$ compared
to the leading term.

Thus it seems
likely that, when the coefficients in $t^{\left( 1 \right)} \left( b \right)$
are all $\leq 0$, a reasonable approximation to the small $b$ attractor
trajectory, generalizing the small $b$ power laws (\ref{small b power law}),
(\ref{power law trajectory for small b}), valid when $t^{\left( 1 \right)}
\left( b \right)$ is a pure power, will, in the limit of large $- t^{\left( 1
\right)} \left( b \right)$, be:
\begin{equation}
  \label{approximate small b attractor trajectory} c \simeq \sqrt{-
  \frac{b^2}{9} \kappa^2 t^{\left( 1 \right)} \left( b \right)}
\end{equation}

Now for any sufficiently large value of $c$, the trajectory passing through
the point $\left( b, c \right)$ will still be of the bulk power law type
(\ref{bulk type power law for upper sign}), even for $b$ in the range
(\ref{small b region}).  However, when all the coefficients in $ t^{\left(1
\right)}\left(b\right) $ are $ \leq 0 $, any power law trajectory, of the type
(\ref{bulk type power law for upper sign}), will intersect the small $b$
attractor trajectory, (\ref{approximate small b attractor trajectory}), for
sufficiently small $b$.  From the perturbative analysis carried out in
connection with (\ref{perturbation of power law trajectory for small b}), it is
clear that what actually happens, when all the coefficients in $ t^{\left(1
\right)}\left(b\right) $ are $ \leq 0 $, is that each bulk power law
trajectory, (\ref{bulk type power law for upper sign}), curves upwards as it
approaches the small $b$ attractor trajectory, (\ref{approximate small b
attractor trajectory}), and then approaches the small $b$ attractor trajectory
gradually, without ever actually crossing it.

Now regarding $a$ as a function of $b$ again, and considering the case where
$C^{\left( 1 \right)}_n$ is only nonzero, for the same $n$ as in (\ref{second
Einstein equation for c in terms of b with upper sign and specific C n}), the
equation (\ref{a dot over a}) for $\frac{\dot{a}}{a}$, with the upper choice of
sign, and dropping the term $\frac{4}{a^2}$ in the square root, becomes:
\begin{equation}
  \label{a dot over a for a in terms of b} \frac{c}{a}  \frac{da}{db} = - 2
  \frac{c}{b} + \frac{1}{2} \sqrt{6 \frac{c^2}{b^2} - \frac{8}{b^2} +
  \frac{2}{3} C^{\left( 1 \right)}_n \frac{\kappa^{\frac{2}{3} \left( n + 2
  \right)}}{b^{8 + 3 n}}}
\end{equation}
On the unique
small $b$ power law trajectory, defined by (\ref{small b power law}) and
(\ref{power law trajectory for small b}), we can neglect the $- \frac{8}{b^2}$
term in the square root in this equation, which then becomes:
\begin{equation}
  \label{small b power law for a} \frac{da}{db} = - \frac{\left( 4 + 2 n
  \right)}{\left( 4 + n \right)}  \frac{a}{b}
\end{equation}
Hence:
\begin{equation}
  \label{small b power law for a in terms of b} a = A_1 \left(
  \frac{\kappa^{2/9}}{b} \right)^{\frac{4 + 2 n}{4 + n}}
\end{equation}
where $A_1$ is another constant of integration.  For $n = 0$, the solution
defined by (\ref{small b power law}), (\ref{power law trajectory for small
b}), and (\ref{small b power law for a in terms of b}), which corresponds to
$b = \beta \kappa^{2/9} \left( \frac{y}{\kappa^{2/9}}
\right)^{\frac{1}{4}}$, $a = \frac{A_1}{\beta} \left(
\frac{\kappa^{2/9}}{y} \right)^{\frac{1}{4}}$, where $\beta$ is a
constant, has the functional form of the supersymmetric solution found by
Lukas, Ovrut, Stelle, and Waldram {\cite{Lukas Ovrut Stelle Waldram}}, for the
case when the compact six-manifold is a Calabi-Yau threefold with $h_{11} =
1$, transformed to the coordinate system where the metric has the form
(\ref{metric ansatz}).

We now have to consider whether these solutions can be self-consistent, when
we recalculate the expansion coefficients $C_n^{\left( i \right)}$ in
(\ref{the t i as functions of b}) for $b \left( y \right)$ corresponding to
these solutions, in accordance with the discussion in the preceding
subsection.  We can no longer assume that $t^{\left( 3 \right)} = t^{\left( 1
\right)}$, but since we are now just considering orders of magnitude, it will
be adequate to consider the case where $t^{\left( 3 \right)} = t^{\left( 1
\right)}$.  Let us suppose that in the quantum region, where
$\frac{b}{\kappa^{2/9}} < \left( \frac{B}{\kappa^{2/9}}
\right)^{0.6551}$, we have a power law, $c = \left( \frac{B_q}{b}
\right)^{\gamma}$, with $\gamma \geq 0$, which joins continuously onto the
bulk power law (\ref{bulk type power law for upper sign}), at
$\frac{b}{\kappa^{2/9}} = \left( \frac{B}{\kappa^{2/9}}
\right)^{0.6551}$.  Then
\begin{equation}
  \label{continuity condition for B sub q} \left(
  \frac{B_q}{\kappa^{2/9}} \right)^{\gamma} = \left(
  \frac{B}{\kappa^{2/9}} \right)^{0.6551 \left( 1 + \gamma \right)}
  \simeq 400^{1 + \gamma},
\end{equation}
where I used that $\frac{B}{\kappa^{2/9}} \simeq 10^4$ for TeV-scale
gravity.  Now by the preceding subsection, we expect a term $\kappa^{-
\frac{22}{9}} C^{\left( i \right)}_n \left( \frac{\kappa^{2/9}}{b}
\right)^{8 + 3 n}$, in (\ref{the t i as functions of b}), to be accompanied by
an additional term $\sim \kappa^{- \frac{22}{9}} C^{\left( i \right)}_n \left(
\frac{\kappa^{2/9}}{b} \right)^{8 + 3 n} c^{8 + 3 n}$.  This now
becomes:
\begin{equation}
   \label{additional term in t i for trial power law}
   \kappa^{- \frac{22}{9}} C^{\left( i \right)}_n \left(
   \frac{\kappa^{2/9}}{b} \right)^{8 + 3 n} \left( \frac{B_q}{b}
   \right)^{\gamma \left( 8 + 3 n \right)} \simeq \kappa^{- \frac{22}{9}}
   C^{\left( i \right)}_n \left( 400 \frac{\kappa^{2/9}}{b}
   \right)^{\left( 1 + \gamma \right) \left( 8 + 3 n \right)}
\end{equation}
and thus contributes expansion coefficients $C^{\left( i \right)}_{\tilde{n}}
\simeq 400^{8 + 3 \tilde{n}} C^{\left( i \right)}_n$ to the recalculated
$t^{\left( i \right)}$, where $\tilde{n} = \left( 1 + \gamma \right) n +
\frac{8}{3} \gamma$.  If we now consider the case where $C^{\left( 1
\right)}_{\tilde{n}}$ is significant only for one value of $\tilde{n}$, and
assume that the significant $C^{\left( 1 \right)}_{\tilde{n}}$ is negative,
then by (\ref{small b power law}) and (\ref{power law trajectory for small
b}), the recalculated $c$, calculated from the recalculated $t^{\left( i
\right)}$, is
\begin{equation}
  \label{recalculated c} \sim 400^{4 + \frac{3}{2} \tilde{n}} \left(
  \frac{\kappa^{2/9}}{b} \right)^{3 + \frac{3}{2} \tilde{n}},
\end{equation}
where I dropped all factors of order $1$.  This is in agreement with the $c$
we started with at the upper limit of the quantum region, where
$\frac{b}{\kappa^{2/9}} \simeq 400$, but for all $\gamma \geq 0$, and
all $n \geq 0$, increases much more rapidly with decreasing $b$ than the $c$
we started with, and for $b \sim \kappa^{2/9}$, is very large compared
to the $c$ we started with.  We would not expect the discrepancy to be any
smaller if more than one $C^{\left( 1 \right)}_{\tilde{n}}$ is significant,
provided all the significant $C^{\left( 1 \right)}_{\tilde{n}}$ are negative.
Thus we cannot obtain a self-consistent solution if all the significant
$C^{\left( 1 \right)}_{\tilde{n}}$ are negative.

Now since the proximity force approximation is not valid in the quantum region
$b_1 \sim \kappa^{2/9} \leq b \leq \kappa^{2/9} \left(
\frac{B}{\kappa^{2/9}} \right)^{0.6551}$, we cannot assume that $ t^{
\left(3\right)} = t^{\left(1\right)} $, but for the purpose of illustration, I
shall continue to consider the case where $ t^{\left(3\right)} = t^{\left(1
\right)} $.  Then by the result above, if a self-consistent solution with $
\frac{B}{\kappa^{2/9}} \simeq 10^4 $ exists, the self-consistent
$t^{\left( 1 \right)}$, in (\ref{T IJ block diagonal structure}) and (\ref{the
t i as functions of b}), must contain at least one significant $C^{\left( 1
\right)}_{\tilde{n}}$ that is positive, which corresponds to a negative
contribution to the energy density $T_{00}$.  This is
expected to be possible for Casimir energy densities, whose sign often depends
on the detailed geometry of a physical situation \cite{0106045 Bordag Mohideen
Mostepanenko}, although recent results of Kenneth and Klich \cite{Kenneth
Klich} and Bachas \cite{Bachas} have shown that Casimir forces are always
attractive in certain circumstances.

If $t^{\left( 1 \right)}$ is dominated by a single term $\kappa^{-
\frac{22}{9}} C^{\left( 1 \right)}_n \left( \frac{\kappa^{2/9}}{b}
\right)^{8 + 3 n}$ in (\ref{the t i as functions of b}), where $C^{\left( 1
\right)}_n$ is positive, then as noted after
(\ref{small b power law}), there is no small $b$ power law solution of
(\ref{second Einstein equation for small b}), for that $C^{\left( 1
\right)}_n$.  Instead, the generic solution of (\ref{second Einstein equation
for small b}), with $C^{\left( 1 \right)}_n > 0$, with $c$ viewed as a
function of $b$, in the quadrant $b > 0$, $c > 0$ of the $\left( b, c \right)$
plane, has a peak at a point $\left( b_p, c_p \right)$ where
\begin{equation}
  \label{maximum value of c on a small b curve with increasing c} c_p =
  \frac{\sqrt{2 \sqrt{6 n^2 + 36 n + 64} + 3 n + 4}}{3 \sqrt{10}}
  \sqrt{C^{\left( 1 \right)}_n} \left( \frac{\kappa^{2/9}}{b_p}
  \right)^{3 + \frac{3 n}{2}}.
\end{equation}

Every point in the quadrant $b > 0$, $c > 0$ of the $\left( b, c
\right)$ plane must now lie on a trajectory that has such a peak, for if
$\frac{dc}{db}$ is positive at the point $\left( b, c \right)$, and we follow
the trajectory in the direction of increasing $b$, the terms in (\ref{second
Einstein equation for small b}) proportional to $\frac{1}{b^{8 + 3 n}}$ will
eventually become negligible, and the trajectory will then take the form
(\ref{bulk type power law for upper sign}), for some $B > 0$, so that
$\frac{dc}{db}$ is now negative.  Now suppose that $\frac{dc}{db}$ is negative
at the point $\left( b, c \right)$, and follow the trajectory in the direction
of decreasing $b$.  If $C^{\left( 1 \right)}_n \frac{\kappa^{\frac{2}{9}
\left( 3 n + 6 \right)}}{b^{8 + 3 n}}$ is small compared to $\frac{c^2}{b^2}$,
then the trajectory has the form (\ref{bulk type power law for upper sign}),
so $C^{\left( 1 \right)}_n \frac{\kappa^{\frac{2}{9} \left( 3 n + 6
\right)}}{b^{8 + 3 n}}$ increases more rapidly than $\frac{c^2}{b^2}$ with
decreasing $b$, and a value of $b > 0$ will be reached where the two terms are
comparable in magnitude.  Then either the two terms continue to be comparable
in magnitude as $b$ decreases further, or $C^{\left( 1 \right)}_n
\frac{\kappa^{\frac{2}{9} \left( 3 n + 6 \right)}}{b^{8 + 3 n}}$ becomes large
compared to $\frac{c^2}{b^2}$, as $b$ decreases further.  But if the two terms
continue to be comparable in magnitude as $b$ decreases further, then we have
$\frac{c^2}{b^2} \simeq \alpha C^{\left( 1 \right)}_n
\frac{\kappa^{\frac{2}{9} \left( 3 n + 6 \right)}}{b^{8 + 3 n}}$, for some
constant $\alpha > 0$, for all $b$ from the value $> 0$ where the two terms
first become comparable in magnitude, down to $b = 0$.  But this is the
characteristic property of the small $b$ power law trajectory (\ref{small b
power law}), (\ref{power law trajectory for small b}), and the trajectories
that asymptotically approach it, in the direction of decreasing $b$, in the
sense described after (\ref{perturbation of power law trajectory for small
b}), and, as noted after (\ref{small b power law}), there is no small $b$
power law trajectory for $C^{\left( 1 \right)}_n > 0$.  Thus $C^{\left( 1
\right)}_n \frac{\kappa^{\frac{2}{9} \left( 3 n + 6 \right)}}{b^{8 + 3 n}}$
must become large compared to $\frac{c^2}{b^2}$, as $b$ decreases further,
beyond the value  $> 0$ where the two terms are comparable.  The trajectory
then tends to the form
\begin{equation}
  \label{small c trajectory for C sub n greater than 0} c = \sqrt{\frac{\left(
  4 + n \right)}{9 \left( 2 + n \right)} C^{\left( 1 \right)}_n \left( \left(
  \frac{\kappa^{2/9}}{b_s} \right)^{6 + 3 n} - \left(
  \frac{\kappa^{2/9}}{b} \right)^{6 + 3 n} \right)},
\end{equation}
where $b_s > 0$ is a constant of integration, so that $\frac{dc}{db}$ is
positive.

Now if such a peak occurs, then for self-consistency, when we include the
Casimir energy density corrections beyond the proximity force approximation, as
discussed in the preceding subsection, the peak must occur at the upper limit
of the quantum region, so $ b_p \sim 400 \kappa^{2/9} $.  This is
because we have the bulk power law (\ref{bulk type power law for upper sign})
to the right of the peak, and from the discussion above, we cannot
self-consistently have any power law $ c \simeq \left(\frac{B_q}{b}\right)^{
\gamma} $, with $ \gamma \geq 0 $, in the quantum region.  The peak will be
broad, with width $ \sim b_p $, so in the region of the peak, we can treat $ c
$ as a constant $ \sim 400 $.
The additional term $\sim \kappa^{- \frac{22}{9}} C^{\left( 1 \right)}_n
\left( \frac{\kappa^{2/9}}{b} \right)^{8 + 3 n} c^{8 + 3 n}$, which by
the preceding subsection, we expect to accompany the term $\kappa^{-
\frac{22}{9}} C^{\left( 1 \right)}_n \left( \frac{\kappa^{2/9}}{b}
\right)^{8 + 3 n}$ in (\ref{the t i as functions of b}), now becomes $\sim
\kappa^{- \frac{22}{9}} C^{\left( 1 \right)}_n \left( 400
\frac{\kappa^{2/9}}{b} \right)^{8 + 3 n}$.  Substituting this into the
right-hand side of (\ref{maximum value of c on a small b curve with increasing
c}), and dropping all factors of order $1$, we see that we have
self-consistency in the region of the peak.  However, from comparison of
(\ref{maximum value of c on a small b curve with increasing c}) and
(\ref{small c trajectory for C sub n greater than 0}), we see that
$\frac{b_p}{b_s}$ cannot be large compared to $1$, because if
$\frac{b_p}{b_s}$ was much larger than $1$, (\ref{small c trajectory for C sub
n greater than 0}) would allow $c$ to become substantially larger than the
maximum value given by (\ref{maximum value of c on a small b curve with
increasing c}), in the region where (\ref{small c trajectory for C sub n
greater than 0}) is still valid.  Thus since $b_1$ cannot be smaller than
$b_s$, we cannot obtain a self-consistent result with $b_1 \sim
\kappa^{2/9}$, and $b_p \sim 400 \kappa^{2/9}$, in this way.

A similar result is also expected when no single term is dominant in
$t^{\left( 1 \right)}$ in (\ref{the t i as functions of b}), because if the
$C^{\left( 1 \right)}_n$ term in $t^{\left( 1 \right)}$ is multiplied by $c^{8
+ 3n}$ for all $n \geq 0$, with $c$ a constant $> 1$, the effect is to multiply
the minimum possible value of $b$, as derived in subsection \ref{The expansion
parameter}, on page \pageref{The expansion parameter}, from the
Giudice-Rattazzi-Wells estimate {\cite{Giudice Rattazzi Wells}} of the
effective expansion parameter for quantum gravity in eleven dimensions, by
$c$.

Since there is no difficulty obtaining self-consistency at the upper limit $b
\sim 400 \kappa^{2/9}$ of the quantum region, but we cannot obtain
consistency inside the quantum region for any power law $c \simeq \left(
\frac{B_q}{b} \right)^{\gamma}$, with $\gamma \geq 0$, we now try for a power
law of this form with $\gamma < 0$.  We see that if $\gamma = - 1$,
corresponding to a linear dependence of $c$ on $b$, then $\tilde{n} = \left( 1
+ \gamma \right) n + \frac{8}{3} \gamma$ is independent of $n$, and equal to
$- \frac{8}{3}$.  We can now simply have $B_q \simeq \kappa^{2/9}$, in
which case, if the magnitude of $C^{\left( 1 \right)}_n$ in (\ref{the t i as
functions of b}) is $\sim 0.2^{3 n} C^{\left( 1 \right)}$, for some constant
$C^{\left( 1 \right)}$ of order $1$, as suggested by the minimum value of $b$
estimated in subsection \ref{The expansion parameter}, the additional terms
$\sim \kappa^{- \frac{22}{9}} C^{\left( 1 \right)}_n \left(
\frac{\kappa^{2/9}}{b} \right)^{8 + 3 n} c^{8 + 3 n}$ sum up to no
more than around $\kappa^{- \frac{22}{9}} \left( 1 - 0.2^3 \right)^{- 1}
C^{\left( 1 \right)} \simeq \kappa^{- \frac{22}{9}} C^{\left( 1 \right)}$.
And considering the equation (\ref{second Einstein equation for small b}) for
$n = - \frac{8}{3}$, we see from (\ref{small b power law}) and (\ref{power law
trajectory for small b}) that we do indeed have a unique linear solution, with
\begin{equation}
  \label{self consistent c in the quantum region} c = \frac{1}{3 }
  \sqrt{\frac{C^{\left( 1 \right)}_{- \frac{8}{3}}}{5}}
  \frac{b}{\kappa^{2/9}},
\end{equation}
provided that the effective $C^{\left( 1 \right)}_{- \frac{8}{3}}$ is found to
be positive.  We see that $C^{\left( 1 \right)}_{- \frac{8}{3}}$ will be
self-consistently determined as a fixed number of order $1$, provided that
this number is positive.  Thus it seems reasonable to expect that for around
fifty percent of all possible choices of a smooth compact quotient
$\mathcal{M}^6$ of $\mathbf{C} \mathbf{H}^3$ or $\mathbf{H}^6$ that is a
spin manifold, a spin structure on $\mathcal{M}^6$, and a topologically
stabilized configuration of vacuum Yang-Mills fields on the inner surface of
the thick pipe, consistent with Witten's topological constraint {\cite{Witten
Constraints on compactification}}, a value of $ B $ larger than $ \kappa^{
\frac{2}{9}} $ will be found by this mechanism.

The actual value of $b$ at which the self-consistent quantum linear relation
(\ref{self consistent c in the quantum region}) transforms into the classical
relation (\ref{bulk type power law for upper sign}), and the corresponding
value of $B$, will be determined by how close to the self-consistent
quantum linear relation (\ref{self consistent c in the quantum region})
the system is set by
the boundary conditions at $b = b_1 \sim \kappa^{2/9}$.
We note that $\eta$, in (\ref{perturbation of power law trajectory for small
b}), is equal to $- 10$ when $n = - \frac{8}{3}$, so the linear solution
(\ref{self consistent c in the quantum region}), of (\ref{second Einstein
equation for small b}) with $n = - \frac{8}{3}$, is a very strong attractor in
the direction of increasing $b$.  However this has not taken into account the
fact that in the presence of deviations from the self-consistent linear
relation (\ref{self consistent c in the quantum region}), the equation to be
solved will no longer be precisely (\ref{second Einstein equation for small
b}), with $n = - \frac{8}{3}$.  We also note,
from the discussion above, that it is consistent for $c$ to be approximately
constant in the region of the peak at $b \sim \kappa^{2/9} \left(
\frac{B}{\kappa^{2/9}} \right)^{0.6551}$, although not for $b \ll
\kappa^{2/9} \left( \frac{B}{\kappa^{2/9}} \right)^{0.6551}$,
so we expect the transition from (\ref{self consistent c in the quantum
region}) to (\ref{bulk type power law for upper sign}) to occur smoothly
across a broad peak of width $\sim \kappa^{2/9} \left(
\frac{B}{\kappa^{2/9}} \right)^{0.6551}$.

The linear relation (\ref{self consistent c in the quantum region}) means that
$b$ depends exponentially on $y$ in the quantum region:
\begin{equation}
  \label{dependence of b on y in the quantum region} b = b_q \exp \left(
  \frac{1}{3} \sqrt{\frac{C^{\left( 1 \right)}_{- \frac{8}{3}}}{5}}
  \frac{\left( y - y_q \right)}{\kappa^{2/9}} \right),
\end{equation}
where $b_q \equiv \kappa^{2/9} \left( \frac{B}{\kappa^{2/9}}
\right)^{0.6551}$ and $y_q \equiv 0.3449 \kappa^{2/9}$, for agreement
with (\ref{bulk type power law dependence of b on y for upper sign}) at $b =
b_q$, for $y_0 = 0$.  The thickness in $y$ of the quantum region is $\sim
\kappa^{2/9} \ln \frac{b_q}{\kappa^{2/9}}$, which for
\mbox{TeV-scale} gravity, with $b_q$ greater than around $ 2000 \kappa^{\frac{2
}{9}} $, is $\sim 8 \kappa^{2/9}$.

We note that if the percentage of
possible choices of $ \mathcal{M}^6 $, its spin structure, and the vacuum
Yang-Mills fields, for which the thickness in $ y $ of the quantum region is
greater than a certain value, decreases roughly exponentially with that value,
then the percentage of possible choices, for which $ \frac{B}{\kappa^{\frac{2}{
9}}} $ is greater than a certain value, will be roughly given by a fixed
negative power of that value.

From (\ref{small b power law for a}) and (\ref{small b power law for a in
terms of b}), with $n = - \frac{8}{3}$, we see that in the quantum region,
where the linear relation (\ref{self consistent c in the quantum region})
applies, $a$ also depends linearly on $b$:
\begin{equation}
  \label{dependence of a on b in the quantum region} a = A_1
  \frac{b}{\kappa^{2/9}},
\end{equation}
where $A_1$ is a constant of integration.
However this linear dependence of $a$ on $b$ in the quantum region, for $n = -
\frac{8}{3}$, is a consequence of the proximity force relation $t^{\left( 3
\right)} = t^{\left( 1 \right)}$, which would apply for compactification on
flat $\mathbf{R}^5$ times the compact six-manifold $\mathcal{M}^6$, and as
noted above, there is no reason to expect this relation to hold when $a$ and
$b$ depend nontrivially on $y$.  Consideration of the special case where this
relation holds was adequate for the order of magnitude studies above, where
only the dependence of $b$ on $y$ was considered, but to determine the
possible dependences of $a$ on $b$ in the quantum region, I shall now assume
that the $t^{\left( i \right)}$, in (\ref{T IJ block diagonal
structure}), on page \pageref{T IJ block diagonal structure}, are constrained
only by the conservation equation (\ref{conservation equation for the t i}).

Considering the region $\kappa^{2/9} \ll b \ll b_q$, only the terms
$\kappa^{- \frac{22}{9}} C^{\left( i \right)}_{- \frac{8}{3}}$, in the
self-consistent versions of the expansions (\ref{the t i as functions of b}),
on page \pageref{the t i as functions of b}, will be significant.  The
relevant equations are now (\ref{a dot over a}) and (\ref{second Einstein
equation without a dot}), on page \pageref{a dot over a}, with the upper
choice of sign, and
\begin{equation}
  \label{t i in the quantum region} t^{\left( i \right)} \simeq \kappa^{-
  \frac{22}{9}} C^{\left( i \right)}_{- \frac{8}{3}},
\end{equation}
where the $C^{\left( i \right)}_{- \frac{8}{3}}$ are numerical constants, to
be determined self-consistently, as discussed above.  The only possible
power-law dependence of $c$ on $b$, with this form of the $t^{\left( i
\right)}$, is again $c = \sigma \frac{b}{\kappa^{2/9}}$, where
$\sigma$ is a numerical constant, and this linear dependence of $c$ on $b$
leads self-consistently to the form (\ref{t i in the quantum region}) of the
$t^{\left( i \right)}$ in this region, as before.  However $a$ no longer has
to depend linearly on $b$ in this region, so we try an ansatz
\begin{equation}
  \label{ansatz for a in the quantum region} a = A_1 \left(
  \frac{b}{\kappa^{2/9}} \right)^{\tau} .
\end{equation}
The conservation equation (\ref{conservation equation for the t i}) then
reduces to:
\begin{equation}
  \label{conservation equation in the quantum region} \left( 4 \tau + 6
  \right) C_{- \frac{8}{3}}^{\left( 3 \right)} - 4 \tau C_{-
  \frac{8}{3}}^{\left( 1 \right)} - 6 C_{- \frac{8}{3}}^{\left( 2 \right)} =
  0.
\end{equation}
Choosing $C_{- \frac{8}{3}}^{\left( 1 \right)}$ and $C_{- \frac{8}{3}}^{\left(
3 \right)}$ as independent, equations (\ref{a dot over a}) and (\ref{second
Einstein equation without a dot}) reduce in this region to:
\begin{equation}
  \label{a dot over a in quantum region} \left( 2 \tau + 4 \right) \sigma =
  \sqrt{6 \sigma^2 + \frac{2}{3} C_{- \frac{8}{3}}^{\left( 3 \right)}}
\end{equation}
\begin{equation}
  \label{second Einstein equation in quantum region} - \sigma^2 + \sigma
  \sqrt{6 \sigma^2 + \frac{2}{3} C_{- \frac{8}{3}}^{\left( 3 \right)}} +
  \frac{1}{9} \left( - \left( \tau + 2 \right) C_{- \frac{8}{3}}^{\left( 1
  \right)} + \left( \tau + 1 \right) C_{- \frac{8}{3}}^{\left( 3 \right)}
  \right) = 0,
\end{equation}
from which we find:
\begin{equation}
  \label{first relation between sigma and tau} \left( \left( 2 \tau + 4
  \right)^2 - 6 \right) \sigma^2 = \frac{2}{3} C_{- \frac{8}{3}}^{\left( 3
  \right)}
\end{equation}
\begin{equation}
  \label{second relation between sigma and tau} \left( 2 \tau + 3 \right)
  \sigma^2 = \frac{1}{9} \left( \left( \tau + 2 \right) C_{-
  \frac{8}{3}}^{\left( 1 \right)} - \left( \tau + 1 \right) C_{-
  \frac{8}{3}}^{\left( 3 \right)} \right) .
\end{equation}
Thus almost any value of $\tau$ can be obtained, if there exists a suitable
smooth compact quotient $\mathcal{M}^6$ of $\mathbf{C} \mathbf{H}^3$ or
$\mathbf{H}^6$ that is a spin manifold, and a choice of a spin structure on
$\mathcal{M}^6$ and a topologically stabilized configuration of the Yang-Mills
gauge fields on the inner surface of the thick pipe, that results
self-consistently in the appropriate values of $C_{- \frac{8}{3}}^{\left( 1
\right)}$ and $C_{- \frac{8}{3}}^{\left( 3 \right)}$.
In particular, the bulk power law value $\tau = - 0.7753$ is one of the two
solutions if $C_{- \frac{8}{3}}^{\left( 1 \right)} > 0$ and $C_{-
\frac{8}{3}}^{\left( 3 \right)} = 0$.  However $\tau = - 2$ would imply that
the square root vanished, so that we could not conclude that all three
Einstein equations would be satisfied.

The calculation of $C_{- \frac{8}{3}}^{\left( 1 \right)}$ and $C_{-
\frac{8}{3}}^{\left( 3 \right)}$ for a particular example requires, in
particular, the calculation of the propagators and heat kernels for all the
CJS fields on a flat $\mathbf{R}^5$ times uncompactified $\mathbf{C}
\mathbf{H}^3$ or $\mathbf{H}^6$ background, as appropriate.  These can be
obtained from the corresponding propagators and heat kernels on a flat
$\mathbf{R}^5$ times $\mathbf{C} \mathbf{P}^3$ or $\mathbf{S}^6$ background,
which can be calculated by using the Salam-Strathdee harmonic expansion
method {\cite{Salam Strathdee}}, and summing the expansions by means of a
generating function.  This calculation is currently in progress for
$\mathbf{C} \mathbf{H}^3$, and the scalar heat kernel on $\mathbf{C}
\mathbf{H}^3$, obtained by this method, is presented in subsection \ref{The
Salam Strathdee harmonic expansion method}, on page \pageref{The Salam
Strathdee harmonic expansion method}.

For $t^{\left( 3 \right)} \neq t^{\left( 1 \right)}$, we can no longer study
trajectories near the self-consistent linear trajectory $c = \sigma
\frac{b}{\kappa^{2/9}}$ by perturbing only the dependence of $c$ on
$b$ as $c = \sigma \frac{b}{\kappa^{2/9}} \left( 1 + Sb^{\eta}
\right)$, where $S$ is a small constant of integration.  The dependence
(\ref{ansatz for a in the quantum region}) of $a$ on $b$ also has to be
perturbed as $a = A_1 \left( \frac{b}{\kappa^{2/9}} \right)^{\tau}
\left( 1 + Ub^{\eta} \right)$, where $U$ is a small constant of integration,
and the $t^{\left( i \right)}$ functions (\ref{t i in the quantum region}) in
the region $\kappa^{2/9} \ll b \ll b_q$ have to be perturbed as
$t^{\left( i \right)} \simeq \kappa^{- \frac{22}{9}} C^{\left( i \right)}_{-
\frac{8}{3}} \left( 1 + V^{\left( i \right)} b^{\eta} \right)$, where the
$V^{\left( i \right)}$ are small constants.  The Einstein equations (\ref{a
dot over a}) and (\ref{second Einstein equation without a dot}) and the
conservation equation (\ref{conservation equation for the t i}) impose three
relations among the six constants describing the perturbation, and we would
now expect the exponent $\eta$ to depend on ratios of the small constants $S$,
$U$, and the
$V^{\left( i \right)}$, rather than having the unique value $- 10$ as for the
case when $t^{\left( 3 \right)} = t^{\left( 1 \right)}$.

The possibility of having both a self-consistent quantum region, in which $c$
increases linearly with $b$ as $\sigma \frac{b}{\kappa^{2/9}}$, with
$\sigma$ a numerical coefficient of order $1$, and a self-consistent classical
region where $c$ satisfies the classical bulk power law (\ref{bulk type power
law for upper sign}), on page \pageref{bulk type power law for upper sign}, is
due to the presence, beyond the proximity force approximation, of additional
terms
\begin{equation}
  \label{additional terms in the t i} \kappa^{- \frac{22}{9}} C_{n, n}^{\left(
  i \right)} \left( \frac{\kappa^{2/9} c}{b} \right)^{8 + 3 n},
\end{equation}
with $n \geq 0$, in the expansions (\ref{the t i as functions of b}), on page
\pageref{the t i as functions of b}, of the $t^{\left( i \right)}$ functions
in (\ref{T IJ block diagonal structure}), on page \pageref{T IJ block diagonal
structure}.  These terms sum to finite constant terms $\kappa^{- \frac{22}{9}}
C_{- \frac{8}{3}}^{\left( i \right)}$ at low orders of perturbation theory in
the quantum region, provided $\sigma$ is not too large, and thus result
self-consistently in the linear dependence of $c$ on $b$ in the quantum
region, provided the $C_{- \frac{8}{3}}^{\left( i \right)}$ are consistent
with $\sigma^2 > 0$, as determined by (\ref{first relation between sigma and
tau}) and (\ref{second relation between sigma and tau}).  While if $c$ is
related to $b$ by the classical bulk power law (\ref{bulk type power law for
upper sign}), and $b$ is larger than $b_q = \kappa^{2/9} \left(
\frac{B}{\kappa^{2/9}} \right)^{0.6551}$, so the value of $c$ given by
(\ref{bulk type power law for upper sign}) is smaller than the value that
would be given by extrapolating the linear relation from the quantum region,
then the terms (\ref{additional terms in the t i}) rapidly decrease in
magnitude with further increase in $b$, and quickly become negligible, so that
the classical bulk power law (\ref{bulk type power law for upper sign})
becomes self-consistent.

Thus it is consistent for the quantum region to transform into the classical
region at any point $b_q > b_1$, and the value of $b_q$ at which the
transition occurs in a particular example, and consequently the value of $B$,
will depend on how close to the self-consistent linear trajectory $c = \sigma
\frac{b}{\kappa^{2/9}}$, with $\sigma$ determined by (\ref{first
relation between sigma and tau}) and (\ref{second relation between sigma and
tau}), the system is set by the boundary conditions at $b = b_1$, and on
whether the self-consistent linear trajectory attracts or repels neighbouring
trajectories, in the direction of increasing $b$, and how strongly it does so.
From the discussion above, we see that for $t^{\left( 3 \right)} \neq
t^{\left( 1 \right)}$, the space of relevant neighbouring trajectories is
three-dimensional, and parametrized, for example, by small quantities $S$,
$V^{\left( 1 \right)}$, and $V^{\left( 3 \right)}$.  The actual transition
from the quantum region to the classical region will take place gradually,
over a broad peak of width around $b_q$, as discussed just before
(\ref{dependence of b on y in the quantum region}).

The presence of the additional terms (\ref{additional terms in the t i}) in
the $t^{\left( i \right)}$ functions, beyond the proximity force
approximation, follows from their presence in the local terms formed from
powers of the Riemann tensor, and the components (\ref{Riemann tensor for the
metric ansatz}), on page \pageref{Riemann tensor for the metric ansatz}, of
the Riemann tensor for the metric ansatz (\ref{metric ansatz}).
In particular, $R_{ABC} \, \!^D$ contains both a
term $R_{ABC} \, \!^D \left( h \right)$, which for a local term in the quantum
effective action $\Gamma$ formed from $4 + 3 m$ powers of the Riemann tensor
leads both for $\mathbf{C} \mathbf{H}^3$, on using the $\mathbf{C}
\mathbf{H}^n$ Riemann tensor components (\ref{CHn Riemann tensor}), on page
\pageref{CHn Riemann tensor}, and also for $\mathbf{H}^6$, to terms in the
$t^{\left( i \right)}$ functions of the form $\kappa^{- \frac{22}{9}} C \left(
\frac{\kappa^{2/9}}{b} \right)^{8 + 6 m}$, in agreement with the even
order terms in (\ref{the t i as functions of b}), and a term $\frac{c^2}{b^2}
G_{AC} \delta_B \, \!^D$, which for the same term in $\Gamma$ leads to even
order terms of the form (\ref{additional terms in the t i}).

We note, furthermore, that since, on a power law trajectory,
$\frac{\dot{a}}{a}$ is equal to $\frac{\dot{b}}{b}$ times a fixed number of
order $1$, the $R_{\mu A \nu} \, \!^B$ and $R_{A \mu B} \, \!^{\nu}$
components, and the $\frac{\dot{a}^2}{a^2}$ terms in $R_{\mu \nu \sigma} \,
\!^{\tau}$, will lead both in the quantum region and the
classical region to terms similar in magnitude to the terms
(\ref{additional terms in the t i}).  And since $\frac{\ddot{a}}{a}$ and
$\frac{\ddot{b}}{b}$ are equal, on a power law trajectory, to
$\frac{\dot{b}^2}{b^2}$ times fixed numbers of order $1$, except that
$\frac{\ddot{b}}{b}$ vanishes on the self-consistent linear trajectory in the
quantum region, the $R_{\mu y \nu} \, \!^y$ and $R_{y \mu y} \, \!^{\nu}$
components will also lead both in the quantum region and the classical region
to terms similar in magnitude to the terms (\ref{additional terms in the t
i}), and the $R_{AyB} \, \!^y$ and $R_{yAy} \, \!^B$ components will lead in
the classical region to terms similar in magnitude to the terms
(\ref{additional terms in the t i}).

Now by definition, the metric $g_{\mu \nu}$, in the metric ansatz (\ref{metric
ansatz}), has de Sitter radius equal to $1$.  Hence the value $a_1$ of $a$, at
the inner surface of the thick pipe, is equal to the observed de Sitter radius
(\ref{de Sitter radius}).  We recall that in subsection \ref{The Yang-Mills
coupling constants in four dimensions}, on page \pageref{The Yang-Mills
coupling constants in four dimensions}, we found, by combining an estimate of
the $d = 4$ Yang-Mills
coupling constants at unification, with the Ho\v{r}ava-Witten relation
(\ref{lambda kappa relation}), that $\frac{b_1}{\kappa^{2/9}} \simeq
\frac{1.2772}{\left| \chi \left( \mathcal{M}^6 \right)
\right|^{\frac{1}{6}}}$, when the compact six-manifold $\mathcal{M}^6$ is a
smooth compact quotient of $\mathbf{C} \mathbf{H}^3$, and
$\frac{b_1}{\kappa^{2/9}} \simeq \frac{1.1809}{\left| \chi \left(
\mathcal{M}^6 \right) \right|^{\frac{1}{6}}}$, when $\mathcal{M}^6$ is a
smooth compact quotient of $\mathbf{H}^6$.  Thus from (\ref{ansatz for a in the
quantum region}), we find:
\begin{equation}
  \label{A sub 1 in terms of de Sitter radius} A_1 = \left( \frac{\left| \chi
  \left( \mathcal{M}^6 \right) \right|^{\frac{1}{6}}}{1.2772} \right)^{\tau}
  \times \textrm{de Sitter radius},
\end{equation}
when $\mathcal{M}^6$ is a smooth compact quotient of $\mathbf{C}
\mathbf{H}^3$, and the same relation, with $1.2772$ replaced by $1.1809$,
when $\mathcal{M}^6$ is a smooth compact quotient of $\mathbf{H}^6$.

And from matching the bulk power law (\ref{bulk power law for a in terms of
b}) for $a$ in terms of $b$ to (\ref{ansatz for a in the quantum
region}), at $b = b_q$, we find:
\begin{equation}
  \label{A in terms of A sub 1 and B} A = A_1 \left(
  \frac{b_q}{\kappa^{2/9}} \right)^{\tau + 0.7753} = A_1 \left(
  \frac{B}{\kappa^{2/9}} \right)^{0.6551 \tau + 0.5079} .
\end{equation}
Thus:
\begin{equation}
  \label{A in terms of de Sitter radius} A = \left( \frac{\left| \chi \left(
  \mathcal{M}^6 \right) \right|^{\frac{1}{6}}}{1.2772} \right)^{\tau} \left(
  \frac{B}{\kappa^{2/9}} \right)^{0.6551 \tau + 0.5079} \times
  \textrm{de Sitter radius}
\end{equation}
for a smooth compact quotient of $\mathbf{C} \mathbf{H}^3$, and the same
relation, with $1.2772$ replaced by $1.1809$, holds for a smooth compact
quotient of $\mathbf{H}^6$.

\subsubsection{The boundary conditions at the inner surface of the thick pipe}
\label{The boundary conditions at the inner surface of the thick pipe}

Now, treating $b$ as the independent variable, the boundary conditions
(\ref{boundary conditions at y1}), on page \pageref{boundary conditions at y1},
become:
\begin{equation}
  \label{boundary conditions at y1 with b as variable} \left. \frac{c}{a}
  \frac{da}{db} \right|_{b = b_{1 +}} = \frac{\kappa^2}{18} \left( - 5
  \tilde{t}^{\left[ 1 \right] \left( 1 \right)} + 6 \tilde{t}^{\left[ 1
  \right] \left( 2 \right)} \right), \quad \left. \frac{c}{b} \right|_{b =
  b_{1 +}} = \frac{\kappa^2}{18} \left( 4 \tilde{t}^{\left[ 1 \right] \left( 1
  \right)} - 3 \tilde{t}^{\left[ 1 \right] \left( 2 \right)} \right)
\end{equation}
where $b_1 \equiv b \left( y_1 \right)$.  The coefficients $t^{\left[ 1
\right] \left( i \right)}$ receive contributions from the
Lovelock-Gauss-Bonnet terms, given by (\ref{Lovelock Gauss Bonnet t i j}), on
page \pageref{Lovelock Gauss Bonnet t i j}, for
quotients of $\mathbf{C} \mathbf{H}^3$, and by (\ref{Lovelock Gauss Bonnet
t i j for H6}) for quotients of $\mathbf{H}^6$; from the leading terms in the
Lukas-Ovrut-Waldram harmonic expansion, on the compact six-manifold $
\mathcal{M}^6 $, of the energy-momentum tensor of topologically stabilized
vacuum Yang-Mills fields on the inner surface of the thick pipe; and from
Casimir effects on the inner surface of the thick pipe.

The terms in (\ref{Lovelock
Gauss Bonnet t i j}) and (\ref{Lovelock Gauss Bonnet t i j for H6}) that
involve negative powers of $a$ are negligible at the inner surface of the
thick pipe, so the Lovelock-Gauss-Bonnet terms, for quotients of $\mathbf{C}
\mathbf{H}^3$, are:
\begin{equation}
  \label{Lovelock Gauss Bonnet t 1 j} \tilde{t}^{\left[ 1 \right] \left( 1
  \right) \mathrm{LGB}} = \frac{36}{\lambda^2 b_1^4}, \hspace{2em} \hspace{2em}
  \tilde{t}^{\left[ 1 \right] \left( 2 \right) \mathrm{LGB}} =
  \frac{12}{\lambda^2 b_1^4}
\end{equation}
and for quotients of $\mathbf{H}^6$, they are:
\begin{equation}
  \label{Lovelock Gauss Bonnet t 1 j for H6} \tilde{t}^{\left[ 1 \right]
  \left( 1 \right) \mathrm{LGB}} = \frac{45}{\lambda^2 b_1^4}, \hspace{2em}
  \hspace{2em} \tilde{t}^{\left[ 1 \right] \left( 2 \right) \mathrm{LGB}} =
  \frac{15}{\lambda^2 b_1^4}
\end{equation}

It would seem reasonable to expect that the contributions to the coefficients
$\tilde{t}^{\left[ 1 \right] \left( i \right)}$, from the leading terms in the
Lukas-Ovrut-Waldram harmonic expansion of the energy-momentum tensor of
topologically stabilized vacuum Yang-Mills fields on the inner surface of the
thick pipe, will be roughly a positive numerical multiple of the contributions
that result from embedding the spin connection in the gauge group for
$\mathbf{C} \mathbf{H}^3$, as given in (\ref{Yang Mills t 2 i}), on page
\pageref{Yang Mills t 2 i}, for the outer surface of the thick pipe.  Thus we
estimate the vacuum Yang-Mills field contribution to the coefficients
$\tilde{t}^{\left[ 1 \right] \left( i \right)}$ as:
\begin{equation}
  \label{vacuum Yang Mills t 1 i} \tilde{t}^{\left[ 1 \right] \left( 1 \right)
  \mathrm{YM}} \simeq - \frac{24}{\lambda^2 b_1^4} N, \quad \! \: \! \quad \!
  \: \! \quad \! \: \! \quad \tilde{t}^{\left[ 1 \right] \left( 2 \right)
  \mathrm{YM}} \simeq - \frac{8}{\lambda^2 b_1^4} N,
\end{equation}
where the numerical constant $N \geq 0$ is given by
\begin{equation}
  \label{numerical constant N in Yang Mills t 1 i} N = \frac{1}{96 V \left(
  \mathcal{M}^6 \right)} \int_{\mathcal{M}^6} d^6 z \sqrt{h} h^{CD} h^{EF}
  \mathrm{\mathrm{tr}} F^{\left[ 1 \right]}_{CE} F^{\left[ 1 \right]}_{DF},
\end{equation}
in terms of the topologically stabilized vacuum Yang-Mills fields on the inner
surface of the thick pipe, with $V \left( \mathcal{M}^6 \right)$ given by
(\ref{volume in terms of Euler number for a smooth compact quotient of CH3}),
on page \pageref{volume in terms of Euler number for a smooth compact quotient
of CH3}, for a smooth compact quotient of $\mathbf{C} \mathbf{H}^3$, and
by (\ref{volume in terms of Euler number for a smooth compact quotient of
H6}), for a smooth compact quotient of $\mathbf{H}^6$.

Now the results (\ref{Lovelock Gauss Bonnet t i j}), (\ref{Lovelock Gauss
Bonnet t i j for H6}), (\ref{Lovelock Gauss Bonnet t 1 j}), and (\ref{Lovelock
Gauss Bonnet t 1 j for H6}), for the Lovelock-Gauss-Bonnet contributions, have
been calculated assuming that the Riemann tensor in the Lovelock-Gauss-Bonnet
term in (\ref{FF RR action substitutions}), on page \pageref{FF RR action
substitutions}, is the $d = 10$ Riemann tensor calculated from the induced
metric $G_{UV}$ on the Ho\v{r}ava-Witten orbifold hyperplanes, and not the
restriction to the orbifold hyperplanes of the $d = 11$ Riemann tensor.  This
would seem to be a reasonable assumption, because it implies that for a
Calabi-Yau compactification {\cite{CHSW}}, with the standard embedding of the
spin connection in the gauge group, the Riemann tensor term in (\ref{FF RR
action substitutions}) is automatically equal to $- \frac{1}{2}$ times the
Yang-Mills term, at an arbitrary point of the Calabi-Yau moduli space.  For
the compactifications considered here, it implies that when we go beyond the
proximity force approximation, there are no related terms with factors of $c =
\frac{db}{dy}$ that can cancel the $\frac{1}{b^4}$ factor in the region
$\kappa^{2/9} < b < b_q = \kappa^{2/9} \left(
\frac{B}{\kappa^{2/9}} \right)^{0.6551}$, where, from the previous
subsection, $c$ is $\sim \frac{b}{\kappa^{2/9}}$.

If we assume, by analogy with this, that when we go beyond the proximity force
approximation, there are also no terms related to the higher order terms in
the expansions (\ref{surface Casimir action density}), on page
\pageref{surface Casimir action density}, with enough powers of $c$ to cancel
all the powers of $\frac{1}{b}$ in those terms in the region
$\kappa^{2/9} < b < b_q$, then the boundary conditions (\ref{boundary
conditions at y1 with b as variable}), at the inner surface of the thick pipe,
cannot be solved for any value of $b_1$
much larger than $\kappa^{2/9}$.  This is in agreement with the result
from subsection \ref{The Yang-Mills coupling constants in four dimensions}, on
page \pageref{The Yang-Mills coupling constants in four dimensions}, that to
fit a reasonable estimate of the unification value of the observed $d = 4$
Yang-Mills coupling constants, the value of $b_1$ cannot be larger than around
$1.2 \kappa^{2/9}$, which corresponds to $\left| \chi \left(
\mathcal{M}^6 \right) \right| \simeq 1$.

On the other hand, it would seem reasonable to expect that for perhaps around
three percent of choices of a smooth compact quotient $\mathcal{M}^6$ of
$\mathbf{C} \mathbf{H}^3$ or $\mathbf{H}^6$ that is a spin manifold, a
spin structure on $\mathcal{M}^6$, and a topologically stabilized
configuration of vacuum Yang-Mills fields tangential to $\mathcal{M}^6$, a
solution of the boundary conditions (\ref{boundary conditions at y1 with b as
variable}) will exist with $B$ larger than around $5 \kappa^{2/9}$,
and $b_1$ no smaller than around twice the minimum value $\simeq 0.2
\kappa^{2/9}$ derived in subsection \ref{The expansion parameter}, on
page \pageref{The expansion parameter}, from the GRW estimate {\cite{Giudice
Rattazzi Wells}} of the expansion parameter of quantum gravity in eleven
dimensions, so that the boundary conditions can be solved perturbatively.

For let us suppose that we have done the one-loop calculation for a trial
classical metric $G_{IJ}$, and have an approximation to the expansions
(\ref{the t i as functions of b}), on page \pageref{the t i as functions of
b}, of the $t^{\left( i \right)}$, that contains terms with at least two
different powers of $b$.  Then from the preceding subsection, we expect there
to be roughly a fifty percent chance of having at least a small region $b_1 <
b < b_q$ in which $c$ increases roughly linearly with $b$, so that the lowest
power of $\frac{1}{b}$ in the self-consistent $t^{\left( i \right)}$ will be
zero, as in (\ref{t i in the quantum region}).  The approximation to the
$t^{\left( i \right)}$ contains only a few terms, so for $b$ somewhat smaller
than $\kappa^{2/9}$, it will be dominated by the terms with the
largest power of $\frac{1}{b}$, and there is around a fifty percent chance
that these will lead to a small $b$ power law trajectory.

And similarly, the perturbative approximation to the expansions (\ref{surface
Casimir action density}), on page \pageref{surface Casimir action density},
will contain only a few terms, so in this approximation, the ratio of the
right-hand sides of the boundary conditions (\ref{boundary conditions at y1
with b as variable}) will have an approximately fixed value for $b$ somewhat
larger than $\kappa^{2/9}$, and generically some other approximately
fixed value for $b$ somewhat smaller than $\kappa^{2/9}$.

Now on any power law trajectory in the bulk, $\frac{da}{db}$ is a fixed
multiple of $\frac{a}{b}$, where the fixed multiple is characteristic of the
trajectory, so $\frac{c}{a} \frac{da}{db}$ is a fixed multiple of
$\frac{c}{b}$ on the trajectory.  Thus in this approximation the boundary
conditions (\ref{surface Casimir action density}) generically have no
simultaneous solution for any value of $b_1$ either much larger or much
smaller than $\kappa^{2/9}$, while there is perhaps a fifty percent
chance there will be a solution in the region with $b_1 \sim
\kappa^{2/9}$ where each term is $\sim 1$ in magnitude, and the ratios
of the left-hand sides and right-hand sides are moving between their limiting
values.

Finally, if there is such a solution of the boundary conditions in this
approximation, we would expect there to be roughly a fifty percent chance that
it will have $b_1$ greater than the minimum value of around $0.2
\kappa^{2/9}$ estimated in subsection \ref{The expansion parameter},
on page \pageref{The expansion parameter}, and perhaps another fifty percent
chance that it will have $b_1$ greater than around twice this value, so that
the one-loop calculation would give a reasonable approximation to the correct
result.

\subsubsection{The classical solutions in the bulk}
\label{The classical solutions in the bulk}

I shall now consider solutions of the Einstein equations in the classical part
$b > b_q = \kappa^{2/9} \left( \frac{B}{\kappa^{2/9}}
\right)^{0.6551}$ of the bulk, that start out in the classical region
on a trajectory of the form (\ref{bulk type power law for upper sign}), on page
\pageref{bulk type power law for upper sign}, with $B$ large compared to $
\kappa^{2/9}$,
and follow such solutions further into the bulk, towards the region where $c$
is no longer large compared to $\sqrt{\frac{4}{3}}$, so that (\ref{approximate
second Einstein equation for c in terms of b with upper sign}) no longer
reduces to (\ref{upper sign at large c}).  From (\ref{bulk power law
trajectory for a with upper sign}) and (\ref{bulk power law for a in terms of
b}), we know that $a$ is decreasing in magnitude, as $b$ increases, in this
region, and, depending on the values of the integration constant $B$, in
(\ref{bulk type power law for upper sign}), which is determined by the
boundary conditions at the inner surface of the thick pipe, and the
integration constant $A$, in (\ref{bulk power law for a in terms of b}), which
is not determined by the boundary conditions at the inner surface of the thick
pipe, and is at present a free parameter, that will eventually be determined
by the boundary conditions at the outer surface of the thick pipe, we may or
may not have to stop neglecting the term $\frac{4}{a^2}$, in the square root
(\ref{definition of the square root R}), as it occurs in the classical
Einstein equation (\ref{approximate second Einstein equation for c in terms of
b with upper sign}), before we reach the region where $c$ is no longer large
compared to $\sqrt{\frac{4}{3}}$.
I shall first consider the case where the term
$\frac{4}{a^2}$, in the square root (\ref{definition of the square root R}),
continues to be negligible, into the region where $c$ is no longer large
compared to $\sqrt{\frac{4}{3}}$, so the equation to study is
(\ref{approximate second Einstein equation for c in terms of b with upper
sign}).

We first note that (\ref{approximate second Einstein equation for c in terms
of b with upper sign}) has the solution $c = \sqrt{\frac{4}{3}}$.  However,
when the terms $\frac{4}{a^2}$ and $\frac{2}{3} \kappa^2 t^{\left( 3 \right)}$
are negligible in the square root $R$, defined in (\ref{definition of the
square root R}), as presently assumed, $R$ vanishes identically for this
solution, hence we cannot conclude, from (\ref{first Einstein equation in
terms of other equations}), that (\ref{a dot over a}), (\ref{second Einstein
equation without a dot}), and (\ref{conservation equation for the t i}) imply
that all three Einstein equations are satisfied.  And indeed, this special
solution, of (\ref{approximate second Einstein equation for c in terms of b
with upper sign}), does {\emph{not}} correspond to a solution of all three
Einstein equations.

Now the second term in the parentheses, in (\ref{approximate second Einstein
equation for c in terms of b with upper sign}), is smaller in magnitude that
$\sqrt{\frac{3}{8}}$ times the first term in the parentheses, for all $c \geq
\sqrt{\frac{4}{3}}$, and tends to $0$ relative to the first term as $c
\rightarrow \sqrt{\frac{4}{3}}$ from above.  Hence for $c \geq
\sqrt{\frac{4}{3}}$, but near $\sqrt{\frac{4}{3}}$, (\ref{approximate second
Einstein equation for c in terms of b with upper sign}) reduces to:
\begin{equation}
  \label{second Einstein equation with upper sign near c squared equals four
  thirds} \frac{dc}{db} = - \frac{2}{b} \sqrt{2 \left( 3 c^2 - 4 \right)}
\end{equation}
The solution of this is:
\begin{equation}
  \label{solution of second Einstein equation near c squared equals four
  thirds} c = \sqrt{\frac{1}{3}} \left( \left( \frac{B_1}{b} \right)^{2
  \sqrt{6}} + \left( \frac{b}{B_1} \right)^{2 \sqrt{6}} \right)
\end{equation}
where $B_1 > 0$ is a constant of integration, different from the constant of
integration, $B$, in (\ref{bulk type power law for upper sign}).  Now
(\ref{solution of second Einstein equation near c squared equals four thirds})
gives:
\begin{equation}
  \label{d c by d b near c squared equals four thirds} \frac{dc}{db} = -
  \frac{2 \sqrt{2}}{b} \left( \left( \frac{B_1}{b} \right)^{2 \sqrt{6}} -
  \left( \frac{b}{B_1} \right)^{2 \sqrt{6}} \right)
\end{equation}
This is negative for $b < B_1$, vanishes for $b = B_1$, at which point $c =
\sqrt{\frac{4}{3}}$, and positive for $b > B_1$.  Thus we see that the
solutions (\ref{solution of second Einstein equation near c squared equals
four thirds}), with different values of $B_1$, all osculate with the line $c =
\sqrt{\frac{4}{3}}$, at different points along this line, and that, moreover,
as each solution (\ref{solution of second Einstein equation near c squared
equals four thirds}) passes the point $b = B_1$, in the direction of
increasing $b$, it moves from the positive sign to the negative sign of the
square root, in (\ref{second Einstein equation with upper sign near c squared
equals four thirds}), or in other words, from the upper sign, to the lower
sign, of the square root, in (\ref{second Einstein equation for c in terms of
b}).  $\frac{dc}{db}$ now becomes positive, so, if the term $\frac{4}{a^2}$,
in the square root, $R$, defined in (\ref{definition of the square root R}),
remains negligible, $c$ now starts increasing without limit, and, when $c$ is
large compared to $\sqrt{\frac{4}{3}}$, we reach another power law region,
where, instead of (\ref{upper sign at large c}), (\ref{bulk type power law for
upper sign}), and (\ref{bulk type power law dependence of b on y for upper
sign}) we have (\ref{lower sign at large c}), and
\begin{equation}
  \label{bulk type power law for lower sign} c = \frac{db}{dy} \simeq \left(
  \frac{b}{B_2} \right)^{7.8990}
\end{equation}
for some $B_2$, which in general will be different from both $B$ and $B_1$.
(\ref{bulk type power law for lower sign}) corresponds to:
\begin{equation}
  \label{bulk type power law dependence of b on y for lower sign} \left(
  \frac{b}{B_2} \right) \simeq \left( \frac{1}{2 \sqrt{6} + 2}
  \right)^{\frac{1}{2 \sqrt{6} + 2}} \left( \frac{B_2}{y_3 - y}
  \right)^{\frac{1}{2 \sqrt{6} + 2}} \simeq 0.7558 \left( \frac{B_2}{y_3 - y}
  \right)^{0.1449}
\end{equation}
for some $y_3 > y_1$.

Now we already found, in the first region of the classical part of the bulk,
where we choose the upper sign in (\ref{a dot over a}), (\ref{second Einstein
equation without a dot}), and (\ref{second Einstein equation for c in terms of
b}), and $c$ is sufficiently large, that we are on a bulk power law trajectory
(\ref{upper sign at large c}), (\ref{bulk type power law for upper sign}), and
(\ref{bulk type power law dependence of b on y for upper sign}), that $a$
depends on $b$ through a power law, (\ref{bulk power law trajectory for a with
upper sign}) and (\ref{bulk power law for a in terms of b}), so that $a$
decreases as $b$ increases.  In the second bulk power law region, where we
have the lower sign in (\ref{a dot over a}), (\ref{second Einstein equation
without a dot}), and (\ref{second Einstein equation for c in terms of b}), and
$c$ is sufficiently large, that we are on a bulk power law trajectory
(\ref{lower sign at large c}), (\ref{bulk type power law for lower sign}), and
(\ref{bulk type power law dependence of b on y for lower sign}), we have:
\begin{equation}
  \label{bulk power law trajectory for a with lower sign} \frac{da}{db} = -
  \left( 2 + \frac{1}{2} \sqrt{6} \right) \frac{a}{b} = - 3.2247 \frac{a}{b}
\end{equation}
so that:
\begin{equation}
  \label{bulk power law for a in terms of b with lower sign} a = A_2 \left(
  \frac{\kappa^{2/9}}{b} \right)^{3.2247}
\end{equation}
where $A_2$ is constant, which will in general be different from $A$, in
(\ref{bulk power law for a in terms of b}), so that $a$ now decreases much
more rapidly, with increasing $b$, than it did in the first bulk power law
region, (\ref{bulk power law trajectory for a with upper sign}) and (\ref{bulk
power law for a in terms of b}).

A convenient interpolating function, that agrees with (\ref{bulk type power
law dependence of b on y for upper sign}) in the first bulk power law region,
when $y_0$ is set to $0$, and agrees in form with (\ref{bulk type power law
dependence of b on y for lower sign}), in the second bulk power law region,
is:
\begin{equation}
  \label{interpolating function for b} b = \left( 2 \sqrt{6} - 2
  \right)^{\frac{1}{2 \sqrt{6} - 2}} \frac{B \left( \frac{y}{B}
  \right)^{\frac{1}{2 \sqrt{6} - 2}}}{\left( 1 - \alpha y \right)^{\frac{1}{2
  \sqrt{6} + 2}}} = 1.4436 \frac{B \left( \frac{y}{B} \right)^{0.3449}}{\left(
  1 - \alpha y \right)^{0.1449}}
\end{equation}
To fix $\alpha$, we note that, from (\ref{approximate second Einstein equation
for c in terms of b with lower sign}) and (\ref{approximate second Einstein
equation for c in terms of b with upper sign}), $c \frac{dc}{db} = \ddot{b}$
should vanish, when $c = \dot{b} = \sqrt{\frac{4}{3}}$.  The zero of
$\ddot{b}$, at $y < \frac{1}{\alpha}$, is at:
\begin{equation}
  \label{zero of b double dot} y = - \frac{5 \left( 157 \sqrt{6} - 387
  \right)}{4 \left( 101 \sqrt{6} - 241 \right) \alpha} = \frac{0.4747}{\alpha}
\end{equation}
Imposing the requirement that $\dot{b} = \sqrt{\frac{4}{3}}$ at this value
of $y$, we find that:
\begin{equation}
  \label{alpha in interpolating function for b} \alpha = \frac{\left( 2
  \sqrt{6} - 3 \right) \left( 7 - 2 \sqrt{6}  \right)^{\frac{6 \sqrt{6} +
  29}{25}}}{\left( 2 \sqrt{6} - 2 \right)^{\frac{2 \sqrt{6} + 3}{15}}
  3^{\frac{\sqrt{6} + 9}{15}} 2^{\frac{18 + 2 \sqrt{6}}{25}} B} =
  \frac{0.9094}{B}
\end{equation}
Thus in terms of $ B $, the zero of $ \ddot{b} $, at $ y < \frac{1}{\alpha} $,
is at:
\begin{equation}
\label{zero of b double dot in terms of B}
y \simeq \frac{0.4747}{\alpha} = 0.5220 B
\end{equation} 
Thus $B_1$, in (\ref{solution of second Einstein equation near c squared
equals four thirds}), which is the value of $ b $, at the zero of $ \ddot{b} $,
is given by:
\begin{equation}
  \label{B sub 1 in terms of B}
  B_1 \simeq 1.2664 B
\end{equation}

In the second bulk power law region, the interpolating function,
(\ref{interpolating function for b}), approximately reduces to:
\begin{equation}
  \label{interpolating function in second bulk power law region} b = \left(
  \frac{1}{2 \sqrt{6} + 2}  \right)^{\frac{1}{2 + 2 \sqrt{6}}} \left(
  \frac{\left( \sqrt{6} - 1 \right)^{\frac{2 \sqrt{6} + 3}{15}}  \left(
  \sqrt{6} + 1 \right)^{\frac{2 \sqrt{6} - 3}{15}} 3^{\frac{2 \sqrt{6}}{15}}
  2^{\frac{32 \sqrt{6}}{75}}}{\left( 7 - 2 \sqrt{6} \right)^{\frac{6 \sqrt{6}
  + 4}{25}}  \left( 2 \sqrt{6} - 3 \right)^{\frac{6 \sqrt{6} - 4}{25}}}
  \right)^{\frac{2 \sqrt{6} + 3}{2 \sqrt{6} + 2}} \frac{\left( B
  \right)^{\frac{2 \sqrt{6} + 3}{2 \sqrt{6} + 2}}}{\left( \frac{1}{\alpha} - y
  \right)^{\frac{1}{2 \sqrt{6} + 2}}}
\end{equation}
Comparing with the bulk power law in the second bulk power law region,
(\ref{bulk type power law dependence of b on y for lower sign}), we see that:
\begin{equation}
  \label{y 3 in terms of B} y_3 = \frac{1}{\alpha} = 1.0996 B
\end{equation}
and:
\begin{equation}
  \label{B 2 in terms of B} B_2 = \frac{\left( \sqrt{6} - 1 \right)^{\frac{2
  \sqrt{6} + 3}{15}}  \left( \sqrt{6} + 1 \right)^{\frac{2 \sqrt{6} - 3}{15}}
  3^{\frac{2 \sqrt{6}}{15}} 2^{\frac{32 \sqrt{6}}{75}}}{\left( 7 - 2 \sqrt{6}
  \right)^{\frac{6 \sqrt{6} + 4}{25}}  \left( 2 \sqrt{6} - 3 \right)^{\frac{6
  \sqrt{6} - 4}{25}}} B = 1.8327 B
\end{equation}
A convenient interpolating function for the dependence of $a$ on $b$, that
agrees with (\ref{bulk power law trajectory for a with upper sign}) in the
first bulk power law region, and agrees in form with (\ref{bulk power law
trajectory for a with lower sign}), in the second bulk power law region, is:
\begin{equation}
  \label{interpolating function for a in terms of b} a = \frac{A \left(
  \frac{\kappa^{2/9}}{b} \right)^{\left( 2 - \frac{1}{2} \sqrt{6}
  \right)}}{\left( 1 + \left( \frac{b}{B_1} \right)^{\sqrt{6}} \right)} =
  \frac{A \left( \frac{\kappa^{2/9}}{b} \right)^{0.7753}}{\left( 1 +
  4.9154 \left( \frac{b}{B} \right)^{2.4495} \right)}
\end{equation}
The coefficient of $b^{\sqrt{6}}$, in the denominator of (\ref{interpolating
function for a in terms of b}), has been chosen so that $\frac{da}{db} = - 2
\frac{a}{b}$, when $c = \sqrt{\frac{4}{3}}$, so that $b = B_1$, as follows
from (\ref{a dot over a for a in terms of b}), on neglecting the Casimir
energy term in the square root.  Comparing with (\ref{bulk power law for a in
terms of b with lower sign}), we see that:
\begin{equation}
  \label{A 1 in terms of A} A_2 = 0.2034 \, A \left(\frac{B}
  {\kappa^{2/9}} \right)^{2.4495}
\end{equation}

From (\ref{a dot over
a}), with the lower choice of sign, together with (\ref{bulk type power law
for lower sign}) and (\ref{bulk power law for a in terms of b with lower
sign}), we find that, in the second bulk power law region, $\dot{a}$ is
related to $a$, by:
\begin{equation}
  \label{bulk power law for a dot in terms of a with lower sign} - \dot{a}
  \simeq \left( \frac{\tilde{A}}{a} \right)^{1.1394}
\end{equation}
where:
\[ \tilde{A} = \left( \left( 2 + \frac{1}{2} \sqrt{6} \right) \left(
   \frac{A_2}{B_2} \right) \left( \frac{\kappa^{2/9}}{B_2} \right)^{2
   \sqrt{6} + 2} \right)^{\frac{4 + \sqrt{6}}{3 \sqrt{6}}} A_2 \]
\begin{equation}
  \label{A tilde in terms of A B and kappa} \simeq 2.7944 \left(
  \frac{A_2}{B_2} \right)^{0.8777} \left( \frac{\kappa^{2/9}}{B_2}
  \right)^{6.0550} A_2 \simeq 0.002109 \left( \frac{A}{B} \right)^{0.8777}
  \left( \frac{\kappa^{2/9}}{B} \right)^{1.4556} A
\end{equation}

The power law (\ref{bulk power law for a dot in terms of a with lower sign})
is analogous to (\ref{bulk type power law for upper sign}), on page
\pageref{bulk type power law for upper sign}, so by analogy with the region
near the inner surface of the thick pipe studied in subsection \ref{The region
near the inner surface of the thick pipe}, on page \pageref{The region near the
inner surface of the thick pipe}, we would expect it to be possible to realize
a large value of $\frac{\tilde{A}}{\kappa^{2/9}}$ by the occurrence of
a quantum region $\tilde{a}_q = \kappa^{2/9} \left(
\frac{\tilde{A}}{\kappa^{2/9}} \right)^{0.5326} > a > a_2 \sim
\kappa^{2/9}$ adjacent to the outer surface, in which $- \dot{a}$
self-consistently grows linearly with $a$, and $- \dot{a}$ and $a$ increase
exponentially with the geodesic distance $\left( y_2 - y \right)$ from the
outer surface.  The boundary conditions at the outer surface will determine
both the value $ a_2 $ of $a$ at the outer surface, and the integration
constant $\tilde{A}$, in
(\ref{bulk power law for a dot in terms of a with lower sign}).  The
integration constant $A$ is then determined in terms of $\tilde{A}$, $B$, and
$\kappa^{2/9}$, by (\ref{A tilde in terms of A B and kappa}), which
then determines the relation between $a$ and $b$, by (\ref{interpolating
function for a in terms of b}), and hence the value of $b$, at the outer
surface of the thick pipe.  This then determines $y_2$, or in other words, the
value of $y$, at the outer surface of the thick pipe, by (\ref{interpolating
function for b}).

We note that for this type of solution, in which $a$ is comparable to
$\kappa^{2/9}$ at the outer surface of the thick pipe, the term
$\frac{4}{a^2}$, in the square root in the Einstein equations (\ref{a dot over
a}) and (\ref{second Einstein equation without a dot}), which is the only term
in (\ref{a dot over a}) and (\ref{second Einstein equation without a dot}) that
depends on the existence and sign of the effective cosmological constant in
the four observed dimensions, is not very important, since it
is negligible except near the outer surface, where there will be Casimir terms
of comparable magnitude.

Before considering this type of solution in more detail, I shall now look for
solutions such that both $a$ and $b$ are classical, or in other words, large
compared to $\kappa^{2/9}$, at the outer surface of the thick pipe.
We will see that in contrast to the solutions where $a$ is comparable to
$\kappa^{2/9}$ at the outer surface, the
$\frac{4}{a^2}$ term, in the square root in (\ref{a dot over a}) and
(\ref{second Einstein equation without a dot}), is essential for obtaining
this type of solution, which thus will exist only when the effective
cosmological constant, in the four observed dimensions, is greater than zero.

\subsection{Solutions with both $a$ and $b$ large compared to
$\kappa^{2/9}$, at the outer surface of the thick pipe}
\label{Solutions with both a and b large at the outer surface}

I shall now look for solutions of the Einstein equations, and the boundary
conditions (\ref{boundary conditions at y2}) at the outer surface of the thick
pipe, such that both $a$ and $b$ are large compared to $\kappa^{2/9}$,
at the outer surface, assuming that the boundary conditions at the inner
surface have already been solved, such that in the first bulk power law
region, we are on a trajectory (\ref{bulk type power law for upper sign}),
(\ref{bulk type power law dependence of b on y for upper sign}), and
(\ref{bulk power law for a in terms of b}), with a large value of
$\frac{B}{\kappa^{2/9}}$, but $A$ not yet determined.  We see from
(\ref{boundary conditions at y2}), (\ref{Yang Mills t 2 i}), (\ref{Lovelock
Gauss Bonnet t i j}), (\ref{Lovelock Gauss Bonnet t i j for H6}), the
expansions (\ref{surface Casimir action density for small a}), and the
relations (\ref{lambda kappa relation}) and (\ref{one over lambda squared}),
that when both $a$ and $b$ are large compared to $\kappa^{2/9}$, at
the outer surface of the thick pipe, the largest terms, in the right-hand
sides of the boundary conditions, (\ref{boundary conditions at y2}), are the
Yang-Mills terms, (\ref{Yang Mills t 2 i}), for $\mathbf{C} \mathbf{H}^3$,
and the Lovelock-Gauss-Bonnet terms, (\ref{Lovelock Gauss Bonnet t i j}), for
$\mathbf{C} \mathbf{H}^3$, and (\ref{Lovelock Gauss Bonnet t i j for H6}),
for $\mathbf{H}^6$, and that these terms are of order $\frac{1}{b}$ or
$\frac{1}{a}$, times $\frac{\kappa^{\frac{2}{3}}}{b^3}$,
$\frac{\kappa^{\frac{2}{3}}}{b^2 a}$, $\frac{\kappa^{\frac{2}{3}}}{b^{} a^2}$,
or $\frac{\kappa^{\frac{2}{3}}}{a^3}$, and thus negligible.  Thus we are now
looking for solutions such that $\frac{\dot{a}}{a}$, and $\frac{\dot{b}}{b}$,
are zero, at the outer surface of the thick pipe.

From (\ref{a dot over a}), we therefore require that $\frac{8}{b^2} =
\frac{4}{a^2}$, at the outer surface of the thick pipe, for this type of
solution, or in other words, $b = \sqrt{2} a$, at the outer surface of the
thick pipe.  Furthermore, the term $\frac{4}{a^2}$, in the square root, in
(\ref{a dot over a}), arose from the Ricci tensor of four-dimensional de
Sitter space, $R_{\mu \nu} \left( g \right)$, in (\ref{Ricci tensor for the
metric ansatz}), and would have been absent, if the effective cosmological
constant, in the four observed dimensions, had been zero, and would have had
the opposite sign, if the effective cosmological constant, in the four
observed dimensions, had been zero.  Thus there will be no solutions, such
that both $a$ and $b$ are large compared to $\kappa^{2/9}$, at the
outer surface of the thick pipe, unless the effective cosmological constant,
in the four observed dimensions, is greater than zero.

We now have to study the coupled equations (\ref{a dot over a}) and
(\ref{second Einstein equation without a dot}), when the $t^{\left( i
\right)}$ are negligible, but the term $\frac{4}{a^2}$, in the square root, is
not negligible.  We can still express the two equations as first order
differential equations for $a$, and $c = \dot{b}$, as functions of $b$, but
the two equations are now coupled.  Equation (\ref{a dot over a}), with the
upper choice of sign, now becomes:
\begin{equation}
  \label{bulk coupled a dot over a} \frac{da}{db} = - 2 \frac{a}{b} +
  \frac{a}{2 bc} \sqrt{6 c^2 - 8 + 4 \frac{b^2}{a^2}}
\end{equation}
And (\ref{second Einstein equation without a dot}), with the upper choice of
sign, now becomes:
\begin{equation}
  \label{bulk coupled second Einstein equation} \frac{dc}{db} = \frac{3 c^2 -
  4}{bc} - \frac{2}{b} \sqrt{6 c^2 - 8 + 4 \frac{b^2}{a^2}}
\end{equation}
Qualitatively, when the $\frac{b^2}{a^2}$ term starts to become significant,
in the square root in the right-hand side of (\ref{bulk coupled second
Einstein equation}) the trajectory, in the $\left( b, c \right)$ plane, starts
to peel off below the $\frac{1}{a^2} = 0$ trajectory.  We are looking for a
solution where $\frac{\dot{a}}{a}$, or in other words, $\frac{c}{a}
\frac{da}{db}$, and $\frac{\dot{b}}{b}$, or in other words, $\frac{c}{b}$,
both tend to zero, at the boundary, while $\frac{b}{a}$ tends to $\sqrt{2}$,
at the boundary, and $b$ tends to a finite nonzero limit.  Thus $c \rightarrow
0$ at the boundary, while $\frac{da}{db}$ must remain finite.  Then $4
\frac{b^2}{a^2}$ needs to increase rapidly enough, to compensate for the
decrease in $6 c^2$, so as to keep $6 c^2 - 8 + 4 \frac{b^2}{a^2} > 0$, as $c$
tends towards $0$.  And as $c \rightarrow 0$, $\frac{dc}{db}$ will be
determined by the $- \frac{4}{bc}$ term, which $\rightarrow \infty$.

Now the equation $\frac{dc}{db} = - \frac{4}{bc}$ has the solution $c^2 = 8
\ln \left( \frac{b_2}{b} \right)$, where $b_2$ is a constant of integration,
that we would like to identify as $ b\left( y_2 \right) $, the value of $ b $
at the outer surface of the thick pipe.
This solution applies in the region $c \rightarrow 0$, $b \simeq b_2$, $b \leq
b_2$, so we can expand the logarithm, to find:
\begin{equation}
  \label{dependence of c on b as c tends to 0} c \simeq \sqrt{8 \left( 1 -
  \frac{b}{b_2} \right)}
\end{equation}
If we now define $u \equiv \frac{b}{a}$, the equations (\ref{bulk coupled a
dot over a}), and (\ref{bulk coupled second Einstein equation}) become:
\begin{equation}
  \label{a dot over a in terms of u} \frac{du}{db} = \frac{u}{b} \left( 3 -
  \frac{1}{2 c} \sqrt{6 c^2 - 8 + 4 u^2} \right)
\end{equation}
\begin{equation}
  \label{second Einstein equation in terms of u} \frac{dc}{db} = \frac{3 c^2 -
  4}{bc} - \frac{2}{b} \sqrt{6 c^2 - 8 + 4 u^2}
\end{equation}
And the boundary conditions, at $b = b_2$, become:
\begin{equation}
  \label{boundary conditions in terms of u} u = \sqrt{2}, \hspace{4ex} c = 0
\end{equation}
We now see that there are two different possible behaviours of $u$ near the
boundary, consistent with (\ref{dependence of c on b as c tends to 0}),
(\ref{a dot over a in terms of u}), and (\ref{second Einstein equation in
terms of u}).  Specifically, expanding $u$ in the small quantity $\left( 1 -
\frac{b}{b_2} \right)$, as $u = \sqrt{2} \left( 1 + \alpha \left( 1 -
\frac{b}{b_2} \right) \right)$, we find, from (\ref{dependence of c on b as c
tends to 0}), and (\ref{a dot over a in terms of u}), that
\begin{equation}
  \label{condition on alpha in u} - \alpha = 3 - \frac{1}{2} \sqrt{6 + 2
  \alpha}
\end{equation}
which has the solutions:
\begin{equation}
  \label{solutions for alpha in u} \alpha = - \frac{5}{2}, \hspace{4ex} \alpha
  = - 3
\end{equation}
We note that the first of these is only a solution, for the particular sign of
the square root in (\ref{condition on alpha in u}), while the second is a
solution for both signs of the square root, since the square root vanishes for
it.  Both the solutions (\ref{solutions for alpha in u}) are consistent with
(\ref{second Einstein equation in terms of u}), and substituting one of them
into (\ref{second Einstein equation in terms of u}), fixes the term in $c^2$,
that is quadratic in $\left( 1 - \frac{b}{b_2} \right)$.  Then substitution
into (\ref{a dot over a in terms of u}) fixes the quadratic term in $u$, and
so on.

Now (\ref{a dot over a in terms of u}) and (\ref{second Einstein equation in
terms of u}) imply that:
\begin{equation}
  \label{6 c squared minus 8 plus 4 u squared} \frac{d}{db} \left( 6 c^2 - 8 +
  4 u^2 \right) = - \frac{1}{b} \sqrt{6 c^2 - 8 + 4 u^2} \left( \frac{4}{c}
  \left( 6 c^2 + u^2 \right) - 6 \sqrt{\left( 6 c^2 - 8 + 4 u^2 \right)}
  \right)
\end{equation}
Hence $6 c^2 - 8 + 4 u^2 = 0$ is a solution of (\ref{a dot over a in terms of
u}) and (\ref{second Einstein equation in terms of u}).  However, the square
root, $R$, defined in (\ref{definition of the square root R}), vanishes
identically for this solution, when the $t^{\left( i \right)}$ are negligible,
so we cannot infer, from (\ref{first Einstein equation in terms of other
equations}), that $6 c^2 - 8 + 4 u^2 = 0$ is a solution of all three Einstein
equations, and, in fact, it does not correspond to a solution of all three
Einstein equations.  It is, in fact, the generalization, to the case where $u
\neq 0$, of the line $c = \sqrt{\frac{4}{3}}$, that the trajectories in the
$\left( b, c \right)$ plane, that corresponded to actual solutions of the
Einstein equations, in the limit $u = 0$, osculated with, as they switched
from the first to the second branch of the square root, in (\ref{a dot over
a}) and (\ref{second Einstein equation without a dot}).

We see, furthermore, that the case $\alpha = - 3$, in (\ref{solutions for
alpha in u}), satisfies $6 c^2 - 8 + 4 u^2 = 0$, to the order given, and thus
is the $c \rightarrow 0$ limit of this special solution of (\ref{a dot over a
in terms of u}) and (\ref{second Einstein equation in terms of u}), that does
not correspond to a solution of all three Einstein equations.  We note,
furthermore, that this special solution, of (\ref{a dot over a in terms of u})
and (\ref{second Einstein equation in terms of u}), never rises above the line
$c = \sqrt{\frac{4}{3}}$, in the $\left( b, c \right)$ plane.  It in fact
approaches this line from below, as $b \rightarrow 0$, since $u \rightarrow
0$, as $b \rightarrow 0$.  Furthermore, when $6 c^2 - 8 + 4 u^2 = 0$, (\ref{a
dot over a in terms of u}) reduces to $\frac{du}{db} = 3 \frac{u}{b}$, hence
$u = \sqrt{2} \left( \frac{b}{b_2} \right)^3$, where, by (\ref{boundary
conditions in terms of u}), $b_2$ is the integration constant in
(\ref{dependence of c on b as c tends to 0}).  Hence $c = \sqrt{\frac{4}{3}
\left( 1 - \left( \frac{b}{b_2} \right)^6 \right)}$, which does, indeed, solve
(\ref{second Einstein equation in terms of u}).

Considering, now, the case $\alpha = - \frac{5}{2}$, in (\ref{solutions for
alpha in u}), we see that $6 c^2 - 8 + 4 u^2 \simeq 8 \left( 1 - \frac{b}{b_2}
\right) \simeq c^2$ near the boundary, hence the square root, $R$, is
nonvanishing, as soon as we move away from the boundary, so, by (\ref{first
Einstein equation in terms of other equations}), this solution will correspond
to a solution of all three Einstein equations.  Furthermore, $\frac{du}{db}$
starts positive, specifically $\frac{du}{db} = \frac{5}{2}  \frac{u}{b}$, at
$b = b_2$, hence $u$ decreases, as $b$ decreases downwards, away from $b =
b_2$, hence, provided $\frac{du}{db}$ never becomes negative, and the square
root stays real, the square root is bounded above, by $\sqrt{6} c \simeq 2.45
c$, hence, by (\ref{a dot over a in terms of u}), we have $3\frac{u}{b}\geq
\frac{du}{db} \geq
\frac{u}{b} \left( 3 - \frac{\sqrt{6}}{2} \right) \simeq 1.78 \frac{u}{b}$,
and, by (\ref{6 c squared minus 8 plus 4 u squared}), we have $\frac{d}{db}
\left( 6 c^2 - 8 + 4 u^2 \right) \leq - \left( 24 - 6 \sqrt{6} \right)
\frac{c}{b} \sqrt{6 c^2 - 8 + 4 u^2} \simeq - 9.30 \frac{c}{b}  \sqrt{6 c^2 -
8 + 4 u^2}$, hence $\frac{du}{db}$ never does become negative, and the square
root does stay real.  We can also confirm directly from (\ref{second Einstein
equation in terms of u}), by considering separately the cases $ c\geq\sqrt{
\frac{4}{3}} $ and $ c\leq\sqrt{\frac{4}{3}} $, that $ \frac{dc}{db} $ is
negative irrespective of the value of $ u $, provided the square root is real.
Furthermore, $\sqrt{2} \left( \frac{b}{b_2} \right)^3 \leq u \leq \sqrt{2}
\left( \frac{b}{b_2}
\right)^{1.77} \leq \sqrt{2} \frac{b}{b_2}$, hence $c \geq \sqrt{\frac{4}{3}
\left( 1 - \left( \frac{b}{b_2} \right)^2 \right)}$, hence
\begin{equation}
  \label{inequality for slope of root} \frac{d \left( 6 c^2 - 8 + 4 u^2
  \right)}{\sqrt{6 c^2 - 8 + 4 u^2}} \leq - 9.30 \frac{c}{b} db \leq -
  \frac{10.73}{b_2} \sqrt{1 - \left( \frac{b}{b_2} \right)^2} db
\end{equation}
Hence
\begin{equation}
  \label{inequality for root} \sqrt{6 c^2 - 8 + 4 u^2} \geq 2.68 \left(
  \frac{\pi}{2} - \arcsin \left( \frac{b}{b_2} \right) - \frac{b}{b_2} \sqrt{1
  - \left( \frac{b}{b_2} \right)^2} \right)
\end{equation}
With the bound $u \leq \sqrt{2} \left( \frac{b}{b_2} \right)^{1.77}$, this
implies that $6 c^2 - 8$ is positive for $\frac{b}{b_2} < 0.61$, and is
greater than $9.86$ for $\frac{b}{b_2} = 0.2$, by which point $u^2 < 0.006$.
Thus this solution merges into a solution of (\ref{approximate second Einstein
equation for c in terms of b with upper sign}), as $b$ continues to decrease,
and for $c$ large compared to $\sqrt{\frac{4}{3}}$, will follow a trajectory
of the form (\ref{bulk type power law for upper sign}), in the $\left( b, c
\right)$ plane, with $\frac{B}{b_2}$ a fixed number of order $1$, that will be
the same for all solutions, of this type.  Thus we do, indeed, have a solution
of the boundary conditions, such that both $a$ and $b$ are large compared to
$\kappa^{2/9}$, at the outer surface of the thick pipe.
And moreover, for solutions of this type, namely with $\alpha =
- \frac{5}{2}$ in (\ref{solutions for alpha in u}), the constant of
integration, $ b_2 $, in (\ref{dependence of c on b as c tends to 0}), can be
identified as $ b_2 = b\left( y_2 \right) $, the value of $ b $ at the outer
surface of the thick pipe.

From the behaviour (\ref{dependence of c on b as c tends to 0}),
of $c$ near the outer
boundary, we see that near the outer boundary, $y_2 - y \simeq b_2
\sqrt{\frac{1}{2} \left( 1 - \frac{b}{b_2} \right)} \simeq \frac{b_2}{4} c$,
so $y$ tends to a finite value, $y_2$, at the outer boundary, even though
$\frac{dy}{db} = \frac{1}{c}$ goes to $\infty$, right at the boundary.  $y_2$
will be equal to a number of order $1$, times the value of $y$ at which
$\ddot{b}$ vanishes for the interpolating function (\ref{interpolating
function for b}), on page \pageref{interpolating function for b}, which by
(\ref{zero of b double dot in terms of B}), is at $y = 0.5220 B$.  Thus the
geodesic distance from the inner surface to the outer surface of the thick
pipe is around $B$.

An alternative method of studying solutions of this type, is to take the ratio
of (\ref{a dot over a in terms of u}) and (\ref{second Einstein equation in
terms of u}).  Then $b$ cancels out, and we get a single first order
differential equation, that expresses $\frac{du}{dc}$, as a function of $u$
and $c$.

\subsubsection{Newton's constant and the cosmological constant for solutions
with the outer surface in the classical region}
\label{G sub N and Lambda for solutions with outer surface in classical region}

We now need to consider whether a solution of this type can fit the observed
values of Newton's constant, (\ref{Newtons constant}), and the cosmological
constant, (\ref{Lambda}).  Considering first the value of Newton's constant,
the current observational limits on extra dimensions in high energy physics
experiments {\cite{limits on extra dimensions from high energy physics}}, and
in measurements of the gravitational force at short distances {\cite{Hoyle et
al}}, imply that the maximum
values of $y$, and $b$, namely $y_2$, and $b_2 = b \left( y_2 \right)$, are
required to be sufficiently small, that a four-dimensional effective field
theory description can be used, for all observations up to the present time.
Assuming, provisionally, that $y_2$ and $b_2$ are, indeed, sufficiently small,
the four-dimensional effective field theory description can be obtained by
following the method of Randall and Sundrum {\cite{Randall Sundrum 1}}.

The first step is to identify the massless gravitational fluctuations about
the classical solution found above.  These provide the
gravitational fields of the effective theory.  They are the zero-modes of the
classical solution, and correspond to replacing the locally de Sitter metric,
$g_{\mu \nu}$, in (\ref{metric ansatz}), on page \pageref{metric ansatz}, by $
\tilde{g}_{\mu \nu} = g_{\mu \nu}
+ h_{\mu \nu}$, where $h_{\mu \nu}$ is a small perturbation, that, like
$g_{\mu \nu}$, depends on position in the four extended dimensions, but not on
$y$, nor on the coordinates of the compact six-manifold.  We note that since
the de Sitter radius of $g_{\mu \nu}$ has been set equal to $1$, $h_{\mu
\nu}$ is allowed to very rapidly, with ``Fourier modes'' of wavelength down to
$\sim 10^{- 29}$ of the de Sitter radius, corresponding to the current short
distance limit of about a millimetre, on short-distance tests of Newton's law,
in units of the observed de Sitter radius (\ref{de Sitter radius}).

The four-dimensional effective theory
follows by substituting the zero modes of the classical solution into the
original Ho\v{r}ava-Witten action, (\ref{upstairs bulk action}) plus (\ref{Yang
Mills action}), plus the analogue of (\ref{Yang Mills action}) for $y_2$.
To determine the value of Newton's constant, we focus on the term, in the
Einstein action term in (\ref{upstairs bulk action}), that produces the
Einstein action, (\ref{Einstein action}), in four dimensions.  The Riemann
tensor for the perturbed metric is still given by (\ref{Riemann tensor for the
metric ansatz}), on page \pageref{Riemann tensor for the metric ansatz}, with $
g_{\mu \nu}$ replaced by $\tilde{g}_{\mu \nu}$, since
the derivation of (\ref{Riemann tensor for the metric ansatz}) did not make
use of the locally de Sitter property of $g_{\mu \nu}$.  I shall denote the
metric in eleven dimensions, with the locally de Sitter metric, $g_{\mu \nu}$,
replaced by the perturbed metric, $\tilde{g}_{\mu \nu}$, by $\tilde{G}_{IJ}$.
The relevant term, in (\ref{upstairs bulk action}), is then:
\[ \frac{2}{\kappa^2} \int d^4 x \int_{\mathcal{M}^6} d^6 z \int^{y_2}_{y_1}
   dy \sqrt{- \tilde{G}} \left( - \frac{1}{2} \tilde{G}^{\mu \nu}
   R^{\hspace{0.4ex} \hspace{0.4ex} \hspace{0.4ex} \hspace{0.4ex}
   \hspace{0.4ex} \hspace{1.2ex} \tau}_{\mu \tau \nu} \left( \tilde{g} \right)
   \right) = \hspace{3cm}  \]
\begin{equation}
  \label{Einstein term in four dimensional effective action} \hspace{2cm} = -
  \frac{1}{\kappa^2} \int d^4 x \sqrt{- \tilde{g}} \tilde{g}^{\mu \nu}
  R^{\hspace{0.4ex} \hspace{0.4ex} \hspace{0.4ex} \hspace{0.4ex}
  \hspace{0.4ex} \hspace{1.2ex} \tau}_{\mu \tau \nu} \left( \tilde{g} \right)
  \int_{\mathcal{M}^6} d^6 z \sqrt{h} \int^{y_2}_{y_1} dya^2 b^6
\end{equation}
where I replaced $\frac{1}{\kappa^2}$ by $\frac{2}{\kappa^2}$, because we are
here working in the downstairs picture, and integrating over only one copy of
the bulk, rather than over two copies, one of which is reflected, as in
(\ref{upstairs bulk action}), and I have denoted the coordinates on the
compact six-manifold by $z^A$.

The factor $V \left( \mathcal{M}^6 \right) = \int_{\mathcal{M}^6} d^6 z
\sqrt{h}$ is given by (\ref{volume in terms of Euler number for a smooth
compact quotient of CH3}), on page \pageref{volume in terms of Euler number
for a smooth compact quotient of CH3}, for a smooth compact quotient of
$\mathbf{C} \mathbf{H}^3$, and by (\ref{volume in terms of Euler number
for a smooth compact quotient of H6}), for a smooth compact quotient of
$\mathbf{H}^6$.  To evaluate the factor $\int^{y_2}_{y_1} dya^2 b^6$, we
note that $b$ is a monotonically increasing function of $y$, for the solutions
considered in this subsection, so this integral is equal to $\int^{b_2}_{b_1}
\frac{db}{c} a^2 b^6$, where $c = \frac{db}{dy}$.

For the region $b_1 \leq b \leq b_q = \kappa^{2/9} \left(
\frac{B}{\kappa^{2/9}} \right)^{0.6551}$, we have $c \sim
\frac{b}{\kappa^{2/9}}$, up to a factor of order $1$, and $a = A_1
\left( \frac{b}{\kappa^{2/9}} \right)^{\tau}$, where $A_1$ is related
to the integration constant $A$ in the classical bulk power law (\ref{bulk
power law for a in terms of b}), on page \pageref{bulk power law for a in
terms of b}, by (\ref{A in terms of A sub 1 and B}), on page \pageref{A in
terms of A sub 1 and B}.  Thus:
\begin{equation}
  \label{contribution from quantum region} \int^{b_q}_{b_1} \frac{db}{c} a^2
  b^6 \sim \left\{ \begin{array}{cc}
    \frac{\kappa^{\frac{14}{9}}}{6 + 2 \tau} A^2_1 \left(
    \frac{B}{\kappa^{2/9}} \right)^{1.3102 \tau + 3.9306} & \left(
    \tau > - 3 \right)\\
    \frac{\kappa^{\frac{14}{9}}}{\left| 6 + 2 \tau \right|} A^2_1 & \left(
    \tau < - 3 \right)
  \end{array} \right.
\end{equation}
up to a factor of order $1$.  While for the region $b_q \leq b \leq b_2$, we
note that the equations (\ref{a dot over a in terms of u}) and (\ref{second
Einstein equation in terms of u}), on page \pageref{a dot over a in terms of
u}, are invariant under rescaling of $b$ by a constant factor, so for the
general solution, $c = f \left( \frac{b}{b_2} \right)$, where $f \left( 1
\right) = 0$.  Thus the integration constant $B$, in the classical bulk power
law (\ref{bulk type power law for upper sign}), on page \pageref{bulk type
power law for upper sign}, is a fixed number times $b_2$, and we see from the
discussion around (\ref{inequality for slope of root}) and (\ref{inequality
for root}) that this fixed number is of order $1$.  We also note that the
classical bulk power law (\ref{bulk power law for a in terms of b}), on page
\pageref{bulk power law for a in terms of b}, for $a$ in terms of $b$, will be
approximately valid, up to a factor of order $1$, right up to the outer
surface, for the solutions considered in this subsection.  Thus we find:
\begin{equation}
  \label{contribution from classical region} \int^{b_2}_{b_q} \frac{db}{c} a^2
  b^6 \sim \frac{\kappa^{\frac{14}{9}}}{7.3} A^2 \left(
  \frac{B}{\kappa^{2/9}} \right)^{5.4494} =
  \frac{\kappa^{\frac{14}{9}}}{7.3} A_1^2 \left(
  \frac{B}{\kappa^{2/9}} \right)^{1.3102 \tau + 6.4652},
\end{equation}
up to a factor of order $1$, where I used (\ref{A in terms of A sub 1 and B}).

Comparing (\ref{contribution from quantum region}) and (\ref{contribution from
classical region}), we see that for $B \gg \kappa^{2/9}$, the
contribution from the classical region is large compared to the contribution
from the quantum region for all $\tau > - 4.9345$, while for $\tau \leq -
4.9345$, there is no enhancement of the integral for large $B$.

We now note that, by definition, the de Sitter radius of the unperturbed
metric, $g_{\mu \nu}$, is equal to $1$, so since $\tilde{g}_{\mu \nu}$ differs
from $g_{\mu \nu}$ only by a small perturbation, the use of the metric
$\tilde{g}_{\mu \nu}$ corresponds to measuring distances in units of the de
Sitter radius.  We therefore define a rescaled metric $\bar{g}_{\mu \nu}$ by
$\bar{g}_{\mu \nu} = \left( \textrm{de Sitter radius} \right)^2
\tilde{g}_{\mu \nu}$, which corresponds to measuring distances in ordinary
units rather than in units of the de Sitter radius.  Then from (\ref{Einstein
term in four dimensional effective action}) and (\ref{contribution from
classical region}), together with (\ref{volume in terms of Euler number for a
smooth compact quotient of CH3}) or (\ref{volume in terms of Euler number for
a smooth compact quotient of H6}), on page \pageref{volume in terms of Euler
number for a smooth compact quotient of CH3}, and (\ref{A sub 1 in terms of de
Sitter radius}), on page \pageref{A sub 1 in terms of de Sitter radius}, we
find that for $\tau > - 4.9345$, the Einstein action term, in the
four-dimensional effective action, is for the solutions considered in the
present subsection, equal to:
\begin{equation}
  \label{Einstein term in four dimensional effective action for solutions
  classical at the outer surface} - \frac{1}{\kappa^{\frac{4}{9}}} \left(
  \frac{B}{\kappa^{2/9}} \right)^{1.3102 \tau + 6.4652} \left| \chi
  \left( \mathcal{M}^6 \right) \right|^{1 + \frac{\tau}{3}} \int d^4 x \sqrt{-
  \bar{g}} \bar{g}^{\mu \nu} R_{\mu \tau \nu} \, \!^{\tau} \left( \bar{g}
  \right)
\end{equation}
up to a factor of order $1$.

Comparing with (\ref{Einstein action}), on page \pageref{Einstein action}, we
find that for the solutions considered in the present subsection, with $\tau >
- 4.9345$:
\begin{equation}
  \label{1 over G Newton for solutions classical at the outer surface}
  \frac{1}{G_N} \sim \frac{50}{\kappa^{\frac{4}{9}}} \left(
  \frac{B}{\kappa^{2/9}} \right)^{1.3102 \tau + 6.4652} \left| \chi
  \left( \mathcal{M}^6 \right) \right|^{1 + \frac{\tau}{3}},
\end{equation}
up to a factor of order $1$.  This is the form taken by the ADD mechanism
{\cite{ADD1, ADD2}}, for the solutions considered in the present subsection.
And for $\tau \leq - 4.9345$, we find the same result, but without the
$B$-dependent factor.  Thus for these solutions, there is no ADD mechanism,
unless $\tau > - 4.9345$.

Considering, now, the case of TeV-scale gravity, I shall take $\kappa^{-
\frac{2}{9}} \simeq 0.2$ TeV, so that $\kappa^{2/9} \simeq 10^{- 18}$
metres, as a representative example, which according to Mirabelli, Perelstein,
and Peskin {\cite{Mirabelli Perelstein Peskin}} will for six flat extra
dimensions be just out of reach at the Tevatron, but comfortably accessible at
the LHC, as I shall review further in subsection \ref{Newtons constant and the
cosmological constant}, on page \pageref{Newtons constant and the cosmological
constant}, and section \ref{E8 vacuum gauge fields and the Standard Model}, on
page \pageref{E8 vacuum gauge fields and the Standard Model}.  Then from
(\ref{1 over G Newton for solutions classical at the outer surface}) and
(\ref{Newtons constant}) we find that
\begin{equation}
  \label{B for TeV scale gravity with classical outer surface} \left(
  \frac{B}{\kappa^{2/9}} \right)^{1.3102 \tau + 6.4652} \left| \chi
  \left( \mathcal{M}^6 \right) \right|^{1 + \frac{\tau}{3}} \sim 10^{32}
\end{equation}
for TeV-scale gravity, up to a factor of order $1$.

Now as noted above, the bulk power law (\ref{bulk power law for a in terms of
b}), on page \pageref{bulk power law for a in terms of b}, is valid up to a
factor of order $1$, for the solutions considered in this subsection, right up
to the outer surface, where $b = \sqrt{2} a$ for these solutions.  We also
noted that $b_2 \sim B$, up to a factor of order $1$.  Thus for these
solutions:
\begin{equation}
  \label{A for solutions classical at the outer surface} \frac{
  A}{\kappa^{2/9}} \sim \left( \frac{B}{\kappa^{2/9}}
  \right)^{1.7753},
\end{equation}
up to a factor of order $1$.  Thus from (\ref{A in terms of de Sitter
radius}), on page \pageref{A in terms of de Sitter radius}, we find that for
TeV-scale gravity, the condition for the solutions considered in this
subsection to fit the observed de Sitter radius (\ref{de Sitter radius}), on
page \pageref{de Sitter radius}, is:
\begin{equation}
  \label{B to fit de Sitter radius with classical outer surface} \left(
  \frac{B}{\kappa^{2/9}} \right)^{1.2674 - 0.6551 \tau} \left| \chi
  \left( \mathcal{M}^6 \right) \right|^{- \frac{\tau}{6}} \sim 10^{44},
\end{equation}
up to a factor of order $1$.  The solution of (\ref{B for TeV scale gravity
with classical outer surface}) and (\ref{B to fit de Sitter radius with
classical outer surface}) for the minimum value $\left| \chi \left(
\mathcal{M}^6 \right) \right| = 1$ is
\begin{equation}
  \label{tau and B for mod chi equals 1} \tau = - 3.103, \quad \! \: \! \quad
  \! \: \! \quad \! \: \! \quad \! \: \! \quad B \sim 10^{13}
  \kappa^{2/9} \sim 10^{- 5} \textrm{ metres},
\end{equation}
and the solution for the maximum value $\left| \chi \left( \mathcal{M}^6
\right) \right| \simeq 7 \times 10^4$ is
\begin{equation}
  \label{tau and B for mod chi equals 7 times 10 to the 4} \tau = - 3.023,
  \quad \! \: \! \quad \! \: \! \quad \! \: \! \quad \! \: \! \quad B \sim
  10^{13} \kappa^{2/9} \sim 10^{- 5} \textrm{ metres} .
\end{equation}
The value of $b_q$ corresponding to (\ref{tau and B for mod chi equals 1}) and
(\ref{tau and B for mod chi equals 7 times 10 to the 4}) is $b_q \sim 10^8
\kappa^{2/9} \sim 10^{- 10}$ metres, so the thickness in $y$ of the
quantum region is $\sim 20 \kappa^{2/9} \sim 10^{- 17}$ metres.

We note that the value of $B$ in (\ref{tau and B for mod chi equals 1}) and
(\ref{tau and B for mod chi equals 7 times 10 to the 4}) is about a factor of
$10$ smaller than the shortest distance so far studied in precision
sub-millimetre tests of Newton's law {\cite{Hoyle et al}}.  Nevertheless, we
cannot directly conclude that the solutions just obtained correctly reproduce
the $d = 4$ Newton's law over any distance even up to the de Sitter radius
$\sim 10^{26}$ metres, because we have not fully satisfied the requirement for
a valid reduction to a four-dimensional effective theory, due to the fact that
$a \left( y \right)$ decreases from $\sim 10^{26}$ metres at the inner surface
of the thick pipe, to $\sim 10^{- 5}$ metres at the outer surface.  Thus any
perturbation of $g_{\mu \nu}$, of wavelength less than the de Sitter radius,
will have wavelength less than around $B$, at the outer surface.

We note that $a^2$ corresponds to the warp factor of the first Randall-Sundrum
model (RS1) {\cite{Randall Sundrum 1}}, and that we live on the ``wrong''
boundary, from the point of view of the RS1 model, because the reverse RS1
effect is outweighed by the ADD effect {\cite{ADD1, ADD2}}, which is absent
from the RS1 model.  Arnowitt and Dent {\cite{0412016 Arnowitt Dent}} have
studied Newtonian forces in the RS1 model, and found that Newton's law is
obtained correctly between point sources on the RS1 ``Planck brane'', which
corresponds to the inner surface of the thick pipe, even though there are
problems with Newton's law between point sources on the RS1 ``TeV brane''.
This suggests there is a possibility that Newton's law might be obtained
correctly for the solutions found in this subsection, even though the
requirement for a valid reduction to a four-dimensional effective theory is
not completely satisfied.  However to check this would require repeating the
analysis of Arnowitt and Dent for the solutions found in this subsection, and
that will not be done in this paper.

I shall now consider two alternative ways in which the outer surface of the
thick pipe might be stabilized, consistent with the observed values of
Newton's constant and the effective $d = 4$ cosmological constant, for which
the value of $\tau$ is not fixed uniquely, and the problem noted above can be
avoided.  In the first alternative, considered in the next subsection, $a
\left( y \right)$ has decreased to around $\kappa^{2/9}$ at the outer
surface, and there are Casimir effects near the outer surface.  However for
$\tau$ around the bulk power law value $- 0.7753$, the main part of the
decrease of $a \left( y \right)$ takes place in a very narrow part of the
classical region near the outer surface, corresponding to $y$ near
$\frac{1}{\alpha} = 1.0996 B$ in the interpolating function
(\ref{interpolating function for b}), on page \pageref{interpolating function
for b}, and in the quantum region near the outer surface, so that only a
fraction $\sim 10^{- 6}$ of the integral that determines Newton's constant
comes from values of $y$ for which $a \left( y \right)$ is smaller than around
$10^{18}$ metres.  And in the second alternative, considered in subsection
\ref{Stiffening by fluxes}, on page \pageref{Stiffening by fluxes}, the outer
surface is stabilized in the classical region by extra fluxes of the
three-form gauge field, whose four-form field strength wraps three-cycles of
$\mathcal{M}^6$ times the radial dimension of the thick pipe, and for $\tau$
around $- 0.7753$, the value of $a \left( y \right)$ at the outer surface is
around $10^{22}$ metres.

\subsection{Solutions with $a$ as small as $\kappa^{2/9}$, at the
outer surface of the thick pipe}
\label{Solutions with a as small as kappa to the two ninths at the outer
surface}

I shall now look for solutions of the Einstein equations (\ref{first Einstein
equation}), (\ref{second Einstein equation}), and (\ref{third Einstein
equation}), on page \pageref{first Einstein equation}, and the boundary
conditions (\ref{boundary conditions at y2}) at
the outer surface of the thick pipe, such that the term $\frac{4}{a^2}$, in
the square root, $R$, defined in (\ref{definition of the square root R}), is
still extremely small, compared to the term $- \frac{8}{b^2}$, when $c =
\frac{db}{dy}$ is no longer large compared to $\sqrt{\frac{4}{3}}$, assuming,
as in the preceding two subsections, that the boundary conditions at the inner
surface of the thick pipe have already been solved, such that in the first
bulk power law region, we are on a trajectory (\ref{bulk type power law for
upper sign}), (\ref{bulk type power law dependence of b on y for upper sign}),
and (\ref{bulk power law for a in terms of b}), on page \pageref{bulk type
power law for upper sign}, with a large value of
$\frac{B}{\kappa^{2/9}}$, and $A$ not yet determined, but such that $A
\left( \frac{\kappa^{2/9}}{B} \right)^{0.7753} \gg B$.  We are
therefore, in the main part of the bulk, where both $a$ and $b$ are large
compared to $\kappa^{2/9}$, and both $\left( y - y_1 \right)$ and
$\left( y_2 - y \right)$ are large compared to $\kappa^{2/9}$, on a
solution of the form studied in subsection \ref{The classical solutions in the
bulk}, on page \pageref{The classical solutions in the
bulk}, so that the interpolating function (\ref{interpolating function for
b}), for $b$ as a function of $y$, with $\alpha$ given by (\ref{alpha in
interpolating function for b}), and the interpolating function
(\ref{interpolating function for a in terms of b}), for $a$ as a function of
$b$, are approximately valid, throughout the main part of the bulk.

There is now no possibility of satisfying the boundary conditions at the outer
surface until $ a $ has become as small as $ \kappa^{2/9} $, so that
there are Casimir effects on and near the outer surface.  So we are now looking
for a solution in which the constant of integration $\tilde{A}$, in (\ref{bulk
power law for a dot in terms of a with lower sign}), whose value is determined
by the boundary conditions at the outer surface, obtains a very large value in
units of $\kappa^{2/9}$, by a mechanism analogous to the way in which
the constant of integration $B$, whose value is determined by the boundary
conditions at the inner surface, can obtain a large value in units of $\kappa^{
\frac{2}{9}}$, as studied in subsection \ref{The region near the inner surface
of the thick pipe}, on page \pageref{The region near the inner surface of the
thick pipe}, but with the roles of $b$ and $a$ now reversed.

We therefore now assume that the three observed spatial dimensions, whose
curvature has become very large at the outer surface, due to
the small size of the warp factor, $a$, there, are compact hyperbolic, so as to
maximize the available range of dependences of the Casimir energy densities, at
the outer surface, on $a$.  This violates rotational invariance and Lorentz
invariance globally, but not locally.  The violation of Lorentz invariance
globally means that the Casimir energy-momentum tensors on and
near the outer surface will not necessarily have the forms (\ref{T IJ block
diagonal structure}) and (\ref{T tilde i UV block diagonal structure}), but I
shall consider the case where they do have these forms.

We will find that a large value of $ \frac{\tilde{A}}{\kappa^{2/9}} $
can be obtained self-consistently in the same way as a large value of $ \frac{
B}{\kappa^{2/9}} $.  By analogy with the region near the inner surface,
I shall first consider the case where $t^{\left( 3 \right)} = t^{\left( 2
\right)}$, as would be appropriate for a $ \mathrm{dS}_4 $ times flat $
\mathbf{R}^7 $ background.  In that case, by analogy with (\ref{conservation
equation when t 3 equals t 1}), on page \pageref{conservation equation when t 3
equals t 1}, the conservation equation (\ref{conservation equation for the t
i}), on page \pageref{conservation equation for the t i}, reduces to:
\begin{equation}
\label{conservation for small a in the proximity force approximation}
  \frac{d t^{\left( 2 \right)}}{da} + \frac{4}{a} t^{\left( 2
  \right)} - \frac{4}{a} t^{\left( 1 \right)} = 0
\end{equation}
Hence, in this case, the expansion coefficients $ \tilde{C}^{\left( i \right)
}_n $, in the expansions (\ref{bulk Casimir action density for small a}), on
page \pageref{bulk Casimir action density for small a}, are related by:
\begin{equation}
  \label{C tilde 1 and C tilde 3 for proximity force} \tilde{C}^{\left( 1
  \right)}_n = - \frac{\left( 4 + 3 n \right)}{4} \tilde{C}^{\left( 2
  \right)}_n, \hspace{4em} \tilde{C}^{\left( 3 \right)}_n = \tilde{C}^{\left(
  2 \right)}_n \hspace{4em} n \geq 0
\end{equation}

By analogy with (\ref{second Einstein equation for c in terms of b with upper
sign and specific C n}), on page \pageref{second Einstein equation for c in
terms of b with upper sign and specific C n}, I shall first consider the case
where all the
$\tilde{C}^{\left( 2 \right)}_n$ are zero, except for a single value of $n$,
and consider the region
\begin{equation}
  \label{small a region} a \ll \left( \frac{1}{6} \left\vert \tilde{C}^{\left(
  2 \right)}_n \right\vert \right)^{\frac{1}{6 + 3 n}} \kappa^{2/9}
\end{equation}
The Einstein equations (\ref{a dot over a}) and (\ref{second Einstein equation
without a dot}), with the lower choice of sign, as appropriate for this
region, and dropping the $- \frac{8}{b^2}$ and $\frac{4}{a^2}$ terms in the
square root, and the $\frac{4}{b^2}$ term in (\ref{second Einstein equation
without a dot}), now become:
\begin{equation}
  \label{a dot over a for small a} \frac{c}{a}  \frac{da}{db} = - 2
  \frac{c}{b} - \frac{1}{2} \sqrt{6 \frac{c^2}{b^2} + \frac{2}{3}
  \tilde{C}^{\left( 2 \right)}_n \frac{\kappa^{\frac{2}{3} \left( n + 2
  \right)}}{a^{8 + 3 n}}}
\end{equation}
\begin{equation}
  \label{second Einstein equation for small a} \frac{c}{b}  \frac{dc}{db} - 3
  \frac{c^2}{b^2} - 2 \frac{c}{b} \sqrt{6 \frac{c^2}{b^2} + \frac{2}{3}
  \tilde{C}^{\left( 2 \right)}_n \frac{\kappa^{\frac{2}{3} \left( n + 2
  \right)}}{a^{8 + 3 n}}} + \left( \frac{2 + n}{3} \right) \tilde{C}^{\left( 2
  \right)}_n \frac{\kappa^{\frac{2}{3} \left( n + 2 \right)}}{a^{8 + 3 n}} = 0
\end{equation}

The natural independent variable in this region would be $a$, and we would
expect a trajectory analogous to (\ref{small b power law}), with $\dot{a} = c
\frac{da}{db}$ being given by a power law as a function of $a$, with a fixed
coefficient, and $b$ being given by a power law as a function of $a$, with an
undetermined coefficient, analogous to (\ref{small b power law for a in terms
of b}).  So we try an ansatz:
\begin{equation}
  \label{small a power law} c \frac{da}{db} = - \tilde{\sigma} \left(
  \frac{a}{\kappa^{2/9}} \right)^{\tilde{\rho}}, \hspace{4em} b =
  \tilde{B} \left( \frac{a}{\kappa^{2/9}} \right)^{\tilde{\tau}}
\end{equation}
This implies $\frac{da}{db} = \frac{\kappa^{2/9}}{\tilde{\tau} b}
\left( \frac{b}{\tilde{B}} \right)^{\frac{1}{\tilde{\tau}}}$, and $\frac{c}{b}
\frac{}{} = - \frac{\tilde{\sigma} \tilde{\tau}}{\kappa^{2/9}} \left(
\frac{b}{\tilde{B}} \right)^{\frac{\tilde{\rho} - 1}{\tilde{\tau}}}$.  We then
find, from (\ref{a dot over a for small a}), and (\ref{second Einstein
equation for small a}), that $\tilde{\rho} = - \frac{6 + 3 n}{2}$, which is
the same as $\rho$, at the inner surface of the thick pipe, and
$\frac{\kappa^{\frac{2}{3} \left( n + 2 \right)}}{a^{8 + 3 n}} =
\frac{1}{\tilde{\sigma}^2 \tilde{\tau}^2}  \frac{c^2}{b^2}$.  Thus (\ref{a dot
over a for small a}) and (\ref{second Einstein equation for small a}) reduce
to
\begin{equation}
  \label{a dot over a for tau tilde} \frac{1}{\tilde{\tau}} = - 2 -
  \frac{1}{2}  \sqrt{6 + \frac{2}{3} \tilde{C}^{\left( 2 \right)}_n
  \frac{1}{\tilde{\sigma}^2 \tilde{\tau}^2} }
\end{equation}
\begin{equation}
  \label{second Einstein equation for tau tilde} - \left( \frac{6 + 3 n}{2
  \tilde{\tau}} \right) - 2 - \frac{1}{\tilde{\tau}} - 2 \sqrt{6 + \frac{2}{3}
  \tilde{C}^{\left( 2 \right)}_n \frac{1}{\tilde{\sigma}^2 \tilde{\tau}^2} } +
  \left( \frac{2 + n}{3} \right) \tilde{C}^{\left( 2 \right)}_n
  \frac{1}{\tilde{\sigma}^2 \tilde{\tau}^2} = 0
\end{equation}
which imply $\frac{2}{3} \tilde{C}^{\left( 2 \right)}_n
\frac{1}{\tilde{\sigma}^2 \tilde{\tau}^2} = 10 + \frac{16}{\tilde{\tau}} +
\frac{4}{\tilde{\tau}^2}$, and
\begin{equation}
  \label{quadratic equation for tau tilde} \left( 32 + 10 n \right)
  \tilde{\tau}^2 + \left( 32 + 13 n \right) \tilde{\tau} + \left( 8 + 4 n
  \right) = 0
\end{equation}
which has the solutions $\tilde{\tau} = - \frac{1}{2}$, and $\tilde{\tau} = -
\frac{4 n + 8}{5 n + 16}$.  However, $\tilde{\tau} = - \frac{1}{2}$ implies,
by (\ref{a dot over a for tau tilde}), that the square root, $R$, defined in
(\ref{definition of the square root R}), vanishes identically, so we cannot
conclude, from (\ref{first Einstein equation in terms of other equations}),
that all three Einstein equations are satisfied, and this also applies to the
solution $\tilde{\tau} = - \frac{4 n + 8}{5 n + 16}$, when $n = 0$.  Thus I
now assume $\tilde{\tau} = - \frac{4 n + 8}{5 n + 16}$, with $n \geq 1$.  We
then find:
\begin{equation}
  \label{power law trajectory for small a} \tilde{\sigma} = \sqrt{-
  \frac{\tilde{C}^{\left( 2 \right)}_n \left( 5 n + 16 \right)^2}{6 \left( 15
  n^2 + 96 n + 96 \right)}}
\end{equation}
We note, from (\ref{C tilde 1 and C tilde 3 for proximity force}), that this
requires $\tilde{C}^{\left( 1 \right)}_n$ to be positive, which corresponds to
a negative contribution to the energy density, $T_{00}$, which is opposite to
the situation at the inner surface of the thick pipe.

These results can be checked by solving (\ref{Einstein equation without double
derivatives}) for $\frac{\dot{b}}{b}$, which gives:
\begin{equation}
  \label{b dot over b} \frac{\dot{b}}{b} = \frac{1}{5} \left( - 4
  \frac{\dot{a}}{a} \pm \sqrt{6 \frac{\dot{a}^2}{a^2} - \frac{20}{b^2} +
  \frac{10}{a^2} + \frac{5}{3} \kappa^2 t^{\left( 3 \right)}} \right)
\end{equation}
and then using this result, to eliminate $\frac{\dot{b}}{b}$ from the first
Einstein equation, (\ref{first Einstein equation}), which gives:
\begin{equation}
  \label{first Einstein equation without b dot} \frac{\ddot{a}}{a} -
  \frac{9}{5}  \frac{\dot{a}^2}{a^2} \pm \frac{6}{5}  \frac{\dot{a}}{a}
  \sqrt{6 \frac{\dot{a}^2}{a^2} - \frac{20}{b^2} + \frac{10}{a^2} +
  \frac{5}{3} \kappa^2 t^{\left( 3 \right)}} - \frac{3}{a^2} +
  \frac{\kappa^2}{9} \left( 5 t^{\left( 1 \right)} - 6 t^{\left( 2 \right)} -
  t^{\left( 3 \right)} \right) = 0
\end{equation}
Comparing with (\ref{bulk power law trajectory for a with lower sign}),
we see that the upper sign in (\ref{b dot over b}) and (\ref{first Einstein
equation without b dot}) corresponds to the lower sign in (\ref{a dot over
a}) and (\ref{second Einstein equation without a dot}), and is thus the
appropriate sign for the region nearer the outer surface, in the case under
consideration in the present subsection, where $ a \sim \kappa^{2/9} $
at the outer surface.

Considering, again, the case where $t^{\left( 3 \right)} = t^{\left( 2
\right)}$, and all the $\tilde{C}^{\left( 2 \right)}_n$ are zero, except for a
single value of $n$, and trying for a solution of the form $\tilde{c} \equiv
\dot{a} = \tilde{c}_0 \left( 1 + s \right)$, where $\tilde{c}_0$ is the small
$a$ power law trajectory found above, and $s$ is a small perturbation, we
find, similarly to the region near the inner surface, that
\begin{equation}
  \label{perturbation of power law trajectory for small a} s =
  Sa^{\tilde{\eta}}, \quad \! \: \! \quad \! \: \! \quad \! \: \! \quad
  \tilde{\eta} = \frac{15 n^2 + 96 n + 96}{10 n + 32}
\end{equation}
where $S$ is a constant of integration.  And since $\tilde{\eta} \geq 3$ for
$n \geq 0$, trajectories near the small $a$ power law trajectory tend to
converge towards it in the direction of decreasing $a$, or in other words, of
increasing $y$, in the sense that $s$ decreases in magnitude in this
direction, so in this sense, the small $a$ power law trajectory is an
attractor in the direction of decreasing $a$, for $n \geq 0$.

Now, by analogy with (\ref{small b attractor trajectory}), the small $a$ power
law trajectories, found above, can all be written as:
\begin{equation}
  \label{small a attractor trajectory} \dot{a}^2 = - \frac{5 a^2}{18} \kappa^2
  t^{\left( 2 \right)} \left( a \right)^{} - \frac{2
  a^{\frac{18}{5}}}{\sqrt{6}} \int^{\infty}_a \frac{dx}{x^{\frac{13}{5}}}
  \left( \left( \frac{x}{a} \right)^{\frac{12}{5} \sqrt{6}} - \left(
  \frac{a}{x} \right)^{\frac{12}{5} \sqrt{6}} \right) \kappa^2 t^{\left( 2
  \right)} \left( x \right)
\end{equation}
And as near the inner surface of the thick pipe, the integral is convergent,
as $x \rightarrow \infty$, because $t^{\left( 2 \right)} \left( x \right)$
decreases at least as rapidly as $x^{- 8}$, as $x \rightarrow \infty$,
although of course $t^{\left( 2 \right)} \left( x \right)$, near the outer
surface of the thick pipe, is not the same function as $t^{\left( 2 \right)}
\left( x \right)$, near the inner surface of the thick pipe, and (\ref{small a
attractor trajectory}) does not give an exact solution of (\ref{first Einstein
equation without b dot}), except when $t^{\left( 2 \right)} \left( a \right)$
is a pure power of $a$.  However, if $t^{\left( 2 \right)} \left( a \right)$
is a linear combination of two different pure powers, say $a^{- \left( 8 + 3 n
\right)}$ and $a^{- \left( 8 + 3 m \right)}$, with $n$ and $m$ large, then the
remainder term, in (\ref{first Einstein equation without b dot}), will be
$\sim \frac{1}{\sqrt{nm}}$, while the leading terms will be $\sim n$ or $m$.
And for a pure power $a^{- \left( 8 + 3 n \right)}$, the integral, in
(\ref{small a attractor trajectory}), is of order $\frac{1}{n^2}$, compared to
the leading term.

Thus, by analogy with the situation near the inner surface
of the thick pipe, we expect that for large $- t^{\left( 2 \right)} \left( a
\right)$, there will be an attractor trajectory in the $\left( a, \dot{a}
\right)$ plane, such that nearby trajectories approach it, in the direction of
decreasing $a$, or in other words, in the direction towards the outer surface
of the thick pipe, in the sense that the relative separation of the two
trajectories decreases, in the direction of decreasing $a$, and this attractor
trajectory will approximately be given by:
\begin{equation}
  \label{approximate small a attractor trajectory} \dot{a} \simeq \sqrt{-
  \frac{5 a^2}{18} \kappa^2 t^{\left( 2 \right)} \left( a \right)^{}}
\end{equation}
This trajectory will eventually intersect every second bulk power law region
bulk power law trajectory (\ref{bulk power law for a dot in terms of a with
lower sign}), and, as near the inner surface of the thick pipe, we expect each
bulk power law trajectory to curve upwards as it approaches the attractor
trajectory, and then approach the attractor trajectory gradually.

The square roots in (\ref{b dot over b}) and (\ref{first Einstein equation
without b dot}) approximately vanish on the approximate small $a$ attractor
trajectory (\ref{approximate small a attractor trajectory}), but for $n \geq
1$, the small $a$ power law trajectories, which are approximately described by
(\ref{approximate small a attractor trajectory}), are already known to be
approximate solutions of all three Einstein equations near the outer surface
of the thick pipe, when all the $ \tilde{C}_n^{\left(2\right)} $ are zero,
except for a single value of $ n $, and that $ \tilde{C}_n^{\left(2\right)} $
is negative, so it seems reasonable to expect that (\ref{approximate small a
attractor trajectory}) will also give an approximate solution of all three
Einstein equations in the more general case, when $ \tilde{C}_n^{\left(2\right)
} \leq 0 $ for all $ n \geq 1 $.

Now in the same way as in the discussion beginning just before
(\ref{continuity condition for B sub q}), on page \pageref{continuity
condition for B sub q}, for the region near the inner surface of the thick
pipe, we have to consider whether these solutions can be self-consistent, when
we recalculate the expansion coefficients $\tilde{C}^{\left( i \right)}_n$ in
(\ref{bulk Casimir action density for small a}), on page \pageref{bulk Casimir
action density for small a}, for $a \left( y \right)$
corresponding to these solutions, in accordance with the discussion in
subsection \ref{Beyond the proximity force approximation}, on page
\pageref{Beyond the proximity force approximation}.  We first recall
that the bulk power law in the second classical power law region, that
corresponds to the bulk power law (\ref{bulk type power law for upper sign}),
on page \pageref{bulk type power law for upper sign}, in the first classical
power law region, is (\ref{bulk power law for a dot in terms of a with lower
sign}), on page \pageref{bulk power law for a dot in terms of a with lower
sign}, as we can confirm from (\ref{first Einstein equation without b dot})
above, with the upper choice of sign.  And from the discussion in subsection
\ref{Beyond the proximity force approximation}, a term $\kappa^{-
\frac{22}{9}} \tilde{C}^{\left( i \right)}_n \left(
\frac{\kappa^{2/9}}{a} \right)^{8 + 3 n}$ in (\ref{bulk Casimir action
density for small a}) will be associated with additional terms
$\kappa^{- \frac{22}{9}} \tilde{C}^{\left( i \right)}_{n, m} \left(
\frac{\kappa^{2/9}}{a} \right)^{8 + 3 n} \tilde{c}^m$, $1 \leq m \leq
n$, as well as terms with factors of higher derivatives of $a$ with respect to
$y$, which can, however, be bounded by constant multiples of the terms without
factors of higher derivatives, when the dependence of $a$ on $y$ is by a power
law.  The constant of integration $\tilde{A}$, in (\ref{bulk power law for a
dot in terms of a with lower sign}), will have to have a very large value, in
units of $\kappa^{2/9}$, in order to fit the observed value (\ref{de
Sitter radius}) of the de Sitter radius, so the largest additional terms will
be those with $m = n$.

Thus the Casimir terms in (\ref{b dot over b}) and (\ref{first Einstein
equation without b dot}), with the upper choice of sign, will be significant
near the outer surface for $\frac{\kappa^{2/9} \tilde{c}}{a} \geq 1$,
which from (\ref{bulk power law for a dot in terms of a with lower sign})
corresponds to $\frac{a}{\kappa^{2/9}} \leq \left(
\frac{\tilde{A}}{\kappa^{2/9}} \right)^{0.5326}$.  Defining
$\tilde{a}_q$ to be the value of $a$ where this is an equality, we then find,
from (\ref{A tilde in terms of A B and kappa}), that:
\begin{equation}
  \label{a sub q} \frac{\tilde{a}_q}{\kappa^{2/9}} = \left(
  \frac{\tilde{A}}{\kappa^{2/9}} \right)^{0.5326} = 0.03757 \left(
  \frac{\kappa^{2/9}}{B} \right)^{1.2427}
  \frac{A}{\kappa^{2/9}}
\end{equation}
And defining $\tilde{b}_q$ to be the corresponding value of $b$, we find from
(\ref{bulk power law for a in terms of b with lower sign}), on page
\pageref{bulk power law for a in terms of b with lower sign}, and (\ref{A 1 in
terms of A}), on page \pageref{A 1 in terms of A}, that
\begin{equation}
  \label{b tilde sub q} \frac{\tilde{b}_q}{\kappa^{2/9}} = 1.6884
  \left( \frac{B}{\kappa^{2/9}} \right)^{1.1450}
\end{equation}

Then in the same way as in subsection \ref{The region near the inner surface
of the thick pipe}, on page \pageref{The region near the inner surface of the
thick pipe}, for the region near the inner surface, we find that the
only self-consistent way to obtain a large value of the integration constant
$\tilde{A}$, is for $\tilde{c}$ to depend linearly on $a$ in the
quantum region $a \leq \tilde{a}_q$, which results in (\ref{a dot over a for
small a}) and (\ref{second Einstein equation for small a}) for $n = -
\frac{8}{3}$, and an effective coefficient $\tilde{C}^{\left( 2 \right)}_{-
\frac{8}{3}}$, for $\tilde{a}_q \geq a \gg \kappa^{2/9}$.  This
results self-consistently in (\ref{small a power law}), with $\tilde{\rho} =
\tilde{\tau} = 1$, and $\tilde{\sigma} = \frac{1}{3}
\sqrt{\frac{\tilde{C}^{\left( 2 \right)}_{- \frac{8}{3}}}{5}}$, so that the
effective coefficient $\tilde{C}^{\left( 2 \right)}_{- \frac{8}{3}}$ has to be
$> 0$ in order to obtain the linear relation.  We note that $\tilde{\eta}$, in
(\ref{perturbation of power law trajectory for small a}), takes the value $-
10$ when $n = - \frac{8}{3}$, so the linear trajectory is a very strong
attractor in the direction of increasing $a$.  However, in the same way as for
the corresponding result for the region near the inner surface, this has not
taken account of the fact that in the presence of deviations from the linear
trajectory, the equations to be solved will no longer be precisely (\ref{a dot
over a for small a}) and (\ref{second Einstein equation for small a}), with $n
= - \frac{8}{3}$.

Now $a$ continues to decrease with increasing $y$ in the quantum region near
the outer surface, since $ \tilde{c} = \frac{da}{dy} $ is still negative in the
quantum region.  Hence since
$\frac{b}{a}$ has the fixed value $\frac{\tilde{B}}{\kappa^{2/9}}$ in
the quantum region, $b$ stops increasing with increasing $y$ at the upper
limit $\tilde{b}_q$ of the classical region, and decreases with increasing $y$
in the quantum region.  Thus a
necessary condition for the existence of a solution to the boundary conditions
at the outer surface is that $b$ must be comparable to or larger than $a$ at
the start of the quantum region, or in other words, $\tilde{a}_q \gg
\tilde{b}_q$ must not hold, for if $\tilde{a}_q$ was $\gg \tilde{b}_q$, the
boundary conditions at the outer surface would not depend significantly on the
integration constant $\tilde{A}$, so that $\tilde{A}$ would be undetermined,
and $B$ would be overdetermined.

In the next subsection, I shall determine the values of $B$ and $\tilde{A}$
required to fit the observed values of Newton's constant and the cosmological
constant, for given values of $ \tau $ and $\kappa^{- \frac{2}{9}}$, assuming
that this consistency condition is satisfied.  We will then find that for $
\tau = 1 $, which follows from assuming that $ t^{\left( 3 \right)} = t^{\left(
1 \right)} $ in the quantum region near the inner surface of the thick pipe,
the consistency condition cannot be satisfied unless $\kappa^{-\frac{2}{9}}$ is
much smaller than the minimum value $ \sim 0.1 $ TeV allowed by current
observations.  However the linear relation between $ b $ and $ a $ in the
quantum region near the outer surface, which follows from (\ref{small a power
law}) and (\ref{quadratic equation for tau tilde}) for $ n = -\frac{8}{3} $, on
rejecting the solution $ \tilde{\tau} = -\frac{1}{2} $, is a consequence of the
assumption that $ t^{\left( 3 \right)} = t^{\left( 2 \right)} $ near the outer
surface, and there is no reason to expect this relation to be valid when $ a $
and $ b $ depend exponentially on $ y $.

Thus in a similar way to the discussion following (\ref{dependence of a on b in
the quantum region}), on page \pageref{dependence of a on b in the quantum
region}, we should discard the assumption that $ t^{\left( 3 \right)} = t^{
\left( 2 \right)} $ near the outer surface, and assume that the $ t^{\left( i
\right)} $, in (\ref{T IJ block diagonal structure}), on page \pageref{T IJ
block diagonal structure}, are constrained only by the conservation equation
(\ref{conservation equation for the t i}).  We would then expect, by analogy
with the situation near the inner surface, that almost any value of $
\tilde{\tau} $ could be obtained, provided there exists a suitable smooth
compact quotient $ \mathcal{M}^3 $ of $ \mathbf{H}^3 $, and a choice of spin
structure on $ \mathcal{M}^3 $, that results self-consistently in the
appropriate values of the independent coefficients $ \tilde{C}^{\left( 2
\right)}_{-\frac{8}{3}} $ and $ \tilde{C}^{\left( 3 \right)}_{-\frac{8}{3}} $.

\subsubsection{Newton's constant and the cosmological constant}
\label{Newtons constant and the cosmological constant}

I shall now determine the values of the integration constants $B$, in
(\ref{interpolating function for b}), on page \pageref{interpolating function
for b}, and $A$, in (\ref{interpolating function for
a in terms of b}), or equivalently, $\tilde{A}$, in (\ref{bulk power law for a
dot in terms of a with lower sign}), and the constants $ \tau $, in
(\ref{ansatz for a in the quantum region}), on page \pageref{ansatz for a in
the quantum region}, and $\kappa^{2/9}$, for which the
solutions found above can fit the observed values of
Newton's constant, (\ref{Newtons constant}), and the cosmological constant,
(\ref{Lambda}), and check that this type of solution is consistent with
observational limits on the existence of large extra dimensions, and can avoid
the possible problem noted in the discussion following (\ref{tau and B for mod
chi equals 7 times 10 to the 4}), on page \pageref{tau and B for mod chi equals
7 times 10 to the 4}, for the solutions studied in subsection \ref{Solutions
with both
a and b large at the outer surface}, on page \pageref{Solutions with both a and
b large at the outer surface}.  I shall then check that this type of solution
is consistent with experimental limits on deviations from Newton's law at
sub-millimetre distances, in subsection \ref{Comparison with sub-millimetre
tests of Newtons law}, on page \pageref{Comparison with sub-millimetre tests of
Newtons law}, and with precision solar system tests of General Relativity, in
subsection \ref{Comparison with precision solar system tests of General
Relativity}, on page \pageref{Comparison with precision solar system tests of
General Relativity}.  Some further consequences of the warp factor decreasing
to a small
value, at the outer surface of the thick pipe, in this type of solution, are
considered briefly in subsection \ref{Further consequences of the warp factor
decreasing to a small value}, on page \pageref{Further consequences of the warp
factor decreasing to a small value}.

We can follow the same method as used in subsection \ref{G sub N and Lambda for
solutions with outer surface in classical region}, on page \pageref{G sub N and
Lambda for solutions with outer surface in classical region}.  The term,
in the Einstein action term in (\ref{upstairs bulk action}), that produces the
Einstein action, (\ref{Einstein action}), in four dimensions, is again given
by (\ref{Einstein term in four dimensional effective action}), where $b \left(
y \right)$ is now given by (\ref{interpolating function for b}), and $a$, as a
function of $b$, is given by (\ref{interpolating function for a in terms of
b}).  Thus we now have:
\begin{equation}
  \label{a squared b to the sixth} a^2 b^6 \simeq A^2 B^6  \left(
  \frac{\kappa^{2/9}}{B} \right)^{1.5506} f \left( \frac{y}{B} \right)
\end{equation}
where $f \left( Y \right)$ is defined by:
\begin{equation}
  \label{definition of f of Y} f \left( Y \right) \equiv \frac{5.1220 \left( 1
  - 0.9094 Y \right)^{0.0651} Y^{1.5346}}{\left( \left( 1 - 0.9094 Y
  \right)^{0.3549} + 12.0816 Y^{0.8448} \right)^2}
\end{equation}

\begin{figure}[t]
\setlength{\unitlength}{1cm}
\begin{picture}(10.18,8.41)(-2.0,0.2)
\put(1.260,1.08){\line(0,1){7.42}}
\put(1.03,0.355){$ 0.0 $}
\put(1.260,1.08){\line(1,0){9.5}}
\put(2.02,0.90){\line(0,1){0.36}}
\put(2.78,0.90){\line(0,1){0.36}}
\put(3.54,0.90){\line(0,1){0.36}}
\put(4.30,0.90){\line(0,1){0.36}}
\put(5.06,0.90){\line(0,1){0.36}}
\put(4.83,0.355){$ 0.5 $}
\put(5.82,0.90){\line(0,1){0.36}}
\put(6.58,0.90){\line(0,1){0.36}}
\put(7.34,0.90){\line(0,1){0.36}}
\put(8.10,0.90){\line(0,1){0.36}}
\put(8.86,0.90){\line(0,1){0.36}}
\put(8.63,0.355){$ 1.0 $}
\put(9.62,0.90){\line(0,1){0.36}}
\put(10.38,0.90){\line(0,1){0.36}}
\put(1.09,3.43){\line(1,0){0.36}}
\put(0.25,3.29){$ 0.01 $}
\put(1.09,5.78){\line(1,0){0.36}}
\put(0.25,5.64){$ 0.02 $}
\put(1.09,8.13){\line(1,0){0.36}}
\put(0.25,7.99){$ 0.03 $}
\qbezier(9.59,6.45)(9.62,6.09)(9.64,0.893)
\qbezier(2.08,6.29)(2.66,8.04)(5.81,8.15)
\qbezier(1.26,1.10)(1.67,5.25)(2.08,6.32)
\qbezier(5.81,8.17)(9.69,7.99)(9.59,6.47)
\end{picture}
\caption{The function $ f\left( Y \right) $ defined in (\ref{definition of f
of Y})}
\label{graph of f of Y}
\end{figure}

The function $f \left( Y \right)$ is illustrated in Figure \ref{graph of f
of Y}.
The peak is at $Y = 0.5777$, at which point the value of the function is
$0.03002$.  The function is $0$ at $Y = 0$, and at $Y = 1.0996$, and by use of
PARI/GP {\cite{PARI GP}}, we find:
\begin{equation}
  \label{integral of f of Y} \int^{1.0996}_0 dYf \left( Y \right) = 0.02967
\end{equation}
The contribution to this integral, from the regions $0 \leq Y \leq
\frac{y_1}{B}$, and $\frac{y_2}{B} \leq Y \leq 1.0996$, will be negligible, to
the accuracy to which we are working, so we now find:
\begin{equation}
  \label{integral of a squared b to the sixth for a small at outer boundary}
  \int^{y_2}_{y_1} dya^2 b^6 \simeq 0.02967 \kappa^{\frac{14}{9}} A^2 \left(
  \frac{B}{\kappa^{2/9}} \right)^{5.4494} = 0.02967
  \kappa^{\frac{14}{9}} A_1^2 \left( \frac{B}{\kappa^{2/9}}
  \right)^{1.3102 \tau + 6.4652}
\end{equation}
instead of (\ref{contribution from classical region}).
The numerical coefficient in (\ref{integral of a squared b to the
sixth for a small at outer boundary}) should now be approximately correct, for
the solutions found in subsection \ref{Solutions with a as small as kappa to
the two ninths at the outer
surface}, on page \pageref{Solutions with a as small as kappa to the two ninths
at the outer surface}, to
the extent that the interpolating functions (\ref{interpolating function for
b}), and (\ref{interpolating function for a in terms of b}), are approximately
valid, whereas the numerical coefficient, in (\ref{contribution from classical
region}), was only valid up to a factor of order $1$.

The result (\ref{integral of a squared b to the sixth for a small at outer
boundary}) is for a smooth compact quotient of $\mathbf{C} \mathbf{H}^3$.
To obtain the corresponding result for a smooth compact quotient of
$\mathbf{H}^6$, we note that throughout the range where they give
significant contributions to the integral, $a$ and $b$ are solutions of the
vacuum Einstein equations, and $a$ is so large that the curvature of the
four-dimensional de Sitter space can be neglected.  We recall that we have
chosen the metric
$h_{AB}$ for $\mathbf{H}^6$ to have radius of curvature equal to $1$, so
that $R_{ABCD} \left( h \right) = h_{AC} h_{BD} - h_{AD} h_{BC}$, and $R_{AB}
\left( h \right) = 5 h_{AB}$, as stated after (\ref{Ricci tensor for the metric
ansatz}), on page \pageref{Ricci tensor for the metric ansatz}.  Then looking
at the Ricci tensor components (\ref{Ricci tensor for the metric ansatz}),
and noting that $R_{AB} \left( h \right) = 4
h_{AB}$ for the standard metric on $\mathbf{C} \mathbf{H}^3$ introduced in
subsection \ref{CH3}, on page \pageref{CH3}, we see that the vacuum Einstein
equations for $\mathbf{C} \mathbf{H}^3$, when $a$ is so large that the
curvature of the $\mathrm{dS}_4$ can be neglected, can be transformed into the
corresponding equations for $\mathbf{H}^6$, by rescaling $y$ by a factor
$\sqrt{\frac{4}{5}}$.  Furthermore, derivatives with respect to $y$ are larger
for $\mathbf{H}^6$ than for $\mathbf{C} \mathbf{H}^3$ by a factor
$\sqrt{\frac{5}{4}}$, so the range of $y$ is smaller for $\mathbf{H}^6$ than
for $\mathbf{C} \mathbf{H}^3$, by a factor $\sqrt{\frac{4}{5}}$.  Thus the
integral $\int^{y_2}_{y_1} dya^2 b^6$ for $\mathbf{H}^6$ is obtained from
the corresponding integral for $\mathbf{C} \mathbf{H}^3$ by multiplying by
a factor $\sqrt{\frac{4}{5}}$, or in other words, replacing the coefficient
$0.02967$, in (\ref{integral of a squared b to the sixth for a small at outer
boundary}), by $0.02654$.

The integral over the compact six-manifold, in terms of the Euler number of
the compact six-manifold, will be the same as before, so we find that when
the compact six-manifold is a smooth compact quotient of $\mathbf{C}
\mathbf{H}^3$, the Einstein action term, in the four-dimensional effective
action, for the solutions considered in subsection \ref{Solutions with a as
small as kappa to the two ninths at the outer surface}, will be equal to:
\begin{equation}
  \label{Einstein term in four dimensional effective action for solutions with
  a small at outer surface} 0.3067 \frac{1}{\kappa^{\frac{4}{9}}} A^2 \left(
  \frac{ B}{\kappa^{2/9}} \right)^{5.4494} \chi \left( \mathcal{M}^6
  \right) \int d^4 x \sqrt{- \tilde{g}} \tilde{g}^{\mu \nu} R_{\mu \tau \nu}
  \, \!^{\tau} \left( \tilde{g} \right)
\end{equation}
And when the compact six-manifold is a smooth compact quotient of
$\mathbf{H}^6$, we get the same result as in (\ref{Einstein term in four
dimensional effective action for solutions with a small at outer surface}),
but with the numerical coefficient replaced by $\sqrt{\frac{4}{5}} \times
\frac{8}{5} \times 0.3067 \simeq 0.4389$.

Thus from the relation (\ref{A in terms of de Sitter radius}), on page
\pageref{A in terms of de Sitter radius}, between $A$, and the observed de
Sitter radius (\ref{de Sitter radius}), and the discussion following
(\ref{Einstein term in four dimensional effective action for solutions
classical at the outer surface}), on page \pageref{Einstein term in four
dimensional effective action for solutions
classical at the outer surface}, we see that when we define the rescaled
metric $\bar{g}_{\mu \nu}$ by $\bar{g}_{\mu \nu} = \left(
\textrm{de Sitter radius} \right)^2 \tilde{g}_{\mu \nu}$ as before,
so as to measure distances in ordinary units, rather than
in units of the de Sitter radius, the Einstein action term, in the
four-dimensional effective action, for the solutions considered in subsection
\ref{Solutions with a as small as kappa to the two ninths at the outer
surface}, will for smooth compact
quotients of $\mathbf{C} \mathbf{H}^3$ be equal to:
\begin{equation}
  \label{Einstein action term in the four dimensional effective action without
  A} - \frac{ 0.3067}{1.2772^{2 \tau}}
  \frac{1}{\kappa^{\frac{4}{9}}} \left( \frac{B}{\kappa^{2/9}}
  \right)^{1.3102 \tau + 6.4652} \left| \chi \left( \mathcal{M}^6 \right)
  \right|^{1 + \frac{\tau}{3}} \int d^4 x \sqrt{- \bar{g}} \bar{g}^{\mu \nu}
  R_{\mu \tau \nu} \, \!^{\tau} \left( \bar{g} \right)
\end{equation}
And for smooth compact quotients of $\mathbf{H}^6$, we get the same result
as in (\ref{Einstein action term in the four dimensional effective action
without A}), but with the numerical coefficient replaced by $\frac{0.4389}{
1.1809^{2 \tau}}$.

Thus, comparing with (\ref{Einstein action}), we find that for smooth compact
quotients of $\mathbf{C} \mathbf{H}^3$:
\begin{equation}
  \label{1 over G Newton for solutions with a small at the outer surface}
  \frac{1}{G_N} \simeq \frac{15.416}{1.2772^{2 \tau}}
  \frac{1}{\kappa^{\frac{4}{9}}}^{} \left( \frac{B}{\kappa^{2/9}}
  \right)^{1.3102 \tau + 6.4652} \left| \chi \left( \mathcal{M}^6 \right)
  \right|^{1 + \frac{\tau}{3}}
\end{equation}
And for smooth compact quotients of $\mathbf{H}^6$, the numerical coefficient
is replaced by $\frac{22.062}{1.1809^{2 \tau}}$.  This is the form taken by the
ADD mechanism
{\cite{ADD1, ADD2}}, for the solutions considered in subsection \ref{Solutions
with a as small as kappa to the two ninths at the outer surface}, on page
\pageref{Solutions with a as small as kappa to the two ninths at
the outer surface}.  We see that in the same way as for the solutions
considered in subsection
\ref{Solutions with both a and b large at the outer surface}, on page
\pageref{Solutions with both a and b large at the outer surface}, there is no
ADD effect unless $ \tau > -4.9345 $.  This is due to the fact that for the
classical region in the bulk, and for $ \tau < 0 $, also for the quantum region
near the inner surface of the thick pipe, we live on the wrong boundary, from
the point of view of the first Randall-Sundrum model \cite{Randall Sundrum 1},
and for $ \tau < -4.9345 $, the reverse RS1 effect outweighs the ADD effect.

For $ \tau = 1 $, we find from (\ref{1 over G Newton for solutions with a small
at the outer surface}) that for smooth compact quotients of $\mathbf{C}
\mathbf{H}^3$:
\begin{equation}
  \label{B for solutions with a small at the outer surface}
  \frac{B}{\kappa^{2/9}} \simeq \frac{0.7491}{\left| \chi \left(
  \mathcal{M}^6 \right) \right|^{0.1715}} \left(
  \frac{\kappa^{\frac{4}{9}}}{G_N} \right)^{0.1286}
\end{equation}

Considering, now, the case of TeV-scale gravity, we will find in section
\ref{E8 vacuum gauge fields and the Standard Model}, on page~\pageref{E8
vacuum gauge fields and the Standard Model}, that $\kappa$ is related to the
gravitational masses $M$, $M_p$, and $M_D$, with $D = 11$, defined
respectively by Mirabelli, Perelstein, and Peskin {\cite{Mirabelli Perelstein
Peskin}}, Giddings and Thomas {\cite{Giddings Thomas}}, and Giudice, Rattazzi,
and Wells {\cite{Giudice Rattazzi Wells}}, by $M = M_p = 2^{\frac{1}{9}} M_D =
2 \pi \left( \frac{1}{\pi \kappa} \right)^{\frac{2}{9}}$.  I shall use the
results of Mirabelli, Perelstein, and Peskin, for six flat extra dimensions,
as an indication of the current experimental limits on $\kappa^{-
\frac{2}{9}}$.  Thus from their Table 1, we see that in 1998, the LEP 2 lower
bound on $\kappa^{- \frac{2}{9}}$ was around 107 GeV, and the
Tevatron lower bound was around 125 GeV.  And the final lower bound on
$\kappa^{- \frac{2}{9}}$ attainable at the Tevatron is expected to be around
166 GeV, and the final lower bound on $\kappa^{- \frac{2}{9}}$ attainable at
the LHC is expected to be around 677 GeV.

As a representative example of TeV-scale gravity, I shall consider the case
where the Giudice, Rattazzi, and Wells gravitational mass $M_D$, for $ D = 11
$, is equal to $1$ TeV, which corresponds to $\kappa^{- \frac{2}{9}} = 0.2217$
TeV, so that $\kappa^{2/9} = 8.899 \times 10^{- 19}$ metres.  We then
find from (\ref{Newtons constant}), on page \pageref{Newtons constant}, that
for $ \tau = 1 $, and smooth compact quotients of $\mathbf{C} \mathbf{H}^3$:
\begin{equation}
  \label{B for TeV scale gravity} B \simeq \frac{1.515 \times 10^4}{\left|
  \chi \left( \mathcal{M}^6 \right) \right|^{0.1715}} \kappa^{2/9}
  \simeq \frac{1.348 \times 10^{- 14} \textrm{ metres}}{\left| \chi \left(
  \mathcal{M}^6 \right) \right|^{0.1715}}
\end{equation}
Thus from (\ref{A in terms of de Sitter radius}), on page \pageref{A in terms
of de Sitter radius}, and (\ref{de Sitter radius}), on page \pageref{de Sitter
radius}, we have for $ \tau = 1 $, and smooth compact quotients of $\mathbf{C}
\mathbf{H}^3$:
\begin{equation}
  \label{A in metres} A \simeq \frac{8.6 \times 10^{30}\textrm{ metres}}{\left|
  \chi \left( \mathcal{M}^6 \right) \right|^{0.0328}} \simeq \frac{5.4 \times
  10^{65}\sqrt{G_N}}{\left| \chi \left( \mathcal{M}^6 \right) \right|^{0.0328}}
\end{equation}
Thus from (\ref{A tilde in terms of A B and kappa}), on page \pageref{A tilde
in terms of A B and kappa}, the integration
constant $\tilde{A}$, in (\ref{bulk power law for a dot in terms of a with
lower sign}), is given for $ \tau = 1 $, and smooth compact quotients of
$\mathbf{C} \mathbf{H}^3$, by:
\begin{equation}
  \label{A tilde for TeV scale gravity} \tilde{A} \simeq 3.15 \times 10^{61}
  \left| \chi \left( \mathcal{M}^6 \right) \right|^{0.3386} \textrm{ metres}
  \simeq 3.54 \times 10^{79} \left| \chi \left( \mathcal{M}^6 \right)
  \right|^{0.3386} \kappa^{2/9}
\end{equation}
Thus from (\ref{a sub q}) and (\ref{b tilde sub q}), on page \pageref{a sub
q}, we find that for $\tau = 1$, and smooth compact quotients of $\mathbf{C}
\mathbf{H}^3$:
\begin{equation}
  \label{a tilde sub q and b tilde sub q for tau equals 1}
  \frac{\tilde{a}_q}{\kappa^{2/9}} \simeq 2.33 \times 10^{42} \left|
  \chi \left( \mathcal{M}^6 \right) \right|^{0.1803}, \quad \! \: \! \quad \!
  \: \! \quad \! \: \! \quad \frac{\tilde{b}_q}{\kappa^{2/9}} \simeq
  \frac{1.033 \times 10^5}{\left| \chi \left( \mathcal{M}^6 \right)
  \right|^{0.1964}}
\end{equation}

Thus the consistency requirement that when the exponent $\tilde{\tau}$ in
(\ref{small a power law}), on page \pageref{small a power law}, is equal to
$1$, $\tilde{a}_q$ must not be large compared to $\tilde{b}_q$, is violated
for $\tau = 1$.  Thus the relation $t^{\left( 3 \right)} = t^{\left( 1
\right)}$ near the inner surface of the thick pipe and the relation $t^{\left(
3 \right)} = t^{\left( 2 \right)}$ near the outer surface
cannot both be satisfied, but as noted in subsections \ref{The region near the
inner surface of the thick pipe}, on page \pageref{The region near the inner
surface of the thick pipe}, and \ref{Solutions with a as small as kappa to the
two ninths at the outer surface}, on page \pageref{Solutions with a as small as
kappa to the two ninths at the outer surface}, there is no reason for either of
these relations to be
satisfied, since $a$ and $b$ depend exponentially on $y$ in the quantum
regions.  If $\tilde{\tau}$ is $< 0$, there is no consistency condition on
$\tilde{a}_q$ and $\tilde{b}_q$, since $b \left( y \right)$ continues to
increase with increasing $y$ in the quantum region near the outer surface.

When $\tau = - 0.7753$, as for the classical power law (\ref{bulk power law
for a in terms of b}), on page \pageref{bulk power law for a in terms of b},
in the first classical region, we find from (\ref{1 over G Newton for
solutions with a small at the outer surface}) that for smooth compact
quotients of $\mathbf{C} \mathbf{H}^3$:
\begin{equation}
  \label{B for solutions with a small at the outer surface and classical tau}
  \frac{B}{\kappa^{2/9}} \simeq \frac{0.5646}{\left| \chi \left(
  \mathcal{M}^6 \right) \right|^{0.1361}} \left(
  \frac{\kappa^{\frac{4}{9}}}{G_N} \right)^{0.1835}
\end{equation}
And for smooth compact quotients of $\mathbf{H}^6$, the coefficient $0.5646$
is replaced by $0.5406$.

Considering, again, the case of TeV-scale gravity, with $\kappa^{-
\frac{2}{9}} = 0.2217$ TeV, we find that for $\tau = - 0.7753$, and smooth
compact quotients of $\mathbf{C} \mathbf{H}^3$:
\begin{equation}
  \label{B for TeV scale gravity and classical tau} B \simeq \frac{7.864
  \times 10^5}{\left| \chi \left( \mathcal{M}^6 \right) \right|^{0.1361}}
  \kappa^{2/9} \simeq \frac{6.998 \times 10^{- 13} \textrm{ metres}}{
  \left| \chi \left( \mathcal{M}^6 \right) \right|^{0.1361}},
\end{equation}
\begin{equation}
  \label{A in metres for classical tau} A = \frac{1.83 \times 10^{26}
  \textrm{ metres}}{\left| \chi \left( \mathcal{M}^6 \right) \right|^{0.1292}}
  = \frac{1.14 \times 10^{61} \sqrt{G_N}}{\left| \chi \left( \mathcal{M}^6
  \right) \right|^{0.1292}},
\end{equation}
and
\begin{equation}
  \label{A tilde for TeV scale gravity and classical tau} \tilde{A} \simeq
  5.29 \times 10^{48} \left| \chi \left( \mathcal{M}^6 \right)
  \right|^{0.0750} \textrm{ metres} = 5.95 \times 10^{66} \left| \chi \left(
  \mathcal{M}^6 \right) \right|^{0.0750} \kappa^{2/9} .
\end{equation}

Comparing (\ref{A tilde for TeV scale gravity}) and (\ref{A tilde for TeV
scale gravity and classical tau}), we see that the cost of decreasing
$\frac{B}{\kappa^{2/9}}$ by a factor of around $50$, by increasing
$\tau$ from $- 0.7753$ to $1$, is to increase $\frac{\tilde{A}}{\kappa^{\frac{2
}{9}}}$ by a factor of around $ 6 \times 10^{12} $.
Thus it does not seem likely that $B$ will be much smaller than the value
(\ref{B for TeV scale gravity and classical tau}) corresponding to $\tau = -
0.7753$.  Thus from the upper bound of around $7 \times 10^4$ on $\left| \chi
\left( \mathcal{M}^6 \right) \right|$ found in subsection \ref{The expansion
parameter}, on page \pageref{The expansion parameter}, it does not seem likely
that $\frac{B}{\kappa^{2/9}}$ will be much smaller than $10^5$, for
TeV-scale gravity.

By decreasing $\tau$ below $- 0.7753$, it will be possible to decrease
$\frac{\tilde{A}}{\kappa^{2/9}}$ at a cost of increasing
$\frac{B}{\kappa^{2/9}}$, until as $\tau$ approaches the values near
$- 3$ in (\ref{tau and B for mod chi equals 1}) and (\ref{tau and B for mod
chi equals 7 times 10 to the 4}), on page \pageref{tau and B for mod chi
equals 1}, the assumption made in subsection \ref{Solutions with a as small as
kappa to the two ninths at the outer surface}, on page \pageref{Solutions with
a as small as kappa to the two ninths at the outer surface}, that the term
$\frac{4}{a^2}$ in the square root $R$ defined in (\ref{definition of the
square root R}), on page \pageref{definition of the square root R}, is still
extremely small compared to the term $- \frac{8}{b^2}$, when $c =
\frac{db}{dy}$ is no longer large compared to $\sqrt{\frac{4}{3}}$, will no
longer be valid, and the type of solution considered in subsection
\ref{Solutions with a as small as kappa to the two ninths at the outer
surface} will resemble the solutions studied in subsection \ref{Solutions with
both a and b large at the outer surface}, on page \pageref{Solutions with both
a and b large at the outer surface}, except in the region close to the outer
surface.  However $a$ will still decrease to around $\kappa^{2/9}$ at
the outer surface for the solutions considered in subsection \ref{Solutions
with a as small as kappa to the two ninths at the outer surface}, because there
are no solutions of the Einstein equations where $c$ goes to zero at a finite
value of $b$ on the second branch of the square root.  For if such a solution
existed, then expanding $u = \frac{b}{a}$ near the boundary as in subsection
\ref{Solutions with both a and b large at the outer surface}, we would find an
equation that is obtained from (\ref{condition on alpha in u}), on page
\pageref{condition on alpha in u}, by reversing the sign of the square root.
This leads to the same quadratic equation as before, with the same solutions,
(\ref{solutions for alpha in u}), as before.  But the solution $\alpha = - 3$
corresponds to the osculating line, which is not a solution of all three
Einstein equations, and the solution $\alpha = - \frac{5}{2}$ no longer solves
the original equation.

The large values of $ \frac{B}{\kappa^{2/9}} $ and $\frac{\tilde{A}}{
\kappa^{2/9}}$ are the large numbers built into the structure of the
universe, that make the universe into the stiff, strong structure that we
observe.  We note that due to the unique properties of smooth compact quotients
of $ \mathbf{H}^3 $, it might be easier to obtain large values of $ \frac{
\tilde{A}}{\kappa^{2/9}} $ than of $ \frac{B}{\kappa^{2/9}} $.
The three-volume $ V\left(\mathcal{M}^3\right) $ of a compact hyperbolic
three-manifold $ \mathcal{M}^3 $ is a topological invariant when the Ricci
scalar has a fixed value, which is usually chosen to be $ 6 $, corresponding to
sectional curvature equal to $ -1 $.  And uniquely to three dimensions, for any
given three-volume $V_1$, there is a finite, larger three-volume $V_2$, such
that there are an infinite number of topologically distinct compact hyperbolic
three-manifolds $ \mathcal{M}^3 $ with Ricci scalar equal to $ 6 $, such that
$V_1 \leq V\left(\mathcal{M}^3\right) \leq V_2$.  The existence of this
property follows from a construction of Thurston {\cite{Thurston}}, and its
uniqueness to three dimensions follows from a theorem of Wang {\cite{Wang}}, as
I shall briefly discuss in section \ref{Smooth compact quotients of CH3 H6 H3
and S3}, on page \pageref{Smooth compact quotients of CH3 H6 H3 and S3}.

There is no
observational upper limit to the topological invariant $ V\left(\mathcal{M}^3
\right) $.  For approximately homogeneous $ \mathcal{M}^3 $ the Casimir terms
in the energy-momentum tensor near the outer surface of the thick pipe may
tend to become independent of the topology of $ \mathcal{M}^3 $ for large $ V
\left(\mathcal{M}^3\right) $, but all but a finite number of the $
\mathcal{M}^3 $ with volumes in a finite range $ V_1 $ to $ V_2 $ produced by
the Thurston construction are significantly inhomogeneous.  The inhomogeneity
takes the form of a finite number of finite length ``spikes'' with smooth
rounded ends, that approximate the infinite length ``cusps'' of the finite
volume non-compact quotients of $ \mathbf{H}^3 $ to which the smooth compact
quotients of $ \mathbf{H}^3 $ produced by the Thurston construction are
related.  The value of $ \frac{\tilde{A}}{\kappa^{2/9}} $ depends only
on the average over $ \mathcal{M}^3 $ of the functions $ t^{\left( i \right)}
$, and it would seem reasonable to expect that for the majority of the smooth
compact $ \mathcal{M}^3 $ produced by the Thurston construction, these averages
will continue to depend on the topology of $ \mathcal{M}^3 $ for arbitrarily
large $ V\left( \mathcal{M}^3 \right) $, and perhaps might tend to populate
some ranges of values densely.

Comparing (\ref{B for TeV scale gravity and classical tau}) with (\ref{A tilde
for TeV scale gravity and classical tau}), we see that when $ \tau $ has the
value $ -0.7753 $, corresponding to the classical power law (\ref{bulk power
law for a in terms of b}) in the first classical region, the value of $\frac{B}
{\kappa^{2/9}}$ required for TeV-scale gravity is relatively small in
comparison to the very large value required for $\frac{\tilde{A}}{\kappa^{
\frac{
2}{9}}}$.  Moreover, from (\ref{A in terms of de Sitter radius}), (\ref{de
Sitter radius}), (\ref{B for solutions with a small at the outer surface and
classical tau}), and (\ref{A tilde in terms of A B and kappa}), we find that
for $ \tau = -0.7753 $ and smooth compact quotients of $\mathbf{C} \mathbf{H}^3
$, with a general value of $\kappa^{2/9}$:
\begin{equation}
  \label{A tilde for general kappa to the two ninths}
  \frac{\tilde{A}}{\kappa^{2/9}} \simeq 3.53 \times 10^{112} \left(
  \frac{G_N}{\kappa^{\frac{4}{9}}} \right)^{1.3670} \left| \chi \left(
  \mathcal{M}^6 \right) \right|^{0.0750}
\end{equation}
Thus for $ \tau = -0.7753 $ the required value of $\frac{\tilde{A}}{\kappa^{
\frac{2}{9}}}$ is minimized by choosing $\kappa^{2/9}$ as large as
possible, which means TeV-scale gravity, provided this is consistent with the
precision tests of Newton's law down to sub-millimetre distances {\cite{Hoyle
et al}}, which I will check in the next subsection.

\subsubsection{Comparison with sub-millimetre tests of Newton's law}
\label{Comparison with sub-millimetre tests of Newtons law}

We now need to check that TeV-scale gravity, in the type of model considered
here, is consistent with the precision tests of Newton's law, down to
sub-millimetre distances.  The shortest distance over which Newton's law has
been tested precisely is currently about $0.2$ millimetres, so to be sure of
the validity of the
four dimensional effective action description, we require that for all $y$
that make a significant contribution to the integral (\ref{integral of a
squared b to the sixth for a small at outer boundary}), $\frac{a \left( y
\right)}{a \left( y_1 \right)}$ times $0.2$ millimetres is large compared to
both $y$ and $b \left( y \right)$.  From subsection \ref{G sub N and Lambda
for solutions with outer surface in classical region}, on page \pageref{G sub N
and Lambda for solutions with outer surface in classical region}, we know that
if $ \tau > -4.9345 $, so that there is an ADD effect, then the dominant
contribution to the integral on the first branch of the square root comes from
the classical region.

I shall consider
the case where $ \tau = -0.7753 $, as in the classical power law (\ref{bulk
power law for a in terms of b}), on page \pageref{bulk power law for a in terms
of b}, in the first part of the classical region, and $ \tilde{\tau} = -0.3101
$, corresponding to the classical power law (\ref{bulk power law for a in terms
of b with lower sign}), on page \pageref{bulk power law for a in terms of b
with lower sign}, in the second part of the classical region.  Then the
interpolating function (\ref{interpolating function for a in terms of b}), on
page \pageref{interpolating function for a in terms of b}, will be
approximately valid throughout the whole range from the inner surface to the
outer surface of the thick pipe.  I shall make the approximation of treating
the interpolating function (\ref{interpolating function for b}), on page
\pageref{interpolating function for b}, as if it was also valid throughout the
whole range from the inner surface to the outer surface.

The condition to be sure of the validity of the four dimensional effective
action description will be strictest as $y$
approaches the outer surface of the thick pipe, at $y \simeq 1.0996 B$, since
$a \left( y \right)$ decreases monotonically with increasing $y$, and $b
\left( y \right)$ increases monotonically with increasing $y$.  Moreover, we
see, from (\ref{interpolating function for b}), that $b \left( y \right)$ is
comparable to $y$ in the mid-region of the thick pipe, but becomes large
compared to $y$, as either boundary of the thick pipe is approached.  Thus it
is sufficient to check the requirement for $b \left( y \right)$, in the region
where $b$ is approaching the outer surface of the thick pipe.  We then have,
from (\ref{interpolating function for b}), (\ref{interpolating function for a
in terms of b}), and (\ref{A in terms of de Sitter radius}), that:
\begin{equation}
  \label{b in second bulk power law region} b \simeq 1.5123 B \left(
  \frac{B}{1.0996 B - y} \right)^{0.1449}
\end{equation}
\begin{equation}
  \label{a in second bulk power law region in terms of de Sitter radius} a
  \simeq \frac{0.2459}{\left| \chi \left( \mathcal{M}^6 \right)
  \right|^{0.1292}}  \left( \frac{B}{\kappa^{2/9}} \right)^{2.4495}
  \left( \frac{\kappa^{2/9}}{b} \right)^{3.2247} \times
  \hspace{0.4ex} \textrm{de Sitter radius}
\end{equation}
Thus, by (\ref{a in second bulk power law region in terms of de Sitter
radius}), the requirement is that for all $y$ that make a significant
contribution to the integral (\ref{integral of a squared b to the sixth for a
small at outer boundary}):
\begin{equation}
  \label{condition for agreement with sub millimetre tests of Newtons law}
  \frac{0.2459}{\left| \chi \left( \mathcal{M}^6 \right) \right|^{0.1292}}
  \left( \frac{B}{\kappa^{2/9}} \right)^{2.4495} \times \frac{0.2
  \textrm{ millimetre}}{\kappa^{2/9}} \gg \left(
  \frac{b}{\kappa^{2/9}} \right)^{4.2247}
\end{equation}
And for TeV-scale gravity, this becomes, by (\ref{B for TeV scale gravity}):
\begin{equation}
  \label{condition for agreement with sub millimetre tests for TeV scale
  gravity} 1.89 \times 10^3  \left| \chi \left( \mathcal{M}^6 \right)
  \right|^{0.1124} \gg \left( \frac{b}{B} \right)^{4.2247}
\end{equation}
And by (\ref{b in second bulk power law region}), this becomes:
\begin{equation}
  \label{upper limit on y for four dimensional effective action} 1.0996 -
  \frac{y}{B} \gg \frac{7.72 \times 10^{- 5}}{\left| \chi \left( \mathcal{M}^6
  \right) \right|^{0.1836}}
\end{equation}
Now $\left| \chi \left( \mathcal{M}^6 \right) \right| \geq 1$, hence
(\ref{upper limit on y for four dimensional effective action}) will be
satisfied, provided:
\begin{equation}
  \label{sufficient condition on y for four dimensional effective action}
  1.0996 - \frac{y}{B} \gg 7.72 \times 10^{- 5}
\end{equation}
Now the contribution to the integral (\ref{integral of f of Y}), from the
region where $Y$ is within $7.72 \times 10^{- 5}$ of the upper limit, is $1.35
\times 10^{- 6}$, which is a fraction $4.54 \times 10^{- 5}$ of the full
integral (\ref{integral of f of Y}).  Thus for tests of Newton's law at
distances around $0.2$ millimetres, we anticipate deviations from Newton's
law, in the shape of a small change in the effective value of Newton's
constant, at the level of about $50$ parts in a million, or $5 \times 10^{- 3}$
percent.

To compare this result with the measurements of Hoyle et al {\cite{Hoyle et
al}}, we
note that one of the ways they expressed their results, was by giving 95\%
confidence level limits on the magnitude of the parameter $\alpha$, as a
function of $\lambda$, in a modified Newtonian potential of the form:
\begin{equation}
  \label{modified Newtonian potential tested by Hoyle et al} V \left( r
  \right) = - G \frac{m_1 m_2}{r} \left[ 1 + \alpha e^{- r / \lambda} \right]
\end{equation}
The 95\%  confidence level limits on $\left| \alpha \right|$, as a function
of $\lambda$, are given in their Table XIII, from which we see that for
$\lambda = 0.10$ millimetre, $\left| \alpha \right| \leq 1.8 \times 10^1$.
For $\lambda = 0.25$ millimetre, $\left| \alpha \right| \leq 4.3 \times 10^{-
1}$.  And for $\lambda$ in the range 1.00 millimetres to 10.0 millimetres, the
upper bound on $\left| \alpha \right|$ is around $10^{- 2}$.

The form of equation (\ref{modified Newtonian potential tested by Hoyle et
al}) is such that for $r$ large compared to $\lambda$, the correction to
Newton's law is negligible, but for $r$ comparable with $\lambda$, or smaller
than $\lambda$, there is effectively a modification of Newton's constant, by a
factor $\sim \left( 1 + \alpha \right)$.  Thus for the form of TeV-scale
gravity considered in the present paper, the expected deviations from Newton's
law, at distances around a millimetre, are around $5 \times 10^{- 3}$ times
smaller than the current best experimental limits of Hoyle et al.

It is interesting to note that the upper bound, (\ref{upper limit on y for
four dimensional effective action}), on $y$, for the four dimensional
reduction to be valid, for submillimetre tests of Newton's law, from the inner
surface of the thick pipe, up to $y$, corresponds, by (\ref{B for TeV scale
gravity}), to:
\begin{equation}
  \label{upper limit on y in metres for the four dimensional reduction to be
  valid} 1.0996 B - y \gg \frac{5.40 \times 10^{- 17} \textrm{ metres}}{\left|
  \chi \left( \mathcal{M}^6 \right) \right|^{0.3197}}
\end{equation}
or in other words, since $\kappa^{2/9} = 8.899 \times 10^{- 19}$
metres, for TeV-scale gravity, to:
\begin{equation}
  \label{upper limit on y in terms of kappa for the four dimensional reduction
  to be valid} 1.0996 B - y \gg \frac{60.7}{\left| \chi \left( \mathcal{M}^6
  \right) \right|^{0.3197}} \kappa^{2/9}
\end{equation}
On the other hand, from (\ref{a sub q}), on page \pageref{a sub q}, the value
of $ \frac{a_q}{\kappa^{2/9}} $ that corresponds to the value (\ref{A
tilde for TeV scale gravity and classical tau}) of $ \tilde{A} $ is $ 3.67
\times 10^{35} \left\vert\chi\left( \mathcal{M}^6 \right)\right\vert^{0.0399}
$, so the thickness in $ y $ of the quantum region near the outer surface is
around $ \kappa^{2/9} \ln\left(\frac{a_q}{\kappa^{2/9}}\right)
\simeq 81.9 \kappa^{2/9} $.  Thus if the precision of the
submillimetre tests of Newton's law could be increased by another three decimal
places, they would be probing the quantum region near the outer surface of the
thick pipe, for the solutions considered in subsection \ref{Solutions with a as
small as kappa to the two ninths at the outer surface}, when $\tau = -0.7753$.

From (\ref{b in second bulk power law region}) and (\ref{B for TeV scale
gravity and classical tau}), the value of $b$, at the value of $y$ where the
inequalities in
(\ref{upper limit on y in metres for the four dimensional reduction to be
valid}) and (\ref{upper limit on y in terms of kappa for the four dimensional
reduction to be valid}) become equality, is:
\begin{equation}
  \label{b at the upper limit of y for the four dimensional reduction} b
  \simeq 5.96 \left| \chi \left( \mathcal{M}^6 \right) \right|^{0.0266} B
  \simeq \frac{4.17 \times 10^{- 12} \textrm{ metres}}{\left| \chi
  \left( \mathcal{M}^6 \right) \right|^{0.1095}} \simeq \frac{4.69 \times
  10^6}{\left| \chi \left( \mathcal{M}^6 \right) \right|^{0.1095}}
  \kappa^{2/9}
\end{equation}
And from (\ref{a in second bulk power law region in terms of de Sitter
radius}) and (\ref{de Sitter radius}), the corresponding value of $a$ is:
\begin{equation}
   \label{a at the upper limit of y for the four dimensional reduction}
   a \simeq \frac{2.09 \times 10^{- 8}}{\left| \chi \left( \mathcal{M}^6
   \right) \right|^{0.1095}} \times \textrm{de Sitter radius} =
   \frac{3.16 \times 10^{18} \textrm{ metres}}{\left| \chi \left( \mathcal{M}^6
   \right) \right|^{0.1095}} = \frac{3.55 \times 10^{36}}{\left| \chi \left(
   \mathcal{M}^6 \right) \right|^{0.1095}} \kappa^{2/9}
\end{equation}

\subsubsection{Comparison with precision solar system tests of General
Relativity}
\label{Comparison with precision solar system tests of General Relativity}

There are also very precise tests of General Relativity, via lunar laser
ranging measurements of the lunar orbit, using reflectors left on the surface
of the moon by Apollo astronauts, and by unmanned Soviet lunar missions
{\cite{Williams Turyshev Boggs 1, Williams Turyshev Boggs 2}}.  In particular,
a test of the equivalence principle, obtained from a fit of lunar laser
ranging data, gives a value for the difference in the ratio of gravitational
mass to inertial mass, $M_G / M_I$, between the Earth ($e$) and the Moon
($m$).  The value quoted in {\cite{Williams Turyshev Boggs 1}}, which has been
corrected for solar radiation pressure, is:
\begin{equation}
  \label{lunar laser ranging test of the equivalence principle} \left[ \left(
  \frac{M_G}{M_I} \right)_e - \left( \frac{M_G}{M_I} \right)_m \right] =
  \left( - 1.0 \pm 1.4 \right) \times 10^{- 13}
\end{equation}
To check the consistency with this measurement, of the models studied here,
we need to decide, if this measurement is interpreted as giving a bound on the
variation of Newton's constant with distance, what the shortest relevant
distance is.  The lunar orbit is determined by the gravitational interaction
between the Moon and the Earth, while both move in the gravitational field of
the Sun ($s$).

From equation (2) of {\cite{Williams Turyshev Boggs 1}}, the effective
acceleration of the Moon with respect to the Earth, $\vec{a} = \vec{a}_m -
\vec{a}_e$, for the three-body Earth-Moon-Sun system, is:
\begin{equation}
  \label{acceleration of the Moon with respect to the Earth} \vec{a} = - G_N
  \left( M_e \left( \frac{M_G}{M_I} \right)_m + M_m \left( \frac{M_G}{M_I}
  \right)_e \right) \frac{\vec{r}_{em}}{r_{em}^3} - G_N M_s \left(
  \frac{M_G}{M_I} \right)_e \frac{\vec{r}_{es}}{r_{es}^3} + G_N M_s \left(
  \frac{M_G}{M_I} \right)_m \frac{\vec{r}_{ms}}{r_{ms}^3}
\end{equation}
The last two terms in
(\ref{acceleration of the Moon with respect to the Earth}) represent the solar
effect on the motion of the Moon with respect to the Earth.  A violation of
the equivalence principle would produce a lunar orbit perturbation
proportional to the difference in the two $M_G / M_I$ ratios.

From the form of equation (\ref{acceleration of the Moon with respect to the
Earth}), it appears that a small percentage difference in $G_N$, between the
first term, and the last two terms, corresponding to a small percentage
difference in $G_N$, for the Earth-Moon distance, and the Earth-Sun distance,
might result in an orbital perturbation different in form, but of the same
order of magnitude, as the perturbation resulting from a similar percentage
difference in the two $M_G / M_I$ ratios.  Thus I shall provisionally
interpret the measurement (\ref{lunar laser ranging test of the equivalence
principle}), as also giving an order of magnitude bound on the percentage
difference of Newton's constant for the Earth-Moon distance, and for the
Earth-Sun distance.  Thus we have to repeat the calculation performed above,
for tests of Newton's law over distances of around $0.2$ millimetres, for
distances around the Earth-Moon distance, which is around $4 \times 10^8$
metres.  Instead of (\ref{sufficient condition on y for four dimensional
effective action}), we now find that a sufficient condition on $y$, for the
four-dimensional reduction to be valid from the inner surface of the thick
pipe up to $y$, is that:
\begin{equation}
  \label{upper limit on y for the Earth-Moon distance} 1.0996 - \frac{y}{B}
  \gg 6.20 \times 10^{- 25}
\end{equation}
It follows immediately from the flat topped shape of the function $f \left( Y
\right)$, Figure \ref{graph of f of Y}, together with the fact that the
peak of the
function is outside the range excluded by (\ref{upper limit on y for the
Earth-Moon distance}), that the contribution to the integral (\ref{integral of
f of Y}), from the region excluded by (\ref{upper limit on y for the
Earth-Moon distance}), is not more than a fraction $\sim 10^{- 23}$ of the
value of the integral, if $\gg$ is interpreted as meaning larger by a factor
of at least $10$.  Thus for the form of TeV-scale gravity considered in the
present paper, the fractional difference of $G_N$ for the Earth-Moon distance,
from $G_N$ for the Earth-Sun distance, will not be more than around $10^{-
23}$, at the most, which is smaller than the bound given by (\ref{lunar laser
ranging test of the equivalence principle}), interpreted as discussed above,
by a factor of around $10^{- 10}$.

Thus, notwithstanding the remarkable precision of the lunar laser ranging
measurements, the submillimetre tests of Newton's law are currently closer to
testing the form of TeV-scale gravity considered in subsection \ref{Solutions
with a as small as kappa to the two ninths at the outer surface}, with $ \tau =
-0.7753 $.

\subsubsection{Further consequences of the warp factor decreasing to a small
value, at the outer surface of the thick pipe}
\label{Further consequences of the warp factor decreasing to a small value}

The fact that the warp factor, $a^2 \left( y \right)$, decreases to a small
value as $y$ approaches $y_2$, in the solutions considered in subsection
\ref{Solutions with a as small as kappa to the two ninths at the outer
surface}, implies that there are short spacelike paths
through the bulk between points that are separated by large distances in the
observed universe.  However, for the solutions considered in the present
paper, it is not possible, even in principle, to send signals through the
bulk to distant parts of the observed universe, at what would appear to be
superluminal speeds, from the point of view of observers on the inner surface
of the thick pipe, because the time dimension scales with exactly the same
scale factor, $a \left( y \right)$, as the three observed spatial dimensions.

It would be interesting to find out whether or not this conclusion could be
modified in cosmological-type solutions, which would require the analysis of
some coupled partial differential equations, with the time, and $y$, as
independent variables.  In particular, it would be interesting to find out
whether or not an effect of this type could provide an alternative to
inflation, for solving the horizon problem of the early universe
{\cite{Guth}}.  Alternative solutions to the horizon problem, of this type,
have been discussed in {\cite{Kaelbermann Halevi, Chung Freese 1,
Kaelbermann, Chung Freese 2, Ishihara, Caldwell Langlois, Stoica, Abdalla
Casali Cuadros-Melgar, Pas Pakvasa Weiler}}.  It would also be interesting to
find out whether or not an effect
of this type would be consistent with the type of causality constraints
recently discussed by Arkani-Hamed et al {\cite{Arkani Hamed et al causality
constraints}}.  However these questions will not be addressed in the present
paper.

\subsection{Stiffening by fluxes wrapping three-cycles of the compact
six-manifold times the radial dimension}
\label{Stiffening by fluxes}

The occurrence of non-vanishing fluxes of form fields in the de Sitter
backgrounds for type IIB superstrings constructed by Kachru, Kallosh, Linde,
and Trivedi \cite{KKLT} suggests that it might also be interesting to consider
solutions with extra fluxes of the four-form field strength of the three-form
gauge field in the present context, so
I shall now consider the possible effects of fluxes wrapping three cycles of
the compact six-manifold, times the Ho\v{r}ava-Witten one-cycle along the eleventh
dimension, in the upstairs picture.  I shall assume, to start with, that there
will not be enough non-vanishing components of the three-form gauge field, for
the non-linear term in the classical field equation for the three-form gauge
field to be non-vanishing, so that we can treat the classical field equation
for the three-form gauge field as a linear equation, and add solutions.  We
then seek a classical solution, such that only the components $G_{ABCy}$ are
non-zero, and $G_{ABCy} \left( z, y \right)$, where $z$ denotes the
coordinates on the compact six-manifold, has the factorized form
\begin{equation}
  \label{factorized ansatz for four form} G_{ABCy} \left( z, y \right) =
  G_{ABC} \left( z \right) f \left( y \right)
\end{equation}
Now the Bianchi identity reads:
\begin{equation}
  \label{Bianchi identity for four form} \partial_I G_{JKLM} + \partial_J
  G_{KLMI} + \partial_K G_{LMIJ} + \partial_L G_{MIJK} + \partial_M G_{IJKL} =
  0
\end{equation}
With the ansatz (\ref{factorized ansatz for four form}), one component of this
reads:
\begin{equation}
  \label{Bianchi identity for three form factor} \left( \partial_A G_{BCD}
  \left( z \right) - \partial_B G_{CDA} \left( z \right) + \partial_C G_{DAB}
  \left( z \right) - \partial_D G_{ABC} \left( z \right) \right) f \left( y
  \right) = 0
\end{equation}
which, since $f \left( y \right) \neq 0$ by assumption, is the Bianchi
identity for the three-form factor $G_{ABC} \left( z \right)$.

Now when the gravitino field vanishes, the classical field equation for the
three-form gauge field $C_{IJK}$, from the action (\ref{upstairs bulk
action}), is:
\begin{equation}
  \label{three form field equation} \partial_I \left( \sqrt{- G} G^{IM} G^{JN}
  G^{KO} G^{LP} G_{MNOP} \right) - \frac{\sqrt{2}}{1152} \sqrt{- G} G^{JKLI_4
  \ldots I_7 I_8 \ldots I_{11}} G_{I_4 \ldots I_7} G_{I_8 \ldots I_{11}} = 0
\end{equation}
where the metric in eleven dimensions is denoted $G_{IJ}$, as in (\ref{metric
ansatz}), so that $G^{I_1 I_2 \ldots I_{11}}$ denotes the tensor
$\frac{1}{\sqrt{- G}} \epsilon^{I_1 I_2 \ldots I_{11}}$.  Let us now assume
that $G_{IJKL}$ is zero, if any component is along the four observed
dimensions.  Then there are at most seven possible values for each index, such
that $G_{IJKL}$ is non-zero, so the term in (\ref{three form field equation})
bilinear in $G_{IJKL}$ vanishes, and the field equation reduces to:
\begin{equation}
  \label{reduced three form field equation} \partial_I \left( \sqrt{- G}
  G^{IM} G^{JN} G^{KO} G^{LP} G_{MNOP} \right) = 0
\end{equation}
Now, bearing in mind the metric ansatz (\ref{metric ansatz}), one set of
components of this equation, for the factorized ansatz (\ref{factorized ansatz
for four form}), reads:
\begin{equation}
  \label{field equation for three form factor} \sqrt{- g} \partial_A \left(
  \sqrt{h} h^{AD} h^{BE} h^{CF} G_{DEF} \left( z \right) \right) a \left( y
  \right)^4 f \left( y \right) = 0
\end{equation}
Now, since $\sqrt{- g}$, $a \left( y \right)$, and $f \left( y \right)$ are
assumed to be non-vanishing, this equation, together with (\ref{Bianchi
identity for three form factor}), implies that $G_{ABC} \left( z \right)$ is a
Hodge - de Rham harmonic three-form on the compact six-manifold.  So by
standard Hodge - de Rham theory, there are $B_3$ linearly independent
solutions $G_{ABC} \left( z \right)$ of (\ref{Bianchi identity for three form
factor}) and (\ref{field equation for three form factor}), where $B_3$ is the
third Betti number of the compact six-manifold.  I shall now assume that
$G_{ABC} \left( z \right)$ is a Hodge - de Rham harmonic three-form on the
compact six-manifold.

The remaining set of components of (\ref{reduced three form field equation}),
that are not satisfied identically for the factorized ansatz (\ref{factorized
ansatz for four form}), are:
\begin{equation}
  \label{equation for f in the factorized ansatz} - \sqrt{- g} \partial_y
  \left( a \left( y \right)^4 f \left( y \right) \right) \sqrt{h} h^{AD}
  h^{BE} h^{CF} G_{DEF} \left( z \right) = 0
\end{equation}
Thus $f \left( y \right)$ is equal to a fixed number, times $a \left( y
\right)^{- 4}$, so, absorbing the fixed number into $G_{ABC} \left( z
\right)$, we find that:
\begin{equation}
  \label{solution of factorized ansatz for four form} G_{ABCy} \left( z, y
  \right) = G_{ABC} \left( z \right) a \left( y \right)^{- 4}
\end{equation}
where $G_{ABC} \left( z \right)$ is a Hodge - de Rham harmonic three-form on
the compact six-manifold.  We note that (\ref{solution of factorized ansatz
for four form}) applies for all $y$, in the upstairs picture, since under
reflection in the orbifold hyperplane at $y = y_1$, we have $G_{UVWy} \left(
x, 2 y_1 - y \right) = G_{UVWy} \left( x, y \right)$, and also $a \left( 2 y_1
- y \right) = a \left( y \right)$, where $x$ here denotes the coordinates on
$\mathcal{M}^{10}$.

Now, following an argument given by Witten, in section 2 of {\cite{9609122
Witten}}, we consider a four-cycle $X$ in $\mathcal{M}^{10}$, on the $y_{1 +}$
side of the orbifold hyperplane at $y = y_1$, and apply the relation (\ref{G
boundary condition}), with the substitution (\ref{FF to FF RR substitutions}).
If the Pontryagin number of $X$ is zero, then the $RR$ term in (\ref{FF to FF
RR substitutions}) will not contribute to the integral of the right-hand side
of (\ref{FF to FF RR substitutions}) over $X$, so we find that
\begin{equation}
  \label{Integral of G over a four cycle} \frac{\sqrt{2}}{4 \pi}  \left(
  \frac{4 \pi}{\kappa} \right)^{\frac{2}{3}} \int_X \left. G \right|_{y = y_{1
  +}} = \frac{\sqrt{2}}{4 \pi}  \left( \frac{4 \pi}{\kappa}
  \right)^{\frac{2}{3}}  \frac{1}{4!}  \int_X dx^U dx^V dx^W dx^X  \left.
  G_{UVWX} \right|_{y = y_{1 +}}
\end{equation}
is equal to $\frac{1}{16 \pi^2} \int_X \mathrm{tr} F^{( 1 )} \wedge F^{( 1 )}$,
which Witten indicates is a four-dimensional characteristic class of the $E_8$
bundle at $y_1$, and is equal to an integer.  However, (\ref{Integral of G
over a four cycle}) is a topological invariant for smoothly varying $G$, and
thus has the same value no matter what value of $y$ it is evaluated at, and,
indeed, has the same value for any four-cycle in $\mathcal{M}^{11}
=\mathcal{M}^{10} \times \mathbf{S}^1 /\mathbf{Z}_2$ that is topologically
equivalent to $X$.

Following Witten's argument, if we now consider
Ho\v{r}ava-Witten theory with a large value of $\left( y_2 - y_1 \right)$,
specifically, much larger than the diameter of $X$, and the integral
(\ref{Integral of G over a four cycle}) at some value of $y$ a long distance
away from both $y_1$ and $y_2$, then it would seem unlikely that the value of
the integral would depend on whether or not there exist orbifold hyperplanes a
very large distance away, at $y_1$ and $y_2$.  Thus we expect that
(\ref{Integral of G over a four cycle}) should be equal to an integer for an
arbitrary four-cycle $X$ with zero Pontryagin number, for smoothly varying
$G$, in supergravity in eleven dimensions.  In other words, (\ref{Integral of
G over a four cycle}) gives a form of Dirac quantization condition on the
integral of the Cremmer-Julia-Scherk four-form field strength $G$, over a
four-cycle with zero Pontryagin number.  Witten gives further arguments
supporting this interpretation, and also, a generalization of the quantization
condition, to four-cycles with non-zero Pontryagin number.

Witten's arguments
do not cover the case of a four-cycle, in the upstairs formulation of
Ho\v{r}ava-Witten theory, that has the form of a three-cycle in
$\mathcal{M}^{10}$, times a one-cycle that wraps the $\mathbf{S}^1$ in the
$y$ direction.  However, since the Ho\v{r}ava-Witten boundary conditions, at the
orbifold fixed-point hyperplanes, imply that $G_{UVWy}$ is continuous across
the orbifold fixed-point hyperplanes, and such a four-cycle automatically has
zero Pontryagin number, I shall assume that (\ref{Integral of G over a four
cycle}) also has an integer value, for such a four-cycle, and that this
applies, in particular, for the factorized ansatz (\ref{factorized ansatz for
four form}).  Thus, from (\ref{solution of factorized ansatz for four form}),
we find that, for any three-cycle, $Z$, of the compact six-manifold:
\begin{equation}
  \label{Integral of G for the factorized ansatz} 8 \frac{\sqrt{2}}{4 \pi}
  \left( \frac{4 \pi}{\kappa} \right)^{\frac{2}{3}}  \frac{1}{4!}  \int_Z dz^A
  dz^B dz^C G_{ABC} \left( z \right) \int^{y_2}_{y_1} dya \left( y \right)^{-
  4}
\end{equation}
must be equal to an integer.

We now have to calculate the modified value of the contribution (\ref{three
form energy momentum tensor}), on page \pageref{three form energy momentum
tensor}, of the three-form gauge field to the
energy-momentum tensor, (\ref{energy momentum tensor}), in the presence of
fluxes wrapping three-cycles of the compact six-manifold, with the ansatz
(\ref{factorized ansatz for four form}).  Since the three-form field
configurations considered in subsection \ref{The three form gauge field} are
only significant, in the energy-momentum tensor, near the inner surface of the
thick pipe, while, from (\ref{solution of factorized ansatz for four form}),
the three-form field configurations considered in the present subsection are
suppressed by the very small factor $a \left( y \right)^{- 4}$, near the inner
surface of the thick pipe, I shall provisionally assume that cross terms in
the energy-momentum tensor, between the three-form field configurations
considered in subsection \ref{The three form gauge field}, and those
considered in the present subsection, can be neglected, for compactifications
on smooth compact quotients of $\mathbf{C} \mathbf{H}^3$, while for
compactifications on smooth compact quotients of $\mathbf{H}^6$, the
three-form field configurations of the type considered in subsection \ref{The
three form gauge field} are absent, since Witten's topological constraint is
satisfied with zero $G$, as noted at the end of subsection \ref{Wittens
topological constraint}.

From the metric ansatz (\ref{metric ansatz}), we find:
\begin{equation}
  \label{G sub A G sub B for fluxes wrapping three cycles of compact sixfold}
  G^{KN} G^{LO} G^{MP} G_{AKLM} G_{BNOP} = \frac{3}{b^4 a^8} h^{CE} h^{DF}
  G_{ACD} \left( z \right) G_{BEF} \left( z \right)
\end{equation}
I shall now assume, as in the study of the Casimir contributions to the
energy-momentum tensor in subsection \ref{The Casimir energy density
corrections}, that the Einstein equations are expanded in harmonics on the
compact six-manifold, following the procedure of Lukas, Ovrut, and Waldram
{\cite{Lukas Ovrut Waldram}}, and I shall consider the Einstein equations in
the approximation of dropping all but the lowest harmonic.  I shall also
assume that $G_{ABC} \left( z \right)$, which is a sum of constant multiples
of $B_3$ linearly independent Hodge - de Rham harmonic three-forms, where
$B_3$ is the third Betti number of the compact six-manifold, has been chosen
such that
\[ \int_{\mathcal{M}^6} d^6 z \sqrt{h} h^{CE} h^{DF} G_{ACD} \left( z \right)
   G_{BEF} \left( z \right) = \hspace{4em} \hspace{4em} \hspace{4em} \]
\begin{equation}
  \label{first condition on G sub A B C} \hspace{6em} = \int_{\mathcal{M}^6}
  d^6 z \sqrt{h}  \frac{1}{6} h_{AB} h^{CF} h^{DG} h^{EH} G_{CDE} \left( z
  \right) G_{FGH} \left( z \right)
\end{equation}
and
\begin{equation}
  \label{second condition on G sub A B C} \int_{\mathcal{M}^6} d^6 z \sqrt{h}
  h_{AB} h^{CF} h^{DG} h^{EH} G_{CDE} \left( z \right) G_{FGH} \left( z
  \right) = G^2 \int_{\mathcal{M}^6} d^6 z \sqrt{h} h_{AB}
\end{equation}
for a suitable real constant $G > 0$.  These conditions (\ref{first condition
on G sub A B C}) and (\ref{second condition on G sub A B C}) constitute at
most $20 + 20$ linearly independent constraints on the $B_3$ independent
coefficients in $G_{ABC} \left( z \right)$, and thus can presumably always be
satisfied, for sufficiently large $B_3$, unless this somehow conflicts with
the requirement that (\ref{Integral of G for the factorized ansatz}) be an
integer for all three-cycles $Z$, which I shall assume does not occur.  Then
in the approximation of dropping all but the lowest harmonic, (\ref{G sub A G
sub B for fluxes wrapping three cycles of compact sixfold}) becomes:
\begin{equation}
  \label{lowest harmonic of G sub A G sub B} G^{KN} G^{LO} G^{MP} G_{AKLM}
  G_{BNOP} = \frac{1}{2 b^6 a^8} G_{AB} G^2
\end{equation}
Similarly, we find:
\begin{equation}
  \label{G 4 sub y G 4 sub y} G^{KN} G^{LO} G^{MP} G_{yKLM} G_{yNOP} =
  \frac{1}{b^6 a^8} h^{AD} h^{BE} h^{CF} G_{ABC} \left( z \right) G_{DEF}
  \left( z \right)
\end{equation}
And in the approximation of dropping all but the lowest harmonic, this
becomes:
\begin{equation}
  \label{lowest harmonic of G 4 sub y G 4 sub y} G^{KN} G^{LO} G^{MP} G_{yKLM}
  G_{yNOP} = \frac{1}{b^6 a^8}  \tilde{G}^2
\end{equation}
where the real constant $\tilde{G} > 0$ is defined by
\begin{equation}
  \label{definition of the constant G tilde} \int_{\mathcal{M}^6} d^6 z
  \sqrt{h} h^{AD} h^{BE} h^{CF} G_{ABC} \left( z \right) G_{DEF} \left( z
  \right) = \tilde{G}^2 \int_{\mathcal{M}^6} d^6 z \sqrt{h}
\end{equation}
We also find:
\begin{equation}
  \label{scalar G G with flux on three cycles of the compact sixfold} G^{QR}
  G^{KN} G^{LO} G^{MP} G_{QKLM} G_{RNOP} = \frac{4}{b^6 a^8} h^{AD} h^{BE}
  h^{CF} G_{ABC} \left( z \right) G_{DEF} \left( z \right)
\end{equation}
Thus, in the approximation of dropping all but the leading harmonic, we have:
\begin{equation}
  \label{G sub A B times scalar G G with three cycles on compact sixfold}
  G_{AB} G^{QR} G^{KN} G^{LO} G^{MP} G_{QKLM} G_{RNOP} = \frac{4}{b^6 a^8} G^2
  G_{AB}
\end{equation}
\begin{equation}
  \label{G sub mu nu times scalar G G with three cycles on compact sixfold}
  G_{\mu \nu} G^{QR} G^{KN} G^{LO} G^{MP} G_{QKLM} G_{RNOP} = \frac{4}{b^6
  a^8}  \tilde{G}^2 G_{\mu \nu}
\end{equation}
\begin{equation}
  \label{G sub y y times scalar G G with three cycles on compact sixfold}
  G_{yy} G^{QR} G^{KN} G^{LO} G^{MP} G_{QKLM} G_{RNOP} = \frac{4}{b^6 a^8}
  \tilde{G}^2
\end{equation}
Hence, from (\ref{three form energy momentum tensor}), we find the following
additional contributions to the energy- \\
momentum tensor of the three-form gauge
field, to be added to (\ref{three form energy momentum tensor components}), in
the approximation of neglecting cross terms between the three-form gauge field
configurations considered in subsection \ref{The three form gauge field}, and
those considered in the present subsection:
\begin{equation}
  \label{three form energy momentum tensor components for three cycles on the
  sixfold} T_{\mu \nu}^{\left( 3 f \right)} = - \frac{\tilde{G}^2}{12 \kappa^2
  b^6 a^8} G_{\mu \nu}, \hspace{4em} T_{AB}^{\left( 3 f \right)} = 0,
  \hspace{4em} T_{yy}^{\left( 3 f \right)} = \frac{\tilde{G}^2}{12 \kappa^2
  b^6 a^8}
\end{equation}
These components satisfy the conservation equation (\ref{conservation equation
for the t i}), with $t^{\left( 1 \right)}$, $t^{\left( 2
\right)}$, and $t^{\left( 3 \right)}$, defined by (\ref{T IJ block diagonal
structure}), now interpreted as unrestricted functions of $y$.

\subsubsection{The region near the outer surface}
\label{The region near the outer surface}

Now comparing with (\ref{T IJ block diagonal structure}) on page \pageref{T IJ
block diagonal structure}, and with (\ref{a dot over a}) and (\ref{second
Einstein equation without a dot}), on page \pageref{a dot over a}, with the
upper choice of sign for the square root, we see that the new term in the
square root, resulting from (\ref{three form energy momentum tensor components
for three cycles on the sixfold}), has the correct sign, namely the same sign
as the $\frac{4}{a^2}$ term, to make possible a solution of the boundary
conditions, at the outer surface of the thick pipe, with both $a$ and $b$
large compared to $\kappa^{2/9}$ at the outer surface, by the same
mechanism as in subsection \ref{Solutions with both a and b large at the outer
surface}, on page \pageref{Solutions with both a and b large at the outer
surface}.  Furthermore, from (\ref{bulk power law for a in terms of b}), on
page \pageref{bulk power law for a in terms of b}, we see that $\frac{1}{b^6
a^8}$ behaves as $b^{0.2024}$ in the first bulk type power law region,
governed by (\ref{upper sign at large c}), on page \pageref{upper sign at
large c}, and (\ref{bulk power law trajectory for a with upper sign}), on page
\pageref{bulk power law trajectory for a with upper sign}.  Thus the
correction terms from (\ref{three form energy momentum tensor components for
three cycles on the sixfold}) do indeed grow in importance with increasing
$b$, or equivalently, with increasing $y$, and thus the most important of the
correction terms, which is the one in the square root, has the correct
qualitative behaviour, for a suitable value of $\tilde{G}$, to make possible a
solution of the boundary conditions at the outer surface in the classical
region, along the lines of subsection \ref{Solutions with both a and b large
at the outer surface}.  Furthermore, for a sufficiently large value of
$\tilde{G}$, it might be possible to arrange for $a \gg b$ at the outer
surface, and thus avoid the problem that prevented the solutions of subsection
\ref{Solutions with both a and b large at the outer surface} from being in
agreement with observation.

To study this possibility in detail, I shall now retrace the steps in
subsection \ref{Solutions with both a and b large at the outer surface}, but
assuming, now, that $\tilde{G}$ is sufficiently large, and the integration
constant $A$, in (\ref{bulk power law for a in terms of b}), is also
sufficiently large, that as $y$ increases, in the first bulk power law region,
(\ref{upper sign at large c}), (\ref{bulk type power law for upper sign}), and
(\ref{bulk type power law dependence of b on y for upper sign}), with the
integration constant, $B$, large compared to $\kappa^{2/9}$, the
correction terms in (\ref{a dot over a}) and (\ref{second Einstein equation
without a dot}), due to (\ref{three form energy momentum tensor components for
three cycles on the sixfold}), first become significant long before the
$\frac{4}{a^2}$ term, in the square root, becomes significant.  This
assumption will be satisfied, if we find a solution in the classical region,
with $a \gg b$ at the outer surface.

With these assumptions, we find from (\ref{T IJ block diagonal structure}),
(\ref{a dot over a}), (\ref{second Einstein equation without a dot}), and
(\ref{three form energy momentum tensor components for three cycles on the
sixfold}), that the relevant equations, away from the inner surface of the
thick pipe, but still on the first branch of the square root, where we take
the upper sign in (\ref{a dot over a}) and (\ref{second Einstein equation
without a dot}), are:
\begin{equation}
  \label{bulk coupled a dot over a with fluxes} \frac{da}{db} = - 2
  \frac{a}{b} + \frac{a}{2 bc} \sqrt{6 c^2 - 8 + \frac{\tilde{G}^2}{18 b^4
  a^8}}
\end{equation}
\begin{equation}
  \label{bulk coupled second Einstein equation with fluxes} \frac{dc}{db} =
  \frac{3 c^2 - 4}{bc} - \frac{2}{b} \sqrt{6 c^2 - 8 + \frac{\tilde{G}^2}{18
  b^4 a^8}} - \frac{\tilde{G}^2}{36 b^5 a^8 c}
\end{equation}
Here $c = \dot{b} = \frac{db}{dy}$, as defined just before (\ref{second
Einstein equation for c in terms of b}), on page \pageref{second Einstein
equation for c in terms of b}.  These two equations replace the equations
(\ref{bulk coupled a dot over a}) and (\ref{bulk coupled second Einstein
equation}) of subsection \ref{Solutions with both a and b large at the outer
surface}.  Moreover, we are again seeking solutions such that both $a$ and $b$
are large compared to $\kappa^{2/9}$, at the outer surface of the
thick pipe, so, as explained at the start of subsection \ref{Solutions with
both a and b large at the outer surface}, on page \pageref{Solutions with both
a and b large at the outer surface}, the boundary conditions are now that both
$\frac{\dot{a}}{a}$, and $\frac{\dot{b}}{b}$, are zero, at the outer surface
of the thick pipe.  Thus $c = 0$ at the outer surface, and from (\ref{bulk
coupled a dot over a with fluxes}), rewritten in its original form, like
(\ref{a dot over a}), on page \pageref{a dot over a}, we see that
$\frac{\tilde{G}^2}{18 b^4 a^8} = 8$, at the outer surface.  The qualitative
difference from subsection \ref{Solutions with both a and b large at the outer
surface}, is that there we had $b = \sqrt{2} a$, $b \gg \kappa^{2/9}$,
at the outer surface, and here we have a new adjustable parameter, namely
$\frac{\tilde{G}}{\kappa^{\frac{4}{3}}}$, related to the extra fluxes, and we
are going to try to choose a sufficiently large value of
$\frac{\tilde{G}}{\kappa^{\frac{4}{3}}}$, that we find a solution with $a \gg
b \gg \kappa^{2/9}$, at the outer surface.

Now the equations (\ref{bulk coupled a dot over a with fluxes}) and (\ref{bulk
coupled second Einstein equation with fluxes}) have the family of solutions
(\ref{bulk type power law for upper sign}), (\ref{bulk power law for a in
terms of b}), for large values of the integration constants $A$ and $B$, in
the region $c \gg \sqrt{\frac{4}{3}}$, which means $b \ll B$, by (\ref{bulk
type power law for upper sign}), provided also that $\frac{\tilde{G}}{12 b^2
a^4} \ll 1$, or in other words, by (\ref{bulk power law for a in terms of b}),
provided also that $\frac{b}{\kappa^{2/9}} \ll \left( \frac{12
\kappa^{\frac{4}{9}} A^4}{\tilde{G}} \right)^{0.9081}$ since, as noted above,
for this class of solutions, $\frac{1}{b^6 a^8}$ grows as $b^{0.2024}$ with
increasing $b$, hence $\frac{1}{b^4 a^8}$ grows as $b^{2.2024}$ with
increasing $b$.

Let us now follow a solution of (\ref{bulk coupled a dot over a with fluxes})
and (\ref{bulk coupled second Einstein equation with fluxes}) in the class
(\ref{bulk type power law for upper sign}), (\ref{bulk power law for a in
terms of b}), from small $b$, in the direction of increasing $b$, and suppose
that the $\tilde{G}^2$ terms start to become significant while the solution of
(\ref{bulk coupled second Einstein equation with fluxes}), which is decoupled
from (\ref{bulk coupled a dot over a with fluxes}) while these terms are
negligible, is still on the first branch of the square root.  Thus the
integration constants $A$ and $B$, in (\ref{bulk type power law for upper
sign}) and (\ref{bulk power law for a in terms of b}), must be such that
$\kappa^{2/9} \left( \frac{12 \kappa^{\frac{4}{9}} A^4}{\tilde{G}}
\right)^{0.9081}$ is not large compared to $B$.  Then, in a similar manner to
the situation in subsection \ref{Solutions with both a and b large at the
outer surface}, when the $\tilde{G}$ terms first start to become significant,
the solution of (\ref{bulk coupled second Einstein equation with fluxes}), in
the $\left( b, c \right)$ plane, starts to peel off below the $\tilde{G} = 0$
trajectory.  Equation (\ref{bulk coupled second Einstein equation with
fluxes}) then starts to become coupled to equation (\ref{bulk coupled a dot
over a with fluxes}), and we are looking for a solution such that the
trajectory, in the $\left( b, c \right)$ plane, curves downwards and meets the
line $c = 0$, at a finite value of $b$, which will be $b_2 = b\left( y_2
\right)$, the value of $ b $ at the outer surface of the thick pipe.
At this point, $a$
will take the value $\left( \frac{\tilde{G}}{12 b_2^2} \right)^{\frac{1}{4}}$.

In the limit $c \rightarrow 0$, the other boundary condition,
$\frac{\tilde{G}^2}{18 b^4 a^8} = 8$, implies that (\ref{bulk coupled second
Einstein equation with fluxes}) reduces to $\frac{dc}{db} = - \frac{8}{bc}$.
This, in turn, reduces to $\frac{dc}{db} = - \frac{8}{b_2 c}$, in the region
of the boundary, so that, in the region of the boundary, we have:
\begin{equation}
  \label{dependence of c on b as c tends to 0 with fluxes} c \simeq \sqrt{16
  \left( 1 - \frac{b}{b_2} \right)}
\end{equation}
This replaces equation (\ref{dependence of c on b as c tends to 0}) of
subsection \ref{Solutions with both a and b large at the outer surface}, in
the present context.

Following the method of subsection \ref{Solutions with both a and b large at
the outer surface}, we now define $v \equiv \frac{1}{b^2 a^4}$.  The above
equations then become:
\begin{equation}
  \label{a dot over a in terms of v} \frac{dv}{db} = \frac{v}{b} \left( 6 -
  \frac{2}{c} \sqrt{6 c^2 - 8 + \frac{\tilde{G}^2}{18} v^2} \right)
\end{equation}
\begin{equation}
  \label{second Einstein equation in terms of v} \frac{dc}{db} = \frac{3 c^2 -
  4}{bc} - \frac{2}{b} \sqrt{6 c^2 - 8 + \frac{\tilde{G}^2}{18} v^2} -
  \frac{\tilde{G}^2}{36 bc} v^2
\end{equation}
Now, as noted shortly after (\ref{inequality for root}), one way of studying a
pair of equations of this type, would be to take the ratio of (\ref{a dot over
a in terms of v}) and (\ref{second Einstein equation in terms of v}).  Then
$b$ cancels out, and we get a single first order differential equation, that
expresses $\frac{dv}{dc}$, as a function of $v$ and $c$.  However, I shall
follow the method of subsection \ref{Solutions with both a and b large at the
outer surface}.  The boundary conditions, at $b = b_2$, are now that:
\begin{equation}
  \label{boundary conditions in terms of v} v = \frac{12}{\tilde{G}},
  \hspace{5em} c = 0
\end{equation}
Near the boundary, we expand $v$ in the small quantity $\left( 1 -
\frac{b}{b_2} \right)$, as \\
$v = \frac{12}{\tilde{G}} \left( 1 + \alpha \left(
1 - \frac{b}{b_2} \right) \right)$.  Then from (\ref{dependence of c on b as c
tends to 0 with fluxes}) and (\ref{a dot over a in terms of v}), we find that:
\begin{equation}
  \label{condition on alpha in v} - \alpha = 6 - 2 \sqrt{6 + \alpha}
\end{equation}
which has the solutions:
\begin{equation}
  \label{solutions for alpha in v} \alpha = - 2, \hspace{3em} \alpha = - 6
\end{equation}
We note that, analogously to the situation in subsection \ref{Solutions with
both a and b large at the outer surface}, the first of these is only a
solution, for the particular sign of the square root in (\ref{condition on
alpha in v}), while the second is a solution for both signs of the square
root, since the square root vanishes for it.  The solutions can be developed
to higher order in $\left( 1 - \frac{b}{b_2} \right)$, by substituting next
into (\ref{second Einstein equation in terms of v}), to fix the next term in
$c^2$, then after that into (\ref{a dot over a in terms of v}) again, to fix
the next term in $v$, and so on, in the same way as in subsection
\ref{Solutions with both a and b large at the outer surface}.

Furthermore, in a similar manner to subsection \ref{Solutions with both a and
b large at the outer surface}, (\ref{a dot over a in terms of v}) and
(\ref{second Einstein equation in terms of v}) imply that:
\begin{equation}
  \label{vanishing square root with fluxes} \frac{d}{db} \left( 6 c^2 - 8 +
  \frac{\tilde{G}^2}{18} v^2 \right) = \frac{6}{b} \sqrt{6 c^2 - 8 +
  \frac{\tilde{G}^2}{18} v^2} \left( \sqrt{6 c^2 - 8 + \frac{\tilde{G}^2}{18}
  v^2} - \frac{1}{c} \left( 4 c^2 + \frac{\tilde{G}^2}{27} v^2 \right) \right)
\end{equation}
Thus $6 c^2 - 8 + \frac{\tilde{G}^2}{18} v^2 = 0$ is a solution of (\ref{a dot
over a in terms of v}) and (\ref{second Einstein equation in terms of v}).
However the square root, $R$, defined in (\ref{definition of the square root
R}), vanishes identically for this solution, so we cannot infer, from
(\ref{first Einstein equation in terms of other equations}), that $6 c^2 - 8 +
\frac{\tilde{G}^2}{18} v^2 = 0$ is a solution of all three Einstein equations,
and by analogy with subsection \ref{Solutions with both a and b large at the
outer surface}, we would expect that it does not correspond to a solution of
all three Einstein equations, but is, rather, the generalization to the case
where $v \neq 0$ of the line $c = \sqrt{\frac{4}{3}}$, which is the line in
the $\left( b, c \right)$ plane that actual solutions of the Einstein
equations, in the limit $v = 0$, osculate with as they switch from the first
to the second branch of the square root.

Moreover, the case $\alpha = - 6$, in (\ref{solutions for alpha in v}),
satisfies $6 c^2 - 8 + \frac{\tilde{G}^2}{18} v^2 = 0$, to the order given,
and is thus the $c \rightarrow 0$ limit of this particular solution of (\ref{a
dot over a in terms of v}) and (\ref{second Einstein equation in terms of v}),
and is thus not expected to correspond to a solution of all three Einstein
equations.  We note that this particular solution of (\ref{a dot over a in
terms of v}) and (\ref{second Einstein equation in terms of v}) satisfies $c
\leq \sqrt{\frac{4}{3}}$, and thus can never rise above the line $c =
\sqrt{\frac{4}{3}}$, in the $\left( b, c \right)$ plane.  Furthermore, when $6
c^2 - 8 + \frac{\tilde{G}^2}{18} v^2 = 0$, (\ref{a dot over a in terms of v})
reduces to $\frac{dv}{db} = 6 \frac{v}{b}$, hence $v = \frac{12}{\tilde{G}}
\left( \frac{b}{b_2} \right)^6$, where, by (\ref{boundary conditions in terms
of v}), $b_2$ is the integration constant in (\ref{dependence of c on b as c
tends to 0 with fluxes}).  Hence $c = \sqrt{\frac{4}{3} \left( 1 - \left(
\frac{b}{b_2} \right)^{12} \right)}$, which does, indeed, also solve
(\ref{second Einstein equation in terms of v}).

Considering now, the case $\alpha = - 2$, in (\ref{solutions for alpha in v}),
and still following the method of subsection \ref{Solutions with both a and b
large at the outer surface}, we see that $6 c^2 - 8 + \frac{\tilde{G}^2}{18}
v^2 \simeq 64 \left( 1 - \frac{b}{b_2} \right) \simeq 4 c^2$ near the
boundary, hence the square root, $R$, is nonvanishing, as soon as we move away
from the boundary, so, by (\ref{first Einstein equation in terms of other
equations}), this solution will correspond to a solution of all three Einstein
equations.  Furthermore, for $c \geq \sqrt{\frac{4}{3}}$, the right-hand side
of (\ref{second Einstein equation in terms of v}) is $\leq$ the right-hand
side of (\ref{approximate second Einstein equation for c in terms of b with
upper sign}), and as noted before (\ref{second Einstein equation with upper
sign near c squared equals four thirds}), the right-hand side of
(\ref{approximate second Einstein equation for c in terms of b with upper
sign}) is $\leq 0$ for all $c \geq \sqrt{\frac{4}{3}}$, so the right-hand side
of (\ref{second Einstein equation in terms of v}) is $\leq 0$ for all $c \geq
\sqrt{\frac{4}{3}}$, and the right-hand side of (\ref{second Einstein equation
in terms of v}) is certainly $\leq 0$ for $c \leq \sqrt{\frac{4}{3}}$ such
that the square root is real, so $\frac{dc}{db}$ is $\leq 0$ for all $c \geq
0$ and $b \geq 0$ such that the square root is real.  Furthermore,
$\frac{dv}{db}$ starts positive, specifically $\frac{dv}{db} = 2 \frac{v}{b}$
at $b = b_2$, hence $v$ decreases, as $b$ decreases downwards, away from $b =
b_2$, hence, provided $\frac{dv}{db}$ never becomes negative, and the square
root stays real, the square root is bounded above by $\sqrt{6} c \simeq 2.45
c$, hence, by (\ref{a dot over a in terms of v}), we have $\frac{dv}{db} \geq
\frac{v}{b} \left( 6 - 2 \sqrt{6} \right) \simeq 1.10 \frac{v}{b}$, and, by
(\ref{vanishing square root with fluxes}), we have $\frac{d}{db} \left( 6 c^2
- 8 + \frac{\tilde{G}^2}{18} v^2 \right) \leq - \left( 24 - 6 \sqrt{6} \right)
\frac{c}{b}  \sqrt{6 c^2 - 8 + \frac{\tilde{G}^2}{18} v^2} \simeq - 9.30
\frac{c}{b}  \sqrt{6 c^2 - 8 + \frac{\tilde{G}^2}{18} v^2}$, hence
$\frac{dv}{db}$ never does become negative, and the square root does stay
real.  Furthermore $v \leq \frac{12}{\tilde{G}}  \left( \frac{b}{b_2}
\right)^{1.10} \leq \frac{12}{\tilde{G}}  \frac{b}{b_2}$, hence $c \geq
\sqrt{\frac{4}{3} \left( 1 - \left( \frac{b}{b_2} \right)^2 \right)}$, hence
\begin{equation}
  \label{inequality for slope of root with fluxes} \frac{d \left( 6 c^2 - 8 +
  \frac{\tilde{G}^2}{18} v^2 \right)}{\sqrt{6 c^2 - 8 + \frac{\tilde{G}^2}{18}
  v^2}} \leq - 9.30 \frac{c}{b} db \leq - \frac{10.73}{b_2} \sqrt{1 - \left(
  \frac{b}{b_2} \right)^2} db
\end{equation}
Hence
\begin{equation}
  \label{inequality for root with fluxes} \sqrt{6 c^2 - 8 +
  \frac{\tilde{G}^2}{18} v^2} \geq 2.68 \left( \frac{\pi}{2} - \arcsin \left(
  \frac{b}{b_2} \right) - \frac{b}{b_2} \sqrt{1 - \left( \frac{b}{b_2}
  \right)^2} \right)
\end{equation}
With the bound $v \leq \frac{12}{\tilde{G}} \left( \frac{b}{b_2}
\right)^{1.10}$, this implies that $6 c^2 - 8$ is positive for $\frac{b}{b_2}
\leq 0.54$, and is greater than $13.45$ for $\frac{b}{b_2} = 0.1$, by which
point $\frac{\tilde{G}^2}{18} v^2 < 0.05$.  Thus this solution merges into a
solution of (\ref{approximate second Einstein equation for c in terms of b
with upper sign}), as $b$ continues to decrease, and for $c$ large compared to
$\sqrt{\frac{4}{3}}$, will follow a trajectory of the form (\ref{bulk type
power law for upper sign}), in the $\left( b, c \right)$ plane, with
$\frac{B}{b_2}$ a fixed number of order 1, that will be the same for all
solutions of this type.
And for solutions of this type, namely with $\alpha = - 2$ in (\ref{solutions
for alpha in v}), the constant of integration, $ b_2 $, in (\ref{dependence of
c on b as c tends to 0 with fluxes}), can be identified as $ b_2 = b\left( y_2
\right) $, the value of $ b $ at the outer surface of the thick pipe.

To estimate the integration constants $B$, in (\ref{bulk type power law for
upper sign}), and $A$, in (\ref{bulk power law for a in terms of b}), in terms
of $b_2$ and $\tilde{G}$, it is convenient to define $\tilde{v} \equiv
\frac{\tilde{G}}{12} v$.  The equations (\ref{a dot over a in terms of v}) and
(\ref{second Einstein equation in terms of v}) then become:
\begin{equation}
  \label{a dot over a in terms of v tilde} \frac{d \tilde{v}}{db} =
  \frac{\tilde{v}}{b} \left( 6 - \frac{2}{c} \sqrt{6 c^2 - 8 + 8 \tilde{v}^2}
  \right)
\end{equation}
\begin{equation}
  \label{second Einstein equation in terms of v tilde} \frac{dc}{db} = \frac{3
  c^2 - 4 - 4 \tilde{v}^2}{bc} - \frac{2}{b} \sqrt{6 c^2 - 8 + 8 \tilde{v}^2}
\end{equation}
The boundary conditions, at $b = b_2$, are now that:
\begin{equation}
  \label{boundary conditions in terms of v tilde} \tilde{v} = 1, \hspace{4em}
  c = 0
\end{equation}
Near the boundary, we have (\ref{dependence of c on b as c tends to 0 with
fluxes}) and $\tilde{v} \simeq 1 - 2 \left( 1 - \frac{b}{b_2} \right) \simeq 1
- \frac{c^2}{8}$.  Moreover, $c$ increases monotonically, and $\tilde{v}$
decreases monotonically, as $b$ decreases downwards from $b_2$.  And as $b$
tends to zero, and $c$ becomes large compared to $1$, $c$ tends to the form
(\ref{bulk type power law for upper sign}), where $B$ will be a fixed multiple
of $b_2$, that we now wish to estimate, and from (\ref{a dot over a in terms
of v tilde}), $\tilde{v}$ tends to the form
\begin{equation}
  \label{bulk power law for v tilde in terms of b} \tilde{v} \simeq V \left(
  \frac{b}{\kappa^{2/9}} \right)^{1.1010} \simeq V \left(
  \frac{B}{\kappa^{2/9}} \right)^{1.1010} c^{- 0.5798}
\end{equation}
where the second form follows from (\ref{bulk type power law for upper sign}),
and the constant $V$ is given, from (\ref{bulk power law for a in terms of
b}), and the relation $\tilde{v} = \frac{\tilde{G}}{12 b^2 a^4}$, by
\begin{equation}
  \label{V in terms of G tilde and kappa and A} V = \frac{\tilde{G}}{12
  \kappa^{\frac{4}{9}} A^4}
\end{equation}
where $A$ is the constant of integration that occurs in (\ref{bulk power law
for a in terms of b}).

A simple estimate of the dependence of $\tilde{v}$ on $c$, with the required
behaviour $\tilde{v} \simeq 1 - \frac{c^2}{8}$ as $c \rightarrow 0$, and the
power law behaviour (\ref{bulk power law for v tilde in terms of b}) as $c
\rightarrow \infty$, is
\begin{equation}
  \label{estimate of the dependence of v tilde on c} \tilde{v} \simeq \left( 1
  + \left( \frac{2 \sqrt{6} - 3}{24 - 8 \sqrt{6}} \right) c^2 \right)^{-
  \left( \frac{6 - 2 \sqrt{6}}{4 \sqrt{6} - 6} \right)} \simeq \left( 1 +
  0.4312 c^2 \right)^{- 0.2899}
\end{equation}
which gives:
\begin{equation}
  \label{estimate of V times a constant involving B and kappa} V \left(
  \frac{B}{\kappa^{2/9}} \right)^{1.1010} \simeq 0.4312^{- 0.2899}
  \simeq 1.2762
\end{equation}
As a check on (\ref{estimate of the dependence of v tilde on c}) and
(\ref{estimate of V times a constant involving B and kappa}), we note that, on
dividing (\ref{a dot over a in terms of v tilde}) by (\ref{second Einstein
equation in terms of v tilde}), we have:
\begin{equation}
  \label{d v tilde by d c} \frac{d \tilde{v}}{dc} = \frac{\tilde{v} \left( 6 c
  - 2 \sqrt{6 c^2 - 8 + 8 \tilde{v}^2} \right)}{3 c^2 - 4 - 4 \tilde{v}^2 - 2
  c \sqrt{6 c^2 - 8 + 8 \tilde{v}^2}}
\end{equation}
And with the help of Maxima {\cite{Maxima}}, we find that the solution of
(\ref{d v tilde by d c}), that behaves as $\tilde{v} \simeq 1 - \frac{c^2}{8}$
near $c = 0$, has the Taylor expansion:
\begin{equation}
  \label{Taylor expansion of v tilde in terms of c} 1 - \frac{c^2}{8} +
  \frac{5 c^4}{384} + \frac{23 c^6}{46080} - \frac{3559 c^8}{10321920} +
  \frac{15167 c^{10}}{3715891200} + \cdots
\end{equation}
where $\ldots$ denotes terms of order $c^{12}$.  And using the graphical
facility of Maxima, we see that (\ref{Taylor expansion of v tilde in terms of
c}) is accurately approximated by its first four terms up to around $c = 1.3$,
at which point (\ref{Taylor expansion of v tilde in terms of c}) is $0.8256$,
while (\ref{estimate of the dependence of v tilde on c}) is $0.8533$, and
(\ref{Taylor expansion of v tilde in terms of c}) is accurately approximated
by its first five terms up to around $c = 2.0$, at which point (\ref{Taylor
expansion of v tilde in terms of c}) is $0.6562$, while (\ref{estimate of the
dependence of v tilde on c}) is $0.7478$.  And (\ref{Taylor expansion of v
tilde in terms of c}) starts to curve rapidly downwards above around $c =
2.0$, and would thus appear likely to depart from the true dependence of
$\tilde{v}$ on $c$, starting at around $c = 2.0$.  Thus it seems likely that
for large $c$, the estimate (\ref{estimate of the dependence of v tilde on c})
of $\tilde{v}$ will be around 15 to 20 percent too large, and the estimate
(\ref{estimate of V times a constant involving B and kappa}), of $V \left(
\frac{B}{\kappa^{2/9}} \right)^{1.1010}$, will also be around 15 to 20
percent too large.  So a better estimate of $V \left(
\frac{B}{\kappa^{2/9}} \right)^{1.1010}$ would perhaps be around
$1.1$.

A simple estimate of the dependence of $c$ on $b$, with the required behaviour
(\ref{dependence of c on b as c tends to 0 with fluxes}) as $c \rightarrow 0$,
and $b \rightarrow b_2$ from below, and the power law behaviour (\ref{bulk
type power law for upper sign}) for $c \gg \sqrt{\frac{4}{3}}$, is:
\begin{equation}
  \label{estimate of the dependence of c on b} c \simeq \left( \frac{b_2}{b}
  \right)^{1.8990} \sqrt{16 \left( 1 - \frac{b}{b_2} \right)}
\end{equation}
which also has the required property that $c$ only depends on $b$, and the
integration constant $b_2$, through the ratio $\frac{b}{b_2}$, as follows from
(\ref{second Einstein equation in terms of v tilde}), after substituting for
$\tilde{v}$ as a function of $c$, with a Taylor expansion that begins as in
(\ref{Taylor expansion of v tilde in terms of c}).  The estimate
(\ref{estimate of the dependence of c on b}) leads to the estimate:
\begin{equation}
  \label{estimate of B in terms of B sub 3} B \simeq 4^{\frac{1}{1.8990}} b_2
  \simeq 2.0751 b_2
\end{equation}
To check (\ref{estimate of the dependence of c on b}) and (\ref{estimate of B
in terms of B sub 3}), we note that, from (\ref{second Einstein equation in
terms of v tilde}), we have:
\begin{equation}
  \label{integral for b as a function of c} \ln \left( \frac{b_2}{b} \right) =
  \int^c_0 \frac{xdx}{\left( 4 + 4 \tilde{v}^2 + 2 x \sqrt{6 x^2 - 8 + 8
  \tilde{v}^2} - 3 x^2 \right)}
\end{equation}
where, in the integrand, $\tilde{v}$ is given as a function of $x$, by
rewriting $c$ as $x$, in the dependence of $\tilde{v}$ on $c$ as above, whose
Taylor expansion begins as in (\ref{Taylor expansion of v tilde in terms of
c}).  To calculate the integral in (\ref{integral for b as a function of c}),
I used the numerical integration facility of PARI/GP {\cite{PARI GP}}, with
the dependence of $\tilde{v}$ on $c$ given by (\ref{Taylor expansion of v
tilde in terms of c}) for $c \leq 2$, and by (\ref{estimate of the dependence
of v tilde on c}), multiplied by $\frac{0.6562}{0.7478} = 0.8775$, so as to
obtain continuity at $c = 2$, for $c \geq 2$.  In this way, writing the
integral in the right hand side of (\ref{integral for b as a function of c})
as $\int^c_0 fdx$, we find the entries in the second column of
Table \ref{Table of the dependence of c on b}.
The entries in the third column are the values of $\frac{b}{b_2}$
which correspond by (\ref{integral for b as a function of c}) to the entries
in the second column, and the entries in the fourth column are the values of
$c$ which the estimate (\ref{estimate of the dependence of c on b}) gives, for
the values of $\frac{b}{b_2}$ in the third column.  The fifth column gives the
ratio of the estimated value of $c$ calculated in the fourth column by the
estimate (\ref{estimate of the dependence of c on b}), to the original value
of $c$ in the first column.

\begin{table}
\begin{center}
\begin{tabular}{|c|c|c|c|c|c|c|}\hline
  $c$ & $\int^c_0 fdx$ & $\frac{b}{b_2}$ by (\ref{integral for b as a function
  of c}) & \begin{tabular}{c}
    $c$ for this\\
    $\frac{b}{b_2}$ by (\ref{estimate of the dependence of c on b})
  \end{tabular} & $\frac{c_{(\ref{estimate of the dependence of c on
  b})}}{c}$ & \begin{tabular}{c}
    $c$ for this\\
    $\frac{b}{b_2}$ by (\ref{improved estimate of the dependence of c on b})
  \end{tabular} & $\frac{c_{(\ref{improved estimate of the dependence of
  c on b})}}{c}$\\ \hline
  0.5 & $0.01561$ & $0.9845$ & $0.5130$ & $1.0260$ & $0.5032$ & $1.0064$\\
  1.0 & $0.06170$ & $0.9402$ & $1.0997$ & $1.0997$ & $1.0209$ & $1.0209$\\
  1.5 & $0.1329$ & $0.8756$ & $1.8156$ & $1.2104$ & $1.5577$ & $1.0385$\\
  2.0 & $0.2177$ & $0.8044$ & $2.6746$ & $1.3373$ & $2.1081$ & $1.0541$\\
  3.0 & $0.3821$ & $0.6824$ & $4.6576$ & $1.5525$ & $3.1991$ & $1.0664$\\
  5.0 & $0.6286$ & $0.5333$ & $9.0170$ & $1.8034$ & $5.3261$ & $1.0652$\\
  10.0 & $0.9866$ & $0.3728$ & $20.631$ & $2.0631$ & $10.621$ & $1.0621$\\
  20.0 & $1.3504$ & $0.2591$ & $44.747$ & $2.2374$ & $21.220$ & $1.0610$\\
  80.0 & $2.0801$ & $0.1249$ & $194.41$ & $2.4301$ & $84.842$ & $1.0605$\\ \hline
\end{tabular}
\caption{\label{Table of the dependence of c on b}
The numerical dependence of $ c $ on $ b $.}
\end{center}
\end{table}

From the form of the discrepancy factor in the fifth column, we would expect
that the estimate (\ref{estimate of the dependence of c on b}) could be
improved by replacing the factor $4 \sqrt{1 - \frac{b}{b_2}}$, in
(\ref{estimate of the dependence of c on b}), by a factor of the form
$\frac{4}{\sqrt{n}} \sqrt{1 - \left( \frac{b}{b_2} \right)^n}$, which has the
same limiting behaviour as $b \rightarrow b_2$ from below, and where $\sqrt{n}
\sim 2.43 \sim \sqrt{6}$.  Thus, taking $n = 6$, we try an estimate:
\begin{equation}
  \label{improved estimate of the dependence of c on b} c \simeq 1.633 \left(
  \frac{b_2}{b} \right)^{1.8990} \sqrt{1 - \left( \frac{b}{b_2} \right)^6}
\end{equation}
The values of $c$ given by the estimate (\ref{improved estimate of the
dependence of c on b}) are listed in the sixth column of the table, and from
the discrepancy factor, in the seventh column of the table, we see that the
error now stays below 7 percent, and is actually slowly decreasing, as
$\frac{b}{b_2}$ continues to decrease below $0.5$.  I shall therefore use
(\ref{improved estimate of the dependence of c on b}) as a reasonable estimate
of the dependence of $c$ on $b$, in the presence of the extra fluxes.  The
corresponding estimate of the integration constant $B$, in (\ref{bulk type
power law for upper sign}), is:
\begin{equation}
  \label{improved estimate of B in terms of B sub 3} B \simeq
  1.633^{\frac{1}{1.8990}} b_2 \simeq 1.2947 b_2
\end{equation}
which now replaces the estimate (\ref{estimate of B in terms of B sub 3}).

Returning, now, to the dependence of $\tilde{v}$ on $c$, I used a standard
fourth-order Runge-Kutta method {\cite{Runge-Kutta in Wikipedia}} to integrate
(\ref{d v tilde by d c}) from $c = 0.3$, where
$\tilde{v}$ is reliably given by (\ref{Taylor expansion of v tilde in terms of
c}) as $\tilde{v} \simeq 0.9889$, into the power law region, where $c \gg
\sqrt{\frac{4}{3}}$.  The same result was obtained with a Runge-Kutta interval
$h = 0.01$ as with $h = 0.00001$, even for $c = 80$.  In fact, to four
significant digits, the same result was also obtained with $h = 0.1$, even for
$c = 80$.  The results are shown in Table \ref{Table of the dependence of v
tilde on c}.

\begin{table}
\begin{center}
\begin{tabular}{|c|c|c|c|c|}\hline
  $c$ & $\tilde{v}$ & \begin{tabular}{c}
    $\tilde{v}$ from Taylor\\
    series (\ref{Taylor expansion of v tilde in terms of c})
  \end{tabular} & \begin{tabular}{c}
    $\tilde{v}$ from\\
    estimate (\ref{estimate of the dependence of v tilde on c})
  \end{tabular} & $\frac{\tilde{v}_{(\ref{estimate of the dependence of
  v tilde on c})}}{\tilde{v}}$\\ \hline
  $0.5$ & $0.9696$ & $0.9696$ & $0.9708$ & $1.0012$\\
  $1.0$ & $0.8882$ & $0.8882$ & $0.9013$ & $1.0147$\\
  $1.5$ & $0.7830$ & $0.7818$ & $0.8215$ & $1.0492$\\
  $2.0$ & $0.6830$ & $0.6562$ & $0.7478$ & $1.0949$\\
  $3.0$ & $0.5383$ & - & $0.6315$ & $1.1731$\\
  $5.0$ & $0.3919$ & - & $0.4892$ & $1.2483$\\
  $10.0$ & $0.2581$ & - & $0.3336$ & $1.2925$\\
  $20.0$ & $0.1719$ & - & $0.2243$ & $1.3048$\\
  $80.0$ & $0.07680$ & - & $0.1006$ & $1.3099$\\ \hline
\end{tabular}
\caption{\label{Table of the dependence of v tilde on c}
The numerical dependence of $ \tilde{v} $ on $ c $.}
\end{center}
\end{table}

Thus the error of the estimate (\ref{estimate of the dependence of v tilde on
c}) stabilizes at about 31 percent in the power law region, and the estimate
(\ref{estimate of V times a constant involving B and kappa}) should be
replaced by:
\begin{equation}
  \label{improved estimate of V times a constant involving B and kappa} V
  \left( \frac{B}{\kappa^{2/9}} \right)^{1.1010} \simeq
  \frac{0.4312^{- 0.2899}}{1.31} \simeq 0.9742
\end{equation}

We next consider the dependence of $b$ on $y$, and note, following the
discussion shortly after (\ref{inequality for root}), in subsection
\ref{Solutions with both a and b large at the outer surface}, that
the behaviour (\ref{dependence of c on b as c tends to 0 with fluxes}), for
$c$ near the boundary, implies that near the boundary, $y_2 - y \simeq
\frac{b_2}{2}  \sqrt{1 - \frac{b}{b_2}} \simeq \frac{b_2}{8} c$, thus $y$
does, indeed, tend to a finite value, $y_2$, at the boundary, even though
$\frac{dy}{db} = \frac{1}{c}$ goes to $\infty$, right at the boundary.  And
using the approximate relation (\ref{improved estimate of B in terms of B sub
3}), we find that near the outer boundary:
\begin{equation}
  \label{approximate dependence of b on y near outer boundary with fluxes}
  \frac{b}{B} \simeq 0.7724 \left( 1 - 6.7050 \left( \frac{y_2 - y}{B}
  \right)^2 \right)
\end{equation}
For a first estimate of $y_2$, we could simply use the form (\ref{bulk type
power law dependence of b on y for upper sign}), with $y_0 = 0$, all the way
from $y_1$ to $y_2$, and determine $y_2$ as the point where this gives $b =
b_2 \simeq 0.7724 B$, which gives $y_2 \simeq \left( \frac{0.7724}{1.4436}
\right)^{\frac{1}{0.3449}} B \simeq 0.1631 B$.  This underestimates $y_2$ by a
factor of order $1$, because, by (\ref{approximate dependence of b on y near
outer boundary with fluxes}), the curve of $b \left( y \right)$, in the
$\left( y, b \right)$ plane, curves to the right as the outer boundary is
approached, so that $b_2$ is not reached until a larger value of $y$ than
would be indicated by (\ref{bulk type power law dependence of b on y for upper
sign}) with $y_0 = 0$.

For a better estimate of $y_2$, a convenient interpolating function would be
\begin{equation}
  \label{interpolating function for dependence of b on y with fluxes} b =
  1.4436 \frac{B \left( \frac{y}{B} \right)^{0.3449}}{\left( 1 + \alpha
  y^{\beta} \right)^{\gamma}}
\end{equation}
with $\beta > 0$, and either $\alpha > 0$ and $\beta \gamma > 0.3449$, or
$\alpha < 0$ and $\gamma < 0$.  This agrees with (\ref{bulk type power law
dependence of b on y for upper sign}) for $y \ll \frac{1}{\alpha}$, and has a
peak at $y = \left( \frac{0.3449}{\alpha \left( \beta \gamma - 0.3449 \right)}
\right)^{\frac{1}{\beta}}$, which we attempt to identify with $y_2$.
Requiring agreement with (\ref{approximate dependence of b on y near outer
boundary with fluxes}) for $b \left( y_2 \right)$ leads to the requirement
that:
\begin{equation}
  \label{first condition for interpolating function for b with fluxes} 1.4436
  \left( \frac{0.3449}{\alpha B^{\beta}} \right)^{\frac{0.3449}{\beta}} \left(
  \beta \gamma - 0.3449 \right)^{\frac{\beta \gamma - 0.3449}{\beta}} = 0.7724
  \left( \beta \gamma \right)^{\gamma}
\end{equation}
This is written for the $\alpha > 0$ case, and for the $\alpha < 0$ case
should be rewritten in the equivalent form with the contents of each of the
three pairs of parentheses multiplied by $- 1$.  And requiring agreement with
(\ref{approximate dependence of b on y near outer boundary with fluxes}) for
$\frac{1}{b}  \frac{d^2 b}{dy^2}$, evaluated at $y = y_2$, leads to the
requirement that:
\begin{equation}
  \label{second condition for interpolating function for b with fluxes} \left(
  \alpha B^{\beta} \right)^{\frac{2}{\beta}}  \left( \frac{\beta \gamma -
  0.3449}{0.3449} \right)^{\frac{2}{\beta} + 1} = 112.73 \gamma
\end{equation}
This is also written for the $\alpha > 0$ case, and for the $\alpha < 0$ case
should be rewritten in the equivalent form with the contents of each of the
two pairs of parentheses multiplied by $- 1$, and the right hand side also
multiplied by $- 1$.  Eliminating $\alpha B^{\beta}$ between (\ref{first
condition for interpolating function for b with fluxes}) and (\ref{second
condition for interpolating function for b with fluxes}), we find:
\begin{equation}
  \label{reduced condition for interpolating function for b with fluxes}
  \left( \frac{\beta \gamma - 0.3449}{\beta \gamma} \right)^{\gamma + 0.1725}
  = \frac{1.0059}{\beta^{0.1725}}
\end{equation}
Trying first $\beta = 1$, there is no solution in the $\alpha > 0$ case, but
there is a solution with $\gamma \simeq - 0.1672$ in the $\alpha < 0$ case.
We note that the improved estimate, (\ref{improved estimate of the dependence
of c on b}), of the dependence of $c$ on $b$, shows that the power law
behaviour remains a good approximation until $b$ is quite close to $b_2$.  For
small $\frac{b}{b_2}$, the correction to the power law behaviour, in
(\ref{improved estimate of the dependence of c on b}), is by a term of
relative size $\left( \frac{b}{b_2} \right)^6$, which is $\sim \left(
\frac{y}{B} \right)^{2.0694}$ in the power law region, which suggests that
$\beta = 2$ might be a good choice in (\ref{interpolating function for
dependence of b on y with fluxes}).  However with $\beta = 2$, there still
appears to be no solution in the $\alpha > 0$ case, while there is a solution
with $\gamma \simeq - 0.6666$ in the $\alpha < 0$ case.  Trying $\beta = 2.5$,
there is a solution with $\gamma \simeq 2.4194$ in the $\alpha > 0$ case, but
apparently no solution in the $\alpha < 0$ case.  And trying $\beta = 4$,
there is a solution with $\gamma \simeq 0.1771$ in the $\alpha > 0$ case, and
apparently no solution, again, in the $\alpha < 0$ case.  These example
solutions, and the corresponding values of $\alpha B^{\beta}$ and
$\frac{y_2}{B}$, are listed in Table \ref{Table of parameters for the
dependence of b on y}.

\begin{table}
\begin{center}
\begin{tabular}{|c|c|c|c|}\hline
  $\beta$ & $\gamma$ & $\alpha B^{\beta}$ & $\frac{y_2}{B}$\\ \hline
  $1$ & $- 0.1672$ & $- 2.3996$ & $0.2807$\\
  $2$ & $- 0.6666$ & $- 3.1744$ & $0.2545$\\
  $2.5$ & $2.4194$ & $2.0098$ & $0.2462$\\
  $4$ & $0.1771$ & $340.47$ & $0.2298$\\ \hline
\end{tabular}
\caption{\label{Table of parameters for the dependence of b on y}
Parameters for the interpolating function (\ref{interpolating function for
dependence of b on y with fluxes}) for $ b $ as a function of $ y $.}
\end{center}
\end{table}

We see that, notwithstanding the substantial differences between the
parameters of the interpolating function, for the different choices of
$\beta$, the corresponding values of $\frac{y_2}{B}$ only differ by around 20
percent, and are around $1.4$ to $1.7$ times larger than the value $0.1631$
given by the uncorrected power law (\ref{bulk type power law dependence of b
on y for upper sign}).  They show a trend towards the uncorrected power law
value with increasing $\beta$, corresponding to a later and more rapid onset
of the corrrections to the power law.

We can also obtain an approximate value of $y_2$ by integrating the
approximate formula (\ref{improved estimate of the dependence of c on b}).
From (\ref{improved estimate of the dependence of c on b}) and (\ref{improved
estimate of B in terms of B sub 3}) we obtain:
\begin{equation}
  \label{integral for y sub 2 with fluxes} \int^{0.7724}_{\frac{b_1}{B}}
  \frac{x^{1.8990} dx}{\sqrt{1 - \left( \frac{x}{0.7724} \right)^6}} \simeq
  \frac{y_2 - y_1}{B}
\end{equation}
But $b_1 \sim \kappa^{2/9}$, which by assumption is small compared to
$B$, so we can extend the lower limit of the integral in the left hand side of
(\ref{integral for y sub 2 with fluxes}) to zero, and choosing, as usual, the
integration constant $y_0$ in (\ref{bulk type power law dependence of b on y
for upper sign}) to be zero, we have $y_1 \ll \kappa^{2/9} \ll B$, so
we can drop the $y_1$ term in the right hand side of (\ref{integral for y sub
2 with fluxes}).  Then by use of the numerical integration facility of PARI-GP
{\cite{PARI GP}}, plus an analytic approximation for the contribution from the
region close to the upper limit, the integral in the left hand side of
(\ref{integral for y sub 2 with fluxes}) is found to be $\simeq 0.2536$.  And
comparing with the estimates of $\frac{y_2}{B}$ as given in
Table \ref{Table of parameters for the dependence of b on y},
for the different choices of $\beta$ in the interpolating function
(\ref{interpolating function for dependence of b on y with fluxes}), we see
that the best agreement is obtained for the choice $\beta = 2$, as expected
from the discussion following (\ref{reduced condition for interpolating
function for b with fluxes}).

\subsubsection{Newton's constant and the cosmological constant in the presence
of the extra fluxes}
\label{Newtons constant and the cosmological constant with fluxes}

Turning now to fitting the observed values of Newton's constant and the
cosmological constant, we again follow the method used in subsection \ref{G sub
N and Lambda for solutions with outer surface in classical region}, on page
\pageref{G sub N and Lambda for solutions with outer surface in classical
region}.  The term, in the Einstein action term in
(\ref{upstairs bulk action}), that produces the Einstein action,
(\ref{Einstein action}), in four dimensions, is again given by (\ref{Einstein
term in four dimensional effective action}), where $b \left( y \right)$ is now
given approximately by (\ref{interpolating function for dependence of b on y
with fluxes}), with $\beta$ preferably chosen as $2$, and $\gamma$ and
$\alpha$ as given by the row corresponding to $\beta = 2$ in the above table,
and $a$, as a function of $b$, is given, as a first approximation, in terms of
the approximate dependence of $\tilde{v}$ on $c$, in (\ref{estimate of the
dependence of v tilde on c}), the approximate dependence of $c$ on $b$, in
(\ref{improved estimate of the dependence of c on b}), and the relation
$\tilde{v} = \frac{\tilde{G}}{12 b^2 a^4}$.  The worst approximation here is
the estimate (\ref{estimate of the dependence of v tilde on c}) of the
dependence of $\tilde{v}$ on $c$, which has a percentage error that stabilizes
at around 31 percent in the power law region, as found above.  We then have:
\begin{equation}
  \label{integral of a squared b to the sixth with fluxes} \int^{y_2}_{y_1}
  dya^2 b^6 = \int^{b_2}_{b_1} \frac{db}{c} a^2 b^6 =
  \sqrt{\frac{\tilde{G}}{12}}  \int^{0.7724 B}_{b_1} \frac{b^5 db}{c
  \sqrt{\tilde{v}}}
\end{equation}
We note that since $\tilde{v}$ occurs in (\ref{integral of a squared b to the
sixth with fluxes}) only through its square root, the contribution to the
error percentage resulting from the use of (\ref{estimate of the dependence of
v tilde on c}) will be roughly halved, to not more than around 16 percent.
And in the power law region, where by (\ref{bulk power law for v tilde in
terms of b}), $\tilde{v} \sim c^{- 0.5798}$, the denominator, in the last
integral in (\ref{integral of a squared b to the sixth with fluxes}), is $\sim
c^{0.7101}$, so since (\ref{improved estimate of the dependence of c on b})
overestimates $c$ by not more than about 7 percent, and (\ref{estimate of the
dependence of v tilde on c}) will overestimate $\sqrt{\tilde{v}}$, as a
function of $c$, by not more than about 16 percent, the use of (\ref{improved
estimate of the dependence of c on b}) and (\ref{estimate of the dependence of
v tilde on c}), in (\ref{integral of a squared b to the sixth with fluxes}),
is expected to give a result that will be smaller than the correct result, but
by not more than about $16 + 0.71 \times 7 \simeq 21$ percent.  Furthermore,
since $c \sim b^{- 1.8990}$ in the power law region, by (\ref{bulk type power
law for upper sign}), the integrand, in the last integral in (\ref{integral of
a squared b to the sixth with fluxes}), is $\sim b^{6.3485}$ in the power law
region, and goes to infinity as $\left( 0.7724 B - b \right)^{- \frac{1}{2}}$
at the upper limit, due to the behaviour (\ref{dependence of c on b as c tends
to 0 with fluxes}) of $c$, so the last integral in (\ref{integral of a squared
b to the sixth with fluxes}) is substantially dominated by the contribution
from the region near the upper limit, where (\ref{improved estimate of the
dependence of c on b}) and (\ref{estimate of the dependence of v tilde on c})
are accurate, so the error is in fact expected to be substantially smaller
than 21 percent.

Inserting the approximate expressions (\ref{improved estimate of the
dependence of c on b}) and (\ref{estimate of the dependence of v tilde on c}),
we find:
\begin{equation}
  \label{approximate integral of a squared b to the sixth with fluxes}
  \int^{y_2}_{y_1} dya^2 b^6 \simeq B^6  \sqrt{\frac{\tilde{G}}{12}}
  \int^{0.7724}_{\frac{b_1}{B}} \frac{x^{6.8990} dx \left( 1 +
  \frac{0.4312}{x^{3.7980}}  \left( 1 - \left( \frac{x}{0.7724} \right)^6
  \right) \right)^{0.1450}}{\sqrt{1 - \left( \frac{x}{0.7724} \right)^6} }
\end{equation}
Now as before, $\frac{b_1}{B} \sim \frac{\kappa^{2/9}}{B}$, so we can
set the lower limit of the integral in the right hand side of
(\ref{approximate integral of a squared b to the sixth with fluxes}) to zero.
Then using again the numerical integration facility of PARI-GP, plus an
analytic approximation for the contribution from the region near the upper
limit of the integration domain in the integral in the right hand side of
(\ref{approximate integral of a squared b to the sixth with fluxes}), we find:
\begin{equation}
  \label{approximate value of integral of a squared b to the sixth with
  fluxes} \int^{y_2}_{y_1} dya^2 b^6 \simeq 0.03896 B^6
  \sqrt{\frac{\tilde{G}}{12}}
\end{equation}
And, as explained above, the coefficient $0.03896$ is expected to be smaller
than the correct value, but the percentage error is expected to be
substantially smaller than 21 percent.

Now from (\ref{V in terms of G tilde and kappa and A}) and (\ref{improved
estimate of V times a constant involving B and kappa}), we have:
\begin{equation}
  \label{estimate of G tilde in terms of A and B and kappa}
  \frac{\tilde{G}}{12} \simeq 0.9742 \kappa^{\frac{4}{9}} A^4  \left(
  \frac{\kappa^{2/9}}{B} \right)^{1.1010}
\end{equation}
Hence from (\ref{approximate value of integral of a squared b to the sixth
with fluxes}), and (\ref{A in terms of A sub 1 and B}), on page \pageref{A in
terms of A sub 1 and B}, we have:
\begin{equation}
  \label{integral of a squared b to the sixth with fluxes in terms of A and B
  and kappa} \int^{y_2}_{y_1} dya^2 b^6 \simeq 0.03845 \kappa^{\frac{14}{9}}
  A^2 \left( \frac{B}{\kappa^{2/9}} \right)^{5.4495} \simeq 0.03845
  \kappa^{\frac{14}{9}} A^2_1 \left( \frac{B}{\kappa^{2/9}}
  \right)^{1.3102 \tau + 6.4653}
\end{equation}
which replaces (\ref{contribution from classical region}), on page
\pageref{contribution from classical region}, and (\ref{integral of a squared b
to the sixth for a small at outer boundary}), on page \pageref{integral of a
squared b to the sixth for a small at outer boundary}, for the present
situation, where the outer boundary is controlled by the extra fluxes.
We see that, as found in subsections \ref{G sub N and Lambda for solutions
with outer surface in classical region} and \ref{Newtons constant and the
cosmological constant}, there is no ADD effect unless $\tau > - 4.9346$.

Continuing to follow subsections \ref{G sub N and Lambda for solutions with
outer surface in classical region} and \ref{Newtons constant and the
cosmological constant}, we
now find that when the compact six-manifold is a smooth compact quotient of
$\mathbf{C} \mathbf{H}^3$, the Einstein action term, in the
four-dimensional effective action, for the solutions considered in the present
subsection, will be equal to:
\begin{equation}
  \label{Einstein term in four dimensional effective action with fluxes}
  0.3974 \frac{1}{\kappa^{\frac{4}{9}}} A^2 \left( \frac{
  B}{\kappa^{2/9}} \right)^{5.4495} \chi \left( \mathcal{M}^6 \right)
  \int d^4 x \sqrt{- \tilde{g}} \tilde{g}^{\mu \nu} R_{\mu \tau \nu} \,
  \!^{\tau} \left( \tilde{g} \right)
\end{equation}

I shall now consider the case where $\tau = - 0.7753$, which corresponds to
the classical power law (\ref{bulk power law for a in terms of b}), on page
\pageref{bulk power law for a in terms of b}.  Then defining the rescaled
metric $\bar{g}_{\mu \nu}$ by $\bar{g}_{\mu \nu} = \left(
\textrm{de Sitter radius} \right)^2 \tilde{g}_{\mu \nu}$, as in subsections
\ref{G sub N and Lambda for solutions with outer surface in classical region}
and \ref{Newtons constant and the cosmological constant}, so as to measure
distances in ordinary units rather than in units of the de Sitter radius, we
find from (\ref{A in terms of de Sitter radius}), on page \pageref{A in terms
of de Sitter radius}, that the Einstein action term in the four-dimensional
effective action, for the solutions considered in the present subsection, will
for $ \tau = -0.7753 $ and smooth compact quotients of $\mathbf{C} \mathbf{H}^3
$ be equal to
\begin{equation}
  \label{Einstein action term in the four dimensional effective action with
  fluxes without A} - 0.5808 \frac{1}{\kappa^{\frac{4}{9}}} \left(
  \frac{B}{\kappa^{2/9}} \right)^{5.4495} \left| \chi \left(
  \mathcal{M}^6 \right) \right|^{0.7416} \int d^4 x \sqrt{- \bar{g}}
  \bar{g}^{\mu \nu} R_{\mu \tau \nu} \, \!^{\tau} \left( \bar{g} \right)
\end{equation}
Thus, comparing with (\ref{Einstein action}), we find that for $ \tau = -0.7753
$ and smooth compact quotients of $\mathbf{C} \mathbf{H}^3$:
\begin{equation}
  \label{1 over G Newton for solutions with fluxes} \frac{1}{G_N} \simeq 29.19
  \frac{1}{\kappa^{\frac{4}{9}}}  \left( \frac{B}{\kappa^{2/9}}
  \right)^{5.4495} \left| \chi \left( \mathcal{M}^6 \right) \right|^{0.7416}
\end{equation}
This is the form taken by the ADD mechanism
{\cite{ADD1, ADD2}}, for the solutions considered in the present subsection
with $ \tau = -0.7753 $.  Thus for $ \tau = -0.7753 $ and smooth compact
quotients of $\mathbf{C} \mathbf{H}^3$:
\begin{equation}
  \label{B for solutions with fluxes} \frac{B}{\kappa^{2/9}} \simeq
  \frac{0.5384}{\left| \chi \left( \mathcal{M}^6 \right) \right|^{0.1361}}
  \left( \frac{\kappa^{\frac{4}{9}}}{G_N} \right)^{0.1835}
\end{equation}

Considering, now, the case of TeV-scale gravity, I shall again consider the
case where $\kappa^{- \frac{2}{9}} = 0.2217$ TeV, so that
$\kappa^{2/9} = 8.899 \times 10^{- 19}$ metres, and the Giudice,
Rattazzi, and Wells {\cite{Giudice Rattazzi Wells}} gravitational mass $M_D$,
for $D = 11$, is equal to $1$ TeV.  We then find, from (\ref{Newtons
constant}), that for $ \tau = -0.7753 $ and smooth compact quotients of $
\mathbf{C} \mathbf{H}^3$:
\begin{equation}
  \label{B for TeV scale gravity with fluxes} B \simeq \frac{7.499 \times
  10^5}{\left| \chi \left( \mathcal{M}^6 \right) \right|^{0.1361}}
  \kappa^{2/9} \simeq \frac{6.673 \times 10^{- 13} \hspace{0.4ex}
  \textrm{metres}}{\left| \chi \left( \mathcal{M}^6 \right) \right|^{0.1361}}
\end{equation}
On the other hand, for $\tau = - 0.7753$, the integration constant $A$ in
(\ref{bulk power law for a in terms of b}), on page \pageref{bulk power law
for a in terms of b}, is from (\ref{A in terms of de Sitter radius}), on page
\pageref{A in terms of de Sitter radius}, fixed directly in terms of $\left|
\chi \left( \mathcal{M}^6 \right) \right|$ and the observed de Sitter radius
(\ref{de Sitter radius}), on page \pageref{de Sitter radius}, and given for
smooth compact quotients of $\mathbf{C} \mathbf{H}^3$ by (\ref{A in metres for
classical tau}), on page \pageref{A in metres for classical tau}. Hence from
(\ref{estimate of G tilde in terms
of A and B and kappa}) we find, for $\tau = - 0.7753$ and smooth compact
quotients of $\mathbf{C} \mathbf{H}^3$, that:
\begin{equation}
  \label{G tilde for TeV scale gravity} \tilde{G} \simeq \frac{3.53 \times
  10^{63} \hspace{0.4ex} \textrm{metres}^6}{\left| \chi \left( \mathcal{M}^6
  \right) \right|^{0.3670}} \simeq \frac{7.11 \times 10^{171}}{\left| \chi
  \left(   \mathcal{M}^6 \right) \right|^{0.3670}} \kappa^{\frac{4}{3}} \simeq
  \frac{\left( 4.39 \times 10^{28} \kappa^{2/9} \right)^6}{\left| \chi
  \left( \mathcal{M}^6 \right) \right|^{0.3670}}
\end{equation}
where $\tilde{G}$ is defined in (\ref{definition of the constant G tilde}).
This is the large constant of integration, not constrained by the field
equations or boundary conditions, that is built into the structure of the
universe, to make it into the stiff, strong structure that we observe, for the
solutions considered in the present subsection with $ \tau = -0.7753 $, in the
case of TeV-scale gravity.

In a similar way to the situation with $\tau$,
$\frac{\tilde{A}}{\kappa^{2/9}}$, and $\frac{B}{\kappa^{2/9}}$
in subsection \ref{Newtons constant and the cosmological constant}, it will be
possible, by decreasing $\tau$ below $- 0.7753$, to decrease
$\frac{\tilde{G}}{\kappa^{\frac{4}{3}}}$ at a cost of increasing
$\frac{B}{\kappa^{2/9}}$, until as $\tau$ approaches the values near
$- 3$ in (\ref{tau and B for mod chi equals 1}) and (\ref{tau and B for mod
chi equals 7 times 10 to the 4}), on page \pageref{tau and B for mod chi
equals 1}, it will no longer be a good approximation to neglect the term
$\frac{4}{a^2}$ in the square root in comparison to the term
$\frac{\tilde{G}^2}{18 b^6 a^8}$, and the solutions considered in this
subsection will tend as $\tilde{G} \rightarrow 0$ to those studied in
subsection \ref{G sub N and Lambda for solutions with outer surface in
classical region}, on page \pageref{G sub N and Lambda for solutions with
outer surface in classical region}.

More generally, from (\ref{A in terms of de Sitter radius}), (\ref{de Sitter
radius}), (\ref{B for solutions with fluxes}), and (\ref{estimate of G tilde in
terms of A and B and kappa}), we find that for $ \tau = -0.7753 $ and smooth
compact quotients of $\mathbf{C} \mathbf{H}^3$, with a general
value of $\kappa^{2/9}$:
\begin{equation}
  \label{G tilde for general kappa to the two ninths}
  \frac{\tilde{G}}{\kappa^{\frac{4}{3}}} \simeq \frac{3.90 \times 10^{245}
  }{\left| \chi \left( \mathcal{M}^6 \right) \right|^{0.3670}}  \left(
  \frac{G_N}{\kappa^{\frac{4}{9}}} \right)^{2.2020}
\end{equation}
Thus the required value of $\frac{\tilde{G}}{\kappa^{\frac{4}{3}}}$ is
minimized by choosing $\kappa^{2/9}$ to be as large as possible, which
means TeV-scale gravity, provided this is consistent with the precision tests
of Newton's law down to sub-millimetre distances {\cite{Hoyle et al}}.  To
check that this requirement is satisfied, we now determine the values of $ b $,
$ y $, and $ a $, at the outer surface of the thick pipe, for the solutions
considered in the present subsection with $ \tau = -0.7753 $, in the case of
TeV-scale gravity, with $ \kappa^{-\frac{2}{9}} = 0.2217 $ TeV.

From (\ref{B for TeV scale gravity with fluxes}) and (\ref{improved estimate of
B in terms of B sub 3}), we find that $b_2$, the value of $b$ at the outer
surface of the thick pipe, is given, for $ \tau = -0.7753 $ and smooth compact
quotients of $\mathbf{C} \mathbf{H}^3$, by:
\begin{equation}
  \label{B sub 3 for TeV scale gravity with fluxes} b_2 \simeq \frac{5.15
  \times 10^{- 13} \hspace{0.4ex} \textrm{metres}}{\left| \chi \left(
  \mathcal{M}^6 \right) \right|^{0.1361}}
\end{equation}
And since, from above, $y_2$, the value of $y$ at the outer surface of the
thick pipe, which is also the ``radius'' of the thick pipe, is approximately
given by $y_2 \simeq 0.254 B$, we find, from (\ref{B for TeV scale gravity
with fluxes}), that for $ \tau = -0.7753 $ and smooth compact quotients of $
\mathbf{C} \mathbf{H}^3$, $y_2$ is approximately given, for TeV-scale gravity,
by:
\begin{equation}
  \label{y sub 2 for TeV scale gravity with fluxes} y_2 \simeq 0.254 B \simeq
  \frac{1.70 \times 10^{- 13} \hspace{0.4ex} \textrm{metres}}{\left| \chi
  \left( \mathcal{M}^6 \right) \right|^{0.1361}}
\end{equation}
Furthermore, from (\ref{G tilde for TeV scale gravity}), (\ref{B sub 3 for TeV
scale gravity with fluxes}), the relation $\tilde{v} = \frac{\tilde{G}}{12 b^2
a^4}$, and the boundary condition (\ref{boundary conditions in terms of v
tilde}) on $\tilde{v}$ at the outer boundary, where $b = b_2$, we find that for
$ \tau = -0.7753 $ and smooth compact quotients of $ \mathbf{CH}^3 $,
$a_2 = a \left( y_2 \right)$, the value of $a \left( y \right)$ at the outer
boundary, is given by:
\begin{equation}
  \label{a sub 2 for TeV scale gravity with fluxes} a_2 = \left(
  \frac{\tilde{G}}{12 b_2^2} \right)^{\frac{1}{4}} \simeq \frac{5.77 \times
  10^{21} \hspace{0.4ex} \textrm{metres}}{\left| \chi \left( \mathcal{M}^6
  \right) \right|^{0.0237}}
\end{equation}
We note that, since $\left| \chi \left( \mathcal{M}^6 \right) \right| \geq 1$,
this is large compared to $b_2$, as assumed near the beginning of this
subsection.
Furthermore, since $a_1 = a \left( y_1 \right)$ is the de Sitter radius
(\ref{de Sitter radius}), and $\left| \chi \left( \mathcal{M}^6 \right)
\right|$ is bounded above by around $7 \times 10^4$, the ratio
$\frac{a_1}{a_2}$ is bounded above by around $3.4 \times 10^4$.  Thus $0.2$
millimetres on the inner surface of the thick pipe corresponds on the outer
surface to a distance no shorter than around $6$ nanometres, so the four
dimensional effective field theory description is certainly valid for
distances down to $0.2$~millimetres, and the realization of TeV-scale gravity
considered in this subsection is for $\tau = - 0.7753$ consistent with the
precision tests of Newton's law at sub-millimetre distances.

Turning now to the flux quantization condition, (\ref{Integral of G for the
factorized ansatz}), we find, from the relation $\tilde{v} =
\frac{\tilde{G}}{12 b^2 a^4}$, and the approximate relation (\ref{improved
estimate of B in terms of B sub 3}) between $b_2 = b \left( y_2 \right)$ and
$B$, that
\begin{equation}
  \label{integral of a to the minus four with fluxes} \int^{y_2}_{y_1} dya
  \left( y \right)^{- 4} \simeq \frac{12}{\tilde{G}} \int^{0.7724 B}_{b_1}
  \frac{db}{c} b^2  \tilde{v}
\end{equation}
Inserting the approximate dependence of $\tilde{v}$ on $c$, in (\ref{estimate
of the dependence of v tilde on c}), and the approximate dependence of $c$ on
$b$, in (\ref{improved estimate of the dependence of c on b}), we find:
\begin{equation}
  \label{approximate integral of a to the minus four with fluxes}
  \int^{y_2}_{y_1} dya \left( y \right)^{- 4} \simeq \frac{12 B^3}{\tilde{G}}
  \int^{0.7724}_{\frac{b_1}{B}} \frac{x^{3.8990} dx}{\sqrt{1 - \left(
  \frac{x}{0.7724} \right)^6}  \left( 1 + \frac{0.4312}{x^{3.7980}}  \left( 1
  - \left( \frac{x}{0.7724} \right)^6 \right) \right)^{0.2899}}
\end{equation}
For small $b$, or equivalently, for large $c$, the integrand in the right-hand
side of (\ref{integral of a to the minus four with fluxes}) behaves as
$\frac{b^2}{c^{1.5798}}$, and thus as $b^5$, so the integral is dominated by
the contribution from the region near the upper limit.  The estimate
(\ref{estimate of the dependence of v tilde on c}) of the dependence of
$\tilde{v}$ on $c$ is accurate near the upper limit, and becomes too large by
about 31 percent at large $c$, and the estimate (\ref{improved estimate of the
dependence of c on b}) of the dependence of $c$ on $b$ is accurate near the
upper limit, and becomes around 6 percent too large at small $b$, so the
integrand in the right-hand side of (\ref{approximate integral of a to the
minus four with fluxes}) is accurate near the upper limit, and too large by
around 22 percent near the lower limit.  Thus we expect (\ref{approximate
integral of a to the minus four with fluxes}) to give a result that is too
large, but by a lot less than 22 percent.  We can again set the lower limit to
zero, since $\frac{b_1}{B} \sim \frac{\kappa^{2/9}}{B}$, and using
again the numerical integration facility of PARI-GP, plus an analytic
approximation near the upper limit, we find:
\begin{equation}
  \label{approximate value of integral of a to the minus four with fluxes}
  \int^{y_2}_{y_1} dya \left( y \right)^{- 4} \simeq 1.0907
  \frac{B^3}{\tilde{G}}
\end{equation}
Thus from the flux quantization condition (\ref{Integral of G for the
factorized ansatz}), the quantity that is required to be an integer, for each
three-cycle, $Z$, of the compact six-manifold, is approximately:
\begin{equation}
  \label{numerical quantized flux} 0.2212 \frac{B^3}{\kappa^{\frac{2}{3}}
  \tilde{G}}  \int_Z dz^A dz^B dz^C G_{ABC} \left( z \right)
\end{equation}
Now comparing with the definition (\ref{definition of the constant G tilde})
of the constant $\tilde{G}$, we see that $\tilde{G}$ cancels out of
(\ref{numerical quantized flux}), which is thus independent of the overall
normalization of $G_{ABC} \left( z \right)$.  Thus the flux quantization
condition (\ref{Integral of G for the factorized ansatz}) does not constrain
the integration constant $A$ in (\ref{bulk power law for a in terms of b}),
the de Sitter radius (\ref{de Sitter radius}), or the effective cosmological
constant in four dimensions, (\ref{Lambda}).  Furthermore, we recall that
$G_{ABC} \left( z \right)$ is a linear combination, with position-independent
coefficients, of the $B_3$ linearly independent Hodge - de Rham harmonic
three-forms on the compact six-manifold $\mathcal{M}^6$, where $B_3$ is the
third Betti number of $\mathcal{M}^6$, that has been assumed to satisfy the
conditions (\ref{first condition on G sub A B C}) and (\ref{second condition
on G sub A B C}), which constitute at most $20 + 20$ linearly independent
constraints on the $B_3$ independent coefficients in $G_{ABC} \left( z
\right)$.

We can always choose a linearly independent set of $B_3$ Hodge - de
Rham harmonic three-forms $g_{ABC}^{\left( i \right)} \left( z \right)$, $1
\leq i \leq B_3$, and a set of $B_3$ three-cycles $Z_{\left( j \right)}$ of
$\mathcal{M}^6$, $1 \leq j \leq B_3$, linearly independent in the sense of
homology, such that $\int_{Z_{\left( j \right)}} dz^A dz^B dz^C g^{\left( i
\right)}_{ABC} \left( z \right) = \delta_{\hspace{0.4ex} \hspace{0.4ex}
\hspace{0.4ex} \hspace{0.4ex} \left( j \right)}^{\left( i \right)}$.  Choosing
a basis of harmonic three-forms and a set of $B_3$ three-cycles that satisfy
this condition, the requirement that (\ref{numerical quantized flux}) be an
integer, for each three-cycle $Z_{\left( j \right)}$, $1 \leq j \leq B_3$,
implies that the $\left( B_3 - 1 \right)$ independent ratios of the
coefficients in $G_{ABC} \left( z \right)$ are rational numbers.  The overall
normalization of the coefficients, which cancels out of (\ref{numerical
quantized flux}), is fixed by (\ref{G tilde for TeV scale gravity}) for
TeV-scale gravity, and by (\ref{G tilde for general kappa to the two ninths})
in general, together with the definition (\ref{definition of the constant G
tilde}) of $\tilde{G}$.  If we now define $\rho_j \equiv \frac{1}{\tilde{G}}
\int_{Z_{\left( j \right)}} dz^A dz^B dz^C G_{ABC} \left( z \right)$, $1 \leq
j \leq B_3$, the requirement that (\ref{numerical quantized flux}) be an
integer for all three-cycles $Z$ of $\mathcal{M}^6$ reduces to the requirement
that $0.2212 \frac{B^3}{\kappa^{\frac{2}{3}}} \rho_j$ be an integer for all $1
\leq j \leq B_3$.

Now the value of $\frac{B}{\kappa^{2/9}}$ has been
assumed to be fixed by the boundary condition at the inner surface of the
thick pipe, with its actual value determined by the Casimir energy densities
on and near the inner surface of the thick pipe, so
$\frac{B}{\kappa^{2/9}}$ would be overconstrained if the flux
quantization conditions significantly restricted its value.  However for
TeV-scale gravity, (\ref{B for TeV scale gravity with fluxes}) implies that
the value of $\frac{B^3}{\kappa^{\frac{2}{3}}}$ is around $10^{16}$, provided
$\left| \chi \left( \mathcal{M}^6 \right) \right|$ is not too large, so
provided none of the nonvanishing $\rho_j$ are too small in magnitude, and the
nonvanishing $\frac{\rho_j}{\rho_k}$ are expressible as ratios of sufficiently
small integers, an alteration of $a \left( y \right)$ by a tiny percentage in
the region near the outer surface, where the alteration would have the
greatest effect on the integral (\ref{integral of a to the minus four with
fluxes}), would be sufficient to satisfy all the flux quantization conditions.

Furthermore we are free to choose the independent ratios of the $\rho_j$, and
thus to set them equal to ratios of small nonvanishing integers, in which case
it seems plausible that the magnitudes of the $\rho_j$ would generally lie
more or less within the range $\frac{1}{B_3}$ to $\frac{1}{\sqrt{B_3}}$.  Thus
provided $\left| \chi \left( \mathcal{M}^6 \right) \right|$ and $B_3$ are not
too large, it seems plausible, at least for the case of TeV-scale gravity,
that the flux quantization conditions, (\ref{Integral of G for the factorized
ansatz}), will not significantly restrict the solutions considered in the
present subsection.

We note that, notwithstanding the large value (\ref{G tilde for TeV scale
gravity}) of $\tilde{G}$ in the case of TeV-scale gravity, and its large value
(\ref{G tilde for general kappa to the two ninths}) in general, the extra
fluxes of the four-form field strength of the three form gauge field
considered in
the present subsection, which wrap three-cycles of the compact six-manifold
times the radial dimension of the thick pipe, never have a large enough field
strength that we would expect them to produce quantum effects.  To estimate
whether we would expect the extra fluxes to produce quantum effects, we note
that we expect quantum gravitational effects when the Ricci scalar has
magnitude $\sim \kappa^{- \frac{4}{9}}$ or larger.  Hence from the
supergravity action (\ref{upstairs bulk action}), we would expect the
four-form field strength $G_{IJKL}$ to produce quantum effects when $G^{IM}
G^{JN} G^{KO} G^{LP} G_{IJKL} G_{MNOP}$ has magnitude $\sim \kappa^{-
\frac{4}{9}}$ or larger.  And, noting that there are no cross terms in $G^{IM}
G^{JN} G^{KO} G^{LP} G_{IJKL} G_{MNOP}$ between the extra fluxes and the
standard Witten fluxes that follow for smooth compact quotients of
$\mathbf{C} \mathbf{H}^3$ from Witten's topological constraint, as studied
in subsections \ref{Wittens topological constraint} and \ref{The three form
gauge field}, we find, from (\ref{lowest harmonic of G 4 sub y G 4 sub y}),
that in the approxiation of dropping all but the leading harmonic, the
contribution to $G^{IM} G^{JN} G^{KO} G^{LP} G_{IJKL} G_{MNOP}$ from the extra
fluxes is given by
\begin{equation}
  \label{kinetic term from the extra fluxes} G^{QR} G^{KN} G^{LO} G^{MP}
  G_{QKLM} G_{RNOP} = \frac{4}{b^6 a^8}  \tilde{G}^2 = 576
  \frac{\tilde{v}^2}{b^2}
\end{equation}
where the relation $\tilde{v} = \frac{\tilde{G}}{12 b^2 a^4}$ was used.  Thus
since $\tilde{v} = 1$ at the outer surface of the thick pipe, by
(\ref{boundary conditions in terms of v tilde}), and $b \sim B$ at the outer
surface of the thick pipe, which for TeV-scale gravity is $\gg
\kappa^{2/9}$ by (\ref{B for TeV scale gravity with fluxes}), unless
$\left| \chi \left( \mathcal{M}^6 \right) \right|$ is extremely large, which
seems very unlikely since it would require a correspondingly small value of
$\frac{b_1}{\kappa^{2/9}}$, by (\ref{b sub 1 in terms of chi}), we see
that for TeV-scale gravity the extra fluxes are not large enough at the outer
surface of the thick pipe that we would expect them to cause quantum effects
there.

Furthermore, from (\ref{bulk power law for a in terms of b}), on page
\pageref{bulk power law for a in terms of b}, $\frac{1}{b^6 a^8}$ behaves as
$b^{0.2024}$ in the bulk power law region, and thus decreases with decreasing
$b$, and from the form of the solutions studied above, neither $\tilde{v}$ nor
$b$ changes significantly in order of magnitude between the power law region
and the outer surface of the thick pipe, hence for TeV-scale gravity
(\ref{kinetic term from the extra fluxes}) is small in magnitude compared to
$\kappa^{- \frac{4}{9}}$ throughout the whole thick pipe.  Thus for TeV-scale
gravity, we do not expect the extra fluxes considered in the present
subsection to produce any significant quantum effects at all, and away from
the inner surface of the thick pipe, the solutions studied in the present
subsection are entirely classical in character.

We note, furthermore, that even though $\tilde{G}$, and the integration
constant, $A$, in (\ref{bulk power law for a in terms of b}), cancel out of
the reduced Einstein equations (\ref{a dot over a in terms of v tilde}) and
(\ref{second Einstein equation in terms of v tilde}) and boundary conditions
(\ref{boundary conditions in terms of v tilde}), and also cancel out of the
contribution of the extra fluxes to \\
$G^{IM} G^{JN} G^{KO} G^{LP} G_{IJKL}
G_{MNOP}$, they are nevertheless physically significant.  For first of all, if
$\tilde{G}$ and $A$ had not been sufficiently large, it would not have been
possible to neglect the term $\frac{4}{a^2}$ in the square root in (\ref{a dot
over a}) and (\ref{second Einstein equation without a dot}), in comparison
with the new term in $\frac{2}{3} \kappa^2 t^{\left( 3 \right)}$ coming from
(\ref{three form energy momentum tensor components for three cycles on the
sixfold}), as explained between (\ref{three form energy momentum tensor
components for three cycles on the sixfold}) and (\ref{bulk coupled a dot over
a with fluxes}), and it would not then have been possible to eliminate
$\tilde{G}$ from the Einstein equations by defining $\tilde{v} =
\frac{\tilde{G}}{12 b^2 a^4}$.  And secondly, from the metric ansatz
(\ref{metric ansatz}), and (\ref{bulk power law for a in terms of b}), the
observed de Sitter radius (\ref{de Sitter radius}), as estimated from
observations of type Ia supernovae {\cite{cosmological constant 1, cosmological
constant 2, cosmological constant 3}}, and
significantly bounded below by a great variety of astronomical observations,
as well as by the approximate flatness of the everyday world, is equal to $A
\left( \frac{\kappa^{2/9}}{b_1} \right)^{0.7753}$, where $b_1 = b
\left( y_1 \right)$ is expected, by (\ref{b sub 1 in terms of chi}), to be
$\sim \kappa^{2/9}$.  It is the large value of $A$, which for the
solutions considered in the present subsection results from the large value of
$\tilde{G}$, that results in the existence of a large and approximately flat
platform at the inner surface of the thick pipe, on which the interesting
processes of intermediate range astronomy, and the everyday world, can take
place.

The value of $\tilde{G}$ is not constrained by the field equations or the
boundary
conditions, since the relevant field equation, (\ref{equation for f in the
factorized ansatz}), is satisfied both in the bulk and on the orbifold
fixed-point hyperplanes, in the upstairs picture, in consequence of the
Ho\v{r}ava-Witten orbifold conditions, as summarized after (\ref{upstairs bulk
action}), which imply that the components $G_{yUVW}$ are even under
reflections in the orbifold hyperplanes.  It seems that the existence of the
arbitrary constant $\tilde{G}$, defined in (\ref{definition of the constant G
tilde}), in compactifications of Ho\v{r}ava-Witten theory of the type studied in
the present paper, is implicit in the field content of supergravity in eleven
dimensions {\cite{Nahm, Cremmer Julia Scherk}}, and the boundary conditions or
orbifold conditions of Ho\v{r}ava-Witten theory, as summarized after
(\ref{upstairs bulk action}).  The question of how $\tilde{G}$ came to have
the large value required to fit the observed small value of the cosmological
constant, and the related question of whether $\tilde{G}$ has any effects on
the dynamics of the early universe, other than preventing the occurrence of a
large cosmological constant, in models of this type, will not be considered in
the present paper.

Finally we should check whether the solutions considered in the present
subsection are consistent with the precision tests of Newton's law down to
submillimetre distances, and the very high precision tests of Newton's law
over solar system distances, as carried out for the solutions of subsection
\ref{Solutions with a as small as kappa to the two ninths at the outer
surface} in subsection \ref{Newtons constant and the cosmological constant},
starting shortly after equation (\ref{A tilde for general kappa to the two
ninths}), on page \pageref{A tilde for general kappa to the two ninths}.
However, it does not seem very likely that the constraints from these tests
will be more stringent for the solutions considered in the present subsection
than for the solutions of subsection \ref{Solutions with a as small as kappa
to the two ninths at the outer surface}, and this will not be considered in
detail in the present paper.

The fact that the value of $\tilde{G}$ does seem to be quantized seems to
suggest that if the metric ansatz (\ref{metric ansatz}) is generalized to a
cosmological ansatz of the form $ds^2_{11} = - a \left( y \right)^2 dt^2 +
\tilde{a} \left( y, t \right)^2 g_{ij} dx^i dx^j + \tilde{b} \left( y, t
\right)^2 h_{AB} dx^A dx^B + \tilde{c} \left( y, t \right)^2 dy^2$, which is
consistent with (\ref{metric ansatz}) if the metric on the four-dimensional de
Sitter space, in (\ref{metric ansatz}), is taken in Friedmann-Robertson-Walker
form, a parameter related to $\tilde{G}$ might evolve with time, as in
quintessence models {\cite{Steinhardt quintessence}}.  This could perhaps be
investigated by studying cosmological solutions that are small perturbations
of the de Sitter solutions studied in this section.

\section{Smooth compact quotients of $ \mathbf{CH}^3 $, $ \mathbf{H}^6 $,
$ \mathbf{H}^3 $ and $ \mathbf{S}^3 $}
\label{Smooth compact quotients of CH3 H6 H3 and S3}

For the compactifications of Ho\v{r}ava-Witten theory considered in the present
paper, the compact six-manifold, $\mathcal{M}^6$, is a smooth compact quotient
of either the symmetric space $\mathbf{C} \mathbf{H}^3$, or the symmetric
space $\mathbf{H}^6$, by a discrete subgroup of the isometry group of the
symmetric space, and for the solutions considered in subsection \ref{Solutions
with a as small as kappa to the two ninths at the outer surface}, on page
\pageref{Solutions with a as small as kappa to the two ninths at the outer
surface}, the three observed spatial dimensions are also compactified to
a smooth compact quotient of either the symmetric space $\mathbf{H}^3$, or
the symmetric space $\mathbf{S}^3$, by a discrete subgroup of the symmetric
space.  I shall first consider smooth compact quotients of the non-compact
symmetric spaces $\mathbf{C} \mathbf{H}^3$, $\mathbf{H}^6$, and
$\mathbf{H}^3$, then briefly consider smooth compact quotients of
$\mathbf{S}^3$, at the end of this section.

Let $G$ be the identity component, or in other words, the connected component
that contains the identity, of either SU($n$,1), for $n \geq 1$, or SO($n$,1),
for $n \geq 2$, and let $K$ be the maximal compact subgroup of $G$, which is
$\mathrm{SU} \left( n \right) \times U \left( 1 \right)$ for SU($n$,1), and
SO($n$), for SO($n$,1).  Then the non-compact symmetric space, $\mathcal{S}
\equiv G / K$, is $\mathbf{C} \mathbf{H}^n$ for SU($n$,1), and
$\mathbf{H}^n$ for SO($n$,1).  I shall assume that the metric of
$\mathbf{C} \mathbf{H}^n$ is normalized such that the Riemann tensor is
given by (\ref{CHn covariant Riemann tensor}), in complex coordinates, and
that the metric of $\mathbf{H}^n$ is normalized such that the Riemann tensor
is given by $R_{\mu \nu \sigma \tau} = g_{\mu \sigma} g_{\nu \tau} - g_{\mu
\tau} g_{\nu \sigma}$, consistent with the choice made between (\ref{Lovelock
Gauss Bonnet t i j}) and (\ref{scalars formed from R tilde for H6}), and
between (\ref{volume in terms of Euler number for a smooth compact quotient of
CH3}) and (\ref{volume in terms of Euler number for a smooth compact quotient
of H6}), so that the sectional curvature (\ref{definition of sectional
curvature}) of $\mathbf{H}^n$ is equal to $- 1$, which is the conventional
value.

We choose a point of $\mathcal{S}$, called $O$, to be the origin of
$\mathcal{S}$.  For example, for $\mathbf{C} \mathbf{H}^n$, we could
choose $O$ to be the origin of the coordinates used for $\mathbf{C}
\mathbf{H}^n$, in subsection \ref{CH3}.  For
any subgroup, $H$, of $G$, let $\mathcal{C} \left( H \right)$ denote the set
of all the set of all the images of $O$, by the action of elements of $H$.  A
discrete subgroup, $\Gamma$, of $G$, is a subgroup of $G$ such that there is a
real number $\rho > 0$, such that for all members $x$ of $\mathcal{C} \left(
\Gamma \right)$ different from $O$, the geodesic distance from $O$ to $x$ is
$\geq \rho$.  For any discrete subgroup, $\Gamma$, of $G$, and any member,
$x$, of $\mathcal{C} \left( \Gamma \right)$, the Wigner-Seitz cell, or Voronoi
cell, $\mathcal{W} \left( \Gamma, x \right)$, is the set of all points of
$\mathcal{S}$, that are closer to $x$, than to any other member of
$\mathcal{C} \left( \Gamma \right)$.  The fundamental domain of the quotient
$\mathcal{S}/ \Gamma$ is $\mathcal{W} \left( \Gamma, O \right)$.  $\Gamma$ is
called a lattice in $G$, if $\mathcal{W} \left( \Gamma, O \right)$ has finite
volume.  When $\Gamma$ is a lattice in $G$, the set $\mathcal{C} \left( \Gamma
\right)$, of all the images of $O$ in $\mathcal{S}$, looks like a hyperbolic
analogue of a crystal lattice.

$G$ always has an infinite family of lattices called ``arithmetic lattices'',
whose existence was demonstrated by Borel and Harish-Chandra {\cite{Borel
Harish-Chandra}}.  A very helpful introduction to arithmetic lattices has been
provided by Morris {\cite{Morris}}.  Some examples of arithmetic lattices in
$G$ are considered in subsection \ref{Smooth compact arithmetic quotients of
CHn and Hn}, on page \pageref{Smooth compact arithmetic quotients of CHn and
Hn}.  For SO($n$,1), SU(2,1), and SU(3,1), there also exist additional
lattices called ``non-arithmetic lattices''.  Non-arithmetic lattices in
SO($n$,1), for $n \leq 5$, were constructed by Makarov and Vinberg
{\cite{Makarov, Vinberg}}, and non-arithmetic lattices in SO($n$,1), for all
$n$, were constructed by Gromov and Piatetski-Shapiro {\cite{Gromov
Piatetski-Shapiro}}.  The construction of Gromov and Piatetski-Shapiro
involves cutting two different quotients of $\mathcal{S}$ into two parts along
totally geodesic $n$-dimensional submanifolds, and smoothly joining together
one part from each of the two different quotients.  There is no analogous
construction for SU($n$,1) for $n \geq 2$, because for $n \geq 2$,
$\mathbf{C} \mathbf{H}^n$ has no totally geodesic $\left( 2 n - 1
\right)$-dimensional submanifolds {\cite{Goldman}}.  Non-arithmetic lattices
in SU(2,1) were constructed by Mostow {\cite{Mostow SU 2 1}}, and
non-arithmetic lattices in SU(3,1) were constructed by Deligne and Mostow
{\cite{Deligne Mostow}}.  To the best of my knowledge, it is not yet known
whether or not any non-arithmetic lattices exist for SU($n$,1), $n \geq 4$.

For the models considered in the present paper, I assume that the quotient
$\mathcal{S}/ \Gamma$ is a smooth manifold, not an orbifold, so $\Gamma$ is
required to act on $\mathcal{S}$ without fixed points, or in other words, no
non-trivial element of $\Gamma$ is allowed to leave any point of $\mathcal{S}$
invariant.  The fact that $\mathcal{S}$ is the quotient of $G$, by its maximal
compact subgroup, $K$, implies that a necessary condition for $\Gamma$ to act
without fixed points, is that $\Gamma$ must have no torsion, in the sense of
discrete group theory, or in other words, $\Gamma$ must not contain any
element $g \neq 1$, such that $g^n = 1$, for some finite $n$.  For the finite
group generated by such a $g$ is a compact subgroup of $G$, and every compact
subgroup of $G$ is contained in a maximal compact subgroup, and all maximal
compact subgroups of $G$ are conjugate.  Furthermore, $K$ is the subgroup of
$G$ that leaves $O$ invariant, and the conjugate $hKh^{- 1}$ of $K$, where $h$
is a fixed element of $G$, leaves the point $hO$ invariant.  Conversely, the
requirement that $\Gamma$ have no torsion is also sufficient to ensure that
$\Gamma$ acts on $\mathcal{S}$ without fixed points.  For suppose an element
$g \neq 1$ of $\Gamma$ leaves a point $x$ of $\mathcal{S}$ invariant.  Then
since $\Gamma$ is an isometry, and maps members of $\mathcal{C} \left( \Gamma
\right)$ to members of $\mathcal{C} \left( \Gamma \right)$, $g$ must permute
the members of $\mathcal{C} \left( \Gamma \right)$, at any given fixed
distance from $x$, among themselves.  In particular, $g$ must permute the
nearest neighbours of $x$, in $\mathcal{C} \left( \Gamma \right)$, amongst
themselves.  But the discreteness of $\Gamma$ implies that the number of
nearest neighbours of $x$, in $\mathcal{C} \left( \Gamma \right)$, is finite.
Hence $g$ is an element of a finite group, hence $g^n = 1$, for some finite
$n$.

Now let $\Gamma$ be a lattice in $G$, that acts without fixed points on
$\mathcal{S}$, so that the quotient $\mathcal{S}/ \Gamma$ is a smooth
manifold, of finite volume.  Let $d$ denote the real dimension of
$\mathcal{S}$, which is $2 n$ for SU($n$,1), and $n$ for SO($n$,1).  Then for
$d \geq 3$, Mostow's rigidity theorem {\cite{Mostow 68, Prasad,
Mostow, Thurston MSRI, Gromov}} implies that the locally symmetric space
$\mathcal{S}/ \Gamma$ is completely determined, up to isometry, by its
fundamental group, which is $\Gamma$.  This result is not true for $d = 2$,
since smooth compact quotients of $\mathbf{C} \mathbf{H}^1$, which differs
from $\mathbf{H}^2$ only in the normalization of its metric, in the
conventions adopted here, have shape moduli, as is well known in superstring
theory.  An orientable smooth compact quotient of $\mathbf{C}
\mathbf{H}^1$ of genus $g \geq 2$, which is topologically equivalent to a
sphere with $g$ handles, has a moduli space of dimension $6 g - 6$.  The
moduli are called Teichm"uller parameters, and are the minimum number of
parameters needed to characterize conformally inequivalent closed Riemann
surfaces.  They correspond to the positions and radii of six circles in the
complex plane, which are identified in pairs to produce the closed surface,
less six parameters that relate conformally equivalent surfaces.

Mostow's rigidity theorem implies, in particular, that for $d \geq 3$, the
volume, $\mathcal{V} \left( \mathcal{S}/ \Gamma \right)$, is a topological
invariant.  For $d$ even, $\mathcal{V} \left( \mathcal{S}/ \Gamma \right)$ is
a fixed multiple of the Euler number, given for $d = 6$ by (\ref{volume in
terms of Euler number for a smooth compact quotient of CH3}) for smooth
compact quotients of $\mathbf{C} \mathbf{H}^3$, and by (\ref{volume in
terms of Euler number for a smooth compact quotient of H6}) for smooth compact
quotients of $\mathbf{H}^6$, but for $d$ odd, the Euler number is zero, and,
at least for $d = 3$, there is no corresponding restriction on the possible
values of $\mathcal{V} \left( \mathcal{S}/ \Gamma \right)$.

If there is a finite upper bound, on the geodesic distance between pairs of
points in the fundamental domain $\mathcal{W} \left( \Gamma, O \right)$, then
the quotient $\mathcal{S}/ \Gamma$ is compact, while if, for any given finite
distance, there exist pairs of points, in the fundamental domain, such that
the geodesic distance between them is greater than that given distance, then
the quotient is non-compact.  In the coordinate system used for $\mathbf{C}
\mathbf{H}^n$, in subsection \ref{CH3}, a
quotient of $\mathbf{C} \mathbf{H}^n$ is compact, if all points in the
closure of its fundamental domain have $z^r z^{\bar{r}} < 1$, and non-compact
if the closure of its fundamental domain has one or more vertices on the
sphere $z^r z^{\bar{r}} = 1$.  If a quotient is non-compact, then the
non-compactness is associated with a finite number of tubular regions, called
cusps, which extend out to infinite distances, but become narrow so rapidly,
that their contribution to the total volume is finite.

Inspection of the list in section (14.4) of {\cite{Deligne Mostow}} shows that
the non-arithmetic quotients of $\mathbf{C} \mathbf{H}^3$ found by Deligne
and Mostow, which correspond to the lattice in $\mathrm{SU} \left( 3, 1 \right)$
denoted $753333$ in their notation, are not compact.  This same lattice is
also the only non-arithmetic lattice in $\mathrm{SU} \left( 3, 1 \right)$ listed
in the Appendix in {\cite{Thurston Shapes of polyhedra}}, where it occurs as
No. 66 in the list.  Thus to the best of my knowledge, it is at present not
known whether or not any compact non-arithmetic quotients of $\mathbf{C}
\mathbf{H}^3$ exist.  However, the non-arithmetic quotients of $\mathbf{C}
\mathbf{H}^2$ found by Mostow in {\cite{Mostow SU 2 1}} are compact.

For the models considered in the present paper, I assume that the quotient is
compact.  However, smooth non-compact finite volume quotients of
$\mathbf{H}^3$ are important, because by a construction of Thurston
{\cite{Thurston MSRI, Thurston 82, Thurston 86, Thurston}}, there exist
infinite sequences of smooth compact quotients of $\mathbf{H}^3$, all with
distinct topology, whose volumes converge to the volumes of smooth non-compact
finite volume quotients of $\mathbf{H}^3$.  This cannot happen in any
dimension larger than $3$, because by a theorem of Wang {\cite{Wang}}, the
number of topologically distinct smooth finite volume quotients of a
non-compact symmetric space of dimension $\geq 4$, whose volume $\mathcal{V}
\left( \mathcal{S}/ \Gamma \right)$ is less than a given volume, is finite.
Moreover, Borel {\cite{Borel arithmetic quotients of H3}} demonstrated that
the number of topologically distinct smooth compact arithmetic quotients of
$\mathbf{H}^3$, whose volume is less than a given volume, is finite, so all
but a finite number of the smooth compact quotients of $\mathbf{H}^3$
resulting from Thurston's construction, whose volume is less than a given
volume, are non-arithmetic.  Each cusp of a non-compact finite volume quotient
of $\mathbf{H}^3$ is topologically equivalent to the Cartesian product of a
two-torus and the infinite half-line.

Thurston's construction makes use of a method of modifying the topology of a
three-manifold $\mathcal{M}^3$, called Dehn surgery {\cite{Wikipedia Dehn
surgery, Wikipedia Hyperbolic Dehn surgery, Springer Dehn surgery}}.  A Dehn
surgery consists of removing a tubular neighbourhood $N$ of an
$\mathbf{S}^1$ embedded in the manifold, then putting it back, in a twisted
fashion.  The surface of $N$ is a two-torus, and twists can be introduced in
two independent senses.  We choose two oriented simple closed curves $m$ and
$l$, called the meridian and the longitude, embedded in the common boundary
torus of $N$ and its complement, that generate the fundamental group of that
torus.  When $\mathcal{M}^3$ is $\mathbf{S}^3$, $l$ is chosen such that it
bounds a surface in the complement of $N$, and $m$ is chosen such that it
crosses $l$ exactly once.  This gives any oriented simple closed curve $c$ on
that torus two coordinates $p$ and $q$, which correspond to the net number of
times $c$ crosses $m$ and $l$ respectively. These coordinates depend only on
the homotopy class of $c$.  The Dehn surgery with slope $u = \frac{p}{q}$,
where $p$ and $q$ are coprime integers, then corresponds to gluing back $N$ by
means of a homeomorphism of its two-torus boundary to the two-torus boundary
of its complement, such that the meridian curve of the boundary of $N$ maps to
a $\left( p, q \right)$ curve in the boundary of its complement.  By a theorem
of Lickorish {\cite{Lickorish}} and Wallace {\cite{Wallace}}, every closed,
orientable, connected three-manifold can be obtained from $\mathbf{S}^3$ by
Dehn surgery around finitely many copies of $\mathbf{S}^1$ embedded
disjointly in $\mathbf{S}^3$.

Now let $\mathcal{M}^3$ a smooth non-compact finite volume quotient of
$\mathbf{H}^3$ with $n$ cusps, where $n \geq 1$.  Then because each cusp of
$\mathcal{M}^3$ is topologically equivalent to the Cartesian product of a
two-torus and the infinite half-line, $\mathcal{M}^3$ is topologically
equivalent to the interior of a compact three-manifold with $n$ connected
boundary components, each of which is topologically equivalent to a two-torus.
We choose a meridian and longitude for each boundary torus, as in the case of
Dehn surgery. Let $\mathcal{M}^3 ( u_1, u_2, \ldots, u_n )$ denote the
manifold obtained from $\mathcal{M}^3$ by filling in each boundary two-torus
with a solid torus using the slopes $u_i = p_i / q_i$, where each pair $p_i$
and $q_i$ are coprime integers.  This is called Dehn filling.  Thurston's
hyperbolic Dehn surgery theorem then states that $\mathcal{M}^3 ( u_1, u_2,
\ldots, u_n )$ is topologically equivalent to a smooth compact quotient
$\mathbf{H}^3 / \Gamma \left( u_1, u_2, \ldots, u_n \right)$ of
$\mathbf{H}^3$, provided a finite set $E_i$ of slopes is avoided for each
$i$.  Furthermore, $\mathcal{V} \left( \mathbf{H}^3 / \Gamma \left( u_1,
u_2, \ldots, u_n \right) \right) <\mathcal{V} \left( \mathcal{M}^3 \right)$,
and the volumes $\mathcal{V} \left( \mathbf{H}^3 / \Gamma \left( u_1, u_2,
\ldots, u_n \right) \right)$ converge to $\mathcal{V} \left( \mathcal{M}^3
\right)$ as all $p_i^2 + q_i^2 \rightarrow \infty$, $p_i \neq 0$, $q_i \neq
0$.  It is also known that only a finite number of topologically distinct
smooth compact quotients of $\mathbf{H}^3$ with any given volume exist
{\cite{Thurston 82}}.

Many smooth non-compact finite volume quotients of $\mathbf{H}^3$ have been
discovered by studying the complement of disjoint tubular neighbourhoods of
finitely many copies of $\mathbf{S}^1$ embedded disjointly in
$\mathbf{S}^3$, to see if it can be constructed by gluing together a small
number of hyperbolic tetrahedra, some of whose vertices stretch out to
infinity as parts of cusps.  A hyperbolic polyhedron, one or more of whose
vertices stretches out to infinity as part of a cusp, is called an ideal
hyperbolic polyhedron.  This method was originally applied by Thurston to show
that the complement of the figure of eight knot was hyperbolic, by
constructing it by gluing together two ideal hyperbolic tetrahedra.  It had
earlier been shown to be hyperbolic by Riley, and by Jorgensen, using other
methods.

Weeks's computer program SnapPea {\cite{Weeks SnapPea}}, which can
perform Dehn surgeries automatically, includes a census of smooth non-compact
finite volume quotients of $\mathbf{H}^3$ constructed by gluing together up
to seven hyperbolic tetrahedra.  The smooth non-compact finite volume quotient
of $\mathbf{H}^3$ that is topologically equivalent to the complement of the
figure of eight knot is designated m004 in SnapPea, and has volume $2.02988
\ldots$.  This has been shown by Cao and Meyerhoff {\cite{Cao Meyerhoff}} to
be the smallest possible volume of an orientable cusped hyperbolic
three-manifold.  There is one other known smooth non-compact finite volume
quotient of $\mathbf{H}^3$ with this volume, which is designated m003 in
SnapPea, and can be obtained by a $\left( 5, 1 \right)$ Dehn filling on the
complement of the Whitehead link.  The Whitehead link is a disjoint embedding
of two copies of $\mathbf{S}^1$ in $\mathbf{S}^3$, such that neither
$\mathbf{S}^1$ is knotted by itself, but the two copies of $\mathbf{S}^1$
are linked such that one $\mathbf{S}^1$ has a loose twist to resemble a
figure of eight, and the other $\mathbf{S}^1$ links both loops of the figure
of eight.

The smooth compact quotient of $\mathbf{H}^3$ of smallest known
volume is called the Weeks manifold or the Fomenko-Matveev-Weeks manifold
{\cite{Weeks thesis, Matveev Fomenko}}, and can be obtained by a (5,2) Dehn
filling on m003 or by a $\left( 3, - 1 \right)$ Dehn filling on m003, and has
volume $0.9427 \ldots$.  The smooth compact quotient of $\mathbf{H}^3$ of
second smallest known volume is called the Thurston manifold, and can be
obtained by a $\left( - 2, 3 \right)$ Dehn filling on m003, and has volume
$0.9814 \ldots$.

By Thurston's hyperbolic Dehn surgery theorem, there are already an infinite
number of topologically distinct smooth compact quotients of $\mathbf{H}^3$
with volume less than the volume $2.02988 \ldots$ of m003 and m004, while as
noted above, when the dimension $d$ of $\mathcal{S}$ is $\geq 4$, the number
of topologically distinct smooth finite volume quotients of $\mathcal{S}$,
whose volume $\mathcal{V} \left( \mathcal{S}/ \Gamma \right)$ is less than a
given volume, is finite.  Let $\rho_{\mathcal{S}} \left( v \right)$ denote the
number of topologically distinct smooth finite volume quotients of
$\mathcal{S}$, whose volume $\mathcal{V} \left( \mathcal{S}/ \Gamma \right)$
is less than $v$.  Then Gelander {\cite{Gelander}} has proved that when the
dimension $d$ of $\mathcal{S}$ is $\geq 4$, there is a constant $c$, depending
on $\mathcal{S}$, such that
\begin{equation}
  \label{Gelanders bound} \log \rho_{\mathcal{S}} \left( v \right) \leq cv
  \log v
\end{equation}
for all $v > 0$.  And for $\mathbf{H}^n$, $n \geq 4$, Burger, Gelander,
Lubotzky, and Mozes (BGLM) {\cite{Burger Gelander Lubotzky Mozes}} have proved
that there exist constants $c_n > b_n > 0$ and $v_n > 0$, such that
\begin{equation}
  \label{Burger Gelander Lubotzky and Mozes bounds} b_n v \log v \leq \log
  \rho_{\mathbf{H}^n} \left( v \right) \leq c_n v \log v
\end{equation}
whenever $v > v_n$.

Thus the number of topologically distinct smooth finite volume quotients of
$\mathbf{H}^6$ with $\left| \chi \left( \mathcal{M}^6 \right) \right| < n$
grows as $n^{cn}$ for sufficiently large $n$, where $c$ is a constant $> 0$.
Furthermore, for both the smooth finite volume quotients of $\mathbf{H}^3$
obtained by Thurston's construction, and for the arithmetic quotients
considered in the following subsection, the vast majority of the smooth finite
volume quotients are in fact compact, so it seems likely that the number of
topologically distinct smooth compact quotients of $\mathbf{H}^6$ with
$\left| \chi \left( \mathcal{M}^6 \right) \right| < n$ also grows as $n^{cn}$
for sufficiently large $n$, with the same constant $c > 0$.

However, since supergravity in eleven dimensions does not contain any
Yang-Mills fields, and the three-form gauge field only enters the
supercovariant derivative on the gravitino through its three-form field
strength, which is well defined globally, there is no
possibility of introducing an analogue of a spin$^c$ structure to compensate
for the compact six-manifold $\mathcal{M}^6$ not being a spin manifold, so
$\mathcal{M}^6$ is required to be a spin manifold.
To the best of my knowledge, none of the known examples of smooth compact
quotients of $\mathbf{C} \mathbf{H}^3$ or $\mathbf{H}^6$ have yet been
shown to be spin manifolds.

For smooth compact quotients of $\mathbf{C}
\mathbf{H}^3$, the fraction of smooth compact quotients that are spin would
be expected to be $\sim 2^{- B_2}$, where $B_2$ is the second Betti number of
the quotient, since the second Steifel-Whitney class is the mod 2 reduction of
the first Chern class, hence the vanishing of the second Steifel-Whitney class
requires that $h_{11} \sim B_2$ integers be even.  Now by a theorem of Gromov
\cite{Gromov Bound on Betti numbers}, there is a number $ \beta $ such that
for all smooth compact quotients $ \mathcal{M}^6 $ of $ \mathbf{CH}^3 $ or
$ \mathbf{H}^6 $ all the Betti numbers of $ \mathcal{M}^6 $ are bounded by
$\beta \left| \chi \left( \mathcal{M}^6 \right) \right|$.  Thus if we suppose
that the fraction of the smooth compact quotients of $\mathbf{H}^6$ that are
spin is also $\sim 2^{- B_2}$, then the result of Burger, Gelander, Lubotzky,
and Mozes stated above implies that for sufficiently large $\left| \chi \left(
\mathcal{M}^6 \right) \right|$, there will be smooth compact quotients $
\mathcal{M}^6 $ of $ \mathbf{H}^6 $ that are spin.

It is known that
all smooth compact orientable manifolds of dimension $\leq 3$ are spin
{\cite{Wikipedia Spin structure}}, and the Davis manifold {\cite{Davis
manifold}}, which is the simplest known smooth compact quotient of
$\mathbf{H}^4$, has been shown to be a spin manifold {\cite{Ratcliffe
Tschantz}}.

We note
that most of the smooth compact hyperbolic threefolds associated by Thurston's
construction with a given smooth non-compact finite volume quotient of
$\mathbf{H}^3$ will be highly inhomogeneous.  In fact, by a theorem
of Cheeger \cite{Cheeger}, summarized recently in \cite{Acharya Douglas},
if a sequence of Riemannian manifolds is such that there is a fixed upper
bound on the magnitudes of all sectional curvatures of all manifolds in the
sequence, a fixed lower bound $ > 0 $ on the volumes of all manifolds in
the sequence, and a fixed upper bound on the diameters of all manifolds in
the sequence, then the sequence contains only a finite number of diffeomorphism
types.  Thus since the members of a Thurston infinite sequence of
topologically distinct smooth compact quotients of $ \mathbf{H}^3 $, whose
volumes tend to the volume of a finite volume cusped hyperbolic threefold,
satisfy the first two requirements of Cheeger's theorem, they must violate the
third requirement, which means that the diameters of the members of the
sequence must increase without limit.  This suggests that the members of such
a Thurston sequence develop longer and longer spikes, that approximate more
and more closely to the infinite cusps of the finite volume cusped hyperbolic
threefold, and that, moreover, the differences in topology between the members
of the sequence might become localized further and further out along these
spikes, where the spikes become narrower and narrower.

\subsection{Smooth compact arithmetic quotients of $\mathbf{C}
\mathbf{H}^n$ and $\mathbf{H}^n$}

\label{Smooth compact arithmetic quotients of CHn and Hn}

I shall now outline the construction of some smooth compact arithmetic
quotients of $\mathbf{C} \mathbf{H}^n$ and $\mathbf{H}^n$.  The first
step is to construct some cocompact arithmetic lattices in SU($n$,1) and
SO($n$,1), whose existence follows from section 12 of {\cite{Borel
Harish-Chandra}}.  I shall then briefly review Selberg's lemma
{\cite{Selberg}} for the case of these arithmetic lattices, which states that
certain finite index subgroups of these discrete groups are torsion-free, or
in other words, have no nontrivial finite subgroups.  A subgroup $\Gamma_1$ of
a discrete group $\Gamma$ is said to have finite index in $\Gamma$, if
$\Gamma_1$ divides $\Gamma$ into a finite number of left cosets.

We recall that an algebraic number field {\cite{Wikipedia algebraic number
field}} is a finite-dimensional (and therefore algebraic) field extension of
the field $\mathbf{Q}$ of rational numbers. That is, it is a field which
contains $\mathbf{Q}$ and has finite dimension, or degree, when considered
as a vector space over $\mathbf{Q}$.  To form an algebraic number field, we
recall that for any field $F$, the ring of polynomials with coefficients in
$F$ is denoted by $F [ x ]$. A polynomial $p ( x )$ in $F [ x ]$ is called
irreducible over $F$ {\cite{Wikipedia irreducible polynomial}} if it is
non-constant and cannot be represented as the product of two or more
non-constant polynomials from $F [ x ]$.  Then if $\alpha$ is a root of some
irreducible polynomial $f \left( x \right)$ in $F \left[ x \right]$, the
extension field $F \left( \alpha \right)$ is the set of all polynomials $g
\left( \alpha \right)$, with two polynomials $g \left( \alpha \right)$ and $h
\left( \alpha \right)$ being defined to be equal, if $f \left( \alpha \right)
= 0$ implies $g \left( \alpha \right) = h \left( \alpha \right)$.  In practice
this means that, if $f \left( x \right)$ is of degree $m$, every polynomial $g
\left( \alpha \right)$, of degree $\geq m$, is equal to some polynomial of
degree $< m$.  The field extension $F \left( \alpha \right)$ will then be of
degree $m$, as a vector space over the field $F$, and a possible basis for $F
\left( \alpha \right)$ is the set of monomials $1, \alpha, \alpha^2, \ldots,
\alpha^{m - 1}$.  If an element $a \neq 0$ of $F \left( \alpha \right)$ is
expressed in this basis as $a = a_1 + a_2 \alpha + a_3 \alpha^2 + \ldots + a_m
\alpha^{m - 1}$, where the coefficients $a_i$ are elements of $F$, then the
reciprocal of $a$, expressed in this basis as $a^{- 1} = b_1 + b_2 \alpha +
b_3 \alpha^2 + \ldots + b_m \alpha^{m - 1}$, where the coefficients $b_i$ are
elements of $F$, can be found by solving the $2 m - 1$ linear equations, that
result from equating coefficients of all powers of $\alpha$, up to and
including $\alpha^{2 m - 2}$, in the equation
\[ \left( a_1 + a_2 \alpha + a_3 \alpha^2 + \ldots + a_m \alpha^{m - 1}
   \right) \left( b_1 + b_2 \alpha + b_3 \alpha^2 + \ldots + b_m \alpha^{m -
   1} \right) = \]
\begin{equation}
  \label{equation for the reciprocal of an element of an extension field} = 1
  + \left( c_1 + c_2 \alpha + c_3 \alpha^2 + \ldots + c_{m - 1} \alpha^{m - 2}
  \right) f \left( \alpha \right)
\end{equation}
for the $m$ coefficients $b_i$, and the $m - 1$ coefficients $c_i$, which are
also elements of $F$.  The extension field $F \left( \alpha \right)$ is
sometimes written $F \left( \left\{ \alpha \right\} \right)$, to allow for the
possibility of adjoining more than one new element to $F$.  In general, if $S$
is a set of elements not in $F$, the extension field $F \left( S \right)$ is
the smallest field that contains $F$ and $S$.

We recall, also, that an algebraic number is a root of a polynomial with
integer coefficients.  For any algebraic number, $\alpha$, there is a unique
polynomial $f \left( x \right)$ in $\mathbf{Q} \left[ x \right]$, such that
$f \left( x \right)$ is irreducible over $\mathbf{Q}$, the coefficient of
the highest power of $x$ in $f \left( x \right)$ is equal to $1$, and $\alpha$
is a root of $f \left( x \right)$.  This is called the minimal polynomial of
$\alpha$, and the degree of this polynomial is called the degree of $\alpha$.
Every polynomial $g \left( x \right)$ in $\mathbf{Q} \left[ x \right]$, such
that $g \left( \alpha \right) = 0$, is a multiple of the minimal polynomial of
$\alpha$.  The roots of the minimal polynomial of $\alpha$, including $\alpha$
itself, are called the {\emph{conjugates}} of $\alpha$, and are all distinct.

By the primitive element theorem {\cite{Wikipedia primitive element theorem}},
every algebraic number field $F$ is of the form $\mathbf{Q} \left( \alpha
\right)$, where $\alpha$ is a root of a polynomial $f \left( x \right)$ in
$\mathbf{Q} \left[ x \right]$, such that $f \left( x \right)$ is irreducible
over $\mathbf{Q}$.  An element $\alpha$ of $F$, such that $F$ is generated
by adjoining $\alpha$ to $F$, is called a {\emph{primitive element}} of $F$.
A primitive element of $F$ can also be characterized by the fact that it does
not belong to any proper subfield of $F$, and it can also be characterized by
the fact that the degree of its minimal polynomial is equal to the degree of
$F$.  An algebraic number field $F$ has only a finite number of subfields $K$
such that $\mathbf{Q} \subseteq K \subseteq F$, and since these correspond
to subspaces of $F$ as a vector space over $\mathbf{Q}$, most elements of
$F$ are in fact primitive elements.

Considering the field $F =\mathbf{Q} \left( \left\{ \sqrt{2}, \sqrt{3}
\right\} \right)$, for example, neither $\sqrt{2}$ nor $\sqrt{3}$ is a
primitive element of $F$, since they are respectively elements of the
subfields $\mathbf{Q} \left( \sqrt{2} \right)$ and $\mathbf{Q} \left(
\sqrt{3} \right)$.  We cannot form $F$ by adjoining to $\mathbf{Q}$ a root
of the polynomial $x^4 - 5 x^2 + 6$, whose roots are $\pm \sqrt{2}$ and $\pm
\sqrt{3}$, because this polynomial factors as $\left( x^2 - 2 \right) \left(
x^2 - 3 \right)$, and is thus not irreducible over $\mathbf{Q}$.  The
\verb#nfinit# function of PARI/GP {\cite{PARI GP}}, for example, simply
rejects an attempt to form an algebraic number field with this polynomial.
However $\alpha = \sqrt{2} + \sqrt{3}$, whose minimal polynomial is $x^4 - 10
x^2 + 1$, is a primitive element of $F$, and, indeed, we have
$\frac{\alpha^3}{2} - \frac{9 \alpha}{2} = \sqrt{2}$, and $-
\frac{\alpha^3}{2} + \frac{11 \alpha}{2} = \sqrt{3}$.  The conjugates of
$\alpha$ are $\pm \sqrt{2} \pm \sqrt{3}$ with all four sign choices allowed.

If $F$ is an algebraic number field of degree $m$ over $\mathbf{Q}$, and
$v_i$, $1 \leq i \leq m$, is a basis for $F$, as a vector space over
$\mathbf{Q}$, then we may associate to each element $x$ of $F$, a square
matrix $x_{ij}$ with rational elements, defined by $xv_i = v_j x_{ji}$, where
the summation convention is used.  We then find, for any elements $x$ and $y$
of $F$, that $xyv_i = xv_j y_{ji} = v_j x_{jk} y_{ki}$.  Thus the matrices
$x_{ij}$ form a matrix representation of the elements of $F$, called the
regular representation for the basis given by the $v_i$.  For example, for the
field $\mathbf{Q} \left( \sqrt{2} \right)$, we find that $1$ is represented
by $\left(\begin{array}{cc}
  1 & 0\\
  0 & 1
\end{array}\right)$, and $\sqrt{2}$ is represented by $\left(\begin{array}{cc}
  0 & 2\\
  1 & 0
\end{array}\right)$.  Invariants of the matrix $x_{ij}$ representing an
element $x$ of $F$, such as its trace, determinant, and characteristic
polynomial, are properties of $x$, and do not depend on the basis.  In
particular, the characteristic polynomial of $x_{ij}$ is a polynomial of
degree $m$, with the coefficient of $\lambda^m$ equal to $1$, and by the
Cayley-Hamilton theorem has $x$ as a root.  If $x$ is a primitive element of
$F$, then the characteristic polynomial if its matrix representation,
$x_{ij}$, is irreducible over $\mathbf{Q}$, and is the minimal polynomial of
$x$.

We recall, also, that an algebraic integer {\cite{Wikipedia algebraic
integer}} is a root of a polynomial with integer coefficients, such that the
coefficient of the highest power of $x$ is equal to $1$.  The sum, difference
and product of two algebraic integers is an algebraic integer.  If $F$ is an
algebraic number field of degree $m$ over $\mathbf{Q}$, then the elements
$x$ of $F$, whose regular representation matrices $x_{ij}$ have a
characteristic polynomial with integer coefficients, are the algebraic
integers in $F$.  The set of all the algebraic integers in $F$ is a ring,
called the ring of algebraic integers of $F$, that is often denoted
$\mathcal{O}_F$.  It is always possible to use a basis for $F$ consisting of
algebraic integers, called an integral basis, in which every algebraic integer
$x$ is represented by a matrix $x_{ij}$ with integer matrix elements.  When we
use an integral basis for $F$, the algebraic integers of $F$ are then
precisely those elements $x$ of $F$ which, when expressed as a linear
combination $x = x_1 v_1 + \ldots + x_m v_m$ of elements of the basis, are
such that the $x_i$ are all integers.  Most of the algebraic integers of $F$
are primitive elements of $F$, since those which are not primitive are one of
the finite number of proper subfields $K$ such that $\mathbf{Q} \subseteq K
\subset F$, and thus lie in one of a finite number of linear spaces over
$\mathbf{Q}$, each of dimension $< m$ over $\mathbf{Q}$.  Thus $F$ can
always be expressed in the form $\mathbf{Q} \left( \alpha \right)$, where
$\alpha$ is an algebraic integer in $F$, of degree equal to $m$.

If $F$ is an algebraic number field of degree $m$ over $\mathbf{Q}$, and
$\alpha$ is an algebraic integer of $F$, of degree equal to $m$, then $F$
always has an integral basis whose first element is $1$, whose second element
has the form $\frac{a_{21} + \alpha}{d_2}$, whose third element has the form
$\frac{a_{31} + a_{32} \alpha + \alpha^2}{d_3}$, and so on, where the $a_{ij}$
and the $d_i$ are ordinary integers in $\mathbf{Z}$.  This is called a
canonical basis of $F$ {\cite{PlanetMath Canonical basis}}.

If $t$ is an integer not divisible by the square of an integer $> 1$, then for
$t = 1 \left( \mathrm{mod} 4 \right)$, an integral basis for the quadratic
number field $\mathbf{Q} \left( \sqrt{t} \right)$ is given by $1,
\frac{1}{2} \left( \sqrt{t} - 1 \right)$, while for $t = 2$ or $3$ (mod 4), an
integral basis for $\mathbf{Q} \left( \sqrt{t} \right)$ is given by $1,
\sqrt{t}$ {\cite{PlanetMath Examples of the ring of integers}}.

If $F$ is an algebraic number field, then an {\emph{embedding}} of $F$ into
the field $\mathbf{C}$ of complex numbers, sometimes called an isomorphism
of $F$ into $\mathbf{C}$, means a one to one map of $F$ into $\mathbf{C}$,
that preserves all the structure of $F$.  In particular, the subfield
$\mathbf{Q}$ of $F$ is mapped by the identity map to the subfield
$\mathbf{Q}$ of $\mathbf{C}$.  The number of distinct embeddings of $F$
into $\mathbf{C}$ is finite.  In particular, if $F$ is defined by adjoining
to $\mathbf{Q}$ a root, $\alpha$, of a polynomial $f \left( x \right)$ in
$\mathbf{Q} \left[ x \right]$, such that $f \left( x \right)$ is irreducible
over $\mathbf{Q}$, then an embedding of $F =\mathbf{Q} \left( \alpha
\right)$ into $\mathbf{C}$ is specified by saying which root of $f \left( x
\right)$, in $\mathbf{C}$, $\alpha$ corresponds to.

An algebraic number field $F$ is called {\emph{totally real}}, if every
embedding of $F$ into $\mathbf{C}$, is equivalent to its complex conjugate.
If $F$ is defined by adjoining to $\mathbf{Q}$ a root, $\alpha$, of a
polynomial $f \left( x \right)$ in $\mathbf{Q} \left[ x \right]$, such that
$f \left( x \right)$ is irreducible over $\mathbf{Q}$, then $F =\mathbf{Q}
\left( \alpha \right)$ is totally real, if all the roots of $f \left( x
\right)$ are real.

If an object $X$, such as a number, matrix, or group, is defined for a
specific embedding, $I$, of an algebraic number field, $F$, into
$\mathbf{C}$, then the object corresponding to $X$, for an embedding
$\sigma$ of $F$ into $\mathbf{C}$, is denoted by $X^{\sigma}$, and called
the {\emph{Galois conjugate}} of $X$ by $\sigma$.  For example, if $F$ is
$\mathbf{Q} \left( \sqrt{2} \right)$, then there is only one embedding of
$F$ into $\mathbf{C}$ different from $I$, and taking the Galois conjugate of
an object by that embedding, corresponds to replacing all occurrences of
$\sqrt{2}$ by $- \sqrt{2}$.

The discreteness of the arithmetic lattices, to be constructed below, will
follow from the fact, noted in section 12 of {\cite{Borel Harish-Chandra}},
that if $F$ is an algebraic number field, $\mathcal{O}_F$ is the ring of
algebraic integers of $F$, and $\Phi_F$ is the set of all distinct embeddings
$\sigma$ of $F$ into $\mathbf{C}$, then for any positive number $r$, there
are only a finite number of elements $b$ of $\mathcal{O}_F$, such that
{\emph{all}} the Galois conjugates $b^{\sigma}$ of $b$, $\sigma \in \Phi_F$,
have magnitude less than $r$.  For example, in the case when $F =\mathbf{Q}
\left( \sqrt{2} \right)$ and $\mathcal{O}_F =\mathbf{Z} \left[ \sqrt{2}
\right]$, there are an infinite number of elements $a + b \sqrt{2}$, $a, b \in
\mathbf{Z}$, of $\mathcal{O}_F$, whose magnitude is less than $1$, but no
elements at all of $\mathcal{O}_F$, such that both $a + b \sqrt{2}$ and $a - b
\sqrt{2}$ have magnitude less than $1$.  To check the result for general $F$,
we note, first, that for any member, $b$, of $F$, the set of all the Galois
conjugates $b^{\sigma}$ of $b$, $\sigma \in \Phi_F$, is the same as the set of
all the conjugates of $b$, as an algebraic number.  This set will have the
same number of members as the degree of $F$, if $b$ is a primitive element of
$F$, and a smaller number, if $b$ is not a primitive element of $F$.
Furthermore, the product of all the conjugates of an algebraic number of
degree $s$ is equal to $\left( - 1 \right)^s$ times the constant term in its
minimal polynomial, which must be non-zero, for otherwise the polynomial would
be reducible over $\mathbf{Q}$.  Thus the product of all the conjugates of
an algebraic integer has magnitude $\geq 1$, hence the result is certainly
true when $r \leq 1$, since there are then no elements of $\mathcal{O}_F$, all
of whose Galois conjugates have magnitude less than $r$.  For general $r$, we
note that if all the conjugates of an algebraic integer of degree $s$ have
magnitude less than $r$, then, denoting this algebraic integer and its
conjugates by $b_1, b_2, \ldots, b_s$, we have $\left| b_1 + b_2 + \ldots +
b_s \right| < sr$, $\left| b_1 b_2 + b_1 b_3 + \ldots + b_{s - 1} b_s \right|
< \frac{s \left( s - 1 \right)}{2} r^2$, $\ldots$, $\left| b_1 b_2 \ldots b_s
\right| < r^s$, hence every coefficient of the minimal polynomial of that
algebraic integer is bounded by a binomial coefficient times a power of $r$,
hence since these coefficients are integers, there are, in fact, only a finite
number of distinct algebraic integers of degree $s$, all of whose conjugates
have magnitude less than $r$.  Hence, since all elements of $\mathcal{O}_F$
are algebraic integers of degree $\leq$ the degree of $F$, there are only a
finite number of elements of $\mathcal{O}_F$, all of whose Galois conjugates,
or equivalently, all of whose conjugates, have magnitude less than $r$.  I
will call this result the bounded conjugates lemma.

We can now construct examples of cocompact lattices in SU($n$,1) and
SO($n$,1), by choosing:
\begin{enumerate}
  \item A totally real algebraic number field $F \neq \mathbf{Q}$;

  \item A specific embedding $I$ of $F$ into $\mathbf{R}$; and

  \item A diagonal $\left( n + 1 \right) \times \left( n + 1 \right)$ matrix
  $B$, with signature $\left( +, +, \ldots, +, - \right)$, and diagonal matrix
  elements in the ring of algebraic integers $\mathcal{O}_F$ of $F$, such that
  for all embeddings $\sigma \neq I$ of $F$ into $\mathbf{R}$, the Galois
  conjugate $B^{\sigma}$ is either positive definite or negative definite.
\end{enumerate}
For example, we could choose $F$ to be $\mathbf{Q} \left( \alpha \right)$,
where $\alpha$ is a root of the polynomial $x^2 - 2$, we could specify $I$ by
choosing $\alpha$ to be $\sqrt{2}$ rather than $- \sqrt{2}$, and we could
choose $B$ to be the diagonal $\left( n + 1 \right) \times \left( n + 1
\right)$ matrix with diagonal matrix elements $\left( 1, 1, \ldots, 1, -
\sqrt{2} \right)$, so that for the one embedding $\sigma$ different from $I$,
the Galois conjugate $B^{\sigma}$ is the positive definite diagonal matrix
with diagonal matrix elements \\
$\left( 1, 1, \ldots, 1, \sqrt{2} \right)$.

We note that $\mathrm{SU} \left( B \right)$, the group of all complex $\left( n
+ 1 \right) \times \left( n + 1 \right)$ matrices with unit determinant, that
preserve the quadratic form $B_{R \bar{S}} z^R z^{\bar{S}}$, is isomorphic to
SU($n$,1), and that the matrix $\eta_{R \bar{S}}$, defined between (\ref{CHn
Ricci tensor}) and (\ref{homogeneous coordinates}), could be transformed to
equal $B_{R \bar{S}}$, by a suitable rescaling of the coordinates, and
similarly, $\mathrm{SO} \left( B \right)$, the group of all real $\left( n + 1
\right) \times \left( n + 1 \right)$ matrices with unit determinant, that
preserve the quadratic form $B_{RS} x^R x^S$, is isomorphic to SO($n$,1), and
that the standard Minkowski metric $\eta_{RS}$ could be transformed to equal
$B_{RS}$, by a suitable rescaling of the coordinates.  Here $z^{\bar{S}} =
\left( z^S \right)^{\ast}$, in accordance with the conventions of subsection
\ref{CH3}.

If we now identify $G$ with the identity component, or in other words, the
connected component containing the identity, of either $\mathrm{SU} \left( B
\right)$ or $\mathrm{SO} \left( B \right)$, then the required cocompact lattice,
$\Gamma$, is in the unitary case, the subgroup $G_{\mathcal{O}_F \left[ i
\right]}$ of $G$, consisting of the elements of $G$, all of whose matrix
elements are in $\mathcal{O}_F \left[ i \right]$, the extension of the ring of
algebraic integers $\mathcal{O}_F$ of $F$, by the square root of $- 1$, and in
the orthogonal case, the subgroup $G_{\mathcal{O}_F}$ of $G$, consisting of
the elements of $G$, all of whose matrix elements are in $\mathcal{O}_F$
{\cite{Borel Harish-Chandra}}.

To check the discreteness of $\Gamma$ I will consider the case of $\mathrm{SU}
\left( n, 1 \right)$ and SU($B$), since the corresponding discussion for
SO($n$,1) and SO($B$) will follow by dropping the extension of $\mathcal{O}_F$
to $\mathcal{O}_F \left[ i \right]$.  We first note that, for all embeddings
$\sigma \neq I$ of $F$ into $\mathbf{R}$, the Galois conjugate group
$G^{\sigma} = \mathrm{SU} \left( B^{\sigma} \right)$ is isomorphic to $\mathrm{SU}
\left( n + 1 \right)$, and thus compact.  Furthermore, for any element
$\gamma$ of $\Gamma$, the Galois conjugate $\gamma^{\sigma}$ will be a member
of $\Gamma^{\sigma}$, and thus of $G^{\sigma}$.  Thus $\gamma^{\sigma}$ could
be transformed to a unitary matrix by a rescaling of the coordinates, so there
is a number $r_{\sigma}$, depending only $B^{\sigma}$, such that every matrix
element of $\gamma^{\sigma}$ has magnitude less than $r_{\sigma}$.  Let $r$ be
any number $\geq$ the maximum of the numbers $r_{\sigma}$, for all the
embeddings $\sigma \neq I$ of $F$ into $\mathbf{R}$.  Then since
$\mathcal{O}_F$ is the ring of algebraic integers of the algebraic number
field $F$, the bounded conjugates lemma implies that there are only a finite
number of elements $b$ of $\mathcal{O}_F$, such that {\emph{all}} the Galois
conjugates $b^{\sigma}$ of $b$, including $b$ itself, have magnitude less than
$r$.  Thus if $\gamma$ is an element of $\Gamma$, such that every matrix
element $\left( a + ib \right)$, $a, b \in \mathcal{O}_F$, of $\gamma$, has
magnitude less than $r$, then there are only a finite number of elements $c$
of $\mathcal{O}_F \left[ i \right]$, that can be matrix elements of $\gamma$.
Thus there are only a finite number of elements $\gamma$ of $\Gamma$, such
that every matrix element of $\gamma$ has magnitude less than $r$.  In
particular, there is a number $r > 0$ such that there are only a finite number
of elements $\gamma$ of $\Gamma$, such that every matrix element of $\gamma -
1$, where $1$ denotes the unit matrix, has magnitude less than $r$.  Hence
$\Gamma$ is discrete {\cite{Borel Harish-Chandra}}.

We note that if $r$ is the maximum of the numbers $r_{\sigma}$, for all the
embeddings $\sigma \neq I$ of $F$ into $\mathbf{R}$, and $\gamma$ is any
element of $\Gamma$, then every matrix element $\left( a + ib \right)$ of
$\gamma$ has the property that each of $a$ and $b$ is an algebraic integer in
$\mathcal{O}_F$, such that all its conjugates, other than itself, have
magnitude less than $r$.  An algebraic integer, such that all its conjugates,
other than itself, have magnitude less than $r$, is called an $r$-Pisot
number.  Pisot numbers are sometimes called Pisot-Vijayaraghavan numbers, or
PV numbers.  Fan and Schmeling {\cite{Fan Schmeling}} have demonstrated that
for any real algebraic number field, and any $r > 0$, there exists a number
$L$ such that for all $x \in \mathbf{R}$, there is at least one $r$-Pisot
number $\eta$ in that algebraic number field, such that $x \leq \eta \leq x +
L$.  This result gives an indication of the distribution of algebraic integers
that can be matrix elements of an element of $\Gamma$.  For example, if $F$ is
the field $\mathbf{Q} \left( \sqrt{2} \right)$, $G$ is isomorphic to
SU(1,1), and $B$ is the diagonal matrix with diagonal entries $\left( 1, -
\sqrt{2} \right)$, then every matrix element of every element of $\Gamma$ is a
$2^{\frac{1}{4}}$-Pisot number.  Now for every integer, $b$, there is an
integer, $a$, such that $\left| a - b \sqrt{2} \right| \leq \frac{1}{2} <
2^{\frac{1}{4}}$, so that $a + b \sqrt{2}$ is a $2^{\frac{1}{4}}$-Pisot
number.  And for that integer, $a$, we have $2 b \sqrt{2} - \frac{1}{2} \leq a
+ b \sqrt{2} \leq 2 b \sqrt{2} + \frac{1}{2}$.  Thus in this case we can take
$L = 2 \sqrt{2} + 1$.  Some examples of elements of $\Gamma$, in this case,
are:
\begin{equation}
  \label{examples of elements of Gamma} \left(\begin{array}{cc}
    3 + 2 \sqrt{2} & 4 + 2 \sqrt{2}\\
    2 + 2 \sqrt{2} & 3 + 2 \sqrt{2}
  \end{array}\right), \hspace{2em} \left( \begin{array}{cc}
    33 + 24 \sqrt{2} & 40 + 28 \sqrt{2}\\
    28 + 20 \sqrt{2} & 33 + 24 \sqrt{2}
  \end{array} \right)
\end{equation}
Let $\mathcal{P}_{F, r}$ denote the set of all the $r$-Pisot numbers in
$\mathcal{O}_F$.  We note that, for all real numbers $s > 0$, there are only a
finite number of elements of $\mathcal{P}_{F, r}$ with magnitude $< s$.  For
it is sufficient to prove this for $s \geq r$.  And for $s \geq r$, every
element of $\mathcal{P}_{F, r}$ is an $s$-Pisot number in $F$.  And by
definition, all the conjugates $\sigma \neq I$, of an $s$-Pisot number, have
magnitude $< s$.  Hence by the bounded conjugates lemma, there are only a
finite number of $s$-Pisot numbers in $F$, whose magnitude is $< s$, hence
there are only a finite number of elements of $\mathcal{P}_{F, r}$, whose
magnitude is $< s$.  Hence, in particular, $\mathcal{P}_{F, r}$ is discrete.

\subsubsection{Compactness of $G / \Gamma$ for the examples of arithmetic
lattices}

The compactness of $G / \Gamma$, for groups such as those in the examples
given above, was originally proved by Borel and Harish-Chandra {\cite{Borel
Harish-Chandra}}, making use of their proof that $\Gamma$ is a lattice in $G$,
and their proof of a compactness criterion that had been conjectured by
Godement.  The following direct proof of compactness is adapted from sections
(6.36) and (6.45) of {\cite{Morris}}, and the proof of Mahler's compactness
theorem {\cite{Mahler compactness theorem}} in section (5.34) of
{\cite{Morris}}.

To check the compactness of the quotient $G / \Gamma$, it is sufficient to
check that, given any infinite sequence $\left\{ g_k \right\}$ of elements of
$G$, there exists a sequence $\left\{ \gamma_k \right\}$ of elements of
$\Gamma$, such that the sequence $\left\{ g_k \gamma_k \right\}$ has a
convergent infinite subsequence.

We first note that if $G = \mathrm{SU} \left( B \right) \cong \mathrm{SU} \left(
n, 1 \right)$, the subgroup $G_{F \left( i \right)}$ of $G$, consisting of the
elements of $G$, all of whose matrix elements are in $F \left( i \right)$, the
extension of $F$, by the square root of $- 1$, is dense in $G$, while if $G =
\mathrm{SO} \left( B \right) \cong \mathrm{SO} \left( n, 1 \right)$, the subgroup
$G_F$ of $G$, consisting of the elements of $G$, all of whose matrix elements
are in $F$, is dense in $G$.  For an arbitrary element $V$ of SU($B$)
satisfies
\begin{equation}
  \label{equation satisfied by V and B} V^{\dag} BV = B
\end{equation}
where $V^{\dag}$ denotes the hermitian conjugate of $V$, and (\ref{equation
satisfied by V and B}) is also satisfied by an arbitrary element $V$ of
$\mathrm{SO} \left( B \right)$, since in that case $V^{\dag} = V^T$, where $V^T$
denotes the transpose of $V$.  And in general, if $B$ is a nonsingular
hermitian matrix, and $V$ is a complex matrix that satisfies (\ref{equation
satisfied by V and B}), then the matrix
\begin{equation}
  \label{definition of A in terms of B and V} A \equiv B \left( V - 1 \right)
  \left( V + 1 \right)^{- 1}
\end{equation}
is antihermitian, and $V$ is expressed rationally in terms of the
antihermitian matrix $A$ by
\begin{equation}
  \label{V in terms of B and A} V = \left( B - A \right)^{- 1} \left( B + A
  \right)
\end{equation}
Furthermore, the matrix $\left( B - A \right) = 2 B \left( V + 1 \right)^{-
1}$ is nonsingular.  Moreover, if $A$ is an arbitrary antihermitian matrix,
such that $\left( B - A \right)$ is nonsingular, and $V$ is defined in terms
of $A$ by (\ref{V in terms of B and A}), then $V$ satisfies $V^{\dag} BV = B$.
Thus, for an arbitrary element $V$ of $\mathrm{SU} \left( B \right)$ or
$\mathrm{SO} \left( B \right)$, such that $\left( V + 1 \right)$ is
non-singular, we can define the antihermitian matrix $A$ by (\ref{definition
of A in terms of B and V}), and then, by choosing an antihermitian matrix
$\tilde{A}$, with matrix elements in $F \left( i \right)$ or $F$, as
appropriate, that approximates $A$ sufficiently closely, and is such that
$\left( B - \tilde{A} \right)$ is nonsingular, we can obtain an element
$\tilde{V}$ of $\mathrm{SU} \left( B \right)_{F \left( i \right)}$ or $\mathrm{SO}
\left( B \right)_F$, as appropriate, such that every matrix element of $\left(
\tilde{V} - V \right)$ has magnitude less than any given number $> 0$.  And if
$\left( V + 1 \right)$ is singular, we can follow the same procedure, for an
element $V_1$ of $\mathrm{SU} \left( B \right)$ or $\mathrm{SO} \left( B \right)$,
as appropriate, such that that matrix elements of $\left( V_1 - V \right)$ are
sufficiently small in magnitude, and $\left( V_1 + 1 \right)$ is nonsingular,
so as to obtain, again, an element $\tilde{V}$ of $\mathrm{SU} \left( B
\right)_{F \left( i \right)}$ or $\mathrm{SO} \left( B \right)_F$, as
appropriate, such that every matrix element of $\left( \tilde{V} - V \right)$
has magnitude less than any given number $> 0$.  Thus it is sufficient to
check that, given any infinite sequence $\left\{ g_k \right\}$ of elements of
$G_{F \left( i \right)}$ or $G_F$, as appropriate, there exists a sequence
$\left\{ \gamma_k \right\}$ of elements of $\Gamma$, such that the sequence
$\left\{ g_k \gamma_k \right\}$ has a infinite Cauchy subsequence, or in other
words, an infinite subsequence $\left\{ g_p \gamma_p \right\}$, such that for
any given $\varepsilon > 0$, there exists an integer $t$, such that for all $p
> t$ and all $q > t$, every matrix element of $\left( g_p \gamma_p - g_q
\gamma_q \right)$ has magnitude less than $\varepsilon$.

The matrices $\gamma_k$ will be constructed as one block of a block diagonal
matrix that includes all the Galois conjugates of $\Gamma$ along its block
diagonal, because we can then transform these block diagonal matrices to
matrices with integer matrix elements, by a similarity transformation that
consists of multiple copies of the inverse of the similarity transformation
that diagonalizes the matrix representations of the elements of $F$ in an
integral basis, discussed above.  Once we are working with matrices with
integer matrix elements, we can construct the sequence $\left\{ \gamma_k
\right\}$ by a method due to Mahler {\cite{Mahler compactness theorem}}.

It is convenient, first, if $G = \mathrm{SU} \left( B \right) \cong \mathrm{SU}
\left( n, 1 \right)$, to embed $G$ in a group of $2 \left( n + 1 \right)
\times 2 \left( n + 1 \right)$ matrices with real matrix elements.  For each
element $g$ of $G$, let $\bar{g}$ denote the $2 \left( n + 1 \right) \times 2
\left( n + 1 \right)$ matrix with real matrix elements, obtained from $g$ by
replacing each complex matrix element $\left( a + ib \right)$ by the real
matrix $\left(\begin{array}{cc}
  a & - b\\
  b & a
\end{array}\right)$.  We note that, by this rule, the hermitian conjugate
$g^{\dag}$ of $g$ corresponds to the transpose $\bar{g}^T$ of $\bar{g}$.  And
let $\bar{B}$ be obtained from $B$ by the same rule.  Thus $\bar{B}$ is a
diagonal matrix whose first and second diagonal matrix elements are equal to
one another, whose third and fourth diagonal matrix elements are equal to one
another, and so on.  $\bar{B}$ has signature $\left( +, +, \ldots, +, -, -
\right)$, so we are embedding $G$ in a group isomorphic to $\mathrm{SO} \left(
n, 2 \right)$.  However the following discussion will not depend on the
detailed signature of $\bar{B}$ or $B$, beyond the fact that $\bar{B}$ or $B$
is indefinite, while its Galois conjugates for $\sigma \neq I$ are either
positive definite or negative definite.

When an element $g$ of $\mathrm{SU} \left( B \right)$ acts on a complex $\left(
n + 1 \right)$-vector, each complex matrix element $\left( c + id \right)$ of
that $\left( n + 1 \right)$-vector is replaced by the real column vector
$\left(\begin{array}{c}
  c\\
  d
\end{array}\right)$, so that $\left(\begin{array}{cc}
  a & - b\\
  b & a
\end{array}\right) \left(\begin{array}{c}
  c\\
  d
\end{array}\right) = \left(\begin{array}{c}
  ad - bc\\
  bc + ad
\end{array}\right)$, and the complex $\left( n + 1 \right)$-vector becomes a
real $2 \left( n + 1 \right)$-vector.  These two ways of representing a
complex number, as a real matrix, or as a real column vector, are an example
of the relation between the representation of an element $x$, of an algebraic
number field, by the matrix $x_{ij}$, and by its components $x_i$, in a
particular basis, as discussed above.

We note that if $\bar{g}$ corresponds to an element $g$ of $\mathrm{SU} \left( B
\right)$ by the transformation described above, then each $2 \times 2$ block
in $\bar{g}$ can be diagonalized by the similarity transformation
$\frac{1}{\sqrt{2}} \left(\begin{array}{cc}
  1 & i\\
  i & 1
\end{array}\right) \left(\begin{array}{cc}
  a & - b\\
  b & a
\end{array}\right) \frac{1}{\sqrt{2}} \left(\begin{array}{cc}
  1 & - i\\
  - i & 1
\end{array}\right) = \left(\begin{array}{cc}
  a + ib & 0\\
  0 & a - ib
\end{array}\right)$, hence all the $2 \times 2$ block matrices in $\bar{g}$
can be diagonalized by applying a block diagonal similarity transformation
with $n + 1$ copies of $\left(\begin{array}{cc}
  1 & - i\\
  - i & 1
\end{array}\right)$ along the block diagonal.  Then by permuting rows and
columns, $\bar{g}$ can be brought to the form of a block diagonal matrix with
two $\left( n + 1 \right) \times \left( n + 1 \right)$ blocks along the block
diagonal, one of which is $g$, and the other of which is $g^{\ast}$, the
complex conjugate of $g$.  Hence $\det \bar{g} = \left| \det g \right|^2$.
But $\det g = 1$, hence $\det \bar{g} = 1$.

Let $\mathcal{J}$ be the block diagonal matix with $n$ copies of
$\left(\begin{array}{cc}
  0 & - 1\\
  1 & 0
\end{array}\right)$ along the block diagonal.  Then the subgroup $\mathrm{SU}
\left( B \right)$ of $\mathrm{SO} \left( \bar{B} \right)$ is the group of all
elements $g$ of $\mathrm{SO} \left( \bar{B} \right)$ such that $\mathcal{J}g =
g\mathcal{J}$.

Hence by renaming $\bar{B}$ as $B$, we can now assume that either $B$ is a
diagonal matrix with $n$ positive diagonal matrix elements and one negative
diagonal matrix element, and $G = \mathrm{SO} \left( B \right)$, or $B$ is a
diagonal matrix with $2 n$ positive diagonal matrix elements and 2 negative
diagonal matrix elements, and $G$ is the subgroup of $\mathrm{SO} \left( B
\right)$ that commutes with $\mathcal{J}$.  And for every Galois conjugate
$\sigma$ of $F$ different from the identity $I$, $B^{\sigma}$ is in both cases
either a positive definite matrix or a negative definite matrix.  We define $p
= \left( n + 1 \right)$ in the orthogonal case, and $p = 2 \left( n + 1
\right)$ in the unitary case.

Let $H$ be the set of all $p \times p$ matrices with matrix elements in $F$,
and $m$ be the degree of $F$.  We now choose a fixed sequence of the $m$
Galois conjugates $\sigma$ of $F$, starting with the identity $\sigma = I$,
and for an arbitrary element $h$ of $H$, we define $\hat{h}$ to be the $pm
\times pm$ block diagonal matrix which has $h$ as its top left $p \times p$
matrix elements, then the first Galois conjugate $\sigma \neq I$ of $h$ as its
second block of $p \times p$ matrix elements on the block diagonal, then the
second Galois conjugate $\sigma \neq I$ of $h$ as its third block of $p \times
p$ matrix elements on the block diagonal, and so on, and all other matrix
elements equal to zero.  We also define $\hat{H}$ to be the set of all the
matrices $\hat{h}$, for $h \in H$, $\hat{G}$ to be the set of all the matrices
$\hat{g}$, for $g \in G_F$, and $\hat{\Gamma}$ to be the set of all the
matrices $\hat{g}$, for $g \in \Gamma$.  We note that if $g$ is any element of
$G_F$, then since $\hat{g}$ is block diagonal, and every block on the block
diagonal of $\hat{g}$ has determinant equal to $1$, $\det \hat{g} = 1$.

For an arbitrary $p$-vector $x$ in $F^p$, we define $\hat{x}$ to be the
$pm$-vector whose first $p$ components are $x$, whose next $p$ components are
the first Galois conjugate $\sigma \neq I$ of $x$, whose third set of $p$
consecutive components are the second Galois conjugate $\sigma \neq I$ of $x$,
and so on.  We also define $\hat{F}^p$ to be the set of all the $pm$-vectors
$\hat{x}$, for $x \in F^p$.  Thus for all $\hat{h} \in \hat{H}$, and all
$\hat{x} \in \hat{F}^p$, the $pm$-vector $\hat{h}  \hat{x}$ is an element of
$\hat{F}^p$.  We also define $\hat{\mathcal{O}}_F^p$ to be the set of all the
$pm$-vectors $\hat{x}$, for $x \in \mathcal{O}_F^p$.

We note that, if $x$ is any nonzero $p$-vector in $F^p$, then the quadratic
form $x^T Bx$ is nonzero.  For by assumption $F \neq \mathbf{Q}$, hence $F$
has at least one Galois conjugate $\sigma$ different from $I$, and by
assumption $\left( x^{\sigma} \right)^T B^{\sigma} x^{\sigma} = \left( x^T Bx
\right)^{\sigma}$ is either positive definite or negative definite.
Furthermore, no nonzero element of $F$ can have any Galois conjugate equal to
$0$, for $0$ is of degree $1$, hence has no conjugates other than itself.
Thus if $\hat{x}$ is any nonzero element of $\hat{F}^p$, then the quadratic
form $\hat{x}^T \hat{B}  \hat{x}$ is nonzero.

Furthermore, if $x$ is any nonzero $p$-vector in $\mathcal{O}^p_F$, then the
value of the quadratic form $\hat{x}^T \hat{B}  \hat{x}$ is an ordinary
integer in $\mathbf{Z}$, and its magnitude is $\geq 1$.  For all the matrix
elements of $B$ are in $\mathcal{O}_F$, hence $x^T Bx$ is an algebraic integer
in $\mathcal{O}_F$.  Furthermore, $\hat{x}^T \hat{B}  \hat{x}$ is the sum of
all the Galois conjugates $\left( x^T Bx \right)^{\sigma}$ of $x^T Bx$, which
if the degree of the algebraic integer $x^T Bx$ is equal to $m$, is $- 1$
times the coefficient of $x^{m - 1}$ in the minimal polynomial of $x^T Bx$,
and thus an integer in $\mathbf{Z}$, while if the degree $k$ of $x^T Bx$ is
less than $m$, it must divide $m$, and $\hat{x}^T \hat{B}  \hat{x}$ is equal
to the integer $\frac{m}{k}$, times $- 1$ times the coefficient of $x^{k - 1}$
in the minimal polynomial of $x^T Bx$, and thus again an integer in
$\mathbf{Z}$.  And furthermore, by the preceding paragraph, the ordinary
integer $\hat{x}^T \hat{B}  \hat{x}$ cannot be equal to zero, hence it has
magnitude $\geq 1$.

We now carry out a similarity transformation $\hat{h} \rightarrow S^{- 1}
\hat{h} S$ on the elements $\hat{h}$ of $\hat{H}$, such that $S$ consists of
$p$ copies of the inverse, $s$, of a similarity transformation that
diagonalizes the matrix representations of the elements of $F$ in an integral
basis, discussed above.  Each of the $p$ copies of $s$ is ``spread out'', so
that, for example, the first copy acts from the right only on the first column
of each of the $m$ Galois conjugates of elements of $G_F$, the second copy
acts from the right only on the second column of each of the $m$ Galois
conjugates of elements of $G_F$, and so on.  For example, if $F =\mathbf{Q}
\left( \sqrt{2} \right)$, we can choose the similarity transformation $S^{- 1}
\hat{h} S$ to consist of $p$ copies of the similarity transformation:
\begin{equation}
  \label{similarity transformation for Q root 2} \frac{1}{2}
  \left(\begin{array}{cc}
    1 & 1\\
    \frac{1}{\sqrt{2}} & - \frac{1}{\sqrt{2}}
  \end{array}\right) \left(\begin{array}{cc}
    a + b \sqrt{2} & 0\\
    0 & a - b \sqrt{2}
  \end{array}\right) \left(\begin{array}{cc}
    1 & \sqrt{2}\\
    1 & - \sqrt{2}
  \end{array}\right) = \left( \begin{array}{cc}
    a & 2 b\\
    b & a
  \end{array} \right)
\end{equation}
And if, in addition, $G \cong \mathrm{SO} \left( 1, 1 \right)$, so that $p = 2$,
the similarity transformation $S^{- 1}  \hat{h} S$ would have the form:
\[ \frac{1}{2}  \left( \begin{array}{cccc}
     1 & 0 & 1 & 0\\
     \frac{1}{\sqrt{2}} & 0 & - \frac{1}{\sqrt{2}} & 0\\
     0 & 1 & 0 & 1\\
     0 & \frac{1}{\sqrt{2}} & 0 & - \frac{1}{\sqrt{2}}
   \end{array} \right) \left(\begin{array}{cccc}
     a + b \sqrt{2} & c + d \sqrt{2} & 0 & 0\\
     e + f \sqrt{2} & g + h \sqrt{2} & 0 & 0\\
     0 & 0 & a - b \sqrt{2} & c - d \sqrt{2}\\
     0 & 0 & e - f \sqrt{2} & g - h \sqrt{2}
   \end{array}\right) \times \hspace{2em} \]
\begin{equation}
  \label{similarity transformation for SO 1 1} \hspace{10em} \times \left(
  \begin{array}{cccc}
    1 & \sqrt{2} & 0 & 0\\
    0 & 0 & 1 & \sqrt{2}\\
    1 & - \sqrt{2} & 0 & 0\\
    0 & 0 & 1 & - \sqrt{2}
  \end{array} \right) = \left( \begin{array}{cccc}
    a & 2 b & c & 2 d\\
    b & a & d & c\\
    e & 2 f & g & 2 h\\
    f & e & h & g
  \end{array} \right)
\end{equation}
We see that for each element $h$ of $H$, the similarity transformation $S^{-
1}  \hat{h} S$ transforms $\hat{h}$ into a $p \times p$ block matrix, each
block of which is the $m \times m$ matrix representation of the corresponding
matrix element of $h$, in the chosen integral basis.  Thus the matrix elements
of $S^{- 1}  \hat{h} S$ are rational numbers.  For each element $h$ of $H$, we
define $\tilde{h} \equiv S^{- 1}  \hat{h} S$, where $\hat{h}$ is the element
of $\hat{H}$ that corresponds to $h$ as above.  We see that the elements of
the subgroup $\hat{\Gamma}$ of $\hat{G}$ are precisely those elements
$\hat{g}$ of $\hat{G}$ for which $\tilde{g} = S^{- 1}  \hat{g} S$ has
integer-valued matrix elements.  We define $\tilde{H}$ to be the set of all
the matrices $\tilde{h}$, for $h \in H$, $\tilde{G}$ to be the set of all the
matrices $\tilde{g}$, for $g \in G_F$, and $\tilde{\Gamma}$ to be the set of
all the matrices $\tilde{g}$, for $g \in \Gamma$.  We note that if $g$ is any
element of $G_F$, then since $\tilde{g}$ is related to $\hat{g}$ by a
similarity transformation, and $\det \hat{g} = 1$, we have $\det \tilde{g} =
1$.

Now if $\alpha$ is a primitive element of $F$, or in other words, an algebraic
number in $F$, whose degree is equal to $m$, and $\mathbf{Q}^{m \times m}$
denotes the set of all $m \times m$ matrices with rational matrix elements,
then the elements of $\mathbf{Q}^{m \times m}$ that are matrix
representations of elements of $F$, in the chosen integral basis, are
precisely those that commute with the matrix representation of $\alpha$ in the
chosen integral basis.  For every element of $F$ commutes with $\alpha$, and
if an element $\chi$ of $\mathbf{Q}^{m \times m}$ commutes with the matrix
representation of $\alpha$, then since $\left( 1, \alpha, \alpha^2, \ldots,
\alpha^{m - 1} \right)$ is a possible basis for $F$, so the matrix
representations of $1, \alpha, \alpha^2, \ldots, \alpha^{m - 1}$ are linearly
independent elements of $\mathbf{Q}^{m \times m}$, and are thus a complete
basis for the elements of $\mathbf{Q}^{m \times m}$ that commute with
$\alpha$, $\chi$ is a linear combination, with coefficients in $\mathbf{Q}$,
of the matrix representations of $1, \alpha, \alpha^2, \ldots, \alpha^{m -
1}$, and is thus the matrix representation of an element of $F$.

We now choose an algebraic integer $\varphi$ of $F$, such that $\varphi$ is
primitive in $F$.  Such an algebraic integer $\varphi$ of $F$ always exists,
because, as noted above, most algebraic integers in $F$ are primitive in $F$.
We define $\mathcal{F}$ to be the element of $H$ that is the $p \times p$
diagonal matrix, with each matrix element on the diagonal equal to $\varphi$,
so that, in other words, $\mathcal{F}$ is equal to $\varphi$ times the $p
\times p$ unit matrix.  The elements $\hat{\mathcal{F}}$ of $\hat{H}$, and
$\tilde{\mathcal{F}}$ of $\tilde{H}$, are then defined in the standard way, as
above.  Thus $\tilde{\mathcal{F}}$ is the $pm \times pm$ block diagonal
matrix, such that each $m \times m$ block on the block diagonal is equal to
the matrix representation of $\varphi$, in the chosen integral basis.  Then if
$\mathbf{Q}^{pm \times pm}$ denotes the set of all $pm \times pm$ matrices
with matrix elements in $\mathbf{Q}$, the elements of $\tilde{H}$ are
precisely the elements of $\mathbf{Q}^{pm \times pm}$ that commute with
$\tilde{\mathcal{F}}$, since an element $\xi$ of $\mathbf{Q}^{pm \times pm}$
commutes with $\tilde{\mathcal{F}}$ if and only if every $m \times m$ block of
$\xi$ commutes with the matrix representation of $\varphi$, in the chosen
integral basis.

We note that $\tilde{B}$ will have integer-valued matrix elements, and be
block diagonal, with each block on the block diagonal being an $m \times m$
matrix which, for at least one block, in the example when $F$ is $\mathbf{Q}
\left( \sqrt{2} \right),$ will not be symmetric.  $\tilde{\mathcal{F}}$ will
also have integer-valued matrix elements, since $\varphi$ is an algebraic
integer.  And in the unitary case, $\tilde{\mathcal{J}}$ will have
integer-valued matrix elements, which will in fact be $+ 1$ or $- 1$, and be
block diagonal, with each block on the block diagonal being a $2 m \times 2 m$
antisymmetric matrix.

We note that in the above example, (\ref{similarity transformation for Q root
2}), $s$ has been chosen such that every matrix element in its first column is
equal to $1$.  This has the consequence that for an arbitrary element $h$ of
$H$, the first column of the matrix $\hat{h} S$ is the element $\hat{x}$ of
$\hat{F}^p$, where $x$ denotes the first column of $h$, and the element
$\hat{x}$ of $\hat{F}^p$ is related to the $p$-vector $x$ in $F^p$, as
described above.  Furthermore, the first $m$ matrix elements of the first
column of $S^{- 1}  \hat{h} S$, or in other words, the first $m$ matrix
elements of $S^{- 1}  \hat{x}$, are the components of the first matrix element
of $x$, with respect to the integral basis $\left( 1, \sqrt{2} \right)$ of
$\mathbf{Q} \left( \sqrt{2} \right)$, and the next $m$ matrix elements of
the first column of $S^{- 1}  \hat{h} S$, or in other words, the next $m$
matrix elements of $S^{- 1}  \hat{x}$, are the components of the second matrix
element of $x$, with respect to the integral basis $\left( 1, \sqrt{2}
\right)$ of $\mathbf{Q} \left( \sqrt{2} \right)$.  Therefore, since for an
arbitrary $p$-vector $x$ in $F^p$, we can write down an element $h$ of $H$,
such that the first column of $h$ is $x$, it follows, for this example, that
for an arbitrary $p$-vector $x$ in $F^p$, the components of the $pm$-vector
$S^{- 1}  \hat{x}$, where the element $\hat{x}$ of $\hat{F}^p$ is related to
the $p$-vector $x$ in $F^p$ as described above, are the $m$ components of the
first matrix element of $x$, with respect to the integral basis $\left( 1,
\sqrt{2} \right)$ of $\mathbf{Q} \left( \sqrt{2} \right)$, followed by the
$m$ components of the second matrix element of $x$, with respect to the
integral basis $\left( 1, \sqrt{2} \right)$ of $\mathbf{Q} \left( \sqrt{2}
\right)$.  Thus, for this example, for an arbitrary $p$-vector $x$ in
$\mathcal{O}^p_F$, the $pm$-vector $S^{- 1}  \hat{x}$ has integer components
in $\mathbf{Z}$, and conversely, for an arbitrary $pm$-vector $\tilde{x}$ in
$\mathbf{Z}^{pm}$, the $pm$-vector $S \tilde{x}$ is an element $\hat{x}$ of
$\hat{\mathcal{O}}^p_F$, that corresponds to an element $x$ of
$\mathcal{O}_F^p$ in the manner described above.  Specifically, we have:
\begin{equation}
  \label{action of S to the minus 1 on an element of F hat to the p}
  \frac{1}{2}  \left( \begin{array}{cccc}
    1 & 0 & 1 & 0\\
    \frac{1}{\sqrt{2}} & 0 & - \frac{1}{\sqrt{2}} & 0\\
    0 & 1 & 0 & 1\\
    0 & \frac{1}{\sqrt{2}} & 0 & - \frac{1}{\sqrt{2}}
  \end{array} \right) \left(\begin{array}{c}
    a + b \sqrt{2}\\
    c + d \sqrt{2}\\
    a - b \sqrt{2}\\
    c - d \sqrt{2}
  \end{array}\right) = \left( \begin{array}{c}
    a\\
    b\\
    c\\
    d
  \end{array} \right)
\end{equation}
This corresponds to the fact that, with $s$ chosen as in the example
(\ref{similarity transformation for Q root 2}), $s^{- 1}$ acts on a column
$m$-vector, that consists of all the Galois conjugates of an element of $F$,
in the standard order, as:
\begin{equation}
  \label{action of s to the minus 1 on an element of F hat} \frac{1}{2}
  \left(\begin{array}{cc}
    1 & 1\\
    \frac{1}{\sqrt{2}} & - \frac{1}{\sqrt{2}}
  \end{array}\right) \left(\begin{array}{c}
    a + b \sqrt{2}\\
    a - b \sqrt{2}
  \end{array}\right) = \left(\begin{array}{c}
    a\\
    b
  \end{array}\right)
\end{equation}
I shall now show that for an arbitrary algebraic number field $F$, and thus,
in particular, for an arbitrary totally real algebraic number field $F$, we
can always choose an integral basis $\left\{ v_i \right\}$ for $F$, such that
the elements of the first column of the matrix representation of each element
of $F$, in the basis $\left\{ v_i \right\}$, are the expansion coefficients of
that element of $F$ in the basis $\left\{ v_i \right\}$, $1 \leq i \leq m$.
And furthermore, in such a basis for $F$, the inverse, $s$, of the similarity
transformation that diagonalizes the matrix representations of elements of $F$
in the integral basis $\left\{ v_i \right\}$, will be such that all the matrix
elements in its first column are nonzero, and moreover, all the matrix
elements in its first column can be chosen equal to $1$.

When we use such a basis for $F$, and choose all the matrix elements in the
first column of $s$ to be equal to $1$, then it immediately follows, as in the
above example, that for an arbitrary $p$-vector $x$ in $\mathcal{O}^p_F$, the
$pm$-vector $S^{- 1}  \hat{x}$ has integer components in $\mathbf{Z}$, which
are in fact the expansion coefficients of the successive matrix elements of
$x$, with respect to the integral basis $\left\{ v_i \right\}$, and
conversely, for an arbitrary $pm$-vector $\tilde{x}$ in $\mathbf{Z}^{pm}$,
the $pm$-vector $S \tilde{x}$ is an element $\hat{x}$ of
$\hat{\mathcal{O}}^p_F$, that corresponds to an element $x$ of
$\mathcal{O}_F^p$ in the manner described above.

We choose an integral basis $v_i$, $1 \leq i \leq m$, for $F$, such that $v_1
= 1$.  For example, we could choose a canonical integral basis, associated
with the algebraic integer $\varphi$, in terms of which we defined the
matrices $\mathcal{F}$, $\hat{\mathcal{F}}$, and $\tilde{\mathcal{F}}$ above.
We recall, from the beginning of this subsection, that the matrix elements
$x_{ij}$ of the representation of an element $x$ of $F$, for the basis given
by the $v_i$, are defined by $xv_i = v_j x_{ji}$, where the summation
convention is used.  Thus for the basis element $v_k$ we find $v_k v_1 = v_k =
v_j \left( v_k \right)_{j 1}$, hence by the linear independence of the basis
elements, we must have $\left( v_k \right)_{j 1} = \delta_{kj}$.  On the other
hand, we can also express a general element $x$ of $F$, in the integral basis
$v_i$, as $x = x_k v_k$.  Hence $x_{i 1} = x_k \left( v_k \right)_{i 1} =
x_i$.  Thus the elements of the first column of $x_{ij}$ are the expansion
coefficients of $x$ in the integral basis $v_i$.

Let $s$ be the inverse of the similarity transformation that diagonalizes the
matrix representations of elements of $F$ in the integral basis $v_i$, so that
$S$ consists of $p$ copies of $s$, each ``spread out'', as described above.
Thus if $x$ is a general element of $F$, and $\hat{x}$ is the diagonal $m
\times m$ matrix, whose matrix elements on the diagonal are the $m$ Galois
conjugates of $x$, taken in the same order as we chose above, then $\left(
s^{- 1}  \hat{x} s \right)_{ij} = x_{ij}$, where $x_{ij}$ are the matrix
elements of $x$ in the basis $v_i$, which by assumption has $v_1 = 1$.  Now
this equation remains true, with the same $\hat{x}$ and $x_{ij}$, if we
pre-multiply $s$ by an arbitrary diagonal matrix, so that $s^{- 1}$ gets
post-multiplied by the inverse of that diagonal matrix.  Thus by
pre-multiplying $s$ by a suitable diagonal matrix, we can assume that every
nonzero matrix element, in the first column of $s$, is equal to $1$.
Furthermore, no matrix element in the first column of $s$ can be zero.  For by
assumption, each matrix element of the first column of $\hat{x} s$ is either
the appropriate Galois conjugate of $x$, or zero.  And by the preceding
paragraph, the set of the first columns of $\left( s^{- 1}  \hat{v}_i s
\right)$, for the $m$ basis elements $v_i$, is a set of $m$ linearly
independent column vectors of real numbers, namely $\left(\begin{array}{c}
  1\\
  0\\
  \vdots\\
  0
\end{array}\right), \left(\begin{array}{c}
  0\\
  1\\
  \vdots\\
  0
\end{array}\right), \ldots, \left(\begin{array}{c}
  0\\
  0\\
  \vdots\\
  1
\end{array}\right)$.  But this would be impossible, if any matrix element of
the first column of $s$ was zero, because the set of the first columns of
$\left( \hat{v}_i s \right)$ would not then be a set of $m$ linearly
independent column vectors of real numbers.  Thus we can assume that every
matrix element, in the first column of $s$, is equal to $1$, as in the example
above.

We now choose all the matrix elements in the first column of $s$ to be equal
to $1$, so that for an arbitrary $p$-vector $x$ in $\mathcal{O}^p_F$, the
$pm$-vector $S^{- 1}  \hat{x}$ has integer components in $\mathbf{Z}$, which
are the expansion coefficients of the successive matrix elements of $x$, with
respect to the integral basis $\left\{ v_i \right\}$, and for an arbitrary
$pm$-vector $\tilde{x}$ in $\mathbf{Z}^{pm}$, the $pm$-vector $S \tilde{x}$
is an element $\hat{x}$ of $\hat{\mathcal{O}}^p_F$, that corresponds to an
element $x$ of $\mathcal{O}_F^p$ in the manner described above.

We next note that by choosing each matrix element in the first column of $s$
to be equal to $1$, we have guaranteed that the matrix $V \equiv s^T s$ has
rational matrix elements.  For by the definition of $s$, we have, for an
arbitrary element $x$ of $F$, that $\tilde{x} s^{- 1} = s^{- 1} \lambda$,
where $\lambda$ is a diagonal matrix.  Hence $\tilde{x} = s^{- 1} \lambda s$,
and $\tilde{x}^T = s^T \lambda \left( s^{- 1} \right)^T = V \tilde{x} V^{-
1}$, hence $\left( V \tilde{x} \right)^T = V \tilde{x}$, or in other words, $V
\tilde{x}$ is a symmetric matrix.  If we now regard $V$ as an independent
symmetric matrix, and impose this condition on $V$, for an arbitrary primitive
element $x$ of $F$, then since all the eigenvalues of $\tilde{x}$ are distinct
for $x$ primitive, we find $\frac{1}{2} m \left( m - 1 \right)$ linearly
independent equations among the matrix elements of $V$, with coefficients
linear in the matrix elements of $\tilde{x}$, and thus rational numbers.  For
if we regard $V$ as an independent symmetric matrix, and express the equation
$\tilde{x}^T V = V \tilde{x}$ in a basis in which $\tilde{x}$ is diagonal, or
in other words, if we write the equation as $\lambda \left( s^{- 1} \right)^T
Vs^{- 1} = \left( s^{- 1} \right)^T Vs^{- 1} \lambda$, and treat this as an
equation for the symmetric matrix $V$, without making use of the relation
between $s$ and $V$, then the fact that $\lambda$ is a diagonal matrix, all of
whose eigenvalues are different, implies that the equation $\tilde{x}^T V = V
\tilde{x}$ is equivalent to $\frac{1}{2} m \left( m - 1 \right)$ linearly
independent relations among the matrix elements of the symmetric matrix $V$,
of the form $\left( \lambda_i - \lambda_j \right) \left( \left( s^{- 1}
\right)^T Vs^{- 1} \right)_{ij} = 0$, $1 \leq i < j \leq m$, where the
summation convention is not applied to $i$ and $j$.  But the number of
linearly independent relations in the matrix equation $\tilde{x}^T V = V
\tilde{x}$ is independent of what basis we express it in, hence for $x$
primitive this matrix equation gives $\frac{1}{2} m \left( m - 1 \right)$
linearly independent linear relations, with rational coefficients, among the
$\frac{1}{2} m \left( m + 1 \right)$ independent matrix elements of the
symmetric matrix $V$, which we can use to express $\frac{1}{2} m \left( m - 1
\right)$ matrix elements of $V$ as linear combinations, with rational
coefficients, of the remaining $m$ independent matrix elements.  Furthermore,
the form of the equation, in the basis in which all the $\tilde{x}$ are
diagonal, which simply states that the symmetric matrix $\left( s^{- 1}
\right)^T Vs^{- 1}$ is diagonal, shows that no further information can be
obtained, by imposing the relation $\tilde{x}^T V = V \tilde{x}$, for any
further $\tilde{x}$.

We choose to use the relation $\tilde{x}^T V = V \tilde{x}$, for one primitive
$x$, to express all the $\frac{1}{2} m \left( m - 1 \right)$ independent
$V_{ij}$, $2 \leq i \leq m$, $2 \leq j \leq m$, as linear combinations, with
rational coefficients, of the $V_{i 1}$, $1 \leq i \leq m$.  And making use,
now, of the definition $V = s^T s$, and the fact that we have set $s_{i 1} =
1$, for all $1 \leq i \leq m$, we find that:
\begin{equation}
  \label{V in terms of s} V = \left( \begin{array}{cccc}
    m & s_{12} + s_{22} + \ldots + s_{m 2} & \ldots & s_{1 m} + s_{2 m} +
    \ldots + s_{mm}\\
    s_{12} + s_{22} + \ldots + s_{m 2} & s_{12}^2 + s_{22}^2 + \ldots +
    s_{32}^2 & \ldots & s_{12} s_{1 m} + \ldots + s_{m 2} s_{mm}\\
    \vdots & \vdots & \ddots & \vdots\\
    s_{1 m} + s_{2 m} + \ldots + s_{mm} & s_{12} s_{1 m} + \ldots + s_{m 2}
    s_{mm} & \ldots & s_{1 m}^2 + s_{2 m}^2 + \ldots + s_{mm}^2
  \end{array} \right)
\end{equation}
Thus it remains to check that the linear combinations $s_{12} + s_{22} +
\ldots + s_{m 2}$, $s_{13} + s_{23} + \ldots + s_{m 3}$, ..., $s_{1 m} + s_{2
m} + \ldots + s_{mm}$, are rational numbers.  To do this, we use the fact that
$s^T$ is the matrix of eigenvectors of the $\tilde{x}$, or in other words, for
all $x$ in $F$, we have $\tilde{x}^T s^T = s^T \lambda$, where $\lambda$ is
the diagonal matrix of eigenvalues of $\tilde{x}^T$.  We have chosen the top
matrix element of each column of $s^T$ to be equal to $1$.  The sums whose
rationality we want to determine, are the sums of the matrix elements across
the rows 2 to $m$ of $s^T$.  In other words, for each row of $s^T$ after the
first, we need to check that the sum of all the matrix elements in that row of
$s^T$ is rational.  We choose a primitive element $x$ of $F$, so that the
eigenvalues $\lambda_i$ of $\tilde{x}^T$ are all distinct, and for each column
$i$ of $s^T$, use rows $2$ to $m$ of the equation $\tilde{x}^T s^T = s^T
\lambda$, to express each matrix element after the first of that column of
$s^T$, or in other words, the $s_{ij}$, $2 \leq j \leq m$, as a ratio of
multinomials formed from the matrix elements of $\tilde{x}^T$, the eigenvalue
$\lambda_i$ for that column of $s^T$, and the top matrix element of that
column of $s^T$, which is $1$.  When we do this for all the columns $i$ of
$s^T$, $1 \leq i \leq m$, we find that for all the $s_{ij}$ in each row $j$ of
$s^T$, we have a formula of the form
\begin{equation}
  \label{formula for the s i j in row j of s T} s_{ij} = \frac{f_j \left(
  \lambda_i \right)}{g_j \left( \lambda_i \right)}
\end{equation}
where $f_j \left( \lambda_i \right)$ and $g_j \left( \lambda_i \right)$ are
multinomials in the matrix elements of $\tilde{x}^T$ and $\lambda_i$, such
that the dependence of $f_j \left( \lambda_i \right)$ and $g_j \left(
\lambda_i \right)$ on the matrix elements of $\tilde{x}^T$ is the same for all
the $s_{ij}$ in the row $j$ of $s^T$.  For example, for $m = 3$ we find, for
row 2 of $s^T$, that:
\begin{eqnarray}
  \label{s i 2 for m equals 3} s_{12} & = & - \frac{\tilde{x}_{12}  \left(
  \lambda_1 - \tilde{x}_{33} \right) + \tilde{x}_{13}
  \tilde{x}_{32}}{\lambda_1  \tilde{x}_{33} + \tilde{x}_{22}  \left( \lambda_1
  - \tilde{x}_{33} \right) + \tilde{x}_{23}  \tilde{x}_{32} - \lambda_1^2}
  \nonumber\\
  s_{22} & = & - \frac{\tilde{x}_{12}  \left( \lambda_2 - \tilde{x}_{33}
  \right) + \tilde{x}_{13}  \tilde{x}_{32}}{\lambda_2  \tilde{x}_{33} +
  \tilde{x}_{22}  \left( \lambda_2 - \tilde{x}_{33} \right) + \tilde{x}_{23}
  \tilde{x}_{32} - \lambda_2^2} \nonumber\\
  s_{32} & = & - \frac{\tilde{x}_{12}  \left( \lambda_3 - \tilde{x}_{33}
  \right) + \tilde{x}_{13}  \tilde{x}_{32}}{\lambda_3  \tilde{x}_{33} +
  \tilde{x}_{22}  \left( \lambda_3 - \tilde{x}_{33} \right) + \tilde{x}_{23}
  \tilde{x}_{32} - \lambda_3^2}
\end{eqnarray}
From the structure of the formula (\ref{formula for the s i j in row j of s
T}), we see that $\sum^m_{i = 1} s_{ij}$ is a symmetric function of the
eigenvalues $\lambda_i$, and can in fact be expressed as the ratio of two
symmetric multinomials in the $\lambda_i$.  Thus it is equal to the ratio of
two multinomials in the matrix elements of $\tilde{x}$, so it is a rational
number.  Thus the matrix elements of the matrix $V = s^T s$ are rational
numbers.

Now the quadratic form $x^T Bx$ is preserved by all elements $g$ of $G_F$, for
by definition of $G$, we have $g^T Bg = B$, for all elements $g$ of $G_F$.
And similarly, we have $\hat{g}^T  \hat{B}  \hat{g} = \hat{B}$, for all
elements $\hat{g}$ of $\hat{G}$, because this equation is block diagonal, with
each of the $m$ blocks on the block diagonal being one of the $m$ Galois
conjugates of the first $p \times p$ block on the block diagonal, which is an
equation of the form $g^T Bg = B$, with $g \in G_F$.  We now define the
symmetric matrix $\breve{B} = S^T  \hat{B} S = S^T S \tilde{B}$, where
$\tilde{B} = S^{- 1}  \hat{B} S$ as above, and the $S$ is constructed from $p$
``spread out'' copies of $s$, as described above.  Then since the matrix
elements of $\tilde{B}$ are ordinary integers, and the matrix elements of $S^T
S$ are rational numbers, the matrix elements of $\breve{B}$ are rational
numbers.  We next note that, for all elements $\tilde{g}$ of $\tilde{G}$, we
have:
\begin{equation}
  \label{B breve is preserved by G tilde} \tilde{g}^T  \breve{B}  \tilde{g} =
  \left( S^{- 1}  \hat{g} S \right)^T  \left( S^T  \hat{B} S \right) S^{- 1}
  \hat{g} S = S^T  \hat{g}^T  \hat{B}  \hat{g} S = S^T  \hat{B} S = \breve{B}
\end{equation}
Now let $\tilde{x}$ be an arbitrary nonzero $pm$-vector with integer
components, or in other words, an arbitrary nonzero element of
$\mathbf{Z}^{pm}$.  Then $\hat{x} = S \tilde{x}$ is a nonzero element of
$\hat{\mathcal{O}}^p_F$, that corresponds to a nonzero element $x$ of
$\mathcal{O}_F^p$ in the manner described above.  Hence, from above, the value
of $\tilde{x}^T  \breve{B}  \tilde{x} = \left( S^{- 1}  \hat{x} \right)^T
\left( S^T  \hat{B} S \right) S^{- 1}  \hat{x} = \hat{x}^T  \hat{B}  \hat{x}$
is an ordinary integer in $\mathbf{Z}$, and its magnitude is $\geq 1$.  The
fact that the set of possible values of $\tilde{x}^T  \breve{B}  \tilde{x}$ is
discrete, and that there is a minimum distance $> 0$ between adjacent possible
values of $\tilde{x}^T  \breve{B}  \tilde{x}$, also follows directly from the
fact that the matrix elements of $\breve{B}$ are rational numbers.

We next note that there is a real number $\lambda > 0$, such that for all $g
\in G_F$, and all nonzero $pm$-vectors $\tilde{x} \in \mathbf{Z}^{pm}$,
$\left| \tilde{g}  \tilde{x} \right| \geq \lambda$.  Here, and throughout the
following, $\left| \tilde{g}  \tilde{x} \right|$ has its usual meaning of
$\left| \tilde{g}  \tilde{x} \right| = \sqrt{\tilde{x}^T \tilde{g}^T
\tilde{g}  \tilde{x}}$.  For if $\left| \tilde{g}  \tilde{x} \right|$ could be
arbitrarily small, then the value of $\tilde{x}^T  \tilde{g}^T  \breve{B}
\tilde{g}  \tilde{x} = \tilde{x}^T  \breve{B}  \tilde{x}$ could be arbitrarily
close to $0$.  But by the preceding paragraph, the magnitude of $\tilde{x}^T
\breve{B}  \tilde{x}$ is $\geq 1$, for arbitrary nonzero $\tilde{x} \in
\mathbf{Z}^{pm}$.  Hence for nonzero $\tilde{x} \in \mathbf{Z}^{pm}$,
$\tilde{x}^T  \breve{B}  \tilde{x}$ cannot be arbitrarily close to $0$, hence
$\left| \tilde{g}  \tilde{x} \right|$ cannot be arbitrarily small.  Let
$\lambda > 0$ be the largest number such that for all $g \in G_F$, and all
nonzero $pm$-vectors $\tilde{x} \in \mathbf{Z}^{pm}$, $\left| \tilde{g}
\tilde{x} \right| \geq \lambda$.

Given an infinite sequence $\left\{ g_k \right\}$ of elements of $G_F$, the
plan now is to find, first, a sequence $\left\{ \beta_k \right\}$ of elements
of $\mathrm{SL} \left( pm,\mathbf{Z} \right)$, such that the sequence $\left\{
\tilde{g}_k \beta_k \right\}$ has an infinite Cauchy subsequence $\left\{
\tilde{g}_j \beta_j \right\}$, and then show that this infinite Cauchy
subsequence itself has an infinite subsequence $\left\{ \tilde{g}_i \beta_i
\right\}$, such that $\beta_i \beta^{- 1}_1$ is an element $\tilde{\gamma}_i$
of $\tilde{\Gamma}$, for all $i$ in this infinite subsequence.  The sequence
$\left\{ \hat{g}_i \hat{\gamma}_i \right\} = \left\{ S \tilde{g}_i
\tilde{\gamma}_i S^{- 1} \right\}$ is then an infinite Cauchy sequence of
block diagonal matrices in $\hat{G}$, such that all the $\hat{\gamma}_i$ are
in $\hat{\Gamma}$, and the sequence $\left\{ g_i \gamma_i \right\}$ of the
first $p \times p$ blocks, on the block diagonal, is an infinite Cauchy
sequence in $G_F$, such that the sequence $\left\{ g_i \right\}$ is an
infinite subsequence of the given sequence $\left\{ g_k \right\}$, and all the
$\gamma_i$ are in $\Gamma$.

To construct the required sequence $\beta_k$ of elements of $\mathrm{SL} \left(
pm,\mathbf{Z} \right)$, we first use a method of Mahler {\cite{Mahler
compactness theorem}} to
construct, for an arbitrary element $g$ of $G_F$, an element $\beta$ of
$\mathrm{SL} \left( pm,\mathbf{Z} \right)$, such that all matrix elements of
$\tilde{g} \beta$ are bounded above in magnitude in terms of $\lambda$, where
$\lambda > 0$ was defined above to be the largest number such that for all $g
\in G_F$, and all nonzero $pm$-vectors $\tilde{x} \in \mathbf{Z}^{pm}$,
$\left| \tilde{g}  \tilde{x} \right| \geq \lambda$.  The following form of
Mahler's construction is adapted from section (5.34) of {\cite{Morris}}.  We
define $l \equiv pm$.

Given an element $g$ of $G_F$, the required element $\beta$ of $\mathrm{SL}
\left( l,\mathbf{Z} \right)$ will be constructed column by column, as a
sequence of nonzero column vectors in $\mathbf{Z}^l$, that I shall call
$v_1, v_2, v_3, \ldots, v_l$.

We first choose $v_1 \in \mathbf{Z}^l \backslash \left\{ 0 \right\}$, where
$\backslash$ means ``outside'', such that $\left| \tilde{g} v_1 \right|$ has
its minimum possible value, for $v \in \mathbf{Z}^l \backslash \left\{ 0
\right\}$.  This is always possible, because $\tilde{g}$ is nonsingular, hence
$v^T  \tilde{g}^T  \tilde{g} v$ is a positive definite quadratic form, with no
flat directions.  Let $\pi_1$ denote the projection to the line $\mathbf{R}
\tilde{g} v_1$, and $\pi^{\perp}_1$ denote the projection to the subspace
orthogonal to this line.

We next choose $v_2 \in \mathbf{Z}^l \backslash \mathbf{R}v_1$, such that
$\left| \pi^{\perp}_1  \tilde{g} v_2 \right|$ has its minimum possible value,
for $v \in \mathbf{Z}^l \backslash \mathbf{R}v_1$.  This is always
possible, for the same reason as before.  Moreover, $\pi^{\perp}_1  \tilde{g}
v_2$ is unaltered by adding a multiple of $v_1$ to $v_2$.  For $k \in
\mathbf{Z}$, the values of $\left| \pi_1  \tilde{g}  \left( v_2 + kv_1
\right) \right| = \left| \frac{\tilde{g} v_1  \left( v^T_1  \tilde{g}^T
\tilde{g} \left( v_2 + kv_1 \right) \right)}{v^T_1  \tilde{g}^T  \tilde{g}
v_1} \right| = \left| \tilde{g} v_1 \right| \left( \frac{v^T_1  \tilde{g}^T
\tilde{g} v_2}{v^T_1  \tilde{g}^T  \tilde{g} v_1} + k \right)$ are spaced by
$\left| \tilde{g} v_1 \right|$, so by replacing $v_2$ by $v_2 + kv_1$, with a
suitable value of $k$, we can assume that $\left| \pi_1  \tilde{g} v_2 \right|
\leq \frac{1}{2}  \left| \tilde{g} v_1 \right|$.  Then from the minimality of
$\left| \tilde{g} v_1 \right|$, we have that:
\begin{equation}
  \label{first Mahler bound} \left| \tilde{g} v_1 \right| \leq \left|
  \tilde{g} v_2 \right| \leq \left| \pi^{\perp}_1  \tilde{g} v_2 \right| +
  \left| \pi_1  \tilde{g} v_2 \right| \leq \left| \pi^{\perp}_1  \tilde{g} v_2
  \right| + \frac{1}{2}  \left| \tilde{g} v_1 \right|
\end{equation}
Hence:
\begin{equation}
  \label{second Mahler bound} \left| \pi^{\perp}_1  \tilde{g} v_2 \right| \geq
  \frac{1}{2}  \left| \tilde{g} v_1 \right|
\end{equation}
Let $\pi_2$ denote the projection to the plane $\mathbf{R} \tilde{g} v_1
+\mathbf{R} \tilde{g} v_2$, and $\pi^{\perp}_2$ denote the projection to the
subspace orthogonal to this plane.

We next choose $v_3 \in \mathbf{Z}^l \backslash \left( \mathbf{R}v_1
+\mathbf{R}v_2 \right)$, such that $\left| \pi^{\perp}_2  \tilde{g} v_3
\right|$ has its minimum possible value, for $v \in \mathbf{Z}^l \backslash
\left( \mathbf{R}v_1 +\mathbf{R}v_2 \right)$.  This is always possible,
for the same reason as before.  Moreover, $\pi^{\perp}_2  \tilde{g} v_3$ is
unaltered by adding multiples of $v_1$ and $v_2$ to $v_3$.  We first add a
suitable integer multiple of $v_2$, to arrange that $\left| \pi_2
\pi^{\perp}_1 \tilde{g} v_3 \right| \leq \frac{1}{2} \left| \pi^{\perp}_1
\tilde{g} v_2 \right|$.  Then, without affecting this bound, we add a suitable
integer multiple of $v_1$, to arrange that $\left| \pi_1  \tilde{g} v_3
\right| \leq \frac{1}{2}  \left| \tilde{g} v_1 \right|$.  The minimality of
$\left| \pi^{\perp}_1  \tilde{g} v_2 \right|$ now implies:
\begin{equation}
  \label{third Mahler bound} \left| \pi^{\perp}_1  \tilde{g} v_2 \right| \leq
  \left| \pi^{\perp}_1  \tilde{g} v_3 \right| \leq \left| \pi^{\perp}_2
  \tilde{g} v_3 \right| + \left| \pi_2 \pi^{\perp}_1 \tilde{g} v_3 \right|
  \leq \left| \pi^{\perp}_2  \tilde{g} v_3 \right| + \frac{1}{2} \left|
  \pi^{\perp}_1  \tilde{g} v_2 \right|
\end{equation}
Hence:
\begin{equation}
  \label{fourth Mahler bound} \left| \pi^{\perp}_2  \tilde{g} v_3 \right| \geq
  \frac{1}{2}  \left| \pi^{\perp}_1  \tilde{g} v_2 \right|
\end{equation}
Then we carry on after this pattern, until we eventually choose \\
$v_l \in
\mathbf{Z}^l \backslash \left( \mathbf{R}v_1 + \ldots +\mathbf{R}v_{l -
1} \right)$, such that $\left| \pi^{\perp}_{l - 1}  \tilde{g} v_l \right|$ has
its minimum possible value, for $v \in \mathbf{Z}^l \backslash \left(
\mathbf{R}v_1 + \ldots +\mathbf{R}v_{l - 1} \right)$.  Then by
successively adding suitable integer multiples of $v_{l - 1}$, $v_{l - 2}$,
..., $v_2$, and $v_1$, we arrange that $\left| \pi_{l - 1} \pi^{\perp}_{l - 2}
\tilde{g} v_l \right| \leq \frac{1}{2} \left| \pi^{\perp}_{l - 2}  \tilde{g}
v_{l - 1} \right|$, $\left| \pi_{l - 2} \pi^{\perp}_{l - 3}  \tilde{g} v_l
\right| \leq \frac{1}{2} \left| \pi^{\perp}_{l - 3}  \tilde{g} v_{l - 2}
\right|$, ..., $\left| \pi_1  \tilde{g} v_l \right| \leq \frac{1}{2}  \left|
\tilde{g} v_1 \right|$.  Then from the minimality of $\left| \pi^{\perp}_{l -
2}  \tilde{g} v_{l - 1} \right|$, we find, in the same way as before, that:
\begin{equation}
  \label{fifth Mahler bound} \left| \pi^{\perp}_{l - 1}  \tilde{g} v_l \right|
  \geq \frac{1}{2}  \left| \pi^{\perp}_{l - 2}  \tilde{g} v_{l - 1} \right|
\end{equation}
We next note that the successive minimality of $\left| \tilde{g} v_1 \right|$,
$\left| \pi^{\perp}_1  \tilde{g} v_2 \right|$, $\left| \pi^{\perp}_2
\tilde{g} v_3 \right|$, ..., $\left| \pi^{\perp}_{l - 1}  \tilde{g} v_l
\right|$, implies in turn that the convex hull of $\left\{ 0, v_1 \right\}$
contains no points of $\mathbf{Z}^l$ other than $0$ and $v_1$, the convex
hull of $\left\{ 0, v_1, v_2 \right\}$ contains no points of $\mathbf{Z}^l$
other than $0$, $v_1$, and $v_2$, ..., and finally that the convex hull of
$\left\{ 0, v_1, v_2, \ldots, v_l \right\}$ contains no points of
$\mathbf{Z}^l$ other than $0$, $v_1$, $v_2$, ..., $v_l$.  Hence the
parallelepiped generated by the vectors $v_1$, $v_2$, ..., $v_l$, whose
vertices are the expressions of the form $k_1 v_1 + k_2 v_2 + \ldots + k_l
v_l$, where each $k_i$ can independently take the values $0$ or $1$, contains
no points of $\mathbf{Z}^l$ in its convex hull, other than its $2^l$
vertices.

Now by considering tesselations of $\mathbf{R}^l$ by lattice
parallelepipeds, the volume of a lattice parallelepiped is given, in terms of
the points of $\mathbf{Z}^l$ in its convex hull, by:
\begin{equation}
  \label{volume of lattice parallelepiped} V = \frac{1}{2^l} f_0 +
  \frac{1}{2^{l - 1}} f_1 + \frac{1}{2^{l - 2}} f_2 + \ldots + \frac{1}{2}
  f_{l - 1} + f_l,
\end{equation}
where $f_0$ is the number of points of $\mathbf{Z}^l$ that are vertices of
the parallelpiped, $f_1$ is the number of points of $\mathbf{Z}^l$ that lie
within the ``interiors'' of edges of the parallepiped, ..., and $f_l$ is the
number of points of $\mathbf{Z}^l$ that lie within the interior of the
$l$-volume of the parallepiped.  Hence in the present instance, the volume of
the parallelepiped generated by the vectors $v_1$, $v_2$, ..., $v_l$, is $1$,
hence the determinant of $\beta$, which is defined to be the $l \times l$
matrix whose columns are $v_1$, $v_2$, ..., $v_l$, is $\pm 1$.  And if $\det
\beta = - 1$, we note that we can replace $v_l$ by $- v_l$, which is also in
$\mathbf{Z}^l \backslash \left( \mathbf{R}v_1 + \ldots +\mathbf{R}v_{l -
1} \right)$, without affecting the minimality of $\left| \pi^{\perp}_{l - 1}
\tilde{g} v_l \right|$, and we therefore replace $v_l$ by $- v_l$, to obtain
$\det \beta = 1$.

Furthermore:
\begin{equation}
  \label{Mahler determinant equation} \left| \pi^{\perp}_{l - 1}  \tilde{g}
  v_l \right|  \left| \pi^{\perp}_{l - 2}  \tilde{g} v_{l - 1} \right|  \left|
  \pi^{\perp}_{l - 3}  \tilde{g} v_{l - 2} \right| \ldots \left| \pi^{\perp}_1
  \tilde{g} v_2 \right|  \left| \tilde{g} v_1 \right| = \left| \det \left(
  \tilde{g} \beta \right) \right| = 1
\end{equation}
Hence from (\ref{second Mahler bound}), (\ref{fourth Mahler bound}), ..., and
(\ref{fifth Mahler bound}), we find that:
\begin{equation}
  \label{sixth Mahler bound} \left| \tilde{g} v_1 \right| \leq 2^{\frac{l -
  1}{2}}
\end{equation}
Furthermore, since $v_1 \neq 0$, we have $\left|
\tilde{g} v_1 \right| \geq \lambda > 0$.  Therefore, returning to (\ref{Mahler
determinant equation}), and the bounds (\ref{second Mahler bound}),
(\ref{fourth Mahler bound}), ..., we find that
\[ \left| \pi^{\perp}_1  \tilde{g} v_2 \right| \leq 2^{\frac{l - 2}{2}}
   \left( \frac{1}{\lambda} \right)^{\frac{1}{l - 1}} \]
\[ \left| \pi^{\perp}_2  \tilde{g} v_3 \right| \leq 2^{\frac{1}{l - 2}}
   2^{\frac{l - 3}{2}}  \left( \frac{1}{\lambda} \right)^{\frac{2}{l - 2}} \]
\[ \ldots \]
\[ \left| \pi^{\perp}_{k - 1}  \tilde{g} v_k \right| \leq 2^{\frac{\left( k -
   2 \right) \left( k - 1 \right)}{2 \left( l + 1 - k \right)}} 2^{\frac{l -
   k}{2}} \left( \frac{1}{\lambda} \right)^{\frac{k - 1}{l + 1 - k}} \]
\[ \ldots \]
\begin{equation}
  \label{Mahler perpendicular bounds} \left| \pi^{\perp}_{l - 1}  \tilde{g}
  v_l \right| \leq 2^{\frac{\left( l - 2 \right) \left( l - 1 \right)}{2}}
  \left( \frac{1}{\lambda} \right)^{l - 1}
\end{equation}
Furthermore, since $\left| \pi_1  \tilde{g} v_2 \right| \leq \frac{1}{2}
\left| \tilde{g} v_1 \right|$, we find:
\begin{equation}
  \label{bound on g hat v2} \left| \tilde{g} v_2 \right| \leq \left|
  \pi^{\perp}_1  \tilde{g} v_2 \right| + \left| \pi_1  \tilde{g} v_2 \right|
  \leq 2^{\frac{l - 2}{2}}  \left( \frac{1}{\lambda} \right)^{\frac{1}{l - 1}}
  + 2^{\frac{l - 3}{2}}
\end{equation}
And similarly:
\[ \left| \tilde{g} v_3 \right| \leq \left| \pi^{\perp}_2  \tilde{g} v_3
   \right| + \left| \pi_2 \pi^{\perp}_1  \tilde{g} v_3 \right| + \left| \pi_1
   \tilde{g} v_3 \right| \leq \hspace{6em} \]
\begin{equation}
  \label{bound on g hat v3} \hspace{2em} \leq 2^{\frac{1}{l - 2}} 2^{\frac{l -
  3}{2}}  \left( \frac{1}{\lambda} \right)^{\frac{2}{l - 2}} + 2^{\frac{l -
  4}{2}}  \left( \frac{1}{\lambda} \right)^{\frac{1}{l - 1}} + 2^{\frac{l -
  3}{2}}
\end{equation}
And so on.  Thus, since $\lambda > 0$, all matrix elements of $\tilde{g}
\beta$ are, indeed, bounded, independently of $\tilde{g}$.

Given an infinite sequence $\left\{ g_k \right\}$ of elements of $G_F$, we now
take, for each $k$, $\beta_k$ to be the matrix $\beta$, as constructed above,
with $\tilde{g}$ taken as $\tilde{g}_k$.  The elements of the sequence
$\left\{ \tilde{g}_k \beta_k \right\}$ are then bounded in terms of $\lambda$
as above, independently of $k$.  Hence this sequence has a Cauchy subsequence.
We can find a Cauchy subsequence by subdividing the bounded $l^2$-dimensional
domain of the matrix elements into a finite number of subsectors, choosing a
subsector in which the sequence has an infinite number of elements,
subdividing that subsector into an finite number of subsectors, choosing one
of them in which the sequence has an infinite number of elements, and so on.
Let $\left\{ \tilde{g}_j \beta_j \right\}$ be an infinite Cauchy subsequence
of the sequence $\left\{ \tilde{g}_k \beta_k \right\}$.

Now we found above that the matrix elements of $\breve{B}$ are rational
numbers.  Hence there is an integer $a \in \mathbf{Z}$ such that all the
matrix elements of $a \breve{B}$ are ordinary integers in $\mathbf{Z}$.  On
the other hand, the fact that $\left\{ \tilde{g}_j \beta_j \right\}$ is a
Cauchy sequence implies that the sequence
\begin{equation}
  \label{Cauchy sequence involving B breve} a \left\{ \beta^T_j  \tilde{g}^T_j
  \breve{B}  \tilde{g}_j \beta_j \right\} = a \left\{ \beta^T_j  \breve{B}
  \beta_j \right\}
\end{equation}
is a Cauchy sequence.  Hence since all the matrix elements of $a \beta^T_j
\breve{B} \beta_j$ are ordinary integers in $\mathbf{Z}$, there must be a
value $q_1$ of $j$ such that for all $s \geq q_1$ and all $t \geq q_1$, $a
\beta^T_s  \breve{B} \beta_s = a \beta^T_t  \breve{B} \beta_t$, hence $\left(
\beta_s \beta^{- 1}_t \right)^T  \breve{B} \beta_s \beta^{- 1}_t = \breve{B}$.
This result can also be obtained without directly using the fact that the
matrix elements of $\breve{B}$ are rational numbers, by using the fact that
for an arbitrary element $\tilde{x}$ of $\mathbf{Z}^l$, the value of
$\tilde{x}^T  \breve{B}  \tilde{x}$ is an ordinary integer in $\mathbf{Z}$,
and considering the Cauchy sequence $\left\{ \tilde{x}^T \beta^T_j
\tilde{g}^T_j  \breve{B}  \tilde{g}_j \beta_j  \tilde{x} \right\} = \left\{
\tilde{x}^T \beta^T_j  \breve{B} \beta_j  \tilde{x} \right\}$ for $\frac{1}{2}
l \left( l + 1 \right)$ suitable choices of $\tilde{x}$, such as the $l$ unit
vectors in the positive coordinate directions, and the $\frac{1}{2} l \left( l
- 1 \right)$ distinct sums of two such unit vectors.  This procedure can also
be used to give an alternative proof that the matrix elements of $\breve{B}$
are rational numbers, without using the fact that the matrix elements of $s^T
s$ are rational numbers.

Furthermore, the fact that $\left\{ \tilde{g}_j \beta_j \right\}$ is a Cauchy
sequence implies that the sequence
\begin{equation}
  \label{Cauchy sequence involving calligraphic F} \left\{ \beta^{- 1}_j
  \tilde{g}^{- 1}_j  \tilde{\mathcal{F}}  \tilde{g}_j \beta_j \right\} =
  \left\{ \beta^{- 1}_j  \tilde{\mathcal{F}} \beta_j \right\}
\end{equation}
is a Cauchy sequence.  Hence since the matrix elements of
$\tilde{\mathcal{F}}$ are ordinary integers in $\mathbf{Z}$, hence the
matrix elements of $\beta^{- 1}_j  \tilde{\mathcal{F}} \beta_j$ are ordinary
integers in $\mathbf{Z}$, there must be a value $q_2$ of $j$ such that for
all $s \geq q_2$ and all $t \geq q_2$, $\beta^{- 1}_s  \tilde{\mathcal{F}}
\beta_s = \beta^{- 1}_t  \tilde{\mathcal{F}} \beta_t$, hence
$\tilde{\mathcal{F}} \beta_s \beta^{- 1}_t = \beta_s \beta^{- 1}_t
\tilde{\mathcal{F}}$.

And finally, in the unitary case, the preceding paragraph is also valid with
$\tilde{\mathcal{F}}$ replaced by $\tilde{\mathcal{J}}$, hence there must be a
value $q_3$ of $j$ such that for all $s \geq q_3$ and all $t \geq q_3$,
$\tilde{\mathcal{J}} \beta_s \beta^{- 1}_t = \beta_s \beta^{- 1}_t
\tilde{\mathcal{J}}$.

Hence there is a value $r$ of $j$, namely the maximum of $q_1$ and $q_2$ in
the orthogonal case, and the maximum of $q_1$, $q_2$, and $q_3$ in the unitary
case, such that for all $s \geq r$ and all $t \geq r$, $\beta_s \beta^{- 1}_t$
is an element of $\tilde{G}$.  Let $\tilde{\gamma}_j \equiv \beta_j \beta^{-
1}_r$ for all $j \geq r$, and let $\left\{ \tilde{g}_i \beta_i \right\}$ be
the infinite Cauchy sequence obtained from $\left\{ \tilde{g}_j \beta_j
\right\}$ by dropping all terms with $j < r$.  Then $\left\{ \tilde{g}_i
\beta_i \beta^{- 1}_r \right\} = \left\{ \tilde{g}_i  \tilde{\gamma}_i
\right\}$ is an infinite Cauchy sequence in $\tilde{G}$, such that $\left\{
\tilde{g}_i \right\}$ is an infinite subsequence of $\left\{ \tilde{g}_k
\right\}$, and $\tilde{\gamma}_i \in \tilde{G}$, for all $i$.  Then as
anticipated above, the sequence $\left\{ \hat{g}_i \hat{\gamma}_i \right\} =
\left\{ S \tilde{g}_i \tilde{\gamma}_i S^{- 1} \right\}$ is an infinite Cauchy
sequence of block diagonal matrices in $\hat{G}$, such that all the
$\hat{\gamma}_i$ are in $\hat{\Gamma}$, and the sequence $\left\{ g_i \gamma_i
\right\}$ of the first $p \times p$ blocks, on the block diagonal, is the
required infinite Cauchy sequence in $G_F$, such that the sequence $\left\{
g_i \right\}$ is an infinite subsequence of the given sequence $\left\{ g_k
\right\}$, and all the $\gamma_i$ are in $\Gamma$.

\subsubsection{Obtaining finite index torsion-free subgroups of $\Gamma$ by
Selberg's lemma}

We recall from above that for compact quotients of $\mathrm{SU} \left( n, 1
\right)$ or $\mathrm{SO} \left( n, 1 \right)$, the requirement that the quotient
be smooth, rather than an orbifold, or in other words, that all elements $\neq
1$ of the discrete subgroup $\Gamma$ act on the symmetric space $\mathbf{C}
\mathbf{H}^n = \mathrm{SU} \left( n, 1 \right) / \left( \mathrm{SU} \left( n
\right) \times U \left( 1 \right) \right)$ or $\mathbf{H}^n = \mathrm{SO}
\left( n, 1 \right) / \mathrm{SO} \left( n \right)$ without fixed points, is
equivalent to the requirement that $\Gamma$ have no torsion, or in other
words, no nontrivial finite subgroups.  Selberg's lemma {\cite{Selberg}}, for
the case of arithmetic lattices such as those in the examples above, states
that certain finite index subgroups of these discrete groups have no torsion.
We recall that a subgroup $\Gamma_1$ of a discrete group $\Gamma$ is said to
have finite index in $\Gamma$, if $\Gamma_1$ divides $\Gamma$ into a finite
number of left cosets.  Thus if $G / \Gamma$ is compact, and $\Gamma_1$ has
finite index in $\Gamma$, then $G / \Gamma_1$ is also compact, so for all the
examples above, we can obtain smooth compact quotients of $\mathbf{C}
\mathbf{H}^n$ or $\mathbf{H}^n$ by using any of the subgroups of $\Gamma$
specified by Selberg's lemma for this case.  I shall briefly review Selberg's
lemma for the case of these arithmetic lattices, following section (5.60) of
{\cite{Morris}}.

As in the previous subsection, we define $p = 2 \left( n + 1 \right)$ if $G =
\mathrm{SU} \left( n, 1 \right)$, and $p = n + 1$ if $G = \mathrm{SO} \left( n, 1
\right)$, and $l \equiv pm$, where $m$ is the degree of $F$.  We choose an
integral basis for $F$, and represent each element $\gamma$ of $\Gamma$ as a
$p \times p$ block matrix, each block of which is the $m \times m$ matrix
representation of the corresponding matrix element of $\gamma$, in the chosen
integral basis.  Thus each element $\gamma$ of $\Gamma$ is represented by an
$l \times l$ matrix, with matrix elements in $\mathbf{Z}$.  In the preceding
subsection, such $l \times l$ matrices representing elements $\gamma$ of
$\Gamma$ were denoted $\tilde{\gamma}$, and the group of all of them was
denoted $\tilde{\Gamma}$, but since $\tilde{\Gamma}$ is isomorphic to
$\Gamma$, and this representation of $\Gamma$, as a group of $l \times l$
matrices, with matrix elements in $\mathbf{Z}$, is the only representation
of $\Gamma$ that will be used in the present subsection, I shall not use the
tildes in this subsection.  Thus we now regard $\Gamma$ as a subgroup of
$\mathrm{SL} \left( l,\mathbf{Z} \right)$.  The following construction is
valid for all subgroups $\Gamma$ of $\mathrm{SL} \left( l,\mathbf{Z} \right)$.

For $k \in \mathbf{Z}$, such that $k \geq 2$, let $\Gamma_k$ denote the set
of all elements of $\Gamma$ of the form $\left( 1 + kT \right)$, where $1$
denotes the $l \times l$ unit matrix, and the matrix elements of $T$ are in
$\mathbf{Z}$.  Then $\Gamma_k$ is a group, and is moreover a normal subgroup
of $\Gamma$, since if $\gamma$ is an element of $\Gamma$, then $\gamma$ is a
matrix with matrix elements in $\mathbf{Z}$, and determinant equal to $1$,
so $\gamma^{- 1} \left( 1 + kT \right) \gamma$ has the form $1 + kT_1$, where
the matrix elements of $T_1$ are in $\mathbf{Z}$.

We next note that elements $\gamma_1$ and $\gamma_2$ of $\Gamma$ are in the
same left left coset of $\Gamma_k$ in $\Gamma$, if and only if corresponding
matrix elements of $\gamma_1$ and $\gamma_2$ are equal, mod $k$.  For if
corresponding matrix elements of $\gamma_1$ and $\gamma_2$ are equal, mod $k$,
then $\gamma_2 = \gamma_1 + kT$, for some matrix $T$ with matrix elements in
$\mathbf{Z}$, hence $\gamma^{- 1}_1 \gamma_2 = 1 + k \gamma^{- 1}_1 T$,
which is in $\Gamma_k$, so $\gamma_1$ and $\gamma_2$ are in the same left
coset of $\Gamma_k$ in $\Gamma$, while if $\gamma_1$ and $\gamma_2$ are in the
same left coset of $\Gamma_k$ in $\Gamma$, then $\gamma_2 = \gamma_1 \left( 1
+ kT \right)$ for some matrix $T$ with matrix elements in $\mathbf{Z}$,
hence corresponding matrix elements of $\gamma_1$ and $\gamma_2$ are equal,
mod $k$.

Thus the quotient group $\Gamma / \Gamma_k$, of $\Gamma$ by its normal
subgroup $\Gamma_k$, is the group obtained from $\Gamma$, by considering its
matrix elements mod $k$.  Thus each matrix element of $\Gamma / \Gamma_k$
takes values in the finite set $\left\{ 0, 1, 2, \ldots, k - 1 \right\}$,
hence $\Gamma / \Gamma_k$ cannot have more than $k^{\left( l^2 \right)}$
elements, and is thus a finite group, and $\Gamma_k$ has finite index in
$\Gamma$.

We now demonstrate that for $k \geq 3$, $\Gamma_k$ has no torsion.  It is
sufficient to demonstrate that for an arbitrary element $\gamma$ of
$\Gamma_k$, such that $\gamma \neq 1$, no integer power $s \geq 1$ of $\gamma$
is equal to $1$, for if $\gamma$ is an element of a finite group, the sequence
$\left\{ 1, \gamma, \gamma^2, \gamma^3, \ldots \right\}$ cannot contain more
distinct elements than the number of elements of that finite group.

We assume now that $k \geq 3$.  Then $k$ is divisible by either $2^2$ or an
odd prime.  Furthermore, $\Gamma_j$ is a subgroup of $\Gamma_k$ whenever $k$
is a divisor of $j$, so it is sufficient to prove that $\Gamma_k$ has no
torsion when $k$ is either $2^2$ or an odd prime.  Thus we now assume $k =
p^r$, where $p$ is prime, and $r = 2$ for $p = 2$, and $r = 1$ for $p \geq 3$.
Furthermore, it is sufficient to prove that for an arbitrary element $\gamma$
of $\Gamma_k$, such that $\gamma \neq 1$, no power $\gamma^s$ is equal to $1$
for $s$ prime, since if $s$ factorizes as $s = tq$, where $q$ is prime, we can
write $\gamma^s = \left( \gamma^t \right)^q$.  Thus we now assume $s$ is
prime, so either $p$ does not divide $s$, or $p = s$.

We can write a general element of $\Gamma_k$ as $\left( 1 + p^u T \right)$,
where $u \geq r \geq 1$, and not every matrix element of $T$ is divisible by
$p$.  If $p$ does not divide $s$, we note that $\left( p^u T \right)^2$ is
equal to $0$, mod $p^{u + 1}$, hence $\left( 1 + p^u T \right)^s$ is equal to
$1 + sp^u T$, mod $p^{u + 1}$, which is $\neq 1$, mod $p^{u + 1}$.  And if $p
= s$, we note that $\left( p^u T \right)^3$ is equal to $0$, mod $p^{u + 2}$,
hence $\left( 1 + p^u T \right)^p = 1 + p^{u + 1} T + \frac{p^{2 u + 1} \left(
p - 1 \right)}{2} T^2$, mod $p^{u + 2}$.  Furthermore, if $p \geq 3$, then
$\left( p - 1 \right)$ is even, hence $\frac{p^{2 u + 1} \left( p - 1
\right)}{2}$ is an integer that is equal to $0$, mod $p^{u + 2}$, hence
$\left( 1 + p^u T \right)^p \neq 1$, mod $p^{u + 2}$, while if $p = 2$, then
$u \geq 2$, hence again $\frac{p^{2 u + 1} \left( p - 1 \right)}{2}$ is an
integer that is equal to $0$, mod $p^{u + 2}$, hence $\left( 1 + p^u T
\right)^p \neq 1$, mod $p^{u + 2}$.

An alternative method of constructing a torsion-free subgroup of $\Gamma$ has
been considered by Everitt and Maclachlan {\cite{Everitt Maclachlan}}, who
applied their method to obtain a construction of the Davis manifold
{\cite{Davis manifold}}, which is the smallest known smooth compact quotient of
$\mathbf{H}^4$.

\subsection{Smooth compact quotients of $S^3$}

\label{Smooth compact quotients of S3}

For even $n$, every real antisymmetric $\left( n + 1 \right) \times \left( n +
1 \right)$ matrix has a zero eigenvalue, hence no element of $\mathrm{SO} \left(
n + 1 \right)$ acts without fixed points on $\mathbf{S}^n$, and the only
smooth compact quotient of $\mathbf{S}^n$ is the non-orientable
$n$-dimensional real projective space, obtained from $\mathbf{S}^n$ by
identifying every point with its antipode.  On the other hand,
$\mathbf{S}^3$ is well known to have smooth compact quotients, which fall
into a small number of families, that were first classified by Seifert and
Threlfall {\cite{Threlfall Seifert 1, Threlfall Seifert 2}}.  Smooth compact
quotients of $\mathbf{S}^3$ have been considered recently as possible
topologies for the three observed spatial dimensions, in consequence of the
current slight preference of astrophysical data for $k = + 1$ rather than $k =
- 1$, as discussed in section \ref{Thick pipe geometries}, and have recently
been reclassified by Gausmann, Lehoucq, Luminet, Uzan, and Weeks
{\cite{Gausmann Lehoucq Luminet Uzan Weeks}}.

\section{The Casimir energy densities}
\label{The Casimir energy densities}

The validity of the realization of TeV-scale gravity by the thick pipe
geometries studied in section \ref{Thick pipe geometries}, for the
compactification of Ho\v{r}ava-Witten theory on a particular smooth compact
quotient of $\mathbf{C} \mathbf{H}^3$ or $\mathbf{H}^6$ that is a spin
manifold, and a particular choice of spin structure on that spin manifold,
depends on the Casimir energy densities on and near the inner surface of the
thick pipe resulting in the integration constant $B$, in (\ref{bulk type power
law for upper sign}), taking the value (\ref{B for TeV scale gravity}), or
(\ref{B for TeV scale gravity with fluxes}), and in the case where the outer
surface of the thick pipe is stabilized by Casimir effects, also on the
Casimir energy densities on and near the outer surface resulting in the
integration constant $\tilde{A}$, in (\ref{bulk power law for a dot in terms
of a with lower sign}), taking the value (\ref{A tilde for TeV scale
gravity}).  The Casimir energy densities are, by definition, the correction
terms in the field equations and boundary conditions for the graviton, when
they are derived by varying the full quantum effective action, or in other
words, the generating functional of the proper vertices, with respect to the
graviton field, rather than by varying the classical Cremmer-Julia-Scherk
action, augmented by supersymmetrized Gibbons-Hawking {\cite{York, Gibbons
Hawking, Luckock Moss, Moss 1, Moss 2, Moss 3}} terms and the semiclassical
Ho\v{r}ava-Witten supersymmetric Yang-Mills actions on the orbifold fixed-point
hyperplanes.

Now, as noted in connection with (\ref{action for calculating quantum
effective action}), on page \pageref{action for calculating quantum effective
action}, the quantum effective action $\Gamma \left( \Phi \right)$, for a
properly gauge-fixed classical action $A \left( \varphi \right)$, where
$\varphi$ denotes all the fields occurring in the gauge invariant classical
action, together with all the Faddeev-Popov ghosts {\cite{Feynman, DeWitt,
Faddeev Popov, t Hooft ghosts}}, and also the Nielsen-Kallosh ghosts
{\cite{Nielsen,
Kallosh ghosts}} if appropriate, can be calculated, for an arbitrary
classical field configuration $\Phi$, as the sum of all the one line
irreducible vacuum diagrams, calculated from the action $A \left( \Phi +
\varphi \right)$, with the term linear in $\varphi$ deleted, where $\varphi$
denotes the quantum fields.  In other words, using DeWitt's compact index
notation {\cite{DeWitt}}, where a single index, $i$,
runs over all combinations of type of field, space-time position, and
coordinate and other indices, the quantum effective action, as a function of
the classical fields, $\Phi$, is given by the sum of all the one line
irreducible vacuum diagrams, calculated with the action $A \left( \Phi +
\varphi \right) - \varphi_i \frac{\delta A \left( \Phi \right)}{\delta
\Phi_i}$, as in (\ref{action for calculating quantum effective action}), where
the summation convention is applied to the index $i$.

To check this, we note
that, with the functional integral defined as $Z \left( J \right) = e^{- iW
\left( J \right)} \hspace{-1.2pt} =
\int \left[ d \varphi \right] \exp i \left( A \left(
\varphi \right) + J_i \varphi_i \right)$, the classical field $\Phi_i$ defined
as $\Phi_i \equiv - \frac{\delta W}{\delta J_i}$, and $\Gamma \left( \Phi
\right)$ defined by a Legendre transformation by the relation $\Gamma \left(
\Phi \right) + J_i \Phi_i = - W \left( J \right)$, we have $J_i = - \iota^{ii}
\frac{\delta \Gamma}{\delta \Phi_i}$, where $\iota^{ij}$ is $- 1$ if both
$\varphi_i$ and $\varphi_j$ are fermionic, and $1$ otherwise, indices on
$\iota^{ij}$ are ignored in applying the summation convention, and all
derivatives act from the left {\cite{Schwinger, DeWitt effective action,
Jona-Lasinio}}.  We then have {\cite{Taylor Veneziano}}:
\begin{equation}
  \label{integro differential equation for Gamma of Phi} e^{i \Gamma \left(
  \Phi \right)} = \int \left[ d \varphi \right] e^{i \left( A \left( \varphi
  \right) + J_i \left( \varphi_i - \Phi_i \right) \right)} = \int \left[ d
  \varphi \right] e^{i \left( A \left( \varphi \right) - \left( \varphi_i -
  \Phi_i \right) \frac{\delta \Gamma}{\delta \Phi_i} \right)}
\end{equation}
which can be regarded as an alternative definition of $\Gamma \left( \Phi
\right)$.  Shifting the integration variables $\varphi_i$ by $\Phi_i$, we
have:
\begin{equation}
  \label{integro differential equation for Gamma of Phi with shifted
  integration variables} e^{i \Gamma \left( \Phi \right)} = \int \left[ d
  \varphi \right] e^{i \left( A \left( \Phi + \varphi \right) - \varphi_i
  \frac{\delta \Gamma}{\delta \Phi_i} \right)} = \int \left[ d \varphi \right]
  e^{i \left( \left( A \left( \Phi + \varphi \right) - \varphi_i \frac{\delta
  A}{\delta \Phi_i} \right) - \left( \varphi_i \frac{\delta \Gamma_1}{\delta
  \Phi_i} + \varphi_i \frac{\delta \Gamma_2}{\delta \Phi_i} + \ldots \right)
  \right)}
\end{equation}
where the loop expansion $\Gamma \left( \Phi \right) = A \left( \Phi \right) +
\Gamma_1 \left( \Phi \right) + \Gamma_2 \left( \Phi \right) + \ldots$ was
introduced.  Now if the term $- i \left( \varphi_i \frac{\delta
\Gamma_1}{\delta \Phi_i} + \varphi_i \frac{\delta \Gamma_2}{\delta \Phi_i} +
\ldots \right)$ in the exponent in the right-hand side of (\ref{integro
differential equation for Gamma of Phi with shifted integration variables})
was neglected, (\ref{integro differential equation for Gamma of Phi with
shifted integration variables}) would express $\Gamma \left( \Phi \right)$ as
the sum of all connected, but not necessarily one line irreducible, vacuum
bubbles, calculated with the action $\left( A \left( \Phi + \varphi \right) -
\varphi_i \frac{\delta A \left( \Phi \right)}{\delta \Phi_i} \right)$.  An
arbitrary such vacuum bubble can be regarded as a tree diagram, such that a
vertex of the tree diagram on which $n$ propagators end corresponds to $i
\frac{\delta^n \Gamma \left( \Phi \right)}{\delta \Phi_{i_1} \ldots \delta
\Phi_{i_n}}$, and a propagator of the tree diagram corresponds to $iG \left(
\Phi \right)_{i_1 i_2}$, where $G \left( \Phi \right)_{ij}$ is the inverse of
the matrix $\frac{\delta^2 A \left( \Phi \right)}{\delta \Phi_i \delta
\Phi_j}$.

And when the effects of the term $- i \left( \varphi_i \frac{\delta
\Gamma_1}{\delta \Phi_i} + \varphi_i \frac{\delta \Gamma_2}{\delta \Phi_i} +
\ldots \right)$ in the exponent are included, the only change is that each
vertex, of a tree diagram, on which precisely {\emph{one}} propagator ends,
can now come from either of two alternative sources, namely either as a one
line irreducible diagram built from the propagators $G \left( \Phi
\right)_{ij}$ and vertices $\frac{\delta^n A \left( \Phi \right)}{\delta
\Phi_{i_1} \ldots \delta \Phi_{i_n}}$, $n \geq 3$, of the action $\left( A
\left( \Phi + \varphi \right) - \varphi_i \frac{\delta A \left( \Phi
\right)}{\delta \Phi_i} \right)$, as before, or alternatively from the term of
appropriate loop order $\geq 1$ in the extra term $- i \left( \varphi_i
\frac{\delta \Gamma_1}{\delta \Phi_i} + \varphi_i \frac{\delta
\Gamma_2}{\delta \Phi_i} + \ldots \right)$.  The result of this is that a tree
diagram that contains $m$ vertices $i \frac{\delta \Gamma}{\delta \Phi_i}$, on
which precisely {\emph{one}} propagator ends, gets a factor $\left( 1 - 1
\right)^m$.  Hence since every tree diagram with more than one vertex contains
at least one such vertex, all the tree diagrams cancel out except for those
with precisely one vertex, and these are the one line irreducible vacuum
bubbles calculated with the action $\left( A \left( \Phi + \varphi \right) -
\varphi_i \frac{\delta A \left( \Phi \right)}{\delta \Phi_i} \right)$, as
stated.

Continuing to use DeWitt's abstract index notation, the one-loop effective
action, $\Gamma_1 \left( \Phi \right)$, is given, by (\ref{integro
differential equation for Gamma of Phi with shifted integration variables}),
by:
\begin{equation}
  \label{one loop effective action in terms of superdet} e^{i \Gamma_1 \left(
  \Phi \right)} = K_1 \int \left[ d \varphi \right] e^{\frac{i}{2}
  \frac{\delta^2 A \left( \Phi \right)}{\delta \Phi_j \delta \Phi_i} \varphi_i
  \varphi_j} = K_2  \frac{1}{\sqrt{\mathrm{superdet} \frac{\delta^2 A \left(
  \Phi \right)}{\delta \Phi_j \delta \Phi_i}}} = K_2 e^{- \frac{1}{2} \:
  \mathrm{supertr} \; \ln \:  \frac{\delta^2 A \left( \Phi \right)}{\delta
  \Phi_j \delta \Phi_i}}
\end{equation}
where $K_1$ and $K_2$ are constants independent of the fields $\Phi_i$, and
the matrix $\frac{\delta^2 A \left( \Phi \right)}{\delta \Phi_j \delta
\Phi_i}$ has been assumed to have a bose-bose part $\left. \frac{\delta^2 A
\left( \Phi \right)}{\delta \Phi_j \delta \Phi_i} \right|_b$, for which the
indices $j$ and $i$ denote bosonic fields, and a fermi-fermi part $\left.
\frac{\delta^2 A \left( \Phi \right)}{\delta \Phi_j \delta \Phi_i} \right|_f$,
for which the indices $j$ and $i$ denote fermionic fields, but no
non-vanishing matrix elements such that one of the indices $i$ and $j$ is
bosonic, and the other fermionic, in which case the superdeterminant
{\cite{Berezin 1, Berezin 2, 0512031 Khudaverdian Voronov}} is defined by:
\begin{equation}
  \label{definition of superdet} \mathrm{superdet} \frac{\delta^2 A \left( \Phi
  \right)}{\delta \Phi_j \delta \Phi_i} \equiv \frac{\det \left.
  \frac{\delta^2 A \left( \Phi \right)}{\delta \Phi_j \delta \Phi_i}
  \right|_b}{\det \left. \frac{\delta^2 A \left( \Phi \right)}{\delta \Phi_j
  \delta \Phi_i} \right|_f}
\end{equation}
and the bose-bose part $\left. \frac{\delta^2 A \left( \Phi \right)}{\delta
\Phi_j \delta \Phi_i} \right|_b$ has been assumed to have an infinitesimal
positive-definite imaginary part.

We assume now that $A \left( \Phi \right)$ has an expansion:
\begin{equation}
  \label{expansion of classical action in DeWitts notation} A \left( \Phi
  \right) = \frac{1}{2} A_{ij} \Phi_j \Phi_i + \frac{1}{6} A_{ijk} \Phi_k
  \Phi_j \Phi_i + \frac{1}{24} A_{ijkl} \Phi_l \Phi_k \Phi_j \Phi_i + \ldots
\end{equation}
and define $G \left( \Phi \right)_{ij}$ to be the inverse of the matrix
$\frac{\delta^2 A \left( \Phi \right)}{\delta \Phi_j \delta \Phi_i}$, and
$G_{ij}$ to be the inverse of the matrix $A_{ij}$.  Then since $A \left( \Phi
\right)$ is bosonic, and the assumed vanishing of all bose-fermi matrix
elements of $\frac{\delta^2 A \left( \Phi \right)}{\delta \Phi_j \delta
\Phi_i}$ thus implies that all non-vanishing matrix elements of
$\frac{\delta^2 A \left( \Phi \right)}{\delta \Phi_j \delta \Phi_i}$ are
bosonic, we have:
\[ G \left( \Phi \right)_{ij} = G_{ij} - G_{ik} \left( \frac{\delta^2 A \left(
   \Phi \right)}{\delta \Phi_k \delta \Phi_m} - A_{km} \right) G_{mj}
   \hspace{16em} \]
\begin{equation}
  \label{expansion of G of Phi} \hspace{8em} + G_{ik} \left( \frac{\delta^2 A
  \left( \Phi \right)}{\delta \Phi_k \delta \Phi_m} - A_{km} \right) G_{mn}
  \left( \frac{\delta^2 A \left( \Phi \right)}{\delta \Phi_n \delta \Phi_p} -
  A_{np} \right) G_{pj} - \ldots
\end{equation}
And from (\ref{one loop effective action in terms of superdet}) we have, up to
an additive constant, independent of all the fields $\Phi_i$:
\begin{displaymath}
\Gamma_1 = \frac{i}{2}
  \left( \iota^{jj} G_{jk} \left( \frac{\delta^2 A \left( \Phi \right)}{\delta
  \Phi_k \delta \Phi_j} - A_{kj} \right) \right. \hspace{45.0ex}
\end{displaymath}
\begin{equation}
  \label{one loop effective action in DeWitts notation}
  \hspace{20.0ex} \left. - \frac{1}{2} \iota^{jj} G_{jk}
  \left( \frac{\delta^2 A \left( \Phi \right)}{\delta \Phi_k \delta \Phi_l} -
  A_{kl} \right) G_{lm} \left( \frac{\delta^2 A \left( \Phi \right)}{\delta
  \Phi_m \delta \Phi_j} - A_{mj} \right) + \ldots \right)
\end{equation}
where the effect of the factors $\iota^{jj}$ is to introduce a $-$ sign when
the field circulating in the loop is fermionic.  The expression (\ref{one loop
effective action in DeWitts notation}) is well-known to be real, in Minkowski
signature.  Checking it is real is simplest for the field equations.  We have:
\[ \frac{\delta \Gamma_1}{\delta \Phi_i} = \frac{i}{2} \left( \frac{\delta^3 A
   \left( \Phi \right)}{\delta \Phi_i \delta \Phi_k \delta \Phi_j} \right)
   \iota^{jj} \left( G_{jk} - G_{jm} \left( \frac{\delta^2 A \left( \Phi
   \right)}{\delta \Phi_m \delta \Phi_n} - A_{mn} \right) G_{nk} + \ldots
   \right) = \hspace{4em} \]
\begin{equation}
  \label{one loop term in field equations in DeWitts notation} \quad =
  \frac{i}{2} \left( \frac{\delta^3 A \left( \Phi \right)}{\delta \Phi_i
  \delta \Phi_k \delta \Phi_j} \right) \iota^{jj} G \left( \Phi \right)_{jk}
\end{equation}
Thus, from the definition (\ref{energy momentum tensor}) of the
energy-momentum tensor, the one-loop Casimir energy density contributions to
the energy-momentum tensor are obtained from this equation, by choosing the
field $\Phi_i$ to be the metric, $g_{\mu \nu}$, and multiplying by
$\frac{2}{\sqrt{- g}}$.  Considering, now, the contribution to the one-loop
Casimir energy densities from a real scalar boson, with the classical action:
\begin{equation}
  \label{classical action for scalar field} A_{\mathrm{scalar}} = - \frac{1}{2}
  \int d^d x \sqrt{- g}  \left( g^{\mu \nu} \left( D_{\mu} \varphi \right)
  \left( D_{\nu} \varphi \right) + m^2 \varphi^2 - \xi R \varphi^2 \right)
\end{equation}
where $R$ is the Ricci scalar, and $\xi$ is a real constant, sometimes called
the conformal coupling, when $d = 4$, we find, by use of the Palatini identity
$\delta R_{\mu \nu} = D_{\mu} \delta \Gamma_{\tau \nu}^{\tau} - D_{\tau}
\delta \Gamma_{\mu \nu}^{\tau}$, and the identity $\delta \Gamma_{\mu
\nu}^{\tau} = \frac{1}{2} g^{\tau \sigma} \left( D_{\mu} \delta g_{\sigma \nu}
+ D_{\nu} \delta g_{\sigma \mu} - D_{\sigma} \delta g_{\mu \nu} \right)$,
that:
\[ \frac{\delta A_{\mathrm{scalar}}}{\delta \Phi_{g_{x \mu \nu}}} = \frac{1}{2}
   \sqrt{- g}  \left( \left( 1 - 2 \xi \right) g^{\mu \sigma} g^{\nu \tau}
   \left( D_{\sigma} \varphi \right) \left( D_{\tau} \varphi \right) - \left(
   \frac{1}{2} - 2 \xi \right) g^{\mu \nu} g^{\sigma \tau} \left( D_{\sigma}
   \varphi \right) \left( D_{\tau} \varphi \right) \right. \hspace{2em} \]
\begin{equation}
  \label{metric variation of scalar field classical action} \left. - 2 \xi
  g^{\mu \sigma} g^{\nu \tau} \varphi D_{\sigma} D_{\tau} \varphi + 2 \xi
  g^{\mu \nu} g^{\sigma \tau} \varphi D_{\sigma} D_{\tau} \varphi - \xi \left(
  R^{\mu \nu} - \frac{1}{2} g^{\mu \nu} R \right) \varphi^2 - \frac{1}{2}
  g^{\mu \nu} m^2 \varphi^2 \right)
\end{equation}
For the particular case $\xi = \frac{1}{6}$, this is in agreement with the
formulae of Muller, Fagundes, and Opher {\cite{Muller Fagundes Opher 1, Muller
Fagundes Opher 2}}, after for allowing for their sign convention for the
Riemann tensor, which results in the opposite sign of the Ricci tensor to
(\ref{Ricci tensor}).

From (\ref{energy momentum tensor}), (\ref{one loop term in field equations in
DeWitts notation}), and (\ref{metric variation of scalar field classical
action}), we find:
\[ T^{\mu \nu}_{1, \mathrm{scalar}} = \frac{2}{\sqrt{- g}}  \frac{\delta
   \Gamma_{1, \mathrm{scalar}}}{\delta \Phi_{g_{x \mu \nu}}} = \frac{i}{2}
   \left. \left( \left( 1 - 2 \xi \right) g^{\mu \sigma} g^{\nu \tau} \left(
   D_{y \sigma} D_{x \tau} + D_{x \sigma} D_{y \tau} \right)
   \rule[-1.5ex]{0pt}{2.5ex} \right. \hspace{6em} \right. \]
\[ - \left( \frac{1}{2} - 2 \xi \right) g^{\mu \nu} g^{\sigma \tau} \left(
   D_{y \sigma} D_{x \tau} + D_{x \sigma} D_{y \tau} \right) - 2 \xi g^{\mu
   \sigma} g^{\nu \tau} \left( D_{x \sigma} D_{x \tau} + D_{y \sigma} D_{y
   \tau} \right) \]
\begin{equation}
  \left. \label{scalar contribution to one loop energy momentum tensor} \left.
  + 2 \xi g^{\mu \nu} g^{\sigma \tau} \left( D_{x \sigma} D_{x \tau} + D_{y
  \sigma} D_{y \tau} \right) - 2 \xi \left( R^{\mu \nu} - \frac{1}{2} g^{\mu
  \nu} R \right) - g^{\mu \nu} m^2 \right) G \left( \Phi \right)_{\varphi_x
  \varphi_y} \right|_{y = x}
\end{equation}
In the particular case of $3 + 1$ dimensional Minkowski space, the scalar
contribution to the one-loop energy density is given by (\ref{scalar
contribution to one loop energy momentum tensor}) as:
\[ T^{00}_{1, \mathrm{scalar}} = \frac{i}{2}  \left( \left( 2 - 4 \xi \right)
   \partial_{y 0} \partial_{x 0} + \left( 1 - 4 \xi \right) \left( -
   \partial_{y 0} \partial_{x 0} + \vec{\partial}_y \vec{\partial}_x \right) -
   2 \xi \left( \partial_{x 0} \partial_{x 0} + \partial_{y 0} \partial_{y 0}
   \right) \right. \hspace{2em} \]
\begin{equation}
  \label{scalar contribution to one loop energy density} \hspace{2em} \left.
  \left. - 2 \xi \left( - \partial^2_{x 0} + \vec{\partial}^2_x -
  \partial^2_{y 0} + \vec{\partial}^2_y \right) + m^2 \right) G \left( \Phi
  \right)_{\varphi_x \varphi_y} \right|_{y = x}
\end{equation}
Furthermore, for the scalar propagator, requiring that the Fresnel integral in
(\ref{one loop effective action in terms of superdet}) be well-defined
uniquely selects the Feynman $i \varepsilon$ prescription for the propagator.
Thus the scalar propagator is:
\begin{equation}
  \label{scalar propagator in Minkowski space} G_{\varphi_x \varphi_y} = -
  \int \frac{d^4 p}{\left( 2 \pi \right)^4}  \frac{e^{i \left( - p_0 \left(
  x_0 - y_0 \right) + \vec{p} . \left( \vec{x} - \vec{y} \right) \right)}}{-
  p^2_0 + \vec{p}^2 + m^2 - i \varepsilon}
\end{equation}
Substituting (\ref{scalar propagator in Minkowski space}) into (\ref{scalar
contribution to one loop energy density}), and taking the limit $y \rightarrow
x$ from either $y_0 > x_0$ or $y_0 < x_0$, we find:
\begin{equation}
  \label{scalar contribution to one loop vacuum energy density in Minkowski
  space} T^{00}_{1, \mathrm{scalar}} = \int \frac{d^3 \vec{p}}{\left( 2 \pi
  \right)^3}  \frac{1}{2}  \sqrt{\vec{p}^2 + m^2}
\end{equation}
which is real, as required, and is the standard divergent expression for the
one-loop vacuum energy density of a real scalar field.  For models that have
unbroken supersymmetry in $3 + 1$ spacetime dimensions, and do not involve
gravitons, the one-loop vacuum energy densities cancel between fermions and
bosons, and, moreover, the vacuum energy density is exactly zero to all orders
in the coupling constants {\cite{Zumino Supersymmetry and the vacuum}}, and the
one-loop vacuum energy
densities have also been found to vanish in some models with broken
supergravity {\cite{Allen Davis, Gibbons Nicolai, Burgess Hoover 1, Burgess
Hoover 2}}, whilst for $d = 11$ supergravity {\cite{Cremmer Julia Scherk}}, it
appears that a cosmological constant is not consistent with supersymmetry
{\cite{Nicolai Townsend van Nieuwenhuizen, Sagnotti Tomaras, Bautier Deser
Henneaux Seminara}}, so that divergences corresponding to a cosmological
constant term would be cancelled unambiguously within the framework of BPHZ
renormalization, to all orders in the semiclassical expansion in the number of
loops in the Feynman diagrams.

Considering the gravitino propagator, for the compactification of
Ho\v{r}ava-Witten theory on $\mathcal{M}^6$, as a sum over images:
\begin{equation}
  \label{gravitino propagator as a sum over images} G \left( \Phi
  |\mathcal{M}^6 \right)_{\psi_{x \mu i} \psi_{y \nu j}} = \sum_{\gamma \in
  \Gamma} G \left( \Phi |\mathbf{C} \mathbf{H}^3 \right)_{\psi_{x \mu i}
  \psi_{\gamma \left( y \right) \nu j}},
\end{equation}
where $\mathcal{M}^6$ is the quotient of $\mathbf{C} \mathbf{H}^3$ by the
cocompact, torsionless, discrete subgroup $\Gamma$ of $\mathrm{SU} \left( 3, 1
\right)$, we see that if the sign of the gravitino field, at the image $\gamma
\left( y \right)$ of $y$ by an element $\gamma$ of $\Gamma$, depends on the
route taken from $y$ to $\gamma \left( y \right)$, then the sum of $G \left(
\Phi |\mathbf{C} \mathbf{H}^3 \right)_{\psi_{x \mu i} \psi_{\gamma \left(
y \right) \nu j}}$, for $y$ close to $x$, over all elements $\gamma$ of
$\Gamma$ different from the identity, will not be well defined, even if it
converges.  But this sum is directly physically significant, because it
determines the finite part of the gravitino contribution to the one-loop
Casimir contribution to the energy-momentum tensor, by a formula analogous to
(\ref{scalar contribution to one loop energy momentum tensor}).  Furthermore,
the three-form gauge field, $C_{IJK}$, only enters the gravitino field
equation through its four-form field strength $G_{IJKL}$, which is globally
well defined, so a background configuration of the three-form gauge field
cannot make any difference to whether or not the sum over $\gamma \in \Gamma$
in $G \left( \Phi |\mathcal{M}^6 \right)_{\psi_{x \mu i} \psi_{y \nu j}}$ well
defined.  Thus it does, indeed, seem that models of this type are not
physically well defined, unless $\mathcal{M}^6$ is a spin manifold.  Of
course, a rotation through $2 \pi$ changes the sign of a spinor field, so it
is natural to wonder whether introducing twists or rotations in the local
Lorentz part of the vielbein between different coordinate patches, which will
cancel out of the relations between the metric on the different coordinate
patches, can cancel the ambiguity, but this is presumably taken into
consideration in determining whether or not a manifold is a spin manifold.  A
direct explanation of why $\mathbf{C} \mathbf{P}^2$ is not a spin manifold
has been given by Hawking and Pope {\cite{Hawking Pope}}, and recently
reviewed in Appendix B of {\cite{Alexanian Balachandran Immirzi Ydri}}.

\subsection{The Salam-Strathee harmonic expansion method}
\label{The Salam Strathdee harmonic expansion method}

For the explicit calculation of the Casimir energy densities for
compactifications on smooth compact quotients of $\mathbf{C} \mathbf{H}^3$
or $\mathbf{H}^6$, by means of the sum over images method of Muller,
Fagundes, and Opher {\cite{Muller Fagundes Opher 1, Muller Fagundes, Muller
Fagundes Opher 2}}, for obtaining the propagators on the quotients, or some
extension of their method if the sums diverge at large distances due to the
masslessness of the fields, the propagators and heat kernels for the $d = 11$
supergravity multiplet are needed for flat $\mathbf{R}^5$, times
$\mathbf{C} \mathbf{H}^3$ or $\mathbf{H}^6$, and the propagators and
heat kernels for the $d = 10$ supersymmetric Yang-Mills multiplet are needed
for flat $\mathbf{R}^4$, times $\mathbf{C} \mathbf{H}^3$ or
$\mathbf{H}^6$.  The propagators and heat kernels for $\mathbf{C}
\mathbf{H}^3$ or $\mathbf{H}^6$ can be obtained from the corresponding
propagators and heat kernels for $\mathbf{C} \mathbf{P}^3$ or
$\mathbf{S}^6$, which can in turn be obtained by a straightforward but
lengthy application of the harmonic expansion method of Salam and Strathdee
{\cite{Salam Strathdee, Strathdee 1, Strathdee 2}}, which is currently in
progress.  The harmonic expansions can be summed by means of a generating
function, and for the heat kernel of a massive scalar, on $\mathbf{C}
\mathbf{H}^3$, we find the integral representation:
\[ H \left( \chi, s \right) = \frac{4 e^{- sm^2} e^{9 s}}{\sqrt{2 \pi s}}
   \left( \left( \frac{d}{d \cosh \left( 2 \chi \right)} \right)^2 + \left(
   \cosh \left( 2 \chi \right) + 1 \right) \left( \frac{d}{d \cosh \left( 2
   \chi \right)} \right)^3 \right) \times \]
\begin{equation}
\label{scalar heat kernel on CP 3}
   \times \int^{\infty}_{\chi} \left( dy \right) e^{- \frac{y^2}{4 s}} \sinh
   \left( 2 y \right) \left( \cosh \left( 2 y \right) - \cosh \left( 2 \chi
   \right) \right)^{- \frac{1}{2}}
\end{equation}
Here $\chi$ is proportional to the geodesic distance between the two position
arguments of the heat kernel.  The same integral, but with different
differential operators acting on it, occurs in the heat kernel of a massive
scalar, on real hyperbolic spaces of all even dimensions $\geq 2$, while for
real hyperbolic spaces of odd dimension $\geq 3$, the heat kernel can be
written in closed form, as found by Muller, Fagundes, and Opher, for $d = 3$.
The application of the Salam-Strathdee method to $\mathbf{C}
\mathbf{P}^3$, which is a spin manifold, was begun by Strathdee
{\cite{Strathdee 2}}, and developed by Sobczyk {\cite{Sobczyk 1, Sobczyk 2}}.

A special effect in a related background was discovered by Gibbons and Nicolai
{\cite{Gibbons Nicolai}}, who calculated the one-loop vacuum energy density of
the Freund-Rubin $\mathrm{AdS}_4 \times \mathbf{S}^7$ compactification of $d =
11$ supergravity {\cite{Freund Rubin}}, and found that it vanished ``floor by
floor'', or in other words, separately for each $N = 8$ supersymmetric
Kaluza-Klein level, whereas to preserve the supersymmetry of the vacuum, it
would have been sufficient for the sum over all the Kaluza-Klein levels to
vanish.  The contribution of the lowest Kaluza-Klein level, namely the $N = 8$
supergravity multiplet, had earlier been found to vanish by Allen and Davis
{\cite{Allen Davis}}.

There is also a
$\mathrm{AdS}_4 \times \mathbf{C} \mathbf{P}^3$ compactification of
type IIA $d = 10$ supergravity, discovered by Watamura {\cite{Watamura}}, that
was shown by Nilsson and Pope \cite{Nilsson Pope} to have $N = 6$
supersymmetry, and also to be
related to the Freund-Rubin compactification of $d = 11$ supergravity, via the
fact that $\mathbf{S}^7$ is a non-trivial $U \left( 1 \right)$ fiber bundle
over $\mathbf{C} \mathbf{P}^3$, called a Hopf fibration.
What this means is that the Watamura $\mathrm{AdS}_4 \times
\mathbf{C} \mathbf{P}^3$ compactification of type IIA $d = 10$
supergravity can be identified with a particular $\mathrm{AdS}_4 \times
\mathbf{C} \mathbf{P}^3 \times \mathbf{S}^1$ compactification of $d =
11$ supergravity, such that the metric ansatz (\ref{metric ansatz}) has been
modified by the replacement
\begin{equation}
  \label{replacement of dy squared}
  dy^2 \to \left( dy - A_A dz^A \right)^2,
\end{equation}
where $A_A$
is proportional to a potential for the K\"ahler form of the $\mathbf{C}
\mathbf{P}^3$, and $y$ is now the coordinate around the $\mathbf{S}^1$.
The Watamura $N = 6$ compactification is then obtained in an appropriate
limit, where the radius of the $\mathbf{S}^1$ tends to $0$, while for
another special case, where the radius of the $\mathbf{S}^1$ is
appropriately related to the diameter of the $\mathbf{C} \mathbf{P}^3$,
the supersymmetry is presumably extended to $N = 8$, and the Freund-Rubin
compactification is obtained.

Nilsson and Pope showed that the complete spectrum of small fluctuations of
the Watamura $N = 6$ compactification can be directly obtained from the known
spectrum of small fluctuations of the Freund-Rubin solution {\cite{Biran
Casher Englert Rooman Spindel, DAuria Fre, Englert Nicolai, Casher Englert
Nicolai Rooman, Sezgin}}.  I shall
now obtain the complete list of the modes by the Salam-Strathdee method, and
check it against the list given by Nilsson and Pope, and then repeat the
Gibbons-Nicolai calculation, for all but the lowest two Kaluza-Klein levels,
for the Watamura $N = 6$ compactification.

The isometry group of $\mathbf{C} \mathbf{P}^3$, with the standard
Fubini-Study metric {\cite{Green Schwarz Witten}}, is $\mathrm{SU} \left( 4
\right)$, and the subgroup of the isometry group, that leaves a chosen point
fixed, which I shall call the tangent space isometry group, is $\mathrm{SU}
\left( 3 \right) \times U \left( 1 \right)$.  The tangent space group of
$\mathbf{C} \mathbf{P}^3$ is $\mathrm{SO} \left( 6 \right)$, because
$\mathbf{C} \mathbf{P}^3$ has six real dimensions, so by accident, the
tangent space group is locally isomorphic to the isometry group, although the
tangent space group and the isometry group are completely distinct, and the
tangent space isometry group, $\mathrm{SU} \left( 3 \right) \times U \left( 1
\right)$, is found to be embedded in the tangent space group, $\mathrm{SO}
\left( 6 \right)$, and the isometry group, $\mathrm{SU} \left( 4 \right)$, in
different ways.  I shall put a tilde above the irreducible representations of
the tangent space group, $\mathrm{SO} \left( 6 \right)$, to distinguish them
from the irreducible representations of the isometry group, $\mathrm{SU} \left(
4 \right)$.

The first step of the Salam-Strathdee method is to decompose all the fields
involved, which are here the metric, the three-form gauge field, and the
gravitino, of $d = 11$ supergravity, into irreducible representations of the
product of the tangent space isometry groups $\mathrm{SO} \left( 3, 1 \right)$,
of the four extended dimensions, and $\mathrm{SU} \left( 3 \right) \times U
\left( 1 \right)$, of $\mathbf{C} \mathbf{P}^3$, and possible components
along the $\mathbf{S}^1$, that does not have a nontrivial continuous tangent
space isometry group.  The next step is then to determine, for each
irreducible represention of the tangent space isometry group $\mathrm{SU} \left(
3 \right) \times U \left( 1 \right)$ of $\mathbf{C} \mathbf{P}^3$ that
arises, the list of all the irreducible representations of $\mathrm{SU} \left( 4
\right)$, the isometry group of $\mathbf{C} \mathbf{P}^3$, that contain
that irreducible representation of $\mathrm{SU} \left( 3 \right) \times U \left(
1 \right)$, under the subgroup decomposition $\mathrm{SU} \left( 4 \right)
\rightarrow \mathrm{SU} \left( 3 \right) \times U \left( 1 \right)$.  This is
then the list of all the harmonics that occur, in the harmonic expansion, on
$\mathbf{C} \mathbf{P}^3$, of that particular irreducible representation
of the tangent space isometry group of $\mathbf{C} \mathbf{P}^3$.

According to Salam and Strathdee's general prescription, {\cite{Salam
Strathdee}}, for harmonic expansions on the quotient space, $G / H$, of a Lie
group, $G$, and a Lie subgroup, $H$, of $G$, the quotient space $\mathbf{C}
\mathbf{P}^3 = \mathrm{SU} \left( 4 \right) / \left( \mathrm{SU} \left( 3
\right) \times U \left( 1 \right) \right)$ is coordinatized by the ``boosts''
generated by the six generators of $\mathrm{SU} \left( 4 \right)$, that are not
generators of its subgroup $\mathrm{SU} \left( 3 \right) \times U \left( 1
\right)$.  Now $\mathrm{SU} \left( 3 \right) \times U \left( 1 \right)$ is
contained in $\mathrm{SU} \left( 4 \right)$ such that the $4$ of $\mathrm{SU}
\left( 4 \right)$ has the $\mathrm{SU} \left( 3 \right) \times U \left( 1
\right)$ content:
\begin{equation}
  \label{SU 3 cross U 1 content of 4 of SU 4} 4 = 3_{\frac{1}{4}} + 1_{-
  \frac{3}{4}}
\end{equation}
where the relative value of the $U \left( 1 \right)$ charges, which are shown
as subscripts, is determined by the tracelessness of the $\mathrm{SU} \left( 4
\right)$ generators, and the overall normalization of the $U \left( 1 \right)$
charges is a convention, that I have chosen to agree with Strathdee,
{\cite{Strathdee 2}}, and Sobczyk, {\cite{Sobczyk 1, Sobczyk 2}}.

From (\ref{SU 3 cross U 1 content of 4 of SU 4}), we find that the $\mathrm{SU}
\left( 3 \right) \times U \left( 1 \right)$ content of the adjoint of
$\mathrm{SU} \left( 4 \right)$ is determined by:
\begin{equation}
  \label{SU 3 cross U 1 content of adjoint and singlet of SU 4} 15 + 1 = 4
  \times \bar{4} = \left( 3_{\frac{1}{4}} + 1_{- \frac{3}{4}} \right) \times
  \left( \bar{3}_{- \frac{1}{4}} + 1_{ \frac{3}{4}} \right) = 8_0 + 1_0 + 3_1
  + \bar{3}_{- 1} + 1_0
\end{equation}
Thus the generators of $\mathrm{SU} \left( 4 \right)$, that are not generators
of $\mathrm{SU} \left( 3 \right) \times U \left( 1 \right)$, have the $\mathrm{SU}
\left( 3 \right) \times U \left( 1 \right)$ content $3_1 + \bar{3}_{- 1}$, so
the tangent space isometry group, $\mathrm{SU} \left( 3 \right) \times U \left(
1 \right)$, of $\mathbf{C} \mathbf{P}^3$, is embedded in the tangent space
group, $\mathrm{SO} \left( 6 \right)$, of $\mathbf{C} \mathbf{P}^3$, such
that the tangent space six-vector, the $\tilde{6}$ of $\mathrm{SO} \left( 6
\right)$, has the $\mathrm{SU} \left( 3 \right) \times U \left( 1 \right)$
content {\cite{Strathdee 2}}:
\begin{equation}
  \label{SU 3 cross U 1 content of 6 of SO 6} \tilde{6} = 3_1 + \bar{3}_{- 1}
\end{equation}
The decomposition (\ref{SU 3 cross U 1 content of 6 of SO 6}) now determines
the decompositions of all the other irreducible representations of $\mathrm{SO}
\left( 6 \right)$.  In particular, if we consider the $\tilde{4}$ of
$\mathrm{SO} \left( 6 \right)$ that contains the $1$ and the $\bar{3}$ of
$\mathrm{SU} \left( 3 \right)$, and write its decomposition as $\tilde{4} = 1_a
+ \bar{3}_b$, where the $U \left( 1 \right)$ charges $a$ and $b$ are to be
determined, we find that:
\begin{equation}
  \label{decomposition of 4 tilde times 4 tilde} \tilde{4} \times \tilde{4} =
  1_{2 a} + \bar{3}_{a + b} + \bar{3}_{a + b} + \bar{6}_{2 b} + 3_{2 b}
\end{equation}
However, we know that the antisymmetric part of $ \tilde{4} \times \tilde{4} $
is the $\tilde{6}$, so for
consistency with (\ref{SU 3 cross U 1 content of 6 of SO 6}), we must have $2
b = 1$, and $a + b = - 1$, so that we find $\tilde{4} = 1_{- \frac{3}{2}} +
\bar{3}_{\frac{1}{2}}$.  The other $\tilde{4}$ of $\mathrm{SO} \left( 6 \right)$
then decomposes as $1_{\frac{3}{2}} + 3_{- \frac{1}{2}}$, so, comparing with
(\ref{SU 3 cross U 1 content of 4 of SU 4}), we see that the tangent space
isometry group, $\mathrm{SU} \left( 3 \right) \times U \left( 1 \right)$, of
$\mathbf{C} \mathbf{P}^3$, is, indeed, embedded differently in the tangent
space group, $\mathrm{SO} \left( 6 \right)$, and the isometry group, $\mathrm{SU}
\left( 4 \right)$, as stated above {\cite{Strathdee 2}}.

To determine which irreducible representations of $\mathrm{SU} \left( 4 \right)$
contain a given irreducible representation of $\mathrm{SU} \left( 3 \right)
\times U \left( 1 \right)$, it will be convenient to use Young tableau
notations for the irreducible representations of $\mathrm{SU} \left( 3 \right)$
and $\mathrm{SU} \left( 4 \right)$.  I shall denote the irreducible
representation of $\mathrm{SU} \left( 3 \right)$, that corresponds to a Young
tableau with rows of lengths $a$, $b$, and $c$, such that $a \geq b \geq c
\geq 0$, by $\left[ a, b, c \right]$, with a corresponding notation for
$\mathrm{SU} \left( 4 \right)$.  Then $\left[ a + n, b + n, c + n \right]$, for
all integer $n$ such that $c + n \geq 0$, all correspond to the same
irreducible representation of $\mathrm{SU} \left( 3 \right)$, whose Dynkin label
is $\left( a - b, b - c \right)$.

It will be very convenient also to allow Young tableaux with negative length
rows, which means that $a$, $b$, and $c$, in the Young tableau $\left[ a, b, c
\right]$, are restricted only by $a \geq b \geq c$, without the restriction to
$c \geq 0$.  Negative length rows are represented by rows of blocks extending
out to the left of what would normally be the left-hand side of the Young
tableau.  The corresponding irreducible representations of $\mathrm{SU} \left( 3
\right)$ are constructed from appropriately symmetrized Kronecker products of
the fundamental and the antifundamental representations, one fundamental
representation factor for each block in a positive length row, and one
antifundamental representation factor for each block in a negative length row,
with all traces that can be formed by contracting the $\mathrm{SU} \left( 3
\right)$ invariant tensor $\delta_{r \bar{s}}$ with an antifundamental index
and a fundamental index, both from among the left-hand indices of the
representation matrix, or both from among the right-hand indices of the
representation matrix, removed.

Analogous constructions also apply for all the other special unitary groups.
For example, the $\left[ p, 0, 0, - p \right]$ representation of $\mathrm{SU}
\left( 4 \right)$, whose Dynkin label is $\left( p, 0, p \right)$, with the
three components of the Dynkin label corresponding to the three vertices of
the $\mathrm{SU} \left( 4 \right)$ Dynkin diagram, written in sequence from end
to end along the line, has the representation matrices:
\[ U_{I_1 I_2 \ldots I_p \bar{J}_1 \bar{J}_2 \ldots \bar{J}_p, \bar{K}_1
   \bar{K}_2 \ldots \bar{K}_p L_1 L_2 \ldots L_p} = \hspace{18em}  \]
\[ = \frac{1}{\left( p! \right)^2} \sum_{S_I S_J S_K S_L} \sum^p_{r = 0}
   \left( \left( - 1 \right)^r \frac{ \left( 2 p + 2 - r \right) !}{\left(
   \left( p - r \right) ! \right)^2 r! \left( 2 p + 2 \right) !} \times
   \right. \hspace{2em} \]
\begin{equation}
  \label{p 0 p representation of SU 4} \hspace{8em}  \left. \times \delta_{L_1
  \bar{K}_1} \delta_{I_1 \bar{J}_1} \ldots \delta_{L_r \bar{K}_r} \delta_{I_r
  \bar{J}_r} U_{I_{r + 1} \bar{K}_{r + 1}} U_{\bar{J}_{r + 1} L_{r + 1}}
  \ldots U_{I_p \bar{K}_p} U_{\bar{J}_p L_p} \right)
\end{equation}
where, in the last line of this expression,
\[ \delta_{L_1 \bar{K}_1} \delta_{I_1 \bar{J}_1} \ldots \delta_{L_r \bar{K}_r}
   \delta_{I_r \bar{J}_r} U_{I_{r + 1} \bar{K}_{r + 1}} U_{\bar{J}_{r + 1}
   L_{r + 1}} \ldots U_{I_p \bar{K}_p} U_{\bar{J}_p L_p} \]
is interpreted as having no $\delta$s, if $r < 1$, and as no $U$s, if $r + 1 >
p$.  In other words, the subscripts, on the subscripts, are to increase from
$1$ to $p$, going from left to right, along this expression.  $U_{I \bar{J}}$
here denotes the fundamental representation of an element of $\mathrm{SU} \left(
4 \right)$, and $U_{\bar{I} J} \equiv \left( U_{I \bar{J}} \right)^{\ast}$
denotes the antifundamental representation of that same element of $\mathrm{SU}
\left( 4 \right)$, in accordance with the conventions of subsection \ref{CH3},
on page \pageref{CH3}, for barred and unbarred indices, and
$\sum_{S_I}$denotes the sum over all permutations of the ``$I$'' subscripts,
and thus contains $p!$ terms, since there are $p$ such subscripts, and so on.
The formula (\ref{p 0 p representation of SU 4}) is used in the derivation of
the scalar heat kernel on $\mathbf{C} \mathbf{P}^3$, (\ref{scalar heat
kernel on CP 3}), by the Salam-Strathdee method.

We then find, using indices $\mu, \nu, \sigma, \ldots$, for the four extended
dimensions, $r, s, t, \ldots$, and $\bar{r}, \bar{s}, \bar{t}, \ldots$, in the
complex coordinate notation of subsection \ref{CH3}, for $\mathbf{C}
\mathbf{P}^3$, and $y$ for the $\mathbf{S}^1$, that the $d = 11$
gravition, $h_{IJ}$, contains the $d = 4$ graviton, $h_{\mu \nu}$, in the
Young tableau representation $\left[ 0, 0, 0 \right]_0$ of $\mathrm{SU} \left( 3
\right) \times U \left( 1 \right)$, where the subscript denotes the $U \left(
1 \right)$ charge, and also a $d = 4$ vector, $h_{\mu y}$, and a $d = 4$
scalar, $h_{yy}$, in the $\left[ 0, 0, 0 \right]_0$ of $\mathrm{SU} \left( 3
\right) \times U \left( 1 \right)$, a $d = 4$ vector, $h_{\mu r}$, and a $d =
4$ scalar, $h_{yr}$, in the $\left[ 1, 0, 0 \right]_1$ of $\mathrm{SU} \left( 3
\right) \times U \left( 1 \right)$, a $d = 4$ vector, $h_{\mu \bar{r}}$, and a
$d = 4$ scalar, $h_{y \bar{r}}$, in the $\left[ 0, 0, - 1 \right]_{- 1}$ of
$\mathrm{SU} \left( 3 \right) \times U \left( 1 \right)$, and $d = 4$ scalars,
$h_{r \bar{s}}$, in the $\left[ 1, 0, - 1 \right]_0$, $h_{rs}$, in the $\left[
2, 0, 0 \right]_2$, and $h_{\bar{r} \bar{s}}$, in the $\left[ 0, 0, - 2
\right]_{- 2}$, of $\mathrm{SU} \left( 3 \right) \times U \left( 1 \right)$.
The decomposition of the three-form gauge field, $C_{IJK}$, is worked out
similarly, bearing in mind that for $d = 4$, a two-form gauge field is
equivalent to a scalar {\cite{Kalb Ramond, Cremmer Scherk, Nambu, Townsend
ghosts for ghosts}}, and a three-form gauge field has no degrees of freedom.

To work out the decomposition of the gravitino, we first determine the
decomposition of a $d = 11$ spinor.  We decompose the 32-valued spinor index
into the Cartesian product of an 8-valued spinor index, for the $\mathbf{C}
\mathbf{P}^3$, and a four-valued spinor index, for the four extended
dimensions and the $\mathbf{S}^1$, considered together as a five-dimensional
space, and consider the decomposition of the 8-valued spinor index, which is
the sum of the two opposite chirality $\tilde{4}$'s of $\mathrm{SO} \left( 6
\right)$.  Thus, from above, the 8-valued spinor index decomposes into the
$1_{- \frac{3}{2}} + \bar{3}_{\frac{1}{2}} + 1_{\frac{3}{2}} + 3_{-
\frac{1}{2}} = \left[ 0, 0, 0 \right]_{- \frac{3}{2}} + \left[ 0, 0, - 1
\right]_{\frac{1}{2}} + \left[ 0, 0, 0 \right]_{\frac{3}{2}} + \left[ 1, 0, 0
\right]_{- \frac{1}{2}}$ of $\mathrm{SU} \left( 3 \right) \times U \left( 1
\right)$.  The $d = 11$ gravitino, $\psi_I$, thus contains $d = 4$ gravitinos,
$\psi_{\mu}$, and $d = 4$ spinors, $\psi_y$, in these four representations of
$\mathrm{SU} \left( 3 \right) \times U \left( 1 \right)$, together with $d = 4$
spinors, $\psi_r$, in the $\mathrm{SU} \left( 3 \right) \times U \left( 1
\right)$ representations that result from forming the Cartesian product of the
$3_1$ with these four representations, namely $3_{_{- \frac{1}{2}}} +
8_{\frac{3}{2}} + 1_{\frac{3}{2}} + 3_{\frac{5}{2}} + 6_{\frac{1}{2}} +
\bar{3}_{\frac{1}{2}} = \left[ 1, 0, 0 \right]_{- \frac{1}{2}} + \left[ 1, 0,
- 1 \right]_{\frac{3}{2}} + \left[ 0, 0, 0 \right]_{\frac{3}{2}} + \left[ 1,
0, 0 \right]_{\frac{5}{2}} + \left[ 2, 0, 0 \right]_{\frac{1}{2}} + \left[ 0,
0, - 1 \right]_{\frac{1}{2}}$, and $d = 4$ spinors, $\psi_{\bar{r}}$, in the
complex conjugates of these six representations.

To determine which irreducible representations of $\mathrm{SU} \left( 4 \right)$
contain these representations of $\mathrm{SU} \left( 3 \right) \times U \left( 1
\right)$, we first recall the general rule for the irreducible representations
of $\mathrm{SU} \left( q - 1 \right)$ contained in an irreducible representation
of $\mathrm{SU} \left( q \right)$, which, for the present case, states that the
irreducible representations of $\mathrm{SU} \left( 3 \right)$, contained in the
irreducible representation of $\mathrm{SU} \left( 4 \right)$ that corresponds to
a Young tableau $\left[ n_1, n_2, n_3, n_4 \right]$, $n_1 \geq n_2 \geq n_3
\geq n_4$, are the irreducible representations of $\mathrm{SU} \left( 3 \right)$
that correspond to all the Young tableaux $\left[ m_1, m_2, m_3 \right]$, such
that $n_1 \geq m_1 \geq n_2 \geq m_2 \geq n_3 \geq m_3 \geq n_4$.  This
general rule is the basis for the Gelfand-Tsetlin patterns that can be used to
label the basis vectors of the irreducible representations of $\mathrm{SU}
\left( q \right)$, via the subgroup chain $U \left( 1 \right) \subset
\mathrm{SU} \left( 2 \right) \subset \mathrm{SU} \left( 3 \right) \subset \ldots
\subset \mathrm{SU} \left( q - 1 \right) \subset \mathrm{SU} \left( q \right)$, as
reviewed, for example, in {\cite{Molev}}.

We next note that if the $\mathrm{SU} \left( 3 \right)$ representation,
corresponding to a Young tableau $\left[ m_1, m_2, m_3 \right]$, is contained
in an $\mathrm{SU} \left( 4 \right)$ representation, corresponding to a Young
tableau $\left[ n_1, n_2, n_3, n_4 \right]$, with $n_4 \geq 0$, then $m_1 +
m_2 + m_3$ of the $n_1 + n_2 + n_3 + n_4$ copies of the $\mathrm{SU} \left( 4
\right)$ fundamental, from which the $\mathrm{SU} \left( 4 \right)$
representation is constructed, branch to the $3$ of $\mathrm{SU} \left( 3
\right)$, and the remaining $n_1 + n_2 + n_3 + n_4 - \left( m_1 + m_2 + m_3
\right)$ copies of the $\mathrm{SU} \left( 4 \right)$ fundamental branch to the
$1$ of $\mathrm{SU} \left( 3 \right)$, so from (\ref{SU 3 cross U 1 content of 4
of SU 4}), the $U \left( 1 \right)$ charge of the $\mathrm{SU} \left( 3 \right)$
representation is
\begin{equation}
  \label{U 1 charge of the SU 3 representation} m_1 + m_2 + m_3 - \frac{3}{4}
  \left( n_1 + n_2 + n_3 + n_4 \right)
\end{equation}
Furthermore, this relation, like the rule for the $\mathrm{SU} \left( 3 \right)$
irreducible representations contained within a given $\mathrm{SU} \left( 4
\right)$ irreducible representation, is unaltered by adding a constant to all
the $n_i$ and all the $m_i$, so it remains true when the assumption that $n_4
\geq 0$ is no longer satisfied.

We now find that the rule (\ref{U 1 charge of the SU 3 representation}), in
combination with the rule that $n_1 \geq m_1 \geq n_2 \geq m_2 \geq n_3 \geq
m_3 \geq n_4$, is very restrictive.  For example, the $\mathrm{SU} \left( 3
\right)$ representation, corresponding to the Young tableau $\left[ 0, 0, 0
\right]$, is contained in all the $\mathrm{SU} \left( 4 \right)$ representations
$\left[ p, 0, 0, - p' \right]$, with $p \geq 0$ and $p' \geq 0$, but the
$\mathrm{SU} \left( 3 \right) \times U \left( 1 \right)$ representation $\left[
0, 0, 0 \right]_0$ is contained only in the representations $\left[ p, 0, 0, -
p \right]$, with $p \geq 0$, whose representation matrices are given in
(\ref{p 0 p representation of SU 4}).  The Dynkin labels of these
representations of $\mathrm{SU} \left( 4 \right)$ are $\left( p, 0, p \right)$,
as already noted, and in general, the Dynkin label of an $\mathrm{SU} \left( 4
\right)$ irreducible representation, that corresponds to a Young tableau
$\left[ a, b, c, d \right]$, is $\left( a - b, b - c, c - d \right)$, with the
three components of the Dynkin label corresponding to the three vertices of
the $\mathrm{SU} \left( 4 \right)$ Dynkin diagram, written in sequence from end
to end along the line.

We next note that each massive $d = 4$ graviton mode $h_{\mu \nu}$,
corresponding to harmonics on $\mathbf{C} \mathbf{P}^3$ with Dynkin labels
$\left( p, 0, p \right)$, $p > 0$, will absorb a $d = 4$ vector with the same
Dynkin label, for which $h_{\mu y}$ is available, and a $d = 4$ scalar with
the same Dynkin label, for which $h_{yy}$ is available.  Many of the other $d
= 4$ massive modes, with a subscript $y$, also get absorbed by higher spin $d
= 4$ massive modes, with matching Dynkin labels, in a similar way, although
for the modes $C_{\mu \nu r}$, $C_{\mu yr}$, $C_{\mu \nu \bar{r}}$, and
$C_{\mu y \bar{r}}$, the situation is reversed, with the $d = 4$ vector fields
$C_{\mu yr}$ and $C_{\mu y \bar{r}}$ absorbing the fields $C_{\mu \nu r}$ and
$C_{\mu \nu \bar{r}}$, which are equivalent to $d = 4$ scalar fields.  In this
way, we find that the unabsorbed $d = 4$ massive modes, and the $\mathrm{SU}
\left( 4 \right)$ irreducible representations that occur in their harmonic
expansions on $\mathbf{C} \mathbf{P}^3$, are as shown in Table \ref{Boson
harmonics on CP 3} for the bosons, and in Table \ref{Fermion harmonics on CP
3} for the fermions.  Most of the $\mathrm{SU} \left( 4 \right)$ irreducible
representations that occur in the harmonic expansions of the metric, and of a
vector field, were listed by Sobczyk, {\cite{Sobczyk 1}}, in the context of a
$\mathbf{C} \mathbf{P}^3$ compactification of a $d = 10$
Einstein-Yang-Mills theory.

\begin{table}
\begin{center}
\begin{tabular}{|c|c|c|c|}\hline
  $\begin{array}{c}
    d \hspace{-0.6pt} = \hspace{-0.6pt} 4\\
    \mathrm{spin}
  \end{array}$ & $\begin{array}{c}
    d = 11\\
    \textrm{compo-}\\
    \mathrm{nents}
  \end{array}$ & $\begin{array}{c}
    \mathrm{SU}\! \left( 3 \right)\! \times\! \mathrm{U}\! \left( 1 \right)\\
    \mathrm{tableau}
  \end{array}$ & $\begin{array}{c}
    \mathrm{SU} \left( 4 \right) \hspace{0.4ex} \textrm{Dynkin labels}\\
    p \geq 0
  \end{array}$\\
  \hline
  $2$ & $h_{\mu \nu}$ & $\left[ 0, 0, 0 \right]_0$ & $\left( p, 0, p
  \right)$\\
  \hline
  $1$ & $h_{\mu r}$ & $\left[ 1, 0, 0 \right]_1$ & $\left( p + 1, 0, p + 1
  \right) + \left( p, 1, p + 2 \right)$\\
  \hline
  $1$ & $h_{\mu \bar{r}}$ & $\left[ 0, 0, - 1 \right]_{- 1}$ & $\left( p + 1,
  0, p + 1 \right) + \left( p + 2, 1, p \right)$\\
  \hline
  $0$ & $h_{r \bar{s}}$ & $\begin{array}{c}
    \left[ 1, 0, - 1 \right]_0\\
    + \left[ 0, 0, 0 \right]_0
  \end{array}$ & $\begin{array}{c}
    \left( p + 1, 0, p + 1 \right) + \left( p, 1, p + 2 \right) + \left( p +
    2, 1, p \right)\\
    + \left( p, 2, p \right) + \left( p, 0, p \right)
  \end{array}$\\
  \hline
  $0$ & $h_{rs}$ & $\left[ 2, 0, 0 \right]_2$ & $\left( p\! +\! 2, 0, p\! +\! 2
  \right) + \left( p\! +\! 1, 1, p\! +\! 3 \right) +
  \left( p, 2, p\! +\! 4 \right)$\\
  \hline
  $0$ & $h_{\bar{r} \bar{s}}$ & $\left[ 0, 0, - 2 \right]_{- 2}$ & $\left( p\!
  +\! 2, 0, p\! +\! 2 \right) + \left( p\! +\! 3, 1, p\! +\! 1 \right) +
  \left( p\! +\! 4, 2, p \right)$\\
  \hline
  $0$ & $C_{\mu \nu y}$ & $\left[ 0, 0, 0 \right]_0$ & $\left( p, 0, p
  \right)$\\
  \hline
  $1$ & $C_{\mu yr}$ & $\left[ 1, 0, 0 \right]_1$ & $\left( p + 1, 0, p + 1
  \right) + \left( p, 1, p + 2 \right)$\\
  \hline
  $1$ & $C_{\mu y \bar{r}}$ & $\left[ 0, 0, - 1 \right]_{- 1}$ & $\left( p +
  1, 0, p + 1 \right) + \left( p + 2, 1, p \right)$\\
  \hline
  $1$ & $C_{\mu r \bar{s}}$ & $\begin{array}{c}
    \left[ 1, 0, - 1 \right]_0\\
    + \left[ 0, 0, 0 \right]_0
  \end{array}$ & $\begin{array}{c}
    \left( p + 1, 0, p + 1 \right) + \left( p, 1, p + 2 \right) + \left( p +
    2, 1, p \right)\\
    + \left( p, 2, p \right) + \left( p, 0, p \right)
  \end{array}$\\
  \hline
  $1$ & $C_{\mu rs}$ & $\left[ 0, 0, - 1 \right]_2$ & $\left( p, 1, p + 2
  \right) + \left( p, 0, p + 4 \right)$\\
  \hline
  $1$ & $C_{\mu \bar{r} \bar{s}}$ & $\left[ 1, 0, 0 \right]_{- 2}$ & $\left( p
  + 2, 1, p \right) + \left( p + 4, 0, p \right)$\\
  \hline
  $0$ & $C_{rst}$ & $\left[ 0, 0, 0 \right]_3$ & $\left( p, 0, p + 4
  \right)$\\
  \hline
  $0$ & $C_{rs \bar{t}}$ & $\begin{array}{c}
    \left[ 0, 0, - 2 \right]_1\\
    + \left[ 1, 0, 0 \right]_1
  \end{array}$ & $\begin{array}{c}
    \left( p, 0, p + 4 \right) + \left( p, 1, p + 2 \right) + \left( p, 2, p
    \right)\\
    + \left( p + 1, 0, p + 1 \right) + \left( p, 1, p + 2 \right)
  \end{array}$\\
  \hline
  $0$ & $C_{r \bar{s} \bar{t}}$ & $\begin{array}{c}
    \left[ 2, 0, 0 \right]_{- 1}\\
    + \left[ 0, 0, - 1 \right]_{- 1}
  \end{array}$ & $\begin{array}{c}
    \left( p + 4, 0, p \right) + \left( p + 2, 1, p \right) + \left( p, 2, p
    \right)\\
    + \left( p + 1, 0, p + 1 \right) + \left( p + 2, 1, p \right)
  \end{array}$\\
  \hline
  $0$ & $C_{\bar{r} \bar{s} \bar{t}}$ & $\left[ 0, 0, 0, \right]_{- 3}$ &
  $\left( p + 4, 0, p \right)$\\
  \hline
\end{tabular}
\caption{\label{Boson harmonics on CP 3}
Boson harmonics for type IIA $ d = 10 $ supergravity compactified on
$ \mathbf{CP}^3 $.}
\end{center}
\end{table}

\begin{table}[t]
\begin{center}
\begin{tabular}{|c|c|c|c|}\hline
  $\begin{array}{c}
    d = 4\\
    \mathrm{spin}
  \end{array}$ & $\begin{array}{c}
    d = 11\\
    \textrm{compo-}\\
    \mathrm{nents}
  \end{array}$ & $\begin{array}{c}
    \mathrm{SU}\! \left( 3 \right)\! \times\! \mathrm{U}\! \left( 1 \right)\\
    \mathrm{tableau}
  \end{array}$ & $\begin{array}{c}
    \mathrm{SU} \left( 4 \right) \hspace{0.4ex} \textrm{Dynkin labels}\\
    p \geq 0
  \end{array}$\\
  \hline
  $\frac{3}{2}$ & $\psi_{\mu}$ & $\left[ 0, 0, 0 \right]_{- \frac{3}{2}}$ &
  $\left( p + 2, 0, p \right)$\\
  \hline
  $\frac{3}{2}$ & $\psi_{\mu}$ & $\left[ 0, 0, - 1 \right]_{\frac{1}{2}}$ &
  $\left( p, 0, p + 2 \right) + \left( p, 1, p \right)$\\
  \hline
  $\frac{3}{2}$ & $\psi_{\mu}$ & $\left[ 0, 0, 0 \right]_{\frac{3}{2}}$ &
  $\left( p, 0, p + 2 \right)$\\
  \hline
  $\frac{3}{2}$ & $\psi_{\mu}$ & $\left[ 1, 0, 0 \right]_{- \frac{1}{2}}$ &
  $\left( p + 2, 0, p \right) + \left( p, 1, p \right)$\\
  \hline
  $\frac{1}{2}$ & $\psi_r$ & $\left[ 1, 0, 0 \right]_{- \frac{1}{2}}$ &
  $\left( p + 2, 0, p \right) + \left( p, 1, p \right)$\\
  \hline
  $\frac{1}{2}$ & $\psi_r$ & $\left[ 1, 0, - 1 \right]_{\frac{3}{2}}$ &
  $\begin{array}{c}
    \left( p\! +\! 1, 0, p\! +\! 3 \right) + \left( p, 1, p\! +\! 4 \right) +
    \left( p\! +\! 1, 1, p\! +\! 1 \right)\\
    + \left( p, 2, p + 2 \right)
  \end{array}$\\
  \hline
  $\frac{1}{2}$ & $\psi_r$ & $\left[ 0, 0, 0 \right]_{\frac{3}{2}}$ & $\left(
  p, 0, p + 2 \right)$\\
  \hline
  $\frac{1}{2}$ & $\psi_r$ & $\left[ 1, 0, 0 \right]_{\frac{5}{2}}$ & $\left(
  p + 1, 0, p + 3 \right) + \left( p, 1, p + 4 \right)$\\
  \hline
  $\frac{1}{2}$ & $\psi_r$ & $\left[ 2, 0, 0 \right]_{\frac{1}{2}}$ & $\left(
  p + 2, 0, p \right) + \left( p + 1, 1, p + 1 \right) + \left( p, 2, p + 2
  \right)$\\
  \hline
  $\frac{1}{2}$ & $\psi_r$ & $\left[ 0, 0, - 1 \right]_{\frac{1}{2}}$ &
  $\left( p, 0, p + 2 \right) + \left( p, 1, p \right)$\\
  \hline
  $\frac{1}{2}$ & $\psi_{\bar{r}}$ & $\left[ 0, 0, - 1 \right]_{-
  \frac{5}{2}}$ & $\left( p + 3, 0, p + 1 \right) + \left( p + 4, 1, p
  \right)$\\
  \hline
  $\frac{1}{2}$ & $\psi_{\bar{r}}$ & $\left[ 0, 0, - 2 \right]_{-
  \frac{1}{2}}$ & $\left( p, 0, p + 2 \right) + \left( p + 1, 1, p + 1 \right)
  + \left( p + 2, 2, p \right)$\\
  \hline
  $\frac{1}{2}$ & $\psi_{\bar{r}}$ & $\left[ 1, 0, 0 \right]_{- \frac{1}{2}}$
  & $\left( p + 2, 0, p \right) + \left( p, 1, p \right)$\\
  \hline
  $\frac{1}{2}$ & $\psi_{\bar{r}}$ & $\left[ 0, 0, - 1 \right]_{\frac{1}{2}}$
  & $\left( p, 0, p + 2 \right) + \left( p, 1, p \right)$\\
  \hline
  $\frac{1}{2}$ & $\psi_{\bar{r}}$ & $\left[ 1, 0, - 1 \right]_{-
  \frac{3}{2}}$ & $\begin{array}{c}
    \left( p\! +\! 3, 0, p\! +\! 1 \right) + \left( p\! +\! 4, 1, p \right) +
    \left( p\! +\! 1, 1, p\! +\! 1 \right)\\
    + \left( p + 2, 2, p \right)
  \end{array}$\\
  \hline
  $\frac{1}{2}$ & $\psi_{\bar{r}}$ & $\left[ 0, 0, 0 \right]_{- \frac{3}{2}}$
  & $\left( p + 2, 0, p \right)$\\
  \hline
\end{tabular}
\caption{\label{Fermion harmonics on CP 3}
Fermion harmonics for type IIA $ d = 10 $ supergravity compactified on
$ \mathbf{CP}^3 $.}
\end{center}
\end{table}

We see that for each $d = 4$ spin, the $\mathrm{SU} \left( 4 \right)$ multiplets
in Tables \ref{Boson harmonics on CP 3} and \ref{Fermion harmonics on CP 3}
are in one to one correspondence with the $\mathrm{SU} \left( 4 \right)$
multiplets listed by Nilsson and Pope {\cite{Nilsson Pope}} for that $d = 4$
spin, except that in some cases, $p$ has to be shifted by a small number.
This was to be expected, because simply listing the harmonics corresponding to
each $d = 4$ state does not determine the corresponding masses, nor does it
determine how the states are organized into $N = 6$ supermultiplets.  I have
listed the $\mathrm{SU} \left( 4 \right)$ multiplets in Tables \ref{Boson
harmonics on CP 3} and \ref{Fermion harmonics on CP 3} so that the complete
set of harmonics entering the expansion, on $\mathbf{C} \mathbf{P}^3$, of
the corresponding $d = 4$ state, is given by the $\mathrm{SU} \left( 4 \right)$
multiplets shown, for all $p \geq 0$, whereas $p$, in Nilsson and Pope's Table
1, identifies the distinct $N = 6$ supermultiplets, with $p = 0$ corresponding
to the standard $N = 6$ supergravity multiplet.  Thus in Nilsson and Pope's
Table 1, many of the $\mathrm{SU} \left( 4 \right)$ Dynkin labels have a
negative component for small values of $p$, and in particular, for $p = 0$,
indicating the absence of an $\mathrm{SU} \left( 4 \right)$ representation in
that $p$-series, in the corresponding low-lying $N = 6$ supermultiplet.

Nilsson and Pope also listed the parities of the $d = 4$ boson states. I have
not calculated the parities of the $d = 4$ boson states by the Salam-Strathdee
method, but we note that if we assume that the parity of a $d = 4$ boson state
is the product of a factor of $- 1$ if the state arises from the $d = 11$
three-form gauge field, a factor of $- 1$ if the $d = 4$ boson state has spin
$1$, a factor of $- 1$ if the second component of the $\mathrm{\mathrm{SU}}
\left( 4 \right)$ Dynkin label is an odd number, and a factor of $- 1$ for
each index $y$ of the $d = 11$ components that the state arises from, then the
$\mathrm{\mathrm{SU}} \left( 4 \right)$ boson multiplets listed in Table
\ref{Boson harmonics on CP 3}, for each $d = 4$ spin and parity, can be paired
one to one with the $\mathrm{\mathrm{SU}} \left( 4 \right)$ boson multiplets
listed by Nilsson and Pope, of the same $d = 4$ spin and parity, up to small
shifts of $p$ in some cases, as before.

We can now calculate the one-loop vacuum energy of the Watamura $N = 6$
compactification of type IIA $d = 10$ supergravity, by the method of Gibbons
and Nicolai {\cite{Gibbons Nicolai}}, which uses the zeta function
regularization method of Hawking {\cite{Hawking Zeta function regularization}}.
We can directly use
Gibbons and Nicolai's formula (9) for the contribution to the vacuum energy
from a spin $s$ state, $s > 0$, and their formula (11) for the contribution to
the vacuum energy from spin $0$ state, except that the last term in their
formula (9), in the scanned version of the preprint from KEK {\cite{Gibbons
Nicolai}}, which seems to be a misprint, has to be replaced by $- \left(
\frac{20 s + 9}{120} \right)$.  I have verified, using Maxima \cite{Maxima},
that the
one-loop vacuum energy of the Freund-Rubin compactification does, indeed,
vanish floor by floor, for all Kaluza-Klein levels above the lowest, when this
replacement is made in their formula (9).  The lowest Kaluza-Klein level,
namely the ordinary supergravity multiplet, requires a separate calculation,
which was carried out by Allen and Davis \cite{Allen Davis},
because the formula for the dimension of an $ \mathrm{SO}\left(8\right) $
irreducible representation does not vanish for some of the $ \mathrm{SO}
\left(8\right) $ Dynkin labels with a negative component that arise in this
case, such as $ \left( -2, 1, 0, 0 \right) $.

The dimensions of the irreducible representations of $\mathrm{SU} \left( 4
\right)$ with Dynkin labels $\left( a, b, c \right)$, where the integers $a
\geq 0$, $b \geq 0$, and $c \geq 0$ are associated with the three verteices of
the $\mathrm{SU} \left( 4 \right)$ Dynkin diagram, taken in sequence from end to
end along the line, can be calculated from Weyl's dimension formula
{\cite{Wikipedia Weyl character formula, Woit}}, or from the combinatorial
result, summarized in section 3.I.(b) of {\cite{Barcelo Ram}}, that the
dimension of the irreducible representation of $\mathrm{GL} \left(
n,\mathbf{C} \right)$ associated with an ordinary Young tableau with $n$
rows, and no negative length rows, is the product over all the boxes $x$ of
the tableau, of $\frac{n + c \left( x \right)}{h_x}$, where $c \left( x
\right)$ is the horizontal position of $x$ minus its vertical position,
counting from left to right and downwards, starting from the box at the top
left-hand corner of the tableau, and $h_x$, the length of the hook whose top
left-hand corner is $x$, is the number of boxes directly under $x$, plus the
number of boxes directly to the right of $x$, plus $1$.  The result is:
\begin{equation}
  \label{dimension of SU 4 irreducible representation} D \left( a, b, c
  \right) = \left( 1 + a \right)  \left( 1 + b \right)  \left( 1 + c \right)
  \left( 1 + \frac{a + b}{2} \right)  \left( 1 + \frac{b + c}{2} \right)
  \left( 1 + \frac{a + b + c}{3} \right)
\end{equation}

Then using Maxima {\cite{Maxima}}, we find, by a formula analogous to Gibbons
and Nicolai's formula (14), with $z = - 1$, but using the entries in Nilsson
and Pope's Table 1, instead of from Gibbons and Nicolai's Table 1, that in
units of $\left( - \frac{1}{3} \Lambda \right)^{\frac{1}{2}}$, where $\Lambda$
is the cosmological constant of the $\mathrm{AdS}_4$, the contributions to the
vacuum energy, from the states in the $N = 6$ supersymmetry multiplet at
Kaluza-Klein level $p$, for $ p \geq 2 $, where $p = 0$ corresponds to the
$N = 6$ supergravity multiplet, are as follows.

Spin $2$:
\[ \frac{1}{576}  \left( - 20 p^9 - 270 p^8 - 1580 p^7 - 5250 p^6 - 10888 p^5
   - 14565 p^4 - 12506 p^3 \right. \]
\begin{equation}
  \label{spin 2 contribution to the vacuum energy} \left. - 6597 p^2 - 1916 p
  - 228 \right)
\end{equation}
Spin $\frac{3}{2}$:
\[ \frac{1}{1440}  \left( 320 p^9 + 4320 p^8 + 24800 p^7 + 78960 p^6 + 152188
   p^5 + 181290 p^4 + 129622 p^3 \right. \]
\begin{equation}
  \label{spin three halves contribution to the vacuum energy} \left. + 50049
  p^2 + 7275 p - 414 \right)
\end{equation}
Spin $1$:
\[ \frac{1}{320}  \left( - 180 p^9 - 2430 p^8 - 13740 p^7 - 42210 p^6 - 76152
   p^5 - 80685 p^4 - 45930 p^3 \right. \]
\begin{equation}
  \label{spin 1 contribution to the vacuum energy} \left. - 9045 p^2 + 2892 p
  + 1260 \right)
\end{equation}
Spin $\frac{1}{2}$:
\[ \frac{1}{1440}  \left( 960 p^9 + 12960 p^8 + 72480 p^7 + 216720 p^6 +
   370644 p^5 + 353070 p^4 \right. \]
\begin{equation}
  \label{spin half contribution to the vacuum energy} \left. + 153770 p^3 -
  7065 p^2 - 28859 p - 6690 \right)
\end{equation}
Spin $0$:
\[ \frac{1}{1440}  \left( - 420 p^9 - 5670 p^8 - 31500 p^7 - 92610 p^6 -
   152928 p^5 - 134865 p^4 - 45442 p^3 \right. \]
\begin{equation}
  \label{spin 0 contribution to the vacuum energy} \left. + 14211 p^2 + 13360
  p + 2004 \right)
\end{equation}
The fact that the contribution of the spin $2$ states is negative is
presumably an artifact of the zeta function regularization used.  The sum of
these contributions is zero, so for all the Kaluza-Klein levels with $ p \geq 2
$, the one-loop vacuum energy vanishes ``floor by floor''
for the Watamura-Nilsson-Pope $N = 6$ $\mathbf{C} \mathbf{P}^3$
compactification of type IIA $d = 10$ supergravity, just as it does for the $N
= 8$ Freund-Rubin compactification of $d = 11$ supergravity.

The cases of $ p
= 0 $ and $ p = 1 $ require separate calculations, because some $ \mathrm{SU}
\left(4\right) $ multiplets occur that should be omitted in these cases, and
some Dynkin labels with a negative component occur, for which the formula
(\ref{dimension of SU 4 irreducible representation}) for the dimension of an
$ \mathrm{SU}\left(4\right) $ irreducible representation does not give zero.
For the $N = 6$ supergravity multiplet, which is the case with $p = 0$, the
one-loop result vacuum energy was found to vanish by Allen and Davis
{\cite{Allen Davis}}.  The case of $ p = 1 $ requires further study, and will
not be considered in this paper.  But it does not seem very likely that the
one-loop vacuum energy would fail to vanish for this one Kaluza-Klein level,
when it does for all the others.

We note that this calculation has not included any Kaluza-Klein excitations
associated with the $ \mathbf{S}^1 $, so in the context of Nilsson and Pope's
compactifications of $ d = 11 $ supergravity, interpolating between the
Freund-Rubin compactification of $ d = 11 $ supergravity, and Watamura's $ N =
6 $ compactification of type IIA $ d = 10 $ supergravity, the $ d = 4 $ states
listed here are appropriate for the limit in which the radius of the $
\mathbf{S}^1 $ tends to zero.  To consider the opposite limit, in which the
radius of the $ \mathbf{S}^1 $ tends to infinity, which is presumably related
to the $ N = 6 $ supergravity in five dimensions listed by Cremmer
\cite{Cremmer}, it would be necessary to repeat the calculation with the extra
modes included.  However, at a certain value of the radius of the $
\mathbf{S}^1 $, the $ N = 6 $ supersymmetry would be extended to the $ N = 8 $
supersymmetry of the Freund-Rubin compactification, for which it is known from
the Gibbons-Nicolai calculation that the one-loop vacuum energy vanishes floor
by floor.  So it is perhaps plausible that the one-loop vacuum energy might
also vanish floor by floor for all values of the radius of the $ \mathbf{S}^1
$, from $ 0 $ to $ \infty $.

If the numbers of fermion and boson helicity states are equal for all the $ N =
6 $ massive multiplets, which I have not explicitly checked, then we would
presumably find that the one-loop vacuum energy would still vanish when the
background is flat four-dimensional Minkowski space times $ \mathbf{CP}^3 $,
even though this background is not a solution of the classical field equations,
and is not supersymmetric,
and, on the basis of relations between the propagators and heat kernels on
Minkowski space times $ \mathbf{CP}^3 $, and on Minkowski space times $
\mathbf{CH}^3 $, it might then also vanish when the background is flat
four-dimensional Minkowski space times $ \mathbf{CH}^3 $.  And for similar
reasons, it seems possible that the one-loop vacuum energy might also vanish
when the background is flat five-dimensional Minkowski space times $
\mathbf{CH}^3 $.

However, in
consequence of the rule, discussed at the beginning of this section, that the
quantum effective action of the BRST-BV gauge-fixed theory, in a background
that is not a solution of the classical field equations, is the sum of all the
one-line-irreducible vacuum bubbles, calculated with an action given by the
BRST-BV gauge-fixed classical action in the presence of the background field,
but with the terms linear in the quantum fields deleted, the action used in the
calculation of the quantum effective action of the BRST-BV gauge-fixed theory,
on a background that is flat four-dimensional or five-dimensional Minkowski
space, times $
\mathbf{CP}^3 $ or $ \mathbf{CH}^3 $, would presumably not satisfy identities
needed to use Zumino's arguments \cite{Zumino Supersymmetry and the vacuum} for
the vanishing of the higher loop vacuum energies.

Nevertheless, if the massive $ N = 6 $ multiplets satisfied Curtright's spin
sum rules \cite{Curtright} for a theory with $ N = 6 $ supersymmetry in $ d =
4 $, which I have not explicitly checked, then some of the ingredients for a
possible cancellation of higher loop vacuum energies, on a flat
four-dimensional
Minkowski space times $ \mathbf{CP}^3 $ or $ \mathbf{CH}^3 $ background, would
be in place, so the possibility that such cancellations might occur is not yet
excluded.  However, the grounds for expecting such higher loop cancellations to
occur are not very strong, and it does not seem very likely that the higher
loop vacuum energies of type IIA $d = 10$ supergravity on a four-dimensional
Minkowski space times uncompactified $\mathbf{C} \mathbf{H}^3$ background, and
of $d = 11$ supergravity on a five-dimensional Minkowski space times
uncompactified $\mathbf{C} \mathbf{H}^3$ background, will vanish,
notwithstanding the special properties of the Watamura $ N = 6 $
compactification of type IIA $ d = 10 $ supergravity, and its oxidation to $ d
= 11 $ by Nilsson and Pope, for three reasons.

Firstly, the lowest
Kaluza-Klein energies of the states in the $N = 6$ supermultiplet at
Kaluza-Klein level $p$, are not all the same.  Instead, they differ by up to
four units of $\left( - \frac{1}{3} \Lambda \right)^{\frac{1}{2}}$ within the
same multiplet, so the energy differences, between the lowest energies of
states within one multiplet, are up to four times greater than the energy
difference between corresponding states within successive multiplets, and
these energy differences, between the states within a multiplet, are likely to
be essential for the cancellation of the vacuum energy of a multiplet, at least
when $ \Lambda $ is nonzero.

The
contribution to the vacuum energy, from a state of lowest energy $ E_0 $, is a
quartic polynomial in $ E_0 $, by Gibbons and Nicolai's equations (9) and (11).
$ E_0 $ is linear in the Kaluza-Klein level number $ p $.
When the relation between the sectional curvature of the
$\mathrm{AdS}_4$, and the minimum sectional curvature of the $\mathbf{C}
\mathbf{P}^3$, is broken, there are two independent units of energy, namely
$\left( - \frac{1}{3} \Lambda \right)^{\frac{1}{2}}$ and $\left( - \frac{1}{3}
\Lambda_{\mathbf{C} \mathbf{P}^3} \right)^{\frac{1}{2}}$, where
$\Lambda_{\mathbf{C} \mathbf{P}^3}$ is defined in terms of the minimum
sectional curvature of the $\mathbf{C} \mathbf{P}^3$.  We would now expect the
lowest energy $ E_0 $ of a state to contain a term $ p \left( - \frac{1}{3}
\Lambda_{\mathbf{C} \mathbf{P}^3} \right)^{\frac{1}{2}}$, associated with its
Kaluza-Klein level number $ p $, and a term $ q \left( - \frac{1}{3} \Lambda
\right)^{\frac{1}{2}}$, where $ q $ is the integer or half integer, such that $
1 \leq q \leq 5 $, that determines the offset of $ E_0 $ from $ p \left( -
\frac{1}{3} \Lambda \right)^{\frac{1}{2}}$, as listed by Nilsson and Pope, for
the case when $ \Lambda_{\mathbf{C} \mathbf{P}^3} = \Lambda $.

Vanishing of
the one-loop vacuum energy floor by floor, for independent $ \Lambda $ and
$\Lambda_{
\mathbf{C} \mathbf{P}^3}$, would then require that the coefficients of the
different powers of $\left( - \frac{1}{3} \Lambda \right)^{\frac{1}{2}}$ and
$\left( - \frac{1}{3} \Lambda_{\mathbf{C} \mathbf{P}^3} \right)^{\frac{1}{2}}$
in the vacuum energy, which are polynomials in $ p $ of degree up to $ 9 $, all
vanish separately, and, although this has not been excluded, there is no
reason to expect it to happen, to the best of my knowledge, except that, if the
numbers of fermion and boson states in each $ N = 6 $ multiplet are equal, we
would expect the coefficients of the terms independent of $ \Lambda $, which
are polynomials in $ p $ of degree $ 9 $, to vanish, since the one-loop vacuum
energy of each $ N = 6 $ multiplet would then vanish in four-dimensional
Minkowski space.

Thus, although the one-loop vacuum energy of each $ N = 6 $
multiplet would vanish for $ \Lambda = 0 $, if the numbers of fermion and boson
states in each $ N = 6 $ multiplet are equal, it does not seem very likely that
the one-loop vacuum energy of each $ N = 6 $ multiplet would vanish for values
of the ratio $ \frac{ \Lambda }{ \Lambda_{\mathbf{C} \mathbf{P}^3} } $ strictly
between $ 0 $ and $ 1 $, so the vanishing of the one-loop vacuum energy of each
$ N = 6 $ multiplet, for $ \Lambda = 0 $, would be an isolated phenomenon, not
continuously connected to the supersymmetric system with $ \Lambda = \Lambda_{
\mathbf{C} \mathbf{P}^3} $, so it does not seem very likely that the
supersymmetric system could result in the vanishing of the higher loop vacuum
energies of type IIA $ d = 10 $ supergravity on a four-dimensional Minkowski
space times $ \mathbf{CP}^3 $ background, or that its $ d = 11 $ oxidation
could result in the vanishing of the higher loop vacuum energies of $ d = 11 $
supergravity on a five-dimensional Minkowski space times $ \mathbf{CP}^3 $
background, when there are no other reasons to expect this to happen.

Secondly, there is a second Watamura-Nilsson-Pope $\mathbf{C}
\mathbf{P}^3$ compactification of type IIA $d = 10$ supergravity, that has
no supersymmetry, but differs from the $N = 6$ compactification only by the
relative orientation of a four-form field strength $F_{\mu \nu \sigma \tau}$,
which is proportional to the $d = 4$ tensor density $\epsilon_{\mu \nu \sigma
\tau}$, and a two-form field strength $F_{AB}$, which is proportional to the
K\"ahler form of the $\mathbf{C} \mathbf{P}^3$.  The relative orientation
of $F_{\mu \nu \sigma \tau}$ and $F_{AB}$ is detected by the supersymmetry
variations of
the fermions.  Now, by the Salam-Strathdee construction, the small fluctuation
modes, about this $N = 0$ compactification, will consist of exactly the same
collection of series of $\mathrm{SU} \left( 4 \right)$ representations as listed
above for the $N = 6$ compactification, but the lowest energies, of the
smallest representations of some of the series, will be shifted up or down, by
a small number of units of $\left( - \frac{1}{3} \Lambda
\right)^{\frac{1}{2}}$, so that the vacuum energy will presumably no longer
vanish.

And to distinguish the two cases, both $F_{\mu \nu \sigma \tau}$ and $F_{AB}$
would still have to be nonzero, when the $\mathrm{AdS}_4$ is replaced by
Minkowski space, and the two cases would still have to be distinguished, when
the $\mathbf{C} \mathbf{P}^3$ is replaced by $\mathbf{C}
\mathbf{H}^3$, so it seems unlikely that the higher loop vacuum energies will
vanish for a four-dimensional Minkowski space times $\mathbf{C}
\mathbf{H}^3$ background, without nonvanishing background fields
corresponding to $F_{\mu \nu \sigma \tau}$ and $F_{AB}$.  And for the
corresponding compactifications of $d = 11$ supergravity, Nilsson and Pope
showed that these two $\mathbf{C} \mathbf{P}^3$ compactifications of type
IIA $d = 10$ supergravity are ``Hopf fibrations'' of the Freund-Rubin
$\mathrm{AdS}_4 \times \mathbf{S}^7$ compactification of $d = 11$
supergravity, which means that the metric ansatz (\ref{metric ansatz}) would
have to be modified by the replacement (\ref{replacement of dy squared}),
where $A_A$ is proportional to a potential for the K\"ahler form of the
$\mathbf{C} \mathbf{P}^3$ or $\mathbf{C} \mathbf{H}^3$.

And thirdly, there is an $\mathrm{AdS}_5 \times \mathbf{C} \mathbf{P}^3$
compactification of $d = 11$ supergravity, such that the only nonvanishing
form field, in the background, has the form of the Lukas-Ovrut-Stelle-Waldram
{\cite{Lukas Ovrut Stelle Waldram}} ansatz (\ref{four form field strength}).
This compactification is investigated in the following subsection \ref{AdS5
times CP3
compactification}, and found to have no supersymmetry.  Its one-loop vacuum
energy will thus presumably be nonvanishing, and $ d = 11 $ supergravity, on
a five-dimensional Minkowski space times $ \mathbf{CP}^3 $ or uncompactified
$ \mathbf{CH}^3 $ background, is as closely related to this compactification,
as it is to Nilsson and Pope's $ d = 11 $ oxidation of Watamura's $ N = 6 $
compactification of type IIA $ d = 10 $ supergravity.
This suggests, again, that if the higher loop vacuum energies of $d = 11$
supergravity were to vanish on any five-dimensional Minkowski space times
$\mathbf{C} \mathbf{H}^3$ background, there would have to be a nonvanishing
field strength $ F_{\mu \nu \sigma \tau} $ in the background, and the metric
ansatz (\ref{metric ansatz}) would have to be modified by the replacement
(\ref{replacement of dy squared}), with $ A_A $ proportional to a potential for
the K\"ahler form of the $ \mathbf{CH}^3 $, in order to relate the background
to the $ d = 11 $ oxidation of the Watamura
$N = 6$ compactification, and distinguish it from a background related
to the $\mathrm{AdS}_5 \times \mathbf{C} \mathbf{P}^3$ compactification.

These arguments do not exclude the possibility that the higher loop vacuum
energies of a four-dimensional Minkowski space times uncompactified
$\mathbf{C} \mathbf{H}^3$
times $ \mathbf{R}^1 $ background for $d = 11$ supergravity, with suitable
dependences of $ a\left(y\right) $ and $ b\left(y\right) $, in the metric
ansatz (\ref{metric ansatz}), on the position $ y $ along the $ \mathbf{R}^1 $,
and a suitable $ y $-dependent value of the field strength $F_{\mu \nu \sigma
\tau}$, proportional to $ \epsilon_{\mu \nu \sigma \tau} $, and the replacement
(\ref{replacement of dy squared}) in the metric ansatz (\ref{metric ansatz}),
with $ A_A $ a suitable $ y $-dependent multiple of a potential for the
K\"ahler form of the $ \mathbf{CH}^3 $, might vanish.  However, the reasons for
expecting such a background to exist, for which the higher loop vacuum energies
vanish, are not very strong, so for the phenomenological estimates in this
paper, I assume that the higher-loop vacuum energies are nonvanishing on
an uncompactified $ \mathbf{CH}^3 $ background, and, moreover, that they have
their typical order of magnitude, in terms of the magnitude of the curvature
of the background, which means that $ \frac{ b_1 }{ \kappa^{
\frac{2}{9} } } $ cannot be smaller than the value $ \sim 0.03 $ to $ 0.2
$ estimated in subsection \ref{The expansion parameter}, on page \pageref{The
expansion parameter}, on the basis of
Giudice, Rattazzi, and Wells's estimate \cite{Giudice Rattazzi Wells} of the
expansion parameter for quantum gravitational corrections in $ d $ dimensions,
so that, in consequence of the relation (\ref{b sub 1 in terms of chi}), on
page \pageref{b sub 1 in terms of chi}, between $ \frac{ b_1 }{ \kappa^{
\frac{2}{9} } } $ and $ \left \vert \chi \left( \mathcal{M}^6 \right) \right
\vert $, which follows from the estimate (\ref{alpha sub U in the Standard
Model}), of the $ d = 4 $ Yang-Mills fine structure constant at unification,
values of $ \left \vert \chi \left( \mathcal{M}^6 \right) \right \vert
$ larger than $ \sim 7 \times 10^4 $ to $ 6 \times 10^9 $ are excluded.

If it turned out that cancellations of higher loop vacuum energies of
Ho\v{r}ava-Witten theory, on a suitable uncompactified $ \mathbf{CH}^3 $
background,
actually did occur, allowing $ \frac{ b_1 }{ \kappa^{ \frac{2}{9} } } $ to be
smaller than $ \sim 0.03 $ to $ 0.2 $, and $ \left \vert \chi \left(
\mathcal{M}^6 \right) \right \vert $ to be larger than $ \sim 7 \times 10^4 $
to $ 6 \times 10^9 $,
when $ \mathcal{M}^6 $ is a smooth compact quotient of $ \mathbf{CH}^3 $ that
is a spin manifold, then the phenomenological estimates in this paper would
presumably still be valid, with minor modifications, for smooth compact
quotients of $ \mathbf{H}^6 $ that are spin manifolds, since, to the best of my
knowledge, there is no reason to expect the vacuum energy of Ho\v{r}ava-Witten
theory to vanish on an uncompactified $ \mathbf{H}^6 $ background.

It would be interesting to find out whether Nilsson and Pope's $ d = 11 $
``oxidation'' of the Watamura $ N = 6 $ $ \mathbf{CP}^3 $ compactification of
type IIA $ d = 10 $ supergravity, as discussed in this subsection, can be
extended by the addition of non-vanishing components $ G_{ABCD} $ of the
four-form field strength of the three-form gauge field, given by the ansatz
(\ref{four form field strength}) of Lukas, Ovrut, Stelle, and Waldram (LOSW)
\cite{Lukas Ovrut Stelle Waldram}, so as to obtain a supersymmetric $
\mathrm{AdS}_4 \times \mathbf{CP}^3 $ compactification of Ho\v{r}ava-Witten
theory, consistent with Witten's topological constraint \cite{Witten
Constraints on compactification}, when the $ \mathrm{SU}\left(3\right) $ part
of the spin connection of the $ \mathbf{CP}^3 $ is embedded in the $ E8 $ on
one of the two orbifold hyperplanes, and the $ \mathrm{U}\left(1\right) $ part
of the spin connection of the $ \mathbf{CP}^3 $ is embedded in the $ E8 \times
E6 $ left unbroken by the $ \mathrm{SU}\left(3\right) $ embedding, in one of
the four ways listed by Pilch and Schellekens, in subsection 4.3 of \cite{Pilch
Schellekens}.

However the components $ G_{\mu \nu \sigma \tau} $ of the four-form field
strength of the three-form gauge field, like the components $ G_{ABCD} $, are
odd under reflection in the Ho\v{r}ava-Witten orbifold hyperplanes, so if they
do not vanish as one or both of the orbifold hyperplanes are approached, they
would have to have discontinuities at the orbifold hyperplanes in the upstairs
picture, which would then, by (\ref{G discontinuity}), require the existence of
non-vanishing components $ F_{\mu \nu} $ of the $ E8 $ Yang-Mills field
strength on the corresponding orbifold hyperplane, which would break invariance
under the $ \mathrm{SO}\left(3,2\right) $ Anti de Sitter group.  Thus to
preserve invariance under the Anti de Sitter group, $ G_{\mu \nu \sigma \tau} $
would have to vanish on both orbifold hyperplanes.  This is not necessarily
inconsistent with the existence of a compactification, since there also exists
an $ \mathrm{AdS}_5 \times \mathbf{CP}^3 $ compactification of $ d = 11 $
supergravity, whose only non-vanishing components of $ G_{IJKL} $ are given by
the LOSW ansatz (\ref{four form field strength}), but I shall show in the next
subsection that this $ \mathrm{AdS}_5 \times \mathbf{CP}^3 $ compactification
has no supersymmetry, so to have a chance of having a supersymmetric $
\mathrm{AdS}_4 \times \mathbf{CP}^3 $ compactification of Ho\v{r}ava-Witten
theory, $ G_{\mu \nu \sigma \tau} $ would have to be nonzero in the bulk, away
from the orbifold hyperplanes.  The boundary conditions, on $ G_{\mu \nu \sigma
\tau} $, would then be that these components vanish on both orbifold
hyperplanes.  A new feature, in the bulk, would be that $ G_{IJKL} $ now has
enough nonvanishing components, in the bulk, to turn on the nonlinear $
G_{IJKL_1 \ldots L_8}G^{L_1 \ldots L_4} G^{L_5 \ldots L_8} $ term in the field
equation for $ C_{IJK} $, where  $ G_{IJKL_1 \ldots L_8} $ denotes the tensor
$ \sqrt{-G} \epsilon_{IJKL_1 \ldots L_8} $.  We would thus expect also to find
some nonvanishing components of $ G_{IJKL} $ that have an index $ y $, and some
nonvanishing components of $ C_{IJK} $, with an index $ y $, were in fact found
in Witten's original investigation of supersymmetric compactifications of
Ho\v{r}ava-Witten theory \cite{Witten Strong coupling expansion}.

Furthermore, the metric components $ G_{Ay} $, which are nonzero in the
Nilsson-Pope $ d = 11 $ oxidation of the Watamura $ N = 6 $ compactification of
type IIA $ d = 10 $ supergravity, due to the replacement (\ref{replacement of
dy squared}) in the metric ansatz (\ref{metric ansatz}), are also odd under
reflection in the Ho\v{r}ava-Witten orbifold hyperplanes, and must thus
presumably vanish on the orbifold hyperplanes, since, to the best of my
knowledge, there is no analogue, for the metric components $ G_{Uy} $, of the
discontinuity equation (\ref{G discontinuity}), for the components $ G_{UVWX} $
of the four-form field strength of the three-form gauge field.  This is, again,
not necessarily inconsistent with the existence of a compactification, due to
the existence of the $ \mathrm{AdS}_5 \times \mathbf{CP}^3 $ compactification
studied in the next subsection, and would give the boundary conditions on $
G_{Ay} $.

In the presence of a boundary, half of the bulk supersymmetry is always broken
\cite{van Nieuwenhuizen Vassilevich}.  However the $ N = 3 $ $ d = 4 $
supergravity supermultiplet contains three vector bosons, which naturally
transform as the adjoint of $ \mathrm{SO}\left(3\right) $, and do not fit
naturally into a $ \mathbf{CP}^3 $ compactification.  However, as noted by
Nilsson and Pope \cite{Nilsson Pope}, the $ \mathrm{SU}\left(4\right) \times
\mathrm{U}\left(1\right) $ gauge bosons, found in the Watamura $ \mathbf{CP}^3
$ compactification of type IIA $ d = 10 $ supergravity, could be consistent
with $ N = 2 $ or $ N = 1 $ supersymmetry, as well as with $ N = 6 $
supersymmetry.  $ N = 2 $, $ d = 4 $ supersymmetry is not consistent with the
existence of chiral fermions, and three of the four embeddings of the $
\mathrm{SU}\left(3\right) \times \mathrm{U}\left(1\right) $ spin connection of
$ \mathrm{CP}^3 $, in $ E8 \times E8 $, found by Pilch and Schellekens
\cite{Pilch Schellekens}, have chiral fermions, so could have at most $ N = 1 $
supersymmetry, whereas the fourth embedding found by Pilch and Schellekens,
their case 4.3.(a), has no chiral fermions for $ \mathbf{CP}^3 $, and thus
might possibly be consistent with $ N = 2 $ supersymmetry.

We note that for gauged $ N $-extended $ d = 4 $ supergravity, with $ N leq 4
$, and not coupled to any matter multiplets, Allen and Davis \cite{Allen Davis}
found that the one-loop vacuum energy, in the $ \mathrm{AdS}_4 $ background, is
nonvanishing, so that there would be no possibility of an analogue of the
Gibbons-Nicolai floor by floor vanishing of the one-loop vacuum energy when the
contributions of the Kaluza-Klein multiplets above the supergravity multiplet
are included.  However, if the Nilsson-Pope $ d = 11 $ oxidation of the
Watamura $ N = 6 $ solution could be modified to obtain a supersymmetric $
\mathbf{CP}^3 $ compactification of Ho\v{r}ava-Witten theory, in the manner
just discussed, there would be additional supersymmetric Yang-Mills multiplets,
together with the Kaluza-Klein multiplets above them, so there would be a
possibility that the floor by floor vanishing of the one-loop vacuum energy
might be restored.

The question of whether or not there exists, in the bulk, a supersymmetric
deformation of the Nilsson-Pope $ d = 11 $ oxidation of the Watamura $ N = 6 $
compactification, whose nonvanishing components of $ G_{IJKL} $ include
components $ G_{ABCD} $ given by the LOSW ansatz (\ref{four form field
strength}), where $ \alpha $ might now depend on $ y $, could perhaps be
investigated, in the first instance, by Witten's method \cite{Witten Strong
coupling expansion}, in which the new components of $ G_{IJKL} $ would be
treated as a perturbation.

\subsection{$\mathrm{AdS}_5 \times \mathbf{C} \mathbf{P}^3$ compactification
of $d = 11$ supergravity}
\label{AdS5 times CP3 compactification}

The value of the integration constant $B$, in (\ref{bulk type power law for
upper sign}), that is required for TeV-scale gravity, is given by (\ref{B for
TeV scale gravity}), when the outer surface of the thick pipe is stabilized in
the quantum region by Casimir effects, and by (\ref{B for TeV scale gravity
with fluxes}), when the outer surface is stabilized in the classical region by
extra fluxes.  From these equations, we see that the value of
$\frac{B}{\kappa^{2/9}}$ required for TeV-scale gravity is reduced if
the Euler number $\chi \left( \mathcal{M}^6 \right)$ of the compact
six-manifold, which is a negative integer for the compact six-manifolds
considered in the present paper, is large in magnitude.  However, $\left| \chi
\left( \mathcal{M}^6 \right) \right|$ is also related to
$\frac{b_1}{\kappa^{2/9}}$ by the relation (\ref{b sub 1 in terms of
chi}), which follows from the value (\ref{alpha sub U in the Standard Model})
of the Yang-Mills fine structure constant assumed at unification, which is the
value of the QCD fine structure constant, $\alpha_3$, as evolved in the
Standard Model to around 150 TeV.  Here $b_1 = b \left( y_1 \right)$ is the
value of $b \left( y \right)$ at the inner surface of the thick pipe, where $b
\left( y \right)$ was introduced in the metric ansatz (\ref{metric ansatz}) as
the scale factor that determines the diameter of the compact six-manifold,
once its topology is fixed by selecting a specific smooth compact quotient of
$\mathbf{C} \mathbf{H}^3$ or $\mathbf{H}^6$.  And
$\frac{b_1}{\kappa^{2/9}}$ is determined by Casimir effects near the
inner surface of the thick pipe, and thus, as discussed in subsection \ref{The
boundary conditions at the inner surface of the thick pipe}, cannot be small
compared to $1$, unless, for some reason, not only are the one-loop
coefficients in the Casimir energy densities (\ref{the t i as functions of b})
and (\ref{surface Casimir action density}) small compared to $1$,
but also the multi-loop coefficients are all suppressed by the appropriate
powers of the small number $\frac{b_1}{\kappa^{2/9}}$, either to all
loop orders, or at least up to some high loop order.  Thus we cannot have a
very large value of $\left| \chi \left( \mathcal{M}^6 \right) \right|$, and
also obtain a reasonable value value of the Yang-Mills fine structure constant
at unification, unless the coefficients in the Casimir energy densities
(\ref{the t i as functions of b}) and (\ref{surface Casimir action
density}), either to all loop orders, or at least to some high loop order, all
tend to zero as the appropriate power of $\frac{b_1}{\kappa^{2/9}}$,
where $\frac{b_1}{\kappa^{2/9}}$ is given by (\ref{b sub 1 in terms of
chi}), as $\left| \chi \left( \mathcal{M}^6 \right) \right|$ becomes very
large.  However there is no reason for this to happen unless some special
effect occurs, because the limit $\left| \chi \left( \mathcal{M}^6 \right)
\right| \rightarrow \infty$ does not correspond to any restoration of
supersymmetry.

A special effect of the required type was, however, discovered by Gibbons and
Nicolai {\cite{Gibbons Nicolai}}, who calculated the one-loop Casimir energy
density of the Freund-Rubin compactification of $d = 11$ supergravity on the
round seven-sphere {\cite{Freund Rubin}}, including the effects of all the
Kaluza-Klein states, and found that not only did the Casimir energy density
vanish, as required to preserve the supersymmetry of the solution at one loop,
but also the contributions to the Casimir energy density vanished ``floor by
floor'', or in other words, at each separate Kaluza-Klein level or
$ \mathrm{Osp}(8|4) $
multiplet, which is not required to preserve the supersymmetry.  This appears
to suggest that the one-loop Casimir energy density of this compactification
would still vanish ``floor by floor'' even if the Freund-Rubin relation
between the $\mathrm{AdS}_4$ radius and the $\mathbf{S}^7$ radius was broken,
in which case the background would no longer satisfy the classical
Cremmer-Julia-Scherk field equations, but the Casimir energy density would
nevertheless still be defined by the general formula for the quantum effective
action, $\Gamma$, as a function of arbitrary background fields, as described
before (\ref{action for calculating quantum effective action}).  Thus the
Gibbons-Nicolai result would seem to imply that the one-loop Casimir energy
density of $d = 11$ supergravity, defined in this way, would vanish ``floor by
floor'' even when the background is flat $\mathbf{R}^4$, times
$\mathbf{S}^7$.  And furthermore, since there will be relations between the
propagators and heat kernels on a flat $\mathbf{R}^4$, times
$\mathbf{S}^7$, background, and the propagators and heat kernels on a flat
$\mathbf{R}^4$, times $\mathbf{H}^7$, background, analogous to those
discussed above for the flat $\mathbf{R}^5$, times $\mathbf{C}
\mathbf{P}^3$, and the flat $\mathbf{R}^5$, times $\mathbf{C}
\mathbf{H}^3$, backgrounds, the Gibbons-Nicolai result would seem to suggest
that the one-loop Casimir energy density of $d = 11$ supergravity, defined by
the quantum effective action, $\Gamma$, as a function of arbitrary background
fields, will also vanish when the background is flat $\mathbf{R}^4$, times
$\mathbf{H}^7$, for arbitrary radius of curvature of the $\mathbf{H}^7$.
In that case, the one-loop Casimir energy density of $d = 11$ supergravity, on
a flat $\mathbf{R}^4$, times $\mathcal{M}^7$, background, where
$\mathcal{M}^7$ is a smooth compact quotient of $\mathbf{H}^7$, would
presumably tend to zero, in the limit as the volume of the $\mathcal{M}^7$ at
fixed Ricci scalar, which is a topological invariant by Mostow's rigidity
theorem even though the Euler number vanishes for a smooth compact manifold of
odd dimension, tends to infinity.

Thus it is appropriate to ask if there exist supersymmetric compactifications
of $d = 11$ supergravity on $\mathrm{AdS}_5 \times \mathbf{C} \mathbf{P}^3$
or $\mathrm{AdS}_5 \times \mathbf{S}^6$, which might lead, by an analogue of
the Gibbons-Nicolai effect, to the vanishing of the one-loop Casimir energy
density of $d = 11$ supergravity, as defined by the quantum effective action,
$\Gamma$, on a flat $\mathbf{R}^5$, times $\mathbf{C} \mathbf{H}^3$,
background, or a flat $\mathbf{R}^5$, times $\mathbf{H}^6$, background.
To the best of my knowledge, there is no classical solution of the
Cremmer-Julia-Scherk field equations on an $\mathrm{AdS}_5 \times
\mathbf{S}^6$ background, that has a maximally symmetric metric on both
factors, because there is no natural ansatz for the four-form field strength
of the three-form gauge field.  However, there is, indeed, a classical
solution of the Cremmer-Julia-Scherk field equations on an $\mathrm{AdS}_5
\times \mathbf{C} \mathbf{P}^3$ background, with the
Lukas-Ovrut-Stelle-Waldram (LOSW) ansatz (\ref{four form field strength}) for
the four-form field strength of the three-form gauge field.  I shall seek a
solution with the metric ansatz (\ref{metric ansatz}), such that
$\mathrm{AdS}_5$ is realized as flat four-dimensional Minkowski space times the
$y$ direction, with $a \left( y \right)$ depending exponentially on $y$, as in
the Randall-Sundrum model {\cite{Randall Sundrum 1}}, and $b \left( y \right)$
independent of $y$.  Comparing the Ricci tensor components (\ref{Ricci tensor
for the metric ansatz}), the energy-momentum tensor components contributed by
the three-form gauge field with the LOSW ansatz (\ref{three form energy
momentum tensor components}), the definition of the $t^{\left( i \right)}
\left( y \right)$ energy-momentum tensor coefficients (\ref{T IJ block
diagonal structure}), and the Einstein equations (\ref{first Einstein
equation}), (\ref{second Einstein equation}), and (\ref{third Einstein
equation}), we see that on replacing $\mathbf{C} \mathbf{H}^3$ by
$\mathbf{C} \mathbf{P}^3$, so that the relation $R_{AB} \left( h \right) =
4 h_{AB}$ is replaced by $R_{AB} \left( h \right) = - 4 h_{AB}$, and replacing
four-dimensional de Sitter space by four dimensional Minkowski space, so that
the relation $R_{\mu \nu} \left( g \right) = - 3 g_{\mu \nu}$ is replaced by
$R_{\mu \nu} \left( g \right) = 0$, and setting the $t^{\left( i \right)}
\left( y \right)$ energy-momentum tensor coefficients to the values given by
the LOSW ansatz, the Einstein equations become:
\begin{equation}
  \label{first Einstein equation for CP 3} \frac{\ddot{a}}{a} + 3
  \frac{\dot{a}^2}{a^2} + 6 \frac{\dot{a} \dot{b}}{ab} - \frac{\alpha^2}{9
  b^8} = 0
\end{equation}
\begin{equation}
  \label{second Einstein equation for CP 3} \frac{\ddot{b}}{b} + 5
  \frac{\dot{b}^2}{b^2} + 4 \frac{\dot{a} \dot{b}}{ab} - \frac{4}{b^2} +
  \frac{\alpha^2}{9 b^8} = 0
\end{equation}
\begin{equation}
  \label{third Einstein equation for CP 3} 4 \frac{\ddot{a}}{a} + 6
  \frac{\ddot{b}}{b} - \frac{\alpha^2}{9 b^8} = 0
\end{equation}
Requiring that $\dot{b} = 0$, $\ddot{b} = 0$, the second of these equations
reduces to
\begin{equation}
  \label{relation between alpha and b for CP 3} \alpha^2 = 36 b^6
\end{equation}
The first and third equations then reduce to:
\begin{equation}
  \label{reduced first Einstein equation for CP 3} \frac{\ddot{a}}{a} + 3
  \frac{\dot{a}^2}{a^2} = \frac{4}{b^2}
\end{equation}
\begin{equation}
  \label{reduced third Einstein equation for CP 3} \frac{\ddot{a}}{a} =
  \frac{1}{b^2}
\end{equation}
which have the solutions $a = Ae^{\frac{y}{b}}$ and $a = Ae^{- \frac{y}{b}}$.
And from the formulae (\ref{Riemann tensor for the metric ansatz}) for the
Riemann tensor components, we see that
\begin{equation}
  \label{AdS 5 Riemann tensor components} R_{\mu \nu \sigma \tau} =
  \frac{1}{b^2} \left( G_{\mu \sigma} G_{\nu \tau} - G_{\nu \sigma} G_{\mu
  \tau} \right), \hspace{4em} R_{\mu y \nu y} = \frac{1}{b^2} G_{\mu \nu}
\end{equation}
hence since $G_{\mu y} = 0$ and $G_{yy} = 1$, we have:
\begin{equation}
  \label{maximally symmetric form of Riemann tensor for AdS 5} R_{\bar{\mu}
  \bar{\nu} \bar{\sigma} \bar{\tau}} = \frac{1}{b^2} \left( G_{\bar{\mu}
  \bar{\sigma}} G_{\bar{\nu} \bar{\tau}} - G_{\bar{\nu} \bar{\sigma}}
  G_{\bar{\mu} \bar{\tau}} \right)
\end{equation}
where the barred Greek indices run over four-dimensional Minkowski space and
$y$.  Thus the five-dimensional space formed from four-dimensional Minkowski
space and the $y$ direction is maximally symmetric, and in consequence of its
$\left( - + + + + \right)$ signature and the relation $R_{\bar{\mu}
\bar{\sigma}} = \frac{4}{b^2} G_{\bar{\mu} \bar{\sigma}}$, is $\mathrm{AdS}_5$.

We now need to determine whether this solution has any supersymmetries.  There
are no Majorana spinors in five dimensions, but a symplectic-Majorana
condition can be imposed on a pair of spinors {\cite{Cremmer}}, in consequence
of which the possible numbers of supersymmetries in five dimensions are even,
and there do, indeed, exist supergravities with 2, 4, 6, and 8 supersymmetries
in five dimensions {\cite{Cremmer}}.  We know from the Figueroa-O'Farrill -
Papadopoulos theorem {\cite{Figueroa-O'Farrill Papadopoulos}} that the
solution cannot have 8 supersymmetries.  The SU(4) isometry group of
$\mathbf{C} \mathbf{P}^3$ with its standard metric implies there will be
15 Yang-Mills vector bosons in the adjoint of SU(4), and looking at the table
of states of the extended supergravities in five dimensions given by Cremmer
{\cite{Cremmer}}, we see that $N = 6$ supergravity in five dimensions has
precisely 15 vector fields, which on toroidal compactification to four
dimensions join the extra vector field coming from the metric, to produce the
standard $15 + 1 = 16$ vector fields of $N = 6$ supergravity in four
dimensions.  Furthermore, Nilsson and Pope {\cite{Nilsson Pope}} found that a
known compactification {\cite{Watamura}} of Type IIA supergravity in ten
dimensions on $\mathrm{AdS}_4 \times \mathbf{C} \mathbf{P}^3$ has either $N
= 6$ supersymmetry or no supersymmetry, depending on the relative sign of form
field fluxes on the $\mathrm{AdS}_4$ and $\mathbf{C} \mathbf{P}^3$ factors.
However, notwithstanding these positive indications, the $\mathrm{AdS}_5 \times
\mathbf{C} \mathbf{P}^3$ compactification of $d = 11$ supergravity
considered above has no supersymmetry.

To check this, I shall use the notations of subsection \ref{Horava-Witten
theory} for supergravity in eleven dimensions, so coordinate indices $I, J, K,
\ldots$ run over all directions on $\mathcal{M}^{11}$.  The Dirac matrices
$\Gamma^I$ satisfy $\left\{ \Gamma^I, \Gamma^J \right\} = 2 G^{IJ}$, and
$\Gamma^{I_1 I_2 \ldots I_n} \equiv \Gamma^{\left[ I_1 \right.} \Gamma^{I_2}
\ldots \Gamma^{\left. I_n \right]}$.  Coordinate indices $\mu, \nu, \sigma,
\ldots$ will now run over all directions on $\mathrm{AdS}_5$, which is a change
from the meaning of the Greek indices used above and in section \ref{Thick
pipe geometries}, and coordinate indices $A, B, C$ will run over all
directions on the compact six-manifold, which is in agreement with section
\ref{Thick pipe geometries}, although the compact six-manifold is now
$\mathbf{C} \mathbf{P}^3$.  Local Lorentz indices will be indicated by
putting a bar over the corresponding coordinate indices, so the meaning of
barred Greek indices is also now changed from their meaning in equation
(\ref{maximally symmetric form of Riemann tensor for AdS 5}) above.  A real
representation of the $\Gamma$ matrices for eleven dimensions does not
decompose neatly into Dirac matrices for the five extended dimensions with
signature $\left( - + + + + \right)$ and Dirac matrices for the six compact
dimensions with signature $\left( + + + + + + \right)$, so I shall instead use
a representation of the form used by Lukas, Ovrut, Stelle, and Waldram
{\cite{Lukas Ovrut Stelle Waldram}}, with $\Gamma^I = \frac{1}{b} \left\{
\gamma^{\mu} \times \lambda, 1 \times \lambda^A \right\}$, where
$\gamma^{\mu}$ and $\lambda^A$ are the five- and six-dimensional Dirac
matrices, respectively.  Here, $\lambda$ is the chiral projection matrix in
six dimensions with $\lambda^2 = 1$.  For a specific representation of the
$\lambda^{\bar{A}}$ we can choose $\lambda^5 = \sigma_1 \times 1 \times 1$,
$\lambda^6 = \sigma_2 \times 1 \times 1$, $\lambda^7 = \sigma_3 \times
\sigma_1 \times 1$, $\lambda^8 = \sigma_3 \times \sigma_2 \times 1$,
$\lambda^9 = \sigma_3 \times \sigma_3 \times \sigma_1$, $\lambda^{10} =
\sigma_3 \times \sigma_3 \times \sigma_2$.  We define $\lambda = i \lambda^5
\lambda^6 \lambda^7 \lambda^8 \lambda^9 \lambda^{10} = \sigma_3 \times
\sigma_3 \times \sigma_3$.  For a specific representation of the
$\gamma^{\bar{\mu}}$ we can choose $\gamma^1 = \sigma_1 \times 1$, $\gamma^2 =
\sigma_2 \times 1$, $\gamma^3 = \sigma_3 \times \sigma_1$, $\gamma^4 =
\sigma_3 \times \sigma_2$, $\gamma^0 = i \sigma_3 \times \sigma_3$.  Then for
the charge conjugation matrix $C$ in eleven dimensions, which satisfies as
usual $\left( \Gamma^I \right)^T C = - C \Gamma^I$, $C^T = - C$, we can take
$C = C_5 \times C_6$, where $C_5 = \sigma_1 \times \sigma_2$ is the charge
conjugation matrix in five dimensions, and satisfies $\left( \gamma^{\mu}
\right)^T C_5 = C_5 \gamma^{\mu}$, $C^T_5 = - C_5$, in agreement with
{\cite{Cremmer}}, and $C_6 = \sigma_2 \times \sigma_1 \times \sigma_2$ is the
charge conjugation matrix in six dimensions, and satisfies $\left( \lambda^A
\right)^T C_6 = - C_6 \lambda^A$, $\lambda^T C_6 = - C_6 \lambda$, $C^T_6 =
C_6$.

Now the gravitino field is zero in the above classical solution, so if it has
any supersymmetries, there must exist supersymmetry variation parameters $\eta
\left( x, z \right)$, where $x^{\mu}$ are coordinates on $\mathrm{AdS}_5$, and
$z^A$ are coordinates on $\mathbf{C} \mathbf{P}^3$, such that the
supersymmetry variation of the gravitino vanishes.  The supersymmetry
variation of the gravitino, about a configuration in which the gravitino field
is zero, is {\cite{Cremmer Julia Scherk, HW2, Witten Strong
coupling expansion}}:
\begin{equation}
  \label{SUSY variation of the gravitino} \delta \psi_I = D_I \eta +
  \frac{\sqrt{2}}{288} \left( \Gamma_{IJKLM} - 8 G_{IJ} \Gamma_{KLM} \right)
  G^{JKLM} \eta
\end{equation}
To study the condition on $\eta$ that results from setting this variation
equal to zero, when $G_{JKLM}$ is given by the LOSW ansatz (\ref{four form
field strength}), I shall follow the method of Nilsson and Pope {\cite{Nilsson
Pope}}.  It is convenient, first, to note the identities:
\begin{equation}
  \label{identity for a term in the gravitino variation} \Gamma_{IJKLM}
  G^{JKLM} = \left( \Gamma_I \Gamma_{JKLM} - 4 G_{IJ} \Gamma_{KLM} \right)
  G^{JKLM}
\end{equation}
and:
\begin{equation}
  \label{another identity for a term in the gravitino variation} 8 G_{IJ}
  \Gamma_{KLM} G^{JKLM} = \left[ \Gamma_I, \Gamma_{JKLM} \right] G^{JKLM}
\end{equation}
Thus the supersymmetry variation of the gravitino, (\ref{SUSY variation of the
gravitino}), can be written:
\begin{equation}
  \label{equivalent form of the SUSY variation of the gravitino} \delta \psi_I
  = D_I \eta + \frac{\sqrt{2}}{576} \left( - \Gamma_I \Gamma_{JKLM} + 3
  \Gamma_{JKLM} \Gamma_I \right) G^{JKLM} \eta
\end{equation}
Now from the definition (\ref{Kahler form}) of the K\"ahler
form, we have:
\begin{equation}
  \label{square of the Kahler form} \omega_{AB} \omega_{\hspace{0.4ex}
  \hspace{0.4ex} \hspace{0.4ex} C}^B = - h_{AC}
\end{equation}
Furthermore, for an arbitrary $2 n \times 2 n$ antisymmetric matrix $M$, with
real matrix elements, we have the identity:
\begin{equation}
  \label{Pfaffian identity with two free indices} \varepsilon_{i_1 i_2 i_3 i_4
  \ldots i_{2 n - 1} j} M_{i_1 i_2} M_{i_3 i_4} \ldots M_{i_{2 n - 1} k} =
  2^{n - 1} \left( n - 1 \right) ! \sqrt{\det M} \delta_{jk}
\end{equation}
This is proved by applying an orthogonal similarity transformation to
transform $M$ to a block diagonal matrix $\tilde{M}$, such that each block in
the block diagonal of $\tilde{M}$ is an antisymmetric $2 \times 2$ matrix with
real matrix elements, then replacing each index $i$ by an index pair $aI$,
where $a$ runs from 1 to 2, and $I$ runs from 1 to $n$, so that $\tilde{M}$
can be expressed as a Kronecker product $\tilde{M}_{aI, bJ} = \varepsilon_{ab}
\bar{M}_{IJ}$, where $\varepsilon_{ab} = \left(\begin{array}{cc}
  0 & 1\\
  - 1 & 0
\end{array}\right)$, and $\bar{M}$ is an $n \times n$ diagonal matrix with
real matrix elements.

Applying this to the K\"ahler form, we have the identity:
\begin{equation}
  \label{Pfaffian identity for Kahler form} \omega_{AB} \omega_{CD} h^{ABCDEF}
  = 8 \omega^{EF}
\end{equation}
Following Nilsson and Pope, it is convenient to define:
\begin{equation}
  \label{definition of Nilsson Pope Q} Q \equiv - i \omega^{AB} \lambda_{AB}
  \lambda
\end{equation}
We note that $\left[ Q, \lambda \right] = 0$.  And from the definition of the
$\Gamma^A$ in terms of the $\lambda^A$, as above, $\left\{ \lambda_A,
\lambda_B \right\} = 2 h_{AB}$.  Thus from the identities
\begin{equation}
  \label{identity for lambda sub AB lambda sub CD} \lambda_{AB} \lambda_{CD} =
  \lambda_{ABCD} - h_{AC} \lambda_{BD} + h_{AD} \lambda_{BC} + h_{BC}
  \lambda_{AD} - h_{BD} \lambda_{AC} - h_{AC} h_{BD} + h_{AD} h_{BC}
\end{equation}
\begin{equation}
  \label{identity for lambda sub ABCD} \lambda_{ABCD} = \frac{i}{2} h_{ABCDEF}
  h^{EG} h^{FH} \lambda_{GH} \lambda
\end{equation}
and (\ref{Pfaffian identity for Kahler form}), we find that:
\begin{equation}
  \label{identity for square of Nilsson Pope Q} Q^2 = 4 Q + 12
\end{equation}
Hence the eigenvalues of $Q$ are $- 2$ and $6$, hence since $Q$ is traceless,
there are six eigenvalues $- 2$ and two eigenvalues $6$.

We now assume that $\eta \left( x, z \right)$ factorizes in the form $\eta
\left( x, z \right) = \varepsilon \left( x \right) \tilde{\eta} \left( z
\right)$, where $\varepsilon \left( x \right)$ is a four component spinor
acted on by the first factor in the Kronecker product expressions for the
$\Gamma^I$, and $\tilde{\eta} \left( z \right)$ is an eight component spinor
acted on by the second factor in the Kronecker product expressions for the
$\Gamma^I$.  Substituting in the LOSW ansatz (\ref{four form field strength}),
and requiring that $\delta \psi_I = 0$, we find from the components of
(\ref{equivalent form of the SUSY variation of the gravitino}) with $I$ along
$\mathbf{C} \mathbf{P}^3$ that:
\begin{equation}
  \label{condition on eta tilde} D_A \tilde{\eta} + \frac{\alpha
  \sqrt{2}}{3456 b^3} \left( - \lambda_A \lambda_{BCDE} + 3 \lambda_{BCDE}
  \lambda_A \right) h^{BCDEFG} \omega_{FG} \tilde{\eta} = 0
\end{equation}
Now from (\ref{definition of Nilsson Pope Q}) and (\ref{identity for lambda
sub ABCD}), we find:
\begin{equation}
  \label{lambda h omega in terms of Q} \lambda_{BCDE} h^{BCDEFG} \omega_{FG} =
  - 24 Q
\end{equation}
Hence (\ref{condition on eta tilde}) reduces to:
\begin{equation}
  \label{reduced condition on eta tilde} D_A \tilde{\eta} \pm
  \frac{\sqrt{2}}{24} \left( \lambda_A Q - 3 Q \lambda_A \right) \tilde{\eta}
  = 0
\end{equation}
where I also used (\ref{relation between alpha and b for CP 3}), and the sign
choice corresponds to $\alpha = \pm 6 b^3$.  And from (\ref{definition of
Nilsson Pope Q}), we also have:
\begin{equation}
  \label{anticommutator of Q and lambda sub A} \left\{ Q, \lambda_A \right\} =
  - 4 i \omega^{\hspace{0.4ex} \hspace{0.4ex} \hspace{0.4ex} B}_A \lambda_B
  \lambda
\end{equation}
Hence (\ref{reduced condition on eta tilde}) is equivalent to:
\begin{equation}
  \label{equivalent reduced condition on eta tilde} D_A \tilde{\eta} \pm
  \frac{\sqrt{2}}{6} \left( \lambda_A Q + 3 i \omega^{\hspace{0.4ex}
  \hspace{0.4ex} \hspace{0.4ex} B}_A \lambda_B \lambda \right) \tilde{\eta} =
  0
\end{equation}
A necessary condition for the existence of solutions of (\ref{equivalent
reduced condition on eta tilde}) is the integrability condition:
\begin{equation}
  \label{integrability condition for eta tilde} \left[ D_A \pm
  \frac{\sqrt{2}}{6} \left( \lambda_A Q + 3 i \omega^{\hspace{0.4ex}
  \hspace{0.4ex} \hspace{0.4ex} C}_A \lambda_C \lambda \right), D_B \pm
  \frac{\sqrt{2}}{6} \left( \lambda_B Q + 3 i \omega^{\hspace{0.4ex}
  \hspace{0.4ex} \hspace{0.4ex} D}_B \lambda_D \lambda \right) \right] = 0
\end{equation}
To evaluate the left-hand side of (\ref{integrability condition for eta
tilde}), we note first that with the convention (\ref{Riemann tensor
definition}) for the Riemann tensor, we have $\left[ D_A, D_B \right] = -
\frac{1}{4} R_{ABCD} \Gamma^{CD} = - \frac{1}{4} \tilde{R}_{ABCD}
\lambda^{CD}$, where $R_{ABCD}$ is the Riemann curvature of $\mathbf{C}
\mathbf{P}^3$ with the metric $G_{AB} = b^2 h_{AB}$, and $\tilde{R}_{ABCD} =
\frac{1}{b^2} R_{ABCD}$ is the Riemann curvature of $\mathbf{C}
\mathbf{P}^3$ with the metric $h_{AB}$.  And secondly, there are no cross
terms between $D_A$ or $D_B$, and the extra terms that came from the
$G^{JKLM}$ term in (\ref{SUSY variation of the gravitino}), because the extra
terms are built from the K\"ahler form and the vielbein, which are covariantly
constant, and the Dirac matrices with local Lorentz indices, and $\lambda$,
which are position-independent invariant tensors with respectively a vector
index and two spinor indices, and two spinor indices, and thus also
covariantly constant.

To evaluate the commutator of the extra terms, we note that:
\begin{equation}
  \label{lambda Q lambda Q commutator} \left[ \lambda_A Q, \lambda_B Q \right]
  = - 4 i \omega^{\hspace{0.4ex} \hspace{0.4ex} \hspace{0.4ex} D}_B \lambda_A
  \lambda_D \lambda Q + 4 i \omega^{\hspace{0.4ex} \hspace{0.4ex}
  \hspace{0.4ex} D}_A \lambda_B \lambda_D \lambda Q + 2 \lambda_{BA} \left( 4
  Q + 12 \right)
\end{equation}
\begin{equation}
  \label{lambda Q lambda lambda commutator} \left[ \lambda_A Q, \lambda_D
  \lambda \right] = - 4 i \omega^{\hspace{0.4ex} \hspace{0.4ex} \hspace{0.4ex}
  \hspace{0.4ex} C}_D \lambda_A \lambda_C + 2 \lambda_{DA} \lambda Q
\end{equation}
where (\ref{identity for square of Nilsson Pope Q}) was used to obtain
(\ref{lambda Q lambda Q commutator}).  Thus we find:
\[ \left[ \pm \frac{\sqrt{2}}{6} \left( \lambda_A Q + 3 i
   \omega_A^{\hspace{0.4ex} \hspace{0.4ex} \hspace{0.4ex} C} \lambda_C \lambda
   \right), \pm \frac{\sqrt{2}}{6} \left( \lambda_B Q + 3 i
   \omega_B^{\hspace{0.4ex} \hspace{0.4ex} \hspace{0.4ex} D} \lambda_D \lambda
   \right) \right] = \hspace{8em}  \]
\begin{equation}
  \label{commutator of the extra terms} \hspace{2em} = \frac{5 i}{9} \left(
  \omega^{\hspace{0.4ex} \hspace{0.4ex} \hspace{0.4ex} D}_B \lambda_{DA} -
  \omega^{\hspace{0.4ex} \hspace{0.4ex} \hspace{0.4ex} D}_A \lambda_{DB}
  \right) \lambda Q + \frac{4 i}{9} \omega_{AB} \lambda Q + \frac{4}{9}
  \lambda_{BA} \left( Q + 6 \right) + \omega_A^{\hspace{0.4ex} \hspace{0.4ex}
  \hspace{0.4ex} C} \omega_B^{\hspace{0.4ex} \hspace{0.4ex} \hspace{0.4ex} D}
  \lambda_{CD}
\end{equation}
Terms of the same structure, but with different coefficients, occurred in
Nilsson and Pope's calculation of the corresponding commutator for the
$\mathrm{AdS}_4 \times \mathbf{C} \mathbf{P}^3$ compactification of Type IIA
supergravity in ten dimensions {\cite{Watamura}}, and in that case, for one of
two alternative choices of a relative sign, the result was that after adding
the Riemann tensor term, each nonvanishing term had a factor of $\left( Q + 2
\right)$ at its right-hand side, so that acting on any linear combination of
the six linearly independent eigenvectors of $Q$ with eigenvalue $- 2$, the
commutator vanished.  That does not happen in the present case, so we have to
check whether there is any further relation between the terms in the left-hand
side of (\ref{integrability condition for eta tilde}) that might result in
(\ref{integrability condition for eta tilde}) being satisfied when acting on
an appropriate eigenvector of $Q$.

It is convenient now to switch to complex coordinates, as in subsection
\ref{CH3}, on page \pageref{CH3}.  Barred Latin indices will now denote
antiholomorphic indices, as in subsection \ref{CH3}.  Then corresponding to
the Riemann tensor (\ref{CHn covariant Riemann tensor}) for $\mathbf{C}
\mathbf{H}^n$, the Riemann tensor for $\mathbf{C} \mathbf{P}^3$ is:
\begin{equation}
  \label{Riemann tensor for CP 3} \tilde{R}_{r \bar{s} t \bar{u}} = h_{r
  \bar{s}} h_{t \bar{u}} + h_{r \bar{u}} h_{t \bar{s}}
\end{equation}
Evaluating the left-hand side of (\ref{integrability condition for eta tilde})
for $A = r$, $B = s$, the Riemann tensor term does not contribute, and the
result is:
\begin{equation}
  \label{non satisfaction of integrability condition for eta tilde}
  \frac{1}{9} \lambda_{sr} \left( 10 \lambda Q + 4 Q + 33 \right)
\end{equation}
which is nonvanishing for any combination of $\lambda = + 1$ or $- 1$, and $Q
= - 2$ or $+ 6$, and thus proves the absence of supersymmetry.  And similarly,
for $A = \bar{r}$, $B = \bar{s}$, the left-hand side of (\ref{integrability
condition for eta tilde}) is $\frac{1}{9} \lambda_{\bar{s}  \bar{r}} \left( -
10 \lambda Q + 4 Q + 33 \right)$.  And for $A = r$, $B = \bar{s}$, the Riemann
tensor term $- \frac{1}{4} \tilde{R}_{ABCD} \lambda^{CD}$ contributes
$\frac{1}{4} \left( i \omega_{r \bar{s}} Q \lambda + 2 \lambda_{r \bar{s}}
\right)$, and the left-hand side of (\ref{integrability condition for eta
tilde}) is $\frac{1}{36} \left( 25 i \omega_{r \bar{s}} Q \lambda +
\lambda_{\bar{s} r} \left( 16 Q + 42 \right) \right)$.  In fact, if the
numerical coefficients of the terms in the parentheses in (\ref{equivalent
reduced condition on eta tilde}) had had the values $\pm \frac{3}{4} i$, $\mp
\frac{3}{2} i$, instead of their actual values $1$ and $3$, the left-hand side
of (\ref{commutator of the extra terms}) would have been equal to $\frac{1}{4}
\lambda_{AB} \left( Q + 2 \right)$, and would thus have been consistent with
$N = 6$ supersymmetry.

About 18 months after version 1 of this article was published on arXiv, I
learned from {\cite{Martin Reall}} that the $\mathrm{AdS}_5 \times
\mathbf{CP}^3$ solution was studied by Pope and van
Nieuwenhuizen in 1989, who showed that it is not supersymmetric {\cite{Pope
van Nieuwenhuizen}}. \ The lack of supersymmetry could also have been deduced
from a general study of supersymmetric $\mathrm{AdS}_5$ solutions of
$M$-theory by Gauntlett, Martelli, Sparks, and Waldram {\cite{Gauntlett
Martelli Sparks Waldram}}.

\section{$ E8 $ vacuum gauge fields and the Standard Model}
\label{E8 vacuum gauge fields and the Standard Model}

In the present paper, we have considered the compactification of Ho\v{r}ava-Witten
theory on a smooth compact quotient of either $\mathbf{C} \mathbf{H}^3$ or
$\mathbf{H}^6$, which breaks supersymmetry completely.  The fact that the
observed gauge coupling constants are $\sim 1$ in magnitude implies that the
six-volume of the inner surface of the thick pipe is $\sim
\kappa^{\frac{4}{3}}$, as discussed in subsection \ref{Newtons constant and
the cosmological constant}, following (\ref{Einstein term in four dimensional
effective action for solutions with a small at outer surface}), on page
\pageref{Einstein term in four dimensional effective action for solutions with
a small at outer surface}.  Thus the energy at which supersymmetry is broken
at the inner surface of the thick pipe will be $\sim \kappa^{- \frac{2}{9}}$.
Thus if $\kappa^{- \frac{2}{9}}$ was large compared to the energy $\sim 174$
GeV at which the electroweak symmetry is broken, we would have a hierarchy
problem of the original type {\cite{Gildener}}, without supersymmetry just
above the electroweak breaking energy, to stabilize the parameters of the
effective electroweak Higgs sector.  Thus in models of the present type we
would expect to find the simplest physical picture if $\kappa^{- \frac{2}{9}}$
is as close above the electroweak breaking energy as allowed by present
experimental constraints, which in practice means TeV-scale gravity
{\cite{ADD1, ADD2}}.  In the present section I shall consider how the Standard
Model {\cite{Rosner}} might be realized in the framework considered in the
preceding sections, if $\kappa^{- \frac{2}{9}}$ is of order a TeV.

No positive experimental evidence for the existence of large extra dimensions
and TeV-scale gravity has yet been reported.  However, in the approximation
that the seven extra dimension are flat, the branching ratio for emitting a
graviton, in any process, is $\sim \left( \kappa^{2/9} E \right)^9$,
where $E$ is the energy available to the graviton {\cite{ADD1}}.  Thus if
quantum gravitational effects are observed at the LHC, the effects will start
very suddenly, as the energy of the beams is gradually increased, with no
detectable effects at all up to a certain energy, and very large effects, with
large amounts of missing energy, at slightly higher beam energies, as
gravitons start radiating into the bulk of the thick pipe.  This is in
agreement with the general expectation that, although new physics is not yet
observed at colliders, it cannot be far away {\cite{Altarelli}}.  The
perturbative contributions of virtual graviton exchange to scattering
amplitudes and cross sections, not yet observed, also increase very rapidly
with increasing beam energies {\cite{Giudice Rattazzi Wells, Mirabelli
Perelstein Peskin}}, and once they become observable above the background, are
expected to saturate rapidly at the nonperturbative rate for production of
short-lived microscopic black holes, whose production cross section increases
much more slowly with increasing energy, specifically as $\kappa^{\frac{4}{9}}
\left( \kappa^{2/9} E \right)^{\frac{1}{4}}$ {\cite{Giddings Thomas,
Dimopoulos Landsberg}}.

To estimate the current experimental limits on $\kappa^{- \frac{2}{9}}$, I
shall use the results of Mirabelli, Perelstein, and Peskin {\cite{Mirabelli
Perelstein Peskin}}, who consider the case of flat extra dimensions.
From the discussion around their equations (3) and (4), we see that their
fundamental gravitational mass $M$ is defined such that for seven flat extra
dimensions, compactified to volume $V_7$, Newton's constant $G_N$ is given by
$\pi^2 \left( \frac{M}{2 \pi} \right)^9 V_7 = \frac{1}{16 \pi G_N}$.  On the
other hand, comparing (\ref{Einstein action}) and (\ref{upstairs bulk
action}), and remembering that for working in the ``downstairs'' picture, on
the manifold with boundary, the coefficient $\frac{1}{\kappa^2}$ in
(\ref{upstairs bulk action}) is to be replaced by $\frac{2}{\kappa^2}$, we see
that $\frac{V_7}{\kappa^2} = \frac{1}{16 \pi G_N}$.  Hence $\kappa$ is related
to Mirabelli, Perelstein, and Peskin's $M$ by $\frac{1}{\kappa^2 \pi^2} =
\left( \frac{M}{2 \pi} \right)^9$.  Hence $\kappa^{- \frac{2}{9}} =
\pi^{\frac{2}{9}} \frac{M}{2 \pi} \simeq 0.2053 M$.  The nearest case to the
models of the present paper, for which they give results, is for six flat
extra dimensions.  Thus from the limits on $M$ in their Table 1, we see that
in 1998, the LEP 2 lower bound on $\kappa^{- \frac{2}{9}}$ was around 107 GeV,
and the Tevatron lower bound was around 125 GeV.  And the final lower bound on
$\kappa^{- \frac{2}{9}}$ attainable at the Tevatron is expected to be around
166 GeV, and the final lower bound on $\kappa^{- \frac{2}{9}}$ attainable at
the LHC is expected to be around 677 GeV.  The relations between Mirabelli,
Perelstein, and Peskin's $M$, and $M_p$, the Planck mass in $D$ dimensions, as
defined by Giddings and Thomas {\cite{Giddings Thomas}}, and $M_D$,
the Planck mass in $D$ dimensions, as defined by Giudice, Rattazzi, and Wells
{\cite{Giudice Rattazzi Wells}}, for the case $D = 11$, and $\kappa$,
are
\begin{equation}
  \label{relations between kappa and definitions of the Planck mass in eleven
  dimensions} M = M_p = 2^{\frac{1}{9}} M_D = 2 \pi \left( \frac{1}{\pi
  \kappa} \right)^{\frac{2}{9}} .
\end{equation}

Considering, now, the massless vector bosons in the effective theory in four
dimensions, we note that a smooth compact Einstein space of negative curvature
cannot have any continuous symmetries.  For a vector field $V^A$ that
generates a continuous symmetry on a differentiable manifold $\mathcal{M}$
must satisfy the Killing vector equation $D_A V_B + D_B V_A = 0$.  Hence $0 =
D^A \left( D_A V_B + D_B V_A \right)$.  But from (\ref{Riemann tensor
definition}), on page \pageref{Riemann tensor definition}, we have $D^A D_B
V_A = D_B D^A V_A - R_{BD} V^D$, and from the Killing vector equation, we have
$D^A V_A = 0$.  And if $\mathcal{M}$ is an Einstein space of negative
curvature, then $R_{BD} = \alpha g_{BD}$, where $\alpha > 0$ is independent of
position by the contracted Bianchi identity.  Thus we find $D^A D_A V_B =
\alpha g_{BD} V^D$, hence $V^B D^A D_A V_B = \alpha V^B g_{BD} V^D$.  Thus if
$\mathcal{M}$ is compact, we find on integrating by parts that:
\begin{equation}
  \label{contradiction for generator of continuous symmetry on M}
  \int_{\mathcal{M}} d^d x \left( D^A V^B \right) \left( D_A V_B \right) = -
  \alpha \int_{\mathcal{M}} d^d xV^B g_{BD} V^D
\end{equation}
The left-hand side of this equation is $\geq 0$, but for nonzero $V^A$, the
right-hand side is $< 0$, so there can be no such nonzero $V^A$.  Thus since
there is certainly no continuous symmetry under translation in the radial
direction of the thick pipe, the only massless vector bosons in four
dimensions, in the models considered in this paper, are those which originate
from the $E_8$ Yang-Mills multiplets on the orbifold fixed-point hyperplanes.

In standard compactifications of the weak coupling $E_8$ heterotic superstring
{\cite{Gross Harvey Rohm Martinec 1, Gross Harvey Rohm Martinec 2}}, the $E_8$
containing the Standard Model {\cite{PDG, Rosner}} is first
broken to $E 6$ by embedding the spin connection in the gauge group
{\cite{CHSW, Green Schwarz Witten}}, and the $E
6$ is then further broken by the Hosotani mechanism {\cite{Hosotani 1,
Hosotani 2, Hosotani 3}}.  However, in the models considered in the present
paper, the Standard Model is contained in the $E_8$ on the inner surface of
the thick pipe, whereas if the compact six-manifold, $\mathcal{M}^6$, is a
smooth compact quotient of $\mathbf{C} \mathbf{H}^3$, the spin connection
is embedded in the $E_8$ on the outer surface of the thick pipe, and if
$\mathcal{M}^6$ is a smooth compact quotient of $\mathbf{H}^6$, the spin
connection is not embedded in either of the two $E_8$'s.

The fundamental group of $\mathcal{M}^6$ necessarily has no torsion in the
sense of group theory, or in other words, has no non-trivial finite subgroup,
so if the vacuum contains Hosotani configurations of the Yang-Mills fields, or
in other words, topologically non-trivial configurations of the Yang-Mills
fields, with identically vanishing Yang-Mills field strengths, they might
have to be stabilized dynamically, by radiative corrections, or partly
dynamically and partly topologically, rather than purely topologically, as in
\cite{CHSW}.  The dynamical Hosotani fields in the Cartan subalgebra of $E_8$,
analogous to the Hosotani modes on a torus \cite{Hosotani 1, Hosotani 2,
Hosotani 3}, would be proportional to harmonic 1-forms on $\mathcal{M}^6$,
which are associated with the non-torsion part of the first homology group
$H_1\left(\mathcal{M}^6;\mathbf{Z}\right)$, while Hosotani fields in the
Cartan subalgebra of $E_8$ that are associated with the torsion part of
$H_1\left(\mathcal{M}^6;\mathbf{Z}\right)$ would be partly topologically
stabilized, and might modify the potential for the dynamical Hosotani fields.

I shall assume that the first stage of breaking the $E_8$ on the
inner surface of the thick pipe is by topologically non-trivial $E_8$ vacuum
gauge fields, localized on Hodge - de Rham harmonic two-forms of
$\mathcal{M}^6$, whose field strengths are topologically stabilized in
magnitude, and also partly in orientation within $E_8$, by a form of Dirac
quantization condition, studied in subsection \ref{Dirac quantization
condition for E8 vacuum gauge fields}.  When these Hodge - de Rham
``monopoles'' are all in the Cartan subalgebra of $E_8$, they break $E_8$
either to $\mathrm{SU} \left( 3 \right) \times \left( \mathrm{SU} \left( 2
\right) \right)^3 \times \left( \mathrm{U} \left( 1 \right) \right)^3$, or to
$\mathrm{SU} \left( 3 \right) \times \left( \mathrm{SU} \left( 2 \right)
\right)^2 \times \left( \mathrm{U}
\left( 1 \right) \right)^4$, or to $\mathrm{SU} \left( 3 \right) \times
\mathrm{SU} \left( 2 \right) \times \left( \mathrm{U} \left( 1 \right)
\right)^5$, and
the $\mathrm{U} \left( 1 \right)$'s, other than $\mathrm{U} \left( 1
\right)_Y$, are also broken
by a form of Higgs mechanism involving the $C_{ABy}$ components of the
three-form gauge field, that was discussed by Witten {\cite{Witten Constraints
on compactification}}, and by Green, Schwarz, and West {\cite{Green Schwarz
West}}.  This arises, in the case of Ho\v{r}ava-Witten theory, from the
redefinition of $ G_{yUVW} $ to include a term $ \frac{\kappa^2}{\sqrt{2}
\lambda^2}\delta\left(y - y_1\right)\omega_{UVW}^{\left(1\right)} $, and an
analogous term involving $ \delta\left(y - y_2\right) $, in order to solve the
modified Bianchi identity (\ref{Bianchi identity with FF only}).  Here $
\omega_{UVW}^{\left(1\right)} $ is the Chern-Simons form constructed from the
$ E8 $ gauge fields at $ y_1 $:
\begin{equation}
\label{Chern Simons form}
\omega_{UVW}^{\left(1\right)}\!=\!\mathrm{tr}\!\left(\!A^{\left(1\right)}_U
\!\left(
\partial_V A^{\left(1\right)}_W\! -\! \partial_W A^{\left(1\right)}_V\right)\!
+ \! \frac{2}{3}A^{\left(1\right)}_U\!\left[A^{\left(1\right)}_V,
A^{\left(1\right)}_W
\right] + \textrm{cyclic permutations of }U,V,W\!\right)
\end{equation}
This redefinition of $ G_{yUVW} $ corresponds to the redefinition of the
three-form field strength of the two-form gauge field of $ N = 1 $ supergravity
in ten dimensions, in the Bergshoeff-de Roo-de Wit-van Nieuwenhuizen
\cite{Bergshoeff de Roo de Wit van Nieuwenhuizen} and Chapline-Manton
\cite{Chapline Manton} couplings of $ N = 1 $, $ d = 10 $ supergravity to
Abelian gauge fields, and Yang-Mills fields, respectively.  $ \omega^{\left(1
\right)}_{\mu AB} $ contains a term $ 2\mathrm{tr}\left(A^{\left(1
\right)}_{\mu} F^{\left(1\right)}_{AB}\right) $, and when $ F^{\left(1\right)
}_{AB} $
has a vacuum expectation value in the Cartan subalgebra of $ E8 $, this leads,
through the kinetic term $ G_{IJKL}G^{IJKL} $ of the three-form gauge field, to
a mass term for the corresponding gauge field in the Cartan subalgebra.
However, when $ G_{yUVW} $ is redefined as above, the resulting $ \omega_{UVW}
\omega^{UVW} $ term in the action is formally infinite, being proportional to
$ \delta\left(0\right) $, so it would presumably be preferable to use Moss's
improved form of Ho\v{r}ava-Witten theory \cite{Moss 1, Moss 2, Moss 3},
mentioned shortly after (\ref{FF to FF RR substitutions}), on page \pageref{FF
to FF RR substitutions}, in which the $ \delta\left(0\right) $ terms are
absent.  It was noted by Witten, and by Green, Schwarz, and West, in the papers
cited above, that if the gauge field of a $ \mathrm{U}\left(1\right) $ subgroup
of $ E8 $ develops a vacuum expectation value, but commutes with the gauge
fields in the vacuum, it can be anomalous, so consistency would require any
such field that is anomalous to be massive also in Moss's form of the theory,
so the $ \delta\left(0\right) $ term would have to be replaced by a finite
term, rather than zero.

The Hodge - de Rham ``monopoles'' have non-vanishing Yang-Mills field
strength, and thus contribute to the vacuum energy on the inner surface of the
thick pipe.  However, in the models considered in the present paper, the
universe is stiffened by effects largely determined by the region near the
outer surface of the thick pipe, and in particular, in the case studied in
subsection \ref{Stiffening by fluxes}, the universe is stiffened by the large
value of the integration constant $\tilde{G}$, defined in (\ref{definition of
the constant G tilde}).  Thus the presence of the Hodge - de Rham monopoles,
on the inner surface of the thick pipe, does not lead to a large value of the
effective cosmological constant in four dimensions.

When the Hodge - de Rham ``monopoles'' in the Cartan subalgebra break $E_8$
directly to $\mathrm{SU} \left( 3 \right) \times \mathrm{SU} \left( 2 \right)
\times \left( \mathrm{U} \left( 1 \right) \right)^5$, there is no need for any
Hodge - de Rham monopoles outside the Cartan subalgebra, but unification of the
Yang-Mills coupling constants then depends entirely on the accelerated
unification mechanism studied by Dienes, Dudas, and Gherghetta {\cite{DDG1,
DDG2}}, and by Arkani-Hamed, Cohen,
and Georgi {\cite{Arkani-Hamed Cohen Georgi}}.  In this case, the Hodge - de
Rham monopoles automatically satisfy the classical Yang-Mills field equations.

When the Hodge - de Rham monopoles in the Cartan subalgebra break $E_8$ to
$\mathrm{SU} \left( 3 \right) \times \left( \mathrm{SU} \left( 2 \right)
\right)^2 \times \left( \mathrm{U} \left( 1 \right) \right)^4$, the $\left(
\mathrm{SU} \left( 2 \right) \right)^2$ must then be broken to the diagonal
subgroup $\mathrm{SU} \left( 2 \right)_{\mathrm{diag}}$ by monopoles outside
the Cartan subalgebra, so that, at unification, the Yang-Mills coupling
constant of $\mathrm{SU} \left( 2
\right)_{\mathrm{diag}}$ is smaller than the Yang-Mills coupling constant of
$\mathrm{SU} \left( 3 \right)$, by a factor of $\frac{1}{\sqrt{2}}$, and the
Yang-Mills coupling constants, as evolved in the Standard Model, approximately
unify at around 150 TeV, so there is still a need for an accelerated
unification effect, to achieve unification at around a TeV.  The study of the
Dirac quantization condition, in subsection \ref{Dirac quantization condition
for E8 vacuum gauge fields}, only covers the case where all the Hodge - de
Rham monopoles are in the Cartan subalgebra, and I do not know whether it is
possible, by topological means, to prevent the Hodge - de Rham monopoles
outside the Cartan subalgebra, that break $\left( \mathrm{SU} \left( 2 \right)
\right)^2$ to $\mathrm{SU} \left( 2 \right)_{\mathrm{diag}}$, from
``rotating'', or ``sliding'', back into the Cartan subalgebra.  In the study of
this case, I shall assume, without proof, that this is possible.

And finally, when the Hodge - de Rham monopoles in the Cartan subalgebra break
$E_8$ to $\mathrm{SU} \left( 3 \right) \times \left( \mathrm{SU} \left( 2
\right) \right)^3 \times \left( \mathrm{U} \left( 1 \right) \right)^3$, the
$\left( \mathrm{SU}
\left( 2 \right) \right)^3$ must also be broken to the diagonal subgroup
$\mathrm{SU} \left( 2 \right)_{\mathrm{diag}}$ by monopoles outside the Cartan
subalgebra, so that, at unification, the Yang-Mills coupling constant of
$\mathrm{SU} \left( 2 \right)_{\mathrm{diag}}$ is smaller than the Yang-Mills
coupling constant of $\mathrm{SU} \left( 3 \right)$, by a factor of
$\frac{1}{\sqrt{3}}$, and the $\mathrm{SU} \left( 3 \right)$ and $\mathrm{SU}
\left( 2 \right)_{\mathrm{diag}}$ coupling constants, as evolved in the
Standard Model, now unify at around 413 GeV.  However, it is not possible to do
this without breaking $\mathrm{SU} \left( 2 \right)_{\mathrm{diag}} \times
\mathrm{U} \left( 1
\right)_Y$, and at the same time, obtain an acceptable value of $\sin^2
\theta_W$, which would have to be close to the value $\simeq 0.23$ observed at
$m_Z$, so this case appears to be excluded.

It is not possible to stabilize the absolute orientation of the Cartan
subalgebra within $E_8$ topologically, and there will therefore, by
Goldstone's theorem {\cite{Goldstone}}, be $248 - 12 = 236$ potentially
massless Goldstone boson fields, corresponding to extra-dimensional Lorentz
components of the Yang-Mills fields, proportional to generators of $E_8$
outside the Standard Model $\mathrm{SU}\left(3)\right)\times\mathrm{SU}
\left(2\right)\times\mathrm{U}\left(1\right)_Y$, that can rotate different
possible choices of the Standard Model $\mathrm{SU}\left(3)\right)\times
\mathrm{SU}\left(2\right)\times\mathrm{U}\left(1\right)_Y$ into one another.
These modes, which are independent of position on $\mathcal{M}^6$, do not
correspond to physical massless Lorentz scalar multiplets, but rather become
the longitudinal degrees of freedom of the massive $E_8$ gauge bosons outside
the Standard Model $\mathrm{SU}\left(3)\right)\times\mathrm{SU}\left(2\right)
\times\mathrm{U}\left(1\right)_Y$ \cite{Englert Brout, Higgs 1, Higgs 2,
Guralnik Hagen Kibble}.

I shall assume that $\mathcal{M}^6$ has first Betti number $B_1 > 0$.  There
are then $B_1$ linearly independent harmonic 1-forms on $\mathcal{M}^6$, so
that before the Dirac-quantized harmonic 2-form Hodge - de Rham monopoles in
the $E_8$ Cartan subalgebra are introduced, there are at tree level $B_1$
physical massless Lorentz scalar multiplets in the $E_8$ fundamental/adjoint,
one for each linearly independent harmonic 1-form.  When $E_8$ is broken by the
Hodge - de Rham monopoles, some of the resulting scalar multiplets have the
quantum numbers of the Standard Model Higgs field.  The Hodge - de Rham
monopoles can also produce a potential for some or all of the scalar
multiplets at tree level, which can result in some of the scalars becoming
tachyonic and developing vacuum expectation values, so that the Standard Model
$\mathrm{SU}\left(3\right)\times\mathrm{SU}\left(2\right)\times\mathrm{U}
\left(1\right)_Y$ is broken as in the ordinary Higgs effect \cite{Randjbar
Daemi Salam Strathdee Higgs, Dvali Randjbar Daemi Tabbash}.

After the
inclusion of radiative corrections, the potential is expected to depend
on all the scalar multiplets originating from harmonic 1-forms on
$\mathcal{M}^6$, including any that are not affected by the Hodge - de
Rham monopoles, by the Coleman-Weinberg mechanism {\cite{Coleman Weinberg,
E Weinberg thesis}}, or equivalently, the Hosotani mechanism \cite{Hosotani 1,
Hosotani 2, Hosotani 3}.  I shall assume that this potential has a minimum in
which a scalar multiplet with the quantum numbers of the Standard Model Higgs
field has a vacuum expectation value, which after integration
over position on $\mathcal{M}^6$,
produces masses for the Standard Model $W^{\pm}$ and $Z$ bosons, equivalent to
the masses produced by the Standard Model Higgs boson with a vacuum
expectation value of 246 GeV $\simeq \sqrt{2} \times 174$ GeV, and breaks the
electroweak $\mathrm{SU} \left( 2 \right) \times \mathrm{U} \left( 1 \right)_Y$
to $ \mathrm{U}
\left( 1 \right)_{e.m.}$, as in the Standard Model.  The original
Coleman-Weinberg mechanism resulted in a Higgs mass that was much smaller than
the current experimental lower bound of around 95 to 120 GeV, but more recent
studies, taking into account the large Yukawa coupling of the top quark, have
found consistent solutions, with a Higgs mass consistent with the current
experimental constraints {\cite{0509122 Chishtie et al, 0508107 Elias Mann
McKeon Steele, 0102090 Arkani-Hamed Hall Nomura Smith Weiner}}.

The vacuum expectation value of the scalar multiplet that serves as the
Standard Model Higgs field is proportional to a harmonic 1-form on
$\mathcal{M}^6$, and is thus expected to depend on position on $\mathcal{M}^6$.
I shall assume that this enables the effective Yukawa couplings of this
scalar multiplet, identified as the Standard Model
Higgs field, to different pairs of chiral fermion zero modes to have
different values, so as to realize the fermion mass hierarchy, and the CKM
{\cite{Cabibbo, Kobayashi Maskawa}} and PMNS {\cite{Pontecorvo, Maki Nakagawa
Sakata}} mixing matrices, by a version of the Arkani-Hamed - Schmaltz
mechanism \cite{Arkani-Hamed Schmaltz}.  I shall also assume that all the other
scalar multiplets that
originate from harmonic 1-forms on $\mathcal{M}^6$ are sufficiently massive at
the minimum of the potential to be consistent with experimental
limits, even though they do not develop vacuum expectation values.

The Hodge - de Rham monopoles are required to satisfy Witten's topological
constraint {\cite{Witten Constraints on compactification}}, that was discussed
in subsection \ref{Wittens topological constraint}.  But since the vacuum field
configuration already satisfies this constraint in the absence of the Hodge -
de Rham monopoles, this means that the configuration of the Hodge - de Rham
monopoles is required to satisfy the requirement that for each closed
four-dimensional submanifold $ \mathcal{Q} $ of the compact six-manifold
$ \mathcal{M}^6 $, the
integral $ \int_{\mathcal{Q}} \mathrm{tr} \left( F\wedge F \right) $ is equal
to zero.  For a given configuration $ F $, of the $ E8 $ gauge fields on the
inner surface of the thick pipe, this integral only depends on the cohomology
class of $ Q $, and thus gives $ B_4 $ constraints, where $ B_4 $ is the fourth
Betti number of $ \mathcal{M}^6 $.  But by
Poincar\'{e}
duality, $ B_4 = B_2 $, where $ B_2 $ is the second Betti number of $
\mathcal{M}^6 $.  Hence there is one constraint per harmonic two-form.
However, the embedding of each harmonic two-form, in the Cartan subalgebra of
$ E8 $, is determined by eight independent numbers, which, as I will show in
subsection \ref{Dirac quantization condition for E8 vacuum gauge fields}, are
constrained only to lie on a certain lattice in the Cartan subalgebra of $ E8
$.  Thus it seems likely that there will be non-trivial solutions of Witten's
topological constraint, even when the Hodge - de Rham monopoles are required
to leave $ \mathrm{SU}\left(3\right) \times \left(\mathrm{SU}\left(2\right)
\right)^n \times \left( \mathrm{U}\left(1\right)\right)^{6-n} $ unbroken, for
the required value 3, 2, or 1, of $ n $, and also to be perpendicular to
$ \mathrm{U}\left(1\right)_Y $, so that the $ \mathrm{U}\left(1\right)_Y $ does
not become massive by Witten's Higgs mechanism.  However, when Witten's
topological constraint is imposed in addition to these requirements, there only
remain $ 4 - n $ degrees of freedom per monopole, for the embedding in the $ E8
$ Cartan subalgebra, so the greatest flexibility is obtained for $ n = 1 $.

The Hodge - de Rham monopoles result in
the existence of chiral fermion zero modes, for chiral fermions in various
irreducible representations of the subgroup of $E_8$ left unbroken by the
monopoles and Witten's Higgs mechanism involving the three-form gauge field,
and the number of chiral fermion zero modes, in each such irreducible
representation, is determined by the Atiyah-Singer index theorem {\cite{Atiyah
Singer}}.  Many of these irreducible representations have the quantum numbers
of a fermion representation in the Standard Model, subject to the necessary
accelerated unification of the Yang-Mills coupling constants.  And, as shown by
Witten \cite{Witten Constraints on compactification}, and Green, Schwarz, and
West \cite{Green Schwarz West}, Witten's topological constraint ensures that
there will be no gauge anomalies involving only the gauge bosons left massless
by the Hodge - de Rham monopoles and Witten's Higgs mechanism.  Green, Schwarz,
and West also state that the anomalies involving the $ \mathrm{U}\left(1\right)
$ gauge bosons that commute with the vacuum Yang-Mills fields, but become
massive by Witten's Higgs mechanism, due to having nonvanishing vacuum
expectation values themselves, are harmless.

For all the breakings of $E_8$ considered in the present paper, there exists a
$\mathrm{U} \left( 1 \right)$ gauge boson $B_{\mu}$ that becomes massive by
Witten's
Higgs mechanism, and one or more irreducible representations with the quantum
numbers of each left-handed fermion representation in the Standard Model, such
that the coupling of $B_{\mu}$ to each of those fermion representations is a
fixed multiple of the baryon number of that fermion representation in the
Standard Model.  Sums of triangle diagrams with one or more external
$B_{\mu}$'s are expected to be anomalous, but as explained by Witten
{\cite{Witten Constraints on compactification}}, this does not matter, due to
the fact that $B_{\mu}$ has become massive by the Higgs mechanism involving the
$C_{ABy}$ components of the three-form gauge field.  Thus there might be a
possibility of stabilizing the proton in a manner similar to the Aranda-Carone
mechanism {\cite{Aranda Carone}}, although Aranda and Carone required the
massive gauge boson, whose couplings to the observed fermions are proportional
to baryon number, to be non-anomalous.

In the case where the Hodge - de Rham monopoles in the Cartan subalgebra break
$E_8$ directly to $\mathrm{SU} \left( 3 \right) \times \mathrm{SU} \left( 2
\right) \times \left( \mathrm{U} \left( 1 \right) \right)^5$, and there are no
Hodge - de Rham monopoles outside the Cartan subalgebra, realizing the Standard
Model requires:
\begin{enumerate}
  \item finding a linear combination of the $\mathrm{U} \left( 1 \right)$'s to
  serve as $\mathrm{U} \left( 1 \right)_Y$, such that there exist $\mathrm{SU}
  \left( 3 \right) \times \mathrm{SU} \left( 2 \right)$ irreducible
  representations in the $E_8$ fundamental, with the correct $\mathrm{SU}
  \left( 3 \right) \times \mathrm{SU} \left( 2 \right)$ quantum numbers and
  $\mathrm{U} \left( 1 \right)_Y$
  charges to be identified as the left-handed fermions of one or more
  generations, and the Higgs boson of the Standard Model;

  \item finding another linear combination of the $\mathrm{U} \left( 1
  \right)$'s to serve as $\mathrm{U} \left( 1 \right)_B$, such that for each of
  the five types of $\mathrm{SU} \left( 3 \right) \times \mathrm{SU} \left( 2
  \right)$ multiplet with non-vanishing $\mathrm{U} \left( 1 \right)_Y$ charge
  in the Standard Model, and
  also for the left-handed antineutrino, if these are required, there exists
  at least one $\mathrm{SU} \left( 3 \right) \times \mathrm{SU} \left( 2
  \right)$ irreducible representation in the $E_8$ fundamental, with those
  $\mathrm{SU} \left( 3 \right) \times \mathrm{SU} \left( 2 \right)$ quantum
  numbers and $ \mathrm{U} \left( 1 \right)_Y$ charge, such that the $
  \mathrm{U} \left( 1 \right)_B $ charge of
  that irreducible representation is a fixed multiple of the baryon number of
  the corresponding fermion; and

  \item finding, for each of the $B_2$ linearly independent Hodge - de Rham
  harmonic two forms of $\mathcal{M}^6$, where $B_2$ is the second Betti
  number of $\mathcal{M}^6$, a point perpendicular to $\mathrm{U} \left( 1
  \right)_Y$,
  in the eight-dimensional lattice of points in the $E_8$ Cartan subalgebra
  that is allowed by the Dirac quantization condition, such that:
  \begin{enumerate}
    \item Witten's topological constraint is satisfied, for all $B_4 = B_2$
    linearly independent harmonic four-forms of $\mathcal{M}^6$, or
    equivalently, for a set of $B_4$ topologically non-trivial closed
    four-dimensional surfaces in $\mathcal{M}^6$, linearly independent in the
    sense of homology; and

    \item for each of the five or six types of $\mathrm{SU} \left( 3 \right)
    \times \mathrm{SU} \left( 2 \right)$ left-handed fermion multiplet in the
    Standard Model, depending on whether or not left-handed antineutrinos are
    required:
    \begin{enumerate}
      \item every occurrence of that multiplet in the $E_8$ fundamental, that
      has the correct $\mathrm{U} \left( 1 \right)_Y$ charge, and $\mathrm{U}
      \left( 1
      \right)_B$ charge equal to the correct multiple of baryon number, has a
      net number of chiral fermion zero modes, as given by the Atiyah-Singer
      index theorem, $\geq 0$; and

      \item the sum, over all occurrences of that multiplet in the $E_8$
      fundamental, that have the correct $\mathrm{U} \left( 1 \right)_Y$
      charge, and $\mathrm{U}
      \left( 1 \right)_B$ charge equal to the correct multiple of baryon
      number, of the net number of chiral fermion zero modes, as given by the
      Atiyah-Singer index theorem, is equal to $3$; and

      \item every occurrence of that multiplet in the $E_8$ fundamental, that
      either has the wrong $\mathrm{U} \left( 1 \right)_Y$ charge, or has
      $\mathrm{U} \left( 1
      \right)_B$ charge equal to the wrong multiple of baryon number, has a
      net number of chiral fermion zero modes, as given by the Atiyah-Singer
      index theorem, equal to $0$;
    \end{enumerate}
    \item for each $\mathrm{SU} \left( 3 \right) \times \mathrm{SU} \left( 2
    \right)$ multiplet in the $E_8$ fundamental, that does not correspond to a
    fermion multiplet in the Standard Model, or the complex conjugate of a
    fermion multiplet in the Standard Model, the net number of chiral fermion
    zero modes, as given by the Atiyah-Singer index theorem, is equal to $0$;
    and

    \item if there are sufficiently many left-handed antineutrinos, a Majorana
    mass matrix, with one or more very light eigenstates by a generalized
    seesaw mechanism, as discussed in subsection \ref{Generalized seesaw
    mechanism} below, is generated for them by the Hodge - de Rham monopoles;
    and

    \item a potential is generated for all the ``Higgs'' bosons, by the
    Coleman-Weinberg mechanism, that has a minimum at which all the ``Higgs''
    bosons are massive, and the electrically neutral component of a ``Higgs''
    boson, with the quantum numbers of the Standard Model Higgs boson, has a
    vacuum expectation value, possibly dependent on position on
    $\mathcal{M}^6$, whose value, averaged over position on $\mathcal{M}^6$,
    produces masses for the Standard Model $W^{\pm}$ and $Z$ bosons,
    equivalent to the masses produced by the Standard Model Higgs boson, with
    a vacuum expectation value of 246 GeV; and

    \item the mass matrices with entries given by the overlap integrals of
    pairs of chiral fermion zero modes, with the vacuum expectation of the
    ``Higgs'' boson, which may depend on position on $\mathcal{M}^6$, produce
    the observed mass spectra of the quarks and the electrically charged
    leptons, and the CKM mixing matrix of the quarks, by a version of the
    Arkani-Hamed - Schmalz mechanism; and

    \item the masses of the Standard Model neutrinos, and the PMNS mixing
    matrix of the Standard Model leptons, arise in some way.

  \end{enumerate}

\end{enumerate}

In the present paper, I will present some solutions to the requirements
$ 1.\hspace{-0.3ex} $ and
2. above, both for the case when the Hodge - de Rham monopoles in the Cartan
subalgebra of $E_8$ break $E_8$ directly to $\mathrm{SU} \left( 3 \right)
\times \mathrm{SU} \left( 2 \right) \times \left( \mathrm{U} \left( 1 \right)
\right)^5$, and for the case when they break $E_8$ directly to $\mathrm{SU}
\left( 3 \right) \times \left( \mathrm{SU} \left( 2 \right) \right)^2 \times
\left( \mathrm{U} \left( 1 \right) \right)^4$.  In the solutions where the
Hodge - de Rham monopoles in the Cartan subalgebra break $E_8$ directly to
$\mathrm{SU} \left( 3 \right) \times \left( \mathrm{SU} \left( 2 \right)
\right)^2 \times \left( \mathrm{U} \left( 1
\right) \right)^4$, there exist components of the $E_8$ fundamental, outside
the Cartan subalgebra, that could break $\left( \mathrm{SU} \left( 2 \right)
\right)^2$ to $\mathrm{SU} \left( 2 \right)_{\mathrm{diag}}$, without breaking
$\mathrm{SU} \left( 3 \right) \times \mathrm{SU} \left( 2
\right)_{\mathrm{diag}} \times \mathrm{U} \left( 1 \right)_Y$, if they could be
given topologically stabilized
vacuum expectation values, as Hodge - de Rham monopoles, but, as mentioned
above, I do not know whether or not there is any topological obstruction to
prevent the orientation in $E_8$, of such Hodge - de Rham monopoles, from
``rotating'', or ``sliding'', back into the Cartan subalgebra.

The necessary first step for studying the requirements $ 3.\hspace{-0.4ex} $
(a) - (g) is to find
explicit examples of smooth compact quotients of $\mathbf{C} \mathbf{H}^3$
or $\mathbf{H}^6$ that are spin manifolds.  This is unavoidable, because
Witten's topological constraint depends on the cohomology cup product of the
manifold {\cite{Wikipedia Cup product, PlanetMath Cup product}}, that
expresses the wedge product of pairs of harmonic two-forms as linear
combinations of harmonic four-forms, and this cohomology cup product is a
topological invariant of the manifold.

I shall now consider the lightest massive modes of the supergravity multiplet,
in the following subsection \ref{The lightest massive modes of the
supergravity multiplet}.  The $\mathrm{SU} \left( 9 \right)$ basis for $E_8$ is
studied in subsection \ref{An SU9 basis for E8}, on page \pageref{An SU9 basis
for E8}.  The Dirac quantization condition on the field strengths of Hodge - de
Rham harmonic two-forms, in the Cartan subalgebra of $E_8$, is studied in
subsection \ref{Dirac quantization condition for E8 vacuum gauge fields}, on
page \pageref{Dirac quantization condition for E8 vacuum gauge fields}.  I show
that there are no models with an acceptable
value of $\sin^2 \theta_W$, such that the Hodge - de Rham monopoles, in the
Cartan subalgebra of $E_8$, break $E_8$ to $\mathrm{SU} \left( 3 \right) \times
\left( \mathrm{SU} \left( 2 \right) \right)^3 \times \left( \mathrm{U}
\left( 1 \right)
\right)^3$, in subsection \ref{SU 3 cross SU 2 cubed cross U 1 cubed}, on page
\pageref{SU 3 cross SU 2 cubed cross U 1 cubed}.  Models where the Hodge - de
Rham monopoles, in the Cartan subalgebra of $E_8$, break $E_8$ to $\mathrm{SU}
\left( 3 \right) \times \left( \mathrm{SU} \left( 2 \right) \right)^2 \times
\left( \mathrm{U}
\left( 1 \right) \right)^4$, are studied in subsection \ref{SU 3
cross SU 2 squared cross U 1 to the fourth}, on page \pageref{SU 3 cross SU 2
squared cross U 1 to the fourth}, and models where they break $E_8$ to
$\mathrm{SU} \left( 3 \right) \times \mathrm{SU} \left( 2 \right) \times \left(
\mathrm{U}
\left( 1 \right) \right)^5$, are studied in subsection \ref{SU 3 cross SU 2
cross U 1 to the fifth}, on page \pageref{SU 3 cross SU 2 cross U 1 to the
fifth}.

\subsection{The lightest massive modes of the supergravity multiplet}

\label{The lightest massive modes of the supergravity multiplet}

From the point of view of the effective theory in four dimensions,
supersymmetry is broken explicitly in the models considered in the present
paper, even though, from the point of view of Ho\v{r}ava-Witten theory in eleven
and ten dimensions, the supersymmetry is broken spontaneously, by the
compactification.  Thus the gravitinos, four of which are allowed, by the
Ho\v{r}ava-Witten boundary conditions, to couple directly to the matter on the
inner surface of the thick pipe, and the associated spin-$\frac{1}{2}$
fermions, and also the vectors and scalars which correspond, in four
dimensions, to the three-form gauge field, couple to ordinary matter with at
most gravitational strength, and there is no enhancement of the coupling of
the gravitino to ordinary matter, as can happen in models where $N = 1$
supersymmetry is broken spontaneously in four dimensions, through the
absorption of the goldstino by the gravitino {\cite{Fayet 1, Fayet 2}}.

To study the Kaluza-Klein modes of the supergravity multiplet we have to
expand the quantum effective action to quadratic order in
small fluctuations about the relevant background solution, which is here one
of the solutions found in subsections \ref{Solutions with both a and b large
at the outer surface}, \ref{Solutions with a as small as kappa to the two
ninths at the outer surface}, and \ref{Stiffening by fluxes}. \ For a
first estimate I
shall instead consider a massless scalar field $\Phi$ in the bulk, which is
intended to represent a small fluctuation of a component of any of the
supergravity fields, and retain only its classical action. \ Dropping
also $R \Phi$ and $H^2 \Phi$ terms, the equation for the small
fluctuation $\Phi$ is then:
\begin{equation}
  \label{small fluctuation equation} - \frac{1}{\sqrt{- G}} \partial_I \left(
  \sqrt{- G} G^{I J} \partial_J \Phi \right) = 0
\end{equation}
Trying an ansatz $\Phi \left( x^{\mu}, x^A, y \right) = \varphi \left( x^{\mu}
\right) \psi \left( x^A, y \right)$, where coordinate indices $\mu, \nu,
\sigma,\dots$ are tangent to the four observed space-time dimensions, and
coordinate indices $A, B,$ $C, \dots$ are tangent to $\mathcal{M}^6$, as in
subsection \ref{The field equations and boundary conditions}, we find from
(\ref{small fluctuation equation}) that:
\begin{dmath}
  \label{separation of variables} - \frac{a^2}{b^2 \psi \left( x^C, y \right)
  \sqrt{h}} \partial_A \left( \sqrt{h} h^{A B} \partial_B \psi \left( x^C, y
  \right) \right) - \frac{a^2}{a^4 b^6 \psi \left( x^C, y \right)} \partial_y
  \left( a^4 b^6 \partial_y \psi \left( x^C, y \right) \right) =
  \frac{1}{\varphi \left( x^{\sigma} \right) \sqrt{- g}} \partial_{\mu} \left(
  \sqrt{- g} g^{\mu \nu} \partial_{\nu} \varphi \left( x^{\sigma} \right)
  \right)
\end{dmath}
The left-hand side of (\ref{separation of variables}) is independent of
$x^{\mu}$ and the right-hand side is independent of $x^A$ and $y$, hence each
side must be a constant. \ The left-hand side is a positive operator on a
compact manifold so must be a non-negative constant $m^2 \geq 0$.

From the metric ansatz \ref{metric ansatz}, on page \pageref{metric ansatz},
the metric $G_{\mu \nu}$ at the inner surface of the thick pipe, where we
live, in the models considered here, is $G_{\mu \nu} = A_{\mathrm{dS}}^2
g_{\mu \nu}$, where $A_{\mathrm{dS}}$ is the observed de Sitter radius
\ref{de Sitter radius}, since by definition the de Sitter radius of
$g_{\mu \nu}$ is 1.
Thus in terms of the metric $G_{\mu \nu}$ at the inner surface of the thick
pipe, the wave equation along the 4 extended dimensions, for a Kaluza-Klein
mode $\psi \left( x^C, y \right)$ for which each side of \ref{separation of
variables} is equal to $m^2$, is:
\begin{equation}
   \label{wave equation for light KK mode} - \frac{1}{\sqrt{- G}}
   \partial_{\mu} \left( \sqrt{- G} G^{\mu \nu} \partial_{\nu} \varphi \left(
   x^{\sigma} \right) \right) +
   \frac{m^2}{A^2_{\mathrm{{{dS}}}}} \varphi \left( x^{\sigma} \right) = 0.
\end{equation}

For the solution found in subsection \ref{Solutions with both a and b large at
the outer surface}, starting on page \pageref{Solutions with both a and b
large at the outer surface}, $a$ and $b$ are roughly constant $\sim B$ over
the main part of the classical region around $y \sim B$, so there are modes
spread in this region for which $- \partial_y^2 \psi \sim \frac{n^2}{B^2}
\psi$, so that $m\sim n$, for all integers $n > 0$.
Thus there are very light Kaluza-Klein modes of the bulk
whose mass, as seen at the inner surface of the thick pipe, is $\sim
\frac{n}{A_{\mathrm{dS}}}$, for all integers $n > 0$.  $\psi \left( x^C, y
\right)$ is suppressed in the region of the inner surface of the thick pipe
for these modes, so the situation is qualitatively similar to the situation
considered by Randall and Sundrum in \cite{Randall Sundrum 2}, where the
modifications to Einstein gravity in the 4 extended dimensions, on the brane
we live on, from modes of this form, were found to be negligibly small.
However the model considered
in \cite{Randall Sundrum 2} did not include the ADD effect, so further
study would be needed to determine whether these very light Kaluza-Klein
modes, localized in the classical region of the bulk, prevent the solution
found in subsection \ref{Solutions with both a and b large at the outer
surface} from being consistent with the precision Solar System tests of
Einstein gravity \cite{Williams Turyshev Boggs 1, Williams Turyshev Boggs 2},
and with the sub-millimetre tests of Newton's law \cite{Hoyle et al}.

For the solution found in subsection \ref{Solutions with a as small as kappa to
the two ninths at the outer surface}, starting on page \pageref{Solutions with
a as small as kappa to the two ninths at the outer surface},
there are modes in the second quantum region, adjacent to the outer surface of
the thick pipe, that oscillate sufficiently rapidly as $y$ increases, that
$a^4 b^6$ is approximately constant over $\sim 10$ or more cycles, and
wavepackets localized in this region can be formed from these modes. \ For
such a wavepacket localized at $a \simeq
a_{\mathrm{{{cent}}}}$ and independent of position on
$\mathcal{M}^6$, the left-hand side of (\ref{separation of variables}) is
approximately $- \frac{a^2_{\mathrm{{{cent}}}}}{\psi
\left( y \right)} \partial^2_y \psi \left( y \right) \simeq m^2$, so a
representative mode is $\psi \left( y \right) = \mathrm{\cos}
\frac{my}{a_{\mathrm{{{cent}}}}}$ times a wavepacket
profile. \ In this region $a$ decreases exponentially with increasing
$\frac{y}{\kappa^{2 / 9}} $, with a coefficient $\sim 1$ in the exponent, and
$b$ is a constant times $a^{\tilde{\tau}}$, where $\tilde{\tau}$ is a constant
of magnitude $\sim 1$. \ Thus the requirement that $a^4 b^6$ changes over one
wavelength by at most a factor close to 1 is that $\frac{\left( 4 + 6
\tilde{\tau} \right) 2 \pi
a_{\mathrm{{{cent}}}}}{\kappa^{2 / 9} m} \ll 1$. \ For
example $m \sim 10^3$ would be adequate, for
$a_{\mathrm{{{cent}}}}$ roughly at the outer boundary
and hence $\sim \kappa^{2 / 9}$. \ Thus from (\ref{wave equation for light KK
mode}), there are very light Kaluza-Klein modes of the bulk whose mass, as
seen at the inner surface of the thick pipe, is $\sim 10^3
\frac{n}{A_{\mathrm{{{dS}}}}}$, for all integers $n >
0$. \ $\psi \left( x^C, y \right)$ is again suppressed in the region of the
inner surface of the thick pipe for these modes, and further study would be
needed to determine whether these modes prevent the solution found in
subsection \ref{Solutions with a as small as kappa to the two ninths at the
outer surface} from being consistent with the precision Solar System tests of
Einstein gravity {\cite{Williams Turyshev Boggs 1, Williams Turyshev Boggs 2}},
and with the sub-millimetre tests of Newton's law {\cite{Hoyle et al}}.

For the solution found in subsection \ref{Stiffening by fluxes}, starting on
page \pageref{Stiffening by fluxes}, where the outer surface is stabilized in
the classical region by fluxes,
$a$ and $b$ are roughly constant, with $a \sim 10^{22}$ metres, from (\ref{a
sub 2 for TeV scale gravity with fluxes}), on page \pageref{a sub 2 for TeV
scale gravity with fluxes}, and $b \sim B$, over the main part of the
classical region around $y \sim B$, so there are modes spread in this region
for which $- \partial^2_y \psi \sim \frac{n^2}{B^2} \psi$, so that from
(\ref{separation of variables}), $m \sim \frac{n}{B} \times 10^{22}$ metres,
for all integers $n > 0$. \ Thus from (\ref{wave equation for light KK mode}),
and (\ref{de Sitter radius}), on page \pageref{de Sitter radius}, the mass of
these modes, as seen from the inner surface of the thick pipe, is $\sim 10^{-
4} \frac{n}{B}$, which from (\ref{B for TeV scale gravity with fluxes}), on
page \pageref{B for TeV scale gravity with fluxes}, is $\sim \frac{n}{10^{- 8}
\hspace{0.2em} \mathrm{{{metres}}}} \sim 10 n$ eV. \
The wavefunctions of these modes are again suppressed in the region of the
inner surface of the thick pipe.

\subsection{An $\mathrm{SU} \left( 9 \right)$ basis for $E_8$}
\label{An SU9 basis for E8}

Throughout this section, I shall use an $\mathrm{\mathrm{SU}} ( 9 )$ basis for
$\mathrm{E} 8$, as in {\cite{NTSG}}. On breaking $\mathrm{E} 8$ to
$\mathrm{\mathrm{SU}} ( 9 )$, the \textbf{248} of $\mathrm{E} 8$ splits to the
\textbf{80}, \textbf{84}, and $\mathbf{\overline{84}}$ of $\mathrm{\mathrm{SU}}
( 9 )$. Here the \textbf{80} is the adjoint of $\mathrm{\mathrm{SU}} ( 9 )$, the
\textbf{84} has three totally antisymmetrized $\mathrm{\mathrm{SU}} ( 9 )$
fundamental subscripts, and the $\mathbf{\overline{84}}$ has three totally
antisymmetrized $\mathrm{\mathrm{SU}} ( 9 )$ antifundamental subscripts. The
fundamental representation generators $\left( t_{\alpha} \right)_{i \bar{j}}$
of $\mathrm{SU} \left( 9 \right)$ are normalized to satisfy {\cite{Rosner}}
\begin{equation}
  \label{normalization} \mathrm{\mathrm{tr}} \left( t_{\alpha} t_{\beta} \right)
  = \frac{\delta_{\alpha \beta}}{2}
\end{equation}
The generators of the required representations are as follows:
\begin{equation}
  \label{Antifundamental} \textrm{\textit{Antifundamental}} \hspace{2em}
  \hspace{1.8ex} \hspace{2em} \hspace{2em}
  \left( T_{\alpha} \right)_{\bar{i} j} = - \left(
  t_{\alpha} \right)_{j \bar{i}} \hspace{2em} \hspace{2em} \hspace{2em}
  \hspace{2em} \hspace{2em} \hspace{2em} \hspace{1.8ex}
\end{equation}
\begin{equation}
  \label{Adjoint} \textrm{\textit{Adjoint}} \hspace{2em} \hspace{2em}
  \hspace{0.5ex}
  \hspace{2em} \hspace{0.5em} \hspace{0.75em} \left( T_{\alpha} \right)_{i
  \bar{j}, \bar{k} m} = \left( t_{\alpha} \right)_{i \bar{k}} \delta_{m
  \bar{j}} - \delta_{i \bar{k}} \left( t_{\alpha} \right)_{m \bar{j}}
  \hspace{2em} \hspace{2em} \hspace{2em} \hspace{2em}
  \hspace{0.5em} \hspace{0.75em} \hspace{0.5ex}
\end{equation}
\begin{equation}
  \label{84} \textrm{\textbf{84}} \hspace{2em} \hspace{2em} \hspace{2em}
  \hspace{0.2ex}
  \left( T_{\alpha} \right)_{ijk, \bar{m}  \bar{p}  \bar{q}} =
  \left( t_{\alpha} \right)_{i \bar{m}} \delta_{j \bar{p}} \delta_{k
  \bar{q}} \pm \textrm{seventeen terms} \hspace{2em} \hspace{2em}
  \hspace{2em}
\end{equation}
\begin{equation}
  \label{84 bar} \mathbf{\overline{84}} \hspace{2em} \hspace{2em} \hspace{2em}
  \hspace{0.6ex}
  \left( T_{\alpha} \right)_{\bar{i}  \bar{j}  \bar{k}, mpq} =
  - \left( t_{\alpha} \right)_{m \bar{i}} \delta_{p \bar{j}} \delta_{q
  \bar{k}} \pm \textrm{seventeen terms} \hspace{2em} \hspace{2.0ex}
  \hspace{2em}
\end{equation}
where the additional terms in (\ref{84}) and (\ref{84 bar}) antisymmetrize
with respect to permutations of $( i, j, k )$, and with respect to
permutations of $( m, p, q )$. We can check directly that these generators
satisfy the same commutation relations as $\left( t_{\alpha} \right)_{i
\bar{j}}$, with the same structure constants.

When we check the commutation relations for the \textbf{84}
or $\mathbf{\overline{84}}$, we have to decide whether or not
each sum over an antisymmetrized triple of indices, using the standard
summation convention, as in $\left( T_{\alpha} \right)_{i j k, \bar{m} \bar{p}
\bar{q}} \left( T_{\beta} \right)_{m p q, \bar{r} \bar{s} \bar{t}}$, will be
multiplied by an explicit factor $\frac{1}{3!} = \frac{1}{6}$. \ If we do not
multiply the sum by an explicit factor $\frac{1}{6}$, then the summation
convention results in each element of the \textbf{84} or
$\mathbf{\overline{84}}$ being summed over 6 times.

When we write a Jacobi identity for the structure constants of a Lie algebra,
each element of the Lie algebra should be summed over precisely once, in each
of the three terms in the Jacobi identity. \ Thus when we write a Jacobi
identity for $E_8$ in the SU(9) basis, and use the standard summation
convention, we have to multiply each sum over elements of the
\textbf{84} or $\mathbf{\overline{84}}$ of
SU(9) by an explicit factor $\frac{1}{6}$, to ensure that each element of the
\textbf{84} or $\mathbf{\overline{84}}$ is
summed over exactly once.

For $E_8$, the \textbf{248} is both the fundamental and the
adjoint, the structure constants give the matrix elements of the
\textbf{248}, and the Jacobi identities for the structure
constants give the commutation relations for the
\textbf{248}. \ So for consistency, I shall use a
convention here, that {\emph{every sum over elements of the}} \textbf{84}
{\emph{or}} $\mathbf{\overline{84}}$ {\emph{of}} $\mathrm{SU} (9)$,
{\emph{using the standard summation convention, is to be multiplied by an
explicit factor}} $\frac{1}{6}$, {\emph{to ensure that each element of the}}
\textbf{84} {\emph{or}} $\mathbf{\overline{84}}$ {\emph{is counted exactly
once}}.

This convention has to be used to verify that the generators of the
\textbf{84} and $\mathbf{\overline{84}}$,
as given in (\ref{84}) and (\ref{84 bar}), satisfy the same commutation
relations as $\left(
t_{\alpha} \right)_{i \bar{j}}$, with the same structure constants. \ This
convention was not used in the first two versions of this article, and was not
used in \cite{NTSG}, so the right-hand sides of equations (\ref{84}) and
(\ref{84 bar}) in
version 2 of this article, and the right-hand sides of the corresponding
equations (8) and (9) of \cite{NTSG}, have an overall factor $\frac{1}{6}$,
which is absent from (\ref{84}) and (\ref{84 bar}) above.

It is convenient to define:
\begin{equation}
  \label{projection operator to antisymmetry on three indices} \delta_{ijk,
  \bar{r} \bar{s} \bar{t}} \equiv \delta_{i \bar{r}}
  \delta_{j \bar{s}} \delta_{k \bar{t}} + \delta_{i \bar{s}} \delta_{j
  \bar{t}} \delta_{k \bar{r}} + \delta_{i \bar{t}} \delta_{j \bar{r}}
  \delta_{k \bar{s}} - \delta_{i \bar{r}} \delta_{j \bar{t}} \delta_{k
  \bar{s}} - \delta_{i \bar{s}} \delta_{j \bar{r}} \delta_{k \bar{t}} -
  \delta_{i \bar{t}} \delta_{j \bar{s}} \delta_{k \bar{r}}
\end{equation}
which is the unit matrix in the space of matrices whose rows and columns are
labelled by antisymmetrized triples of indices, when the above convention is
used, and projects expressions with
three indices to their antisymmetric part.  Then we have:
\[ \left( T_{\alpha} \right)_{ijk, \bar{m}  \bar{p}  \bar{q}} = \delta_{ijk,
   \bar{r} \bar{s} \bar{t}} \left( \left( t_{\alpha} \right)_{r \bar{m}}
   \delta_{s \bar{p}} \delta_{t \bar{q}} + \delta_{r \bar{m}} \left(
   t_{\alpha} \right)_{s \bar{p}} \delta_{t \bar{q}} + \delta_{r \bar{m}}
   \delta_{s \bar{p}} \left( t_{\alpha} \right)_{t \bar{q}} \right) = \]
\[  \hspace{4em} =
  \left( \left( t_{\alpha} \right)_{i \bar{r}} \delta_{j \bar{s}} \delta_{k
  \bar{t}} + \delta_{i \bar{r}} \left( t_{\alpha} \right)_{j \bar{s}}
  \delta_{k \bar{t}} + \delta_{i \bar{r}} \delta_{j \bar{s}} \left( t_{\alpha}
  \right)_{k \bar{t}} \right) \delta_{rst, \bar{m} \bar{p} \bar{q}} = \]
\begin{equation}
  \label{generators of 84 in terms of projection operator} = \frac{1}{2}
\delta_{i j k, \bar{r} \bar{s} \bar{t}} \left( t_{\alpha} \right)_{r \bar{u}}
\delta_{s \bar{v}} \delta_{t \bar{w}} \delta_{u v w, \bar{m} \bar{p} \bar{q}}
\hspace{6em}
\end{equation}
\[ \left( T_{\alpha} \right)_{\bar{i}  \bar{j}  \bar{k}, mpq} = - \delta_{rst,
   \bar{i} \bar{j} \bar{k}} \left( \left( t_{\alpha} \right)_{m \bar{r}}
   \delta_{p \bar{s}} \delta_{q \bar{t}} + \delta_{m \bar{r}} \left(
   t_{\alpha} \right)_{p \bar{s}} \delta_{q \bar{t}} + \delta_{m \bar{r}}
   \delta_{p \bar{s}} \left( t_{\alpha} \right)_{q \bar{t}} \right) = \]
\[ \hspace{4.2em} =
  - \left( \left( t_{\alpha} \right)_{r \bar{i}} \delta_{s \bar{j}} \delta_{t
  \bar{k}} + \delta_{r \bar{i}} \left( t_{\alpha} \right)_{s \bar{j}}
  \delta_{t \bar{k}} + \delta_{r \bar{i}} \delta_{s \bar{j}} \left( t_{\alpha}
  \right)_{t \bar{k}} \right) \delta_{mpq, \bar{r} \bar{s} \bar{t}} = \]
\begin{equation}
  \label{generators of 84 bar in terms of projection operator} =
- \frac{1}{2} \delta_{r s t, \bar{i} \bar{j} \bar{k}} \left( t_{\alpha}
\right)_{u \bar{r}} \delta_{v \bar{s}} \delta_{w \bar{t}} \delta_{m p q,
\bar{u} \bar{v} \bar{w}}
\hspace{6em}
\end{equation}
We define the totally antisymmetric $\mathrm{SU} \left( 9 \right)$ structure
constants $f_{\alpha \beta \gamma}$ by $\left[ t_{\alpha}, t_{\beta} \right] =
if_{\alpha \beta \gamma} t_{\gamma}$, noting, from (\ref{normalization}), that
the $\mathrm{SU} \left( 9 \right)$ generators $t_{\alpha}$, in the $\mathrm{SU}
\left( 9 \right)$ fundamental representation, have been chosen to be
hermitian.  The generators of $E_8$ are now the 80 generators $T_{\alpha}$ of
$\mathrm{SU} \left( 9 \right)$, together with 84 generators $T_{rst}$,
antisymmetric in $rst$, whose label, $rst$, runs over the \textbf{84} of
$\mathrm{SU} \left( 9 \right)$, and 84 generators $T_{\bar{r} \bar{s} \bar{t}}$,
antisymmetric in $\bar{r} \bar{s} \bar{t}$, whose label, $\bar{r} \bar{s}
\bar{t}$, runs over the $\mathbf{\overline{84}}$ of $\mathrm{SU} \left( 9
\right)$.  Indices $\mathcal{A},\mathcal{B},\mathcal{C}, \ldots$ will run over
all 248 generators of $E_8$, as in the discussion of the $\mathrm{SO} \left( 16
\right)$ basis, in subsection \ref{Horava-Witten theory}.  The $E_8$ structure
constants will be written $F_{\mathcal{A} \mathcal{B} \mathcal{C}}$, and
defined such that $\left[ T_{\mathcal{A}}, T_{\mathcal{B}} \right] =
iF_{\mathcal{A} \mathcal{B} \mathcal{C}} T_{\mathcal{C}}$.  The convention
stated after (\ref{84 bar}) means, for example, that
when $\mathcal{C}$, in $iF_{\mathcal{A} \mathcal{B}
\mathcal{C}} T_{\mathcal{C}}$, refers to $\bar{r} \bar{s} \bar{t}$ on
$F_{\mathcal{A} \mathcal{B} \mathcal{C}}$, and to $rst$ on $T_{\mathcal{C}}$,
the contribution is $\frac{1}{6}iF_{\mathcal{A} \mathcal{B}, \bar{r} \bar{s}
\bar{t}} T_{rst}$, with the normal summation convention, so that each of the
84 distinct generators $T_{rst}$, $1 \leq r < s < t \leq 9$, is effectively
counted once, instead of 6 times.  This is in contrast to the convention used
in the discussion of the
$\mathrm{SO} \left( 16 \right)$ basis in subsection \ref{Horava-Witten theory},
where the definition (\ref{orthogonal group structure constants}), of the
orthogonal group structure constants, means that the orthogonal group
commutation relation (\ref{JJ commutation relation}) takes the form $\left[
J_{ij}, J_{kl} \right] = f_{ij, kl, rs} J_{rs}$, so that each of the distinct
generators $J_{ij}$, $1 \leq i < j \leq 16$, actually gets counted twice in
the sum.

The structure constants $F_{\mathcal{A} \mathcal{B} \mathcal{C}}$ are totally
antisymmetric under permutations of $\mathcal{A},\mathcal{B},\mathcal{C}$, and
the non-vanishing matrix elements of the $E_8$ generators are:
\begin{equation}
  \label{alpha beta gamma matrix elements} \left( T_{\alpha} \right)_{\beta
  \gamma} = if_{\beta \alpha \gamma} = iF_{\beta \alpha \gamma}
\end{equation}
\[ \left( T_{\alpha} \right)_{ijk, \bar{m} \bar{p} \bar{q}} = - \left(
   T_{\alpha} \right)_{\bar{m} \bar{p} \bar{q}, ijk} = - \left( T_{ijk}
   \right)_{\alpha, \bar{m} \bar{p} \bar{q}} = \left( T_{ijk} \right)_{\bar{m}
   \bar{p} \bar{q}, \alpha} = \hspace{8em} \]
\begin{equation}
  \label{alpha i j k m bar p bar q bar matrix elements} \hspace{14em} = \left(
  T_{\bar{m} \bar{p} \bar{q}} \right)_{\alpha, ijk} = - \left( T_{\bar{m}
  \bar{p} \bar{q}} \right)_{ijk, \alpha} = iF_{ijk, \alpha, \bar{m} \bar{p}
  \bar{q}}
\end{equation}
\begin{equation}
  \label{r s t i j k m p q matrix elements} \left( T_{rst} \right)_{ijk, mpq}
  = \frac{i}{\sqrt{2}} \epsilon_{rstijkmpq} = iF_{ijk, rst, mpq}
\end{equation}
\begin{equation}
  \label{r bar s bar t bar i bar j bar k bar m bar p bar q bar matrix
  elements} \left( T_{\bar{r} \bar{s} \bar{t}} \right)_{\bar{i} \bar{j}
  \bar{k}, \bar{m} \bar{p} \bar{q}} = \frac{i}{\sqrt{2}} \epsilon_{\bar{r}
  \bar{s} \bar{t} \bar{i} \bar{j} \bar{k} \bar{m} \bar{p} \bar{q}} =
  iF_{\bar{i} \bar{j} \bar{k}, \bar{r} \bar{s} \bar{t}, \bar{m} \bar{p}
  \bar{q}}
\end{equation}
The matrix representations of the generators are not antisymmetric in this
basis, even though the structure constants are totally antisymmetric, because
it is necessary to take the three types of index group in a different order
for rows and columns, to ensure that $\mathrm{SU} \left( 9 \right)$
anti-fundamental indices contract with $\mathrm{SU} \left( 9 \right)$
fundamental indices.  The matrix representations of the generators can be
written as:
\begin{equation}
  \label{T sub alpha} \begin{array}{c}
    \\
    T_{\alpha} \quad =
  \end{array} \begin{array}{cc}
    & \begin{array}{ccc}
      {\scriptstyle \bar{m} \bar{p} \bar{q}} \hspace{0.5em} \quad &
      {\scriptstyle \hspace{0.8em} \gamma \hspace{0.8em}} & \quad
      {\scriptstyle  \hspace{0.6em} mpq}
    \end{array}\\
    \begin{array}{c}
      {\scriptstyle ijk}\\
      {\scriptstyle \beta}\\
      {\scriptstyle \bar{i} \bar{j} \bar{k}}
    \end{array} & \left(\begin{array}{ccc}
      iF_{ijk, \alpha, \bar{m} \bar{p} \bar{q}} & 0 & 0\\
      0 & iF_{\beta \alpha \gamma} & 0\\
      0 & 0 & iF_{\bar{i} \bar{j} \bar{k}, \alpha, mpq}
    \end{array}\right)
  \end{array}
  \hspace{3.0ex}
\end{equation}
\begin{equation}
  \label{T sub r s t} \begin{array}{c}
    \\
    T_{rst} \quad =
  \end{array} \begin{array}{cc}
    & \begin{array}{ccc}
      {\scriptstyle \bar{m} \bar{p} \bar{q} \hspace{1.3em}}
      \quad & {\scriptstyle
      \hspace{0.8em} \gamma \hspace{0.8em}} & \quad {\scriptstyle
      \hspace{1.5em} mpq \hspace{0.2em}}
    \end{array}\\
    \begin{array}{c}
      {\scriptstyle ijk}\\
      {\scriptstyle \beta}\\
      {\scriptstyle \bar{i} \bar{j} \bar{k}}
    \end{array} & \left(\begin{array}{ccc}
      0 & 0 & iF_{ijk, rst, mpq}\\
      iF_{\beta, rst, \bar{m} \bar{p} \bar{q}} & 0 & 0\\
      0 & iF_{\bar{i} \bar{j} \bar{k}, rst, \gamma} & 0
    \end{array}\right)
  \end{array}
  \hspace{3.0ex}
\end{equation}
\begin{equation}
  \label{T sub r bar s bar t bar} \begin{array}{c}
    \\
    T_{\bar{r} \bar{s} \bar{t}} \quad =
  \end{array} \begin{array}{cc}
    & \begin{array}{ccc}
      {\scriptstyle \bar{m} \bar{p} \bar{q} \hspace{1.4em}}
      \quad & {\scriptstyle
      \hspace{0.8em} \gamma \hspace{0.8em}} & \quad {\scriptstyle
      \hspace{1.3em} mpq}
    \end{array}\\
    \begin{array}{c}
      {\scriptstyle ijk}\\
      {\scriptstyle \beta}\\
      {\scriptstyle \bar{i} \bar{j} \bar{k}}
    \end{array} & \left(\begin{array}{ccc}
      0 & iF_{ijk, \bar{r} \bar{s} \bar{t}, \gamma} & 0\\
      0 & 0 & iF_{\beta, \bar{r} \bar{s} \bar{t}, mpq}\\
      iF_{\bar{i} \bar{j} \bar{k}, \bar{r} \bar{s} \bar{t}, \bar{m} \bar{p}
      \bar{q}} & 0 & 0
    \end{array}\right)
  \end{array}
  \hspace{3.0ex}
\end{equation}
To check the Jacobi identities, we first note that from the $\mathrm{SU} \left(
9 \right)$ commutation relation for $\left( T_{\alpha} \right)_{ijk, \bar{m}
\bar{p} \bar{q}}$, we have:
\begin{equation}
  \label{first Jacobi identity} \frac{1}{6} F_{rst, \alpha, \bar{m} \bar{p}
\bar{q}}  F_{mpq, \beta, \bar{i} \bar{j} \bar{k}} + \frac{1}{6}  F_{\bar{i}
\bar{j} \bar{k},
  \alpha, mpq} F_{\bar{m} \bar{p} \bar{q}, rst, \beta} + F_{\beta \alpha
  \gamma} F_{\gamma, \bar{i} \bar{j} \bar{k}, rst} = 0
\end{equation}
We next note that:
\[ \frac{1}{6} F_{ijk, \alpha, \bar{u} \bar{v} \bar{w}} F_{uvw, rst, mpq}
+ \frac{1}{6} F_{mpq, \alpha, \bar{u} \bar{v} \bar{w}} F_{uvw, ijk, rst}
+ \frac{1}{6} F_{rst, \alpha,
   \bar{u} \bar{v} \bar{w}} F_{uvw, mpq, ijk} = \hspace{4em}  \]
\[ = - \frac{i}{6 \sqrt{2}} \left( \left( t_{\alpha} \right)_{i \bar{u}}
   \epsilon_{rstujkmpq} + \left( t_{\alpha} \right)_{j \bar{u}}
   \epsilon_{rstukimpq} + \left( t_{\alpha} \right)_{k \bar{u}}
   \epsilon_{rstuijmpq} \right. \]
\[ \hspace{4em}  + \left( t_{\alpha} \right)_{m \bar{u}} \epsilon_{rstijkupq}
   + \left( t_{\alpha} \right)_{p \bar{u}} \epsilon_{rstijkuqm} + \left(
   t_{\alpha} \right)_{q \bar{u}} \epsilon_{rstijkump} \]
\begin{equation}
  \label{second Jacobi identity} \hspace{6em} \left. + \left( t_{\alpha}
  \right)_{r \bar{u}} \epsilon_{ustijkmpq} + \left( t_{\alpha} \right)_{s
  \bar{u}} \epsilon_{utrijkmpq} + \left( t_{\alpha} \right)_{t \bar{u}}
  \epsilon_{ursijkmpq} \rule{0pt}{2.0ex} \right)
\end{equation}
The right-hand side is totally antisymmetric in $rstijkmpq$, for each value of
$\alpha$, and is thus equal to an $\alpha$-dependent multiple of
$\epsilon_{rstijkmpq}$.  The $\alpha$-dependent coefficient of
$\epsilon_{rstijkmpq}$ is found by contracting with $\epsilon_{\bar{r} \bar{s}
\bar{t} \bar{i} \bar{j} \bar{k} \bar{m} \bar{p} \bar{q}}$, which gives zero,
due to the tracelessness of $t_{\alpha}$.

Similarly, we find:
\begin{equation}
  \label{third Jacobi identity} \frac{1}{6} F_{\bar{i} \bar{j} \bar{k},
\alpha, uvw}  F_{\bar{u} \bar{v} \bar{w}, \bar{r} \bar{s} \bar{t}, \bar{m}
\bar{p} \bar{q}} + \frac{1}{6} F_{\bar{m} \bar{p} \bar{q}, \alpha, uvw}
F_{\bar{u} \bar{v} \bar{w}, \bar{i} \bar{j} \bar{k}, \bar{r} \bar{s} \bar{t}}
+ \frac{1}{6} F_{\bar{r}
  \bar{s} \bar{t}, \alpha, uvw} F_{\bar{u} \bar{v} \bar{w}, \bar{m} \bar{p}
  \bar{q}, \bar{i} \bar{j} \bar{k}} = 0
\end{equation}
We next note that, due to the tracelessness and the normalization
(\ref{normalization}) of the $\mathrm{SU} \left( 9 \right)$ generators, we have:
\begin{equation}
  \label{summed Kronecker square of SU 9 generators} \left( t_{\alpha}
  \right)_{r \bar{i}} \left( t_{\alpha} \right)_{s \bar{j}} = \frac{1}{2}
  \left( \delta_{r \bar{j}} \delta_{s \bar{i}} - \frac{1}{9} \delta_{r
  \bar{i}} \delta_{s \bar{j}} \right)
\end{equation}
We now consider the expression:
\[ F_{uvw, \bar{i} \bar{j} \bar{k}, \alpha} F_{\alpha, \bar{m} \bar{p}
   \bar{q}, rst} + F_{rst, \bar{i} \bar{j} \bar{k}, \alpha} F_{\alpha, uvw,
   \bar{m} \bar{p} \bar{q}} = \hspace{14em}  \]
\[ \hspace{0.6em}  = \left( \left( t_{\alpha} \right)_{u \bar{i}}
   \delta_{v \bar{j}} \delta_{w \bar{k}} \pm \textrm{seventeen terms}
   \right) \left( \left( t_{\alpha} \right)_{r \bar{m}}
   \delta_{s \bar{p}} \delta_{t \bar{q}} \pm \textrm{seventeen terms}
   \right) \]
\begin{equation}
  \label{first two terms of fourth Jacobi identity} \hspace{3em}  -
  \left( \left( t_{\alpha} \right)_{r \bar{i}} \delta_{s \bar{j}}
  \delta_{t \bar{k}} \pm \textrm{seventeen terms}
  \textrm{} \right)_{} \left( \left( t_{\alpha} \right)_{u \bar{m}} \delta_{v
  \bar{p}} \delta_{w \bar{q}} \pm \textrm{seventeen terms}
  \textrm{} \right)
\end{equation}
I will show that this is equal to:
\[ \frac{1}{2} \left( \delta_{r \bar{i}} \delta_{s \bar{j}} \delta_{t
   \bar{k}} \delta_{u \bar{m}} \delta_{v \bar{p}} \delta_{w \bar{q}} \pm 719
   \hspace{0.8em} \mathrm{terms} \right) = \hspace{12em}  \]
\begin{equation}
  \label{minus the third term of the fourth Jacobi identity} \hspace{6em}  =
\frac{1}{12} \epsilon_{\bar{i} \bar{j} \bar{k} \bar{m} \bar{p} \bar{q} \bar{x}
\bar{y} \bar{z}} \epsilon_{rstuvwxyz} = - \frac{1}{6} F_{\bar{m} \bar{p}
  \bar{q}, \bar{i} \bar{j} \bar{k}, \bar{x} \bar{y} \bar{z}} F_{xyz, rst, uvw}
\end{equation}
where the additional terms in the first line of (\ref{minus the third term of
the fourth Jacobi identity}) antisymmetrize with respect to permutations of
$\left( \overline{i,} \bar{j}, \bar{k}, \bar{m}, \overline{p,} \bar{q}
\right)$.  We first note that the terms in the first line of (\ref{minus the
third term of the fourth Jacobi identity}) can be classified by the number $n$
of elements of $\left\{ u, v, w \right\}$ that are joined by Kronecker deltas
to elements of $\left\{ \bar{m}, \overline{p,} \bar{q} \right\}$.  We see that
when we use (\ref{summed Kronecker square of SU 9 generators}) in the
right-hand side of (\ref{first two terms of fourth Jacobi identity}), terms
with $n = 0$ can only come from the second term in the right-hand side of
(\ref{summed Kronecker square of SU 9 generators}), used in the first term in
the right-hand side of (\ref{first two terms of fourth Jacobi identity}),
terms with $n = 1$ can only come from the first term in the right-hand side of
(\ref{summed Kronecker square of SU 9 generators}), used in the first term in
the right-hand side of (\ref{first two terms of fourth Jacobi identity}),
terms with $n = 2$ can only come from the first term in the right-hand side of
(\ref{summed Kronecker square of SU 9 generators}), used in the second term in
the right-hand side of (\ref{first two terms of fourth Jacobi identity}), and
terms with $n = 3$ can only come from the second term in the right-hand side
of (\ref{summed Kronecker square of SU 9 generators}), used in the second term
in the right-hand side of (\ref{first two terms of fourth Jacobi identity}).
Thus the first term in the first line of (\ref{minus the third term of the
fourth Jacobi identity}), which has $n = 3$, comes from the second term in the
right-hand side of (\ref{summed Kronecker square of SU 9 generators}), used in
the second term in the right-hand side of (\ref{first two terms of fourth
Jacobi identity}).

Considering, now, the coefficient of the first term in the first line of
(\ref{minus the third term of the fourth Jacobi identity}), we see that it
gets contributions from three terms in the first factor of the second term in
the right-hand side of (\ref{first two terms of fourth Jacobi identity}),
namely $\left( t_{\alpha} \right)_{r \bar{i}} \delta_{s \bar{j}} \delta_{t
\bar{k}} + \delta_{r \bar{i}} \left( t_{\alpha} \right)_{s \bar{j}} \delta_{t
\bar{k}} + \delta_{r \bar{i}} \delta_{s \bar{j}} \left( t_{\alpha} \right)_{t
\bar{k}}$, times three terms in the second factor of the second term in the
right-hand side of (\ref{first two terms of fourth Jacobi identity}), namely
$\left( t_{\alpha} \right)_{u \bar{m}} \delta_{v \bar{p}} \delta_{w \bar{q}} +
\delta_{u \bar{m}} \left( t_{\alpha} \right)_{v \bar{p}} \delta_{w \bar{q}} +
\delta_{u \bar{m}} \delta_{v \bar{p}} \left( t_{\alpha} \right)_{w \bar{q}}$.
Hence its coefficient is $- \frac{1}{2} \times \left( -
\frac{1}{9} \right) \times 9 = \frac{1}{2}$, as required.

Considering, next, the coefficient of a term with $n = 2$, namely $\delta_{r
\bar{m}} \delta_{s \bar{j}} \delta_{t \bar{k}} \delta_{u \bar{i}} \delta_{v
\bar{p}} \delta_{w \bar{q}}$, in the first line of (\ref{minus the third term
of the fourth Jacobi identity}), we see that the locations of the
$t_{\alpha}$'s are now fixed, and this term only gets a contribution from the
first term in the first factor of the second term in the right-hand side of
(\ref{first two terms of fourth Jacobi identity}), times the first term in the
second factor of the second term in the right-hand side of (\ref{first two
terms of fourth Jacobi identity}).  Hence its coefficient is $- \frac{1}{2}$,
as required.

And in a similar manner, we confirm the coefficient of a term with $n = 1$,
namely $\delta_{u \bar{m}} \delta_{v \bar{j}} \delta_{w \bar{k}} \delta_{r
\bar{i}} \delta_{s \bar{p}} \delta_{t \bar{q}}$, as $\frac{1}{2}$, and the
coefficient of a term with $n = 0$, namely \\
$\delta_{u \bar{i}} \delta_{v
\bar{j}} \delta_{w \bar{k}} \delta_{r \bar{m}} \delta_{s \bar{p}} \delta_{t
\bar{q}}$, as $- \frac{1}{2}$.  The coefficients of the remaining $716$ terms
in the first line of (\ref{minus the third term of the fourth Jacobi
identity}), of which $6^2 - 1$ have $n = 3$, $\left( 3^4 \times 2^2 \right) -
1$ have $n = 2$, $\left( 3^4 \times 2^2 \right) - 1$ have $n = 1$ and $6^2 -
1$ have $n = 0$, are then determined by the separate antisymmetries of the
left-hand side of (\ref{first two terms of fourth Jacobi identity}) in $\left(
u, v, w \right)$, $\left( r, s, t \right)$, $\left( \overline{i,} \bar{j},
\bar{k} \right)$, and $\left( \bar{m}, \overline{p,} \bar{q} \right)$.  And
furthermore, all $18^2 + 18^2$ terms in the right-hand side of (\ref{first two
terms of fourth Jacobi identity}) have now been accounted for.  Thus we find
the final Jacobi identity:
\begin{equation}
  \label{fourth Jacobi identity} F_{uvw, \bar{i} \bar{j} \bar{k}, \alpha}
  F_{\alpha, \bar{m} \bar{p} \bar{q}, rst} + F_{rst, \bar{i} \bar{j} \bar{k},
  \alpha} F_{\alpha, uvw, \bar{m} \bar{p} \bar{q}} + \frac{1}{6}
F_{\bar{m} \bar{p}
  \bar{q}, \bar{i} \bar{j} \bar{k}, \bar{x} \bar{y} \bar{z}} F_{xyz, rst, uvw}
  = 0
\end{equation}

The non-zero structure constants of $E_8$ in the $\mathrm{SU} ( 9 )$ basis
were effectively presented in the equation array (41) in section 18 of
Chapter V of Cartan's thesis \cite{CartanThesis}, using a notation where
distinct index letters designate distinct values of the indices, and repeated
indices do {\emph{not}} indicate summations.  Cartan's normalization of the
$\mathrm{SU} ( 9 )$ generators is effectively such that the factor
$\frac{1}{2}$ in the right-hand side of (\ref{normalization}) is absent, which
has the consequence, from (\ref{fourth Jacobi identity}), that the factors
$\frac{1}{\sqrt{2}}$ multiplying the 9-index $\epsilon$ symbols in
(\ref{r s t i j k m p q matrix elements}) and
(\ref{r bar s bar t bar i bar j bar k bar m bar p bar q bar matrix elements})
are absent.  After allowing for that, Cartan's values for the structure
constants are consistent with those given here.

If we had used the alternative convention, as in \cite{NTSG} and the first two
versions of this article, that sums over antisymmetrized triples of indices
are not multiplied by a compensating factor $\frac{1}{6}$, then the changes to
the structure constants are that they are multiplied by a factor
$\frac{1}{\sqrt{6}}$ for each $E_8$ index that is an \textbf{84} or
$\mathbf{\overline{84}}$ index.  Thus in calculations involving contractions
of products of $E_8$ structure constants, the effects of the changed
summation rule cancel against the changes to the structure constants, for each
summation over a contracted pair of $E_8$ indices.  The generators of the
\textbf{84} in (\ref{84}) and the $\mathbf{\overline{84}}$ in (\ref{84 bar})
are multiplied by $\frac{1}{6}$, and the coefficients
$\frac{i}{\sqrt{2}}$ multiplying the 9-index $\epsilon$ symbols in
(\ref{r s t i j k m p q matrix elements}) and
(\ref{r bar s bar t bar i bar j bar k bar m bar p bar q bar matrix elements})
become $\frac{i}{12 \sqrt{3}}$.

We next calculate $\mathrm{Tr} \left( T_{\mathcal{A}} T_{\mathcal{B}} \right)$,
where we recall, from just after (\ref{Yang Mills action}), on page
\pageref{Yang Mills action}, that we are using Ho\v{r}ava and Witten's notation
for traces in $E_8$, so that for $E_8$, ``$\mathrm{tr}$'' denotes $\frac{1}{30}$
of the trace in the adjoint representation, which is denoted by
``$\mathrm{Tr}$''.  We also recall our convention, stated after
(\ref{84 bar}), that each index in
a sum over an antisymmetrized triple of indices, as for example in $\left(
T_{\alpha} \right)_{i j k, \bar{m} \bar{p} \bar{q}} \left( T_{\beta}
\right)_{m p q, \bar{r} \bar{s} \bar{t}}$, is to be summed over its full
range, in accordance with the standard summation convention, and the sum over
the antisymmetrized triple of indices is to be multiplied by a compensating
factor $\frac{1}{6}$, so that each of the 84 distinct elements of the
\textbf{84} or the $\mathbf{\overline{84}}$ of $\mathrm{SU} ( 9 )$ is
effectively counted once, instead of 6 times.

We first note that, from (\ref{normalization}), and the definition, $\left[
t_{\alpha}, t_{\beta} \right] = if_{\alpha \beta \gamma} t_{\gamma}$, the
$\mathrm{SU} \left( 9 \right)$ structure constants $f_{\alpha \beta \gamma}$ are
given by $f_{\alpha \beta \gamma} = - 2 i \mathrm{tr} \left( \left[ t_{\alpha},
t_{\beta} \right] t_{\gamma} \right)$.  Hence we find, using
(\ref{summed Kronecker square of SU 9 generators}):
\begin{equation}
  \label{f sub delta alpha gamma f sub delta beta gamma} f_{\delta \alpha
  \gamma} f_{\delta \beta \gamma} = - 2 \mathrm{tr} \left( \left[ t_{\delta},
  t_{\alpha} \right] \left[ t_{\delta}, t_{\beta} \right] \right) = 9
  \delta_{\alpha \beta}
\end{equation}
We next note that:
\[ \frac{1}{36}
\left( T_{\alpha} \right)_{ijk, \bar{m} \bar{p} \bar{q}} \left( T_{\beta}
   \right)_{mpq, \bar{i} \bar{j} \bar{k}} = \hspace{26em}  \]
\begin{equation}
  \label{trace of 84 84} = \frac{1}{2} \left( t_{\alpha} \right)_{r \bar{x}}
  \delta_{s \bar{y}} \delta_{t \bar{z}} \left( \left( t_{\beta} \right)_{x
  \bar{u}} \delta_{y \bar{v}} \delta_{z \bar{w}} + \delta_{x \bar{u}} \left(
  t_{\beta} \right)_{y \bar{v}} \delta_{z \bar{w}} + \delta_{x \bar{u}}
  \delta_{y \bar{v}} \left( t_{\beta} \right)_{z \bar{w}} \right) \delta_{uvw,
  \bar{r} \bar{s} \bar{t}} = \frac{21}{2} \delta_{\alpha \beta}
\end{equation}
Thus:
\[ \mathrm{Tr} \left( T_{\alpha} T_{\beta} \right) = \left( T_{\alpha}
   \right)_{\gamma \delta} \left( T_{\beta} \right)_{\delta \gamma}
+ \frac{1}{36} \left(
   T_{\alpha} \right)_{ijk, \bar{m} \bar{p} \bar{q}} \left( T_{\beta}
   \right)_{mpq, \bar{i} \bar{j} \bar{k}}
+ \frac{1}{36} \left( T_{\alpha} \right)_{\bar{i}
   \bar{j} \bar{k}, mpq} \left( T_{\beta} \right)_{\bar{m} \bar{p} \bar{q},
   ijk} = \]
\begin{equation}
  \label{Trace T sub alpha T sub beta} = 9 \delta_{\alpha \beta} +
  \frac{21}{2} \delta_{\alpha \beta} + \frac{21}{2} \delta_{\alpha \beta} = 30
  \delta_{\alpha \beta}
\end{equation}
We next note that:
\begin{equation}
  \label{F sub 9 F sub 9 with six indices contracted} - \frac{1}{36}
F_{ijk, rst, mpq}
  F_{\bar{m} \bar{p} \bar{q}, \bar{u} \bar{v} \bar{w}, \bar{i} \bar{j}
  \bar{k}} = - \frac{1}{72} \epsilon_{rstijkmpq} \epsilon_{\bar{u} \bar{v}
  \bar{w} \bar{m} \bar{p} \bar{q} \bar{i} \bar{j} \bar{k}} = 10 \delta_{rst,
  \bar{u} \bar{v} \bar{w}}
\end{equation}
And from (\ref{summed Kronecker square of SU 9 generators}), we find that:
\[ - \frac{1}{6} F_{\alpha, rst, \bar{m} \bar{p} \bar{q}} F_{mpq, \bar{u}
\bar{v} \bar{w}, \alpha} = \frac{1}{2} \delta_{rst, \bar{f} \bar{g} \bar{h}}
\left( t_{\alpha}
   \right)_{f \bar{x}} \delta_{g \bar{y}} \delta_{h \bar{z}} \left( T_{\alpha}
   \right)_{xyz, \bar{u}  \bar{v}  \bar{w}} = \hspace{6em}  \]
\begin{equation}
  \label{F sub alpha 3 3 F sub 3 3 alpha} \hspace{8em}  = \frac{3}{2} \left( 9
  - \frac{1}{9} - 1 - \frac{1}{9} - 1 - \frac{1}{9} \right) \delta_{rst,
  \bar{u} \bar{v} \bar{w}} = 10 \delta_{rst, \bar{u} \bar{v} \bar{w}}
\end{equation}
Thus:
\[ \mathrm{Tr} \left( T_{rst} T_{\bar{u} \bar{v} \bar{w}} \right) =
- \frac{1}{36} F_{ijk, rst, mpq} F_{\bar{m} \bar{p} \bar{q}, \bar{u} \bar{v}
\bar{w}, \bar{i} \bar{j} \bar{k}} - \frac{1}{6} F_{\alpha, rst, \bar{m}
\bar{p} \bar{q}} F_{mpq, \bar{u} \bar{v} \bar{w}, \alpha} - \frac{1}{6}
F_{\bar{i} \bar{j} \bar{k}, rst, \beta}
   F_{\beta, \bar{u} \bar{v} \bar{w}, ijk} = \]
\begin{equation}
  \label{Trace T sub r s t T sub u bar v bar w bar} = \left( 10 + 10 + 10
  \right) \delta_{rst, \bar{u} \bar{v} \bar{w}} = 30 \delta_{rst, \bar{u}
  \bar{v} \bar{w}}
\end{equation}
And from the block matrix structure of the generators (\ref{T sub alpha}),
(\ref{T sub r s t}), and (\ref{T sub r bar s bar t bar}), we see that
\begin{equation}
  \label{the Traces which are zero} \mathrm{Tr} \left( T_{\alpha} T_{rst}
  \right) = \mathrm{Tr} \left( T_{\alpha} T_{\bar{r} \bar{s} \bar{t}} \right) =
  \mathrm{Tr} \left( T_{rst} T_{uvw} \right) = \mathrm{Tr} \left( T_{\bar{r}
  \bar{s} \bar{t}} T_{\bar{u} \bar{v} \bar{w}} \right) = 0
\end{equation}

We note that we can choose a set of generators for the $\mathrm{SU} \left( 9
\right)$ Cartan subalgebra, such that in the $\mathrm{SU} \left( 9 \right)$
fundamental, the generators of the Cartan subalgebra are diagonal matrices,
and their nonzero matrix elements are equal to integers, times an overall
normalization factor that depends on the generator, and that there is an
infinite variety of such choices of the generators of the $\mathrm{SU} \left( 9
\right)$ Cartan subalgebra, consistent with (\ref{normalization}).  And from
(\ref{Adjoint}), (\ref{84}), (\ref{84 bar}), and (\ref{T sub alpha}), we see
that for any such set of generators of the $\mathrm{SU} \left( 9 \right)$ Cartan
subalgebra, each generator of the $E_8$ Cartan subalgebra, in the $E_8$
fundamental / adjoint, will be a $248 \times 248$ diagonal matrix, whose
nonzero matrix elements are equal to integers, times an overall normalization
factor that depends on the generator.

The occurrence of the \textbf{84} and $\overline{\mathbf{8} \mathbf{4}}$
of $\mathrm{SU} \left( 9 \right)$, in the fundamental of $E_8$, is connected to
the presence of the three-form gauge field in $d = 11$ supergravity
{\cite{Nahm, Cremmer Julia Scherk}}, through the embedding of $\mathrm{SO}
\left( 9 \right)$, the little group of the $d = 11$ Poincare group, in $E_8$,
by the subgroup chain $\mathrm{SO} \left( 9 \right) \subset \mathrm{SU} \left( 9
\right) \subset E 8$.  For, as reviewed in subsection 2.2 of {\cite{0212245 de
Wit Supergravity}}, half of the 32 supercharges of $d = 11$ supergravity
vanish on the mass shell, and the representation space of the 16 nonvanishing
supercharges decomposes into the two chiral spinor representations of
$\mathrm{SO} \left( 16 \right)$, one of which corresponds to the bosonic states,
and the other to the fermionic states.  The 16 nonvanishing spinor charges
transform according to a single spinor representation of the little group,
$\mathrm{SO} \left( 9 \right)$, and the helicity content of the bosonic and
fermion states is determined by the branching of the two different
$ \mathbf{128} $'s of $\mathrm{SO} \left( 16 \right)$, when $\mathrm{SO} \left( 9
\right)$ is embedded into $\mathrm{SO} \left( 16 \right)$ such that the spinor
of $\mathrm{SO} \left( 9 \right)$ becomes the vector of $\mathrm{SO} \left( 16
\right)$.  This results in one of the $ \mathbf{128} $'s of $\mathrm{SO} ( 16 )$
branching into the $\mathbf{4} \mathbf{4}+\mathbf{8} \mathbf{4}$ of
$\mathrm{SO} \left( 9 \right)$, corresponding to the graviton and the three-form
gauge field, while the other $\mathbf{1} \mathbf{2} \mathbf{8}$ of
$\mathrm{SO} \left( 16 \right)$ becomes the $\mathbf{1} \mathbf{2}
\mathbf{8}$ vector-spinor of $\mathrm{SO} \left( 9 \right)$, corresponding to
the gravitino, as can be checked by studying weight diagrams.  On the other
hand, the adjoint of $\mathrm{SO} \left( 16 \right)$ branches into the
antisymmetrized square of the spinor of $\mathrm{SO} \left( 9 \right)$, which
contains the $\mathbf{3} \mathbf{6}$ of $\mathrm{SO} \left( 9 \right)$,
which is the adjoint, and the $\mathbf{8} \mathbf{4}$ of $\mathrm{SO} \left(
9 \right)$, which is the three-form.  And under the embedding $\mathrm{SO}
\left( 9 \right) \subset \mathrm{SU} \left( 9 \right)$, the adjoint of
$\mathrm{SU} \left( 9 \right)$ branches to the adjoint and the $\mathbf{4}
\mathbf{4}$ of $\mathrm{SO} \left( 9 \right)$, and the $\mathbf{8}
\mathbf{4}$ and $\overline{\mathbf{8} \mathbf{4}}$ of $\mathrm{SU} \left(
9 \right)$ both become the $\mathbf{8} \mathbf{4}$ of $\mathrm{SO} \left( 9
\right)$.  Thus the decomposition of the adjoint of $E_8$ into irreducible
representations of $\mathrm{SO} \left( 9 \right)$ is the same, when $\mathrm{SO}
\left( 9 \right)$ is embedded into $E_8$ according to the subgroup chains
$\mathrm{SO} \left( 9 \right) \subset \mathrm{SU} \left( 9 \right) \subset E 8$
and $\mathrm{SO} \left( 9 \right) \subset \mathrm{SO} \left( 16 \right) \subset E
8$, provided $\mathrm{SO} \left( 9 \right)$ is embedded into $\mathrm{SO} \left(
16 \right)$ in the manner that determines the helicity content of the $d = 11$
supergravity states on the mass shell, and the spinor of $\mathrm{SO} \left( 16
\right)$, in the fundamental of $E_8$, is the one which branches to the
$\mathbf{4} \mathbf{4}+\mathbf{8} \mathbf{4}$ of $\mathrm{SO} \left( 9
\right)$.

\subsection{Dirac quantization condition for $ E8 $ vacuum gauge fields}
\label{Dirac quantization condition for E8 vacuum gauge fields}

In this subsection, I will show that the field strengths of the Hodge - de
Rham monopoles are restricted in their possible magnitudes, and partly also in
their possible orientations within $E_8$, by a form of Dirac quantization
condition.  In particular, if the configuration of the Yang-Mills fields is
gauge equivalent to a configuration where they lie everywhere within the
Cartan subalgebra of $E_8$, then for an arbitrary closed smooth orientable
two-dimensional surface $\mathcal{S}$ in the compact six-manifold
$\mathcal{M}^6$, the integral of the field strengths, over $\mathcal{S}$, in a
gauge where the Yang-Mills fields lie everywhere within the Cartan subalgebra
of $E_8$, is a $248 \times 248$ diagonal matrix, that must be a lattice point
of a certain discrete lattice in the eight dimensional Cartan subalgebra of $E
8$.

We recall that for an arbitrary Yang-Mills gauge group, and for matter fields
$\psi$ transforming in an arbitrary representation of the gauge group, with
hermitian generators $T_{\alpha}$ satisfying $\left[ T_{\alpha}, T_{\beta}
\right] = if_{\alpha \beta \gamma} T_{\gamma}$, with totally antisymmetric
structure constants $f_{\alpha \beta \gamma}$, the covariant derivative is
$D_{\mu} \psi = \left( \partial_{\mu} - igA_{\mu \alpha} T_{\alpha} \right)
\psi$, where $A_{\mu \alpha}$ are the Yang-Mills fields and $g$ is the
coupling constant, and the Wilson line, or gauge covariant path ordered phase
factor, for a continuous path $x \left( s \right)$, $s_{\min} \leq s \leq
s_{\max}$, differentiable except at a finite number of values of $s$, is:
\[ W\! \left( \left\{ A \right\}\! ,\! \left\{ x \left( s \right) \right\}
   \right)_{i
   \bar{j}} = \sum_{n = 0}^{\infty} \left( - ig \right)^n\! \int\! ds_1 \ldots
   \int\! ds_n \theta \left( s_1 - s_{\min} \right) \theta \left( s_2 - s_1
   \right) \ldots \theta \left( s_n - s_{n - 1} \right) \times \]
\begin{equation}
  \label{Wilson line} \hspace{2em} \hspace{-2.9pt} \times \theta
  \left( s_{\max} - s_n \right)
  \frac{dx^{\mu_1} \left( s_1 \right)}{ds_1} \ldots \frac{dx^{\mu_n} \left(
  s_n \right)}{ds_n} A_{\mu_1 \alpha_1} \left( x \left( s_1 \right) \right)
  \ldots A_{\mu_n \alpha_n} \left( x \left( s_n \right) \right) \left(
  T_{\alpha_1} \ldots T_{\alpha_n} \right)_{i \bar{j}},
\end{equation}
where $\theta \left( s \right)$ is the step function, $\theta \left( s \right)
= 1$ for $s \geq 0$, and $\theta \left( s \right) = 0$ for $s < 0$.  For paths
$x_1 \left( s \right)$, $s_{\min} \leq s \leq s_{\mathrm{mid}}$, and $x_2 \left(
s \right)$, $s_{\mathrm{mid}} \leq s \leq s_{\max}$, such that $x_1 \left(
s_{\mathrm{mid}} \right) = x_2 \left( s_{\mathrm{mid}} \right)$,  $W \left(
\left\{ A \right\}, \left\{ x \left( s \right) \right\} \right)_{i \bar{j}}$
satisfies the product formula:
\begin{equation}
  \label{product formula for Wilson lines} W \left( \left\{ A \right\},
  \left\{ x_1 \left( s \right) \right\} \right)_{i \bar{k}} W \left( \left\{ A
  \right\}, \left\{ x_2 \left( s \right) \right\} \right)_{k \bar{j}} = W
  \left( \left\{ A \right\}, \left\{ x_1 \left( s \right) \right\} \cup
  \left\{ x_2 \left( s \right) \right\} \right)_{i \bar{j}}
\end{equation}
where $\left\{ x_1 \left( s \right) \right\} \cup \left\{ x_2 \left( s \right)
\right\}$ denotes the union of the two paths, which is a map with domain
$s_{\min} \leq s \leq s_{\max}$.

We now consider the transformations of $A_{\mu \alpha}$, $\psi$, and $W \left(
\left\{ A \right\}, \left\{ x \left( s \right) \right\} \right)$, under finite
gauge transformations, that might be topologically non-trivial, and might not
be connected to the identity.  I shall assume that the gauge transformation
parameters $\Lambda_{\alpha} \left( x \right)$ are continuous and
differentiable on each coordinate patch, and that the gauge transformation
acts on $\psi$ by $\psi \left( x \right) \rightarrow U \left( x \right) \psi
\left( x \right)$, where $U \left( x \right) = e^{i \Lambda \left( x \right)}
= e^{i \Lambda_{\alpha} \left( x \right) T_{\alpha}}$.  Then the
gauge-transformed Yang-Mills fields $A'_{\mu \alpha}$ are required to satisfy
$D_{\mu} \left( A' \right) U \psi = UD_{\mu} \left( A \right) \psi$.  Thus we
require:
\begin{equation}
  \label{required formula for gauge transformed Yang Mills field} A'_{\mu
  \alpha} T_{\alpha} = A_{\mu \alpha} UT_{\alpha} U^{\dag} - \frac{i}{g}
  \left( \partial_{\mu} U \right) U^{\dag}
\end{equation}
Using the Baker-Campbell-Hausdorff formula {\cite{PlanetMath
Baker-Campbell-Hausdorff}} $e^A Be^{- A}\hspace{-1.0pt} =\hspace{-1.0pt}
B + \left[ A, B \right] + \frac{1}{2!} \hspace{-1.0pt} \left[ A, \left[ A, B
\right] \right] + \frac{1}{3!}
\left[ A, \left[ A, \left[ A, B \right] \right] \right] + \ldots$, and also,
for expanding $e^{i \Lambda \left( x + \delta x \right)} e^{- i \Lambda \left(
x \right)}$ to first order in $\delta x$, the relation $e^{s \left( A + B
\right)} e^{- sA} = 1 + \int^s_0 e^{tA} Be^{- tA} dt + O \left( B^2 \right)$,
we find that (\ref{required formula for gauge transformed Yang Mills field})
is satisfied for an arbitrary representation with generators $T_{\alpha}$, if:
\begin{equation}
  \label{finite gauge transformation of Yang Mills field} A'_{\mu \alpha} =
  A_{\mu \beta} \left( e^{- i \breve{\Lambda} \left( x \right)} \right)_{\beta
  \alpha} + \frac{1}{g} \left( \partial_{\mu} \Lambda_{\beta} \left( x \right)
  \right) \left( \frac{e^{- i \breve{\Lambda} \left( x \right)} - 1}{- i
  \breve{\Lambda} \left( x \right)} \right)_{\beta \alpha}
\end{equation}
where the matrix $\breve{\Lambda}_{\beta \gamma} \left( x \right)$ is defined
in terms of the generators $\left( \breve{T}_{\alpha} \right)_{\beta \gamma} =
- if_{\alpha \beta \gamma}$ of the adjoint representation, by
$\breve{\Lambda}_{\beta \gamma} \left( x \right) = \Lambda_{\alpha} \left( x
\right) \left( \breve{T}_{\alpha} \right)_{\beta \gamma}$.  The Wilson line
(\ref{Wilson line}) then transforms as:
\begin{equation}
  \label{gauge transformation of Wilson line} W\! \left( \left\{ A \right\}\!
  ,\! \left\{ x \left( s \right) \right\} \right)_{i \bar{j}} \rightarrow W\!
  \left( \left\{ A' \right\}\! ,\! \left\{ x \left( s \right) \right\}
  \right)_{i \bar{j}} = U_{i \bar{k}} \left( x \left( s_{\min} \right) \right)
  W\! \left( \left\{ A
  \right\}\! ,\! \left\{ x \left( s \right) \right\} \right)_{k \bar{l}}
  U^{\dag}_{l \bar{j}} \left( x \left( s_{\max} \right) \right)
\end{equation}

Now on a topologically non-trivial manifold, such as the compact six-manifolds
$\mathcal{M}^6$ considered in the present paper, the Yang-Mills fields can
also be topologically non-trivial.  This means that $A_{\mu \alpha} \left( x
\right)$ is not well-defined globally as a continuous and differentiable
function of the coordinates, which are themselves not defined globally.
Instead $A_{\mu \alpha} \left( x \right)$ is a continuous and differentiable
function of the coordinates on each coordinate patch, and where two patches
$i$ and $j$ overlap, $A^{\left( i \right)}_{\mu \alpha} \left( x^{\left( i
\right)} \right)$ is related to $A^{\left( j \right)}_{\nu \beta} \left(
x^{\left( j \right)} \left( x^{\left( i \right)} \right) \right)$, by both a
general coordinate transformation, and a finite gauge transformation.  This is
the case, for example, when the Yang-Mills fields are in the Cartan subalgebra
of the gauge group, and their field strengths are nonzero and proportional to
Hodge - de Rham harmonic two-forms.

The simplest example of this is a two-sphere centred on a Dirac magnetic
monopole {\cite{Dirac magnetic monopole}} in the Wu-Yang gauge {\cite{Wu Yang,
Eguchi Gilkey Hanson}}.  The vector potential is tangential to the two-sphere,
and is well defined on two coordinate patches, one of which covers the
northern hemisphere, and a strip of the southern hemisphere along the equator,
and the other of which covers the southern hemisphere, and a strip of the
northern hemisphere along the equator.  More generally, there will be three or
more coordinate patches, and at any point where three coordinate patches $i$,
$j$, and $k$ overlap, the gauge transformations $U^{i \rightarrow j}$, $U^{j
\rightarrow k}$, and $U^{k \rightarrow i}$ are required to satisfy $U^{i
\rightarrow j} U^{j \rightarrow k} U^{k \rightarrow i} = 1$.

When a Wilson line crosses from a patch $i$ to a patch $j$, we choose a point
$x$ on the line in the overlap region between the two patches, at which to
make the transition from patch $i$ to patch $j$, and the Wilson line is then
defined to be the matrix product of the segment of the line in patch $i$, and
the segment of the line in patch $j$, as in (\ref{product formula for Wilson
lines}), but with the gauge transformation matrix $U^{i \rightarrow j} \left(
x \right)$ inserted between the two segments.  If we consider two different
choices of the point $x$ on the line in the overlap region, at which to make
the transition between the two patches, we find, from the gauge transformation
(\ref{gauge transformation of Wilson line}) of the segment of the Wilson line
between the two different choices of the transition point, that the Wilson
line is independent of the choice of the transition point.

Now if a Wilson line doubles back on itself like a hairpin, and exactly
retraces its path back to its starting point, then it is identically equal to
the unit matrix, even if the hairpin path crosses between several coordinate
patches.  And this is also true for a hairpin path that has ``branches'', that
are themselves hairpins.  Furthermore, by the result just noted, this is also
true if, for a segment of the hairpin path that lies in the overlap region of
two coordinate patches, we make the transition between the two coordinate
patches, at different points on the ``outward'' and ``return'' sections of the
hairpin path.

Let us now consider a configuration of the Yang-Mills fields that is gauge
equivalent to a configuration where the gauge fields are everywhere in the
Cartan subalgebra of the Lie algebra, and choose a gauge where the gauge
fields are everywhere in the Cartan subalgebra.  Let us also assume that the
manifold has non-vanishing second Betti number, and that the gauge field
configuration is topologically non-trivial, due for example to including Hodge
- de Rham harmonic two forms.

We now consider an arbitrary closed orientable two-dimensional surface in the
manifold, that is embedded in the manifold in a topologically non-trivial
manner, in the sense that it cannot be contracted to a point.  Such surfaces
exist due to the assumption that the manifold has non-vanishing second Betti
number.  We do not know what the intrinsic topology of the surface is, but it
was shown by Seifert and Threlfall that the most general closed orientable
two-dimensional manifold is topologically equivalent to a sphere with $n$
handles, $n \geq 0$.  I shall consider a particular Wilson line that has the
form of a branched hairpin, and is thus equal to the identity matrix.
However, the hairpin branches will loop round and meet at their tips, in such
a way that, due to the assumption that the field configuration is Abelian, we
can also express the Wilson line as a diagonal matrix, such that each matrix
element on the diagonal has the form $e^{- ig \int F}$, where $\int F$ denotes
the integral over the closed two-dimensional surface, of the two-form field
strength of the corresponding diagonal matrix element of $A_{\mu \alpha}
T_{\alpha}$.  This will be non-zero, if the field configuration includes a
Hodge - de Rham harmonic two-form, with non-zero coefficient in that matrix
element, that has non-zero integral over that surface.  Thus $g \int F$ must
be an integer multiple of $2 \pi$.

Considering, first, the case when the intrinsic topology of the
two-dimensional surface is an ordinary two-sphere, the intersections of the
coordinate patches of the manifold will define coordinate patches on the
two-dimensional surface.  Let us suppose, first, that the coordinate patches
on the two-dimensional surface are topologically equivalent to the northern
hemisphere, plus a strip of the southern hemisphere, and the southern
hemisphere, plus a strip of the northern hemisphere, as in the case of the
Wu-Yang gauge for the Dirac monopole.  Then we choose a simple hairpin that
starts at a point on the equator, and wraps once round the equator, so that
the point where the hairpin doubles back on itself is the same as the point
where it started.  We choose the hairpin to start on the northern hemisphere
patch, and remain on the northern hemisphere patch all the way around the
equator to the point where it doubles back on itself, and it makes the
transition to the southern hemisphere patch at the point where it doubles back
on itself, and it remains on the southern hemisphere patch for the entire
``return'' section of the hairpin, until it reaches the starting point, where
it finally makes the transition back to the northern hemisphere patch again.
Then due to the Abelian nature of the gauge field, each of the two transitions
from one patch to the other simply introduces a phase factor, and the two
phase factors cancel one another because the two transitions occurred at the
same point.  Furthermore, for an Abelian field configuration, whose only
non-vanishing matrix elements are on the diagonal, each non-vanishing matrix
element of the Wilson line has the form $\exp \left( - ig \int ds
\frac{dx^{\mu} \left( s \right)}{ds} A_{\mu} \left( x \left( s \right) \right)
\right)$, where $A_{\mu}$ denotes the corresponding diagonal matrix element of
$A_{\mu \alpha} T_{\alpha}$.  We then uses Stokes's theorem to equate the line
integral in the exponent, for the ``outward'' section of the hairpin path, to
the integral of $F$ over the northern hemisphere, and the line integral in the
exponent, for the ``return'' section of the hairpin path, to the integral of
$F$ over the southern hemisphere.

And if the coordinate patches on the two-dimensional surface, topologically
equivalent to a two-sphere, are not topologically equivalent to the northern
hemisphere, plus a strip of the southern hemisphere, and the southern
hemisphere, plus a strip of the northern hemisphere, we can introduce two new
coordinate patches in the manifold, whose intersections with the
two-dimensional surface do have this form, and choose suitable gauges on these
two coordinate patches, such that we can use the intersections of these two
coordinate patches with the two-dimensional surface, as the coordinate patches
on the two-dimensional surface, and then use the argument as above.

\begin{figure}[!t]
\setlength{\unitlength}{1cm}
\begin{picture}(14.77,17.47)(-0.3,1.53)
\put(0.039,16.17){\line(0,1){0.809}}
\put(0.964,15.83){\circle{0.910}}
\put(4.84,15.83){\circle{0.869}}
\put(8.93,15.83){\circle{0.894}}
\put(5.29,15.83){\line(1,0){3.20}}
\put(11.31,14.52){\line(-6,-1){0.736}}
\put(10.56,14.4){\line(-1,0){0.924}}
\put(9.64,14.4){\line(-4,1){1.02}}
\put(8.61,14.65){\line(-2,1){0.845}}
\put(7.77,15.07){\line(-6,5){0.865}}
\put(11.35,17.19){\line(-5,3){0.984}}
\put(13.56,14.23){\line(-2,-1){0.834}}
\put(12.73,13.82){\line(-4,-1){1.29}}
\put(11.44,13.49){\line(-1,0){1.29}}
\put(8.69,13.85){\line(-5,3){1.02}}
\put(7.67,14.46){\line(-3,4){0.478}}
\put(7.20,15.09){\line(-1,3){0.242}}
\put(10.13,13.49){\line(-4,1){1.43}}
\put(4.76,14.72){\line(6,1){0.883}}
\put(5.65,14.87){\line(4,3){1.24}}
\put(1.44,15.8){\line(1,0){0.282}}
\put(1.72,15.8){\line(2,-3){0.276}}
\put(2.00,15.38){\line(2,-5){0.286}}
\put(2.30,14.67){\line(1,-1){0.679}}
\put(2.97,13.99){\line(6,-1){1.20}}
\put(4.18,13.79){\line(6,1){1.09}}
\put(5.27,13.97){\line(2,1){0.871}}
\put(6.15,14.41){\line(2,3){0.411}}
\put(6.96,15.87){\line(-3,5){1.04}}
\put(5.90,17.61){\line(-3,2){1.09}}
\put(4.80,18.34){\line(-6,1){1.42}}
\put(6.96,15.83){\line(3,2){0.940}}
\put(7.89,16.45){\line(2,1){0.705}}
\put(8.59,16.8){\line(1,0){0.721}}
\put(6.96,15.83){\line(1,2){0.553}}
\put(7.51,16.93){\line(1,1){0.529}}
\put(8.04,17.46){\line(5,2){0.739}}
\put(8.77,17.76){\line(6,1){0.859}}
\put(9.62,17.9){\line(6,-1){0.730}}
\put(13.04,15.83){\circle{0.905}}
\put(12.54,15.83){\line(1,0){0.076}}
\put(12.61,15.83){\line(-1,0){0.231}}
\put(13.50,15.8){\line(1,0){0.257}}
\put(14.03,14.74){\line(-1,-1){0.495}}
\put(11.50,18.56){\line(4,-1){1.27}}
\put(12.76,18.24){\line(2,-1){1.07}}
\put(13.84,17.7){\line(1,-1){0.539}}
\put(14.38,17.16){\line(2,-5){0.254}}
\put(14.63,16.53){\line(0,-1){2.26}}
\put(14.63,14.28){\line(-2,-5){0.162}}
\put(14.47,13.88){\line(-3,-2){0.936}}
\put(13.53,13.25){\line(-6,-1){1.86}}
\put(11.67,12.94){\line(-1,0){2.39}}
\put(9.28,12.94){\line(-5,2){1.26}}
\put(8.03,13.45){\line(-1,1){0.974}}
\put(7.06,14.42){\line(-1,6){0.087}}
\put(6.96,15.87){\line(0,-1){0.907}}
\put(7.75,18.09){\line(-4,-5){0.481}}
\put(6.98,16.73){\line(0,-1){0.874}}
\put(8.93,18.58){\line(-5,-2){1.17}}
\put(12.89,15.68){$ \mathrm{C} $}
\put(0.582,17.91){\line(2,1){0.947}}
\put(1.52,18.38){\line(6,1){0.997}}
\put(2.54,18.55){\line(1,0){0.849}}
\put(5.30,15.35){$ \mathrm{P} $}
\put(6.01,15.91){$ \mathrm{h} $}
\put(7.05,16.59){$ \mathrm{e} $}
\put(7.45,16.26){$ \mathrm{l} $}
\put(6.69,14.61){$ \mathrm{k} $}
\put(12.24,16.01){$ \mathrm{T} $}
\put(8.93,18.58){\line(1,0){2.57}}
\put(12.38,15.83){\line(-4,3){0.377}}
\put(12.00,16.11){\line(-2,5){0.280}}
\put(11.72,16.81){\line(-1,1){0.381}}
\put(14.22,15.22){\line(-2,-5){0.192}}
\put(0.039,16.98){\line(3,5){0.553}}
\put(3.04,15.84){\circle{0.876}}
\put(2.27,15.35){$ \mathrm{Q} $}
\put(2.27,16.01){$ \mathrm{Q'} $}
\put(2.60,15.83){\line(-1,0){0.282}}
\put(3.46,15.8){\line(1,0){0.282}}
\put(2.93,15.7){$ \mathrm{B} $}
\put(3.75,15.77){\line(1,-1){0.397}}
\put(4.09,15.4){\line(1,-2){0.195}}
\put(4.26,15){\line(2,-1){0.558}}
\put(5.43,16.86){\line(3,-2){1.50}}
\put(5.43,16.86){\line(-5,2){1.00}}
\put(4.43,17.26){\line(-1,0){1.28}}
\put(2.32,15.83){\line(-4,3){0.362}}
\put(2.57,17.15){\line(5,1){0.576}}
\put(1.95,16.1){\line(-1,6){0.071}}
\put(1.88,16.53){\line(2,5){0.153}}
\put(2.04,16.91){\line(2,1){0.525}}
\put(13.76,15.8){\line(2,-1){0.405}}
\put(14.16,15.59){\line(1,-6){0.061}}
\put(10.91,15.83){\circle{0.892}}
\put(10.47,15.8){\line(-1,0){0.203}}
\put(9.91,15.93){\line(5,-2){0.359}}
\put(9.78,16.46){\line(1,-4){0.134}}
\put(9.34,16.77){\line(3,-2){0.475}}
\put(10.78,15.7){$ \mathrm{B} $}
\put(11.34,15.8){\line(1,0){0.203}}
\put(11.57,15.8){\line(5,-3){0.402}}
\put(12.00,15.56){\line(1,-5){0.071}}
\put(11.83,14.84){\line(2,3){0.265}}
\put(11.83,14.82){\line(-2,-1){0.544}}
\put(12.24,15.38){$ \mathrm{T'} $}
\put(6.59,15){\line(2,5){0.317}}
\put(14.30,15.88){$ \mathrm{S} $}
\put(14.73,15.9){$ \mathrm{S'} $}
\put(11.37,15.93){$ \mathrm{Q'} $}
\put(11.33,15.35){$ \mathrm{Q} $}
\put(10.10,15.96){$ \mathrm{U'} $}
\put(10.15,15.35){$ \mathrm{U} $}
\put(8.16,15.93){$ \mathrm{P'} $}
\put(8.19,15.4){$ \mathrm{P} $}
\put(7.62,15.98){$ \mathrm{g} $}
\put(7.58,15.37){$ \mathrm{b} $}
\put(7.10,14.56){$ \mathrm{d} $}
\put(6.30,14.94){$ \mathrm{f} $}
\put(6.04,15.44){$ \mathrm{a} $}
\put(7.28,17.51){\line(-2,-5){0.301}}
\put(5.32,15.94){$ \mathrm{P'} $}
\put(3.55,15.91){$ \mathrm{U'} $}
\put(3.53,15.29){$ \mathrm{U} $}
\put(1.42,15.98){$ \mathrm{T} $}
\put(1.45,15.22){$ \mathrm{T'} $}
\put(0.124,15.27){$ \mathrm{R'} $}
\put(6.74,16.59){$ \mathrm{j} $}
\put(6.17,16.46){$ \mathrm{c} $}
\put(7.42,14.91){$ \mathrm{i} $}
\put(13.52,15.88){$ \mathrm{R} $}
\put(13.54,15.32){$ \mathrm{R'} $}
\put(4.74,15.7){$ \mathrm{A} $}
\put(8.83,15.7){$ \mathrm{A} $}
\put(0.809,15.65){$ \mathrm{C} $}
\put(0.206,16.08){$ \mathrm{R} $}
\put(0.039,16.17){\line(3,-5){0.219}}
\put(0.487,15.81){\line(-1,0){0.203}}
\put(0.01,18.58){$ \mathrm{(i)} $}
\put(11.69,6.63){\line(-1,-3){0.647}}
\put(9.60,2.99){\line(-3,-1){2.42}}
\put(7.24,2.20){\line(-3,1){2.27}}
\put(4.85,10.55){\line(-4,-5){1.56}}
\put(3.30,8.58){\line(-1,-4){0.485}}
\put(2.82,6.63){\line(1,-4){0.493}}
\put(3.31,4.65){\line(1,-1){1.68}}
\put(9.24,10.42){\line(-3,-5){1.50}}
\put(7.00,11.02){\line(0,-1){3.19}}
\put(7.45,7.84){\line(0,1){3.19}}
\put(6.03,7.46){\line(-2,1){2.44}}
\put(3.13,6.87){\line(1,0){2.75}}
\put(3.09,6.41){\line(1,0){2.74}}
\put(3.46,4.98){\line(2,1){2.48}}
\put(7.45,9.80){\vector(0,-1){0.437}}
\put(7.00,9.31){\vector(0,1){0.387}}
\put(5.72,9.49){\vector(2,-3){0.279}}
\put(4.47,8.26){\vector(2,-1){0.493}}
\put(4.88,7.50){\vector(-2,1){0.442}}
\put(4.32,6.87){\vector(1,0){0.234}}
\put(4.55,6.41){\vector(-1,0){0.361}}
\put(5.11,5.30){\vector(-2,-1){0.453}}
\put(5.31,3.02){\line(3,5){1.35}}
\put(4.97,3.34){\line(3,5){1.38}}
\put(7.45,2.48){\line(0,1){2.82}}
\put(9.26,3.11){\line(-2,3){1.53}}
\put(10.74,4.61){\line(-2,1){2.49}}
\put(8.42,6.85){\line(1,0){3.00}}
\put(10.89,8.68){\line(-2,-1){2.60}}
\put(5.49,4.22){\vector(2,3){0.274}}
\put(6.99,3.51){\vector(0,1){0.672}}
\put(7.45,4.28){\vector(0,-1){0.437}}
\put(8.77,3.86){\vector(-2,3){0.420}}
\put(8.65,4.87){\vector(2,-3){0.275}}
\put(10.06,4.92){\vector(-2,1){0.702}}
\put(9.61,6.85){\vector(1,0){0.463}}
\put(10.41,8.02){\vector(-2,-1){0.773}}
\put(9.11,7.81){\vector(2,1){0.651}}
\put(10.56,8.10){\line(2,1){0.642}}
\put(9.39,9.85){\line(3,5){0.306}}
\put(8.68,10.61){\vector(-3,1){0.417}}
\put(7.22,11.32){\line(3,-1){2.38}}
\put(7.45,10.56){\line(0,1){0.648}}
\put(5.35,10.13){\line(-3,5){0.294}}
\put(4.09,8.46){\line(-2,1){0.649}}
\put(3.56,6.87){\line(-1,0){0.698}}
\put(3.61,6.41){\line(-1,0){0.764}}
\put(7.45,2.68){\line(0,-1){0.446}}
\put(4.14,4.81){\line(-2,-1){0.688}}
\put(5.63,3.57){\line(-3,-5){0.420}}
\put(11.06,4.69){\line(-5,-6){1.40}}
\put(9.22,4.02){\line(2,-3){0.547}}
\put(4.37,4.93){\line(2,1){0.513}}
\put(4.42,5.47){\vector(2,1){0.371}}
\put(2.94,7.14){\line(3,4){0.312}}
\put(7.19,11.32){\line(-3,-1){2.37}}
\put(7.00,7.85){\line(2,-5){0.202}}
\put(7.45,7.84){\line(-1,-2){0.253}}
\put(3.61,8.95){\line(2,1){0.552}}
\put(2.95,6.19){\line(5,-6){0.317}}
\put(11.29,8.20){\line(-1,-5){0.104}}
\put(5.46,2.78){\line(5,1){0.418}}
\put(3.69,4.33){\line(5,-1){0.451}}
\put(5.32,10.68){\line(5,-1){0.455}}
\put(5.77,10.59){\line(3,1){1.19}}
\put(6.72,10.89){\vector(-3,-1){0.515}}
\put(4.16,9.22){\line(3,4){0.675}}
\put(4.71,9.97){\vector(-3,-4){0.325}}
\put(3.25,7.56){\line(1,4){0.167}}
\put(3.42,8.24){\vector(-1,-4){0.167}}
\put(9.97,9.67){\line(5,-6){0.833}}
\put(3.46,4.98){\line(-1,4){0.202}}
\put(3.26,5.81){\vector(1,-4){0.129}}
\put(4.19,4.17){\vector(1,-1){0.399}}
\put(4.19,4.17){\line(1,-1){0.399}}
\put(4.13,4.24){\line(1,-1){0.874}}
\put(5.89,2.89){\line(3,-1){0.902}}
\put(6.99,2.51){\line(0,1){2.87}}
\put(5.89,2.89){\vector(3,-1){0.681}}
\put(7.68,2.38){\line(5,4){0.480}}
\put(8.25,2.80){\vector(3,1){0.493}}
\put(8.15,2.76){\line(3,1){1.10}}
\put(7.73,5.44){\line(-1,6){0.079}}
\put(9.90,3.39){\line(1,3){0.157}}
\put(10.05,3.87){\line(5,6){0.630}}
\put(10.12,3.95){\vector(3,4){0.304}}
\put(11.14,5.71){\line(1,3){0.237}}
\put(11.24,5.14){\line(-1,5){0.105}}
\put(11.14,5.68){\line(1,3){0.254}}
\put(11.14,5.71){\vector(1,3){0.168}}
\put(11.42,6.85){\line(-1,4){0.233}}
\put(11.42,6.85){\vector(-1,4){0.178}}
\put(6.50,5.85){\line(2,1){0.371}}
\put(6.86,5.98){\line(0,-1){0.437}}
\put(6.04,5.16){\line(3,5){0.450}}
\put(7.45,5.44){\line(-1,2){0.228}}
\put(6.99,5.50){\line(3,5){0.234}}
\put(6.99,4.39){\line(0,1){1.09}}
\put(7.45,4.28){\line(0,1){1.14}}
\put(6.18,5.89){\line(3,4){0.326}}
\put(5.80,5.66){\line(5,3){0.390}}
\put(5.93,6.84){\line(5,-2){0.533}}
\put(5.85,6.41){\line(5,2){0.589}}
\put(5.48,6.87){\line(1,0){0.451}}
\put(5.77,6.41){\line(1,0){0.076}}
\put(6.14,7.38){\line(3,-4){0.318}}
\put(6.76,7.80){\line(1,-6){0.091}}
\put(7.88,6.95){\line(6,1){0.535}}
\put(7.69,7.78){\line(-1,-6){0.085}}
\put(8.08,7.64){\line(-4,-3){0.466}}
\put(8.19,8.64){\line(-3,-5){0.511}}
\put(8.44,8.28){\line(-3,-5){0.379}}
\put(9.31,5.81){\vector(2,-1){0.513}}
\put(10.65,5.13){\line(2,-1){0.473}}
\put(8.25,5.85){\line(-1,1){0.400}}
\put(5.94,6.98){\line(1,0){0.569}}
\put(5.08,7.41){\line(2,-1){0.830}}
\put(3.41,8.26){\line(2,-1){2.45}}
\put(8.29,7.39){\line(-1,-1){0.425}}
\put(8.87,7.21){\line(-5,-2){0.436}}
\put(9.51,10.53){\line(-3,-5){0.245}}
\put(9.16,9.89){\line(-3,-5){0.262}}
\put(8.75,9.24){\line(-1,-2){0.233}}
\put(8.40,8.61){\line(-3,-5){0.274}}
\put(8.07,8.02){\line(-2,-3){0.290}}
\put(5.23,3.34){\line(-1,-2){0.203}}
\put(5.61,3.96){\line(-2,-3){0.259}}
\put(5.99,4.62){\line(-3,-5){0.262}}
\put(6.29,5.18){\line(-1,-2){0.188}}
\put(10.79,8.67){\vector(-3,4){0.528}}
\put(8.13,7.75){\line(3,5){1.45}}
\put(9.13,9.38){\vector(-2,-3){0.460}}
\put(8.16,8.66){\vector(2,3){0.364}}
\put(2.86,6.66){\line(1,0){0.334}}
\put(3.47,6.64){\line(1,0){0.410}}
\put(4.18,6.64){\line(1,0){0.487}}
\put(4.91,6.64){\line(1,0){0.435}}
\put(5.57,6.64){\line(1,0){0.385}}
\put(11.26,6.64){\line(1,0){0.435}}
\put(10.63,6.61){\line(1,0){0.385}}
\put(4.53,7.95){\line(2,-1){0.380}}
\put(5.72,7.36){\line(2,-1){0.380}}
\put(9.27,5.56){\line(2,-1){0.410}}
\put(8.66,5.90){\line(2,-1){0.390}}
\put(8.12,6.16){\line(2,-1){0.380}}
\put(4.86,10.53){\line(3,-5){0.232}}
\put(5.17,9.98){\line(2,-3){0.238}}
\put(5.93,8.71){\line(3,-5){0.244}}
\put(6.33,8.03){\line(3,-5){0.232}}
\put(9.62,3.44){\line(-2,3){1.53}}
\put(9.37,3.37){\line(3,-5){0.232}}
\put(8.63,4.54){\line(3,-5){0.244}}
\put(8.23,5.07){\line(3,-5){0.250}}
\put(7.84,5.67){\line(3,-5){0.261}}
\put(3.31,4.64){\line(2,1){0.390}}
\put(3.90,4.94){\line(2,1){0.359}}
\put(4.48,5.22){\line(2,1){0.430}}
\put(5.17,5.60){\line(5,3){0.355}}
\put(5.75,5.90){\line(2,1){0.410}}
\put(8.40,7.22){\line(2,1){0.359}}
\put(7.21,11.30){\line(0,-1){0.385}}
\put(7.21,10.71){\line(0,-1){0.461}}
\put(7.24,10.00){\line(0,-1){0.461}}
\put(7.21,9.22){\line(0,-1){0.461}}
\put(7.21,8.43){\line(0,-1){0.461}}
\put(7.21,2.59){\line(0,-1){0.359}}
\put(7.21,3.28){\line(0,-1){0.410}}
\put(7.21,3.93){\line(0,-1){0.411}}
\put(7.21,4.61){\line(0,-1){0.435}}
\put(7.21,5.33){\line(0,-1){0.385}}
\put(8.68,6.64){\line(1,0){0.359}}
\put(9.27,6.64){\line(1,0){0.385}}
\put(9.87,6.61){\line(1,0){0.411}}
\put(11.39,6.43){\line(-1,0){2.93}}
\put(10.73,6.44){\vector(-1,0){0.925}}
\put(8.95,5.96){\line(-2,1){0.578}}
\put(11.01,4.95){\line(-2,1){2.51}}
\put(9.82,10.15){\line(1,-3){0.163}}
\put(5.99,6.27){\line(1,0){0.499}}
\put(3.85,5.17){\line(2,1){2.17}}
\put(5.72,7.60){\line(2,-1){0.453}}
\put(8.37,6.26){\line(-1,0){0.468}}
\put(6.49,4.90){\vector(-2,-3){0.550}}
\put(1.78,11.03){$ \mathrm{(ii)} $}
\put(11.38,8.58){$ \mathrm{b} $}
\put(10.47,9.58){$ \mathrm{Q} $}
\put(11.88,6.52){$ \mathrm{a} $}
\put(9.01,3.95){\line(3,-5){0.250}}
\put(11.68,7.49){$ \mathrm{P} $}
\put(9.61,10.65){$ \mathrm{c} $}
\put(8.41,11.08){$ \mathrm{R} $}
\put(7.09,11.46){$ \mathrm{d} $}
\put(5.83,11.13){$ \mathrm{S} $}
\put(4.59,10.67){$ \mathrm{e} $}
\put(3.80,9.73){$ \mathrm{T} $}
\put(3.01,8.58){$ \mathrm{f} $}
\put(2.61,7.59){$ \mathrm{U} $}
\put(2.41,6.54){$ \mathrm{g} $}
\put(2.56,5.35){$ \mathrm{P'} $}
\put(2.94,4.40){$ \mathrm{h} $}
\put(3.72,3.38){$ \mathrm{Q'} $}
\put(4.82,2.49){$ \mathrm{i} $}
\put(5.91,2.03){$ \mathrm{R'} $}
\put(7.17,1.73){$ \mathrm{j} $}
\put(8.49,2.16){$ \mathrm{S'} $}
\put(10.55,3.51){$ \mathrm{T'} $}
\put(9.73,2.51){$ \mathrm{k} $}
\put(11.22,4.37){$ \mathrm{l} $}
\put(11.58,5.42){$ \mathrm{U'} $}
\put(7.67,5.93){\line(2,-1){0.460}}
\put(6.44,7.48){\line(2,-1){0.424}}
\put(5.52,9.40){\line(3,-5){0.261}}
\put(5.94,8.28){\vector(-2,3){0.520}}
\put(4.85,10.09){\line(3,-5){1.51}}
\put(6.05,8.08){\line(3,-5){0.365}}
\put(9.96,5.23){\line(2,-1){0.390}}
\put(10.59,4.93){\line(2,-1){0.458}}
\put(5.12,7.70){\line(2,-1){0.410}}
\put(3.31,8.57){\line(2,-1){0.407}}
\put(3.93,8.28){\line(5,-3){0.403}}
\put(9.65,7.87){\line(5,2){0.363}}
\put(6.43,2.71){\line(5,-2){0.603}}
\put(6.99,4.39){\line(0,-1){1.91}}
\put(7.67,11.17){\line(6,-5){0.466}}
\put(9.24,10.42){\line(-3,1){1.11}}
\put(9.07,7.56){\line(2,1){0.380}}
\put(10.18,8.12){\line(2,1){0.386}}
\put(10.76,8.45){\line(2,1){0.376}}
\put(9.57,10.53){\line(5,-6){1.60}}
\put(11.18,8.61){\line(1,-4){0.490}}
\put(11.04,8.33){\line(-2,-1){2.50}}
\put(6.72,5.78){\line(-3,-5){0.233}}
\put(6.38,4.80){\line(3,5){0.488}}
\put(6.66,7.96){\line(-3,5){1.46}}
\put(6.44,8.33){\line(3,-5){0.342}}
\put(3.67,4.59){\line(2,1){2.41}}
\end{picture}
\caption{(i) A sphere with three handles, cut so as to transform it into a
twelve-sided polygon with opposite sides identified.  (ii) A multi-hairpin
Wilson line for a sphere with three handles.  Opposite sides of the polygon,
for example $ \mathrm{P} $ and  $ \mathrm{P'} $, are identified.}
\label{figure for Dirac quantization condition}
\end{figure}

Considering, now, the case where the intrinsic topology of the two-dimensional
surface is a sphere with $n$ handles, $n \geq 1$, it will be sufficient to
show that we can always find a suitable branched hairpin, that divides the
surface into suitable sectors, so that we can use the same arguments as above.
We note, first, that we can always cut a sphere with $n$ handles, $n \geq 1$,
in such as way as to transform it into a polygon with $4 n$ sides, such that
opposite sides are identified.  Figure \ref{figure for Dirac quantization
condition} (i) shows a way of doing this for $n = 3$, that extends directly to
all $n \geq 1$.  In this diagram, paired circles $AA$, $BB$, and $CC$ are
identified by reflection in the vertical midline of the diagram, to form
handles, and the remaining lines are the cuts.  Figure \ref{figure for Dirac
quantization condition} (ii) shows a branched hairpin dividing the sphere with
three handles into twelve triangular regions, which we can assume correspond
to the main parts of the coordinate patches on the two-dimensional surface in
this case.  We make the transitions between the coordinate patches, such that
the three sections of the Wilson line directly surrounding each triangle, are
on the coordinate patch corresponding to that triangle.

The individual branches of the hairpin all branch out of the Wilson line at a
single point, which is the central point of Figure \ref{figure for Dirac
quantization condition} (i), and corresponds to all twelve vertices of the
polygon in Figure \ref{figure for Dirac quantization condition} (ii).  Six of
the $6 + 12 - 1 = 17$ hairpins that branch out of this point loop round and
meet this point again at their tips.  These are the hairpins $PP'$, $QQ'$,
$RR'$, $SS'$, $TT'$, and $UU'$, along the edges of the polygon in Figure
\ref{figure for Dirac quantization condition} (ii).  The Wilson line starts
and ends at a different point, corresponding to the centre of the polygon in
Figure \ref{figure for Dirac quantization condition} (ii), which could be any
other point of the sphere with three handles shown in Figure \ref{figure for
Dirac quantization condition} (i), and the remaining $12 - 1 = 11$ hairpins,
which are the hairpins running from vertices $b$ to $l$ of the polygon in
Figure \ref{figure for Dirac quantization condition} (ii), to the centre of
that polygon, also loop round to meet that point at their tips.  These
hairpins reach that point in Figure \ref{figure for Dirac quantization
condition} (i), by passing along the handles, as necessary.  For example, if
the Wilson line starts and ends at a point somewhere in the external region of
Figure \ref{figure for Dirac quantization condition} (i), the hairpin that
runs from vertex $b$ of the polygon in Figure \ref{figure for Dirac
quantization condition} (ii), to the centre of that polygon, reaches that
point from $b$ in Figure \ref{figure for Dirac quantization condition} (i), by
first passing along handle $A$, then along handle $B$, and finally along
handle $C$.

If we label a hairpin that runs from a vertex of the polygon in Figure
\ref{figure for Dirac quantization condition} (ii) to the centre of that
polygon, by the letter of the corresponding vertex, then after the initial
section from the centre of the polygon to vertex $a$, the Wilson line runs
along the hairpins in the sequence $PP', h, c, RR', j, e, TT', l, g, b, QQ',
i, d, SS', k, f, UU'$, then finally along the final section from vertex $a$
back to the centre of the polygon.  We see that each transition, from one
coordinate patch to another, that occurs across a side of the polygon in
Figure \ref{figure for Dirac quantization condition} (ii), is matched by a
reverse transition through the same point, so that all the phase factors
associated with these transitions cancel out.  While for the transitions at
the centre of the polygon in Figure \ref{figure for Dirac quantization
condition} (ii), we see that, since the Wilson line must end with a transition
back to the coordinate patch it started on, we have transitions corresponding
to diagonal matrices $U^{1 \rightarrow 2}, U^{2 \rightarrow 3}, \ldots, U^{12
\rightarrow 1}$, all at the same point, where the patches are labelled 1 to 12
anticlockwise around the polygon, and the product of all these is equal to
$1$.  Furthermore, each of the twelve triangular regions is circled
anticlockwise by the Wilson line sections around its edge, which are the
sections of the Wilson line on the coordinate patch corresponding to that
triangle, so we can use Stokes's theorem for each triangle.

Considering, now, how this works for general $n \geq 1$, we draw the
corresponding $4 n$-sided polygon with an opposite pair of its vertices
pointing east and west.  We draw a $T$, consisting of the initial and final
sections of the Wilson line, and two half hairpins, with the centre of its top
at the easternmost vertex, as in Figure \ref{figure for Dirac quantization
condition} (ii).  And for each of the remaining $2 n - 1$ sides of the upper
half of the polygon, we draw an $L$, consisting of one and a half hairpins,
with the foot of the $L$ pointing anticlockwise as in Figure \ref{figure for
Dirac quantization condition} (ii).  And for each of the remaining $2 n - 1$
sides of the lower half of the polygon, we draw an $L$, consisting of one and
a half hairpins, with the foot of the $L$ pointing clockwise, as in Figure
\ref{figure for Dirac quantization condition} (ii).  And finally we draw an
$I$, consisting of a single hairpin, with its foot at the westernmost vertex,
as in Figure \ref{figure for Dirac quantization condition} (ii).  We draw an
arrow pointing anticlockwise on every Wilson line section running along an
edge of the polygon, as in Figure \ref{figure for Dirac quantization
condition} (ii), and add arrows to the Wilson line sections directly joined to
these sections, consistent with these arrows, so that every triangular section
is circled anticlockwise by the three Wilson line sections around its edge.

The cancellation of the phase factors associated with the transitions between
coordinate patches, and the use of Stokes's theorem, will now work exactly as
for the $n = 3$ case, so it remains to check that, starting at the start of
the Wilson line, we pass along each Wilson line section exactly once, and in
the correct direction.  To check this, we number the Wilson line sections
running along the perimeter of the top half of the polygon $0, 1, \ldots,
\left( 2 n - 1 \right)$ in sequence anticlockwise, starting at the easternmost
section, which is half the top of the $T$, and labelled $P$ in Figure
\ref{figure for Dirac quantization condition} (ii).  And we number the Wilson
line sections running along the perimeter of the lower half of the polygon
$0', 1', \ldots, \left( 2 n - 1 \right)'$ in sequence anticlockwise, starting
at the westernmost section, which is labelled $P'$ in Figure \ref{figure for
Dirac quantization condition} (ii).  Thus the $L$'s in the top half of the
polygon are numbered $1, 2, \ldots \left( 2 n - 1 \right)$, and the $L$'s in
the lower half of the polygon are numbered $0', 1', \ldots, \left( 2 n - 2
\right)'$.

We observe that, due to the directions of the arrows on the Wilson line
sections, each pair of opposite $L$'s of the form $mm'$, $1 \leq m \leq \left(
2 n - 2 \right)$, is traversed in the sequence: first $m$, then $m'$.
Furthermore, the upper half of the top of the $T$, labelled $P$ in Figure
\ref{figure for Dirac quantization condition} (ii), and $0$ in the general
numbering scheme, is traversed immediately after the initial section of the
Wilson line, and immediately before the $L$ labelled $0'$, which is labelled
$P'$ in Figure \ref{figure for Dirac quantization condition} (ii), and the
lower half of the top of the $T$, labelled $U'$ in Figure \ref{figure for
Dirac quantization condition} (ii), and $\left( 2 n - 1 \right)'$ in the
general numbering scheme, is traversed immediately after the $L$ labelled
$\left( 2 n - 1 \right)$, which is labelled $U$ in Figure \ref{figure for
Dirac quantization condition} (ii), and immediately before the final section
of the Wilson line.  Furthermore, the hairpin based at the westernmost vertex
of the polygon, labelled $g$ in Figure \ref{figure for Dirac quantization
condition} (ii), is traversed immediately after the $L$ labelled $\left( 2 n -
2 \right)'$, which is labelled $T'$ in Figure \ref{figure for Dirac
quantization condition} (ii), and immediately before the $L$ labelled $1$,
which is labelled $Q$ in Figure \ref{figure for Dirac quantization condition}
(ii).  And finally, for $0 \leq m \leq \left( 2 n - 3 \right)$, $L$ number
$m'$, in the lower half of the polygon, is immediately followed by $L$ number
$\left( m + 2 \right)$, in the upper half of the polygon.

Thus the $4 n$ Wilson line sections running along the perimeter of the
polygon, and the Wilson line sections directly connected to them in the
diagram, and the hairpin based at the westernmost vertex of the polygon, which
together comprise the $4 n + 1$ pieces of Wilson line that are directly
connected in the diagram, are traversed in the sequence: $0, 0', 2, 2', 4, 4',
\ldots, \left( 2 n - 2 \right), \left( 2 n - 2 \right)'$, then the hairpin
based at the westernmost vertex of the polygon, then $1, 1', 3, 3', 5, 5',
\ldots, \left( 2 n - 1 \right), \left( 2 n - 1 \right)'$.

If there is just one coordinate patch, as is natural when a compact hyperbolic
manifold is specified by giving a Dirichlet domain for it in uncompactified
hyperbolic space, together with the face-pairing maps for the Dirichlet
domain, a simpler tree of hairpins can be obtained from the one shown in
Figure \ref{figure for Dirac quantization condition} (ii), by moving the start
and end point to just inside the 12-sided polygon at $a$, and shrinking the
eleven hairpins that meet at the
centre of the polygon, back to the perimeter of the polygon, so that
all that remains are the hairpin halves around the perimeter of the polygon,
which are traversed in the sequence $P P' R R' T T' Q Q' S S' U U'$.

Thus we have shown that if the configuration of the Yang-Mills fields lies
entirely within the Cartan subalgebra of the gauge group, then for an
arbitrary representation of the gauge group, with generators $T_{\alpha}$,
such that matter fields exist that transform under that representation of the
gauge group, and for each matrix element on the leading diagonal of that
representation of the gauge group, and for an arbitrary closed orientable
two-dimensional surface embedded smoothly in the manifold, the integral $g
\int F$ must be an integer multiple of $2 \pi$, where $\int F$ denotes the
integral over the closed two-dimensional surface, of the two-form field
strength of the corresponding diagonal matrix element of $A_{\mu \alpha}
T_{\alpha}$.  And we noted that this integral will be non-zero, if the field
configuration includes a Hodge - de Rham harmonic two-form, with non-zero
coefficient in that matrix element, that has non-zero integral over that
surface.

Now for the fundamental / adjoint representation of $E_8$, in the $\mathrm{SU}
\left( 9 \right)$ basis used in this section, each of the eight generators of
the Cartan subalgebra of $E_8$, which are the eight generators of the Cartan
subalgebra of $\mathrm{SU} \left( 9 \right)$, in a reducible representation of
$\mathrm{SU} \left( 9 \right)$ that comprises the \textbf{80}, \textbf{84}, and
$\mathbf{\overline{84}}$ of $\mathrm{SU} \left( 9 \right)$, is such that its
nonzero matrix elements are integer multiples of an overall normalization
factor, specific to that generator.  Let us now consider $A_{\mu \alpha}$ such
that $\alpha$ denotes a fixed one of the eight generators of the Cartan
subalgebra of $E_8$.  Let $B_2$ denote the second Betti number of the
manifold, which by assumption is $> 0$.  Then there are $B_2$ linearly
independent Hodge - de Rham harmonic two-forms, and there are also just $B_2$
non-contractible closed two-dimensional surfaces in the manifold, that are
linearly independent in the sense of homology.  Thus we can choose a basis of
$B_2$ non-contractible closed two-dimensional surfaces in the manifold, such
that for an arbitrary closed two-form $F$, or in other words, for an arbitrary
two-form $F$ that satisfies the Bianchi identity $dF = 0$, or in components,
$\partial_{\left[ \mu \right.} F_{\left. \nu \sigma \right]} = 0$, and an
arbitrary closed two-dimensional surface in the manifold, the integral $\int
F$, of $F$ over the surface, is equal to a linear combination of the
corresponding integrals for the $B_2$ surfaces in the basis.

Thus if we consider one particular matrix element in the diagonal of $T_{\mu
\alpha}$, for the particular $\alpha$ in the Cartan subalgebra we are
considering, and restrict $A_{\mu \alpha}$ to be a linear combination of the
one-form vector potentials of the $B_2$ harmonic two-forms, so that the
Yang-Mills field equations will automatically be satisfied, for this Abelian
field configuration, there are just $B_2$ linearly independent quantization
conditions, to be satisfied by $B_2$ independent coefficients.  And if we now
extend the consideration to all 248 matrix elements on the leading diagonal of
$T_{\mu \alpha}$, but still for the fixed value of $\alpha$ in the Cartan
subalgebra, we see that, because the ratios of the matrix elements are fixed
rational numbers, there will be a finite integer $p$, such that if $A_{\mu
\alpha}$ satisfies the quantization condition for one particular matrix
element on the diagonal, such that that matrix element of $T_{\mu \alpha}$ is
nonzero, then $pA_{\mu \alpha}$ will satisfy the quantization condition for
all the nonzero matrix elements of the diagonal matrix $T_{\mu \alpha}$.

Thus, still considering $A_{\mu \alpha}$ for just one fixed value of $\alpha$
in the Cartan subalgebra, the quantization condition can be satisfied by an
infinite number of non-trivial field configurations, and for field
configurations that satisfy the Yang-Mills field equations, and are thus a
linear combination of the $B_2$ Hodge - de Rham harmonic two-forms, the
allowed values of the coefficients of the $B_2$ Hodge - de Rham harmonic
two-forms will lie on a discrete $B_2$-dimensional lattice, because $B_2$
linearly independent linear combinations of the coefficients have quantized
values, so that after a suitable change of basis, each coefficient would be
quantized independently.  And when we choose such a basis for the $B_2$ Hodge
- de Rham harmonic two-forms, so that we can consider each Hodge - de Rham
harmonic two-form in the basis independently, the allowed values of the
$A_{\mu \alpha}$, associated with any one Hodge - de Rham harmonic two form,
will be integer multiples of a basic ``monopole''.

Let us now choose such a basis for the $B_2$ Hodge - de Rham harmonic
two-forms, and consider one Hodge - de Rham harmonic two-form in the basis, so
that the allowed values of $A_{\mu \alpha}$ will be integer multiples of a
basic ``monopole''.  We now allow $A_{\mu \alpha}$ to be nonzero for all the
eight values of $\alpha$ in the Cartan subalgebra.  Then the solutions of the
quantization condition will include, in particular, a discrete
eight-dimensional lattice, in the Cartan subalgebra of $E_8$, whose lattice
points correspond to Yang-Mills fields of the form of the sum over the Cartan
subalgebra of $q_{\alpha} A_{\mu \alpha} T_{\alpha}$, where the eight
$q_{\alpha}$ are the integers that define the lattice point, and $A_{\mu
\alpha} \left( x \right)$ is the Hodge - de Rham harmonic two-form under
consideration, times a normalization factor, dependent on $\alpha$, that makes
it into the correspnonding basic ``monopole'', for the element $\alpha$ of the
Cartan subalgebra.  There may now be additional solutions of the quantization
conditions, such that some or all of the $q_{\alpha}$ are non-integer rational
numbers, but the number of such additional solutions, in each unit cell of the
lattice defined by integer $q_{\alpha}$, will be finite, since 248 linear
combinations of the eight $q_{\alpha}$, not necessarily all distinct, are
required to satisfy quantization conditions, which are, however, mutually
consistent, and among these 248 linear combinations, there are eight that are
linearly independent.  Furthermore, given a point in the eight-dimensional
space of the $q_{\alpha}$, that satisfies all the quantization conditions, and
such that not all eight of the $q_{\alpha}$ are integers, other non-integer
solutions of the quantization conditions can be obtained by adding arbitrary
integers to the $q_{\alpha}$.  Thus the solutions of the quantization
conditions form an infinite discrete lattice in the space of the $q_{\alpha}$,
which is, however, not necessarily hypercubic.

Thus, for each separate Hodge - de Rham harmonic two-form, in a basis in which
we can apply the quantization conditions to each separate Hodge - de Rham
harmonic two-form independently, we can have Abelian configurations of the $E
8$ Yang-Mills fields, that solve the classical Yang-Mills field equations,
and, within the Cartan subalgebra, are topologically stabilized, and whose
field strengths have the spatial dependence of the Hodge - de Rham harmonic
two form, and an embedding within $E_8$, that lies on any lattice point of an
infinite eight-dimensional lattice in the eight-dimensional Cartan subalgebra
of $E_8$.  Thus, provided the different lattice points are not connected to
one another by orbits within $E_8$ that go outside the Cartan subalgebra, we
can break $E_8$ to a wide variety of subgroups, in a topologically stabilized
manner, by introducing such Hodge - de Rham harmonic two-forms in the vacuum,
embedded in $E_8$ on suitable lattice points of this infinite
eight-dimensional lattice in the eight-dimensional Cartan subalgebra of $E_8$.

Furthermore, for breakings to smaller subgroups of $E_8$, such as the
subgroups $\mathrm{SU} \left( 3 \right) \times \left( \mathrm{SU} \left( 2 \right)
\right)^3 \times \left( \mathrm{U}
\left( 1 \right) \right)^3$, $\mathrm{SU} \left( 3
\right) \times \left( \mathrm{SU} \left( 2 \right) \right)^2 \times \left(
\mathrm{U}
\left( 1 \right) \right)^4$, and $\mathrm{SU} \left( 3 \right) \times \mathrm{SU}
\left( 2 \right) \times \left( \mathrm{U}
\left( 1 \right) \right)^5$, considered in
this paper, there is a multi-dimensional space of embeddings in the Cartan
subalgebra of $E_8$, that achieve the required breaking.  Thus we can choose a
different embedding, consistent with the required breaking, for each
independent Hodge - de Rham harmonic two-form, subject to the requirement of
satisfying Witten's topological constraint {\cite{Witten Constraints on
compactification}}, as discussed in subsection \ref{Wittens topological
constraint}, and thus seek to satisfy the conditions 3. (a) to (g), in the
list above.  Specifically, for breaking to  $\mathrm{SU} \left( 3 \right) \times
\left( \mathrm{SU} \left( 2 \right) \right)^3 \times \left( \mathrm{U}
\left( 1 \right)
\right)^3$, the space of embeddings that achieve the required breaking is
three-dimensional, while for breaking to $\mathrm{SU} \left( 3 \right) \times
\left( \mathrm{SU} \left( 2 \right) \right)^2 \times \left( \mathrm{U}
\left( 1 \right)
\right)^4$, it is four-dimensional, and for breaking to $\mathrm{SU} \left( 3
\right) \times \mathrm{SU} \left( 2 \right) \times \left( \mathrm{U}
\left( 1 \right)
\right)^5$, it is five-dimensional.  However, in each case, we also need to
ensure that the embeddings of all the monopoles are perpendicular to $U \left(
1 \right)_Y$, so that $U \left( 1 \right)_Y$ is not broken by Witten's Higgs
mechanism involving the components $C_{ABy}$ of the three-form gauge field
{\cite{Witten Constraints on compactification}}, which reduces the
dimensionalities of the spaces of available embeddings to two, three, and
four, respectively.  And if we want to make the unwanted $U \left( 1
\right)$'s massive by Witten's Higgs mechanism, rather than by monopoles
outside the Cartan subalgebra, we also have to ensure that the embeddings of
at least some of the monopoles are not perpendicular to the unwanted $U \left(
1 \right)$'s.

We can ensure that we really do get the expected multi-dimensional lattice of
embeddings within the $E_8$ Cartan subalgebra, consistent with the required
breaking, by choosing a basis for the Cartan subalgebra such that a certain
subset of the generators automatically preserve the required subgroup.  For
example, the subgroup $\mathrm{SU} \left( 3 \right) \times \left( \mathrm{SU}
\left( 2 \right) \right)^3 \times \left( \mathrm{U}
\left( 1 \right) \right)^3$ is
preserved by an arbitrary element of the Cartan subalgebra of $E_8$, which in
the basis used in the present section, is also the Cartan subalgebra of
$\mathrm{SU} \left( 9 \right)$, whose diagonal matrix elements, in the
$\mathrm{SU} \left( 9 \right)$ fundamental, are $\left( \sigma_1, \sigma_1,
\sigma_1, \sigma_2, \sigma_2, \sigma_3, \sigma_3, \sigma_4, \sigma_4 \right)$,
with $3 \sigma_1 + 2 \sigma_2 + 2 \sigma_3 + 2 \sigma_4 = 0$.  Of course, for
certain values of the $\sigma_i$, a larger subgroup is preserved.  For
example, $\left( \sigma_1, \sigma_2, \sigma_3, \sigma_4 \right) = \left( 2, -
1, - 1, - 1 \right)$ preserves $E 7$, $\left( 0, 0, 1, - 1 \right)$ preserves
$\mathrm{SO} \left( 14 \right)$, $\left( 4, 4, - 5, - 5 \right)$ preserves
$\mathrm{SU} \left( 5 \right) \times \mathrm{SU} \left( 4 \right)$, $\left( 2, 2,
2, - 7 \right)$ preserves $\mathrm{SU} \left( 7 \right) \times \mathrm{SU} \left(
2 \right)$, $\left( 2, 2, - 1, - 4 \right)$ preserves $E 6 \times \mathrm{SU}
\left( 2 \right)$, $\left( 4, - 5, 1, - 2 \right)$ preserves $\mathrm{SO} \left(
10 \right) \times \mathrm{SU} \left( 3 \right)$, and $\left( 0, 1, - 2, 1
\right)$ preserves $\mathrm{SU} \left( 3 \right) \times \mathrm{SO} \left( 10
\right)$.  However, for most of the points of the lattice, which in this case
is three-dimensional, the required breaking is obtained.

This partial topological stabilization no longer applies for continuous
variations of the gauge field configuration that are allowed to go outside the
Cartan subalgebra. \ For a generic path that has the same start and end point
as the tree of hairpins, and is homotopic to the tree of hairpins, the
logarithm $F\equiv\mathrm{\ln} W$, of the Wilson phase factor $W$ for the
path, can generically be defined by continuity under
continuous variations of the path and of the gauge field configuration. \ For
an assumed small variation $\delta$ of $\mathrm{\ln} W$, we have:
\[ W' = e^{i (F + \delta)} = e^{iF} + \int_0^1 dse^{iFs} i \delta e^{iF (1 -
   s)} + \ldots = \]
\begin{equation}
  \label{variation of ln W} = e^{iF} + i \left( \left( \frac{e^{i \breve{F}} -
  1}{i \breve{F}} \right)_{\beta \gamma} \delta_{\gamma} \right) T_{\beta}
  e^{iF} + \ldots,
\end{equation}
where $\breve{F}_{\beta \gamma} = F_{\alpha} ( \breve{T}_{\alpha})_{\beta
\gamma} = - iF_{\alpha} f_{\alpha \beta \gamma}$ is in the adjoint
representation, and the Baker-Campbell-Hausdorff formula has been used as in
the derivation of (\ref{finite gauge transformation of Yang Mills field}). \
Thus $\delta$ could fail to be determined by $W'$ if $\breve{F}$ has one or
more eigenvalues equal to non-zero multiples of $2 \pi$, and this must
inevitably happen for variations of a path that transform it into the tree of
hairpins, for we can transform the tree of hairpins continuously to the
trivial path by continuously retracting the hairpins. \ Thus $\mathrm{\ln} W$
is undefined for the tree of hairpins, for general variations of the gauge
field configuration that go outside the Cartan subalgebra.

The above discussion has involved only the components of the gauge field
tangential to a particular closed smooth orientable two-dimensional surface
$\mathcal{S}$ embedded in $\mathcal{M}^6$, and for this restricted system, the
question of the existence of any possible absolute topological stabilization
of a non-trivial configuration of the gauge field reduces to the corresponding
question for $\mathcal{S}$. \ From subsection 4.1 of {\cite{9204083 Witten}},
if the gauge group $G$ is connected and simply connected, then a $G$ bundle
over a two-dimensional surface is trivial. \ From {\cite{Wikipedia E8}}, the
compact Lie group $E_8$ is simply connected and appears also to be connected.
\ Furthermore, it has trivial centre, so it is not a covering group of any
other connected Lie group {\cite{9911217 Kubyshin}}. \ Thus the
Dirac-quantized harmonic 2-form monopoles considered in this subsection are
apparently not absolutely stabilized topologically, although they might be
separated by potential energy barriers from other solutions of the classical
Yang-Mills equations, including pure gauge configurations.

\subsection{Nonexistence of models where the Abelian Hodge - de Rham monopoles
break $E_8$ to $\mathrm{SU} \left( 3 \right) \times \left( \mathrm{SU} \left( 2
\right) \right)^3 \times \left( \mathrm{U} \left( 1 \right) \right)^3$}
\label{SU 3 cross SU 2 cubed cross U 1 cubed}

In models of TeV-scale gravity based on Ho\v{r}ava-Witten theory, such as those
considered in the present paper, the gauge couplings have to unify at around a
TeV, rather than
at $10^{16}$ GeV.  One way this could work is if the running of the
coupling constants somehow accelerates, so that the couplings run to
their conventional unification values at the TeV scale, rather than
at
$10^{16}$ GeV.  This possibility was studied by Dienes, Dudas, and
Gherghetta \cite{DDG1, DDG2}, and by Arkani-Hamed, Cohen, and Georgi
\cite{Arkani-Hamed Cohen Georgi}.

An alternative possibility, considered in \cite{NTSG}, is to embed
$\mathrm{SU}(3)\times
\mathrm{SU}(2)\times\mathrm{U}(1)$ into $ E8 $
in an unusual way, so that the values of the coupling constants, at
unification, are equal to their observed values, as evolved
conventionally to the TeV scale.  Usually the coupling constant of
a simple non-Abelian subgroup of a Grand Unification group, at
unification, is equal to the coupling constant of the Grand
Unification group, irrespective of how the subgroup is embedded
in the Grand Unification group.  An exception occurs \cite{Benakli,
Arkani-Hamed Cohen Georgi} if
the initial breaking of the Grand Unification group produces $N$
copies of the simple subgroup, and the $N$ copies of the
simple subgroup then break into their ``diagonal'' subgroup.  In
this case, after the second stage of the breaking, the coupling
constant of the ``diagonal'' subgroup is equal to
$\frac{1}{\sqrt{N}}$ times the coupling constant of the Grand
Unification group.  Effectively, the gauge field, in each of the
$N$ copies of the simple non-Abelian subgroup, becomes equal to
$\frac{1}{\sqrt{N}}$ times the ``diagonal'' gauge field, plus
massive vector terms that can be ignored at low energies.  The sum
of the $N$ copies of the Yang-Mills action, of the simple
non-Abelian subgroup, then becomes equal to the Yang-Mills action of
the ``diagonal'' subgroup, whose coupling constant is
$\frac{1}{\sqrt{N}}$ times the coupling constant of the Grand
Unification group.

Looking at the observed values of the reciprocals of the
$\mathrm{SU}(3)\times\mathrm{SU}(2)\times\mathrm{U}(1)$ fine
structure constants, at $M_Z$, normalized so as to meet at
unification in $\mathrm{SU}(5)$ Grand Unification, \cite{Georgi Glashow},
(Mohapatra \cite{Mohapatra}, page 22):
\begin{equation}\label{fine structure constant reciprocals}
\begin{array}{ccc}
\alpha_3^{-1}(M_Z) & = & 8.47\pm.22 \\
\alpha_2^{-1}(M_Z) & = & 29.61\pm.05 \\
\alpha_1^{-1}(M_Z) & = & 58.97\pm.05
\end{array}
\end{equation}
we see that they are quite close to being in the ratios 1, 3, 6.

If we evolve them in the Standard Model,
\cite{Rosner}, then $\alpha_3^{-1}$ and $\alpha_2^{-1}$ reach an
exact ratio of 1, 3, at 413 GeV, at which point $\alpha_3^{-1}$ is
equal to 10.12.  At this point, $\alpha_1^{-1}$ is equal to 58.00,
which is 4\% off being 6 times $\alpha_3^{-1}$, and $\sin^2 \theta_W
\simeq 0.239$.

Thus it is natural to consider the breaking of $\mathrm{E}8$ to
$\mathrm{SU}(3) \times \left(\mathrm{SU}(2)\right)^3
\times\mathrm{U}(1)_Y$,
followed by the breaking of $(\mathrm{SU}(2))^3$ to
$\mathrm{SU}(2)_{\mathrm{diag}}$, and seek an embedding of
$\mathrm{U}(1)_Y$ that gives the correct hypercharges at
unification.  I have summarized the required left-handed fermions of
the first generation, together with their hypercharges, $Y$,
\cite{Rosner}, the coefficients of their $\mathrm{U}(1)_Y$ couplings
in $\mathrm{SU}(5)$ Grand Unification, and the required coefficients
of their $\mathrm{U}(1)_Y$ couplings, in Table \ref{T1}.  Here I
have assumed that $\alpha_3^{-1}$ and $\alpha_1^{-1}$ are in the
ratio 1, 6, at unification, but it would be useful to study models
that achieve this within a few percent, since the correct form of
running to unification is not yet known.  Since the running of the
coupling constants is always by small amounts, the additional states
in these models, not yet observed experimentally, will not alter the
unification mass, or the value of the $\mathrm{SU}(3)$ coupling
constant at unification, which is equal to the $\mathrm{E}8$
coupling constant at unification, by a large amount.  Thus this
class of models generically predicts that the unification mass is
about a TeV, and the $\mathrm{E}8$ fine structure constant, at
unification, is about $\frac{1}{10}$.

\begin{table}
\begin{center}
\begin{tabular}{|c|c|c|c|c|}\hline
\multicolumn{5}{|c|}{First generation LH states} \\ \hline
Multiplet & Y &
$\begin{array}{c}\mathrm{SU}(3)\times\mathrm{SU}(2) \\
\mathrm{content} \end{array}$ &
SU(5) coefficient & $\begin{array}{c}\mathrm{required} \\
\mathrm{coefficient} \end{array}$ \\
\hline
$\left( \begin{array}{ccc}u_R & u_G & u_B \\
d_R & d_G & d_B \end{array} \right)$         & $\frac{1}{3}$ &
(3,2) & $\frac{1}{\sqrt{60}}$ & $\frac{1}{\sqrt{360}}$ \\ \hline
$\left( \begin{array}{ccc}\bar{u}_R & \bar{u}_G & \bar{u}_B
 \end{array} \right)$ & $-\frac{4}{3}$ & $(\bar{3},1)$ &
$\frac{-4}{\sqrt{60}}$ & $\frac{-4}{\sqrt{360}}$ \\ \hline
$\left( \begin{array}{ccc}\bar{d}_R & \bar{d}_G & \bar{d}_B
 \end{array} \right)$ & $\frac{2}{3}$ & $(\bar{3},1)$ &
$\frac{2}{\sqrt{60}}$ & $\frac{2}{\sqrt{360}}$ \\ \hline
$\left( \begin{array}{c}\nu_e \\ e^- \end{array} \right)$ & $-1$ &
(1,2) & $\frac{-3}{\sqrt{60}}$ & $\frac{-3}{\sqrt{360}}$ \\ \hline
$\left( \begin{array}{c} e^+ \end{array} \right)$ & $2$ &
(1,1) & $\frac{6}{\sqrt{60}}$ & $\frac{6}{\sqrt{360}}$ \\ \hline
$\left( \begin{array}{c} \bar{\nu}_e \end{array} \right)$ & $0$ &
(1,1) & absent & $0$ \\ \hline
\end{tabular}
\caption{\label{T1}
Weak hypercharge, $\mathrm{SU}(3)\times\mathrm{SU}(2)$ assignments,
coefficient of the coupling to the $\mathrm{U}(1)_Y$ vector boson in
$\mathrm{SU}(5)$, and the required coefficient of the coupling to
the $\mathrm{U}(1)_Y$ vector boson, for the left-handed fermions of the first
generation.}
\end{center}
\end{table}

The breaking of $\mathrm{E}8$ to $\mathrm{SU}(3) \times \left(\mathrm{SU}(2)
\right)^3 \times \mathrm{U}(1)_Y$ can
be studied, following \cite{NTSG}, by analyzing the breaking of
$\mathrm{SU}(9)$ to $\mathrm{SU}(3) \times \left(\mathrm{SU}(2)\right)^3
\times\mathrm{U}(1)_Y$.  It
is convenient to use block matrix notation.
Each $\mathrm{SU}(9)$ fundamental index is replaced by a pair of
indexes, an upper-case letter and a lower-case letter.  The
upper-case letter runs from 1 to 4, and indicates which subgroup in
the sequence $\mathrm{SU}(3)\times\mathrm{SU}(2)
\times\mathrm{SU}(2)\times\mathrm{SU}(2)$ the block belongs to.
Thus $B=1$ denotes the $\mathrm{SU}(3)$, $B=2$ denotes the first
$SU(2)$, $B=3$ denotes the second $SU(2)$, and $B=4$ denotes the
third $SU(2)$.  The lower-case index is a fundamental index for the
subgroup identified by the upper-case index it belongs to.  It is
important to note that the range of a lower-case index depends on
the value of the upper-case index it belongs to, so we have to keep
track of which lower-case indexes belong to which upper-case
indexes.  Each $\mathrm{SU}(9)$ antifundamental index is treated in
the
same way, except that the lower-case index is now an antifundamental
index for the appropriate subgroup.  The summation convention is
applied to both upper-case letters and lower-case letters that
derive from an $\mathrm{SU}(9)$ fundamental or antifundamental
index, but we have to remember that lower-case indexes are to be
summed over first, because their ranges of summation depend on the
values of the upper-case indexes they belong to.  Each
$\mathrm{SU}(9)$ adjoint representation index, which in the notation
above, is a lower-case Greek letter, is replaced by a pair of
indexes, an
upper-case letter and a lower-case letter, where the upper-case
letter runs from 1 to 5, and identifies which subgroup in the
sequence $\mathrm{SU}(3)\times\mathrm{SU}(2)
\times\mathrm{SU}(2)\times\mathrm{SU}(2)\times\mathrm{U}(1)_Y$ a
generator belongs to, and the lower-case
letter runs over all the
generators of the subgroup identified by the upper-case letter it
belongs to.  When an upper-case adjoint representation index takes
the value 5, the associated lower-case index takes a single
value, 1.  The summation convention is applied to a lower-case
letter that derives from an $\mathrm{SU}(9)$ adjoint representation
index, but \emph{not} to an upper-case letter that derives from an
$\mathrm{SU}(9)$ adjoint representation index.

\begin{table}
\begin{center}
\begin{tabular}{|c|c|c|c|c|}\hline
\multicolumn{5}{|c|}{States in the \textbf{80}} \\ \hline
Blocks & $\begin{array}{c}\textrm{Number of} \\
\mathrm{distinct} \\ \mathrm{blocks} \end{array}$ &
$\begin{array}{c}\mathrm{SU}(3)\times\mathrm{SU}(2) \\
\mathrm{content} \end{array}$ &
$\begin{array}{c}\textrm{Number of} \\
\mathrm{states} \end{array}$ &
$\begin{array}{c}\textrm{coefficient} \\
\textrm{of coupling} \\ \textrm{to U(1)} \end{array}$ \\
\hline
$\psi_{1\bar{1}}$ & 1 & (8,1) & 8 & 0 \\ \hline
$\psi_{2\bar{2}}$ & 1 & (1,3) & 3 & 0 \\
$\psi_{3\bar{3}}$ & 1 & (1,3) & 3 & 0 \\
$\psi_{4\bar{4}}$ & 1 & (1,3) & 3 & 0 \\ \hline
$\begin{array}{c}\psi_{\mathrm{diag}} \\ \psi_{\mathrm{diag}} \\
\psi_{\mathrm{diag}} \end{array}$ &
$\begin{array}{c}\textrm{not} \\ \textrm{applicable} \end{array}$ &
$\begin{array}{c}(1,1)+(1,1)+ \\ +(1,1) \end{array}$ &
3 & 0 \\  \hline
$\psi_{1\bar{2}}$ & 1 & (3,2) & 6 &
$\frac{\sigma_1-\sigma_2}{\theta}$ \\
$\psi_{1\bar{3}}$ & 1 & (3,2) & 6 &
$\frac{\sigma_1-\sigma_3}{\theta}$ \\
$\psi_{1\bar{4}}$ & 1 & (3,2) & 6 &
$\frac{\sigma_1-\sigma_4}{\theta}$ \\ \hline
$\psi_{2\bar{1}}$ & 1 & $(\bar{3},2)$ & 6 &
$\frac{-\sigma_1+\sigma_2}{\theta}$ \\
$\psi_{3\bar{1}}$ & 1 & $(\bar{3},2)$ & 6 &
$\frac{-\sigma_1+\sigma_3}{\theta}$ \\
$\psi_{4\bar{1}}$ & 1 & $(\bar{3},2)$ & 6 &
$\frac{-\sigma_1+\sigma_4}{\theta}$ \\ \hline
$\psi_{2\bar{3}}$ & 1 & $(1,3)+(1,1)$ & 4 &
$\frac{\sigma_2-\sigma_3}{\theta}$ \\
$\psi_{2\bar{4}}$ & 1 & $(1,3)+(1,1)$ & 4 &
$\frac{\sigma_2-\sigma_4}{\theta}$ \\
$\psi_{3\bar{4}}$ & 1 & $(1,3)+(1,1)$ & 4 &
$\frac{\sigma_3-\sigma_4}{\theta}$ \\ \hline
$\psi_{3\bar{2}}$ & 1 & $(1,3)+(1,1)$ & 4 &
$\frac{-\sigma_2+\sigma_3}{\theta}$ \\
$\psi_{4\bar{2}}$ & 1 & $(1,3)+(1,1)$ & 4 &
$\frac{-\sigma_2+\sigma_4}{\theta}$ \\
$\psi_{4\bar{3}}$ & 1 & $(1,3)+(1,1)$ & 4 &
$\frac{-\sigma_3+\sigma_4}{\theta}$ \\ \hline
\end{tabular}
\caption{\label{T2}
The states in the \textbf{80}, organized by their
$\mathrm{SU}(3)\times\mathrm{SU}(2)\times\mathrm{SU}(2)\times
\mathrm{SU}(2)$ assignments, showing their $\mathrm{SU}(3)\times
\mathrm{SU}(2)_{\mathrm{diag}}$ content, and the coefficients of
their couplings to a $\mathrm{U}(1)$ gauge field, parametrized as in equation
(\ref{Abelian generator}).}
\end{center}
\end{table}

We can now list all the blocks in the \textbf{80}, the \textbf{84},
and the $\mathbf{\overline{84}}$, and display their $\mathrm{SU}(3)\times
\mathrm{SU}(2)$ content.  This is displayed in Table \ref{T2} for
the \textbf{80}, and in Table \ref{T3} for the $\mathbf{\overline{84}}$,
with all the lower-case indexes suppressed.

\begin{table}
\begin{center}
\begin{tabular}{|c|c|c|c|c|}\hline
\multicolumn{5}{|c|}{States in the $\mathbf{\overline{84}}$} \\
\hline
Blocks & $\begin{array}{c}\textrm{Number of} \\
\mathrm{distinct} \\ \mathrm{blocks} \end{array}$ &
$\begin{array}{c}\mathrm{SU}(3)\times\mathrm{SU}(2) \\
\mathrm{content} \end{array}$ &
$\begin{array}{c}\textrm{Number of} \\
\mathrm{states} \end{array}$ &
$\begin{array}{c}\textrm{coefficient} \\
\textrm{of coupling} \\ \textrm{to U(1)} \end{array}$ \\
\hline
$\psi_{\bar{1}\bar{1}\bar{1}}$ & 1 & (1,1) & 1 & $\frac{-3\sigma_1}{\theta}$ \\
\hline
$\begin{array}{ccc}\psi_{\bar{2}\bar{1}\bar{1}} &
\psi_{\bar{1}\bar{2}\bar{1}} &
\psi_{\bar{1}\bar{1}\bar{2}} \end{array}$ & 1 & $(3,2)$ &
6 & $\frac{-2\sigma_1-\sigma_2}{\theta}$ \\
$\begin{array}{ccc}\psi_{\bar{3}\bar{1}\bar{1}} &
\psi_{\bar{1}\bar{3}\bar{1}} &
\psi_{\bar{1}\bar{1}\bar{3}} \end{array}$ & 1 & $(3,2)$ &
6 & $\frac{-2\sigma_1-\sigma_3}{\theta}$ \\
$\begin{array}{ccc}\psi_{\bar{4}\bar{1}\bar{1}} &
\psi_{\bar{1}\bar{4}\bar{1}} &
\psi_{\bar{1}\bar{1}\bar{4}} \end{array}$ & 1 & $(3,2)$ &
6 & $\frac{-2\sigma_1-\sigma_4}{\theta}$ \\ \hline
$\begin{array}{ccc}\psi_{\bar{2}\bar{2}\bar{1}} &
\psi_{\bar{2}\bar{1}\bar{2}} &
\psi_{\bar{1}\bar{2}\bar{2}} \end{array}$ & 1 & $(\bar{3},1)$ &
3 & $\frac{-\sigma_1-2\sigma_2}{\theta}$ \\
$\begin{array}{ccc}\psi_{\bar{3}\bar{3}\bar{1}} &
\psi_{\bar{3}\bar{1}\bar{3}} &
\psi_{\bar{1}\bar{3}\bar{3}} \end{array}$ & 1 & $(\bar{3},1)$ &
3 & $\frac{-\sigma_1-2\sigma_3}{\theta}$ \\
$\begin{array}{ccc}\psi_{\bar{4}\bar{4}\bar{1}} &
\psi_{\bar{4}\bar{1}\bar{4}} &
\psi_{\bar{1}\bar{4}\bar{4}} \end{array}$ & 1 & $(\bar{3},1)$ &
3 & $\frac{-\sigma_1-2\sigma_4}{\theta}$ \\ \hline
$\begin{array}{ccc}
\psi_{\bar{1}\bar{2}\bar{3}} & \psi_{\bar{2}\bar{1}\bar{3}} &
\psi_{\bar{2}\bar{3}\bar{1}} \\
\psi_{\bar{1}\bar{3}\bar{2}} & \psi_{\bar{3}\bar{1}\bar{2}} &
\psi_{\bar{3}\bar{2}\bar{1}}
\end{array}$ & 1 & $(\bar{3},3)+(\bar{3},1)$ & 12 & $\frac{-\sigma_1-\sigma_2
-\sigma_3}{\theta}$ \\ \hline
$\begin{array}{ccc}
\psi_{\bar{1}\bar{2}\bar{4}} & \psi_{\bar{2}\bar{1}\bar{4}} &
\psi_{\bar{2}\bar{4}\bar{1}} \\
\psi_{\bar{1}\bar{4}\bar{2}} & \psi_{\bar{4}\bar{1}\bar{2}} &
\psi_{\bar{4}\bar{2}\bar{1}}
\end{array}$ & 1 & $(\bar{3},3)+(\bar{3},1)$ & 12 & $\frac{-\sigma_1-\sigma_2
-\sigma_4}{\theta}$ \\ \hline
$\begin{array}{ccc}
\psi_{\bar{1}\bar{3}\bar{4}} & \psi_{\bar{3}\bar{1}\bar{4}} &
\psi_{\bar{3}\bar{4}\bar{1}} \\
\psi_{\bar{1}\bar{4}\bar{3}} & \psi_{\bar{4}\bar{1}\bar{3}} &
\psi_{\bar{4}\bar{3}\bar{1}}
\end{array}$ & 1 & $(\bar{3},3)+(\bar{3},1)$ & 12 & $\frac{-\sigma_1-\sigma_3
-\sigma_4}{\theta}$ \\ \hline
$\begin{array}{ccc}
\psi_{\bar{2}\bar{2}\bar{2}} & \psi_{\bar{3}\bar{3}\bar{3}} &
\psi_{\bar{4}\bar{4}\bar{4}}
\end{array}$ &
\multicolumn{4}{|c|}{these three blocks are empty} \\ \hline
$\begin{array}{ccc} \psi_{\bar{2}\bar{2}\bar{3}} &
\psi_{\bar{2}\bar{3}\bar{2}} & \psi_{\bar{3}\bar{2}\bar{2}} \end{array}$ & 1 &
(1,2) & 2 & $\frac{-2\sigma_2-\sigma_3}{\theta}$ \\
$\begin{array}{ccc} \psi_{\bar{2}\bar{2}\bar{4}} &
\psi_{\bar{2}\bar{4}\bar{2}} & \psi_{\bar{4}\bar{2}\bar{2}} \end{array}$ & 1 &
(1,2) & 2 & $\frac{-2\sigma_2-\sigma_4}{\theta}$ \\
$\begin{array}{ccc} \psi_{\bar{3}\bar{3}\bar{2}} &
\psi_{\bar{3}\bar{2}\bar{3}} & \psi_{\bar{2}\bar{3}\bar{3}} \end{array}$ & 1 &
(1,2) & 2 & $\frac{-\sigma_2-2\sigma_3}{\theta}$ \\
$\begin{array}{ccc} \psi_{\bar{3}\bar{3}\bar{4}} &
\psi_{\bar{3}\bar{4}\bar{3}} & \psi_{\bar{4}\bar{3}\bar{3}} \end{array}$ & 1 &
(1,2) & 2 & $\frac{-2\sigma_3-\sigma_4}{\theta}$ \\
$\begin{array}{ccc} \psi_{\bar{4}\bar{4}\bar{2}} &
\psi_{\bar{4}\bar{2}\bar{4}} & \psi_{\bar{2}\bar{4}\bar{4}} \end{array}$ & 1 &
(1,2) & 2 & $\frac{-\sigma_2-2\sigma_4}{\theta}$ \\
$\begin{array}{ccc} \psi_{\bar{4}\bar{4}\bar{3}} &
\psi_{\bar{4}\bar{3}\bar{4}} & \psi_{\bar{3}\bar{4}\bar{4}} \end{array}$ & 1 &
(1,2) & 2 & $\frac{-\sigma_3-2\sigma_4}{\theta}$ \\ \hline
$\begin{array}{ccc}
\psi_{\bar{2}\bar{3}\bar{4}} & \psi_{\bar{3}\bar{2}\bar{4}} &
\psi_{\bar{3}\bar{4}\bar{2}} \\
\psi_{\bar{4}\bar{3}\bar{2}} & \psi_{\bar{4}\bar{2}\bar{3}} &
\psi_{\bar{2}\bar{4}\bar{3}}
\end{array}$ & 1 &
$\begin{array}{c} (1,4)+(1,2)+ \\ +(1,2)\end{array}$ & 8 &
$\frac{-\sigma_2-\sigma_3-\sigma_4}{\theta}$ \\ \hline
\end{tabular}
\caption{\label{T3}
The states in the $\mathbf{\overline{84}}$, organized by their
$\mathrm{SU}(3)\times\mathrm{SU}(2)\times\mathrm{SU}(2)\times
\mathrm{SU}(2)$ assignments, showing their $\mathrm{SU}(3)\times
\mathrm{SU}(2)_{\mathrm{diag}}$ content, and the coefficients of
their couplings to a $\mathrm{U}(1)$ gauge field, parametrized as in equation
(\ref{Abelian generator}).}
\end{center}
\end{table}

The $\mathrm{SU}(9)$ generators, in the $\mathrm{SU}(3)\times
\mathrm{SU}(2)\times\mathrm{SU}(2)\times\mathrm{SU}(2)\times
\mathrm{U}(1)_Y$ subgroup, may be taken as follows, in the block
matrix notation.
\begin{equation}\label{non-Abelian generators}
\left(t_{Aa}^{(9)}\right)_{Bi\bar{C}\bar{j}}\ \ =\ \ \delta_{AB}
\delta_{A\bar{C}}
\left(t_{Aa}\right)_{i\bar{j}}\qquad\qquad\qquad(1\leq A\leq4)
\qquad\qquad\qquad\quad
\end{equation}
\begin{equation}\label{Abelian generator}
\begin{array}{ccl}
\left(t_{Y}^{(9)}\right)_{Bi\bar{C}\bar{j}} & = & \frac{1}{\theta}\left(
\sigma_1\delta_{1B}\delta_{1\bar{C}}\delta_{i\bar{j}}+
\sigma_2\delta_{2B}\delta_{2\bar{C}}\delta_{i\bar{j}}+
\sigma_3\delta_{3B}\delta_{3\bar{C}}\delta_{i\bar{j}}+
\sigma_4\delta_{4B}\delta_{4\bar{C}}\delta_{i\bar{j}}\right) \\
 & = & \frac{1}{\theta}
\left(\displaystyle\sum_{A=1}^4\sigma_A\delta_{AB}\delta_{A\bar{C}}
\delta_{i\bar{j}}\right)
\end{array}
\end{equation}

Here $\left(t_{Aa}\right)_{i\bar{j}}$ denotes the fundamental
representation generator number $a$, of non-Abelian subgroup number
$A$, in the list above.  Thus for $A=1$, the subgroup is
$\mathrm{SU}(3)$, $a$ runs from 1 to 8, and $i$ and $j$ each run
from 1 to 3, while for $A=2,\textrm{ 3, or 4}$, the subgroup is
$\mathrm{SU}(2)$, $a$ runs from 1 to 3, and $i$ and $j$ each run
from 1 to 2.

$\sigma_1$, $\sigma_2$, $\sigma_3$, and $\sigma_4$ are real numbers
parametrizing the embedding of the $\mathrm{U}(1)_Y$ subgroup in
$\mathrm{SU}(9)$, and thus in $\mathrm{E}8$, and $\theta$ is a
normalization factor.

In using this notation, we have to take sensible precautions, such
as grouping within brackets, to keep track of which lower-case
indexes belong to which upper-case indexes.  In equation
(\ref{Abelian generator}), it would be wrong to ``factor out'' the
$\delta_{i\bar{j}}$, because it represents a \mbox{3 by 3} matrix in one
term, and a 2 by 2 matrix in the other three terms.

The tracelessness of $\left(t_{Y}^{(9)}\right)_{Bi\bar{C}\bar{j}}$ implies:
\begin{equation}\label{tracelessness}
0 = 3\sigma_1+2\left(\sigma_2+\sigma_3+\sigma_4\right)
\end{equation}
and the normalization condition (\ref{normalization}) implies:
\begin{equation}\label{theta squared}
\theta^2 =
6\sigma_1^2+4\left(\sigma_2^2+\sigma_3^2+\sigma_4^2\right)
\end{equation}

As an example, I consider the states in the left-handed
$\mathbf{\overline{84}}$.  The covariant derivative is \cite{Rosner}
\begin{equation}\label{covariant derivative}
\mathcal{D}_\mu = \partial_\mu - i g A_{\mu\alpha}T_\alpha
\end{equation}
so, for unbroken $\mathrm{SU}(9)$, and with $ \left(+,+,+,1\right) $ metric,
and $ \left\{\gamma_{\mu},\gamma_{\nu}\right\} = 2 g_{\mu \nu} $, the massless
Dirac action in this case is \cite{Rosner}:
\begin{displaymath}
\bar{\psi} \gamma^\mu\mathcal{D}_\mu\psi\quad = \quad
\bar{\psi} \gamma^\mu\left(\partial_\mu
- i g A_{\mu\alpha}T_\alpha\right)\psi\quad =
\qquad\qquad\qquad\qquad\qquad\qquad\qquad\qquad
\end{displaymath}
\begin{displaymath}
=\quad\bar{\psi}_{ijk} \gamma^\mu\left(\partial_\mu
\frac{1}{6}\left(\delta_{m\bar{i}}\delta_{p\bar{j}}\delta_{q\bar{k}}\pm
\textrm{five terms}\right)\right.
\qquad\qquad\qquad\qquad\qquad\qquad\qquad\qquad
\end{displaymath}
\begin{displaymath}
\qquad\qquad\qquad\qquad
\left. - i g A_{\mu\alpha}\frac{1}{6}\left(-\left(t_\alpha
\right)_{m\bar{i}}\delta_{p\bar{j}}\delta_{q\bar{k}}\pm\textrm{seventeen
terms}\right)\right)\psi_{\bar{m}\bar{p}\bar{q}} =
\end{displaymath}
\begin{equation}\label{massless Dirac action}
\qquad\qquad\qquad\qquad
=\quad\bar{\psi}_{ijk} \gamma^\mu\partial_\mu\psi_{\bar{i}\bar{j}\bar{k}}
+ 3 i g A_{\mu\alpha}\bar{\psi}_{ijk}\gamma^\mu \left(t_\alpha
\right)_{m\bar{i}}\psi_{\bar{m}\bar{j}\bar{k}}
\end{equation}
where I used (\ref{84 bar}), the antisymmetry of $\bar{\psi}_{ijk}$
and $\psi_{\bar{m}\bar{p}\bar{q}}$ in their indexes, and the relabelling of
dummy indexes.
$\bar{\psi}_{ijk}$ are the right-handed \textbf{84} states, and
$\psi_{mpq}$ are the left-handed $\mathbf{\overline{84}}$ states.

Breaking $\mathrm{SU}(9)$ to $\mathrm{SU}(3) \times \left(\mathrm{SU}(2)
\right)^3 \times\mathrm{U}(1)_Y$, and
using the block matrix notation, this becomes:
\begin{displaymath}
\bar{\psi}_{BiCjDk} \gamma^\mu\partial_\mu\psi_{\bar{B}\bar{i}\bar{C}\bar{j}
\bar{D}\bar{k}}
+ 3 i g \sum_{A=1}^5 A_{\mu Aa}\bar{\psi}_{BiCjDk}\gamma^\mu
\left(t_{Aa}^{(9)}\right)_{Em\bar{B}\bar{i}}\psi_{\bar{E}\bar{m}\bar{C}\bar{j}
\bar{D}\bar{k}}\quad =\qquad\qquad
\qquad
\end{displaymath}
\begin{displaymath}
=\quad\left(\bar{\psi}_{BiCjDk} \gamma^\mu\partial_\mu\psi_{\bar{B}\bar{i}
\bar{C}\bar{j}\bar{D}\bar{k}}
+ 3 i g \sum_{A=1}^4 A_{\mu Aa}\bar{\psi}_{AiCjDk}\gamma^\mu
\left(t_{Aa}\right)_{m\bar{i}}\psi_{\bar{A}\bar{m}\bar{C}\bar{j}\bar{D}\bar{k}}
\right.\quad\qquad\quad\quad
\end{displaymath}
\begin{equation}\label{after breaking}
\qquad\qquad\qquad\qquad\qquad\qquad\qquad\qquad\quad
\left. + 3 i g A_{\mu Y}
\frac{1}{\theta}
\displaystyle\sum_{A=1}^4\sigma_A
\bar{\psi}_{AiCjDk}\gamma^\mu\psi_{\bar{A}\bar{i}\bar{C}\bar{j}\bar{D}\bar{k}}
\right)\quad
\end{equation}
where I used (\ref{non-Abelian generators}) and
(\ref{Abelian generator}).  We can now extract the covariant
derivative Dirac action terms for the various entries in Table
\ref{T3}, and thus determine the coefficients of their couplings to
$A_{\mu Y}$.  For example, a block in $\psi_{\bar{B}\bar{i}\bar{C}\bar{j}
\bar{D}\bar{k}}$, where two
upper-case indexes take the value 1, and the remaining upper-case
index takes the value 2, 3, or 4, is a candidate to be a (3,2) quark
multiplet.  The sum of all terms in (\ref{after breaking}), where
two upper-case indexes take the value 1, and the remaining
upper-case index takes the value 2, is:
\begin{displaymath}
\bigg(3\bar{\psi}_{1i1j2k} \gamma^\mu\partial_\mu\psi_{\bar{1}\bar{i}\bar{1}
\bar{j}\bar{2}\bar{k}}
+ 6 i g A_{\mu 1a}\bar{\psi}_{1i1j2k}\gamma^\mu\!
\left(t_{1a}\right)_{m\bar{i}}\psi_{\bar{1}\bar{m}\bar{1}\bar{j}\bar{2}\bar{k}}
+ 3 i g A_{\mu 2a}\bar{\psi}_{2i1j1k}
\gamma^\mu\!
\left(t_{2a}\right)_{m\bar{i}}\psi_{\bar{2}\bar{m}\bar{1}\bar{j}\bar{1}\bar{k}}
\end{displaymath}
\begin{equation}\label{112 after breaking}
\qquad\left.+ 6 i g A_{\mu Y} \frac{1}{\theta}\sigma_1
\bar{\psi}_{1i1j2k}\gamma^\mu\psi_{\bar{1}\bar{i}\bar{1}\bar{j}\bar{2}\bar{k}}
+ 3 i g A_{\mu Y} \frac{1}{\theta}\sigma_2
\bar{\psi}_{2i1j1k}\gamma^\mu\psi_{\bar{2}\bar{i}\bar{1}\bar{j}\bar{1}\bar{k}}
\right)\quad
\end{equation}

Now $\psi_{\bar{1}\bar{i}\bar{1}\bar{j}\bar{2}\bar{k}}$ is antisymmetric under
swapping the two
$\mathrm{SU}(3)$ antifundamental indexes $i$ and $j$, so that we may
write:
\begin{equation}\label{112 as quarks}
\psi_{\bar{1}\bar{i}\bar{1}\bar{j}\bar{2}\bar{k}}=\varepsilon_{\bar{i}\bar{j}
\bar{m}}\phi_{m\bar{k}}
\end{equation}
and analogously:
\begin{equation}\label{RH 112 as antiquarks}
\bar{\psi}_{1i1j2k}=\varepsilon_{ijm}\bar{\phi}_{\bar{m}k}
\end{equation}

We can then use relations such as
\begin{equation}\label{112 without inserted SU3 t}
\bar{\psi}_{1i1j2k}\gamma^\mu\psi_{\bar{1}\bar{i}\bar{1}\bar{j}\bar{2}\bar{k}}
=\varepsilon_{ijp}\bar{\phi}_{\bar{p}k}\gamma^\mu
\varepsilon_{\bar{i}\bar{j}\bar{m}}\phi_{m\bar{k}}
=2\delta_{p\bar{m}}\bar{\phi}_{\bar{p}k}\gamma^\mu\phi_{m\bar{k}}
=2\bar{\phi}_{\bar{m}k}\gamma^\mu\phi_{m\bar{k}}
\end{equation}
and
\begin{displaymath}
\bar{\psi}_{1i1j2k}\gamma^\mu\left(t_{1a}\right)_{m\bar{i}}\psi_{\bar{1}\bar{m}
\bar{1}\bar{j}\bar{2}\bar{k}}
=\varepsilon_{ijp}\bar{\phi}_{\bar{p}k}\gamma^\mu\left(t_{1a}\right)_{m\bar{i}}
\varepsilon_{\bar{m}\bar{j}\bar{q}}\phi_{q\bar{k}}=\qquad\qquad\qquad\qquad
\qquad\qquad
\end{displaymath}
\begin{displaymath}
=\left(\delta_{i\bar{m}}\delta_{p\bar{q}}-\delta_{i\bar{q}}
\delta_{p\bar{m}}\right)\bar{\phi}_{\bar{p}k}\gamma^\mu\left(t_{1a}
\right)_{m\bar{i}} \phi_{q\bar{k}}=
\bar{\phi}_{\bar{p}k}\gamma^\mu\left(t_{1a}\right)_{i\bar{i}}\phi_{p\bar{k}}-
\bar{\phi}_{\bar{m}k}\gamma^\mu\left(t_{1a}\right)_{m\bar{i}}\phi_{i\bar{k}}=
\end{displaymath}
\begin{equation}\label{112 with inserted SU3 t}
\qquad\qquad\qquad\qquad\qquad\qquad\qquad\qquad\qquad
=-\bar{\phi}_{\bar{m}k}\gamma^\mu\left(t_{1a}\right)_{m\bar{i}}\phi_{i\bar{k}}
\end{equation}
to express (\ref{112 after breaking}) as:
\begin{displaymath}
6\bigg(\bar{\phi}_{\bar{i}j} \gamma^\mu\partial_\mu\phi_{i\bar{j}}
- i g A_{\mu 1a}
\bar{\phi}_{\bar{i}j}\gamma^\mu\left(t_{1a}\right)_{i\bar{k}}\phi_{k\bar{j}}
+ i g A_{\mu 2a}\bar{\phi}_{\bar{i}j}\gamma^\mu\!
\left(t_{2a}\right)_{m\bar{j}}\phi_{i\bar{m}}\qquad\qquad\qquad
\end{displaymath}
\begin{equation}\label{112 after breaking and transformation}
\qquad\qquad\qquad\qquad\qquad\qquad\qquad\qquad
\qquad\left.- i g A_{\mu Y} \frac{1}{\theta}\left( - 2 \sigma_1
- \sigma_2\right)\bar{\phi}_{\bar{i}j}\gamma^\mu\phi_{i\bar{j}}\right)\quad
\end{equation}

Thus we see that the index $i$ of $\phi_{i\bar{j}}$ is an $\mathrm{SU}(3)$
fundamental index.  The $\mathrm{SU}(2)$ antifundamental is
equivalent to the fundamental, the relation being given by matrix
multiplication by $\varepsilon_{jk}$, and we could, if we wished,
make a further transformation to replace the $\mathrm{SU}(2)$
antifundamental index $\bar{j}$ of $\phi_{i\bar{j}}$, by an index that is
manifestly in the $\mathrm{SU}(2)$ fundamental.  When $(\mathrm{SU}
(2))^3$ is broken to $\mathrm{SU}(2)_{\mathrm{diag}}$, the $A_{\mu2a}$, in
the third term in (\ref{112 after breaking and transformation}),
will be replaced, at low energy, by $\frac{1}{\sqrt{3}}B_{\mu a}$, where
$B_{\mu a}$ is the gauge field of $\mathrm{SU}(2)_{\mathrm{diag}}$.  The
overall factor of 6 can be absorbed into the normalizations of
$\phi_{i\bar{j}}$ and $\bar{\phi}_{\bar{i}j}$, so from the fourth term in
(\ref{112 after breaking and transformation}), we can read off what
the coefficient of $g A_{\mu Y}\bar{\phi}_{\bar{i}j}\gamma^\mu\phi_{i\bar{j}}$
would be, if the
$\bar{\phi}_{\bar{i}ij}i\gamma^\mu\partial_\mu\phi_{i\bar{j}}$ term had
standard normalization, and thus complete the entries in the second row of
Table \ref{T3}.

The entries in the third column of Tables \ref{T2} and \ref{T3} can
be completed by similar methods.  The entries in the fifth column
of Table \ref{T3}
can be completed by a simple mnemonic: for each upper-case index, of
the untransformed $\psi_{\bar{B}\bar{i}\bar{C}\bar{j}\bar{D}\bar{k}}$, that
takes the value $A$,
$1\leq A\leq 4$, include a term $-\frac{1}{\theta}\sigma_A$.  For
Table \ref{T2}, the mnemonic is that when the index $B$ of
$\psi_{Bi\bar{C}\bar{j}}$ takes the value $A$, $1\leq A\leq 4$, so that $i$ is
in the fundamental of non-Abelian subgroup number $A$, include a
term $\frac{1}{\theta}\sigma_A$, and when the index $\bar{C}$ of
$\psi_{Bi\bar{C}\bar{j}}$ takes the value $A$, $1\leq A\leq 4$, so that $j$ is
in the antifundamental of non-Abelian subgroup number $A$, include
a term $-\frac{1}{\theta}\sigma_A$.

Indeed, suppose we extract all terms from
(\ref{after breaking}) such that $\psi_{\bar{B}\bar{i}\bar{C}\bar{j}\bar{D}
\bar{k}}$ has $n_A$
upper-case indexes with the value $A$, $1\leq A\leq 4$, so that
$n_1+n_2+n_3+n_4=3$.  We get $\frac{3!}{n_1!n_2!n_3!n_4!}$
contributions from the
$\bar{\psi}_{BiCjDk} \gamma^\mu\partial_\mu\psi_{\bar{B}\bar{i}\bar{C}\bar{j}
\bar{D}\bar{k}}$ term.  The
number of times we get $\sigma_A$, from the third term in
(\ref{after breaking}), is
$\frac{2!}{\tilde{n}_1!\tilde{n}_2!\tilde{n}_3!\tilde{n}_4!}$, where
$\tilde{n}_B=n_B$ if $B\neq A$, and $\tilde{n}_A=n_A-1$.  But this
is equal to $\frac{2!n_A}{n_1!n_2!n_3!n_4!}$.  The factor
$\frac{2!}{n_1!n_2!n_3!n_4!}$ combines with the explicit factor of
3, in the third term in (\ref{after breaking}), to produce the same
overall factor of $\frac{3!}{n_1!n_2!n_3!n_4!}$ as found for the
first term, so the coefficient of the contributions from the third
term, if the contributions from the first term had standard
normalization, would be $-\frac{1}{\theta}\displaystyle\sum_{A=1}^4
n_A\sigma_A$.  The mnemonic for Table \ref{T2} can be justified in a
similar manner.

We know that we have to find couplings of the observed fermions, to
the $\mathrm{U}(1)_Y$ gauge field, that are smaller than those found
in the $\mathrm{SU}(5)$ model \cite{Georgi Glashow, Mohapatra}, by an overall
factor that is within a few percent of $\frac{1}{\sqrt{6}}$, so it
is useful to apply the same techniques to calculate the
corresponding coefficients in the $\mathrm{SU}(5)$ model.  In this
case, the relations (\ref{normalization}) and (\ref{tracelessness})
completely determine the $\mathrm{U}(1)$ generator, up to sign, and
we find the entries in the fourth column of Table \ref{T1}.  The
entries in the fifth column have been filled in, assuming the
overall factor is exactly $\frac{1}{\sqrt{6}}$.

If we now choose $\left( \sigma_1, \sigma_2, \sigma_3, \sigma_4 \right) =
\left( 4, 3, - 3, - 6 \right)$, so that $\theta^2 = 312$, we can identify
$\psi_{1 \bar{2}}$ as a $q$, the $\left( \bar{3}, 1 \right)$ state in
$\psi_{\bar{1} \bar{2} \bar{3}}$ as a $\bar{u}$, $\psi_{\bar{3} \bar{3}
\bar{1}}$ as a $\bar{d}$, $\psi_{\bar{2} \bar{2} \bar{3}}$ as an $l$, and the
$\left( 1, 1 \right)$ state in $\psi_{2 \bar{3}}$ as an $e^+$.  We note that
another $U \left( 1 \right)$, defined by $\left( \sigma_{B 1},
\sigma_{B 2}, \sigma_{B 3}, \sigma_{B 4} \right) = \left( 2, 0, 0, - 3
\right)$, couples to these states in proportion to their baryon number.
Furthermore,
$\psi_{\bar{3} \bar{3} \bar{2}}$, which has $Y = 1$, can be identified as the
Standard Model Higgs field.

To determine the value of $\sin^2 \theta_W$ at unification, for this choice of
the $\sigma_i$, let us denote the Higgs field, $\psi_{\bar{3} \bar{3}
\bar{2}}$, by $\phi_i$.  Then by the methods above, we find that at low
energies, its covariant derivative, times $i$, becomes:
\begin{equation}
  \label{i times covariant derivative of Higgs field in the first case} i
  \partial_{\mu} \phi_i + g \frac{1}{\sqrt{3}} B_{\mu a} \frac{1}{2} \left(
  \sigma_a \right)_{ik} \phi_k + g \frac{3}{\sqrt{312}} A_{\mu Y} \phi_i,
\end{equation}
where $B_{\mu a}$ is the gauge field of $\mathrm{SU} \left( 2
\right)_{\mathrm{diag}}$.  While from equation (117) on page 33 of Rosner's
review {\cite{Rosner}}, the standard covariant derivative, times $i$, on the
Standard Model Higgs field is:
\begin{equation}
  \label{standard covariant derivative for first case} i \partial_{\mu} \phi_i
  + g_{\mathrm{Rosner}} B_{\mu a} \frac{1}{2} \left( \sigma_a \right)_{ik}
  \phi_k + g'_{\mathrm{Rosner}} \frac{Y}{2} A_{\mu Y} \phi_i
\end{equation}
Thus since the Standard Model Higgs field has $Y = 1$, we see that:
\begin{equation}
  \label{standard g and g primed for the first case} g_{\mathrm{Rosner}} = g
  \frac{1}{\sqrt{3}}, \hspace{4em} g'_{\mathrm{Rosner}} = g
  \frac{6}{\sqrt{312}}
\end{equation}
Now by definition, $\tan \theta_W =
\frac{g'_{\mathrm{Rosner}}}{g_{\mathrm{Rosner}}}$.  Thus we find that, for
this choice of the $\sigma_i$, $\sin^2 \theta_W = \frac{9}{35} \simeq 0.257$
at unification.  This is the closest I have found to the required value of
$\sin^2 \theta_W \simeq 0.239$, when the Hodge - de Rham harmonic two-forms,
in the Cartan subalgebra, break $E_8$ to $\mathrm{SU} \left( 3 \right) \times
\left( \mathrm{SU} \left( 2 \right) \right)^3 \times \left( \mathrm{U}
\left( 1 \right) \right)^3$.

However, even this value of $\sin^2 \theta_W$ cannot actually be realized.
For to break $\left( \mathrm{SU} \left( 2 \right) \right)^3$ to $\mathrm{SU}
\left( 2 \right)_{\mathrm{diag}}$, without breaking $\mathrm{SU} \left( 3 \right)
\times \mathrm{SU} \left( 2 \right)_{\mathrm{diag}} \times U \left( 1 \right)_Y$,
we need to find states in the $E_8$ fundamental / adjoint, that transform
nontrivially under $\left( \mathrm{SU} \left( 2 \right) \right)^3$, but are
singlets of $\mathrm{SU} \left( 3 \right) \times \mathrm{SU} \left( 2
\right)_{\mathrm{diag}}$, and have $Y = 0$.  The only states which transform
nontrivially under $\left( \mathrm{SU} \left( 2 \right) \right)^3$, but are
singlets of $\mathrm{SU} \left( 3 \right) \times \mathrm{SU} \left( 2
\right)_{\mathrm{diag}}$, are the $\left( 1, 1 \right)$ states in $\psi_{2
\bar{3}}$, $\psi_{2 \bar{4}}$, $\psi_{3 \bar{4}}$, and their complex
conjugates.  Looking at the $U \left( 1 \right)$ couplings of these states, in
Table \ref{T2}, we see that none of them have $Y = 0$, for $\left( \sigma_1,
\sigma_2, \sigma_3, \sigma_4 \right) = \left( 4, 3, - 3, - 6 \right)$.
Furthermore, to ensure that $\mathrm{SU} \left( 2 \right)_{\mathrm{diag}}$ is the
diagonal subgroup of all three $\mathrm{SU} \left( 2 \right)$'s, and not just
two of them, we need at least two of $\psi_{2 \bar{3}}$, $\psi_{2 \bar{4}}$,
and $\psi_{3 \bar{4}}$, to have $Y = 0$.  That means we require $\sigma_1 =
\sigma_2 = \sigma_3$, which means it is impossible to have $\bar{u}$ and
$\bar{d}$ states with different values of $Y$.  Thus we cannot realize the
Standard Model, if the Hodge - de Rham harmonic two-forms, in the Cartan
subalgebra of $E_8$, break $E_8$ to $\mathrm{SU} \left( 3 \right) \times \left(
\mathrm{SU} \left( 2 \right) \right)^3 \times \left( \mathrm{U}
\left( 1 \right) \right)^3$.

\subsection{Models where the Abelian Hodge - de Rham monopoles break $E_8$ to
$\mathrm{SU} \left( 3 \right) \times \left( \mathrm{SU} \left( 2 \right)
\right)^2 \times \left( \mathrm{U} \left( 1 \right) \right)^4$}
\label{SU 3 cross SU 2 squared cross U 1 to the fourth}

In this subsection, I shall consider some models where the Hodge - de Rham
harmonic two-forms break $E_8$ to $\mathrm{SU} \left( 3 \right) \times \left(
\mathrm{SU} \left( 2 \right) \right)^2 \times \left( \mathrm{U}
\left( 1 \right)
\right)^4$.  We will find that there are states of $E_8$ that transform
nontrivially under $\left( \mathrm{SU} \left( 2 \right) \right)^2$, and can
break $\left( \mathrm{SU} \left( 2 \right) \right)^2$ to $\mathrm{SU} \left( 2
\right)_{\mathrm{diag}}$, without breaking $\mathrm{SU} \left( 3 \right) \times
\mathrm{SU} \left( 2 \right)_{\mathrm{diag}} \times U \left( 1 \right)_Y$, for a
reasonable value of the $U \left( 1 \right)_Y$ coupling constant at
unification.  The $\mathrm{SU} \left( 2 \right)_{\mathrm{diag}}$ coupling constant
is now $\frac{g}{\sqrt{2}}$ at unification, so the $\mathrm{SU} \left( 3
\right)$ and $\mathrm{SU} \left( 2 \right)_{\mathrm{diag}}$ coupling constants, as
evolved in the Standard Model, now meet at around 145 TeV, although this could
presumably be reduced to around a TeV by the accelerated unification mechanism
{\cite{DDG1, DDG2, Arkani-Hamed Cohen Georgi}}.  We will find two distinct
types of solution for $U \left( 1 \right)_Y$, both of which give $\sin^2
\theta_W = \frac{3}{10} = 0.300$ at unification, roughly halfway between the
observed value $\simeq 0.23$ at $m_Z$, and the value $\frac{3}{8} = 0.375$
found in conventional grand unification {\cite{Georgi Glashow}}.
The observed value of $\sin^2 \theta_W$ evolves to $\simeq 0.270$ at around
145 TeV, in the Standard Model.

An element of the $E_8$ Cartan subalgebra, and hence of the $\mathrm{SU} \left(
9 \right)$ Cartan subalgebra, that can have a vacuum expectation value without
breaking this subgroup of $E_8$, is, in the $\mathrm{SU} \left( 9 \right)$
fundamental, a diagonal matrix, with diagonal matrix elements
\begin{equation}
  \label{Abelian vacuum gauge fields in second case} \left( \sigma_1,
  \sigma_1, \sigma_1, \sigma_2, \sigma_2, \sigma_3, \sigma_3, \sigma_4,
  \sigma_5 \right)
\end{equation}
such that:
\begin{equation}
  \label{tracelessness in second case} 3 \sigma_1 + 2 \sigma_2 + 2 \sigma_3 +
  \sigma_4 + \sigma_5 = 0
\end{equation}
The normalization condition is now:
\begin{equation}
  \label{normalization condition for the second case} \theta^2 = 6 \sigma^2_1
  + 4 \left( \sigma^2_2 + \sigma^2_3 \right) + 2 \left( \sigma^2_4 +
  \sigma^2_5 \right)
\end{equation}
The states in the $\mathbf{8} \mathbf{0}$  are shown in Table \ref{states
in the 80 in the second case}, and the states in the $\overline{\mathbf{8}
\mathbf{4}}$ are shown in Table \ref{the states in the 84 bar in the second
case}.  To break $\left( \mathrm{SU} \left( 2 \right) \right)^2$ to $\mathrm{SU}
\left( 2 \right)_{\mathrm{diag}}$, without breaking $\mathrm{SU} \left( 3 \right)
\times \mathrm{SU} \left( 2 \right)_{\mathrm{diag}} \times U \left( 1 \right)_Y$,
we need an $\mathrm{SU} \left( 3 \right)$ singlet state that transforms
non-trivially under $\left( \mathrm{SU} \left( 2 \right) \right)^2$, but
contains a singlet of $\mathrm{SU} \left( 2 \right)_{\mathrm{diag}}$, to have $Y =
0$, so that the singlet of $\mathrm{SU} \left( 2 \right)_{\mathrm{diag}}$ can have
a non-vanishing vacuum expectation value, without breaking $U \left( 1
\right)_Y$.  Thus at least one of $\psi_{2 \bar{3}}$, $\psi_{\bar{2} \bar{3}
\bar{4}}$, and $\psi_{\bar{2} \bar{3} \bar{5}}$ is required to have vanishing
$U \left( 1 \right)_Y$ charge, so at least one of $\left( \sigma_2 - \sigma_3
\right)$, $\left( - \sigma_2 - \sigma_3 - \sigma_4 \right)$, and $\left( -
\sigma_2 - \sigma_3 - \sigma_5 \right)$ is required to be zero.

\begin{table}
\begin{center}
\begin{tabular}{|c|c|c|c|c|c|c|c|c|c|c|}
  \hline
  Block & $\mathrm{SU} \left( 3 \right) \times \mathrm{SU} \left( 2
  \right)_{\mathrm{diag}}$ &  & Coupling & $3 Y_1$ &  & $3 Y_2$ &  & $3 Y_3$ &
  & $3 B_3$\\
  \hline
  $\psi_{1 \bar{1}}$ &  $( 8, 1 )$  &  8  &  0  & 0 &  & 0 &  & 0 &  & $0$\\
  \hline
  $\psi_{2 \bar{2}}$ &  $( 1, 3 )$  &  3  &  0  & 0 &  & 0 &  & 0 &  & $0$\\
  $\psi_{3 \bar{3}}$ &  $( 1, 3 )$  &  3  &  0  & 0 &  & 0 &  & 0 &  & $0$\\
  \hline
  $\psi_S$ & $\left( 1, 1 \right)$ & 1 & 0 & 0 &  & 0 &  & 0 &  & $0$\\
  $\psi_T$ &  $( 1, 1 )$  &  1  &  0  & 0 &  & 0 &  & 0 &  & $0$\\
  $\psi_X$ & $( 1, 1 )$ & 1 & 0 & 0 &  & 0 &  & 0 &  & $0$\\
  $\psi_Y$ & $( 1, 1 )$ &  1  &  0  & 0 &  & 0 &  & 0 &  & $0$\\
  \hline
  $\psi_{1 \bar{2}}$ &  $( 3, 2 )$  &  6  &  $\frac{\sigma_1 -
  \sigma_2}{\theta}$  & $1$ & $q$ & $1$ & $q$ & $1$ & $q$ & $1$\\
  $\psi_{1 \bar{3}}$ &  $( 3, 2 )$  &  6  &  $\frac{\sigma_1 -
  \sigma_3}{\theta}$  & $1$ & $q$ & $1$ & $q$ & $- 5$ &  & $1$\\
  \hline
  $\psi_{1 \bar{4}}$ & $( 3, 1 )$ & 3 & $\frac{\sigma_1 - \sigma_4}{\theta}$ &
  $1$ &  & $4$ & $u$ & $- 2$ & $d$ & $1$\\
  $\psi_{1 \bar{5}}$ &  $( 3, 1 )$  &  3  &  $\frac{\sigma_1 -
  \sigma_5}{\theta}$  & $- 5$ &  & $10$ &  & $- 8$ &  & $4$\\
  \hline
  $\psi_{2 \bar{1}}$ &  $( \bar{3}, 2 )$  &  6  &  $\frac{- \sigma_1 +
  \sigma_2}{\theta}$  & $- 1$ & $\bar{q}$ & $- 1$ & $\bar{q}$ & $- 1$ &
  $\bar{q}$ & $- 1$\\
  $\psi_{3 \bar{1}}$ &  $( \bar{3}, 2 )$  &  6  &  $\frac{- \sigma_1 +
  \sigma_3}{\theta}$  & $- 1$ & $\bar{q}$ & $- 1$ & $\bar{q}$ & $5$ &   & $-
  1$\\
  \hline
  $\psi_{4 \bar{1}}$ &  $( \bar{3}, 1 )$  & 3 & $\frac{- \sigma_1 +
  \sigma_4}{\theta}$ & $- 1$ &  & $- 4$ & $\bar{u}$ & $2$ & $\bar{d}$ & $-
  1$\\
  $\psi_{5 \bar{1}}$ &  $( \bar{3}, 1 )$  &  3  &  $\frac{- \sigma_1 +
  \sigma_5}{\theta}$  & $5$ &  & $- 10$ &  & $8$ &  & $- 4$\\
  \hline
  $\psi_{2 \bar{3}}$ &  $( 1, 3 ) + ( 1, 1 )$  &  4  &  $\frac{\sigma_2 -
  \sigma_3}{\theta}$  & $0$ & $\bar{\nu}$ & $0$ & $\bar{\nu}$ & $- 6$ & $e^-$
  & $0$\\
  \hline
  $\psi_{2 \bar{4}}$ & $( 1, 2 )$ & 2 & $\frac{\sigma_2 - \sigma_4}{\theta}$ &
  $0$ &  & $3$ & $\bar{l}$ & $- 3$ & $l$ & $0$\\
  $\psi_{2 \bar{5}}$ &  $( 1, 2 )$  &  2  &  $\frac{\sigma_2 -
  \sigma_5}{\theta}$  & $- 6$ &  & $9$ &  & $- 9$ &  & $3$\\
  $\psi_{3 \bar{4}}$ & $( 1, 2 )$ & 2 & $\frac{\sigma_3 - \sigma_4}{\theta}$ &
  $0$ &  & $3$ & $\bar{l}$ & $3$ & $\bar{l}$ & $0$\\
  $\psi_{3 \bar{5}}$ & $( 1, 2 )$ &  2  & $\frac{\sigma_3 - \sigma_5}{\theta}$
  & $- 6$ &  & $9$ &  & $- 3$ & $l$ & $3$\\
  \hline
  $\psi_{4 \bar{5}}$ & $( 1, 1 )$ &  1  &  $\frac{\sigma_4 -
  \sigma_5}{\theta}$  & $- 6$ & $e^-$ & $6$ & $e^+$ & $- 6$ & $e^-$ & $3$\\
  \hline
  $\psi_{3 \bar{2}}$ &  $( 1, 3 ) + ( 1, 1 )$  &  4  &  $\frac{- \sigma_2 +
  \sigma_3}{\theta}$  & $0$ & $\bar{\nu}$ & $0$ & $\bar{\nu}$ & $6$ & $e^+$ &
  $0$\\
  \hline
  $\psi_{4 \bar{2}}$ & $( 1, 2 )$ & 2 & $\frac{- \sigma_2 + \sigma_4}{\theta}$
  & $0$ &  & $- 3$ & $l$ & $3$ & $\bar{l}$ & $0$\\
  $\psi_{5 \bar{2}}$ &  $\left( 1, 2 \right)$  &  2  &  $\frac{- \sigma_2 +
  \sigma_5}{\theta}$  & $6$ &  & $- 9$ &  & $9$ &  & $- 3$\\
  $\psi_{4 \bar{3}}$ & $( 1, 2 )$ & 2 & $\frac{- \sigma_3 + \sigma_4}{\theta}$
  & $0$ &  & $- 3$ & $l$ & $- 3$ & $l$ & $0$\\
  $\psi_{5 \bar{3}}$ & $( 1, 2 )$ & 2 & $\frac{- \sigma_3 + \sigma_5}{\theta}$
  & $6$ &  & $- 9$ &  & $3$ & $\bar{l}$ & $- 3$\\
  \hline
  $\psi_{5 \bar{4}}$ & $( 1, 1 )$ & 1 &  $\frac{- \sigma_4 +
  \sigma_5}{\theta}$  & $6$ & $e^+$ & $- 6$ & $e^-$ & $6$ & $e^+$ & $- 3$\\
  \hline
\end{tabular}
\caption{\label{states in the 80 in the second case}
The states in the $ \mathbf{80} $ for the $ \mathrm{SU}\left(3\right)
\times \left(\mathrm{SU}\left(2\right)\right)^2 \times \left( \mathrm{U}\left(
1\right)\right)^4 $ case.}
\end{center}
\end{table}

\begin{table}
\begin{center}
\begin{tabular}{|c|c|c|c|c|c|c|c|c|c|c|}
  \hline
  Block & $\mathrm{SU} \left( 3 \right) \times \mathrm{SU} \left( 2
  \right)_{\mathrm{diag}}$  &   & Coupling  & $3 Y_1$ &  & $3 Y_2$ &  & $3 Y_3$
  &  & $3 B_3$\\
  \hline
  $\psi_{\bar{1} \bar{1} \bar{1}}$  &  $( 1, 1 )$  &  1  &  $\frac{- 3
  \sigma_1}{\theta}$  & $0$ & $\bar{\nu}$ & $- 6$ & $e^-$ & $6$ & $e^+$ & $-
  3$\\
  \hline
  $\psi_{\bar{1} \bar{1} \bar{2}}$ &  $( 3, 2 )$  &  6  &  $\frac{- 2 \sigma_1
  - \sigma_2}{\theta}$  & $1$ & $q$ & $- 5$ &  & $7$ &  & $- 2$\\
  $\psi_{\bar{1} \bar{1} \bar{3}}$ &  $( 3, 2 )$  &  6  &  $\frac{- 2 \sigma_1
  - \sigma_3}{\theta}$  & $1$ & $q$ & $- 5$ &  & $1$ & $q$ & $- 2$\\
  \hline
  $\psi_{\bar{1} \bar{1} \bar{4}}$ & $( 3, 1 )$ & 3 & $\frac{- 2 \sigma_1 -
  \sigma_4}{\theta}$ & $1$ &  & $- 2$ & $d$ & $4$ & $u$ & $- 2$\\
  $\psi_{\bar{1} \bar{1} \bar{5}}$ &  $( 3, 1 )$  &  3  &  $\frac{- 2 \sigma_1
  - \sigma_5}{\theta}$  & $- 5$ &  & $4$ & $u$ & $- 2$ & $d$ & $1$\\
  \hline
  $\psi_{\bar{1} \bar{2} \bar{2}}$ &  $( \bar{3}, 1 )$  &  3  &  $\frac{-
  \sigma_1 - 2 \sigma_2}{\theta}$  & $2$ & $\bar{d}$ & $- 4$ & $\bar{u}$ & $8$
  &  & $- 1$\\
  $\psi_{\bar{1} \bar{3} \bar{3}}$ &  $( \bar{3}, 1 )$  &  3  &  $\frac{-
  \sigma_1 - 2 \sigma_3}{\theta}$  & $2$ & $\bar{d}$ & $- 4$ & $\bar{u}$ & $-
  4$ & $\bar{u}$ & $- 1$\\
  \hline
  $\psi_{\bar{1} \bar{2} \bar{3}}$ &  $( \bar{3}, 3 ) + ( \bar{3}, 1 )$  &  12
  &  $\frac{- \sigma_1 - \sigma_2 - \sigma_3}{\theta}$  & $2$ & $\bar{d}$ &
  $- 4$ & $\bar{u}$ & $2$ & $\bar{d}$ & $- 1$\\
  \hline
  $\psi_{\bar{1} \bar{2} \bar{4}}$ & $\left( \bar{3}, 2 \right)$ & 6 &
  $\frac{- \sigma_1 - \sigma_2 - \sigma_4}{\theta}$ & $2$ &  & $- 1$ &
  $\bar{q}$ & $5$ &  & $- 1$\\
  $\psi_{\bar{1} \bar{2} \bar{5}}$ & $\left( \bar{3}, 2 \right)$ &  6  &
  $\frac{- \sigma_1 - \sigma_2 - \sigma_5}{\theta}$  & $- 4$ &  & $5$ &  & $-
  1$ & $\bar{q}$ & $2$\\
  $\psi_{\bar{1} \bar{3} \bar{4}}$ & $\left( \bar{3}, 2 \right)$ & 6 &
  $\frac{- \sigma_1 - \sigma_3 - \sigma_4}{\theta}$ & $2$ &  & $- 1$ &
  $\bar{q}$ & $- 1$ & $\bar{q}$ & $- 1$\\
  $\psi_{\bar{1} \bar{3} \bar{5}}$ & $\left( \bar{3}, 2 \right)$ &  6  &
  $\frac{- \sigma_1 - \sigma_3 - \sigma_5}{\theta}$  & $- 4$ &  & $5$ &  & $-
  7$ &  & $2$\\
  \hline
  $\psi_{\bar{1} \bar{4} \bar{5}}$ &  $( \bar{3}, 1 )$  &  3  &  $\frac{-
  \sigma_1 - \sigma_4 - \sigma_5}{\theta}$  & $- 4$ & $\bar{u}$ & $8$ &  & $-
  4$ & $\bar{u}$ & $2$\\
  \hline
  $\psi_{\bar{2} \bar{2} \bar{3}}$ &  $( 1, 2 )$  &  2  &  $\frac{- 2 \sigma_2
  - \sigma_3}{\theta}$  & $3$ & $\bar{l}$ & $- 3$ & $l$ & $3$ & $\bar{l}$ &
  $0$\\
  $\psi_{\bar{2} \bar{3} \bar{3}}$ &  $( 1, 2 )$  &  2  &  $\frac{- \sigma_2 -
  2 \sigma_3}{\theta}$  & $3$ & $\bar{l}$ & $- 3$ & $l$ & $- 3$ & $l$ & $0$\\
  \hline
  $\psi_{\bar{2} \bar{2} \bar{4}}$ & $\left( 1, 1 \right)$ & 1 & $\frac{- 2
  \sigma_2 - \sigma_4}{\theta}$ & $3$ &  & $0$ & $\bar{\nu}$ & $6$ & $e^+$ &
  $0$\\
  $\psi_{\bar{2} \bar{2} \bar{5}}$ &  $( 1, 1 )$  &  1  &  $\frac{- 2 \sigma_2
  - \sigma_5}{\theta}$  & $- 3$ &  & $6$ & $e^+$ & $0$ & $\bar{\nu}$ & $3$\\
  $\psi_{\bar{3} \bar{3} \bar{4}}$ & $\left( 1, 1 \right)$ & 1 & $\frac{- 2
  \sigma_3 - \sigma_4}{\theta}$ & $3$ &  & $0$ & $\bar{\nu}$ & $- 6$ & $e^-$ &
  $0$\\
  $\psi_{\bar{3} \bar{3} \bar{5}}$ & $\left( 1, 1 \right)$ &  1  &  $\frac{- 2
  \sigma_3 - \sigma_5}{\theta}$  & $- 3$ &  & $6$ & $e^+$ & $- 12$ &  & $3$\\
  \hline
  $\psi_{\bar{2} \bar{3} \bar{4}}$ & $\left( 1, 3 \right) + \left( 1, 1
  \right)$ & 4 & $\frac{- \sigma_2 - \sigma_3 - \sigma_4}{\theta}$ & $3$ &  &
  $0$ & $\bar{\nu}$ & $0$ & $\bar{\nu}$ & $0$\\
  $\psi_{\bar{2} \bar{3} \bar{5}}$ & $\left( 1, 3 \right) + \left( 1, 1
  \right)$ &  4  &  $\frac{- \sigma_2 - \sigma_3 - \sigma_5}{\theta}$  & $- 3$
  &  & $6$ & $e^+$ & $- 6$ & $e^-$ & $3$\\
  \hline
  $\psi_{\bar{2} \bar{4} \bar{5}}$ &  $( 1, 2 )$  &  2  &  $\frac{- \sigma_2 -
  \sigma_4 - \sigma_5}{\theta}$  & $- 3$ & $l$ & $9$ &  & $- 3$ & $l$ & $3$\\
  $\psi_{\bar{3} \bar{4} \bar{5}}$ &  $( 1, 2 )$  &  2  &  $\frac{- \sigma_3 -
  \sigma_4 - \sigma_5}{\theta}$  & $- 3$ & $l$ & $9$ &  & $- 9$ &  & $3$\\
  \hline
\end{tabular}
\caption{\label{the states in the 84 bar in the second case}
The states in the $ \overline{\mathbf{84}} $ for the $ \mathrm{SU}\left(3
\right) \times \left(\mathrm{SU}\left(2\right)\right)^2 \times \left(
\mathrm{U}\left(1\right)\right)^4 $ case.}
\end{center}
\end{table}

There are nine $\mathrm{SU} \left( 3 \right) \times \mathrm{SU} \left( 2
\right)_{\mathrm{diag}}$ singlets, plus their complex conjugates, whose $U
\left( 1 \right)$ charges do not vanish identically.  However, only three of
the $U \left( 1 \right)$ charges of these nine $\mathrm{SU} \left( 3 \right)
\times \mathrm{SU} \left( 2 \right)_{\mathrm{diag}}$ singlets are linearly
independent, so it is not possible to raise the masses of more than two of the
three unwanted $U \left( 1 \right)$'s as much as required, without breaking
$\mathrm{SU} \left( 3 \right) \times \mathrm{SU} \left( 2 \right)_{\mathrm{diag}}
\times U \left( 1 \right)_Y$, and without relying on Witten's Higgs mechanism.

I did a computer search to determine whether the number of distinct choices of
$U \left( 1 \right)_Y$, such that there is at least one set of $q$, $u$, $d$,
$l$, and $e$ states with the correct relative $Y$ values, and such that two
$\left( 1, 1 \right)$ states of $\mathrm{SU} \left( 3 \right) \times \mathrm{SU}
\left( 2 \right)_{\mathrm{diag}}$, with independent $U \left( 1 \right)_Y$
charges, have $Y = 0$, is finite or infinite.  Specifically, I generated all
sets of integer-valued $\left( \sigma_1, \sigma_2, \sigma_3, \sigma_4
\right)$, in order of increasing $\left| \sigma_1 \right| + \left| \sigma_2
\right| + \left| \sigma_3 \right| + \left| \sigma_4 \right|$, up to $\left|
\sigma_1 \right| + \left| \sigma_2 \right| + \left| \sigma_3 \right| + \left|
\sigma_4 \right| = 300$, with $\sigma_5$ fixed by (\ref{tracelessness in
second case}), and tested for the existence of at least one set of $\left( 3,
2 \right)$, $\left( 3, 1 \right)$, $\left( 3, 1 \right)$, $\left( 1, 2
\right)$, and $\left( 1, 1 \right)$ states of $\mathrm{SU} \left( 3 \right)
\times \mathrm{SU} \left( 2 \right)_{\mathrm{diag}}$, with $U \left( 1
\right)_Y$ charges in the ratios $1, 4, - 2, \pm 3, \pm 6$, respectively.  The
result was that, excluding $\left( \sigma_1, \sigma_2, \sigma_3, \sigma_4
\right)$ with greatest common divisor $> 1$, and solutions related to
solutions already found, by multiplying by $- 1$, or by swapping $\sigma_2$
and $\sigma_3$, or by swapping $\sigma_4$ and $\sigma_5$, thirteen distinct
solutions were found with $\left| \sigma_1 \right| + \left| \sigma_2 \right| +
\left| \sigma_3 \right| + \left| \sigma_4 \right| \leq 10$, and no new
solutions were found with $11 \leq \left| \sigma_1 \right| + \left| \sigma_2
\right| + \left| \sigma_3 \right| + \left| \sigma_4 \right| \leq 300$.  Thus
it looks likely that the thirteen distinct solutions, found with $\left|
\sigma_1 \right| + \left| \sigma_2 \right| + \left| \sigma_3 \right| + \left|
\sigma_4 \right| \leq 10$, are the only distinct solutions.

All thirteen solutions were found to satisfy the requirement that at least one
of $\left( \sigma_2 - \sigma_3 \right)$, $\left( - \sigma_2 - \sigma_3 -
\sigma_4 \right)$, and $\left( - \sigma_2 - \sigma_3 - \sigma_5 \right)$ is
zero, so that $\left( \mathrm{SU} \left( 2 \right) \right)^2$ can be broken to
$\mathrm{SU} \left( 2 \right)_{\mathrm{diag}}$, without breaking $\mathrm{SU} \left(
3 \right) \times \mathrm{SU} \left( 2 \right)_{\mathrm{diag}} \times U \left( 1
\right)_Y$.  Furthermore, all thirteen solutions were found to admit a choice
of a set of $q$, $u$, $d$, $l$, and $e$ states with the correct relative $Y$
values, such that there exists a $U \left( 1 \right)_B$, defined by a
different set of $\sigma_i$, whose couplings to that set of $q$, $u$, $d$,
$l$, and $e$ states are proportional to their baryon number, so that there is
a chance of stabilizing the proton by a version of the Aranda-Carone
mechanism.  For a given set of $q$, $u$, $d$, $l$, and $e$ states, the
requirement for such a $U \left( 1 \right)_B$ to exist is four homogeneous
linear equations for $\left( \sigma_1, \sigma_2, \sigma_3, \sigma_4 \right)$,
and is thus equivalent to the vanishing of the determinant of the matrix of
the coefficients of these equations.

To try to find out if any of the thirteen solutions might be physically
equivalent to one another, I calculated several numerical properties of each
solution, to see if they distinguished between the solutions.  Specifically, I
made an arbitrary, but fixed, choice of one of each charge conjugate pair of
$\left( 1, 2 \right)$ states, to include in the tests, and an arbitrary, but
fixed, choice of one of each charge conjugate pair of $\left( 1, 1 \right)$
states, not in the $\mathrm{SU} \left( 3 \right) \times \left( \mathrm{SU} \left(
2 \right) \right)^2 \times \left( \mathrm{U}
\left( 1 \right) \right)^4$ subgroup, to
include in the tests, and then calculated $N$, the number of distinct possible
choices of a set of $q$, $u$, $d$, $l$, and $e$ states with the correct
relative $Y$ values, and $N_B$, the number of distinct possible choices of a
set of $q$, $u$, $d$, $l$, and $e$ states with the correct relative $Y$
values, that admit the existence of a $U \left( 1 \right)_B$ coupling to their
baryon number.  And, defining integer-valued charges, for these integer-valued
$\left( \sigma_1, \sigma_2, \sigma_3, \sigma_4, \sigma_5 \right)$, by the
numerators in the fourth columns of Tables \ref{states in the 80 in the second
case} and \ref{the states in the 84 bar in the second case}, I calculated
$n_q$, the number of $\left( 3, 2 \right)$ states with $Y = 1$; $x_q$, the
number of $\left( 3, 2 \right)$ states with $Y = - 1$; $n_u$, the number of
$\left( 3, 1 \right)$ states with $Y = 4$; $x_u$, the number of $\left( 3, 1
\right)$ states with $Y = - 4$; $n_d$, the number of $\left( 3, 1 \right)$
states with $Y = - 2$; $x_d$, the number of $\left( 3, 1 \right)$ states with
$Y = 2$; $n_l$, the number of $\left( 1, 2 \right)$ states tested with $Y =
\pm 3$; $n_e$, the number of $\left( 1, 1 \right)$ states tested with $Y = \pm
6$; $n_{\nu}$, the number of $\left( 1, 1 \right)$ states tested with $Y = 0$;
and $n_{\mathrm{diag}}$, the number of $\left( \sigma_2 - \sigma_3 \right)$,
$\left( - \sigma_2 - \sigma_3 - \sigma_4 \right)$, and $\left( - \sigma_2 -
\sigma_3 - \sigma_5 \right)$ that are zero.

The result was that the thirteen solutions fell into three groups, with all
these numerical quantities, and also $\theta^2$, having the same values, for
all the members of each group.  Thus it seems possible that there might be
just three physically distinct solutions, one from each group.  I have
tabulated the $Y$ values for one representative solution from each group, in
Tables \ref{states in the 80 in the second case} and \ref{the states in the 84
bar in the second case}.

The solutions in the first group are $\left( \sigma_1, \sigma_2, \sigma_3,
\sigma_4, \sigma_5 \right) = ( 0, - 1, - 1, - 1, 5 )$, \\
$( 1, 0, - 3, 0, 3 )$,
$( - 2, 0, 0, 3, 3 )$, $( - 1, 1, - 2, 1, 4 )$, and $( 2, - 2, - 2, 1, 1 )$.
They all have $N = 48$, $N_B = 20$, $\theta^2 = 60$, $n_q = 4$, $x_q = 0$,
$n_u = 1$, $x_u = 0$, $n_d = 3$, $x_d = 0$, $n_l = 4$, $n_e = 1$, $n_{\nu} =
2$, and $n_{\mathrm{diag}} = 1$.  The $Y$ values for $( 0, - 1, - 1, - 1, 5 )$
are tabulated in Tables \ref{states in the 80 in the second case} and \ref{the
states in the 84 bar in the second case} as $Y_1$.

The solutions in the second group are $\left( \sigma_1, \sigma_2, \sigma_3,
\sigma_4, \sigma_5 \right) = ( 2, 1, 1, - 2, - 8 )$, \\
$( - 2, 3, 3, 0, - 6 )$, and
$( 0, - 1, 5, - 4, - 4 )$.  They all have $N = 400$, $N_B = 208$, $\theta^2 =
168$, $n_q = 4$, $x_q = 0$, $n_u = 5$, $x_u = 0$, $n_d = 1$, $x_d = 0$, $n_l =
4$, $n_e = 5$, $n_{\nu} = 4$, and $n_{\mathrm{diag}} = 2$.  The $Y$ values for
$( 2, 1, 1, - 2, - 8 )$ are tabulated in Tables \ref{states in the 80 in the
second case} and \ref{the states in the 84 bar in the second case} as $Y_2$.

The solutions in the third group are $\left( \sigma_1, \sigma_2, \sigma_3,
\sigma_4, \sigma_5 \right) = \left( - 2, - 3, 3, 0, 6 \right)$, \\
$( 0, - 1, -
1, - 4, 8 )$, $( - 4, 1, 1, 4, 4 )$, $( 2, 1, - 5, - 2, 4 )$, and $( 4, - 3, -
3, 0, 0 )$.  They all have $N = 1296$, $N_B = 592$, $\theta^2 = 168$, $n_q =
4$, $x_q = 0$, $n_u = 3$, $x_u = 0$, $n_d = 3$, $x_d = 0$, $n_l = 6$, $n_e =
6$, $n_{\nu} = 2$, and $n_{\mathrm{diag}} = 1$.  The $Y$ values for $\left( - 2,
- 3, 3, 0, 6 \right)$ are tabulated in Tables \ref{states in the 80 in the
second case} and \ref{the states in the 84 bar in the second case} as $Y_3$.
For this example, we can choose $3 B = \left( 1, 0, 0, 0, - 3 \right)$, which
gives the correct baryon number, except for the first four states of the
$\overline{\mathbf{8} \mathbf{4}}$, and states involving $\sigma_5$, other
than the fifth state of the $\overline{\mathbf{8} \mathbf{4}}$.  The $B$
values for this choice of $B$ are tabulated in Tables \ref{states in the 80 in
the second case} and \ref{the states in the 84 bar in the second case} as
$B_3$.

To determine $\sin^2 \theta_W$ at unification, for the three groups of models,
we recall that the Standard Model Higgs field is a $\left( 1, 2 \right)$ state
of $\mathrm{SU} \left( 3 \right) \times \mathrm{SU} \left( 2
\right)_{\mathrm{diag}}$, with $Y = 1$.  In the examples in Tables \ref{states
in the 80 in the second case} and \ref{the states in the 84 bar in the second
case}, this could, for example, be an extra-dimensional component $A_{A 245}$
of $A_{U 245}$, for the $Y_1$ case, $A_{A 2 \bar{4}}$ of $A_{U 2 \bar{4}}$,
for the $Y_2$ case, and $A_{A 3 \bar{4}}$ of $A_{U 3 \bar{4}}$, for the $Y_3$
case.  Denoting this field by $\phi_i$, we find, by the methods of the
preceding subsection, that for the $Y_2$ and $Y_3$ cases, its covariant
derivative, times $i$, becomes, at low energies:
\begin{equation}
  \label{covariant derivative of Higgs field in the second case} i
  \partial_{\mu} \phi_i + g \frac{1}{\sqrt{2}} B_{\mu a} \frac{1}{2} \left(
  \sigma_a \right)_{ik} \phi_k + g \frac{3}{\sqrt{168}} A_{\mu 4} \phi_i
\end{equation}
where $A_{\mu 4}$ is the gauge field of $U \left( 1 \right)_Y$, in a notation
similar to the previous subsection, and $B_{\mu a} = \frac{1}{\sqrt{2}} \left(
A_{\mu 2 a} + A_{\mu 3 a} \right)$ is the gauge field of $\mathrm{SU} \left( 2
\right)_{\mathrm{diag}}$.  While from equation (117) of Rosner's review of the
Standard Model {\cite{Rosner}}, the standard covariant derivative, times $i$,
on the Standard Model Higgs field is:
\begin{equation}
  \label{covariant derivative on Standard Model Higgs field for the second
  case} i \partial_{\mu} \phi_i + g_{\mathrm{Rosner}} B_{\mu a} \frac{1}{2}
  \left( \sigma_a \right)_{ik} \phi_k + g'_{\mathrm{Rosner}} \frac{Y}{2}
  A_{\mu 4} \phi_i
\end{equation}
Now, as noted above, the Standard Model Higgs field has $Y = 1$. Thus we see
that, for the $Y_2$ and $Y_3$ cases:
\begin{equation}
  \label{Rosners coupling constants for the second case} g_{\mathrm{Rosner}} =
  g \frac{1}{\sqrt{2}}, \hspace{4em} g'_{\mathrm{Rosner}} = g
  \frac{6}{\sqrt{168}}
\end{equation}
And by definition, $\tan \theta_W =
\frac{g'_{\mathrm{Rosner}}}{g_{\mathrm{Rosner}}}$.  Hence we find that, for
the $Y_2$ and $Y_3$ cases, $\sin^2 \theta_W = \frac{3}{10} = 0.300$ at
unification, which is roughly halfway between the value $\simeq 0.23$ observed
at $m_Z$, and the value $\frac{3}{8} = 0.375$ found in conventional grand
unification {\cite{Georgi Glashow}}, and reasonably consistent with the
unification of the $\mathrm{SU} \left( 3 \right)$ and $\mathrm{SU} \left( 2
\right)_{\mathrm{diag}}$ coupling constants at around 145 TeV, in the absence of
accelerated unification.  On the other hand, for the $Y_1$ case, the
$\sqrt{168}$, in (\ref{covariant derivative of Higgs field in the second
case}) and (\ref{Rosners coupling constants for the second case}), gets
replaced by $\sqrt{60}$, which gives $\sin^2 \theta_W = \frac{6}{11} \simeq
0.545$ at unification, so the $Y_1$ case does not seem very likely.

Let us now suppose that we have found a smooth compact quotient of
$\mathbf{C} \mathbf{H}^3$ or $\mathbf{H}^6$, and a set of Hodge - de
Rham harmonic two-forms embedded in the $E_8$ Cartan subalgebra as above, such
that the net number of chiral zero modes of each of the left-handed states of
one generation of the Standard Model, as in Table \ref{T1}, is three, and the
net number of chiral zero modes of each fermion state not in the Standard
Model, is zero, and that $\mathrm{SU} \left( 2 \right)^2$ can be broken to
$\mathrm{SU} \left( 2 \right)_{\mathrm{diag}}$, in a topologically stabilized
manner, by a ``monopole'', embedded in $E_8$ in one or more of whichever of
the states $\psi_{2 \bar{3}}$, $\psi_{\bar{2} \bar{3} \bar{4}}$, and
$\psi_{\bar{2} \bar{3} \bar{5}}$ have vanishing $U \left( 1 \right)_Y$ charge,
in the example under consideration, without spoiling this.  Then it seems
reasonable to expect that the Hodge - de Rham harmonic two-forms will lead to
masses $\sim$ a TeV for all chiral zero modes that can be matched in
left-handed and right-handed pairs, so that the only light fermions will be
the three generations of Standard Model fermions, except possibly for one or
more light singlet neutrino states, which could obtain very small masses by
the generalized seesaw mechanism to be discussed in subsection
\ref{Generalized seesaw mechanism}.

Let us now consider an arbitrary proton decay process, proceeding via a
dimension 6 term in the Standard Model effective action, such as
$\frac{qqql}{\Lambda^2}$, $\frac{d^c d^c u^c e^c}{\Lambda^2}$,
$\frac{\overline{e^c} \overline{u^c} qq}{\Lambda^2}$, or $\frac{\overline{d^c}
\overline{u^c} ql}{\Lambda^2}$ {\cite{Wikipedia Proton decay}}, with the
$\mathrm{SU} \left( 3 \right)$ and $\mathrm{SU} \left( 2 \right)_{\mathrm{diag}}$
indices contracted in an appropriate manner, where $\Lambda$ is an effective
cutoff, that determines the size of the term.  Then for any four specific
states from Tables \ref{states in the 80 in the second case} and \ref{the
states in the 84 bar in the second case}, that have nonvanishing amplitudes in
those four types of Standard Model state, the condition for the existence of a
$U \left( 1 \right)_B$, that couples as a nonzero multiple of baryon number,
just on those four states, is three homogeneous linear equations on the four
linearly independent $\sigma_i$, so is always satisfied.  Thus those parts of
the arguments of Aranda and Carone {\cite{Aranda Carone}}, that depend only on
the existence of such a $U \left( 1 \right)_B$, would seem to suggest that the
contribution of those four states, to the corresponding term in the Standard
Model effective action, will be suppressed.  And since this argument applies
to all sets of states from Tables \ref{states in the 80 in the second case}
and \ref{the states in the 84 bar in the second case}, that have nonvanishing
amplitudes in the Standard Model fermion fields in the effective action term
concerned, we expect the same suppression to apply to the overall coefficient
of that term in the effective action, leading to a large value of the
effective cutoff $\Lambda$, even though the relevant $U \left( 1 \right)_B$
may be different, for different relevant sets of states from Tables
\ref{states in the 80 in the second case} and \ref{the states in the 84 bar in
the second case}.

Of course, it was not necessary to require that two of the $\mathrm{SU} \left( 3
\right) \times \mathrm{SU} \left( 2 \right)$ singlets, with independent $U
\left( 1 \right)$ charges, have $Y = 0$, since the unwanted $U \left( 1
\right)$'s will become massive by Witten's Higgs mechanism, provided that none
of them is orthogonal to all the Hodge - de Rham monopoles in the $E_8$ Cartan
subalgebra.  So additional solutions might exist, such that the largest number
of $\mathrm{SU} \left( 3 \right) \times \mathrm{SU} \left( 2 \right)$ singlets,
with linearly independent $U \left( 1 \right)$ charges, that have $Y = 0$, is
less than two.

\subsection{Models where the Abelian Hodge - de Rham monopoles break $E_8$ to
$\mathrm{SU} \left( 3 \right) \times \mathrm{SU} \left( 2 \right) \times
\left( \mathrm{U} \left( 1 \right) \right)^5$}
\label{SU 3 cross SU 2 cross U 1 to the fifth}

I shall now consider some models where the Hodge - de Rham harmonic two-forms
break $E
8$ to $\mathrm{SU} \left( 3 \right) \times \mathrm{SU} \left( 2 \right) \times
\left( \mathrm{U} \left( 1 \right) \right)^5$.  An element of the $E_8$ Cartan
subalgebra, and hence of the $\mathrm{SU} \left( 9 \right)$ Cartan subalgebra,
that can have a vacuum expectation value without breaking this subgroup of $E
8$, is, in the $\mathrm{SU} \left( 9 \right)$ fundamental, a diagonal matrix,
with diagonal matrix elements
\begin{equation}
  \label{Abelian vacuum gauge fields in third case} \left( \sigma_1, \sigma_1,
  \sigma_1, \sigma_2, \sigma_2, \sigma_3, \sigma_4, \sigma_5, \sigma_6 \right),
\end{equation}
such that:
\begin{equation}
  \label{tracelessness in third case} 3 \sigma_1 + 2 \sigma_2 + \sigma_3 +
  \sigma_4 + \sigma_5 + \sigma_6 = 0
\end{equation}

The states in the $\mathbf{8} \mathbf{0}$, omitting the states in the
unbroken $\mathrm{SU} \left( 3 \right) \times \mathrm{SU} \left( 2 \right) \times
\left( \mathrm{U} \left( 1 \right) \right)^5$, whose $U \left( 1 \right)$ charges
vanish identically, are shown in Table \ref{states in the 80 in the third
case}, and the states in the $\overline{\mathbf{8} \mathbf{4}}$ are shown
in Table \ref{the states in the 84 bar in the third case}.  There are now
fifteen $\mathrm{SU} \left( 3 \right) \times \mathrm{SU} \left( 2 \right)$
singlets, plus their complex conjugates, whose $U \left( 1 \right)$ charges do
not vanish identically, and the $U \left( 1 \right)$ charges of five of these
fifteen $\mathrm{SU} \left( 3 \right) \times \mathrm{SU} \left( 2 \right)$
singlets are linearly independent, so there is now a possibility of raising
the masses of all four unwanted $U \left( 1 \right)$'s as much as required,
without breaking $\mathrm{SU} \left( 3 \right) \times \mathrm{SU} \left( 2 \right)
\times U \left( 1 \right)_Y$, and without relying on Witten's Higgs mechanism,
by choosing the $\sigma_i$ such that four $\mathrm{SU} \left( 3 \right) \times
\mathrm{SU} \left( 2 \right)$ singlets outside the $E_8$ Cartan subalgebra, with
linearly independent $U \left( 1 \right)$ charges, all have vanishing $U
\left( 1 \right)_Y$ charge, and could thus have vacuum expectation values
without breaking $\mathrm{SU} \left( 3 \right) \times \mathrm{SU} \left( 2 \right)
\times U \left( 1 \right)_Y$.

\begin{table}
\begin{center}
\begin{tabular}{|c|c|c|c|c|c|c|c|c|c|c|}
  \hline
  Block & Multiplet &  & Coupling & 3$Y_1$ &  & $3 B_1$ & $3 Y_2$ &  & $3 Y_3$
  & \\
  \hline
  $\psi_{1 \bar{2}}$ &  (3,2)  &  6  &  $\frac{\sigma_1 - \sigma_2}{\theta}$
  & $- 5$ &  & $1$ & 1 & $q$ & 1 & $q$\\
  \hline
  $\psi_{1 \bar{3}}$ &  (3,1)  &  3  &  $\frac{\sigma_1 - \sigma_3}{\theta}$
  & $- 2$ & $d$ & $1$ & 4 & $u$ & 4 & $u$\\
  $\psi_{1 \bar{4}}$ & (3,1) & 3 & $\frac{\sigma_1 - \sigma_4}{\theta}$ & $-
  2$ & $d$ & $1$ & 4 & $u$ & $- 2$ & $d$\\
  $\psi_{1 \bar{5}}$ & (3,1) & 3 & $\frac{\sigma_1 - \sigma_5}{\theta}$ & $-
  2$ & $d$ & $1$ & 4 & $u$ & $- 2$ & $d$\\
  $\psi_{1 \bar{6}}$ &  (3,1)  &  3  &  $\frac{\sigma_1 - \sigma_6}{\theta}$
  & $- 2$ & $d$ & $4$ & 4 & $u$ & $- 2$ & $d$\\
  \hline
  $\psi_{2 \bar{1}}$ &  $( \bar{3}, 2 )$  &  6  &  $\frac{- \sigma_1 +
  \sigma_2}{\theta}$  & $5$ &  & $- 1$ & $- 1$ & $\bar{q}$ & $- 1$ &
  $\bar{q}$\\
  \hline
  $\psi_{3 \bar{1}}$ &  $( \bar{3}, 1 )$  &  3  &  $\frac{- \sigma_1 +
  \sigma_3}{\theta}$  & $2$ & $\bar{d}$ & $- 1$ & $- 4$ & $\bar{u}$ & $- 4$ &
  $\bar{u}$\\
  $\psi_{4 \bar{1}}$ &  $( \bar{3}, 1 )$  & 3 & $\frac{- \sigma_1 +
  \sigma_4}{\theta}$ & $2$ & $\bar{d}$ & $- 1$ & $- 4$ & $\bar{u}$ & $2$ &
  $\bar{d}$\\
  $\psi_{5 \bar{1}}$ &  $( \bar{3}, 1 )$  & 3 & $\frac{- \sigma_1 +
  \sigma_5}{\theta}$ & $2$ & $\bar{d}$ & $- 1$ & $- 4$ & $\bar{u}$ & $2$ &
  $\bar{d}$\\
  $\psi_{6 \bar{1}}$ &  $( \bar{3}, 1 )$  &  3  &  $\frac{- \sigma_1 +
  \sigma_6}{\theta}$  & $2$ & $\bar{d}$ & $- 4$ & $- 4$ & $\bar{u}$ & $2$ &
  $\bar{d}$\\
  \hline
  $\psi_{2 \bar{3}}$ &  $( 1, 2 )$  &  2  &  $\frac{\sigma_2 -
  \sigma_3}{\theta}$  & $3$ & $\bar{l}$ & $0$ & $3$ & $\bar{l}$ & $3$ &
  $\bar{l}$\\
  $\psi_{2 \bar{4}}$ & $( 1, 2 )$ & 2 & $\frac{\sigma_2 - \sigma_4}{\theta}$ &
  $3$ & $\bar{l}$ & $0$ & $3$ & $\bar{l}$ & $- 3$ & $l$\\
  $\psi_{2 \bar{5}}$ &  $( 1, 2 )$  &  2  &  $\frac{\sigma_2 -
  \sigma_5}{\theta}$  & $3$ & $\bar{l}$ & $0$ & $3$ & $\bar{l}$ & $- 3$ &
  $l$\\
  $\psi_{2 \bar{6}}$ &  $( 1, 2 )$  &  2  &  $\frac{\sigma_2 -
  \sigma_6}{\theta}$  & $3$ & $\bar{l}$ & $3$ & $3$ & $\bar{l}$ & $- 3$ &
  $l$\\
  \hline
  $\psi_{3 \bar{4}}$ & $( 1, 1 )$ & 1 & $\frac{\sigma_3 - \sigma_4}{\theta}$ &
  0 &  & $0$ & 0 &  & $- 6$ & $e^-$\\
  $\psi_{3 \bar{5}}$ & $( 1, 1 )$ &  1  & $\frac{\sigma_3 - \sigma_5}{\theta}$
  & 0 &  & $0$ & 0 &  & $- 6$ & $e^-$\\
  $\psi_{3 \bar{6}}$ & $( 1, 1 )$ &  1  & $\frac{\sigma_3 - \sigma_6}{\theta}$
  & 0 &  & $3$ & 0 &  & $- 6$ & $e^-$\\
  $\psi_{4 \bar{5}}$ & $( 1, 1 )$ &  1  & $\frac{\sigma_4 - \sigma_5}{\theta}$
  & 0 &  & $0$ & 0 &  & 0 & \\
  $\psi_{4 \bar{6}}$ & $( 1, 1 )$ & 1 & $\frac{\sigma_4 - \sigma_6}{\theta}$ &
  0 &  & 3 & 0 &  & 0 & \\
  $\psi_{5 \bar{6}}$ & $( 1, 1 )$ &  1  &  $\frac{\sigma_5 -
  \sigma_6}{\theta}$  & 0 &  & $3$ & 0 &  & 0 & \\
  \hline
  $\psi_{3 \bar{2}}$ &  $( 1, 2 )$  &  2  &  $\frac{- \sigma_2 +
  \sigma_3}{\theta}$  & $- 3$ & $l$ & $0$ & $- 3$ & $l$ & $- 3$ & $l$\\
  $\psi_{4 \bar{2}}$ & $( 1, 2 )$ & 2 & $\frac{- \sigma_2 + \sigma_4}{\theta}$
  & $- 3$ & $l$ & $0$ & $- 3$ & $l$ & 3 & $\bar{l}$\\
  $\psi_{5 \bar{2}}$ &  $\left( 1, 2 \right)$  &  2  &  $\frac{- \sigma_2 +
  \sigma_5}{\theta}$  & $- 3$ & $l$ & $0$ & $- 3$ & $l$ & 3 & $\bar{l}$\\
  $\psi_{6 \bar{2}}$ &  $\left( 1, 2 \right)$  &  2  &  $\frac{- \sigma_2 +
  \sigma_6}{\theta}$  & $- 3$ & $l$ & $- 3$ & $- 3$ & $l$ & 3 & $\bar{l}$\\
  \hline
  $\psi_{4 \bar{3}}$ & $( 1, 1 )$ & 1 & $\frac{- \sigma_3 + \sigma_4}{\theta}$
  & 0 &  & $0$ & 0 &  & 6 & $e^+$\\
  $\psi_{5 \bar{3}}$ & $( 1, 1 )$ & 1 & $\frac{- \sigma_3 + \sigma_5}{\theta}$
  & 0 &  & $0$ & 0 &  & 6 & $e^+$\\
  $\psi_{6 \bar{3}}$ & $( 1, 1 )$ & 1 & $\frac{- \sigma_3 + \sigma_6}{\theta}$
  & 0 &  & $- 3$ & 0 &  & 6 & $e^+$\\
  $\psi_{5 \bar{4}}$ & $( 1, 1 )$ & 1 & $\frac{- \sigma_4 + \sigma_5}{\theta}$
  & 0 &  & $0$ & 0 &  & 0 & \\
  $\psi_{6 \bar{4}}$ & $( 1, 1 )$ & 1 & $\frac{- \sigma_4 + \sigma_6}{\theta}$
  & 0 &  & $- 3$ & 0 &  & 0 & \\
  $\psi_{6 \bar{5}}$ & $( 1, 1 )$ & 1 &  $\frac{- \sigma_5 +
  \sigma_6}{\theta}$  & 0 &  & $- 3$ & 0 &  & 0 & \\
  \hline
\end{tabular}
\vspace{-5.1pt}
\caption{\label{states in the 80 in the third case}
The states in the $ \mathbf{80} $ for the $ \mathrm{SU}\left(3\right)
\times \mathrm{SU}\left(2\right) \times \left( \mathrm{U}\left(1\right)
\right)^5 $ case, omitting the states in the $ \mathrm{SU}\left(3\right)
\times \mathrm{SU}\left(2\right) \times \left( \mathrm{U}\left(1\right)
\right)^5 $ subgroup.}
\end{center}
\end{table}

\begin{table}
\begin{center}
\begin{tabular}{|c|c|c|c|c|c|c|c|c|c|c|}
  \hline
  Block & Multiplet &   & Coupling & 3$Y_1$ &  & $3 B_1$ & $3 Y_2$ &  &
  $3 Y_3$ & \\
  \hline
  $\psi_{\bar{1} \bar{1} \bar{1}}$ &  $( 1, 1 )$  &  1  &  $\frac{- 3
  \sigma_1}{\theta}$  & $6$ & $e^+$ & $- 3$ & $- 6$ & $e^-$ & 0 & \\
  \hline
  $\psi_{\bar{1} \bar{1} \bar{2}}$ &  $( 3, 2 )$  &  6  &  $\frac{- 2 \sigma_1
  - \sigma_2}{\theta}$  & $1$ & $q$ & $- 2$ & $- 5$ &  & $1$ & $q$\\
  \hline
  $\psi_{\bar{1} \bar{1} \bar{3}}$ &  $( 3, 1 )$  &  3  &  $\frac{- 2 \sigma_1
  - \sigma_3}{\theta}$  & 4 & $u$ & $- 2$ & $- 2$ & $d$ & 4 & $u$\\
  $\psi_{\bar{1} \bar{1} \bar{4}}$ & $( 3, 1 )$ & 3 & $\frac{- 2 \sigma_1 -
  \sigma_4}{\theta}$ & 4 & $u$ & $- 2$ & $- 2$ & $d$ & $- 2$ & $d$\\
  $\psi_{\bar{1} \bar{1} \bar{5}}$ & $( 3, 1 )$ & 3 & $\frac{- 2 \sigma_1 -
  \sigma_5}{\theta}$ & 4 & $u$ & $- 2$ & $- 2$ & $d$ & $- 2$ & $d$\\
  $\psi_{\bar{1} \bar{1} \bar{6}}$ &  $( 3, 1 )$  &  3  &  $\frac{- 2 \sigma_1
  - \sigma_6}{\theta}$  & 4 & $u$ & $1$ & $- 2$ & $d$ & $- 2$ & $d$\\
  \hline
  $\psi_{\bar{1} \bar{2} \bar{2}}$ &  $( \bar{3}, 1 )$  &  3  &  $\frac{-
  \sigma_1 - 2 \sigma_2}{\theta}$  & $- 4$ & $\bar{u}$ & $- 1$ & $- 4$ &
  $\bar{u}$ & 2 & $\bar{d}$\\
  \hline
  $\psi_{\bar{1} \bar{3} \bar{4}}$ &  $( \bar{3}, 1 )$  &  3  &  $\frac{-
  \sigma_1 - \sigma_3 - \sigma_4}{\theta}$  & 2 & $\bar{d}$ & $- 1$ & 2 &
  $\bar{d}$ & 2 & $\bar{d}$\\
  $\psi_{\bar{1} \bar{3} \bar{5}}$ &  $( \bar{3}, 1 )$  &  3  &  $\frac{-
  \sigma_1 - \sigma_3 - \sigma_5}{\theta}$  & 2 & $\bar{d}$ & $- 1$ & 2 &
  $\bar{d}$ & 2 & $\bar{d}$\\
  $\psi_{\bar{1} \bar{3} \bar{6}}$ &  $( \bar{3}, 1 )$  &  3  &  $\frac{-
  \sigma_1 - \sigma_3 - \sigma_6}{\theta}$  & 2 & $\bar{d}$ & $2$ & 2 &
  $\bar{d}$ & 2 & $\bar{d}$\\
  $\psi_{\bar{1} \bar{4} \bar{5}}$ & $( \bar{3}, 1 )$ & 3 & $\frac{- \sigma_1
  - \sigma_4 - \sigma_5}{\theta}$ & 2 & $\bar{d}$ & $- 1$ & 2 & $\bar{d}$ & $-
  4$ & $\bar{u}$\\
  $\psi_{\bar{1} \bar{4} \bar{6}}$ & $( \bar{3}, 1 )$ & 3 & $\frac{- \sigma_1
  - \sigma_4 - \sigma_6}{\theta}$ & 2 & $\bar{d}$ & $2$ & 2 & $\bar{d}$ & $-
  4$ & $\bar{u}$\\
  $\psi_{\bar{1} \bar{5} \bar{6}}$ &  $( \bar{3}, 1 )$  &  3  &  $\frac{-
  \sigma_1 - \sigma_5 - \sigma_6}{\theta}$  & 2 & $\bar{d}$ & $2$ & 2 &
  $\bar{d}$ & $- 4$ & $\bar{u}$\\
  \hline
  $\psi_{\bar{1} \bar{2} \bar{3}}$ &  $( \bar{3}, 2 )$  &  6  &  $\frac{-
  \sigma_1 - \sigma_2 - \sigma_3}{\theta}$  & $- 1$ & $\bar{q}$ & $- 1$ & $-
  1$ & $\bar{q}$ & 5 & \\
  $\psi_{\bar{1} \bar{2} \bar{4}}$ & $\left( \bar{3}, 2 \right)$ & 6 &
  $\frac{- \sigma_1 - \sigma_2 - \sigma_4}{\theta}$ & $- 1$ & $\bar{q}$ & $-
  1$ & $- 1$ & $\bar{q}$ & $- 1$ & $\bar{q}$\\
  $\psi_{\bar{1} \bar{2} \bar{5}}$ & $\left( \bar{3}, 2 \right)$ & 6 &
  $\frac{- \sigma_1 - \sigma_2 - \sigma_5}{\theta}$ & $- 1$ & $\bar{q}$ & $-
  1$ & $- 1$ & $\bar{q}$ & $- 1$ & $\bar{q}$\\
  $\psi_{\bar{1} \bar{2} \bar{6}}$ & $\left( \bar{3}, 2 \right)$ &  6  &
  $\frac{- \sigma_1 - \sigma_2 - \sigma_6}{\theta}$  & $- 1$ & $\bar{q}$ & $2$
  & $- 1$ & $\bar{q}$ & $- 1$ & $\bar{q}$\\
  \hline
  $\psi_{\bar{2} \bar{2} \bar{3}}$ &  $( 1, 1 )$  &  1  &  $\frac{- 2 \sigma_2
  - \sigma_3}{\theta}$  & $- 6$ & $e^-$ & $0$ & 0 &  & 6 & $e^+$\\
  $\psi_{\bar{2} \bar{2} \bar{4}}$ & $\left( 1, 1 \right)$ & 1 & $\frac{- 2
  \sigma_2 - \sigma_4}{\theta}$ & $- 6$ & $e^-$ & $0$ & 0 &  & 0 & \\
  $\psi_{\bar{2} \bar{2} \bar{5}}$ &  $( 1, 1 )$  &  1  &  $\frac{- 2 \sigma_2
  - \sigma_5}{\theta}$  & $- 6$ & $e^-$ & $0$ & 0 &  & 0 & \\
  $\psi_{\bar{2} \bar{2} \bar{6}}$ &  $( 1, 1 )$  &  1  &  $\frac{- 2 \sigma_2
  - \sigma_6}{\theta}$  & $- 6$ & $e^-$ & $3$ & 0 &  & 0 & \\
  \hline
  $\psi_{\bar{2} \bar{3} \bar{4}}$ & $\left( 1, 2 \right)$ & 2 & $\frac{-
  \sigma_2 - \sigma_3 - \sigma_4}{\theta}$ & $- 3$ & $l$ & $0$ & 3 & $\bar{l}$
  & 3 & $\bar{l}$\\
  $\psi_{\bar{2} \bar{3} \bar{5}}$ & $\left( 1, 2 \right)$ & 2 & $\frac{-
  \sigma_2 - \sigma_3 - \sigma_5}{\theta}$ & $- 3$ & $l$ & $0$ & 3 & $\bar{l}$
  & 3 & $\bar{l}$\\
  $\psi_{\bar{2} \bar{3} \bar{6}}$ & $\left( 1, 2 \right)$ & 2 & $\frac{-
  \sigma_2 - \sigma_3 - \sigma_6}{\theta}$ & $- 3$ & $l$ & $3$ & 3 & $\bar{l}$
  & 3 & $\bar{l}$\\
  $\psi_{\bar{2} \bar{4} \bar{5}}$ & $\left( 1, 2 \right)$ & 2 & $\frac{-
  \sigma_2 - \sigma_4 - \sigma_5}{\theta}$ & $- 3$ & $l$ & $0$ & 3 & $\bar{l}$
  & $- 3$ & $l$\\
  $\psi_{\bar{2} \bar{4} \bar{6}}$ & $\left( 1, 2 \right)$ & 2 & $\frac{-
  \sigma_2 - \sigma_4 - \sigma_6}{\theta}$ & $- 3$ & $l$ & $3$ & 3 & $\bar{l}$
  & $- 3$ & $l$\\
  $\psi_{\bar{2} \bar{5} \bar{6}}$ & $\left( 1, 2 \right)$ &  2  &  $\frac{-
  \sigma_2 - \sigma_5 - \sigma_6}{\theta}$  & $- 3$ & $l$ & $3$ & 3 &
  $\bar{l}$ & $- 3$ & $l$\\
  \hline
  $\psi_{\bar{3} \bar{4} \bar{5}}$ & $( 1, 1 )$ & 1 & $\frac{- \sigma_3 -
  \sigma_4 - \sigma_5}{\theta}$ & 0 &  & $0$ & 6 & $e^+$ & 0 & \\
  $\psi_{\bar{3} \bar{4} \bar{6}}$ & $( 1, 1 )$ & 1 & $\frac{- \sigma_3 -
  \sigma_4 - \sigma_6}{\theta}$ & 0 &  & $3$ & 6 & $e^+$ & 0 & \\
  $\psi_{\bar{3} \bar{5} \bar{6}}$ & $( 1, 1 )$ & 1 & $\frac{- \sigma_3 -
  \sigma_5 - \sigma_6}{\theta}$ & 0 &  & $3$ & 6 & $e^+$ & 0 & \\
  $\psi_{\bar{4} \bar{5} \bar{6}}$ &  $( 1, 1 )$  &  1  &  $\frac{- \sigma_4 -
  \sigma_5 - \sigma_6}{\theta}$  & 0 &  & $3$ & 6 & $e^+$ & $- 6$ & $e^-$ \\
  \hline
\end{tabular}
\vspace{-5.5pt}
\caption{\label{the states in the 84 bar in the third case}
The states in the $ \overline{\mathbf{84}} $ for the $ \mathrm{SU}\left(3
\right) \times \mathrm{SU}\left(2\right) \times \left(\mathrm{U}\left(1
\right)\right)^5 $ case.}
\end{center}
\end{table}

I did a computer search through all
$\frac{15!}{4!11!} = 1365$ choices of which four of the fifteen $\mathrm{SU}
\left( 3 \right) \times \mathrm{SU} \left( 2 \right)$ singlets should be set to
have $Y = 0$, to determine which choices led to the existence of at least one
set of $q$, $u$, $d$, $l$, and $e$ states with the correct $Y$ values, such
that there exists a $U \left( 1 \right)_B$, defined by a different set of
$\sigma_i$, whose couplings to at least one set of these states are
proportional to their baryon number, so that there is a chance of stabilizing
the proton by a version of the Aranda-Carone mechanism.  There were only six
distinct solutions, four of which are related by permutations of $\sigma_3$,
$\sigma_4$, $\sigma_5$, and $\sigma_6$.  Taking only one of these four, the
three solutions are:
\begin{equation}
  3 \label{Y sub 1 for third case} Y_1 =  \left( - 2, 3, 0, 0, 0, 0 \right)
\end{equation}
\begin{equation}
  3 \label{Y sub 2 for third case} Y_2 =  \left( 2, 1, - 2, - 2, - 2, - 2
  \right)
\end{equation}
\begin{equation}
  3 \label{Y sub 3 for third case} Y_3 =  \left( 0, - 1, - 4, 2, 2, 2 \right)
\end{equation}
All three of these have $\theta^2 = 60$, so by the same method as in the
previous two subsections, we find $\sin^2 \theta_W = \frac{3}{8}$ at
unification, as for SU(5) grand unification, so unification depends
entirely on the accelerated unification mechanism {\cite{DDG2, DDG2,
Arkani-Hamed Cohen Georgi}}.  For $Y_1$, we could choose $3 B = \left( 1, 0,
0, 0, 0, - 3 \right)$, and the resulting values of $ B $ are tabulated as $ B_1
$ in Tables \ref{states in the 80 in the third case} and \ref{the states in the
84 bar in the third case}.

The number of states of each type, for each of the three choices of $Y$, are
given in Table \ref{number of states of each type for the third case}.  The
total number of states of each type is the same for all three choices, so it
seems possible that the three different choices of $Y$ might be physically
equivalent.

\begin{table}
\begin{center}
\begin{tabular}{|c|c|c|c|c|c|c|c|c|c|c|c|c|c|c|c|c|c|}
  \hline
  & $q$ & $\bar{q}$ & $u$ & $\bar{u}$ & $d$ & $\bar{d}$ & $l$ & $\bar{l}$ &
  $e^+$ & $e^-$ & $\nu$ & $q^5$ & $\bar{q}^5$ & $8$ & $3$ & $Y$ & total\\
  \hline
  $Y_1$ in 80 & 0 & 0 & 0 & 0 & 4 & 4 & 4 & 4 & 0 & 0 & 12 & 1 & 1 & 1 & 1 & 5
  & 80\\
  $Y_1$ in $\overline{84}$ & 1 & 4 & 4 & 1 & 0 & 6 & 6 & 0 & 1 & 4 & 4 & 0 & 0
  & 0 & 0 & 0 & 84\\
  $Y_1$ in 84 & 4 & 1 & 1 & 4 & 6 & 0 & 0 & 6 & 4 & 1 & 4 & 0 & 0 & 0 & 0 & 0
  & 84\\
  $Y_1$ total & 5 & 5 & 5 & 5 & 10 & 10 & 10 & 10 & 5 & 5 & 20 & 1 & 1 & 1 & 1
  & 5 & 248\\
  \hline
  $Y_2$ in 80 & 1 & 1 & 4 & 4 & 0 & 0 & 4 & 4 & 0 & 0 & 12 & 0 & 0 & 1 & 1 & 5
  & 80\\
  $Y_2$ in $\overline{84}$ & 0 & 4 & 0 & 1 & 4 & 6 & 0 & 6 & 4 & 1 & 4 & 0 & 1
  & 0 & 0 & 0 & 84\\
  $Y_2$ in 84 & 4 & 0 & 1 & 0 & 6 & 4 & 6 & 0 & 1 & 4 & 4 & 1 & 0 & 0 & 0 & 0
  & 84\\
  $Y_2$ total & 5 & 5 & 5 & 5 & 10 & 10 & 10 & 10 & 5 & 5 & 20 & 1 & 1 & 1 & 1
  & 5 & 248\\
  \hline
  $Y_3$ in 80 & 1 & 1 & 1 & 1 & 3 & 3 & 4 & 4 & 3 & 3 & 6 & 0 & 0 & 1 & 1 & 5
  & 80\\
  $Y_3$ in $\overline{84}$ & 1 & 3 & 1 & 3 & 3 & 4 & 3 & 3 & 1 & 1 & 7 & 0 & 1
  & 0 & 0 & 0 & 84\\
  $Y_3$ in 84 & 3 & 1 & 3 & 1 & 4 & 3 & 3 & 3 & 1 & 1 & 7 & 1 & 0 & 0 & 0 & 0
  & 84\\
  $Y_3$ total & 5 & 5 & 5 & 5 & 10 & 10 & 10 & 10 & 5 & 5 & 20 & 1 & 1 & 1 & 1
  & 5 & 248\\
  \hline
\end{tabular}
\caption{\label{number of states of each type for the third case}
The numbers of each type of state, for the three choices of $ Y $ in the
$ \mathrm{SU}\left(3\right) \times \mathrm{SU}\left(2\right) \times \left(
\mathrm{U}\left(1\right)\right)^5 $ case.}
\end{center}
\end{table}

We note that all the fermion states in the $E_8$ fundamental, that are
not in the $\mathrm{SU} \left( 3 \right) \times \mathrm{SU} \left( 2 \right)
\times \left( \mathrm{U}
\left( 1 \right) \right)^5$ subgroup, and can thus be given a
nonzero net number of chiral zero modes by the Hodge - de Rham harmonic
two-forms in the $E_8$ Cartan subalgebra, are now either Standard Model
fermions, as in Table \ref{T1}, or singlet neutrinos, apart from the single
$q^5$ state with $Y = - \frac{5}{3}$, and the single $\bar{q}^5$ state with $Y
= \frac{5}{3}$.  It is well known that the possible sets of left-handed chiral
fermions, in four dimensions, are very strongly constrained by the requirement
of the absence of anomalies \cite{Bouchiat Iliopoulos Meyer, Gross Jackiw,
Ibanez, Font Ibanez Quevedo, Geng Marshak, Minahan Ramond Warner, Foot Joshi
Lew Volkas}, and we will now find that an arbitrary
set of Hodge - de Rham harmonic
two-forms, of a smooth compact quotient of $\mathbf{C} \mathbf{H}^3$ or
$\mathbf{H}^6$ that is a spin manifold, embedded in the $E_8$ Cartan
subalgebra as above, such that Witten's topological constraint is satisfied,
will result in a set of chiral zero modes that is simply a number of Standard
Model generations.

If the net numbers of left-handed chiral zero modes are $n_q$ $q$'s, $n_u$
$u$'s, $n_d$ $d$'s, $n_l$ $l$'s, $n_e$ $e$'s, and $n_5$ $q^5$'s, then the
conditions for the absence of gauge anomalies \cite{Bouchiat Iliopoulos Meyer,
Gross Jackiw}, and mixed gauge-gravitational anomalies \cite{Delbourgo Salam,
Eguchi Freund, Alvarez-Gaume Witten}, in four dimensions, are as follows.

From a triangle diagram with three external $ \mathrm{SU}\left(3\right) $ gauge
bosons:
\begin{equation}
\label{SU 3 SU 3 SU 3 anomaly cancellation condition}
2 n_q + n_u + n_d + 2 n_5 = 0
\end{equation}

From a triangle diagram with two external $ \mathrm{SU}\left(3\right) $ gauge
bosons, and one external $ \mathrm{U} \left( 1 \right)_Y$ gauge boson:
\begin{equation}
\label{SU 3 SU 3 U 1 Y anomaly cancellation condition}
n_q + 2 n_u - n_d - 5 n_5 = 0
\end{equation}

From a triangle diagram with two external $ \mathrm{SU}\left(2\right) $ gauge
bosons, and one external $ \mathrm{U} \left( 1 \right)_Y$ gauge boson:
\begin{equation}
\label{SU 2 SU 2 U 1 Y anomaly cancellation condition}
n_q - n_l - 5 n_5 = 0
\end{equation} 

From a triangle diagram with three external $ \mathrm{U} \left( 1 \right)_Y$
gauge bosons:
\begin{equation}
  \label{cubic anomaly cancellation condition} n_q + 32 n_u - 4 n_d - 9 n_l +
  36 n_e - 125 n_5 = 0
\end{equation}

And from a triangle diagram with two external gravitons, and one external
$U \left( 1 \right)_Y$ gauge boson \cite{Delbourgo Salam, Eguchi Freund,
Alvarez-Gaume Witten}:
\begin{equation}
  \label{linear anomaly cancellation condition} n_q + 2 n_u - n_d - n_l + n_e
  - 5 n_5 = 0
\end{equation}

The five equations (\ref{SU 3 SU 3 SU 3 anomaly cancellation condition}),
(\ref{SU 3 SU 3 U 1 Y anomaly cancellation condition}), (\ref{SU 2 SU 2 U 1 Y
anomaly cancellation condition}), (\ref{cubic anomaly cancellation condition}),
and (\ref{linear anomaly cancellation condition}), are linearly independent,
and the general solution, with integer values for the $ n_i $, is an integer
multiple of one Standard Model generation, which has $\left( n_q, n_u, n_d,
n_l, n_e, n_5 \right) = \left( 1, - 1, - 1, 1, 1, 0 \right)$.  Thus for an
arbitrary set of Hodge - de Rham harmonic two-forms, in the Cartan subalgebra
of $ E8 $, that break $ E8 $ to $ \mathrm{SU}\left(3\right) \times \mathrm{SU}
\left(2\right) \times \mathrm{U}\left(1\right)_Y $ as considered in this
subsection, and satisfy Witten's topological constraint, the chiral fermions
will consist of an integer number of Standard Model generations.

Let us now consider the case where $U \left( 1 \right)_Y$ is $\left( - 2, - 2,
- 2, 3, 3, 0, 0, 0, 0 \right)$.  To ensure that $U \left( 1 \right)_Y$ does
not get a mass by Witten's Higgs mechanism {\cite{Witten Constraints on
compactification}}, the Abelian vacuum gauge fields $\left( \sigma_1,
\sigma_1, \sigma_1, \sigma_2, \sigma_2, \sigma_3, \sigma_4, \sigma_5, \sigma_6
\right)$ with non-vanishing field strength must be perpendicular to $U \left(
1 \right)_Y$.  But this implies that $\sigma_1 = \sigma_2$, so that the
Abelian vacuum gauge fields with non-vanishing field strength must actually
leave $\mathrm{SU} \left( 5 \right)$ unbroken.  Nevertheless, we would still be
able to break $E_8$ to the Standard Model by topologically stabilized vacuum
gauge fields in the $E_8$ Cartan subalgebra, if we could topologically
stabilize a Hosotani vacuum gauge field with vanishing field strength
{\cite{Hosotani 1, Hosotani 2, Hosotani 3}} that is in the Cartan subalgebra
but not perpendicular to $U \left( 1 \right)_Y$.  This could be achieved if
the fundamental group of the compact six-manifold $\mathcal{M}^6$ included a
non-trivial element $a$ such that $a^n = 1$ for some finite integer $n$,
because an Abelian Wilson line looping once round the closed path
corresponding to $a$ will then be a phase factor $f$ satisfying $f^n = 1$.
But as noted in section \ref{Smooth compact quotients of CH3 H6 H3 and S3}, on
page \pageref{Smooth compact quotients of CH3 H6 H3 and S3}, the smooth
compact quotients $\mathcal{M}^6$ considered in this paper have no such
non-trivial elements $a$, called torsion elements.

Nevertheless, examples in three dimensions show that it is possible for the
first homology group $H_1 \left( \mathcal{M},\mathbf{Z} \right)$ of a
hyperbolic manifold $\mathcal{M}$ to have torsion even though the fundamental
group of $\mathcal{M}$ has no torsion.  For example, using Weeks's program
SnapPea {\cite{Weeks SnapPea}}, the Weeks manifold, which is the compact
hyperbolic three-manifold of smallest known volume, and designated m003(-3,1)
by SnapPea, is found to have first homology group $\mathbf{Z}/ 5
+\mathbf{Z}/ 5$.  This can be checked using the presentation of the
fundamental group given by SnapPea, which has generators $a$, $b$, and
relations $a^2 b^2 a^2 b^{- 1} ab^{- 1} = 1$ and $a^2 b^2 a^{- 1} ba^{- 1} b^2
= 1$.  We obtain the first homology group from the fundamental group by
treating the generators as commuting in the relations, which then collapse to
$a^5 = 1$ and $b^5 = 1$.  SnapPea also confirms that the Weeks manifold is
oriented.

Thus it seems reasonable to expect that there may exist smooth compact
quotients $\mathcal{M}^6$ of $\mathbf{C} \mathbf{H}^3$ or $\mathbf{H}^6$
such that $H_1 \left( \mathcal{M}^6,\mathbf{Z} \right)$ has torsion.  This
would be sufficient to obtain a topologically stabilized Hosotani Abelian
vacuum gauge field with vanishing field strength, even though the fundamental
group of $\mathcal{M}^6$ has no torsion.  For suppose there exists a one-cycle
$l$ that is not a boundary, such that $nl$, for some finite integer $n$, is a
boundary.  We consider an $E_8$ Wilson line $w$ that loops once around $l$.
Suppose there is a Hosotani $U \left( 1 \right)$ vacuum field that is locally
pure gauge, but for which $w$ is a non-trivial phase factor.  Then $w^n$ is a
phase factor along a one-cycle that is a boundary.  Thus since the Hosotani
field is locally pure gauge, we find $w^n = 1$ by Stokes's theorem.

As in the preceding section, it seems reasonable to expect that the Hodge - de
Rham harmonic two-forms will lead to masses $\sim$ a TeV for all chiral zero
modes that can be matched in left-handed and right-handed pairs, so that the
only light fermions will be the three generations of Standard Model fermions,
except possibly for one or more light singlet neutrino states, which could
obtain very small masses by the generalized seesaw mechanism to be discussed
in the following subsection.

And as in the preceding subsection, let us now consider an arbitrary proton
decay process, proceeding via a dimension 6 term in the Standard Model
effective action, such as $\frac{qqql}{\Lambda^2}$, $\frac{d^c d^c u^c
e^c}{\Lambda^2}$, $\frac{\overline{e^c} \overline{u^c} qq}{\Lambda^2}$, or
$\frac{\overline{d^c} \overline{u^c} ql}{\Lambda^2}$ {\cite{Wikipedia Proton
decay}}, with the $\mathrm{SU} \left( 3 \right)$ and $\mathrm{SU} \left( 2
\right)_{\mathrm{diag}}$ indices contracted in an appropriate manner, where
$\Lambda$ is an effective cutoff, that determines the size of the term.  Then
for any four specific states from Tables \ref{states in the 80 in the third
case} and \ref{the states in the 84 bar in the third case}, that have
nonvanishing amplitudes in those four types of Standard Model state, the
condition for the existence of a $U \left( 1 \right)_B$, that couples as a
nonzero multiple of baryon number, just on those four states, is three
homogeneous linear equations on the five linearly independent $\sigma_i$, so
is always satisfied.  Thus those parts of the arguments of Aranda and Carone
{\cite{Aranda Carone}}, that depend only on the existence of such a $U \left(
1 \right)_B$, would seem to suggest that the contribution of those four
states, to the corresponding term in the Standard Model effective action, will
be suppressed.  And since this argument applies to all sets of states from
Tables \ref{states in the 80 in the third case} and \ref{the states in the 84
bar in the third case}, that have nonvanishing amplitudes in the Standard
Model fermion fields in the effective action term concerned, we expect the
same suppression to apply to the overall coefficient of that term in the
effective action, leading to a large value of the effective cutoff $\Lambda$,
even though the relevant $U \left( 1 \right)_B$ may be different, for
different relevant sets of states from Tables \ref{states in the 80 in the
third case} and \ref{the states in the 84 bar in the third case}.

To find out whether the mass hierarchy of the observed quarks and charged
leptons could occur by a version of the Arkani-Hamed - Schmaltz mechanism
{\cite{Arkani-Hamed Schmaltz}}, in the type of model considered here, it would
be necessary to find the explicit form of the Hodge - de Rham harmonic
two-forms, for examples of smooth compact quotients of $\mathbf{C}
\mathbf{H}^3$ or $\mathbf{H}^6$ that are spin manifolds, and the
corresponding chiral fermion zero modes, to find out how spread out or
localized they are.  However, we note that in the examples considered by
Arkani-Hamed and Schmaltz {\cite{Arkani-Hamed Schmaltz}}, and by Acharya and
Witten {\cite{Acharya Witten}}, the chiral fermion modes have a Gaussian
shape, even though the fermion ``mass terms'' only depend linearly on
position.  The explicit forms of the chiral fermion zero modes in monopole
backgrounds on the two-sphere have been given by Deguchi and Kitsukawa
{\cite{0512063 Deguchi Kitsukawa}}.

Of course, it was not necessary to require that four of the $\mathrm{SU} \left(
3 \right) \times \mathrm{SU} \left( 2 \right)$ singlets, with linearly
independent $U \left( 1 \right)$ charges, have $Y = 0$, since the unwanted $U
\left( 1 \right)$'s will become massive by Witten's Higgs mechanism, provided
that none of them is orthogonal to all the Hodge - de Rham monopoles in the $E
8$ Cartan subalgebra.  So additional solutions might exist, such that the
largest number of $\mathrm{SU} \left( 3 \right) \times \mathrm{SU} \left( 2
\right)$ singlets, with linearly independent $U \left( 1 \right)$ charges,
that have $Y = 0$, is three or less.

\subsection{Generalized seesaw mechanism}
\label{Generalized seesaw mechanism}

With regard to how small neutrino masses, $\sim 1$ eV or smaller, might arise
in models of the type considered in this paper, it seems to be possible to
produce a very small, but nonzero, eigenvalue, from a matrix whose matrix
elements are
integers in the range $0$ to $10$, if all the matrix elements in the lower
right triangle, below the lower left to upper right diagonal, are zero, the
matrix elements on the lower left to upper right diagonal are $1$, and the
matrix elements in the upper left triangle, above the lower left to upper
right diagonal, are $\sim 10$.  For example the matrix:
\begin{equation}
  \label{example for generalized seesaw mechanism} \left(\begin{array}{cccc}
    10 & 10 & 10 & 1\\
    10 & 10 & 1 & 0\\
    10 & 1 & 0 & 0\\
    1 & 0 & 0 & 0
  \end{array}\right)
\end{equation}
has eigenvalues $22.891$, $- 7.6024$, $4.7127$, and $- 0.001219$.  If this
effect occurs because all but one of the eigenvalues tend to be comparable to
the large matrix elements in the upper left triangle, but the determinant, and
hence the product of the eigenvalues, is equal to $1$, it would presumably be
possible to obtain an eigenvalue as small as required, by considering larger
matrices with this structure.  We note that in the models considered in the
preceding subsection, it might be natural to find a number $\sim 10$ or more
of singlet neutrinos, which could perhaps sometimes have a mass matrix of this
type.  To obtain the required small eigenvalue, the matrix elements in the
lower right triangle would presumably have to be exactly zero.  This would
presumably be possible, if the matrix elements were integer multiples of an
overall factor, but I do not know of a reason why this should be so.

\begin{center}
{\bf Acknowledgements}
\end{center}

\noindent I would like to thank Savas Dimopoulos, David E. Kaplan, and Karin
Slinger for
organizing a very enjoyable and helpful visit to Stanford University ITP, where
part of the work that led to this paper was carried out, Nima Arkani-Hamed,
Savas Dimopoulos, Michal Fabinger, Simeon Hellerman, Veronika Hubeny, Shamit
Kachru, Nemanja Kaloper, Renata Kallosh, David E. Kaplan, Matt Kleban, Albion
Lawrence, Andrei Linde, John McGreevy, Michael Peskin, Steve Shenker, Eva
Silverstein, Matt
Strassler, and Lenny Susskind for helpful discussions or comments, and Fyodor
Tkachov and Kasper Peeters for helpful emails.

At an early stage of this work, some of the calculations were carried out using
\mbox{TeXaide} \cite{TeXaide} and TeXnic Center \cite{TeXnic Center}, rather
than by
using pen and paper.  Subsequently, after migrating to Debian GNU/Linux
\cite{Debian}, the work was done almost entirely by means of GNU TeXmacs
\cite{GNU TeXmacs}, without using pen and paper at all.  The paper was largely
written using GNU TeXmacs, and ported to KTeXmaker2, now renamed to Kile
\cite{KTeXmaker2}, for
completion.  Some of the calculations were done with PARI/GP \cite{PARI GP} and
Maxima \cite{Maxima}, initially freestanding, and subsequently run from within
GNU TeXmacs.  The \LaTeX \hspace{-0.1ex} pictures were prepared with TexPict
\cite{TexPict}.
The online Wolfram Integrator \cite{Wolfram Integrator} was used to perform
some integrals.
The bibliography of version 2 was sequenced with help from Ordercite
\cite{Ordercite}.

\end{document}